%% file: main.tex
\documentclass[12pt]{report}
\usepackage[utf8]{inputenc}
\usepackage{charter}
\usepackage{fullpage}
\usepackage{hyperref}
\usepackage{graphicx}
\usepackage{lipsum}
\usepackage[sort&compress,numbers]{natbib}
\usepackage{hyperref}
\hypersetup{hidelinks}
\usepackage[total={6.5in,9in},top=1in,headsep=0.1in,headheight=1in]{geometry}

\usepackage{amsmath}
\usepackage{amssymb}
\usepackage{xcolor}
\usepackage{verbatim}
\usepackage{multirow}
\usepackage{tiffany}

\newcommand{\psim}{\lower.5ex\hbox{$\; \buildrel \propto \over\sim \;$}}
\newcommand{\lbar}{\lower.0ex\hbox{$\; \buildrel
{\lower0.0ex \hbox{-}} \over\lambda  \;$}}

\newcommand{\fermi}{{\em Fermi}}

\definecolor{hookgreen}{rgb}{0.0,0.44,0.0}
\definecolor{periwinkle}{RGB}{104,127,243}
\definecolor{eggplant}{RGB}{83,27,147}
\definecolor{olive}{RGB}{74,80,7}

\newcommand\snowmass{
\begin{center}
  \rule[-0.2in]{\hsize}{0.01in}\\
  \rule{\hsize}{0.01in}\\
  Submitted to the Proceedings of the US Community Study\\ 
  on the Future of Particle Physics (Snowmass 2021)\\
  \vskip 0.1in
  \rule{\hsize}{0.01in}\\
  \rule[+0.2in]{\hsize}{0.01in}\\[-2em]
\end{center}
}

\usepackage[firstpage=true]{background}
\backgroundsetup{contents={\parbox{6.5in}{\snowmass\editors\sponsors}}, scale=1,placement=top,opacity=1,color=black,position={3.25in,1.2in}}

\usepackage{fancyhdr}
\fancypagestyle{plain}{%
  \fancyhf{}%
  \fancyhead[C]{}
  \fancyfoot[C]{\thepage}
}

\fancypagestyle{empty}{%
  \fancyhf{}%
  \fancyhead[C]{{\it Snowmass2021 CF07 Gamma-Ray Experiments}}
  \fancyfoot[C]{\thepage}
}
\title{The Future of Gamma-Ray Experiments in the MeV--EeV Range}
\date{}
\usepackage{authblk}

\renewcommand\Affilfont{\scriptsize}

\author[1,2]{K.~Engel \orcidlink{0000-0001-5737-1820}}
\author[1]{J.~Goodman \orcidlink{0000-0002-9790-1299}}
\author[3]{P.~Huentemeyer \orcidlink{0000-0002-3302-7897}}
\author[4]{C.~Kierans \orcidlink{0000-0001-6677-914X}}
\author[4,5,$\ddagger$]{T.R.~Lewis\orcidlink{0000-0002-9854-1432}}
\author[4,6,7]{M.~Negro \orcidlink{0000-0002-6548-5622}}
\author[8]{M.~Santander}
\author[9]{D.A.~Williams \orcidlink{0000-0003-2740-9714}}
 
\affil[1]{University of Maryland, College Park, College Park, MD, 20742, USA}
\affil[2]{Physics Division, Los Alamos National Laboratory, Los Alamos, NM, 87545, USA}
\affil[3]{Michigan Technological University, Houghton, MI 49931, USA}
\affil[4]{Astroparticle Physics Laboratory, NASA Goddard Space Flight Center, Greenbelt, MD 20771, USA}
\affil[5]{NASA Postdoctoral Program Fellow}
\affil[6]{University of Maryland, Baltimore County, Baltimore, MD 21250, USA}
\affil[7]{Center for Research and Exploration in Space Science and Technology, NASA/GSFC, Greenbelt, MD 20771, USA}
\affil[8]{University of Alabama, Tuscaloosa, AL 35487, USA}
\affil[9]{Santa Cruz Institute for Particle Physics and Department of Physics, University of California, Santa Cruz, Santa Cruz, CA 95064, USA}
\email{ }
\email{$\ddagger$ tiffany.lewis-1@nasa.gov}

\begin{document}

\maketitle

\begin{abstract}
Naturally occurring particle accelerators shine brightly throughout the universe, inviting us to discover fundamental laws and hone our theories if we look in their directions with the right detectors. Gamma-rays, the most energetic photons, carry information from the far reaches of extragalactic space with minimal interaction or loss of information. They bring messages about particle acceleration in environments so extreme they cannot be reproduced on earth for a closer look. Gamma-ray astrophysics is so complementary with collider work that particle physicists and astroparticle physicists are often one in the same. 

Gamma-ray instruments, especially the {\it Fermi Gamma-ray Space Telescope}, have been pivotal in major multi-messenger discoveries over the past decade. There is presently a great deal of interest and scientific expertise available to push forward new technologies, to plan and build space- and ground-based gamma-ray facilities, and to build multi-messenger networks with gamma rays at their core. It is therefore concerning that before the community comes together for planning exercises again, much of that infrastructure could be lost to a lack of long-term planning for support of gamma-ray astrophysics.
Gamma-rays with energies from the MeV to the EeV band are therefore central to multiwavelength and multi-messenger studies to everything from astroparticle physics with compact objects, to dark matter studies with diffuse large scale structure.

These science goals and the excitement of new discoveries have generated a wave of new gamma-ray facility proposals and programs. Since the legacy of existing facilities is well covered in many other places, this paper highlights new and proposed gamma-ray technologies and facilities that have each been designed to address specific needs in the measurement of extreme astrophysical sources that probe some of the most pressing questions in fundamental physics for the next decade.
The proposed instrumentation would also address the priorities laid out in the recent Decadal Survey of Astronomy and Astrophysics (Astro2020), a complementary study by the astrophysics community that provides opportunities also relevant to Snowmass.

\end{abstract}

\clearpage

\setcounter{secnumdepth}{4}
\setcounter{tocdepth}{4}
\tableofcontents

\chapter{Introduction: High-Energy Astrophysics in Support of Fundamental Physics}

Nature provides particle accelerators and conditions of extreme fields that we can only dream of reproducing in the laboratory.  Cosmic rays have been detected with energy exceeding $10^{20}$ eV.  The magnetic fields of magnetars are estimated to exceed $10^{12}$ G, and black holes with masses from a few to billions of times the mass of our Sun produce strong gravitational fields. 
Evidence for dark matter was first found in astrophysical systems, and whatever dark matter turns out to be, we know these systems must contain it and in roughly what quantity. 
Observations of astrophysical gamma rays give a unique view of the high-energy processes in these extreme environments.
Furthermore, the enormous distances to many of these objects allow subtle effects on gamma-ray propagation to be studied, which are sensitive to the underlying physics and cosmology.  High-energy astrophysics, and gamma-ray observations in particular, presents opportunities to study fundamental physics and cosmology that cannot be found anywhere else.

Gamma rays are also a crucial point of contact between photon observations and multi-messenger astrophysics:  observations with gravitational waves, neutrinos, and cosmic rays (charged hadrons and leptons).  The first confirmed electromagnetic counterpart of a gravitational wave event was the gamma-ray burst GRB\,170817 \cite{LIGOScientific:2017zic}.  The strongest association of neutrino emission with an astrophysical source comes from the detection of a neutrino during a gamma-ray flare of the active galaxy
TXS\,0506+056 \cite{IceCube:2018dnn}.  Production of non-photon messengers generally involves high-energy processes that also can produce gamma rays.

Astrophysical environments, while offering unparalleled opportunity, are not without their challenges.  They do not offer the same control of the experimental conditions that is possible in terrestrial laboratories.  Information about fundamental processes and cosmology can be entangled with the astrophysical understanding of the systems being used.  Therefore, astrophysical studies which might be viewed as outside the scope of an agency's portfolio---regardless of how interesting {\it per se}---can be essential in the long run to achieving the agency's goals and should be evaluated from that perspective.  
Conversely, many observations and instruments can be well motivated on the basis of the astrophysics they can do.  As a result, scientific investments in the fundamental physics and cosmology capabilities of these same projects are highly leveraged and the goals of all participants are cost-effectively advanced when agencies work together towards their varied goals.  From this perspective, the results from the complementary study on the opportunities in astrophysics, the Decadal Survey of Astronomy and Astrophysics (Astro2020)~\cite{NAP26141}, are germane to understanding opportunities relevant for Snowmass at the rich interface between astrophysics and fundamental physics and cosmology.

The Astro2020 Decadal Survey was charged to \textit{``generate consensus recommendations to implement a comprehensive strategy and vision for a decade of transformative science at the frontiers of astronomy and astrophysics.''} Astro2020 was sponsored by the National Aeronautics and Space Administration (NASA), the National Science Foundation (NSF), the Department of Energy (DOE) Office of High Energy Physics, and the Air Force Office of Space Research (AFOSR). One of the prime topics for the report was multi-messenger astrophysics. In the main report they say: \textit{``The confluence of decades of work in theory, numerical relativity, nuclear astrophysics, gravitational wave detectors and analysis methods, combined with measurements across the electromagnetic spectrum from space and ground, building on the international network for rapid follow up originally developed for the study of gamma ray bursts, produced what has become the archetype for multi-messenger astronomy, a field destined to blossom in the coming decade.''}

\begin{figure}[htb]
\centering
	\includegraphics[width=0.8\textwidth]{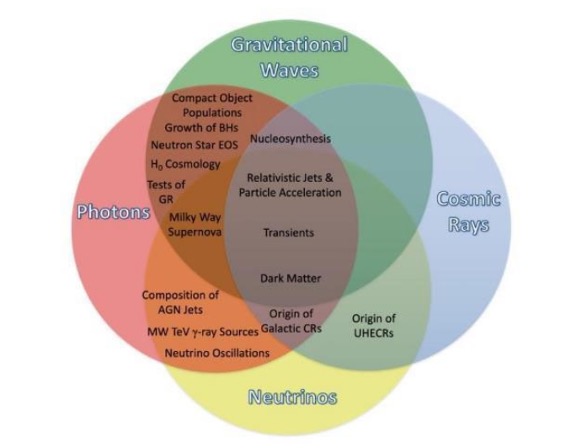}
	\caption{The Astro2020 Decadal Survey report names New Windows on the Dynamic Universe, {\it i.e.}\ multi-messenger astrophysics, as a Priority Science Area. They call for \textit{``higher sensitivity all-sky monitoring of the high-energy sky.''} Figure from Ref.~\citenum{NAP26141}.}	
	\label{fig:decadal-venn}
\end{figure}

The Astro2020 Panel on Compact Objects and Energetic Phenomena points out the need for post-\textit{Swift}~\cite{2004ApJ...611.1005G}, post-\textit{Fermi}~\cite{lat,Meegan2009}, and post-HAWC~\cite{Abeysekara_2019}/VERITAS~\cite{Holder:2006gi} coverage of the gamma-ray sky. 
\textit{``For rapid follow-up and correlation studies, continued support
of Swift and Fermi spacecraft operations will be crucial until newer missions replace them...
wide-field MeV-GeV gamma-ray facilities will be essential for identifying counterparts to high-energy
neutrino and UHECR sources, along with wide-field and follow-up capabilities in the GeV-TeV band to
extend these observations to higher energies for nearby sources.''}
The Panel on Cosmology report points out that wide-field gamma-ray instruments can study dark matter annihilation in dwarf galaxies and in the bulge around the Galactic center. 

The Panel on Particle Astrophysics and Gravitation (PAG) of Astro2020 describes how \textit{``Astronomy is now also being revolutionized by observations with new messengers---gravitational waves, neutrinos, gamma rays, and cosmic rays---that greatly complement and are leveraged by observations in conventional astronomy. The hallmark of this work, for which the projects have largely been developed, funded, and carried out as part of physics programs, is its ability to probe extremes of energy, fields, and density.''} They also point out: \textit{``For gamma rays, there have been dramatic advances in the energy range, angular resolution, and number of sources.''} 

The panel reports that there is a \textit{``Compelling opportunity to dramatically open the discovery space of astronomy through a bold, broad multi-messenger program, with three components.''} They describe a program including neutrinos, gravitational waves, and gamma rays (see Figure~\ref{fig:decadal-venn}). For NASA, they recommend a broad gamma-ray program in space while on the ground they prioritize \textit{``U.S. participation in TeV-range ground-based experiments for precision studies--- e.g., the Cherenkov Telescope Array (CTA) and the Southern Wide-Field Gamma-Ray Observatory (SWGO)''} as medium-scale projects and conclude that \textit{``In the TeV--PeV range, there are exciting developments worldwide, but without U.S. support for and involvement in these activities, the United States will lose its leadership role.''} CTA~\cite{CTAConsortium:2018tzg} and SWGO~\cite{Albert:2019afb,Abreu:2019ahw,Hinton:2021rvp,Schoorlemmer:2019gee} are described in more detail in Sections~\ref{sec:cta} and \ref{sec:swgo}, respectively.

Some of the principal science drivers connecting fundamental physics and gamma-ray astrophysics are presented in Chapter~\ref{sec-fundamental}. An overview of experimental techniques in 
gamma-ray astrophysics is provided in Chapter~\ref{sec-techniques}. A survey of future gamma-ray facilities and experiments from funded projects to proposed missions is given in Chapter~\ref{sec-facilities}. An array of new and developing technologies  in software and instrumentation 
is described in Chapter \ref{sec-technology}. Some broader ideas on community infrastructure and models for dissemination of software and data beyond individual collaborations are discussed in Chapter \ref{sec-synergies}. 

\chapter{Fundamental Physics Questions}
\label{sec-fundamental}


Science goals drive development of instrumentation, technology, and collaborative methods. Of the most fundamental questions in the physical universe, this chapter highlights several that cannot be answered without astrophysical gamma-ray observations and investment in the instruments and infrastructure that facilitate them. 

Some of the greatest fundamental questions of this scientific era relate to the dark sector. Dark matter and dark energy form the vast majority of the universe, and while their nature has eluded us for decades, steady, incremental progress over the last 10 years has significantly narrowed the parameter space available to models. Much of this work has been buoyed by multimessenger discoveries driven by surveys of the gamma-ray sky. Therefore continued investment in gamma-ray astrophysics experiments, with coverage in all energy bands is pivotal to continued discoveries.


High-energy physics and high-energy astrophysics are interconnected fields, sharing instrumentation, personnel, and fundamental questions about how the universe works. The observations of laboratories provided by nature are complementary to the work done in terrestrial facilities. Our understanding of fundamental particles, their interactions and origins depends on the flow of knowledge in both directions. Section \ref{sec:futurecolliders} reviews some key areas of complimentarity between colliders and gamma-ray observatories, especially with regard to electroweak emission and dark matter searches.

The co-detection of gravitational waves with a gamma-ray burst ushered the dawn of the multimessenger era of astronomy. It also raised the question of the speed of gravity because there was a ~2s delay between the arrival of the gamma-rays and the arrival of the gravitational waves.  Future observations from new facilities in the gravity sector will push to test General Relativity and the Cosmological Principle. These tests of well established theories that are incompatible with the Standard Model will be the pivotal questions in the next generation, but will rely on as yet unplanned gamma-ray facilities in the MeV and GeV bands. The importance of gamma-rays to multimessenger topics are further discussed in Section \ref{sec:speedofgravity} and Section \ref{sec:GWlocalization}.

The multimessenger era was further established through the co-detection of neutrinos and gamma-rays during a blazar flare. So, the emergent field of  multimessenger astrophysics is presently held together through space-based gamma-ray all-sky surveying of the {\it Fermi Gamma-ray Space Telescope}. While {\it Fermi} is currently in good condition and expected to make many more distinguished observations, given the length of time associated with planning a probe-class mission it would be a disservice to multimessenger investments in the gravity and particle sectors to allow this decade to pass without addressing the ongoing need for MeV and GeV surveys. In particular, Section \ref{sec:neutrinos} addresses the complimentarity between gamma-ray and neutrino observations to address the decades long question of the origin of cosmic-rays as well as the broader question of large scale structure in the universe. 

In addition to multimessenger studies on particles, the advent of high-energy polarimetry is poised to shed light on fundamental fields. Gamma-ray polarimetry can probe radiation mechanisms and hadronic signatures in cosmic jets. The composition of cosmic jets, especially for blazars is unknown, but could provide a significant source of cosmic-rays if hadronic. These ideas are further explored in Section \ref{sec:polarimetry}.  Gamma-ray polarization measurements are also important to the ongoing study of magnetars. Magnetars exist in the strong field regime, and provide a test of standard quantum electrodynamics while being a potential source for axion-like particles, that is not generally accessible via terrestrial methods. This is further described in Section \ref{sec:magnetars}. 

The problems of missing matter is written all over observations of galaxies. If non-luminous matter exists with gravitational evidence, it should also be explained in particle physics. There are several proposed extensions to the Standard Model that explain the observed behaviors of dark matter, in addition to many predicted behaviors as yet unobserved. Very high-energy and ultra high-energy gamma-ray observatories probe the relevant energies for dark matter annihilation and decay processes that existing models predict. Existing gamma-ray facilities continue to constrain the models, but as these facilities sunset, it is important to be thinking about the next generation of gamma-ray surveys. These topics are further expounded upon in Section~\ref{sec:darkmatter}. Primordial black holes are some of the best studied dark matter candidates and the details of how they are constrained by observations in the MeV and GeV bands are enumerated in Section~\ref{sec:primordealBH}. 

Cosmology and large scale structure probes of dark matter are woven into the fabric of gamma-ray observations. The extragalactic background light, diffuse infrared through ultraviolet radiation that permeates the universe, creates electron-positron pairs in the presence of gamma-rays, and is thus an increasingly significant source of absorption for distant gamma-ray sources, like blazars observed in the GeV and TeV regimes. This is further discussed in Section~\ref{sec:EBLcosmology}.  The evidence of dark matter filaments between galaxies and clusters is traced by catalogues of extragalactic gamma-ray observations. Models currently predict that the unresolved gamma-ray background is produced either by active galactic nuclei or annihilation and decay in dark matter filaments. A probe-scale mission to complement {\it Fermi}-LAT and complete the multiwavelength fleet over the coming decades will be important to identifying new sources and removing them from the background. This is discussed further in Section \ref{sec:largescalestructure}.

 \section{The Unbroken Standard Model at Future Colliders and Gamma-Ray Observatories}
 \label{sec:futurecolliders}

\input{UnbrokenSM-Rodd}

\section{The Speed of Gravity and Lorentz Invariance}
\label{sec:speedofgravity}

\input{SpeedGravity-Burns}

\section{Gravitational Wave Sources Localization}
\label{sec:GWlocalization}

\input{GWLocalization-Judy}

\section{Neutrino Gamma-Ray connection}
\label{sec:neutrinos}
 
 \input{NeutrinoGammaConnection-Karwin}

\section{Gamma-Ray Polarimetry}
\label{sec:polarimetry}

\input{Polarimetry-Haocheng}

\clearpage
\section{Fundamental Physics in Gamma Rays with Magnetars}
\label{sec:magnetars}

\input{Magnetars-Zorawar}

\section{Dark Matter Annihilation \& Decay Signatures}
\label{sec:darkmatter}

\input{DarkMatter-Tansu}

\section{Primordial Black Holes}
\label{sec:primordealBH}

\input{PrimordialBH-Ranjan}

\section{The Extragalactic Background Light and Gamma-Ray Cosmology}
\label{sec:EBLcosmology}

\input{ExtragalacticBackground-Meyer}

\section{Large Scale Structures and the Gamma-Ray Background}
\label{sec:largescalestructure}

\input{LargeScale-Michela}

\chapter{Techniques in Gamma-Ray Facilities}
\label{sec-techniques}

We present herein a brief overview of the major detection techniques used in gamma-ray astrophysics by current facilities and those future facilities detailed below in Chapter~\ref{sec-facilities}. Facilities may use one or multiple of these techniques to perform the gamma-ray observations that enable them to illuminate the mysteries of the Fundamental Physics topics detailed in Chapter~\ref{sec-fundamental}. Applications of these observation strategies have been applied to space-based and ground-based instruments, covering wide-field, pointed, and hybrid facilities. 

These techniques are not limited to gamma-ray observations and may be applied for multimessenger studies (e.g., for the detection of neutrinos and cosmic rays) as well. The core strength of gamma-ray astrophysics comes from the complementarity of these varied technologies and the seamless way in which the strengths of one instrument supports an assortment of others to glean as much information as possible about the many cosmic gamma-ray sources.

\section{Non-Imaging Transient Detectors}
\label{sec:non-imaging}

\input{NonImagingDetectors-Isreal}

\section{Compton Telescopes}
\label{sec:compton_tel}
\input{ComptonTelescopes-Reshmi}

\clearpage
\section{Coded Mask Telescopes}
\label{sec:codedmask_tel}
\input{CodedMask-Brad}


\section{Pair Creation Telescopes}
\label{sec:pair}
\input{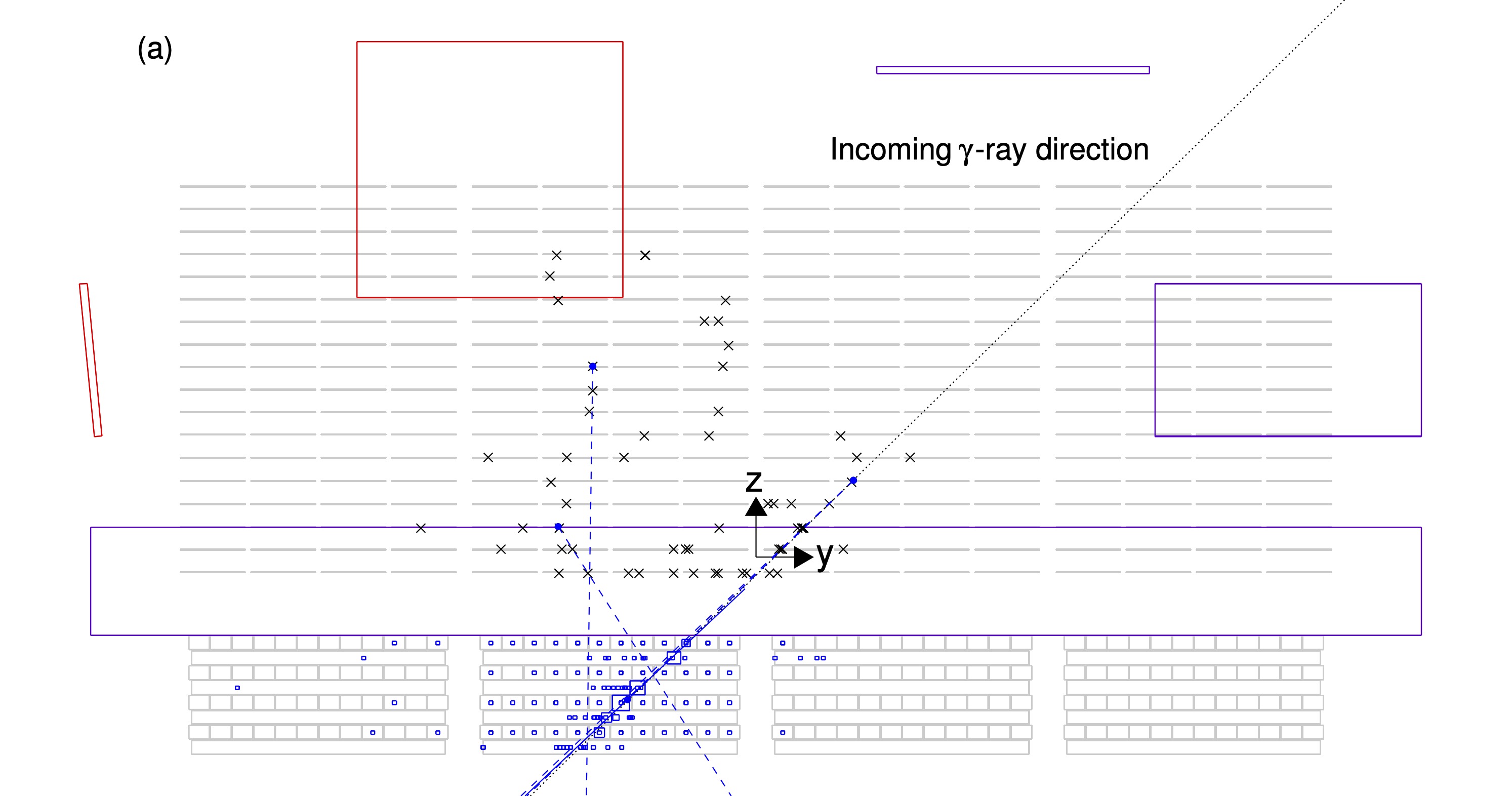}

\section{Ground-Based Air Shower Particle Detectors}
\label{sec:ground-based} 
\input{WaterCherenkov-Henrike}

\section{Imaging Atmospheric Cherenkov Telescopes}
\label{sec:atmospheric_cherenkov}
\input{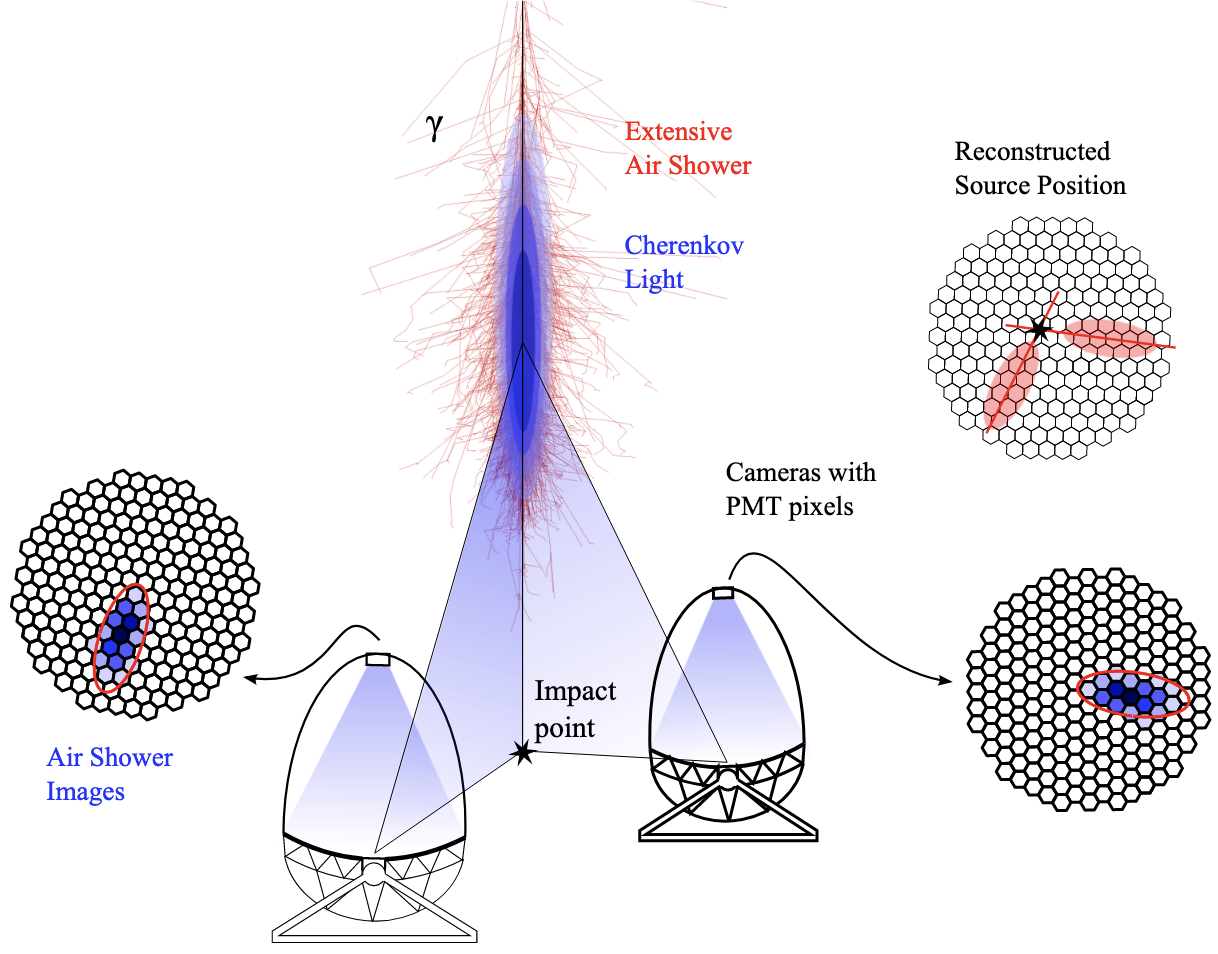}


\chapter{Future Facilities}
\label{sec-facilities}

\begin{figure}[h!]
\begin{minipage}[c]{0.62\textwidth}
\centering
\includegraphics[width=0.99\textwidth]{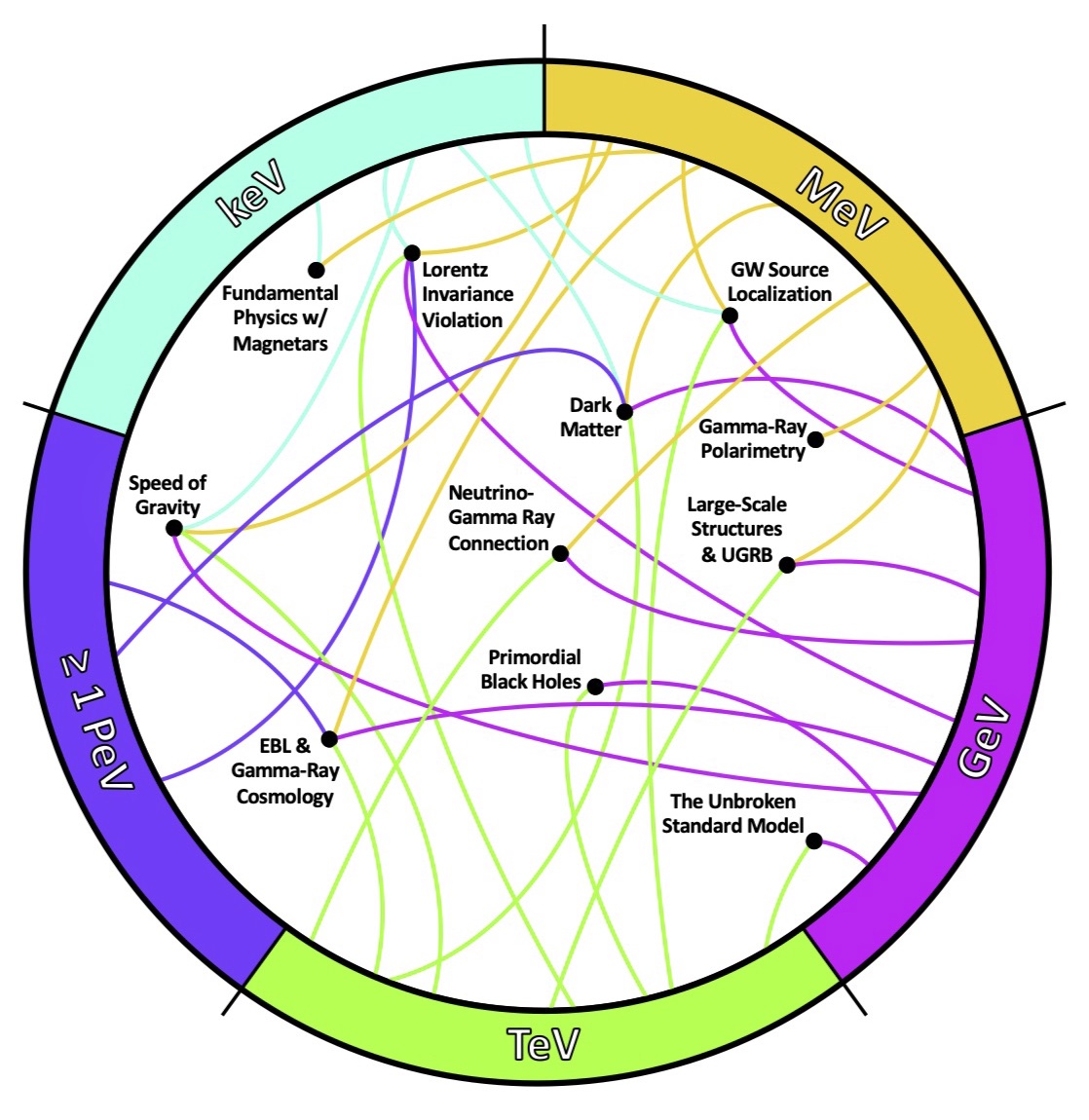}
\end{minipage}\hfill
\begin{minipage}[c]{0.37\textwidth}
\caption{The fundamental physics topics covered in Chapter~\ref{sec-fundamental} are each addressed by gamma rays across multiple energy regimes. Messengers at different energies provide different pieces to each puzzle, with no image being able to be completed with only a single input. Figure by K. Engel.}
\label{fig:chord-neon}
\end{minipage}
\end{figure}

Many key opportunities for investment in the future of Particle Physics in the U.S.  lie in the development, construction, and operation of the next-generation gamma-ray astrophysics facilities that apply the observational techniques covered in Chapter~\ref{sec-techniques}. These experiments, consortiums, and collaborations, detecting gamma-ray energies from MeV to EeV, allow the observation of cosmic accelerators reaching speeds greater than those achievable on Earth and unique probes of the Standard Model and beyond-the-Standard-Model Physics, as well as being promising ground for the discovery of new physics and new particles by providing access to an abundance of mysterious, exotic, and violent environments (see Figure~\ref{fig:chord-neon}).


As previously stated, the power of the Cosmic Frontier's and gamma-ray astrophysics' contributions to elucidating the many mysteries of fundamental physics (Chapter~\ref{sec-fundamental}) comes from the complementarity of these facilities. The specialization of each observatory (in energy range or wide-field, pointed, or instrumentation) as well as their locations on the globe allow them to provide unique contributions, be it through the discovery of new sources and source classes or the notifications or follow-ups to generate more data on a specific observation. Without this interplay of facilities, it would be difficult---if not impossible---to determine many source properties with high accuracy such as localization, transient start time, spectra, etc. 

As pointed out in the Decadal Survey of Astronomy and Astrophysics (Astro 2020)~\cite{NAP26141}, there have been exciting developments worldwide in the field of gamma-ray astrophysics through the development of next-generation facilities, \textit{``but without U.S. support for and involvement in these activities, the United States will lose its leadership role.''} Support of the facilities herein would be a decisive action towards unraveling the mysteries of a wide variety of fundamental physics topics at the forefront of what we hope to learn in the coming decade.

Submissions to this Chapter were open to any in the community seeking U.S. support for their gamma-ray facility, with solicitations going out to those facilities who originally submitted Letters of Interest to the Snowmass 2021 Process. These received submissions are categorized below by energy into those observing in the MeV range (Section~\ref{sec:mev_telescopes}) and those observing in the $\geq$GeV range (Section~\ref{sec:vhe_telescopes}).

\section{Medium Energy Gamma-Ray Telescopes}
\label{sec:mev_telescopes}

\subsection{BurstCube}
\label{sec:burstcube}
\input{BurstCube-Perkins}

\subsection{Glowbug}
\label{sec:glowbug}
\input{Glowbug-Eric}

\subsection{StarBurst}
\label{sec:starburst}
\input{StarBurst-Kocevski}

\subsection{MoonBEAM--- The Moon Bursts Energetics All-sky Monitor}
\label{sec:moonbeam}
\input{MoonBEAM-Hui}

\subsection{LEAP--- The LargE Area burst Polarimeter}
\label{sec:leap}
\input{LEAP_McConnell}

\subsection{COSI--- The Compton Spectrometer and Imager}
\label{sec:cosi}
\input{COSI-Tomsick}

\clearpage
\subsection{AMEGO and AMEGO-X--- The All-sky Medium Energy Gamma-ray Observatory}
\label{sec:amego}
\input{AMEGO-Kierans}

\subsection{GECCO--- The Galactic Explorer with a Coded Aperture Mask Compton Telescope}
\label{sec:gecco}
\input{GECCO-Moiseev}

\clearpage
\subsection{GRAMS---  The Gamma-Ray and AntiMatter Survey}
\label{sec:grams}
\input{GRAMS-Tsuguo}

\clearpage
\subsection{GammaTPC}
\label{sec:gammatpc}
\input{GammaTPC-Shutt}


\section{Very-High-Energy Gamma-Ray Telescopes}
\label{sec:vhe_telescopes}

\subsection{CTA--- The Cherenkov Telescope Array}
\label{sec:cta}
\input{CTA-Williams}

\subsection{SWGO--- The Southern Wide-field Gamma-ray Observatory}
\label{sec:swgo}
\input{SWGO-Engel}

\subsection{GRAND--- The Giant Radio Array for Neutrino Detection}
\label{sec:grand}
\input{GRAND-Batista}

\subsection{POEMMA--- The Probe Of Extreme Multi-Messenger Astrophysics}
\label{sec:poemma}
\input{POEMMA-JohnK}
\chapter{Technology Driven by Science}
\label{sec-technology}

The development of new and improved hardware and software are the cornerstones of the future of gamma-ray astrophysics. Many of the next-generation missions described in Section~\ref{sec-facilities} are enabled by technology advancements that have occurred within the last decade and, often times, the scientific potential is not only validated, but enhanced through the development of accompanying software tools.  

In Section~\ref{sec:detectors} we highlight a subset of the on-going detector development efforts in the gamma-ray community: in Section~\ref{sec:trackers} we list a number of competing technologies for Tracking detectors; in Section~\ref{sec:calorimeters} we give examples of on-going development for novel scintillators and the accompanying readout; and in Section~\ref{sec:VFG-CZT} we give an example of a novel detector technology that can work as a stand-alone Compton telescope or as the back-plane to a coded-mask instrument.

In Section~\ref{sec:pipelines} we outline the current state-of-the-art data pipeline and analysis tools that are being developed by the community, and that will be used by the future missions listed in Section~\ref{sec-facilities}. Section~\ref{sec:modeling} gives an overview of the detector modeling and simulation tools that are key to predicting instrument performance. Section~\ref{sec:reconstruction} describes the computationally intensive process of event reconstruction for Compton and pair gamma-ray telescopes. Finally, Section~\ref{sec:analysistools} gives an overview of higher-level analysis tools in the modern era of big-data and collaborative development. Relatedly, Section~\ref{sec-synergies} covers the data management for the broader community and between missions.

\section{Detector Technology Development}
\label{sec:detectors}

Many of the missions described in Section~\ref{sec-facilities} are enabled by recent advancements in technology. With improved spectral and spatial resolution, better background rejection, and/or improved efficiency, these novel detector approaches enable the next-generation of gamma-ray telescopes to achieve the unprecedented sensitivity required to answer many of the outstanding questions in gamma-ray astrophysics posed in Section~\ref{sec-fundamental}. 

\subsection{Tracking Detectors}
\label{sec:trackers}

Much like the technology for terrestrial particle physics experiments, gamma-ray telescopes rely on the tracking of charged particles and high-energy photon interactions. 
From the first gamma-ray telescope which used a spark chamber, SAS-2~\cite{1975ApJ...198..163F} launched in 1972, to the modern day \textit{Fermi}-Large Area Telescope (LAT)~\cite{fermi-lat}, which uses single-sided silicon semiconductor detectors, advancements in gamma-ray instrumentation follow the progress in particle tracker technology. 
\textit{Fermi}-LAT, which is sensitive from $\sim$50~MeV to over 300 GeV, has made some of the most significant gamma-ray contributions to astroparticle physics to date~\cite{2020ApJS..247...33A, 2016PhR...636....1C}.
Bringing LAT-like sensitivity down to the MeV regime (0.1-100 MeV) with new detector development is a goal being pursued by many groups to advance the science motivated in Section~\ref{sec-fundamental}.

\subsubsection{Double-Sided Silicon Strip Detectors for Next-Generation Gamma-Ray Telescopes}
\label{sec:dssd}

\input{DSSD-Kierans}

\subsubsection{Gamma-Ray Scintillator Fiber Tracker}
\label{sec:fibertracker}
\input{FiberTracker-Nicola}

\subsubsection{A Next-Generation LAr TPC-Based MeV Gamma-Ray Instrument}
\label{sec:LAr_tech_Shutt}
\input{LArTPC-Shutt}

\subsubsection{Silicon Pixel Detectors in Space}
\label{sec:astropix}
\input{SiliconPixel-Regina}

\subsubsection{Artificial Diamond Detectors}
\label{sec:diamonds}
\input{Diamond-Bloser}

\subsection{Scintillators and Calorimeters}
\label{sec:calorimeters}


\subsubsection{Next-Generation Scintillators}
\label{sec:scintillators}
\input{Scintillators-Rich}

\subsubsection{Modern Phoswich Detectors for Wide-Field-of-View Gamma-Ray Astronomy}
\label{sec:phoswich}
\input{Phoswich-Wood}

\subsubsection{Radiation Tolerant Silicon Photo-Multipliers for Next-Generation Particle Space Telescopes}
\label{sec:rad_tol_sipms}
\input{SiliconPhotomultipliers-Perkins}

\subsubsection{SiPMs for Extensive Air Shower Detectors}
\label{sec:eas_sipm}
\input{SiPMEAS-Krizmanic}

\subsection{Virtual Frisch-Grid CZT Detectors}
\label{sec:VFG-CZT}
\input{CZT-Moiseev}

\section{Software Development}
\label{sec:pipelines}
Novel software development can be as mission-enabling as detector development. As the hardware becomes more sophisticated, so too must the  pipeline and analysis software to maximize the full capabilities of the instrument. In the following sections we give an overview of data pipeline and software analysis tools that are developed alongside the hardware for the next-generation gamma-ray missions.

\subsection{Detector Optimization and Modeling Software}
\label{sec:modeling}
\input{ModelingSoftware-Wade}

\subsection{Event Reconstruction Tools}
\label{sec:reconstruction}
\input{EventReconstruction-Andreas}

\subsection{High-Level Analysis Tools}
\label{sec:analysistools}
\input{AnalysisTools-Israel}


\chapter{Synergies and Data Management}
\label{sec-synergies}

Chapter 6 focuses on data management for the broader community and especially synergistic infrastructure that supports the science goals of multiple missions or experiments. It is imperative that future facilities be concerned with ease of accessibility for information that is classed as public in order increase their science output through broader community engagement.  In the simultaneous eras of Big Data and  Multi-messenger Astrophysics, at which gamma-ray facilities are a cornerstone, support for data storage, cloud computing analysis and user accessibility must be commonplace.  This virtual infrastructure across experiments reduces the barrier to entry and time required to find data sets for multi-wavelength and multi-messenger projects where gamma-rays thrive.

\section{Data Infrastructure}

Current experiments produce a lot of data and future experiments will produce even more.  Beyond the increase in data for individual missions, experiments, and collaborations, there is an increasing need to access multiwavelength and multimessenger data sets in order to address the biggest and most exciting questions in cutting edge physics and astrophysics today. Therefore, it is important that funding be set aside in each experiment build to construct data pipelines, storage and an accessible user interface tailored to the needs of the people who will use the data, especially those outside of the collaboration providing it. 

It is also essential that data for multiple experiments be easy to find together and compare. The science return on investment will be larger the easier it is for scientists to acquire the data and sort through it. This may require upgrades to existing systems or new systems of cataloguing and pre-sorting.  Astronomy is already in the era of Big Data and Big Surveys are only getting bigger as technology for data storage and transfer from observing facilities improves. 

Some of this work is already underway at the High Energy Astrophysics Science Archive Research Center at NASA and the Space Science Data Center of the Italian Space Agency. The High Energy Astrophysics Science Archive Research Center (see Section \ref{sec:heasarc}) is the primary archive for high-energy photon data from NASA missions and is invested in facilitating closer collaboration between all multi-wavelength and multi-messenger fields in astronomy and astroparticle research. They were one of the first to enable access and analysis of high-energy data via web browser, and are moving toward enabling additional features through cloud computing with standard interfaces.  The Space Science Data Center shares many of the same goals and is particularly dedicated to providing a user-centric interface and suite of functionality without requiring specialized expertise in individual missions or source data sets (Section \ref{sec:ssdc}). 

While national agencies play an indispensable role in facilitating broad collaboration, individual missions and working groups can be attuned to the upcoming needs as well as agile and innovative in their solutions. For example, a group based at the George Washington University has begun development of An Automated Multiwavelength Machine Learning Classification Pipeline (see Section \ref{sec:muwclass}) for use in identifying sources across wavelengths. Their concept of crowd-sourced data and dynamic web-based user interface are a model for the utility that can be provided to the community by smaller groups of innovators and it will be important to support similar efforts over the next decade.

\subsection{High Energy Astrophysics Science Archive Research Center}
\label{sec:heasarc}

\input{HEASARC-Smale}

\subsection{Space Science Data Center}
\label{sec:ssdc}

\input{SSDC-Gianluca}

\subsection{MUWCLASS: An Automated Multiwavelength Machine Learning Classification Pipeline}
\label{sec:muwclass}

\input{MUWCLASS-Hare}

\section{Software Infrastructure}
\label{sec:softwareinfrastructure}

As modern astronomy moves toward larger data sets and more collaboration between missions, there is a growing need for especially reduction and analysis codes to be available to scientists who want to work with publicly available data without the hurdle of spending a lot of time recreating and validating codes that already exist.  It is especially important to highlight tools and services that facilitate collaboration between missions and experiments. Accessibility of code requires user-oriented search capabilities, documentation, and generally infrastructure for uploading, validation, downloading, and storing code. An important aspect of supporting the production of publicly available software is the standardization of software citations and making sure that users cite developers is key. From the perspective of a funding agency, it is important to invest in the developers, to value listing software in this way, to maintain funding streams for the people that make this system function. 
 
The Astrophysics Source Code Library (see Section \ref{sec:ascl}) is a public-facing scientific software repository which provides individual code citations alongside links to relevant publications, giving users the ability to use and cite software in a standardized way. Interest in the library and its hosted codes has increased exponentially in recent years, demonstrating profound interest in and utility for repositories that support individual developers.  Additionally, the Open Astronomy Consortium (Section \ref{sec:openastronomy}) is a collection of analysis suites developed to support specific missions and analysis goals. It also forms a community of developers through online forums and physical meetings.   

Software suites and analysis packages for high-energy data reduction and analysis have been extensively developed for some applications and mission collaborations. Gammapy is an open source Python package for gamma-ray astronomy described in Section \ref{sec:gammapy}.  Fermipy is a standardized Python wrapper for high-level analysis of gamma-ray data across missions and astrophysical source types further described in Section \ref{sec:fermipy}).  Additionally, ThreeML is a python-based, multi-mission framework and software package for multi-wavelength astronomical data. These open source analysis packages have been invaluable to the gamma-ray astrophysics community and future missions and experiments should plan to support similar collaborative software and analysis infrastructure for the benefit of their science output and the high-energy community.

\subsection{Astrophysics Source Code Library}
\label{sec:ascl}

\input{ASCL-AliceAllen}

\subsection{OpenAstronomy}
\label{sec:openastronomy}

The Open Astronomy Consortium is a collective of reduction and analysis software suites in addition to a forum for the communities of developers that provide them. More information can be found on their website at \url{https://openastronomy.org}.

\subsection{Gammapy}
\label{sec:gammapy}

\input{GammaPy-Axel}

\subsection{Fermipy}
\label{sec:fermipy}

\input{FermiPy-Giacomo}

\clearpage
\subsection{ThreeML}
\label{sec:3ml}

\input{threeML-HenrikeMichael}




\bibliographystyle{unsrt}
\bibliography{main,SiliconPixel-Regina,FiberTracker-Nicola,NeutrinoGammaConnection-Karwin,threeML-HenrikeMichael,VHEgamma-Henrike,ExtragalacticBackground-Meyer,FermiPy-Giacomo,SpeedGravity-Burns,Polarimetry-Haocheng,SiliconPhotomultipliers-Perkins,BurstCube-Perkins,MoonBEAM-Hui,UnbrokenSM-Rodd,LargeScale-Michela,LEAP_McConnell,MUWCLASS-Hare,NonImagingDetectors-Isreal,AnalysisTools-Israel,DSSD-Kierans,Phoswich-Wood,GammaPy-Axel,PrimordialBH-Ranjan,ModelingSoftware-Wade,Magnetars-Zorawar,DarkMatter-Tansu,GWLocalization-Judy,ComptonTelescopes-Reshmi,CodedMask-Brad,Glowbug-Eric,POEMMA-JohnK,ASCL-AliceAllen,SSDC-Gianluca,SiPMEAS-Krizmanic,GammaTPC-Shutt,COSI-Tomsick,SWGO-Engel,GECCO-Moiseev,CZT-Moiseev,Diamond-Bloser,Scintillators-Rich,GRAMS-Tsuguo,AMEGO-Kierans,GRAND-Batista,CTA-Williams}

\end{document}

%% file: UnbrokenSM-Rodd.tex
\noindent
\chapterauthor[]{Nicholas L. Rodd $\quad \ $}\orcidlink{0000-0003-3472-7606}
 \\
 \begin{affils}
   \chapteraffil[]{Theoretical Physics Department, CERN, 1 Esplanade des Particules, CH-1211 Geneva 23, Switzerland}
 \end{affils}

At center-of-mass energies well above the electroweak scale ($\sim$100~GeV), the appropriate description for interactions of the known particles is the unbroken Standard Model (SM).
The transition to the unbroken theory carries with it dramatic changes.
Just as an electron can radiate photons and a quark gluons, at high enough energies, a neutrino can radiate $Z$-bosons.
As a result, neutrinos can initiate jets at a 100~TeV collider, and the decay of heavy dark-matter to neutrinos can produce a sizable gamma-ray signature.
To fully describe this physics, future colliders, and also gamma-ray observatories searching for heavy dark-matter, will require a deeper understanding of the unbroken SM.
In this way, experimental progress on both fronts will drive forward our knowledge of the SM at the highest energies.

While there are conceptual similarities between electroweak emissions and the equivalent processes for photons or gluons, there are important differences.
At low energies the electroweak theory is broken, and as a result we can work with states, like the neutrino, which are not gauge singlets.
(Recall in the SM the neutrino appears in a doublet with the left-handed electron.)
The non-singlet nature of such states generates electroweak double logarithms~\cite{Ciafaloni:2000df}, of the form $\alpha_2 L^2$, where $\alpha_2$ is the SU(2) fine-structure constant, and $L = \ln (\sqrt{s}/m_{\scriptscriptstyle W})$, written in terms of the center-of-mass energy, $\sqrt{s}$, and the $W$ boson mass, $m_{\scriptscriptstyle W} \simeq 80~$GeV.
As $\sqrt{s}$ increases, eventually $\alpha_2 L^2 \sim 1$, and such terms must be resummed into expressions of the form $e^{\alpha_2 L^2}$, which qualitatively capture effects such as the emission of multiple electroweak bosons.
The importance of these terms grows rapidly above the electroweak scale, and the full set of consequences continues to be determined.

The emission of a single electroweak bosons has already been observed at the LHC, for example in final states involving a $W$-boson and a jet~\cite{ATLAS:2016jbu,CMS:2017gbl}.
At a 100 TeV collider, multiple emissions must be accounted for, and the electroweak effects can impact cross-sections by a factor of a few \cite[e.g.][]{Manohar:2014vxa}.
With this in mind, theoretical frameworks to perform such calculations are being developed~\cite[e.g.][]{Chen:2016wkt,Bauer:2017isx,Manohar:2018kfx,Bauer:2018xag}.
Beyond an improved theoretical understanding, such effects must also be incorporated into software used to simulate events at colliders, such as \href{https://pythia.org/}{\texttt{Pythia}}~\cite{Sjostrand:2006za,Sjostrand:2007gs,Sjostrand:2014zea}.
Single electroweak emissions from fermions have been incorporated~\cite{Sjostrand:2014zea}, and extending such approaches to include a broader set of interactions, as well as allowing for multiple emissions, is actively underway~\cite[e.g.][]{Kleiss:2020rcg,Masouminia:2021kne,Brooks:2021kji}.

The importance of single electroweak emissions for the indirect detection of dark matter is well known~\cite{Ciafaloni:2010ti}, and is incorporated in the widely used public code \texttt{PPPC4DMID}~\cite{Cirelli:2010xx}.
A clear example is in scenarios where the dark-matter annihilates or decays to a pair of neutrinos.
The naive expectation of a 2-body final state of neutrinos is dramatically corrected with the emission of just a single electroweak boson: the spectrum now includes photons, positrons, and antiprotons.
Even at TeV energies where canonical WIMP dark-matter candidates exist, such as the wino or higgsino, electroweak corrections are crucial for accurate determinations of the signal~\cite{Bauer:2014ula,Ovanesyan:2014fwa,Baumgart:2014saa,Baumgart:2015bpa,Ovanesyan:2016vkk,Baumgart:2017nsr,Beneke:2018ssm,Baumgart:2018yed,Beneke:2019vhz,Beneke:2019gtg}, and will impact whether the higgsino is discovered or excluded by CTA~\cite{Rinchiuso:2020skh}.
The future experimental landscape will allow one to probe significantly heavier DM candidates, where the full effects of multiple emissions must be accounted for.
The first results computing general dark-matter spectra to arbitrarily high energies appeared recently~\cite{Bauer:2020jay}, and were distributed in the public code \href{https://github.com/nickrodd/HDMSpectra}{\texttt{HDMSpectra}}.
However, there are a number of effects that remain to be added to these results.
For instance, fixed order corrections will be required to smoothly interpolate \texttt{HDMSpectra} onto \texttt{PPPC4DMID} at lower masses.
More broadly, improvements in the dark-matter spectra will continue to occur in parallel with progress at colliders.

%% file: SpeedGravity-Burns.tex
\vspace{0.1in}
 \noindent
 \chapterauthor[1]{Eric Burns}\orcidlink{0000-0002-2942-3379} 
 \chapterauthor[2]{\& Jay Tasson}
 \\
 \begin{affils}
   \chapteraffil[1]{Department of Physics and Astronomy, Louisiana State University, Baton Rouge, LA, USA}
   \chapteraffil[2]{Physics and Astronomy Department, Carleton College, Northfield, MN, USA}
 \end{affils}

The Standard Model encompasses quantum theories for all known forces except gravity; General Relativity is a classical theory of gravity. A Grand Unified Theory may require a quantum theory of gravity, which would be built upon a fundamental length scale of the Universe, which requires breaking of Lorentz Symmetry. As Lorentz Symmetry underlies relativity our two core theories of forces are fundamentally incompatible. Thus, observational evidence of Lorentz Invariance Violation (LIV) would prove a great advance towards unifying the fundamental forces.

Given the breadth of possible LIV, and the plethora of alternative theories of gravity, observers and theorists often communicate results and predictions through the use of effective field theories, where a comprehensive example and updated results are contained in the Standard Model Extension (SME) framework \citep{SME_summary}. The SME allows for violation in specie-specific ways, i.e. separately considering the photon, gravity, neutrino, and matter sectors. Often the photon sector has LIV constraints far more precise than the other sectors. Multimessenger detections allow the existing limits in one sector to place new constraints in another. 

A key example is measurement of the fractional deviation of the speed of gravity from the speed of light, where General Relativity requires them to be exactly the same, and this comparison is the lowest order term in the SME. The test relies on using the relative arrival times of gravity and light from the same event, removal of intrinsic time delay, and division by the distance of the event \citep{GR_mg_Will_1998}. The most precise tests occur for events with small and/or well-known intrinsic time delay and occurring at great distances \citep[accounting for cosmological effects,][]{jacob2008lorentz}.

The key transient for these studies are neutron star mergers \cite{nishizawa_2014_speed_of_gravity}. The time offset between these events is $\mathcal{O}(1)$\,s and they occur out to a redshift of a few. Indeed the first precise measure of the speed of gravity came with the joint detection of a binary neutron star merger in gravitational waves as GW170817 and GRB 170817A, achieving a constraint on the fractional difference of the speed of gravity against the speed of light of $c_{GW}<10^{-15}$ \citep{GW170817-GRB170817A}. This same event additionally provided orders of magnitude improvement in constraints on other gravity-sector LIV terms in the SME. These results conclusively ruled out alternative theories of gravity \citep{cgw_GW170817_dark_matter_Boran_2018,cgw_beyond_GR_constraints_Creminelli_2017, cgw_beyond_GR_constraints_Baker_2017,cgw_beyond_GR_constraints_Ezquiaga_2017,sakstein2017implications}, including several that attempted to explain away dark matter or dark energy.

Future observations will provide orders of magnitude improved sensitivity to LIV in the gravity sector \citep{burns2020neutron}. Improvements will arise for several reasons. GW170817 occurred at only 40 Mpc. The horizon of the A+ gravitational-wave network will be more than an order of magnitude greater than this value, and 3rd generation intereferometers enable an similarly large jump in detection distance. Detection of a population of events allows for joint constraints on propagation delays and the intrinsic time offset of these events. Constraining the intrinsic time offset will have knock-on effects on greater understanding of jet formation and propagation, the emission mechanism of gamma-ray bursts, and studies of hot supranuclear matter by constraining the lifetime of metastable neutron stars. A population of events also allows for observations of sources at different distances and positions on the sky, excluding some remaining finely-tuned alternative theories of gravity and providing the first precise sensitivity to position-dependent LIV in the gravity sector. Should the latter occur it would not only invalidate General Relativity but it would also prove the Cosmological Principle False.

These increases are entirely dependent on continued improvements in the sensitivity of gravitational-wave interferometers and joint observations with gamma-ray burst monitors. These tests require the luminosity distance as measured through gravitational waves; they are improved, but not reliant on, follow-up observations providing a direct redshift determination. The capability of existing gamma-ray burst monitors is sufficient to greatly advance the sensitivity of these tests. The tests can be improved with a greater population of joint detections, enabled by true all-sky gamma-ray burst monitors with improved sensitivity.

%% file: GWLocalization-Judy.tex
 \noindent
 \chapterauthor[]{Judith L. Racusin}\orcidlink{0000-0002-4744-9898}
 \\
 \begin{affils}
   \chapteraffil[]{Astroparticle Physics Laboratory, NASA Goddard Space Flight Center, Greenbelt, MD, USA}
 \end{affils}

The coincident detection of gamma-ray bursts (GRBs) with gravitational waves (GWs) from neutron star mergers provides a wealth of information on both the physics of the merger and subsequent emission components (prompt emission, afterglow, kilonova), allowing for their use as probes of fundamental physics \cite{GW170817-GRB170817A} (Section \ref{sec:speedofgravity}). It is only via wide-band gamma-ray observations of the GRB that one can measure the total energy release in the explosion and put it in context with the population of cosmological short duration GRBs (sGRBs).

By the nature of wide field-of-view (FOV) GRB monitors, they independently trigger within seconds of binary neutron star (BNS) mergers, and possibly also neutron star black hole (NSBH) mergers. GRB monitors (e.g. {\it Swift}-BAT, {\it Fermi}-GBM) autonomously trigger onboard, and rapidly transmit data on the properties of the triggers including their localizations, which are within a few tens of seconds processed by ground pipelines and distributed to the worldwide network of follow-up observers via the Gamma-ray Coordinates Network\footnote{\href{gcn.gsfc.nasa.gov}{gcn.gsfc.nasa.gov}}.

The ideal observational scenario for a nearby on-axis BNS or NSBH would be for it to be within the FOVs of both {\it Swift}-BAT and {\it Fermi}-GBM, providing accurate localization and wide-band spectral energy coverage. {\it Swift} would then autonomously repoint XRT and UVOT providing an incredible data set to probe the physics of both the prompt and afterglow emission, as well as the transition to the kilonova evolution, and an arcsecond position sent to telescopes around the world to begin follow-up observations within minutes providing redshift and high-resolution spectroscopy. However, the probability of this occurrence is low. Roughly 10\% of GW-detections of NS mergers with jets will be oriented such that we can detect the sGRB, and the FOV of the BAT is $\sim$11\% of the sky over the range 15-150 keV. Swift has active programs to localize potential GW counterparts initiated by external triggers via GUANO \cite{SwiftGUANO}, as well as XRT \cite{xrt03} and UVOT \cite{uvoto3} follow-up observations.

The current most prolific detector of sGRBs is {\it Fermi}-GBM, with $\sim$40 onboard sGRB triggers per year, and additional sGRB candidates found via sub-threshold searches \cite{GBMLVCO1O2,gbmtargeted,gbmsingleIFO}. GBM observes the entire unocculted sky from 8 keV to 30 MeV, giving it an instantaneous FOV of $\sim$70\% of the sky. While its localizations are degrees to tens of degrees radius, they are useful for follow up observations. The GW localizations are long arcs, which in some cases will be reduced in upcoming observing runs for nearby events with 3 or more interferometers operating. However, as the GW detectors become more sensitive with upgrades, the population of events will be on-average more distant, with most events detected near threshold \cite{petrov}. These distant events will only be seen by the most sensitive detectors in the network, making their localization arcs longer. The combination of a localization from GBM and the long GW arc, can potentially reduce the search area for follow-up observations significantly. In the case of GW170817, initially only the 2-detector LIGO localization was available, and the combination of GBM and LIGO shrunk the search area.  Although with a several hour time delay, GBM and INTEGRAL SPI-ACS were also combined via the Interplanetary Network (IPN) providing a complementary reduction of the localization region \cite{GW170817MMA}. Low and medium energy gamma-ray survey instruments are important to this ongoing work and several upcoming and proposed examples are discussed in Section \ref{sec:mev_telescopes}.

%% file: NeutrinoGammaConnection-Karwin.tex
  \chapterauthor[]{Christopher M. Karwin}\orcidlink{0000-0002-6774-3111}
 \\
 \begin{affils}
   \chapteraffil[]{Department of Physics and Astronomy, Clemson University, Clemson, SC, USA}
 \end{affils}

Extragalactic cosmic rays (CRs) and high-energy astrophysical neutrinos can reach energies far exceeding those obtained by even the Large Hadron Collider, and thus they serve as probes of fundamental physics at energies unattainable in terrestrial experiments. Their origin, however, still remains an open question~\citep{Hillas:1985is,Fang:2017zjf}. Many astrophysical sources suspected of having the conditions necessary for CR acceleration also contain intense matter and radiation fields with which CRs interact, ultimately producing both neutrinos and gamma-rays. Thus the best approach for answering these fundamental questions is through multi-messenger campaigns (Snowmass2021 white paper: Advancing the Landscape of Multimessenger Science in the Next Decade), which leverage both neutrinos and photons.

Neutrinos have long garnered interest as the unfailing messengers of hadronic interactions in the Universe. With a low interaction cross section and being electrically neutral, they travel virtually unhindered through their sources and over cosmological distances, carrying information about regions from which neither high-energy gamma-rays nor CRs can escape. The ratios of their three flavors ($\nu_e$, $\nu_\mu$, $\nu_\tau$) measured at Earth may encode information about their production within their sources~\citep{Bustamante:2019sdb}. High-energy cosmic neutrinos also present the opportunity to test symmetries in the Standard Model, search for dark matter (Section \ref{sec:darkmatter}), and study neutrino oscillations and cross sections~\citep{Ackermann:2019cxh}. 

Gamma-rays provide complementary information to the neutrinos. The strikingly similar intensities of the extragalactic gamma-ray background and the CR and diffuse neutrino spectra may hint at a common origin (at least in part) for all of these phenomena~\citep{Fang:2017zjf}. Among the possibilities,  blazars--a subclass of active galactic nuclei (AGN) whose jets are directed very close to our line of sight with strong gamma-ray variability--have long been of interest as possible sources of extragalactic CRs and high-energy astrophysical neutrinos.

The IceCube Neutrino Observatory has reported the detection of an isotropic flux of high-energy astrophysical \ neutrinos~\citep{Aartsen:2013jdh,Aartsen:2014gkd}, the origin of which remains unknown, as well-established point sources have yet to be identified~\citep{Aartsen:2015wto,Aartsen:2016oji}. Recently, the track-like $\nu_\mu$ event IceCube-170922A was found to be coincident in direction and time with a gamma-ray flare from the blazar TXS 0506+056~\citep{IceCube:2018dnn}, which lends support to the possibility that relativistic blazar jets may be the first confirmed source of cosmic neutrinos. Moreover, this event also implies that protons can be accelerated in blazar jets, as neutrinos are a unique signature of proton acceleration and interaction. However, the mechanisms and extent of proton acceleration in AGN still remains unclear.

A later analysis showed that roughly three years before the IceCube-170922A event, $\sim$13 neutrinos were detected by IceCube from the same direction in the sky~\citep{IceCube:2018cha}. This earlier `neutrino flare' had no associated gamma-ray flare detected by \textit{Fermi}-LAT, and the neutrino flux was $\sim$5 times higher than the average gamma-ray flux, likely implying a strong absorption of GeV photons by an intense X-ray radiation field. These observational constraints pose a significant difficulty for interpreting the neutrino flares in terms of conventional one-zone models~\citep{Murase:2018iyl,2019ApJ...881...46R,2019ApJ...874L..29R}, and thus numerous models that go beyond the framework of the conventional one-zone model have been proposed~\citep[e.g.][]{2019PhRvD..99f3008L,2019ApJ...886...23X,2020ApJ...889..118Z,2021ApJ...906...51X,2019ApJ...880...40I,2020PhRvL.125a1101M}. In many physical scenarios it is expected that the conditions in blazar jets that are instrumental for efficient neutrino production via photo-hadronic (p$\gamma$) interactions do not allow GeV gamma-rays to escape, due to enhanced $\gamma\gamma$ optical depths~\citep{Murase:2015ndr}, resulting in GeV gamma-rays being reprocessed to the MeV band. 

Observationally, MeV gamma-rays are thus far one of the least explored bands in the electromagnetic (EM) spectrum, whereas theoretical models suggest that this band is the key to identify neutrino production and CR acceleration in blazar jets~\citep{Rani:2019ber,Rani:2019cba,Ojha:2019xan}. To fully understand the corresponding EM and neutrino signatures, support for both observational and theoretical multi-messenger studies will be essential. As one particular example of a prospective MeV mission, we can consider the All-Sky Medium Energy Gamma-ray Observatory Explorer (AMEGO-X)~\cite[Section \ref{sec:amego};][]{2021arXiv210802860F}--an explorer-class mission concept that combines high sensitivity with a wide field of view, good spectral resolution, and polarization capability. In particular, the MeV sensitivity of AMEGO-X would improve upon previous MeV missions by over an order-of-magnitude. 

\begin{figure}[t]
\centering
\includegraphics[width=0.8\textwidth]{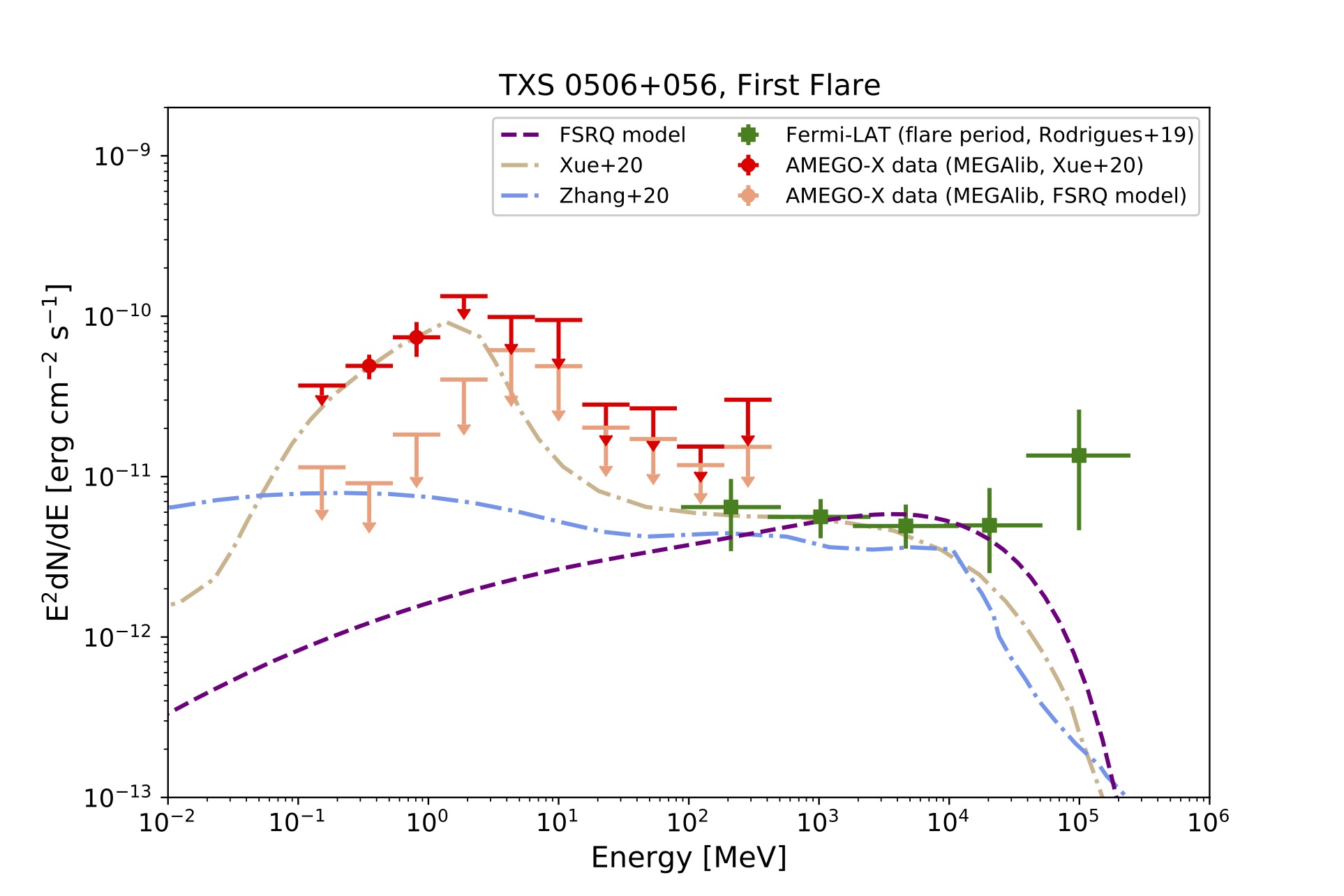}
\caption{Different models and simulated data of the expected MeV flux for the 2014-2015 IceCube neutrino flare, taken from~\cite{Lewis:2021roc}. Note that these models are by no means inclusive of the total body of work currently found in the literature, but rather, they give an estimate of the range for the expected flux in the MeV band corresponding to the neutrino flare. See text for more details.} 
\label{fig:SED1}
\end{figure}

In Ref.~\citenum{Lewis:2021roc}, MEGAlib simulations were used to make inferences about the kinds of MeV observations that would have been possible with AMEGO-X during the 2014-2015 neutrino flare of TXS 0506+056. Figure~\ref{fig:SED1} shows three different models that were considered. Note that these models are by no means inclusive of the total body of work currently found in the literature, but rather, they give an estimate of the uncertainty range for the expected flux in the MeV band corresponding to the neutrino flare. The purple dashed curve shows a one-zone FSRQ model, as described in~\cite{Lewis:2021roc}, and the simulated data is shown with peach markers. The blue dash-dot curve shows the predicted flux from~\cite{2020ApJ...889..118Z} for a neutral beam model. The tan dash-dot curve shows a two-zone radiation model with an inner and outer blob from~\cite{2021ApJ...906...51X}, and the simulated data is shown with red markers. The green markers show {\it Fermi}-LAT date during the flare period from~\cite{2019ApJ...874L..29R}. As can be seen, the different model predictions for the MeV flux cover a broad range. In the most extreme case of bright MeV emission from cascaded photons, AMEGO-X would have been able to significantly detect the EM flare, which went undetected by {\it Fermi}-LAT. More generally, however, AMEGO-X would have provided crucial upper limits on the MeV flux. 

In summary, the origin of high-energy astrophysical neutrinos detected by IceCube remains unknown, and the sources of extragalactic CRs have been a mystery for over a century. Obtaining a better understanding of these fundamental questions will require a multi-messenger approach, for which observations of EM counterparts, particularly in the MeV band, are crucial. The detection of IceCube-170922A coincident in direction and time with a \textit{Fermi}-LAT--detected flare in blazar TXS 0506+056 supports the notion that blazars may be a source of astrophysical neutrinos and CR acceleration. However, an earlier event of $\sim$13 neutrinos from the same source did not coincide with an enhanced flux of GeV gamma-rays. These observational constraints seem to imply an emission mechanism that goes beyond the conventional one-zone model. Future MeV telescopes (Section \ref{sec:mev_telescopes}) such as AMEGO-X will be crucial when it comes to searching for the EM counterparts of neutrino flares and constraining different possible models.

%% file: Polarimetry-Haocheng.tex
\noindent
\chapterauthor[1,2]{Haocheng Zhang}
\\
 \begin{affils}
   \chapteraffil[1]{NASA Postdoctoral Program Fellow}
   \chapteraffil[2]{Astroparticle Physics Laboratory, NASA Goddard Space Flight Center, Greenbelt, MD, USA}
\end{affils}

Magnetic field plays an important role in high-energy astrophysical systems. Polarimetry can directly probe the magnetic field morphology and evolution in these systems. Radio and optical polarimetry have been successful in studying the magnetic field structure in many high-energy astrophysical systems, including  active galactic nuclei (AGN), gamma-ray bursts (GRBs), and  pulsar wind nebulae (PWNe)~\cite{Marscher2008,Mundell2007,Slowikowska2009}. Previous high-energy polarimeter experiments, such as {\it POLAR} and {\it PoGO+}, have revealed very interesting and promising scientific results in X-ray bands~\cite{Chauvin2017,ZhangSN2019}. The recent launch of {\it IXPE} promises significant improvement in X-ray polarization capability and great scientific potential~\cite{Weisskopf2016}. A regime that is not yet explored is the gamma-ray polarization. At even higher energy, gamma-ray polarization can probe more energetic phenomena in extreme physical environments than X-rays. In particular, MeV gamma-ray polarization offers insights into the acceleration of cosmic rays and neutrino production, which are crucial to the multi-messenger astronomy that will be mature in the next decade.

Gamma-ray polarimetry can disentangle the radiation mechanisms in relativistic jets and probe hadronic signatures. The high-energy spectral component of blazars, which are AGN jets close to our line of sight, is often interpreted as the Compton scattering by the same electron population that generates the low-energy synchrotron spectral component~\cite{marscher1985,maraschi1992,dermer1992,sikora1994}. But hadronic processes, including proton synchrotron and hadronic cascades, can explain the high-energy component equally well~\cite{Mannheim1993,mucke2001,Boettcher2013,Petropoulou2015}. The recent detection of a very high energy neutrino event by {\it IceCube} in coincidence with the blazar TXS~0506+056 flare by {\it Fermi} gives strong support to the hadronic model~\cite{IceCube2018}. A similar issue in radiation mechanisms exists for GRBs, where the prompt phase emission can be interpreted by one or a combination of the synchrotron emission, Compton scattering, Compton drag, or thermal emission~\cite{Rees1994,Shaviv1995,Ioka2007,Panitescu2000}. MeV gamma-ray polarimetry can distinguish radiation mechanisms in both systems, because the above radiation mechanisms yield very different polarization degree. For blazars, Compton scattering is typically  unpolarized or has a very low polarization degree (a few to ten percent) in a partially ordered magnetic field, while the hadronic models usually predict $\gtrsim 20\%$ polarization degree in the MeV band~\cite{Krawczynski2012,Zhang2013,Paliya2018,Zhang2019}.  Similarly for GRBs, the gamma-ray polarization degree can distinguish between radiation mechanisms and probe the magnetic field morphology~\cite{Toma2009,Gill2020}. In addition, by synergizing with X-ray polarimetry, MeV gamma-ray polarization can unambiguously identify the acceleration of cosmic ray and neutrino production in AGN jets~\cite{Zhang2019}. Furthermore, dedicated gamma-ray polarimeters with good temporal resolution can diagnose the particle acceleration mechanisms in relativistic jets~\cite{Zhang2016,Zhang2019,Deng2016}. This is because different particle acceleration mechanisms involve distinct magnetic field evolution, leading to distinguishable patterns in time-dependent polarization signatures.

Synergies between observational and theoretical efforts are necessary to address the above scientific issues. Future gamma-ray missions with dedicated polarimetry capability will be optimal for studying gamma-ray polarization in high-energy astrophysical systems. For instance,  AMEGO (Section \ref{sec:amego}) can obtain minimal detectable polarization degree of a few percent for the famous blazar 3C~279 at a typical flaring state~\cite{Rani2019,McEnery2019}. This means that in a five-year mission, such a dedicated gamma-ray polarimeter can detect MeV polarization in $\sim68$ blazars, unambiguously identifying potential hadronic processes in blazars. And during flaring states of bright blazars, such an instrument may probe temporal polarization variations, which are crucial to explore the cosmic ray acceleration mechanisms. This level of sensitivity can also detect prompt gamma-ray polarization for several GRBs per year to probe the emission mechanism, particle acceleration, and jet dynamics. Theoretically, we need comprehensive models under first principles to unveil the dynamical fluid and particle co-evolution in high-energy astrophysical systems, including the radiative transfer and feedback. Although full particle-in-cell (PIC) simulations have been successful in studying the particle acceleration processes, they can only simulate evolution on particle kinetic scales~\cite{Spitkovsky2008,Sironi2009,Sironi2014,Guo2016,Zhdankin2017,Comisso2018}. Several multi-scale hybrid methods have been introduced to push simulations to larger scales that are relevant to observations, but further studies are needed to determine the necessary kinetic-scale physics for particle acceleration and feedback to the fluid dynamics~\cite{Achterberg2001,Drake2019,Li2018,Bai2015}. Additionally, the powerful gamma-ray emission typically found in high-energy astrophysical systems requires a thorough treatment of radiation transfer and feedback to the nonthermal particles~\cite{Zhang2016,Zhang2018}. This involves how cosmic rays propagate and escape from the high-energy astrophysical system, and their interaction with the particle and photon field that may lead to neutrino production. Support of numerical methods on first-principle integrated radiative transfer simulations will synergize with gamma-ray polarimetry to obtain the best scientific return.


%% file: Magnetars-Zorawar.tex
\noindent
 \chapterauthor[1,2,3]{Zorawar Wadiasingh}\orcidlink{0000-0002-9249-0515}
 \\
 \begin{affils}
    \chapteraffil[1]{University of Maryland, College Park, Maryland 20742, USA}
    \chapteraffil[2]{Astroparticle Physics Laboratory, NASA Goddard Space Flight Center, Greenbelt, MD, USA}
    \chapteraffil[3]{Center for Research and Exploration in Space Science and Technology, NASA/GSFC, Greenbelt, Maryland 20771, USA}
 \end{affils}

Magnetars are the most highly magnetized neutron stars, with surface fields exceeding $10^{10}$~Tesla ($10^{14}$~Gauss for astronomers) and are an important and topical subclass of neutron stars in astrophysics~\citep{1995MNRAS.275..255T,2008A&ARv..15..225M,turolla15:mag}. In magnetars, the magnetic field is in the quantumelectrodynamic (QED) domain where $\hbar \omega_B \sim m_e c^2$ ($\hbar \omega_B$ the energy scale of non-relativistic electron Landau states; $\omega_B$ the cyclotron frequency). This defines the Schwinger or critical field $ m_e^2c^3/(\hbar q_e)  \equiv B_{\rm cr} \approx
4.413\times 10^{9}$~T and is a regime where exotic aspects~\citep{1966RvMP...38..626E,2006RPPh...69.2631H,2013LNP...854.....M} of standard (but nonlinear and nonperturbative) QED are important. Beyond standard model physics or particle states (e.g., axion-like particles, that may be a component of Dark Matter) also benefit from intense fields~\citep{PhysRevD.37.1237} for potential conversion into detectable photons. 

The output of nearby magnetars is largely observed through their X-ray/gamma-ray emission via bursts and persistent signals. In particular, the persistent output of many magnetars is thought to peak in the 0.1--1~MeV spectral range~\citep{Kuiper-2004-ApJ,Kuiper-2006-ApJ,Enoto-2017-ApJS,2019BAAS...51c.292W}--- this spectral range is largely unexplored and greater understanding requires the deployment of a sensitive MeV telescope, e.g. AMEGO~\citep[Section \ref{sec:amego};][]{2019BAAS...51g.245M,2021arXiv210802860F}. The preferred standard astrophysical radiative process to produce persistent non-thermal emission in magnetars is resonant Compton scattering of soft surface thermal photons into the hard X-rays and soft gamma-rays by populations relativistic electrons/positrons in the ground Landau state (in the inner magnetospheres of magnetars)~\citep{1986ApJ...309..362D,2005ApJ...630..430B,2007Ap&SS.308..109B,2011ApJ...733...61B,2014PhRvD..90d3014G,2016PhRvD..93j5003M,2019BAAS...51c.292W,2018ApJ...854...98W}. This process produces photon states predominantly in the extraordinary mode (electric field perpendicular to plane formed by momentum and local field) that are inclined to undergo single-photon splitting (see below). The resonant Compton emission, and its propagation effects, constitutes an astrophysical background which must be properly understood to probe Beyond Standard Model processes.

Although there is a strong effort to access some aspects of high-field nonlinear QED processes with powerful lasers, the standard Lorentz invariants of the electromagnetic field tensor (locally defined) can be very different in magnetars environments than in laser-plasma interactions. Magnetars access a strong-field regime where the $e^+e^-$ population generally occupies the ground Landau state and $B^2\gg E^2$ (in Gaussian units). Moreover, magnetars may be the only astrophysical laboratory available to test the high-field QED domain on length scales much larger the reduced Compton wavelength $\hbar/(m_e c)$ (i.e. macroscopic scales where $B > B_{\rm cr}$). In particular, magnetar magnetospheres are opaque to high energy photons~\citep{1983ApJ...273..761D,2001ApJ...547..929B,2019MNRAS.486.3327H}, so that above the pair threshold around 1~MeV, magnetic single-photon pair creation strongly dominates the photon opacity (in a polarization state dependent way). Dispersive influences of the magnetized quantum vacuum introduce birefringence,
especially for different refractive indices for the elliptical polarization eigenstates~\cite{1975PhRvD..12.1132T,2006RPPh...69.2631H}; 
dispersion is small for $\sim1\,$keV photons. Below the
pair threshold, photon splitting is the dominant attenuation mechanism in a strong magnetic field; this
is a $3^{\rm rd}$ order QED process arising from vacuum polarization
(virtual pairs) radiating when interacting with the field~\citep{1971AnPhy..67..599A,1997ApJ...482..372B}. The rate of
splitting is a strong function of photon energy and the local direction of propagation relative to the field. In the
weakly dispersive limit, only extraordinary mode photons may split due to kinematic selection rules~\citep{1971AnPhy..67..599A}. However, splitting of both photon polarizations (modes) does not violate charge-parity (CP) symmetry. It is still an open question if both modes may split in the strongly dispersive nonlinear regime of QED. If both polarizations are permitted to split, then a different polarization and spectral signature may be distinguishable~\citep{2019BAAS...51c.292W}. Thus soft gamma-ray observations may be decisive to clarify this fundamental QED issue. 

Beyond QED, axion-like particles may be produced in the hot cores of magnetars via an axion-nucleon coupling and convert into photons in high in the magnetosphere~\citep{2019JHEP...01..163F,2018JHEP...06..048F,2019PhRvD.100f3005L,2020arXiv200110849L,Fortin-2021-arxiv,Fortin-2021-review}. This is a two-step process, and probes the product of the axion-nucleon and axion-photon couplings. It is best probed in the soft gamma-rays, commensurate with the $\sim 10^8-10^9$~K core temperature energy scale of young magnetars (the photon spectrum traces the core's thermal nucleon spectrum). To this end, understanding how axion-conversion emission models are distinguishable from the ``astrophysical background" emission from magnetospheric resonant Compton scattering is an open research problem. Future diagnosis of pulsed polarized soft gamma-rays from magnetars may discriminate new physics from the exotic (but standard) QED by considering the energy, polarization and pulsation dependence of different models. Thus, deep soft gamma-ray observations of magnetars may offer the prospect of constraining new physics not generally accessible by terrestrial methods. Thus, ongoing monitoring of the MeV sky would open opportunities for new discoveries in fundamental physics of QED.

%% file: DarkMatter-Tansu.tex
 \noindent
 \chapterauthor[1,2]{Tansu Daylan}\orcidlink{0000-0002-6939-9211}
 \\
 \begin{affils}
    \chapteraffil[1]{Department of Physics and Kavli Institute for Astrophysics and Space Research, Massachusetts Institute of Technology, Cambridge, MA 02139, USA}
    \chapteraffil[2]{Department of Astrophysical Sciences, Princeton University, Peyton Hall, Princeton, NJ 08544, USA}
 \end{affils}
 

Astronomical observations point toward the existence of a missing matter problem in the Universe as strongly supported by independent observables such as velocity dispersion in clusters of galaxies, galactic rotation curves, kinematics of gas in galaxy clusters, galaxy collisions, gravitational lensing, and the cosmic microwave background (CMB)~\cite{Zwicky1933, Rubin1983, Bosma1981a, Bosma1981b}. However, the particle nature of dark matter is still unexplored despite extensive research on its potential astrophysical signatures, as well as potential interactions with visible matter and production in particle colliders at high energies~\cite{Bertone2005}. In order to provide a solution to the missing matter problem in the Universe, dark matter is constrained to be at most weakly interacting, cold (i.e., non-relativistic at decoupling from primordial plasma), nearly stable over cosmological time scales, and made of neutral particles with nonzero mass. Extensions of the Standard Model such as Universal Extra Dimensions and Supersymmetry provide candidate particles that can resolve this missing matter problem~\cite{Arbey2021}.

Even though dark matter could be perfectly stable and non-interacting, it can also decay or self-annihilate at a rate that would maintain its dominance over the matter budget of the Universe. For example, dark matter can decay over time scales much longer than the age of the Universe. Equivalently, a small fraction of dark matter may have been decaying within the age of the Universe. Sterile neutrinos with masses in the keV range~\cite{Boyarsky2019} and supersymmetric particles such as neutralino and gravitino in R-parity breaking vacua~\cite{Bertone2007}, which can decay into photon and active neutrino pairs, constitute examples of decaying dark matter. Dark matter may also be self-annihilating into visible particles, whose cross-section sets its relic abundance in the Weakly Interacting Massive Particle (WIMP) paradigm. Because annihilation of dark matter is a two-body process, the annihilation rate scales with the square of the dark matter density, producing a relatively more anisotropic distribution of products. Even when one assumes that dark matter decays into invisible dark radiation, the product of the decay rate and decaying fraction can be constrained by the Cosmic Microwave Background (CMB) and Baryonic Acoustic Oscillation (BAO) measurements to be below $\sim0.01\,$Gyr$^{-1}$~\cite{Poulin2016, Nygaard2021}.

In order to reproduce the observed abundance of dark matter, WIMPs that thermally freeze out are expected to have mass lower than $\sim$340~TeV~\cite{Griest1990}. However, WIMPs can also have much higher masses if they evade overproducing dark matter by having a super weak, i.e., Planck-suppressed, interaction with the Standard Model particles~\cite{Chung1998, Kolb2017}. If dark matter has any mass higher than MeV scale, its annihilation or decay can produce gamma rays among other visible Standard Model products, via either a hadronic or leptonic channel followed by final state radiation, subsequent electromagnetic cascades, and inverse Compton scattering of the microwave, optical, and infrared radiation field~\cite{Stecker1978}. Under this assumption, the lifetime and rate of decaying dark matter can be more strongly constrained based on their visible products~\cite{Ibarra2013, Ando2015}. Gamma rays retain spatial and spectral information at production in contrast to charged dark matter annihilation or decay products, which deviate due to interaction with magnetic fields, undergo diffusion, or otherwise suffer undergo energy losses. The Energetic Gamma Ray Experiment Telescope (EGRET)~\cite{Fichtel1993} and the {\it Fermi} Large Area Telescope (LAT)~\cite{lat} have been our main facilities to measure these gamma rays (Section~\ref{sec:compton_tel}). Furthermore, if WIMPs have mass above $\sim$1~TeV, our \textit{only} hope of detecting dark matter could be such gamma rays since current particle colliders can only probe potential annihilation products of dark matter with mass up to $\sim$1~TeV~\cite{Viana2019}. Ground-based Imaging Atmospheric Cherenkov Telescopes (IACTs; Section~\ref{sec:atmospheric_cherenkov}) and Water Cherenkov Detectors (WCDs; Section~\ref{sec:water_cherenkov}) such as the High Energy Stereoscopic System (H.E.S.S.), Very Energetic Radiation Imaging Telescope Array System (VERITAS), High Altitude Water Cherenkov (HAWC), Major Atmospheric Gamma Imaging Cherenkov Telescopes (MAGIC), First G-APD Cherenkov Telescope (FACT)~\cite{Anderhub2013}, and the expected Southern Wide-field Gamma-ray Observatory (SWGO; Section~\ref{sec:swgo})~\cite{Albert:2019afb,Abreu:2019ahw,Hinton:2021rvp,Schoorlemmer:2019gee} and Cherenkov Telescope Array (CTA; Section~\ref{sec:cta})~\cite{CTAConsortium:2018tzg,Knodlseder2011} are the main experiments providing complementary sensitivity to these very-high-energy gamma-ray products.

Being a two-body process, annihilation products tend to have stronger spatial anisotropies compared to decay products. Therefore, constraints on dark matter decay predominantly come from extragalactic sources making up the isotropic gamma-ray sky~\cite{Bertone2007}, whereas annihilation cross-sections are typically probed by emission from regions such as the Galactic Center or nearby dwarf spheriodal galaxies that are expected to be dominated by dark matter. Active galactic nuclei and star-forming galaxies constitute the dominant background sources in the former searchers, whereas the galactic diffuse gamma ray acts as the dominant background for searches in the inner galaxy. Furthermore, our Sun may also have accreted a dark matter halo, which can act as a source of gamma rays. However, gamma rays produced within the Sun would interact with the plasma and not be observable~\cite{Sivertsson2012}. Instead, a long-lived mediator is a more observationally accessible dark matter annihilation product, if it decays after escaping the Sun~\cite{Bell2011}. Such searches using HAWC~\cite{Albert2018} and {\it Fermi}-LAT~\cite{Ng2016} have produced constraints of $\sim10^{-45}$~cm$^{-2}$ on the mediator decay cross-section for WIMP masses between 4 and $10^6$~GeV, which is competitive compared to the direct detection experiments.

Based on the isotropic gamma-ray background as measured by the {\it Fermi}-LAT~\cite{Ackermann2015}, a lower limit of $(1-5)\times10^{28}$~s was placed on the lifetime of decaying dark matter~\cite{Blanco2019}. Similarly, CTA, SWGO, and HAWC have been forecasted to be able to probe lifetimes up to $10^{27}-10^{28}$~s for the decay of WIMPs with mass between 200~TeV and 20~PeV into $b\bar{b}$ based on gamma rays from dwarf spheroidal galaxies~\cite{Ando2021}. Stronger constraints are obtained on the lifetime of heavier dark matter, with the constraints getting as strong as $\tau \sim10^{30}$~s in the range of $10^{11}-10^{14}$~GeV range~\cite{Ishiwata2020}. The constraints are stronger below $\sim$1~PeV for $\chi \to \tau \bar{\tau}$, where they get stronger with mass for $\chi \to b \bar{b}$. An intriguing gamma-ray observation that could potentially originate from the annihilation of dark matter is the extended gamma-ray excess peaking at 1--2~GeV in the inner regions of the Milky Way~\cite{Hooper2011a, Hooper2011b}. The GeV excess has the amplitude as well as the spectral and morphological properties expected from WIMP annihilation~\cite{Daylan2016}. However, the absence of a consistent signal from dwarf spheroidal galaxies and the ability of millisecond pulsars to also explain the GeV excess currently challenge this interpretation~\cite{Hoof2020}. Hence, improvements to gamma-ray facilities and development of future telescopes (Chapter~\ref{sec-facilities}) are strongly motivated by the anticipated ability of improved observational constraints to distinguish explanations of gamma-ray signals.

%% file: PrimordialBH-Ranjan.tex
 \noindent
 \chapterauthor[]{Ranjan Laha}\orcidlink{0000-0001-7104-5730}
 \\
 \begin{affils}
   \chapteraffil[]{Centre for High Energy Physics, Indian Institute of Science, C.\,V.\,Raman Avenue, Bengaluru 560012, India}
 \end{affils}

\noindent

What is dark matter (DM)?  This is one of the most profound and deep scientific questions that human beings have even attempted to answer. Answering it will help us understand $\sim$ 80\% of the matter energy density of the Universe.  Numerous DM candidates have been proposed in the literature, and it is important to thoroughly test all the well-motivated proposals in order to precisely probe DM. For a majority of these proposals, gamma-ray telescopes will play a decisive role in determining the nature of DM.

Primordial black holes (PBHs) are one of the oldest and well-motivated DM candidates~\cite{1966AZh....43..758Z, Hawking:1971ei, Carr:1974nx, Chapline:1975ojl}. BHs evaporate via Hawking radiation (HR)~\cite{Hawking:1971ei}, and it can be shown that PBHs with masses $\gtrsim$ 6 $\times$ 10$^{14}~$g have a lifetime greater than the current age of the Universe, setting a lower limit on the PBH DM mass~\cite{Page:1976df, Page:1976ki, MacGibbon:2007yq, Arbey:2019jmj}. Angular momentum has a small effect on this lower limit~\cite{Arbey:2019jmj}. Extremal PBHs do not evaporate via HR, and thus lower mass extremal PBH can also be the DM candidates ~\cite{Bai:2019zcd, Chongchitnan:2021ehn}. HR produces all types of particles, and the evaporated Standard Model particles can either decay or hadronize to produce gamma-rays.  The HR flux from non-spinning and spinning PBHs with masses between $\sim$ 6 $\times$ 10$^{14}~$g--$\mathcal{O}$(10$^{18}~$g) is potentially detectable using current and future generation gamma-ray telescopes. Gamma-ray telescopes provide the strongest constraints on PBH DM abundance in the mass range $\sim$6 $\times$ 10$^{14}~$g-- 4 $\times$ 10$^{17}~$g. It is possible that near future gamma-ray data will be able to discover PBH DM in this mass range even if they make up an extremely small fraction, $\sim$10$^{-7}$, of the DM density. We first discuss the various constraints on PBH DM from current gamma-ray observations, and then move on to discussing the future prospects from various planned telescopes.

An electrically neutral BH of mass, $M_{\rm BH}$, and dimensionless spin parameter, $a_*$, has a temperature~\cite{Page:1976df, Page:1976ki, MacGibbon:2007yq, MacGibbon:1990zk, MacGibbon:1991tj}
\begin{eqnarray}
T_{\rm BH} \,=\, \dfrac{1}{4\pi G_N M_{\rm BH}}\,\dfrac{\sqrt{1 - a_*^2}}{1+\sqrt{1 - a_*^2}}\,,
\label{eq:BH temperature}
\end{eqnarray}
where $G_N$ denotes the Newton's gravitational constant.  The number of emitted particles per unit energy interval $dE$ and time interval $dt$ is~\cite{Page:1976df,Page:1976ki,Hawking:1971ei,MacGibbon:2007yq,MacGibbon:1990zk,MacGibbon:1991tj}
\begin{eqnarray}
\dfrac{d^2 N}{dE dt} \,=\, \dfrac{1}{2\pi} \,\dfrac{\Gamma_s (E, M_{\rm BH}, a_*, \mu)}{{\rm exp}(E'/T_{\rm BH}) - (-1)^{2s}} \,,
\label{eq:emitted particle spectrum}
\end{eqnarray}
where $\Gamma_s$ denotes the graybody factor, and the effective energy of the Hawking radiated particles (of mass $\mu$ and spin $s$) is denoted by $E'$. One can calculate the spectrum to great accuracy using the publicly available code {\tt BlackHawk}~\cite{Arbey:2019mbc, Arbey:2021mbl, Auffinger:2022dic}.  The emission of particles from a BH is most prominent at particle energies of the order of $T_{\rm BH}$, for example, the emission of photons peaks at $E \approx$ 5.77 $T_{\rm BH}$~\cite{MacGibbon:1990zk,MacGibbon:2007yq}.

Detection of HR from a low-mass PBH (masses in between $\sim$6 $\times$ 10$^{14}~$g to $\sim$a few times 10$^{18}$~g) will be a smoking-gun signature of their existence.  Various observations have been used to constrain the abundance of low-mass PBH DM~\cite{Poulin:2016anj, Boudaud:2018hqb, Arbey:2019vqx, Ballesteros:2019exr, Iguaz:2021irx, Chen:2021ngo, DeRocco:2019fjq, Laha:2019ssq, Dasgupta:2019cae, Laha:2020ivk, Coogan:2020tuf, Laha:2020vhg, Kim:2020ngi, Siegert:2021upf, Lee:2021qhe, Clark:2018ghm, Mittal:2021egv, Natwariya:2021xki, Cang:2021owu, Halder:2021jiv, Saha:2021pqf, Berteaud:2022tws}. We will first briefly discuss the constraints on low-mass PBH DM arising from the observation of the isotropic diffuse gamma-ray background (IGRB), 511~keV and its associated low-energy continuum photons, and the galactic soft gamma-ray spectrum. (See Section \ref{sec:largescalestructure} for further discussion of the IGRB.)  

The IGRB arises from the emission of all unresolved astrophysical sources in our Universe. In the energy-range $\sim$0.1$~$MeV to 10$^5$~MeV, the IGRB have been observed by Nagoya ballon experiments~\cite{1975Ap&SS..32L...1F}, SMM~\cite{1997AIPC..410.1223W}, COMPTEL~\cite{1995AAS...187.5801K}, EGRET~\cite{1997ApJ...481..205H}, and Fermi-LAT~\cite{2015ApJ...799...86A}. Although, many astrophysical sources have been proposed to be the progenitor of the IGRB, yet we still do not fully understand the origin of this radiation. Low-mass PBH DM will evaporate to produce gamma-rays which can contribute to IGRB.  Multiple analyses (assuming various  astrophysical contributions) have been carried out in order to determine the contribution of low-mass PBH DM to the IGRB, and these have resulted in some of the strongest constraints on PBHs with masses $\sim$6 $\times$ 10$^{14}$~g to 10$^{17}$~g~\cite{Arbey:2019vqx, Ballesteros:2019exr, Iguaz:2021irx, Chen:2021ngo}.

\begin{figure}
\centering
	\includegraphics[angle=0.0,width=0.48\textwidth]{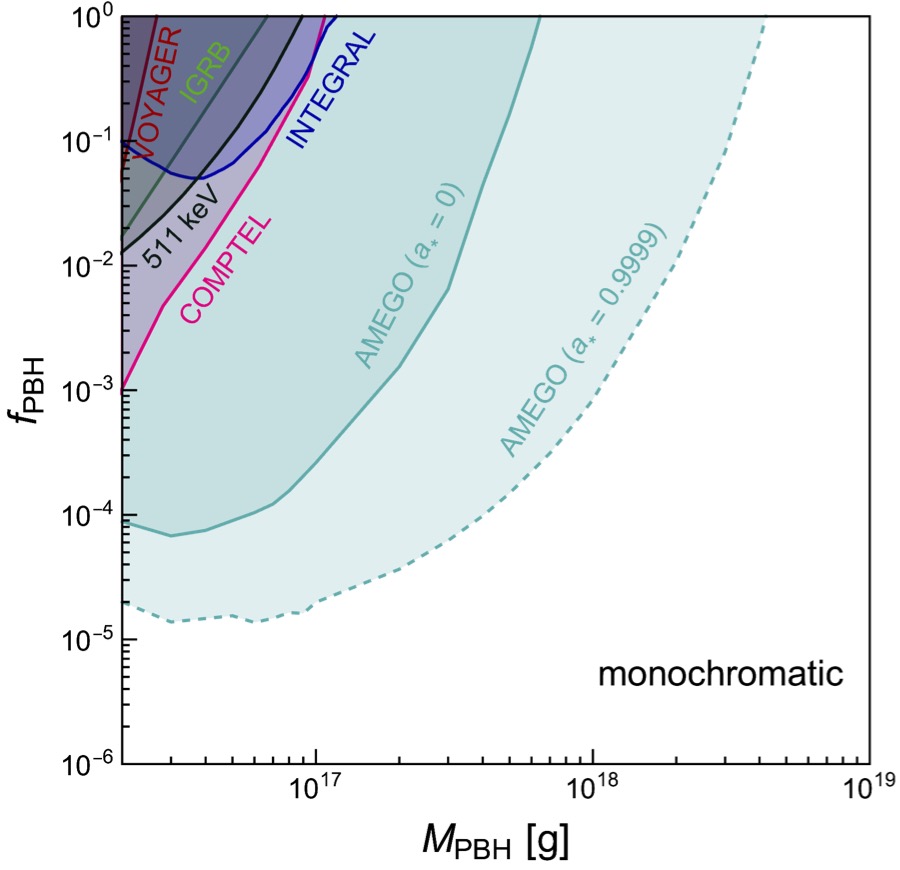}
	\caption{Low-mass PBH DM parameter space (for monochromatic mass distribution) displaying the current constraints on $f_{\rm PBH}$ (the fraction of DM in the form of PBHs) from various observations and future projections from AMEGO for different values of the dimensionless spin parameter ($a_*$ = 0 and $a_*$ = 0.9999).  The current constraints displayed are from the measurement of the cosmic-ray flux by Voyager-1~\cite{Boudaud:2018hqb}, IGRB measurements~\cite{Ballesteros:2019exr, Arbey:2019vqx}, galactic bulge low-energy positron measurements~\cite{DeRocco:2019fjq,Laha:2019ssq,Dasgupta:2019cae}, and from the galactic soft gamma-ray flux~\cite{Laha:2020ivk,Coogan:2020tuf}. The projection is shown for AMEGO observation of the galactic Center. This figure is taken from Ref.~\citenum{Ray:2021mxu}.}	
	\label{fig:low-mass-PBH-DM}
\end{figure}

Observations of the galactic 511~keV line, its associated low-energy continuum photons, and the measurement of the galactic higher energy photons confirm the presence of low-energy positrons in the galactic bulge~\cite{Prantzos:2010wi, Kierans:2019pkh, Siegert:2019tus, Kierans:2019aqz}. Although, we do not yet know the origin of these positrons, one can obtain one of the strongest constraints on low-mass PBH DM by demanding that the evaporated positron flux be less than that required to reproduce the galactic 511 keV and its associated low-energy continuum photon excess~\cite{1980A&A....81..263O, 1991ApJ...371..447M, Bambi:2008kx, DeRocco:2019fjq, Laha:2019ssq, Dasgupta:2019cae}.  Ref.~\citenum{Keith:2021guq} proposes that low-mass PBHs are the source of the low-energy positrons which give rise to the galactic 511 keV and its associated low-energy continuum photon excess.  The measurement of the galactic soft gamma-ray spectrum by the INTEGRAL and COMPTEL satellites also provide one of the strongest constraints on low-mass PBH DM~\cite{Laha:2020ivk, Coogan:2020tuf, Berteaud:2022tws}.  Several next-generation gamma-ray telescopes have been proposed, for 
example, AMEGO~\cite{AMEGO:2019gny} (Section~\ref{sec:amego}), 
GECCO~\cite{2021arXiv211207190O} (Section~\ref{sec:gecco}), 
AdEPT~\cite{2010SPIE.7732E..21H}, MAST~\cite{2019APh...112....1D}, PANGU~\cite{2016SPIE.9905E..6EW}, 
GRAMS~\cite{2020APh...114..107A} (Section~\ref{sec:grams}), and XGIS-THESEUS~\cite{2021arXiv210208701L, 2021ExA...tmp..137A}. 
The parameter space of low-mass PBH DM (for monochromatic mass distribution) with some current constraints and future projections from the AMEGO telescope is shown in Figure~\ref{fig:low-mass-PBH-DM}. The observation of the Milky Way Galactic Center by these telescopes can probe completely new parts of low-mass PBH DM parameter space~\cite{Coogan:2020tuf, Ray:2021mxu, Ghosh:2021gfa}. In large regions of this parameter space, low-mass PBHs can constitute 100\% of the DM density.
Their gamma-ray signatures show the importance of upcoming gamma-ray telescopes to new physics discoveries. 

%% file: ExtragalacticBackground-Meyer.tex
 \noindent
 \chapterauthor[]{Manuel Meyer}\orcidlink{0000-0002-0738-7581}
 \\
 \begin{affils}
   \chapteraffil[]{Institute for Experimental Physics, University of Hamburg, Luruper Chaussee 149, 22671 Hamburg, Germany}
 \end{affils}

The diffuse electromagnetic radiation from ultra-violet to infrared wavelengths integrated from the epoch of reionizaton until today is collectively referred to as the extragalactic background light (EBL)~\cite{2018ApSpe..72..663H}. The main source of this background radiation field is the combined emission of stars and their reprocessed emission by dust in galaxies~\cite{2001ARA&A..39..249H}.
As such, measuring the EBL intensity enables us to infer cosmological parameters, e.g., the cosmic star formation history. However, direct measurements of the EBL are often impeded due to bright foreground sources, in particular zodiacal light. These foregrounds are one to two orders of magnitude brighter than the EBL itself. 
On the other hand, galaxy number counts provide a strict lower limit on the EBL intensity. Deep-field survey galaxy number counts are used to measure the integrated galaxy light (IGL) contribution to the EBL. However, this method misses low surface brightness galaxies below the detection threshold, as well as truly diffuse sources
~\cite{2021arXiv210212089D}. 
Direct EBL measurements and measurements of the IGL can differ by up to an order of magnitude~\cite{2019ConPh..60...23M}. 

Measurements of extragalactic sources at gamma-ray energies provide an independent probe of the EBL photon density. 
Already in the 1960s it was found that gamma-rays should be absorbed during their propagation to Earth due to electron-positron pair production on photon fields~\cite{Nikishov1962,1967PhRv..155.1408G}.
As it turns out, the EBL is particularly important for this interaction since the pair-production cross section peaks for background photons with a wavelength $\lambda \approx 1.2~\mu\mathrm{m}(E / \mathrm{TeV})$, where $E$ is the gamma-ray energy~\cite{1967PhRv..155.1404G}. The absorption leads to an exponential dimming of the emitted gamma-ray flux, $F_\mathrm{observed} = \exp(-\tau_{\gamma\gamma})F_\mathrm{emitted}$, which scales with the optical depth $\tau_{\gamma\gamma}$. The optical depth is given by a line-of-sight integral to the gamma-ray source over the inverse of the mean free path for gamma-rays to pair production. The mean free path decreases with increasing gamma-ray energy and increasing EBL photon density. 

The gamma-ray absorption on the EBL has been unambiguously measured in the gamma-ray spectra of blazars\footnote{Blazars are active galactic nuclei with their relativistic jets pointed close to the line-of-sight to the observer.} obtained with the {\it Fermi}-LAT~\cite{2012Sci...338.1190A,2018Sci...362.1031F} as well as with imaging air Cherenkov telescopes (IACTs)~\cite[Section \ref{sec:atmospheric_cherenkov};][]{2013A&A...550A...4H,2015ApJ...812...60B,2017A&A...606A..59H,2019ApJ...885..150A,2019MNRAS.486.4233A} and the combination of the two types of instruments\cite{2019ApJ...874L...7D}. These achievements mark the beginning of the field of gamma-ray cosmology, recently reviewed in Ref.~\citenum{BiteauMeyer2022}. The IACT measurements, which probe higher energies than the LAT and therefore a regime of stronger absorption, are now able to measure the EBL within 10--20\% (20--30\%) statistical (systematic) uncertainty, although a wavelength-resolved measurement of the EBL comes at lower accuracy. The measurements are broadly consistent with an EBL solely given by IGL. Unfortunately, they do not reach the sensitivity to probe the mild tension recently reported by direct EBL measurements obtained with the New Horizons mission beyond Pluto's orbit~\cite{2021ApJ...906...77L} at a wavelength of $0.6~\mu\mathrm{m}$. The New Horizon observations yield an EBL level roughly a factor of 2 higher than the latest IGL measurements extracted from deep-field surveys~\cite{2021MNRAS.503.2033K}.
The two measurements are inconsistent at the $\sim2~\sigma$ level. 

With the large available data sets from {\it Fermi}-LAT and IACTs it now becomes possible to constrain cosmological parameters that enter the calculation of the EBL intensity and the optical depth. 
For instance, a first indirect measurement of the cosmic star formation rate has been obtained from {\it Fermi}-LAT data~\cite{2018Sci...362.1031F}. Recently, the first independent estimates of the Hubble constant $H_0$ through gamma-ray absorption have also been performed with a reported accuracy of  3--6~km~s$^{-1}$~Mpc$^{-1}$~\cite{2019ApJ...885..137D}. 

Since the gamma-ray absorption depends on the integrated EBL photon density, it is sensitive to all kinds of contributions to the EBL. 
Gamma-ray observations have therefore been used to constrain models of the earliest massive (population III) stars~\cite{2012MNRAS.420..800G} as well as searching for physics beyond the standard model (PBSM) like  dark-matter powered stars~\cite{2012ApJ...745..166M} or the decay of dark matter in the form of hypothetical axions into optical and infrared photons~\cite{2012JCAP...02..032C,2019PhRvD..99b3002K,2020JCAP...03..064K,2020A&A...633A..74K}. Effects of PBSM could also include a modification of the gamma-ray opacity of the Universe. For example, instead of pair production, gamma-rays could oscillate into axions or axion-like particles in the presence of external magnetic fields~\cite{2007PhRvD..76b3001M, 2007PhRvL..99w1102H}. This oscillation could lead to energy-dependent distortions and a reduced absorption in gamma-ray spectra. Claims for such a reduced opacity interpreted as evidence for axion-like particles~\cite{2007PhRvD..76l1301D,2009PhRvD..79l3511S,2011PhRvD..84j5030D,2012JCAP...02..033H,2011JCAP...11..020D,2014JETPL.100..355R,2017PhRvD..96e1701K,2020MNRAS.493.1553G} remain debated in the literature~\cite{2015ApJ...812...60B,2015ApJ...813L..34D}. 
Alternatively, a violation of Lorentz invariance (Section \ref{sec:speedofgravity}) could lead to a modified dispersion relation of photons which leads to a modification of the threshold energy for pair production~\cite{1998Natur.393..763A,2009NJPh...11h5003S,2009ApJ...691L..91S}. This modification could result in a complete suppression of pair production above a certain gamma-ray energy. 
The absence of this suppression has been used to place constraints on the energy scale at which Lorentz invariance is broken~\cite{2015ApJ...812...60B,2019ApJ...870...93A,2019PhRvD..99d3015L}. 

The study of the EBL and its evolution, together with the more general question about which effects influence gamma-ray propagation, will greatly benefit from continued observations with {\it Fermi}-LAT as well as with future IACTs (Section \ref{sec:atmospheric_cherenkov}). In particular, the future Cherenkov Telescope Array (CTA), with its high point-source sensitivity as well as high angular and spectral resolution~\cite{2019scta.book.....C}, promises a sensitivity for the EBL absorption feature two to three times better than that of current IACTs~\cite{2021JCAP...02..048A}. It is also expected that CTA will detect the absorption in spectra of blazars up to redshift of $z\approx 2$. In conjuction with optical observations with the James Webb Space Telescope and the \textit{Euclid} Satellite, CTA observations will probe and possibly resolve the above-mentioned tension at optical wavelengths. Together with water Cherenkov detectors (Section \ref{sec:water_cherenkov}) such as LHAASO, observation of nearby blazars beyond tens of TeV will lead to measurent of the EBL at far infrared wavelenghts with unprecedented accuracy. Furthermore, gamma-ray bursts (GRBs) have been recently detected with IACTs for the first time~\cite{2019Natur.575..455M,2021Sci...372.1081H,2021ApJ...908...90A}. This raises the hopes that CTA observations of GRBs can further be used to test gamma-ray absorption~\cite{2013ExA....35..413G}. 
Observations with CTA will also probe so-far unexplored regions axion-like-particle parameter space (the mass and coupling to photons)~\cite{2021JCAP...02..048A,2014JCAP...12..016M,2017JCAP...01..024K} and to the energy scale of a possible Lorentz Invariance Violation~\cite{2021JCAP...02..048A}. Similar values of this energy scale will also be explored with LHAASO observations~\cite{2019arXiv190502773B}.

%% file: LargeScale-Michela.tex
  \chapterauthor[1,2,3]{Michela Negro}\orcidlink{0000-0002-6548-5622}
 \\
 \begin{affils}
    \chapteraffil[1]{University of Maryland, Baltimore County, Baltimore, MD 21250, USA}
    \chapteraffil[2]{Astroparticle Physics Laboratory, NASA Goddard Space Flight Center, Greenbelt, MD, USA}
    \chapteraffil[3]{Center for Research and Exploration in Space Science and Technology, NASA/GSFC, Greenbelt, MD 20771, USA}
 \end{affils}

The matter in our universe appears to be distributed along large scale structures (LSS), a network of nodes  and filaments, often called \textit{the cosmic web}. These structures can be  reproduced theoretically by admitting some level of inhomogeneity in the matter density field at the early stages of the universe. Tiny primordial fluctuations produced gravitational instabilities that evolved in time and space: under- and over-densities of matter eventually turned into cosmic voids and structures, from galaxies to galaxy clusters and filaments.  

Detailed N-body simulations (see, e.g., The Millennium Simulation Project; \cite{MillenniumProject}) could reproduce the LSS of the universe by assuming that most of the mass in the universe ($\sim$85\%) consists of cold dark matter, while the baryonic matter---the ordinary matter---is only a small fraction ($\sim$4\%) of the total mass-energy content of the universe. Hence, the distributions of LSS tracers, such as galaxies, galaxy clusters or weak lensing maps, directly outline the dark matter distribution, and the distributions at different redshifts are the snapshots of different cosmological epochs. 

A wide range of possible dark matter (DM) candidates exists in literature (e.g.\cite{Taoso_2008} for review). One of the most investigated family of candidates are the Weakly Interacting Massive Particles (WIMPs), characterized by weak-scale interactions and a mass of the order of the GeV--TeV. A generic prediction of WIMP candidates is that they can either annihilate or decay into Standard Model particles, including gamma-rays. The specific mechanisms behind the production of gamma-rays (whether direct production, neutral pion decay, secondary bremsstrahlung or inverse Compton from primarily produced leptons, or other) depend on the DM candidate considered. Despite the extensive campaigns looking for sources of annihilating/decaying DM from known astrophysical gamma-ray emitters, only upper limits have been set so far: it is  reasonable to admit that any DM signature in the gamma-ray regime is unresolved (below the detection threshold of current instruments.  As future missions become sensitive to the regime currently unresolved by {\it Fermi}-LAT it will open new scenarios for DM searches. (See also Section \ref{sec:darkmatter} for additional discussion of DM.)

\begin{figure}[htb]
	\centering
	\includegraphics[height=0.48\textwidth]{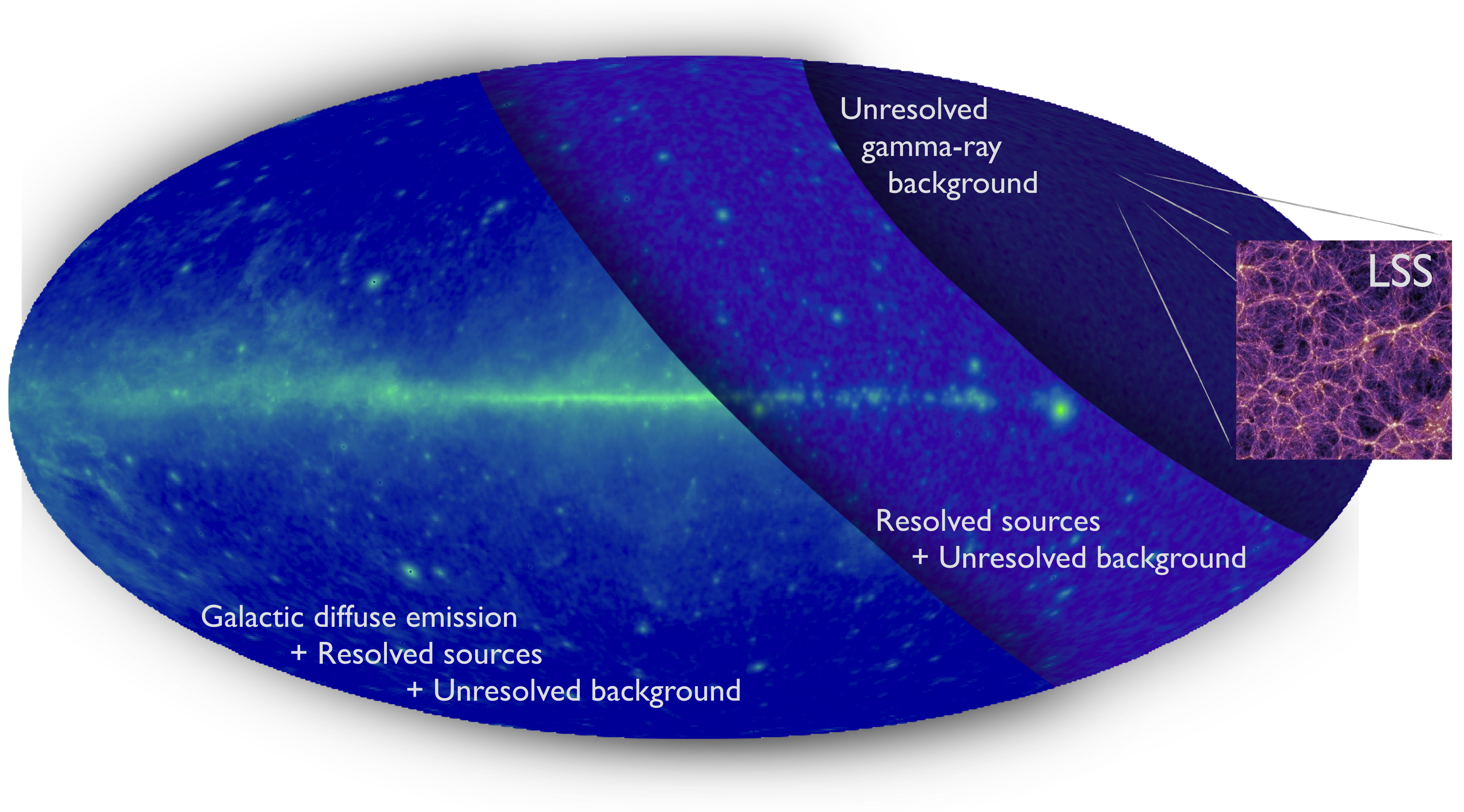}
	\caption{The gamma-ray sky has been revealed in the last decade by the {\it Fermi}-LAT, offering an outstanding picture of our Universe at the highest energies. A great part of this gamma-ray emission has been resolved, studied and attributed to known astrophysical objects and processes. A non negligible fraction (about 20\%) of the total gamma-ray emission is the unresolved gamma-ray background (UGRB), an almost isotropic emission extragalactic in origin. Cross-correlation studies between the UGRB and tracers of the large scale structures of the universe can be carried out to investigate the link between the dark matter distribution and the UGRB anisotropy field. Significant steps forward for these studies would require a probe-scale gamma-ray mission to substitute/complement the {\it Fermi}-LAT with better sensitivity and, most importantly, better angular resolutions.}
	\label{fig:LargeScale-Michela}
\end{figure}

The unresolved gamma-ray background (UGRB) is defined as the smooth residual component (about the 20\% of the total gamma-ray emission as observed by the \textit{Fermi}-LAT), which remains after subtracting all the known sources of gamma-rays (Galactic diffuse emission, point-like and extended detected sources, see Figure \ref{fig:LargeScale-Michela}). For more details about the UGRB see Sec. 2.6.1 of the Snowmass2021 white paper: \textit{Advancing the Landscape of Multimessenger Science in the Next Decade}. Being mostly extragalactic in origin, the objects contributing to the UGRB (whether astrophysical sources or DM halos) trace the LSS of the Universe (at least up to a certain redshift which depends on the attenuation due to the extragalactic background light). Therefore, a certain level of cross-correlation is expected with any LSS tracer, such as galaxies~\citep{xia11, Cuoco:2017bpv, Ammazzalorso:2018evf}, galaxy cluster catalogs~\citep{Branchini:2016glc, 2017arXiv170809385L, 2017arXiv170900416L, 2018PASJ...70S..25M}, weak lensing from cosmic shear~\citep{Camera:2012cj, Camera:2014rja, Shirasaki:2014noa, AmmazzalorsoDES}, and lensing potential of the cosmic microwave background~\citep{Fornengo:2014cya}.

The typical formalism used to derive the cosmological gamma-ray flux produced by DM annihilation over all redshifts can be found, e.g., in the review article by~\cite{Fornasa:2015qua}, where the two main approaches to describe the statistical clustering of DM in the Universe (the Halo Model~\citep{1974ApJ...187..425P, 2002PhR...372....1C} and the Power Spectrum method~\cite{Serpico2012, Sefusatti2014}) are also discussed. Here, we briefly summarize the main results of some most recent cross-correlation analyses involving the UGRB and LSS tracers.

The UGRB is clearly detected on angular scales smaller than 1$^\circ$, in a cross correlation study considering several galaxy catalogs, with significance varying depending on the statistics of the catalog \cite{Cuoco:2017bpv}. 
Interestingly mild evidence is also found of redshift evolution of the cross-correlation signal, which can be interpreted as a change over redshift in the spectral and clustering behavior of the gamma-ray sources contributing to the UGRB. Similar results are obtained by focusing on the local universe (z$<$0.2) exploiting the 2MASS Photometric Redshift catalog (2MPZ; \cite{Ammazzalorso:2018evf}). In this latter study, hints of DM particle annihilation signal was found leading to interesting 95\% CL upper bounds on the annihilation rate vs. DM mass.

There is a positive correlation between the UGRB and galaxy clusters, with different amplitudes depending on the average mass of the objects in each catalog, and rather large angular scales extending to a few to tens of megaparsecs \cite{Branchini:2016glc}. 
 It is likely that the  signal is coming from the cumulative emission of active galactic nuclei associated with the filamentary structures tracing the high-density peaks of the matter field. Another intriguing scenario is that the cross-correlation signal (or a fraction of it) is due to a diffuse gamma-ray emission from the intra-cluster medium (interpretable as signature of annihilating DM particle). However, at the moment of the study, it was not possible to distinguish between the two interpretations. 

Recently, a $>4\sigma$ cross-correlation signal was detected between the UGRB and the cosmic shear from weak lensing as measured by the Dark Energy Survey (DES) \cite{AmmazzalorsoDES}.
Such a correlation was predicted as a novel and relevant channel of DM investigation \cite{Camera:2014rja}.
The signal is mostly localised at small angular scales and high gamma-ray energies (above 5 GeV), so it also likely originates from unresolved blazar emission, which correlates with the mass distribution \cite{AmmazzalorsoDES}. 
However, a hint of correlation at extended separation was also observed and investigated both in terms of astrophysical sources and particle dark matter emission: the inclusion of an annihilating DM component seems to improve the modelling of the measured cross-correlation signal. 
The DM particle parameters are at tension with the bounds provided by dwarf spheroidal galaxies   (e.g. \cite{AmmazzalorsoDES, PhysRevLett.115.231301, 2017ApJ...834..110A}), although uncertainties in the J-factor modeling can weaken the bound of the dSphs by a sizeable factor (e.g. \cite{Calore2019}). 


In conclusion, the gamma-ray background plays a key role in the study of the evolution and content of the large scale structures of the universe. The UGRB measurement evolves with time as more and more gamma-ray emission becomes resolved, revealing fainter (and possibly more exotic) contributors. It is crucial therefore to keep observing the gamma-ray sky. Improved angular resolution is extremely important to study spatial correlation.  Significant steps forward for these studies require a probe-scale gamma-ray mission to complement and eventually replace \textit{Fermi}-LAT with improved sensitivity and, most importantly, better angular resolution. With a prosperous future of big surveys in front of us ( including the James Webb Space Telescope, Vera Rubin Observatory, Euclid and the Nancy Roman Space Telescope), the need for real-time gamma-ray observations and updated measurements acquires a renewed relevance which must not be neglected. The \textit{Fermi}-LAT, whose design and realization took more than a decade, is now approaching its 15$^{th}$ year of operation and, even if there is no sign of malfunction, it is unreasonable to believe that it will still be so in ten years from now: we need to start thinking about new designs and technologies for a future high-energy gamma-ray survey mission.

%% file: NonImagingDetectors-Isreal.tex
\noindent
\chapterauthor[1,2,3]{Israel Martinez-Castellanos}\orcidlink{0000-0002-2471-8696}
\\
 \begin{affils}
   \chapteraffil[1]{University of Maryland, College Park, MD, USA}
   \chapteraffil[2]{Astroparticle Physics Laboratory, NASA Goddard Space Flight Center, Greenbelt, MD, USA}
   \chapteraffil[3]{Center for Research and Exploration in Space Science and Technology, NASA/GSFC, Greenbelt, MD, USA}
\end{affils}

Gamma-ray detectors typically have the ability to detect individual particles. We can potentially measure their energy, incoming direction, and polarization. Although there are ongoing efforts to develop focusing gamma-ray detectors~\cite{10.1117/12.2308289}, current instruments infer the photon direction by tracking the secondary particles produced by the interaction of the primary gamma ray with either the detector or a naturally occurring external medium, e.g., the atmosphere. 

Some instruments, however, do not have this ability, and are called “non-imaging” detectors\footnote{On this section we refer as “non-imaging detectors” strictly to those which cannot reconstruct the direction of the primary gamma ray and do not include in this definition Extensive Air Shower Arrays (see Sections~\ref{sec:water_cherenkov} and \ref{sec:atmospheric_cherenkov}), which can determine the direction of the primary particle but that are sometimes titled “non-imaging” as they sample the air shower at ground level rather than imaging its evolution throughout the atmosphere like Imaging Air Cherenkov Telescopes (IACTs) do (see Section~\ref{sec:atmospheric_cherenkov}).}. They are usually space-based missions and are sensitive to the low end of the gamma-ray spectrum--- below a few MeV to tens of MeV. In these kinds of instruments, the interaction of a gamma ray generally results in a single energy deposit in one of the active elements of the detector that has no tracking capabilities, such as a scintillator. They work in a regime where the efficiency of photoelectric absorption is sufficiently high compared to Compton scattering, such that the energy that escapes the detector is minimized. 

Although this class of detectors is not able to map the sky, they excel in the search for transient events, such as gamma-ray bursts. They have a wide field of view, usually only limited by the portion of the sky occulted by the Earth, and the short-duration nature of the signal counteracts the large background counts caused by their inability to do imaging. Additionally, although they are not able to reconstruct the direction of individual photons, they use the aggregate signal to localize the source that produced it. These characteristics allow them to alert other facilities and guide their observation, making them a crucial component of the multi-wavelength and multi-messenger astronomy effort. 

The position in the sky of a transient source can be estimated from the relative number of counts registered by each active element. These are matched to the expected signal---from simulations or calibration---given a hypothetical source at a given location. The number of counts vary as a function of the direction of the source due to the intrinsic effective area of each active element depending on their line of sight. For example, a relatively flat detector will have an effective area that is approximately proportional to the cosine of the angle with respect to its normal vector. By having multiple such detectors pointing towards different directions, we can deduce the location of the source. The attenuation caused by other active or passive elements also contributes to the change of expected counts as a function of the incoming direction. The resulting shadowing is an additional source of information on the source location.

Non-imaging techniques are very flexible, and are not used exclusively by instruments designed for this purpose. Frequently, some detector elements, such as anti-coincidence detectors, can be repurposed to perform detection and localization of transients. This allows them to increase their field of view, extend their energy range, and use it in combination with imaging techniques to improve their sensitivity.

An interesting possibility is to perform a joint localization utilizing the data from multiple non-imaging detectors. Doing a coherent analysis can decrease the localization uncertainty and even allow the inclusion off information from instruments that do not have localization capabilities by themselves--- e.g., detectors with a single active element. One possibility is to consider all detectors as different components of a single instrument and use the relative number of counts of each detector. Alternatively, if the baseline between instruments is sufficiently long, the source location can be triangulated using timing information. This is regularly used by the Interplanetary Gamma-Ray Burst Timing Network (IPN)~\cite{refId0} and dedicated missions for this purpose are being considered~\cite{pal2020grbalpha}. 

Advances in technology (e.g., Silicon Photomultipliers, see Section~\ref{sec:rad_tol_sipms}) have made the development of SmallSats and CubeSats cheaper, faster, and more accessible---  see some examples in Section~\ref{sec:mev_telescopes}. Within the next decade, multiple such detectors will be operational, which calls for their respective teams to collaborate and analyze their data jointly in order to maximize the science returns.

%% file: ComptonTelescopes-Reshmi.tex
 \noindent
    \chapterauthor[1]{Reshmi Mukherjee}
    \chapterauthor[2]{\& Thomas Shutt}
 \\
 \begin{affils}
    \chapteraffil[1]{Columbia University, New York, NY 10027, USA}
    \chapteraffil[2]{Stanford University, Stanford, CA 94305, USA}
 \end{affils}

In the energy regime between photoabsorption and pair production (approximately in the band 250~keV to 6~MeV), gamma rays primarily interact with matter via the Compton scattering process. The most promising approach for observations of  gamma rays in the low and medium energy range (up to about 30~MeV) is by using ``Compton Telescopes"--- instruments sensitive to measuring the interactions of photons with the detector medium via the Compton scattering process.

\begin{figure} [tb]
\centering
\includegraphics[height=0.43\textwidth]
 {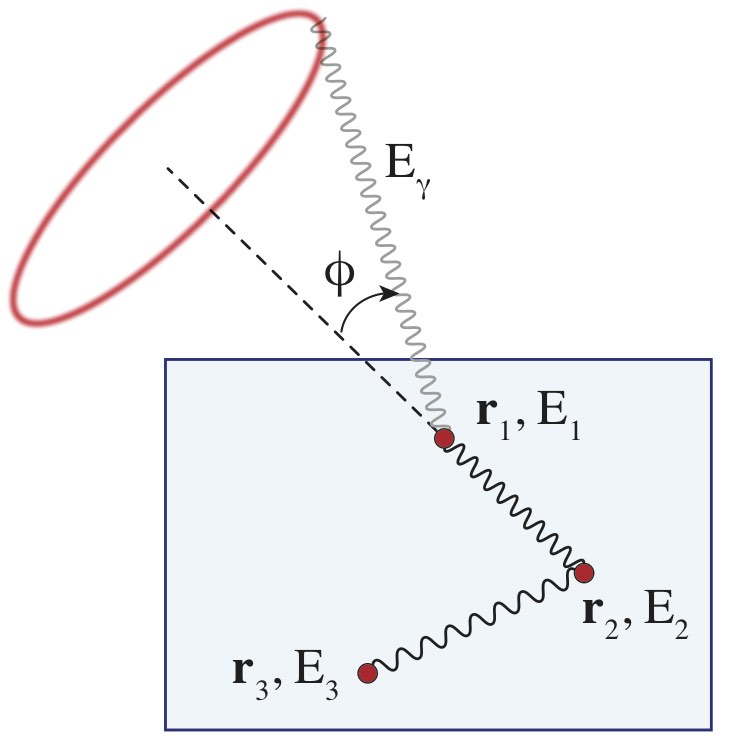}
\includegraphics[height=0.43\textwidth]
 {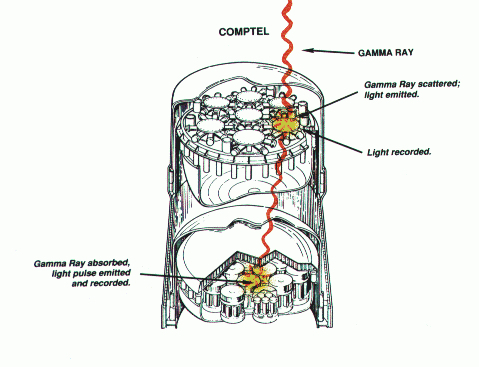}
 \vskip -0.1in
  \caption{ \footnotesize  
Left: Schematic of a Compton scattering event demonstrating the principle of operation of a Compton Telescope. The red ``event circle'' corresponds to the annulus on the sky that constrains the origin of the incident photon, based on the reconstruction of the first interaction in the detector volume shown in blue. Figure from Ref.~\citenum{kieransthesis}. Right: Image of the Imaging Compton Telescope (COMPTEL) on the Compton Gamma-Ray Observatory (CGRO), the first successful Compton Telescope to make astrophysical measurements and provide an all-sky survey in the energy range 0.75 to 30~MeV.~\cite{2000A&AS..143..145S}. }
\label{fig:schematic_compton}
\end{figure}

Figure~\ref{fig:schematic_compton}~(left) shows the principle of operation of a Compton telescope. The figure shows the incident gamma ray of energy $E_\gamma$ being identified by three successive interactions in the active volume. In the first interaction---a Compton scattering event---the primary photon is scattered by an angle $\phi$, having transferred an energy $E_1$ to an electron. The secondary photon subsequently scatters further and may be identified in a second or third interaction. From the measured values of the deposited energies and the positions of the interactions, the original direction of the photon may be inferred. With a single Compton scattering event, the direction of the incoming photon may be constrained within an annulus on the sky by determining the relative order of scatters in addition to the deposited energies and the interaction positions. With three or more Compton scattering events, the intersection of the event circles provides the direction of the gamma-ray source with more precision. The energy and cone angle in a Compton telescope scattering event (for example, for two scatters, followed by photo-absorption) may be calculated as,

$$E_\gamma = E_1 + E_2 + E_3 ,$$

$$\cos\phi = 1 - m_ec^2 \biggl({{1\over{E_2+E_3}} + {1\over{E_1+E_2+E_3}}}\biggr)\enspace ,$$
where $m_e$ is the electron mass, $E_\gamma$ is the energy of the incident photon, $E_1$, $E_2$, and $E_3$ are the deposited energies by the Compton scatterings and photoabsorption, and $\phi$ is the first Compton scattering angle. More than three Compton scattering events may also be reconstructed based on the Compton equations~\cite{KAMAE1987254,DOGAN1990501}. 

A sensitive detector of Compton-scattering gamma rays must have excellent spatial and energy resolution, as well as minimal dead material. Better spatial and energy resolution directly lead to a higher resolution map (i.e., better point-spread function), and, in combination with minimal dead material, enhance the efficiency of event reconstruction. The detector must be sufficiently large to contain the gamma rays while at the same time having fine-grained spatial readout. In addition, if the initial direction of the electron recoil in the first scatter is measured, the cone of possible incoming photon directions is reduced to an arc, enhancing the overall efficiency. The detector must also contend with a background of upward-going gamma rays from cosmic-ray interactions in the atmosphere, gamma rays from activation of detector materials, neutrons, and charged particles.  The rate of all these exceeds the rate of astrophysical gamma rays for all but the lowest energies.  

The first successful Compton telescope to carry out an all-sky survey in the MeV regime was the Imaging Compton Telescope (COMPTEL)~\cite{2000A&AS..143..145S}, flown on NASA's Compton Gamma-Ray Observatory, in operation from 1991 to 2000.  
Current examples of Compton telescopes include COSI (the Compton Spectrometer and Imager; see Section~\ref{sec:cosi})~\cite{2021arXiv210910403T}, recently selected by NASA to continue its development as a small explorer mission (SMEX)~\cite{cosinasa}. Liquid noble gases have been proposed as attractive detector medium for Compton telescopes. An early example is the LXeGRIT (Liquid Xenon Gamma-Ray Imaging Telescope)~\cite{10.1117/12.962588}. Recently, two detectors have been proposed as a Compton telescope using liquid argon (LAr) as the detector medium, GRAMS (the Gamma Ray and AntiMatter Survey; see Section~\ref{sec:grams})~\cite{2020APh...114..107A}, and GammaTPC (see Section~\ref{sec:gammatpc})~\cite{gammatpc_loi}.

%% file: CodedMask-Brad.tex
 \noindent
 \chapterauthor[1]{S. Bradley Cenko}\orcidlink{0000-0003-1673-970X}
 \chapterauthor[2,3]{\& A.A. Moiseev}
 \\
 \begin{affils}
   \chapteraffil[1]{Astroparticle Physics Laboratory, NASA Goddard Space Flight Center, Greenbelt, MD 20771, USA}
   \chapteraffil[2]{University of Maryland, College Park, MD 20742, USA}
   \chapteraffil[3]{NASA Goddard Space Flight Center, Greenbelt, MD 20771, USA}
 \end{affils}

\begin{figure}
  \centerline{\includegraphics[width=8cm]{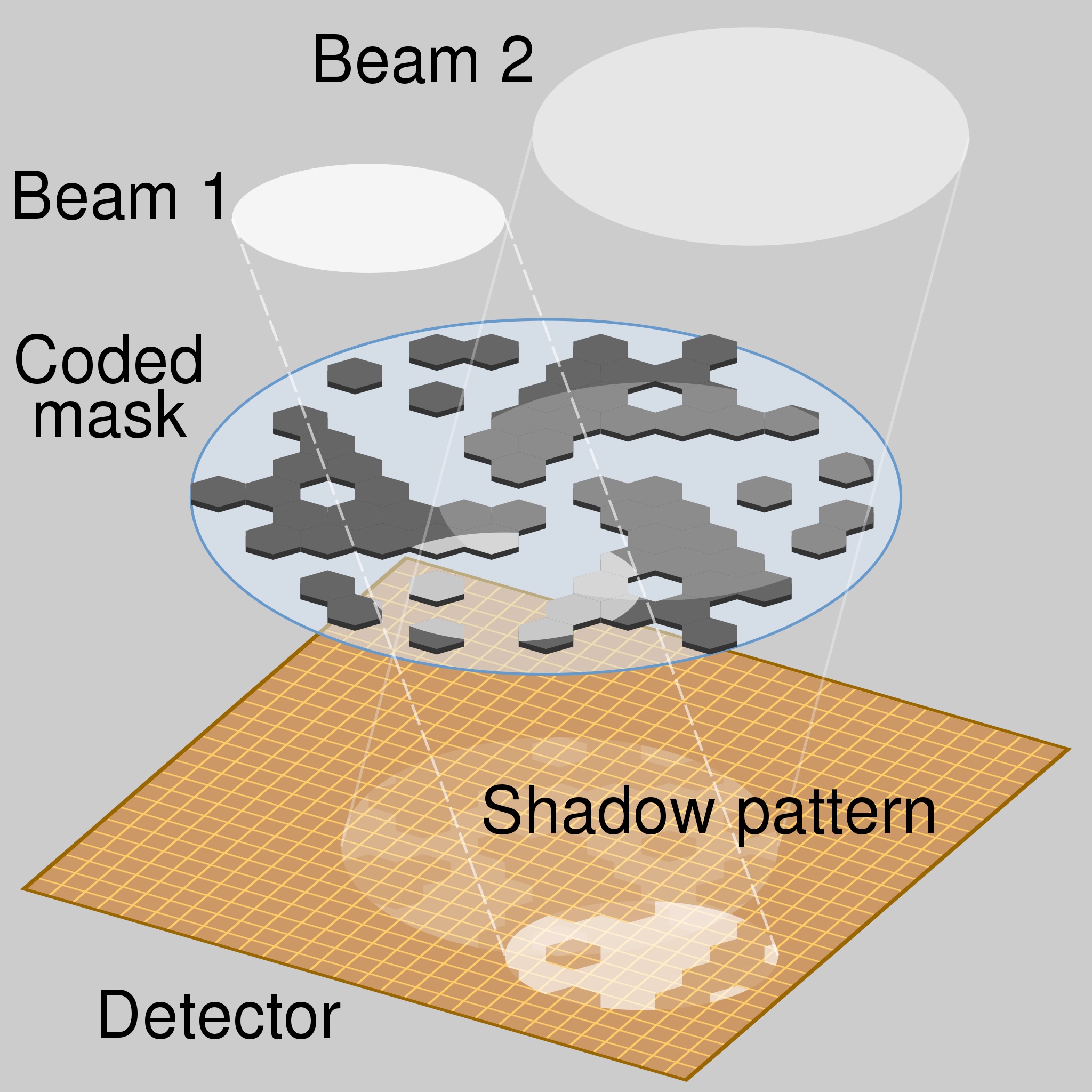}}
  \caption[]
  {A coded-mask telescope uses opaque material to generate ``shadow''
  patterns on the detector plane. This technique enables reconstruction
  of source position (i.e., imaging), often over a large 
  field of view, at wavelengths where normal incidence focusing is not
  possible. [\textit{Source:} Wikimedia Commons.]}
  \label{fig:codedmask}
\end{figure}


Utilization of focusing optics (grazing incidence plus multi-layer optics) provides excellent angular resolution of the order of arcsec in X-rays up to $\sim200$~keV, but is limited to a fraction of degree field of view (as seen in the experiments Chandra~\cite{Weisskopf:1999hm}, NuSTAR~\cite{2010arXiv1008.1362H}, etc.). Alternatively, a coded mask, or coded aperture, instrument employs a pattern of opaque material above the detector plane to generate a ``shadow'', thus enabling imaging at wavelengths where normal incidence optics (i.e., refractive lenses, reflective mirrors) are ineffective (Figure~\ref{fig:codedmask}). This spatial modulation of the incident flux by a pattern of opaque material (called coded-aperture mask) and deconvolution with its measured image in the focal-plane detector is an established method for imaging with fine angular resolution. Such systems have a long history of use in X-ray and gamma-ray astronomy (Table~\ref{tab:codedmask}).

\begin{table}[t]
\centering
\caption{Examples of Coded Mask Instruments}
\footnotesize
\begin{tabular}{|lcccc|}
 \hline
 Instrument & Bandpass & Field of View & Resolution & Lifetime \\
            &  (keV)   &               & (arcminutes) &          \\
 \hline
 \textit{RXTE}-ASM & 2--12 & $6^{\circ} \times 90^{\circ}$ & $3 \times 15$ & 1995--2012 \\
 \textit{INTEGRAL}-IBIS & 15--10,000 & $29^{\circ} \times 29^{\circ}$
 & $12$ & $>$2002 \\
 \textit{Swift}-BAT & 15--150 & $100^{\circ} \times 60^{\circ}$ & 17 & 
 $>$2004 \\
 \textit{AstroSat}-CZTI & 10--150 & $4.6^{\circ} \times 4.6^{\circ}$ &
 $8$ & $>$2015 \\
 \hline
\end{tabular}
\label{tab:codedmask}
\end{table}

As an illustrative example, consider the Burst Alert Telescope 
(BAT~\cite{2005SSRv..120..143B}) on-board the \textit{Neil Gehrels 
Swift Observatory} (\textit{Swift}~\cite{2004ApJ...611.1005G}). The BAT coded mask comprises $\sim$52,000 lead (Pb) tiles, each 5.0~mm square in size and 1.0~mm thick, with a 50\% filling factor. The tiles are oriented in a random pattern in the 2.4~m by 1.2~m D-shaped aperture, and located 1~m above the detector plane. As a result, the BAT instrument provides a field of view of 100$^{\circ}\times~$60$^{\circ}$ (1.4~sr) field of view (half-coded), with an angular resolution of 17~arcminutes. 

Coded mask instruments are typically used for applications where a large field of view and good angular resolution are required~\cite{Caroli, Skinner}, particularly at hard X-ray energies. However, they are limited to imaging at energies where the mask material is opaque--- for example with \textit{Swift}-BAT, the aperture becomes transparent above 150~keV as high-energy limit is constrained by the mask opaqueness range provided by the mask thickness. Additionally, while very large effective areas are achievable with coded mask instruments, the sensitivity is ultimately limited by background in the (physically large) detector plane--- thus, focusing telescopes typically achieve greater sensitivity (at the expense of field of view).

At higher energies, where the dominating photo-absorption interaction cross-section yields to the Compton cross-section, the provision of high angular resolution becomes a problem: Doppler broadening fundamentally limits angular resolution to $\sim$1~degree if Compton reconstruction is used. For MeV energies, Compton scattering is a dominant process of photon interaction with matter, and photon detection using the Compton effect is a well-established observation method. With the Doppler broadening limitation, the coded mask is the only feasible approach to provide arcmin angular resolution in the Compton-dominated energy range (300~keV--10~MeV). The IBIS telescope onboard the \textit{INTEGRAL} space telescope~\cite{IBIS, INTEGRAL} extended the coded mask energy range to 10~MeV by increasing the mask thickness to 16~mm of tungsten, which provides $\sim$70\% of the mask opaqueness (Compton interaction cross-section has a minimum at 4--6~MeV depending on the mask material). The IBIS angular resolution is 12’ with the mask-detector separation 3.7~m--- the fundamental angular resolution of the system is determined by the ratio of the mask pixel size to the distance from the mask to the focal plane detector. Although the angular resolution can, in theory, be made as fine as possible, there are constraints based on other performance requirements of the system. The system signal-to-noise ratio strongly depends on the ratio between the focal plane detector position resolution and the mask pixel size, with the optimal ratio being around 0.5. This driver for IBIS' design yielded its field of view of 8.3$^{\circ}\times~$8.0$^{\circ}$ and angular resolution of 12’.

Development of highly efficient focal plane detectors with fine position resolution the order of $\sim$mm would allow the creation of more compact coded aperture mask systems with higher angular resolution. A feasible and attractive option to improve the angular resolution is to increase the distance between the mask and the detector by deploying the coded mask after reaching orbit. However, in this configuration, the instrument aperture will be exposed to side-entering background radiation, usually protected by either active or passive shielding (e.g., imager IBIS~\cite{IBIS} and spectrometer SPI~\cite{SPI} on-board \textit{INTEGRAL}). Excessive side-entering background will significantly deteriorate the signal-to-noise ratio, and therefore also the instrument sensitivity.  This background can be effectively suppressed if the coded-aperture mask telescope is combined with a Compton telescope (see Section~\ref{sec:compton_tel}), which simultaneously serves as the focal plane detector by selecting for analysis-only events that might have originated from the coded mask location according to their measured Compton-scattered directions.  As an example, if the coded mask is deployed at 20 meters, employing a similar mask as is used in NuSTAR~\cite{2010arXiv1008.1362H} and a focal plane detector based on the virtual Frisch grid drift CZT bar Imaging calorimeter (see Section~\ref{sec:VFG-CZT}) with its $\sim$1~mm 3D position resolution~\cite{bolNIM}, the instrument angular resolution will be $\sim$0.5' with a $\sim$5$^{\circ}\times\,$5$^{\circ}$ fully coded field of view. The combined Compton/Coded mask approach has been demonstrated in simulations~\cite{Aprile, Galloway} and tested with INTEGRAL/IBIS data~\cite{Forot}, but the mature concept has never been implemented as the central motivation for a telescope design. Presently this approach is being proposed for Galactic Explorer with a Coded-Aperture Mask Compton Telescope~\cite{gecco1} (GECCO; see Section~\ref{sec:gecco}).

%% file: PairCreation-HenrikeEric.tex
\noindent
\chapterauthor[1,2,3,]{Henrike Fleischhack}\asteriskfootnote{H.F. acknowledges support by NASA under award number 80GSFC21M0002. Any opinions, findings, and conclusions or recommendations expressed in this material are those of the author(s) and do not necessarily reflect the views of the National Aeronautics and Space Administration.}\orcidlink{0000-0002-0794-8780}
\chapterauthor[4]{\& Eric Charles}\orcidlink{0000-0002-3925-7802}
\\
 \begin{affils}
   \chapteraffil[1]{Catholic University of America, Washington, D.C. 20064, USA}
   \chapteraffil[2]{Astroparticle Physics Laboratory, NASA Goddard Space Flight Center, Greenbelt, MD 20771, USA}
   \chapteraffil[3]{Center for Research and Exploration in Space Science and Technology, NASA/GSFC, Greenbelt, MD 20771, USA}
   \chapteraffil[4]{SLAC National Accelerator Laboratory, Menlo Park, CA 94025, USA}
\end{affils}

\begin{figure}[htb]
\begin{minipage}[c]{1.0\textwidth}
\includegraphics[width=1.0\linewidth]{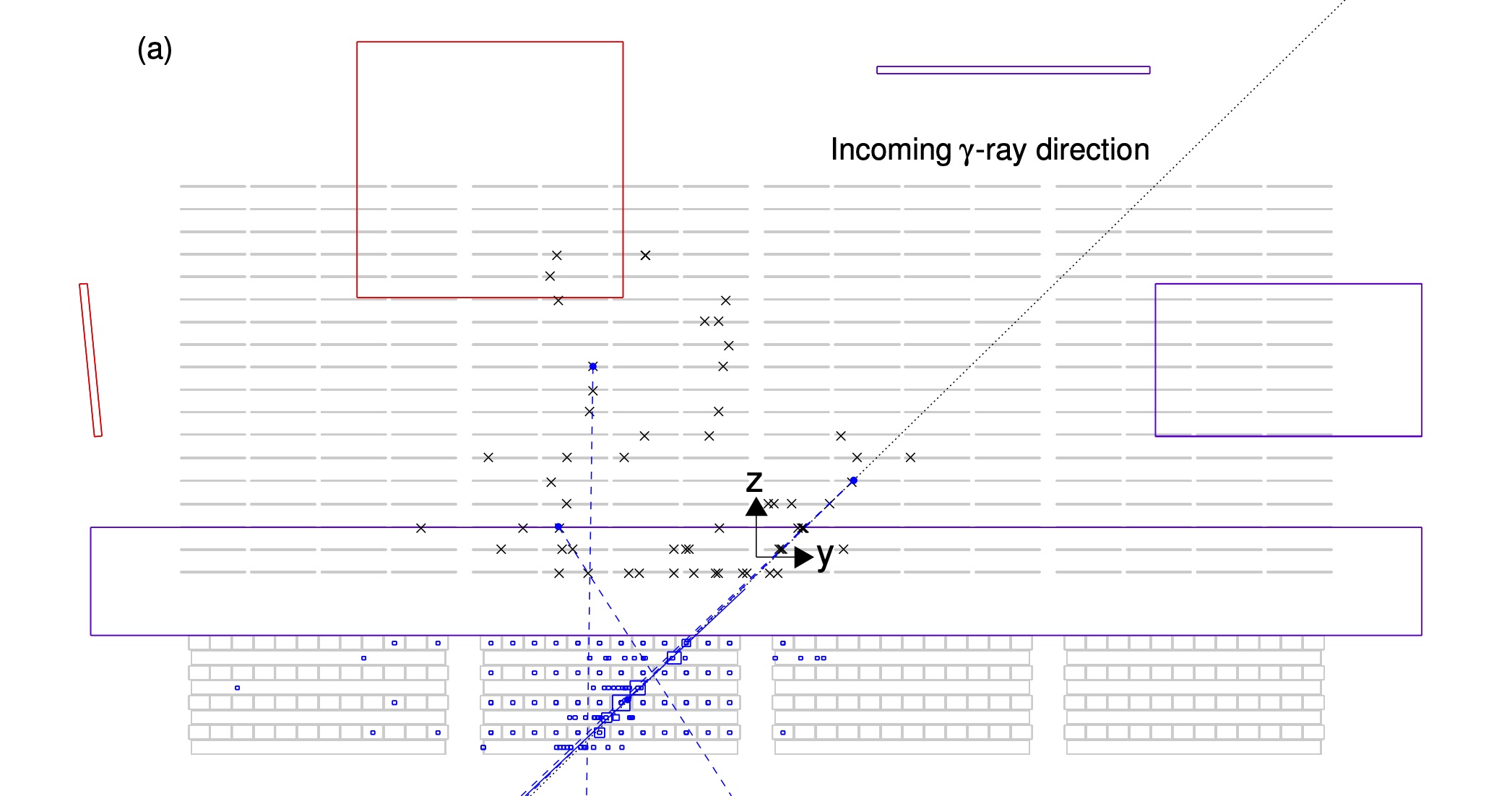}
\end{minipage}
 \caption{A simulated {\it Fermi}-LAT Event. The simulated gamma-ray had 27~GeV of energy.
The small crosses represent the clusters in the silicon tracker, while the variable-size squares indicate the reconstructed location and magnitude of the energy deposition for every hit crystal in the calorimeter. The dotted line represents the true gamma-ray direction. The ``backsplash'' from the calorimeter shower generates tens of hits in the tracker and a few hits in the anti-coincidence detector (colored boxes) which, however, are away from the direction extrapolation and therefore do not compromise our ability to correctly classify the event as a gamma ray. This figure appeared as Figure~12(a) of Ref.~\citenum{2012ApJS..203....4A}
.
}
\label{fig:fermi_event}

\end{figure}

Gamma-ray photons with energies above some tens of GeV predominantly interact with matter through pair production, meaning the original photon is destroyed and an electron-positron pair is produced instead. Pair creation telescopes are space-based (typically satellite-borne) instruments that detect astrophysical gamma rays in this way. Such detectors typically comprise the following components:
\begin{itemize}
 \setlength\itemsep{0mm}
    \item A tracking detector to determine the directions of the secondary electron and positron,
    \item a segmented calorimeter for the electron and positron to deposit (part of) their energy,
    \item a plastic scintillator or similar detector surrounding the rest of the instrument to veto signals produced by charged particles (cosmic rays) crossing the detector, and 
    \item (optionally) non-instrumented conversion layers made from material with a high atomic number, Z, to increase the interaction cross section for gamma rays passing through the detector.
\end{itemize}

The direction of the primary photon can be determined from the directions of the secondary electron and positron in the tracker, and its energy can be reconstructed from the energy deposited in the tracker and calorimeter. A segmented calorimeter enables the estimation of the amount of energy deposited in non-instrumented volumes or escaping the detector. 

Pair telescopes typically have large fields of view and are well suited for all-sky surveys, finding and characterizing new gamma-ray sources and transients. Over time, they accumulate large exposures on the entire gamma-ray sky, allowing the study and characterization of gamma-ray sources including their energy spectra and temporal variability. 

Pair creation telescopes suffer from both reducible (charged cosmic rays) and irreducible (gamma rays) background. Charged cosmic rays (electrons, protons, and heavier nuclei) produce both a prompt signal of ionization tracks and electromagnetic showers, as well as a delayed signal due to activation of the instrument material and ensuing radioactive decays. The prompt cosmic-ray background can be effectively suppressed at the trigger level by employing an anti-coincidence shield which detects charged particles entering the detector. Its flux varies with the solar cycle and the position of the spacecraft and typically increases sharply above the South Atlantic Anomaly (SAA). Spacecraft may choose to pause data taking in close proximity of the SAA. The gamma-ray background is caused by gamma rays from the Earth's atmosphere, the Sun, the (isotropic) extragalactic gamma-ray background, and the Galactic diffuse gamma-ray emission (tracing the cosmic-ray distribution, matter, and radiation fields in the Galaxy).

While the first gamma photons of cosmic origin in this energy regime were detected in 1961 by Explorer XI~\cite{PhysRevLett.8.106}, the pair telescope technique was pioneered by NASA's SAS-2~\cite{1975ApJ...198..163F} in 1972/73, with a 32-level wire spark chamber providing the directions of the electron/positron pairs. Currently operating high-energy gamma-ray pair production telescopes are NASA's \textit{Fermi}-LAT~\cite{lat}, the Chinese mission DAMPE~\cite{CHANG20176}, and the Italian-led \textit{AGILE}-GRID~\cite{2019A&A...627A..13B}.

%% file: WaterCherenkov-Henrike.tex
\chapterauthor[ ]{ }

 \addtocontents{toc}{
     \leftskip3cm
    \scshape\small
    \parbox{5in}{\raggedleft Henrike Fleischhack, J. Patrick Harding, et al. }
    \upshape\normalsize
    \string\par
    \raggedright
    \vskip -0.19in
    }

\noindent
\nocontentsline\chapterauthor[]{Henrike Fleischhack$^{1,2,3,}$}\asteriskfootnote{H.F. acknowledges support by NASA under award number 80GSFC21M0002. Any opinions, findings, and conclusions or recommendations expressed in this material are those of the author(s) and do not necessarily reflect the views of the National Aeronautics and Space Administration.}\orcidlink{0000-0002-0794-8780}
\nocontentsline\chapterauthor[]{J. Patrick Harding$^{4,5}$}\orcidlink{0000-0001-9844-2648}
\nocontentsline\chapterauthor[]{Jordan Goodman$^6$}\orcidlink{0000-0002-9790-1299}
\nocontentsline\chapterauthor[]{Marcos Santander$^7$}\orcidlink{0000-0001-7297-8217}
\\
 \begin{affils}
   \chapteraffil[1]{Catholic University of America, Washington, D.C. 20064, USA}
   \chapteraffil[2]{Astroparticle Physics Laboratory, NASA Goddard Space Flight Center, Greenbelt, MD 20771, USA}
   \chapteraffil[3]{Center for Research and Exploration in Space Science and Technology, NASA/GSFC, Greenbelt, MD 20771, USA}
   \chapteraffil[4]{Physics Division, Los Alamos National Laboratory, Los Alamos, NM 87545, USA}
   \chapteraffil[5]{Michigan State University, East Lansing, MI, 48824, USA}
   \chapteraffil[6]{Department of Physics, University of Maryland, College Park, MD 20742, USA}
   \chapteraffil[7]{Department of Physics and Astronomy, University of Alabama, Tuscaloosa, AL 35487, USA}
\end{affils}

When a gamma-ray photon with an energy of tens of GeV or more impinges upon the atmosphere, it will produce an electron-positron pair within the electric field of an atmospheric nucleus. The electron and positron in turn will emit more gamma-ray photons via bremsstrahlung, which can produce more electron-positron pairs and so on. This cascade of electrons, positrons, and photons is referred to as an extensive air shower. The particles propagate through the atmosphere, with the number of particles approximately doubling after each radiation length (about 37~g/cm$^2$). The shower dissipates when the energy of the electrons and positrons falls below the critical energy, $E_c \approx 70$~MeV, where energy loss via ionization is favored over bremsstrahlung. Charged cosmic rays (electrons, protons, and nuclei) also induce atmospheric air showers. For hadronic particles, the pair-production and bremsstrahlung processes are replaced by strong interactions with nucleons in atmospheric nuclei, typically producing a handful of pions and other mesons~\cite{gaisser_engel_resconi_2016}.

The extended nature of air showers, where the particle shower front can cover areas of $\mathcal{O}(10^{5}~\mathrm{m}^2)$, poses a significant challenge to their detection in a cost-effective manner. The particle detection technology has to therefore be relatively inexpensive per unit surface area and have a high efficiency. The two main approaches that satisfy both requirements are water Cherenkov and scintillator detectors, which constitute most of the air shower arrays currently in operation or under planning. We describe both techniques in the subsections below. These detectors are built at high altitudes (typically above 4,000 m above sea level, see Figure~\ref{fig:ground_vhe_gamma}), as the shower particles are detected directly before they completely attenuate in the atmosphere.

\subsection{Water Cherenkov Detectors}\label{sec:water_cherenkov}

The water Cherenkov method, suitable for both gamma-ray astronomy and cosmic-ray physics, detects the charged component of air showers via Cherenkov light emitted in water. Water Cherenkov Detectors (WCDs) typically comprise one or multiple covered pools filled with several meters of water and instrumented with photo-multiplier tubes (PMTs) or other sensitive optical sensors. The PMTs may be separated by curtains or other optical barriers to avoid light leakage across the pool. Many smaller, discrete water tanks may be used instead of larger pools to ensure optical separation and allow for modular construction and commissioning while reducing the risk of catastrophic water loss. 

Event reconstruction in WCDs uses both the timing and the magnitude of the Cherenkov light signals. The arrival direction of the primary gamma-ray photon is determined by the timing gradient of the shower front detection across the PMT array. The energy is determined by the total signal (i.e. the amount of light) observed across the array. The spatial distribution of signal magnitudes across the array can be used to distinguish gamma-ray-induced air showers from  those induced by cosmic rays (the primary source of backgrounds for gamma-ray searches). Gamma-ray-induced showers tend to have a radially symmetric, smooth profile with a bright core and fainter tails, while hadronic air showers tend to be less regular in nature. The rejection power for cosmic-ray showers can also be improved by constructing vertically-segmented WCDs (or by burying some of them) to enable the identification of deeply penetrating muons, which are abundant in hadronic showers and scarce in gamma-ray ones.

Water Cherenkov telescopes are stationary and observe the sky as it passes overhead. The low-end energy threshold is typically hundreds of GeV, though this depends on the altitude of the detector above sea level. Detectors at lower altitudes above sea level have higher energy thresholds because there is more atmosphere above them in which the electrons can reach the critical ionization energy. Similarly, particles from a gamma-ray shower that has a large zenith angle are less likely to reach the WCDs than those observed directly overhead as they have to traverse less atmosphere. The WCD technique is capable of reaching gamma-ray energies greater than 1~PeV.

Because the water pools or tanks are enclosed by light-tight covers, WCDs can operate continuously, day and night, under most weather conditions. Their performance is not affected by clouds, rain, or moonlight. They tend to have duty cycles close to 100\%.


The use of water Cherenkov detection for gamma-ray astronomy was pioneered by the Cygnus~\cite{CYGNUS:1986dze} and Milagro~\cite{2004ApJ...608..680A} experiments. Currently, there are two active gamma-ray water Cherenkov detectors: the High-Altitude Water Cherenkov (HAWC) Observatory~\cite{3HWC_2020} in Mexico and the Large High-Altitude Air-Shower Observatory (LHAASO)~\cite{Vernetto_2016} in China. An upcoming WCD observatory, the Southern Wide-field Gamma-ray Observatory (SWGO)~\cite{Albert:2019afb,Abreu:2019ahw,Hinton:2021rvp,Schoorlemmer:2019gee}, is planned to be built in South America (see Section~\ref{sec:swgo}).

\subsection{Scintillator Detectors}\label{sec:scint_det}

Similarly to Water Cherenkov Detectors (WCDs), scintillator-based detectors can reconstruct the incoming direction, energy, and type of the primary particle that originated the air shower by recording the time of arrival, number, and spatial distribution of the particles arriving on the ground. This is accomplished by light-tight plastic scintillator paddles with a typical area of $\mathcal{O}(1~\mathrm{m}^2)$ that emit light once an energetic particle traverses it. This light signal is recorded by a PMT and digitized by fast data-acquisition electronics. Scintillator arrays can be used on their own (e.g., Tibet AS-$\gamma$~\cite{Amenomori:2019rjd}, ALPACA~\cite{Sako:2021fyf}) or complement WCD detectors (e.g., LHAASO~\cite{Vernetto_2016}).

Scintillator arrays benefit from being light-weight which simplifies their deployment. Given that most detectors are flat and relatively thin, their sensitivity drops quickly with the zenith angle of the incoming air shower as their cross-sectional area projected onto the shower front decreases.

%% file: ImagingAtmospheric-Henrike.tex
\chapterauthor[1,2,3,]{Henrike Fleischhack}\asteriskfootnote{H.F. acknowledges support by NASA under award number 80GSFC21M0002. Any opinions, findings, and conclusions or recommendations expressed in this material are those of the author(s) and do not necessarily reflect the views of the National Aeronautics and Space Administration.}\orcidlink{0000-0002-0794-8780}
\chapterauthor[4]{\& Marcos Santander}\orcidlink{0000-0001-7297-8217}
\\
 \begin{affils}
   \chapteraffil[1]{Catholic University of America, Washington, D.C. 20064, USA}
   \chapteraffil[2]{Astroparticle Physics Laboratory, NASA Goddard Space Flight Center, Greenbelt, MD 20771, USA}
   \chapteraffil[3]{Center for Research and Exploration in Space Science and Technology, NASA/GSFC, Greenbelt, MD 20771, USA}
    \chapteraffil[4]{Department of Physics and Astronomy, University of Alabama, Tuscaloosa, AL 35487, USA}

 \end{affils}


\begin{figure}[hbt]
\centering
    \includegraphics[width=0.9\textwidth]{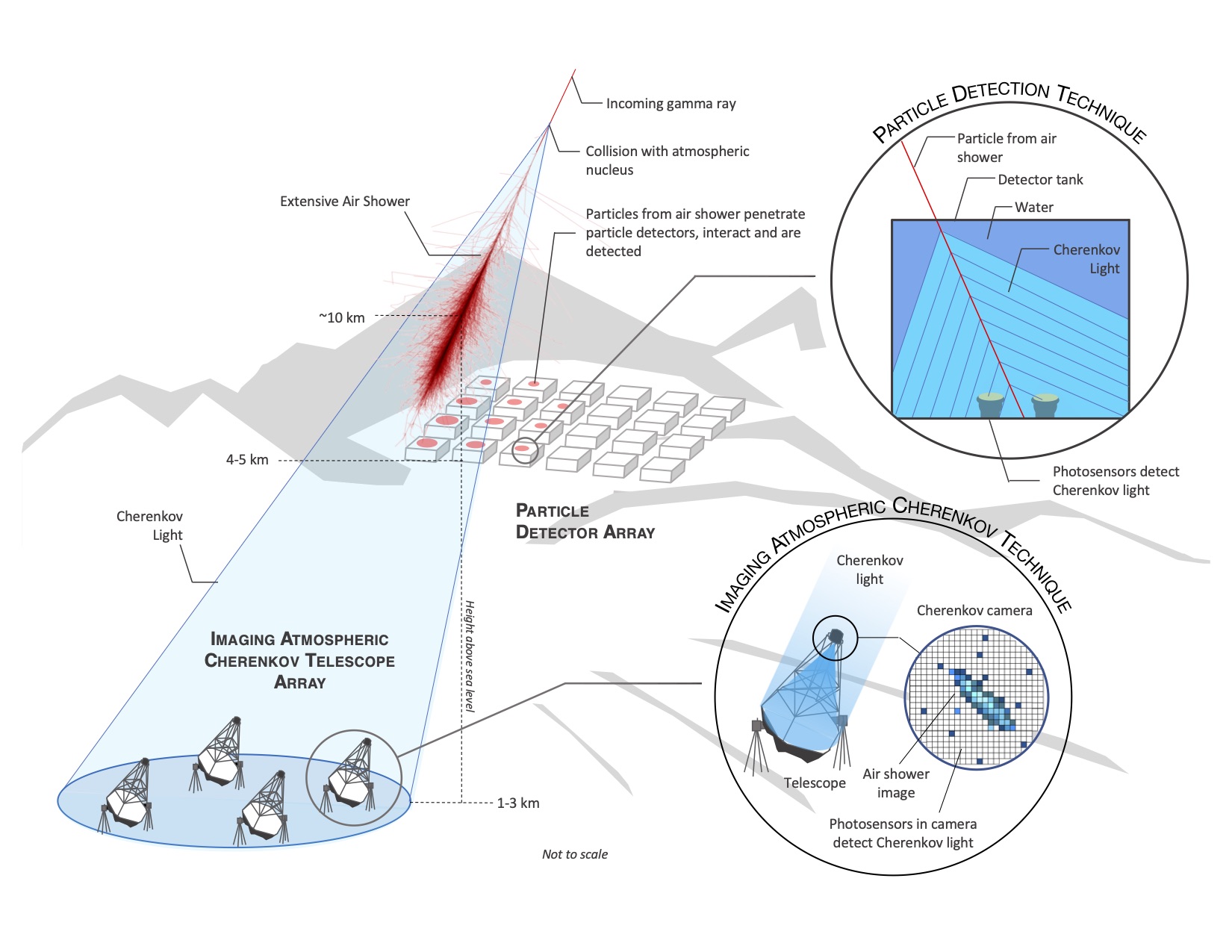}
    \caption{Schematic representation of air-shower detection at ground level with a air shower detector array (in this case illustrated as a water Cherenkov detector) and the imaging atmospheric Cherenkov telescopes (adapted from a figure by Richard White). }
    \label{fig:ground_vhe_gamma}
\end{figure}

Like water Cherenkov detectors (WCDs; Section~\ref{sec:water_cherenkov}), Imaging Air Cherenkov Telescopes (IACTs) detect Cherenkov emission from charged particles in extensive air showers produced by gamma-ray photons and charged cosmic rays. However, IACTs measure Cherenkov light emitted in the air above the detector, not in the detector volume itself. The differences between IACTs and WCDs enable complementary capabilities for observing the gamma-ray sky, as detailed in Table~\ref{tab:comparison}.

\begin{table}[!thp]
\caption{Comparison of typical performance of current and planned IACT arrays and ground particle arrays for gamma-ray astrophysics. Table from Ref.~\citenum{Albert:2019afb}.}
\begin{center}
\begin{tabular}{|l|c|c|}
\hline
 & IACT Arrays & Ground-Particle Arrays \\
\hline
Field-of-view diameter & 3$^{\circ}$--10$^{\circ}$  & 90$^{\circ}$ \\
Duty cycle & 10\%--30\% & $>$95\%\\
Energy range & 30~GeV -- $>$100~TeV& $\sim$500~GeV -- $>$100~TeV \\
Angular resolution & 0.05$^{\circ}$--0.02$^{\circ}$ & 0.4$^{\circ}$--0.1$^{\circ}$ \\
Energy resolution & $\sim$7\% & 60\%--20\%\\
Background rejection & $>$95\% & 90\%--99.8\%\\
\hline
\end{tabular}
\end{center}
\label{tab:comparison}
\end{table}

IACTs generally have a large, typically multi-faceted primary mirror focusing the Cherenkov light onto a camera plane instrumented with sensitive photon detectors (see Figure~\ref{fig:ground_vhe_gamma}) such as photo-multiplier tubes (PMTs) or silicon photo-multipliers (SiPMs). Typical dimensions are 4~m--28~m (mirror diameter) and 2~m--28~m (focal lengths). Instruments with dual- (primary and secondary) mirror designs are under development (e.g.~\cite{Adams:2021hiq,2017ICRC...35..855M}). 

Event reconstruction in IACTs is based on the projected distribution and detected magnitude of the Cherenkov light in the camera. Images of gamma-ray-induced air showers typically have an elliptical shape, with the major axis of the ellipse passing through the projected direction of the primary gamma ray on the sky. Cherenkov light emitted near the top of the air shower, where the atmosphere is less dense, is emitted at a smaller angle and hence seen closer to the source position than the light emitted at a larger angle at the bottom of the air shower. Classical IACT event reconstruction uses geometric methods: if the same shower is observed by two or more telescopes (stereo observation), the incoming direction of the gamma-ray shower (and hence the position of the gamma-ray source in the sky) is determined by the intersection of the major axes of the image ellipses, cf., Ref.~\citenum{Hofmann1999ComparisonOT}, for details on the image reconstruction). If only a single telescope detects the air shower, the source position relative to the shower image can be estimated from the shape of the image itself. The energy of the primary photon can be determined from the total light intensity of the image after correcting for atmospheric absorption and the distance between the telescope and the shower axis (impact distance). Various parameters related to the image shape are used to differentiate between hadronic and photon-induced air showers. Modern reconstruction methods use a template likelihood fitting approach, where a library of shower image templates is first built from simulations for a given combination of event parameters (photon energy, impact distance, etc.), and then the recorded signal in each pixel from a given event is compared to the predictions to determine the set of event parameters that best matches the recorded data (see, e.g., Ref.~\citenum{Impact2014}).

IACTs are typically background-limited at lower energies, especially for longer observing time scales. The two dominant background components for gamma-ray astronomy with IACTs are the night sky background (NSB) due to starlight, moonlight, and light pollution by anthroprogenic sources, and Cherenkov light from cosmic-ray air showers. The magnitude of the NSB depends on weather conditions, moon phase, and the observing field. NSB can cause both spurious triggers as well as artifacts on top of ``real'' air-shower images. The effects can be suppressed at the trigger level (requiring multiple contiguous pixels to trigger, excluding pixels with known bright stars from the trigger logic and/or requiring multiple telescopes to trigger on the same event) and during the ``image cleaning'' stage, e.g., by not considering pixels with bright, isolated signals for the reconstruction. Cosmic-ray air showers produce ``real'' images in the camera, but this background can be partially suppressed by cuts on the image shape parameters (or goodness-of-fit, if using a template likelihood fit). The rate of background events remaining after cuts depends on the observing conditions (e.g., NSB level and pointing direction) and generally has to be measured---rather than derived entirely from simulations---either from dedicated background runs pointing away from known gamma-ray sources or by considering gamma-ray candidates with reconstructed directions far away enough from the source region. 

An IACT's energy threshold depends on the instrument itself as well as on the observing conditions. Typical values are between tens of GeV and several TeV. IACTs are, in principle, sensitive to gamma-ray photons up to PeV energies or more, but few sources have significant emission at those energies. IACTs require (reasonably) dark and clear skies to operate, restricting their duty cycles to around 10\%. Observations under partial moonlight or partial cloud coverage are possible, albeit with reduced sensitivity. 

IACTs are pointing instruments. The field of view is limited by the camera size, typically a few degrees in diameter. IACTs are thus best suited for in-depth studies and monitoring of known or predicted gamma-ray sources. The detection of a transient event such as a gamma-ray burst (GRB) requires luck and/or fast slewing to the position provided by an external alert. Current-generation IACTs can typically slew anywhere on the sky within one to a few minutes. 

The IACT technique was pioneered by Weekes et al. at the Whipple Observatory 10~m reflector, who detected a significant TeV gamma-ray signal from the Crab nebula in 1989~\cite{1989ApJ...342..379W}. Currently operating IACT facilities are the High-Energy Spectroscopic System (H.E.S.S.)\footnote{\url{https://www.mpi-hd.mpg.de/hfm/HESS/}} array~\cite{Aharonian:2006pe} in Namibia, the Major Atmospheric Gamma Imaging Cherenkov (MAGIC) telescope\footnote{\url{https://magic.mpp.mpg.de/}}~\cite{2016APh....72...61A} and the First G-APD Cherenkov Telescope (FACT)\footnote{\url{https://fact-project.org/}}~\cite{Anderhub:2013cqa} on La Palma (Spain), and the Very Energetic Radiation Imaging Telescope Array System (VERITAS)\footnote{\url{https://veritas.sao.arizona.edu/}}~\cite{Holder:2006gi} in the US. The next-generation IACT array will be the Cherenkov Telescope Array (CTA)~\cite{CTAConsortium:2018tzg} (see Section~\ref{sec:cta}) with locations in both hemispheres, the southern one near Cerro Paranal, Chile and the northern one in the Canary Island of La Palma, Spain.

%% file: BurstCube-Perkins.tex
\chapterauthor[1]{Jeremy S. Perkins}\orcidlink{0000-0001-9608-4023}
\chapterauthor[1]{Judith L. Racusin}\orcidlink{0000-0002-4744-9898}
\\ 
 \begin{affils}
   \chapteraffil[1]{Astroparticle Physics Laboratory, NASA Goddard Space Flight Center, Greenbelt, MD 20771, USA}
 \end{affils}

The first joint detection of Gravitational Waves (GWs) by the Laser Interferometer GW Observatory (LIGO) of GW170817 and the short gamma-ray burst (sGRB) GRB 170817A by the Gamma-ray Burst Monitor (GBM) on-board the \textit{Fermi} Gamma-ray Space Telescope~\cite{Meegan2009} and the INTErnational Gamma-Ray Astrophysics Laboratory (INTEGRAL) SPectrometer on INTEGRAL Anti-Coincidence Shield (SPI-ACS)~\cite{spiacs} confirmed that the progenitors of sGRBs are Binary Neutron Star (BNS) mergers~\cite{Abbott2017}. Since that discovery, other BNS mergers have been detected by the GW observatories~\cite{Abbott2020} without electromagnetic (EM) counterparts. LIGO, Virgo, and KAGRA, the Kamioka Gravitational Wave Detector (hereafter, LVK), are commissioning major upgrades~\cite{Abbott2018}, scheduled to begin their next observing run (O4) in mid-December 2022\footnote{\url{https://www.ligo.org/scientists/GWEMalerts.php}}. The  discovery of GW and EM signatures requires dedicated and coordinated observations by large communities of both ground- and space-based facilities. Existing sensitive gamma-ray burst (GRB) instruments cover only $\sim70\%$ of the sky at any one time, and any increase in sky coverage improves both the likelihood of coincident detection and the number of sGRBs that can be correlated with GWs.

\begin{figure}[t!]
   \begin{center}
    \includegraphics[height=3.9cm]{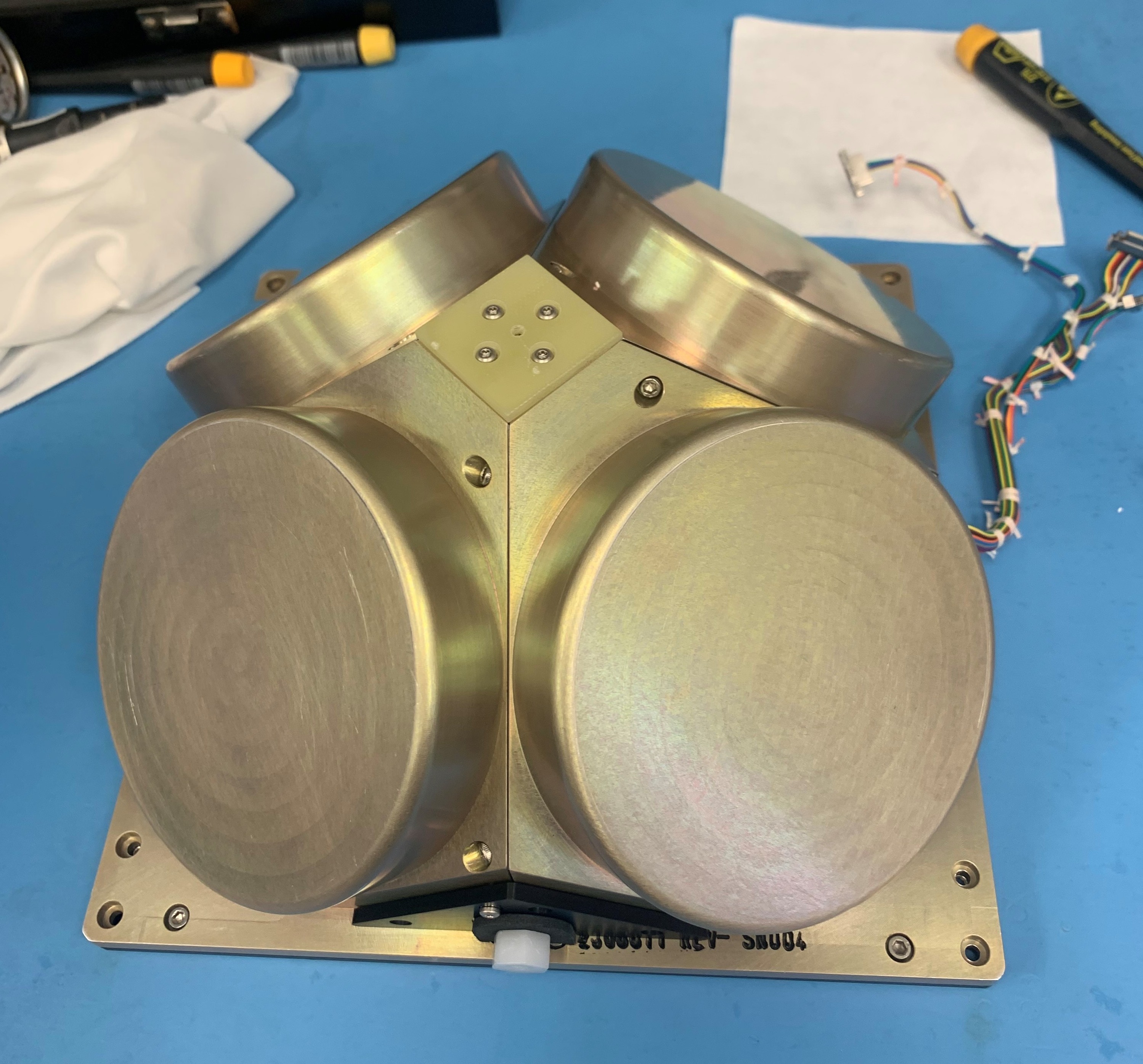}
    \includegraphics[height=3.9cm]{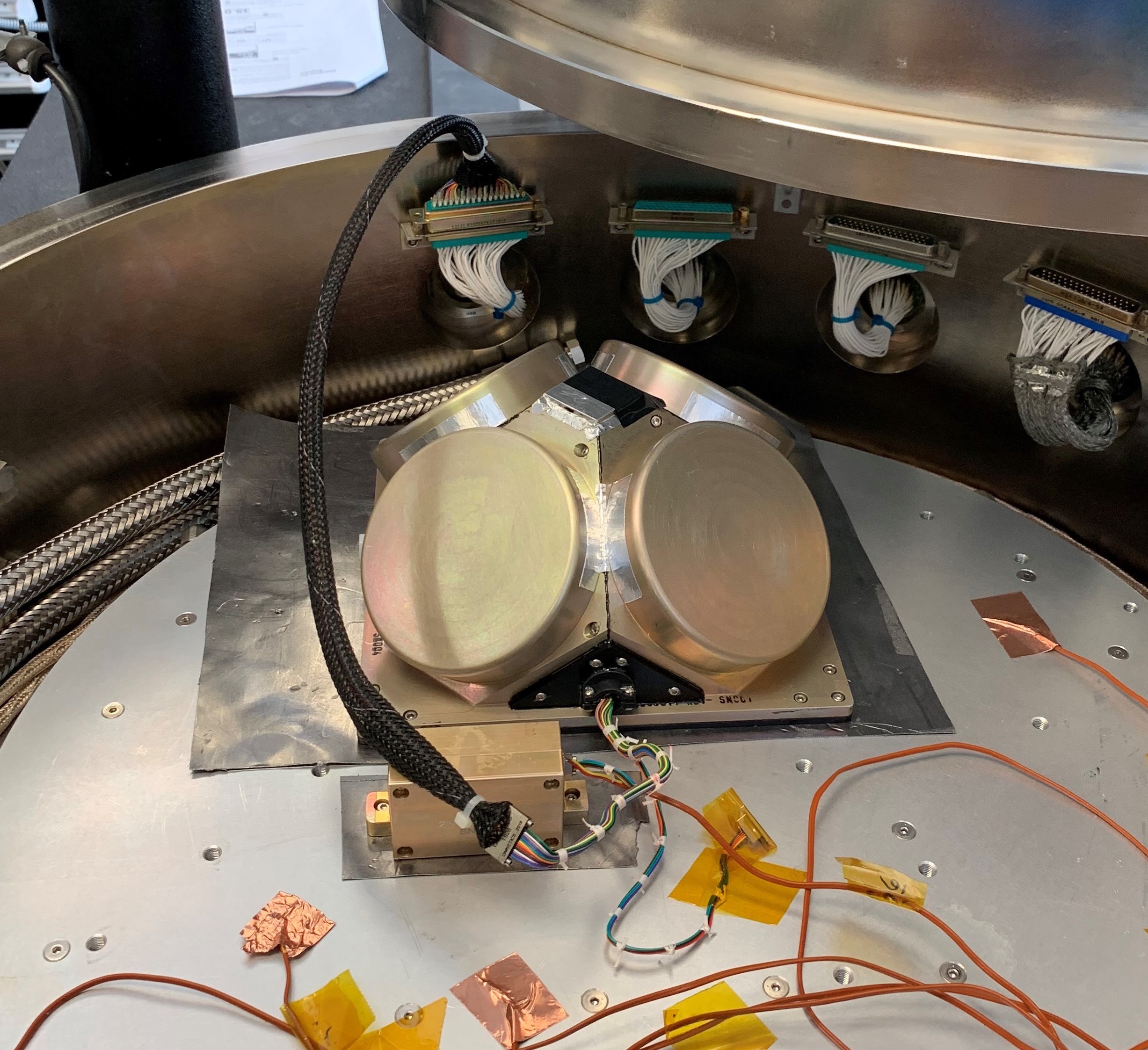}
    \includegraphics[height=3.9cm]{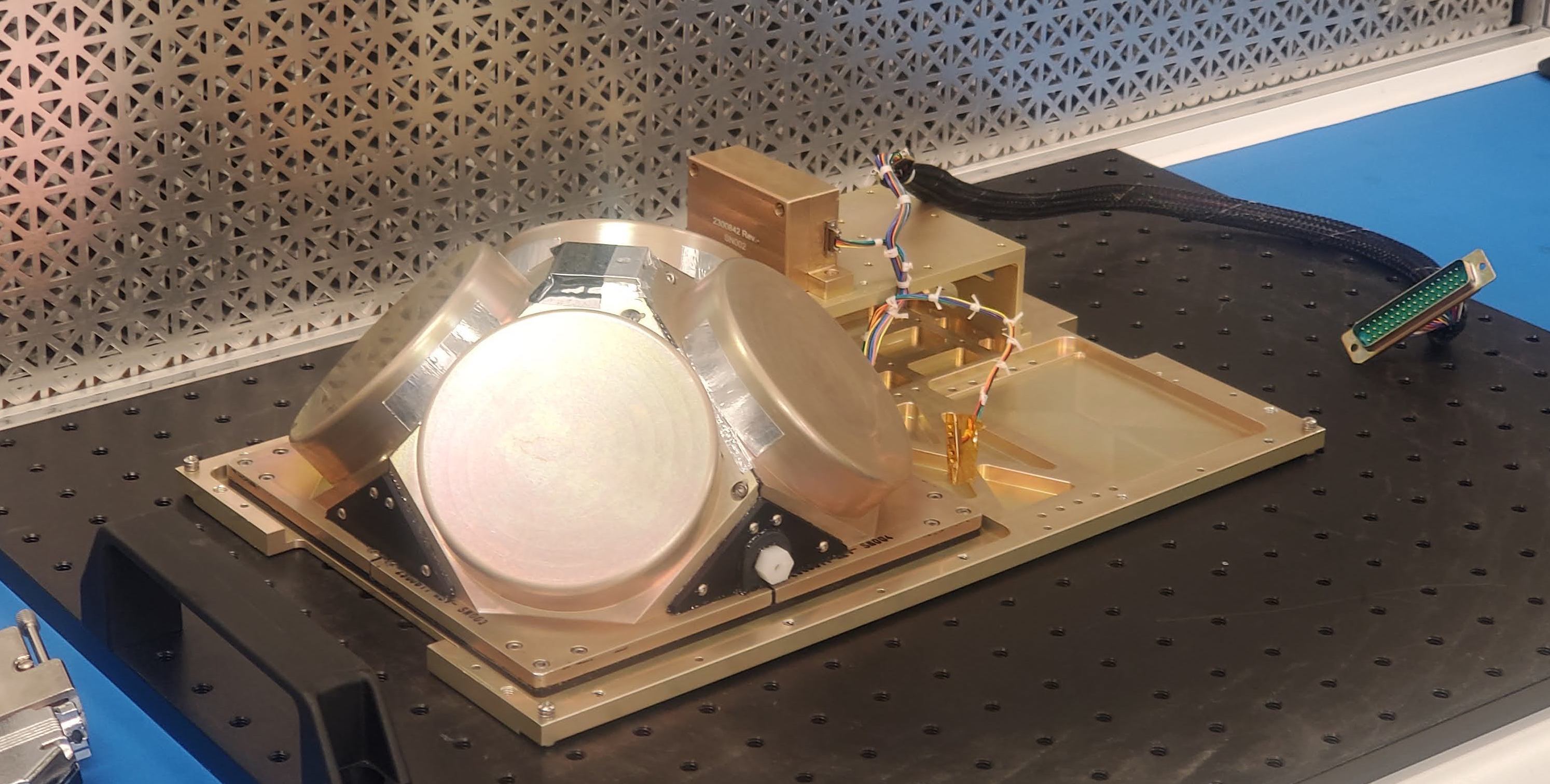}
    
   \end{center}
   \caption[example] 
   { \label{fig:instrument} \textit{Left:} The BurstCube instrument is fully assembled and calibrated. \textit{Center:} In November 2021, the flight instrument and bias supply underwent thermal vacuum testing.  No anomalies were seen. \textit{ Right:} As of December 2021, the instrument and bias supply have now been installed on the spacecraft baseplate and are awaiting integration.}
   \end{figure}

BurstCube is a CubeSat to detect and characterize sGRBs that are counterparts of GW sources. {\it BurstCube addresses a key Astro2020 decadal recommendation~\cite{NAP26141}, studying new messengers and physics by detecting the astrophysical counterparts to GW events.} BurstCube is a `6U' CubeSat (each `U' is  $\sim10$~cm~$\times~10$~cm~$\times~10$~cm, so a `6U' is ${\sim10}$~cm~$\times~20$~cm~$\times~30$~cm) with a `4U' instrument package. The instrument consists of four Cesium Iodide (CsI) scintillators read out by arrays of Silicon Photomultipliers (SiPMs). It will detect photons from $\leq50$~keV to $>1$~MeV and roughly localize gamma-ray transients on the sky, rapidly sending alerts to the ground to enable follow-up at other wavelengths and greater sensitivity. We have completed all design reviews and the instrument is assembled, calibrated, and tested, and will be integrated with the spacecraft soon (see Figure~\ref{fig:instrument}). BurstCube will reach orbit in late 2022 or early 2023, in line with the O4 GW observing run. 

{\it The BurstCube Mission has Three Objectives:} Provide astrophysical context and rapid localizations for high-significance LVK detections coincident with sGRBs; correlate sGRBs with LVK sub-threshold signals, effectively increasing the GW detection volume; and monitor the gamma-ray sky in search of transient and variable sources including Solar flares, flaring Galactic binaries, and magnetar outbursts.

{\it The BurstCube Mission Provides:} Rapid alerts of onboard GRB detections via the Tracking and Data Relay Satellite System (TDRSS); light curves, spectra, and raw data of $>20$ short and $>100$ long GRBs per year; and a long-term test ($\geq1$ year) of SiPMs in low Earth orbit.

The BurstCube instrument consists of four CsI scintillators coupled to arrays of SiPMs~\cite{perkins2020}. It detects photons from $\leq50$~keV to $>1$~MeV and roughly localizes gamma-ray transients on the sky, rapidly sending alerts to the ground via TDRSS which will enable follow-up at other wavelengths and at greater sensitivity~\cite{Martinez2021}. Each CsI crystal is 19~mm thick and 90~mm in diameter, housed in an aluminum can (see Figure~\ref{fig:instrument}). The detailed mechanical design is based on the {\it Fermi}-GBM~\cite{Meegan2009}. Optical pads are used to couple the crystal to a quartz window and the quartz window to the SiPM arrays. These pads provide optical coupling and mechanical buffering. The housing holds each detector at an approximately 45$^\circ$ angle to zenith, allowing for all-sky coverage and localization. Four detectors are mounted together to form the instrument, localizing gamma-ray sources by comparing relative photon rates in each detector.  

Gamma rays pass through the aluminum housing into the CsI crystal and produce scintillation light. This light is collected by an array of 116 SiPMs (Hamamatsu 13360-6050), which convert it into an electrical signal. The SiPMs are mounted on carrier boards (two SiPMs per board) and these carriers are mounted to an instrument detector analog board (IDAB), which combines the signals from 19--20 SiPMs. Each of these groups is individually amplified then summed to provide a single analog output signal per detector. The four IDAB boards are connected to an instrument digital processing unit (IDPU) that provides power to the IDABs (line and bias) and continuously digitizes the signals. The IDPU interfaces with the command and data handling (C\&DH) unit on the spacecraft. The IDPU determines peak times and amplitudes as well as measures the background signal. 

BurstCube is well aligned with the Astro2020 decadal survey~\cite{NAP26141}--- studying new messengers and new physics by detecting the astrophysical counterparts to GW events. BurstCube is part of a suite of missions that will provide electromagnetic context via a small mission. 
The BurstCube team is involved in an international group\footnote{https://asd.gsfc.nasa.gov/conferences/grb\_nanosats/} in sharing data and expertise, as well as pooling resources. This effort is the start of providing a network of GRB-detecting SmallSats in orbit at low cost and low risk. The BurstCube software (pipelines and data analysis tools; e.g., \texttt{bc-tools}) are all being released open-source and developed such that they can be used by future missions. The algorithms and structures developed are already being planned to be used on upcoming missions like StarBurst~\cite{StarBurst} (see Section~\ref{sec:starburst}) and GlowBug~\cite{Grove2020} (see Section~\ref{sec:glowbug}). The \texttt{bc-tools} rely heavily on the \texttt{gbm-data-tools}~\cite{2020ascl.soft10002F} and the software developed for BurstCube are either released on their own or added back into the prior tools.

%% file: Glowbug-Eric.tex
 \noindent
 \chapterauthor[]{J. Eric Grove}
 \\
 \begin{affils}
   \chapteraffil[]{Naval Research Laboratory, Washington, D.C. 20375, USA}
 \end{affils}

Glowbug is a gamma-ray telescope for Gamma-Ray Bursts (GRBs) and other transients in the $\sim$30~keV to 2~MeV band~\cite{2020grbg.conf...57G}. It is one of nine experiments on the DoD Space Test Program STP-H9 payload, which is scheduled to launch to the Japanese Experiment Module - Exposed Facility (JEM-EF) on the International Space Station (ISS) in January 2023. Glowbug was developed and built by the U.S. Naval Research Laboratory (NRL) with funding from the NASA Astrophysics Research and Analysis (APRA) program.

Glowbug will perform forefront GRB research---detecting an estimated hundreds of long GRBs and dozens of short GRBs per year---providing burst spectra, lightcurves, and positions. Its primary science objective is the detection and localization of short GRBs, which are the result of mergers of compact binaries~\cite{PhysRevLett.119.161101, Abbott_2017} involving a neutron star with either another neutron star (NS-NS) or a black hole (NS-BH). The instrument is designed to complement existing (but mature) GRB detection systems (\textit{Fermi} Gamma-ray Burst Monitor (GBM) and \textit{Swift} Burst Alert Telescope (BAT)), and it will join future networks of small GRB instruments to provide all-sky coverage and improved localization of such events. Of immediate importance are the binary neutron star (BNS) systems within the gravitational-wave detection horizon of 200~Mpc expected from the planned upgrades to the Advanced Laser Interferometer Gravitational-wave Observatory (LIGO)-Virgo detectors in the early 2020s~\cite{2020LRR....23....3A}.

\begin{figure}
    \centering
    \includegraphics[width=0.5\textwidth]{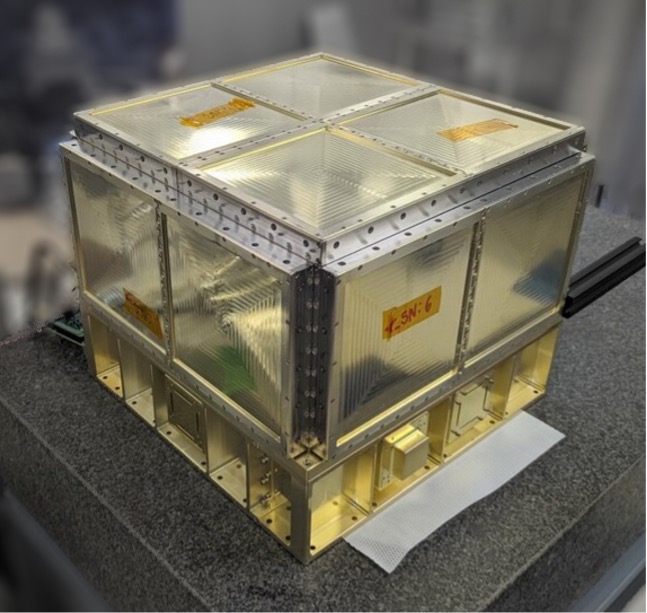}
    \caption{Assembled Glowbug instrument during functional checkout and calibration prior to environmental testing. Eight of 12 CsI(Tl) detectors are visible. The readout electronics and data acquisition system are housed in the electronics pedestal below the detector array.}
    \label{fig:glowbug}
\end{figure}

The instrument design is simple, with an assembly of 12 relatively large Thallium-doped Cesium Iodide (CsI(Tl)) scintillator panels (each 15~cm~$\times$~15~cm~$\times$~1~cm) arrayed on a half cube (see Figure~\ref{fig:glowbug}). This array provides full coverage of the unocculted sky with good sensitivity and modest localization capability. Each crystal is read out on one edge by a custom linear array of ON Semiconductor J-series silicon photomultipliers (SiPMs). SiPMs provide ample gain in a small, low-mass, low-voltage form factor and enable efficient packaging of scintillator detectors. To shield against the diffuse X-ray background, the scintillator panels view the sky through 1~mm aluminum covers.  Each scintillator is backed by 6~mm stainless steel and 1~mm tantalum to attenuate gamma rays that are partially absorbed in the scintillator and increase imaging contrast for the detector array.

Mounted within the half cube are two hexagonal Cs$_2$LiLaBr$_6$(Ce) (CLLB) scintillators, each 10~cm in length and 5~cm in diameter, read out at one end by a custom hexagonal array of ON Semiconductor J-series SiPMs. CLLB scintillators provide excellent spectral resolution ($<$5\% FWHM at 662~keV) and good sensitivity for thermal neutrons, which can readily be distinguished from gamma rays through pulse-shape discrimination. These detectors are included as a secondary instrument and are intended for a study to characterize the performance and activation of CLLB in the space radiation environment. While their neutron sensitivity provides no scientific benefit to the prime GRB science mission, these detectors do provide sensitivity to GRB emission above 2~MeV.

The data acquisition system comprises largely commercial, off-the-shelf electronics. The front end for each scintillator is a Bridgeport Instruments\textsuperscript{TM} slimMorpho multi-channel analyzer (MCA). A Raspberry Pi4\textsuperscript{TM} single-board computer (SBC) is the instrument processor, controlling the MCAs, storing and processing detector data, and providing command and data interface to the STP-H9 payload bus. Instrument flight software provides autonomous burst detection and localization. The processing power of the commercial SBC allows Glowbug flight software to execute more sophisticated detection and localization algorithms, based on a likelihood analysis, than previous GRB instruments. Enabled by the high downlink bandwidth of the ISS, the primary science data product is a continuous, time-tagged event stream.  When a burst is detected, to facilitate rapid notification to the community, a burst alert message is containing the time, intensity, and location derived by the on-board software.

Development of Glowbug has served as a pathfinder for the StarBurst~\cite{StarBurst} (see Section~\ref{sec:starburst}) Pioneer mission, led by NASA Marshall Space Flight Center and recently selected as one of the first four astrophysics Pioneer-class missions. StarBurst is, in essence, an enlarged version of Glowbug, using the same half-cube geometry and 12 larger CsI(Tl) scintillator panels. NRL will design and build the detectors and front-end electronics for StarBurst using designs and processes developed with Glowbug.

%% file: StarBurst-Kocevski.tex
\chapterauthor[ ]{ }

 \addtocontents{toc}{
     \leftskip3cm
    \scshape\small
    \parbox{5in}{\raggedleft Daniel Kocevski, Eric Grove, Michael S. Briggs,  et al.}
    \upshape\normalsize
    \string\par
    \raggedright
    \vskip -0.19in
    }
 
 \noindent
 \nocontentsline\chapterauthor[]{Daniel Kocevski$^{1}$}\orcidlink{0000-0001-9201-4706}
 \nocontentsline\chapterauthor[]{Eric Grove$^{2}$}
 \nocontentsline\chapterauthor[]{Michael S. Briggs$^{3}$}\orcidlink{0000-0003-2105-7711}  
 \nocontentsline\chapterauthor[]{Adam Goldstein$^{5}$}\orcidlink{0000-0002-0587-7042}
 \nocontentsline\chapterauthor[]{Eric Burns$^{4}$}\orcidlink{0000-0002-2942-3379}  
 \nocontentsline\chapterauthor[]{Richard Woolf$^{2}$}\orcidlink{0000-0003-4859-1711}
 \nocontentsline\chapterauthor[]{Michelle Hui$^{1}$}\orcidlink{0000-0002-0468-6025}  
 \nocontentsline\chapterauthor[]{Judith L. Racusin$^{6}$}\orcidlink{0000-0002-4744-9898}
 \nocontentsline\chapterauthor[]{Oliver Roberts$^{5}$}\orcidlink{0000-0002-7150-9061}
 \nocontentsline\chapterauthor[]{Peter Shawhan$^{7}$}\orcidlink{0000-0002-8249-8070}
 \nocontentsline\chapterauthor[]{Jacob R. Smith$^{8,9}$}\orcidlink{0000-0002-3594-6133}
 \nocontentsline\chapterauthor[]{Colleen A. Wilson-Hodge$^{1}$}\orcidlink{0000-0002-8585-0084}
 \nocontentsline\chapterauthor[]{Pete Jenke$^{3}$} 
 \nocontentsline\chapterauthor[]{Rachel Hamburg$^{3}$}\orcidlink{0000-0003-0761-6388} 
 \nocontentsline\chapterauthor[]{Boyan Hristov$^{3}$}\orcidlink{0000-0001-9556-7576}
 \nocontentsline\chapterauthor[]{Matthew Kerr$^{2}$}\orcidlink{0000-0002-0893-4073}
 \nocontentsline\chapterauthor[]{Corinne Fletcher$^{5}$}\orcidlink{0000-0002-0186-3313}
 \nocontentsline\chapterauthor[]{Peter Veres$^{5}$}\orcidlink{0000-0002-2149-9846}
 \nocontentsline\chapterauthor[]{Robert Preece$^{3}$}\orcidlink{0000-0003-1626-7335}
 \nocontentsline\chapterauthor[]{Wen Fe Fong$^{10}$}\orcidlink{0000-0002-7374-935X}
 \nocontentsline\chapterauthor[]{Bing Zhang$^{11}$}\orcidlink{0000-0002-9725-2524}
\\ 
\begin{affils}
   \chapteraffil[1]{NASA Marshall Space Flight Center, Huntsville, AL 35808, USA}
   \chapteraffil[2]{Naval Research Laboratory, Washington, D.C. 20375, USA}
   \chapteraffil[3]{University of Alabama in Huntsville, Huntsville, AL 35899, USA}
   \chapteraffil[4]{Louisiana State University, Baton Rouge, LA. 70803, USA}
   \chapteraffil[5]{Universities Space Research Association, Columbia, MD 21046, USA}
   \chapteraffil[6]{NASA Goddard Space Flight Center, Greenbelt, MD 20771, USA}
   \chapteraffil[7]{University of Maryland, College Park, College Park, MD 20742, USA}
   \chapteraffil[8]{Center for Research and Exploration in Space Science and Technology, NASA/GSFC, Greenbelt, MD 20771, USA}
   \chapteraffil[9]{University of Maryland, Baltimore County, Baltimore, MD 21250, USA}
   \chapteraffil[10]{Northwestern University, Evanston, IL 60208, USA}
   \chapteraffil[11]{University of Nevada, Las Vegas, Las Vegas, NV 89154, USA}
 \end{affils}

The StarBurst Multimessenger Pioneer is a highly sensitive and wide-field gamma-ray monitor designed to detect the prompt emission of short gamma-ray bursts (SGRBs), a key electromagnetic (EM) signature of neutron star mergers. In conjunction with gravitational wave (GW) and follow-up observations across the EM spectrum, StarBurst seeks to understand neutron star mergers through multimessenger observations and use these studies to address four primary science objectives: 1) constrain the progenitors of SGRBs, 2) probe the remnants of neutron star mergers, 3) constrain the neutron star equation of state, and 4) probe the structure of relativistic outflows produced in neutron star mergers. These objectives directly address the time-domain and multimessenger recommendations put forth by the Astro2020 decadal review~\cite{NAP26141}.

StarBurst is designed to capitalize on the new era of multimessenger astronomy following the detection of GW170817/GRB170817A by using the advancements in gamma-ray detectors made over the last decade. With over 500$\%$ the effective area of the \textit{Fermi} Gamma-ray Burst Monitor (GBM) and full coverage of the unocculted sky, StarBurst will make highly sensitive observations of EM counterparts to neutron star mergers and be a key partner to the GW network in discovering these mergers at a fraction of the cost of currently operating gamma-ray missions. StarBurst is designed as a SmallSat to be deployed to Low Earth orbit (LEO) as a secondary instrument using the Evolved Expendable Launch Vehicle (EELV) Secondary Payload Adapter (ESPA) Grande interface for a nominal one-year mission starting in 2025. Assuming a similar orbit and duty cycle as GBM, StarBurst will observe an estimated 200 SGRBs per year, with a median rate of 9.8 joint GW-SGRBs detections per year.

\begin{figure}[t!]
   \begin{center}
    \includegraphics[height=5.3cm]{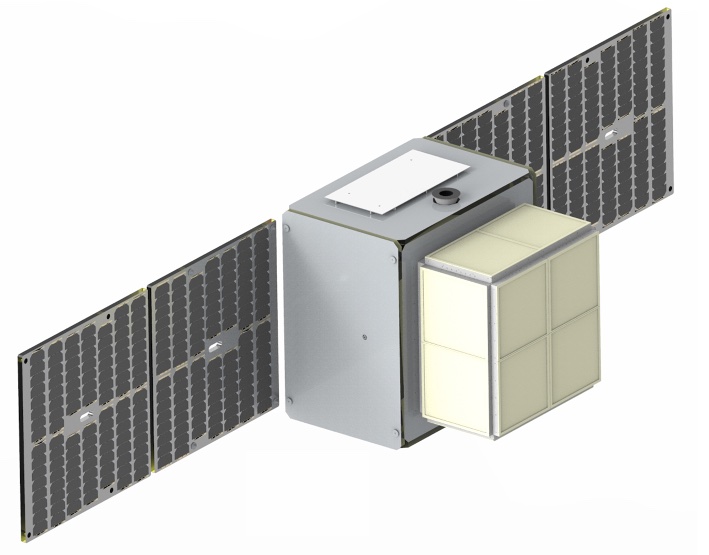}
    \includegraphics[height=5.3cm]{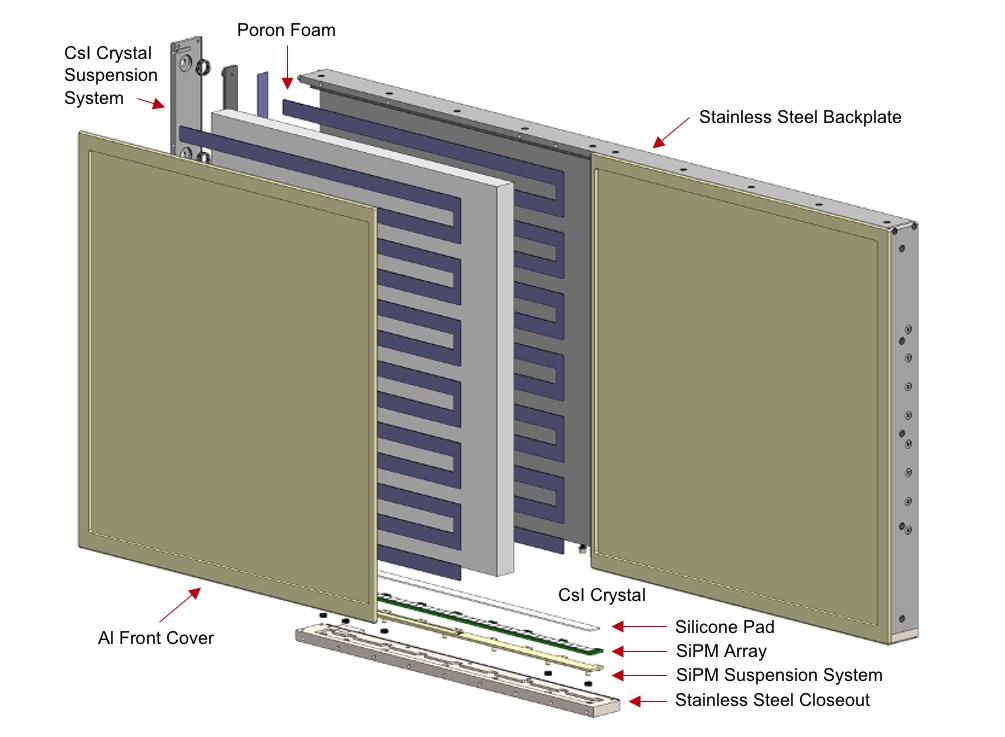}
   \end{center}
   \caption[example] 
   { \label{fig:StarBurst_Figures} \textit{Left:} The integrated StarBurst observatory. \textit{Right:} An exploded view of the StarBurst Bi-Pack detectors.}
   \end{figure}

StarBurst relies heavily on the heritage of the \textit{Fermi}-GBM, Strontium Iodide Radiation Instrumentation (SIRI)~\cite{2019arXiv190711364M}, BurstCube~\cite{Smith:2019zra} (see Section~\ref{sec:burstcube}), and Glowbug~\cite{Grove2020} (see Section~\ref{sec:glowbug}) instruments. Consisting of an array of 12 Thallium-doped Cesium Iodide (CsI(Tl)) scintillator arranged to form a five-sided cube, the instrument achieves coverage of the entire unocculted sky, while preserving forward directionality by shielding the inside face of each detector. The individual StarBurst detector consists of a 27~cm~$\times$~27~cm $\times$~1.5~cm CsI(Tl) scintillator enclosed in a housing with a thin (1~mm) aluminum window transparent down to 30~keV and a stainless steel back shield to provide strong attenuation below $\sim$200~keV. The detector assemblies, referred to as Bi-Packs because they house two detectors side-by-side, are directly based on the assemblies designed for use in Glowbug, with dimensions modified to accommodate the larger StarBurst crystals. An exploded view of the Bi-Pack design is shown in Figure~\ref{fig:StarBurst_Figures}.

The scintillation light is read out by a $2\times40$ linear array of 6~mm~$\times$~6~mm J-Series SiPMs from ON Semiconductor (formerly SensL) optically coupled through an elastomeric silicone optical pad to a single edge of the CsI(Tl) crystal. Edge readout of the CsI(Tl) exploits the planar geometry of the crystal to pipe scintillation light to the SiPM array and dramatically reduces (by as much as a factor of 20) the number of SiPMs that would otherwise be required to cover the larger crystal surface areas to collect the scintillation light. The SiPM array will use a summation amplifier to isolate each SiPM’s capacitance contribution, providing an approximately linear summation of the response of each SiPM. The array is subdivided into modules of 16 SiPMs in which each SiPM is provided with cathode bias and a closely integrated decoupling capacitor. 

The combined instrument concept has a size of roughly 35~cm~$\times$~62~cm~$\times$~62~cm and an estimated mass of 155~kg, including contingency, and achieves an azimuthal averaged effective area of 3,500~cm$^2$, or over 500$\%$ the effective area of \textit{Fermi}-GBM.  This large collecting area should allow StarBurst to localize a SGRB of comparable intensity to GRB170817 to within 3~degrees on the sky.  

StarBurst will use the DAringly UNcommon Technical Leadership in Smaller Satellites (DAUNTLESS) ESPA/ESPA-Grande class bus designed and built by the Space Flight Laboratory (SFL) at the University of Toronto. SFL has extensive experience building and exporting satellites internationally, having accumulated more than 128 years of on-orbit heritage. The essential flight equipment of the DAUNTLESS bus, including the avionics core and attitude control system, are common to all SFL missions, providing a high heritage and well-understood system. Combined with the instrument, the integrated observatory has an estimated mass of 300~kg (including contingency) and stowed dimensions of 90~cm~$\times$~80~cm~$\times$~92~cm.
 
The StarBurst concept of operations closely follows the successful model employed by \textit{Fermi}-GBM, with slight modifications to accommodate telemetry volume constraints. When a transient is detected by the onboard flight software, StarBurst produces a burst alert message that is transmitted with low-latency using NASA’s Tracking and Data Relay Satellite (TDRS) system.  The data contained within this message is received by the science operations center and are formatted for near-real-time dissemination via the Gamma-ray Coordinates Network (GCN). Time tagged event (TTE) data associated with each detection is then automatically downlinked at the next available ground contact. TTE data covering periods not associated with an onboard detection are stored for a period of three days and can be requested by ground command, thereby facilitating joint sub-threshold searches of data surrounding multimessenger events of interest. 

The StarBurst mission’s approach of building a straightforward, low-risk, and high-heritage instrument and spacecraft bus aligns with the goal of NASA’s new Astrophysics Pioneers Program of finding cost-effective, new ways to conduct astrophysics research. By re-using the electrical design, detector readout, and significant portions of the mechanical design from the SIRI and Glowbug instruments, the ground analysis pipeline and flight software developed for GBM and BurstCube, and a commercially available spacecraft bus, StarBurst can be built and launched using mature technologies with significant savings in cost and an equally significant reduction in risk. The resulting mission will significantly outperform existing gamma-ray facilities and substantially increase the sample of joint GW-SGRB detections with which to advance our understanding of NS mergers and the SGRBs they produce.

%% file: MoonBEAM-Hui.tex
 \noindent
 \chapterauthor[]{C. M. Hui}\orcidlink{0000-0002-0468-6025}
 \\
 \begin{affils}
   \chapteraffil[]{NASA Marshall Space Flight Center, Huntsville, AL 35808, USA}
 \end{affils}

Moon Burst Energetics All-sky Monitor (MoonBEAM) is a proposed gamma-ray mission to observe the entire sky instantaneously for relativistic astrophysical explosions from a cislunar orbit. It is designed to explore the behavior of matter and energy under extreme conditions by observing the prompt emission from gamma-ray bursts (GRBs), identifying the conditions capable of launching transient relativistic jets and the origins of high-energy radiation from the relativistic outflows. MoonBEAM provides essential gamma-ray observations for multi-messenger astrophysics by reporting on the prompt emission and providing the critical first alerts to the community for contemporaneous and follow-up observations.

MoonBEAM achieves instantaneous all-sky coverage with a near-continuous ($>$96\%) observing duty cycle based on its instrument design and its cislunar orbit. The instrument consists of six detector modules positioned on the corners of the spacecraft to achieve all-sky coverage. Each detector module consists of a NaI(Tl)/CsI(Na) phoswich scintillator coupled to flat panel photomultiplier tubes and is sensitive to 10--5000~keV photons. The phoswich design allows simultaneous dual-mode analysis: refined localizations from the NaI(Tl) crystal with the thicker CsI(Na) crystal serving as an active shield to reduce the background, while increasing the effective area over a wide field of view (FOV) and extending the sensitivity to higher energies using the signals from both crystals. 

\begin{figure}
    \centering
    \includegraphics[width=\textwidth]{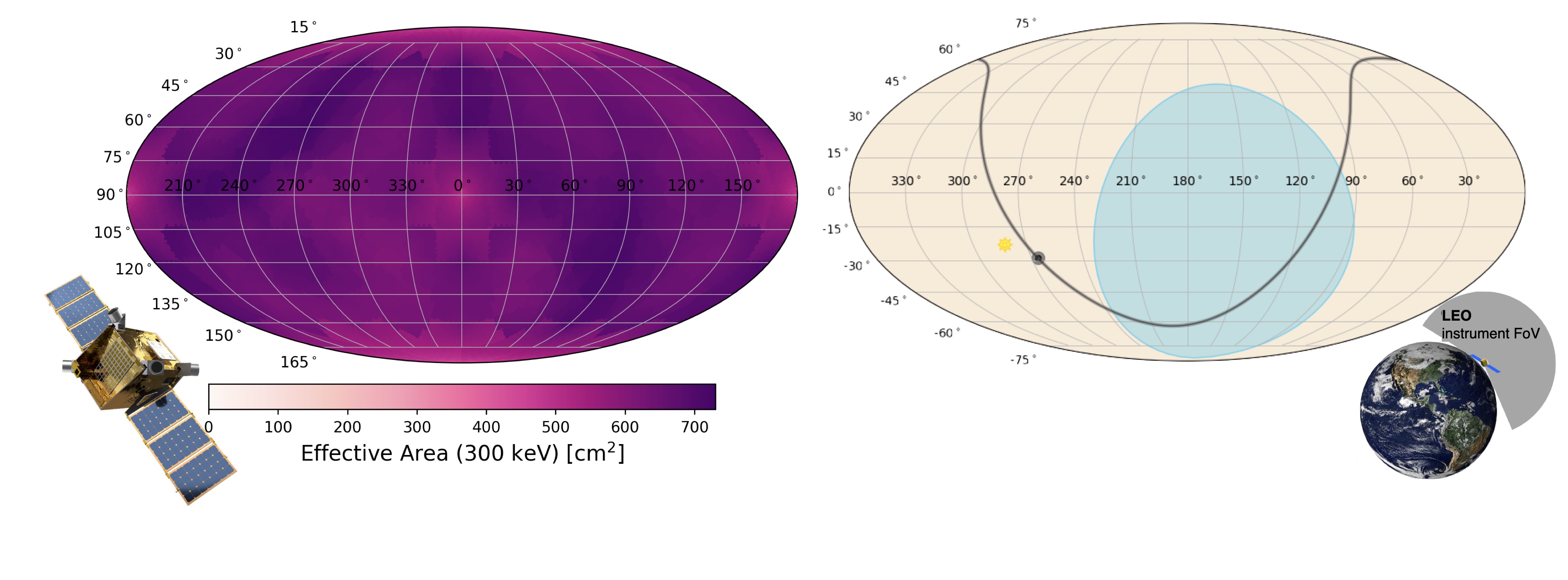}
    \caption{\textit{Left}: MoonBEAM instantaneous effective area at 300~keV, covering the entire sky. \textit{Right}: an equivalent instrument in Low Earth Orbit with $\sim 30\%$ Earth blockage (blue shaded region) in its field of view.}
    \label{fig:moonbeam}
\end{figure}

The science orbit of MoonBEAM is a lunar-resonant orbit at 22,000 to 460,000~km from the Earth. This orbit minimizes Earth blockage to $<$1\% of the all-sky FOV and variation in the background radiation environment. Figure~\ref{fig:moonbeam} highlights the instantaneous all-sky effective area provided by MoonBEAM compared to an equivalent instrument in Low Earth Orbit (LEO) with Earth blockage. The background rate in the cislunar orbit is stable to longer timescales because MoonBEAM is not subjected to the Earth’s atmospheric scattering and only infrequently to trapped radiation. This background stability, coupled with the high duty cycle, allows MoonBEAM to observe weaker and longer-duration transients with an uninterrupted observational period of 13.7~days. 

The intrinsic localization capability of MoonBEAM is comparable to the \textit{Fermi} Gamma-ray Burst Monitor, with a median localization radius of 5~degrees. The cislunar orbit also provides a longer baseline than LEO-only instruments for localization refinement using the timing-triangulation technique~\cite{MB:Hurley2013}. When a gamma-ray transient is detected by multiple instruments in different orbits, the difference in detection times will constrain the arrival direction to an annulus on the sky and aid in reducing the localization area. The median joint localization annulus width is 1.4~degrees with MoonBEAM and a LEO gamma-ray instrument, which can significantly reduce the localization region of compact mergers jointly detected with gravitational waves~\cite{MB:Petrov2022}.

During its three-year primary mission, MoonBEAM is expected to observe 1600 binary compact object mergers~\cite{MB:Petrov2022}, 5900 optically discovered core-collapse supernovae~\cite{MB:Feindt2019}, and 140 magnetar giant flares~\cite{MB:Burns2021} while enabling follow-up observations in other wavelengths and messengers. MoonBEAM will serve as a necessary piece of multimessenger astrophysics infrastructure to study the central engines that power stellar explosions, provide insights into the composition of relativistic outflows, and enable strict constraints on the timescales for jet formation and propagation.

%% file: LEAP_McConnell.tex
 \noindent
 \chapterauthor[1,2]{Mark McConnell}\orcidlink{0000-0001-8186-5978}
 \\
 \begin{affils}
   \chapteraffil[1]{Space Science Center, University of New Hampshire, Durham, NH 03824, USA}
    \chapteraffil[2]{Department of Earth, Oceans, and Space, Southwest Research Institute, Hanover, MD 21076, USA}
 \end{affils}

The LargE Area burst Polarimeter (LEAP) is designed to make the highest fidelity polarization measurements to date of the prompt gamma-ray emission from a large sample of Gamma-Ray Bursts (GRBs). The LEAP science objectives are met with a single instrument--- a wide-field-of-view (FOV) Compton polarimeter that measures GRB polarization over the energy range from 50--500~keV and performs GRB spectroscopy from 20~keV to 5~MeV.  If approved, it will be deployed as an external payload on the International Space Station (ISS) in 2027 for a three year mission~\cite{McConnell.2021,Onate-Melecio.2021}. LEAP measures polarization using seven independent polarimeter modules, each with a 12~$\times$~12 array of optically isolated high-Z and low-Z scintillation detectors read out by individual photomultiplier tubes (PMTs). LEAP was proposed in 2019 to the NASA Astrophysics Mission of Opportunity program and was subsequently selected for a Phase A study that was completed in 2021. After not being selected for flight, the LEAP mission was proposed again to the NASA Astrophysics  Mission of Opportunity program and we are now awaiting the outcome of that review.

The LEAP science investigation is based on the ability to distinguish between three classes of GRB models~\cite{Toma.20097gfh}. The baseline science investigation requires the observation of 65 GRBs with a Minimum Detectable Polarization (MDP) of 30\% or better. Evidence of polarized gamma-ray emission in GRBs ($>$100~keV) has been accumulated in recent years, but the limited sensitivity of these measurements does not yet render a clear picture of the underlying physics~\cite{McConnell.2017,Tatischeff.2019,Chattopadhyay.2021ny}. A sensitive and systematic study of GRB polarization, such as that will be provided by LEAP, is needed to remedy this situation.

Although the focus of the LEAP mission is on our understanding of GRBs through polarization measurements, important secondary science will also be enabled by LEAP. The discovery of gravitational waves (GW) associated with merging compact objects---and their association with GRBs---renders the detection of short GRBs of the utmost importance, especially as the GW detection efficiency continues to improve. With its large detection area, LEAP will be very effective in detecting and studying short GRBs and for correlating with other observations. Other science topics include magnetars (as represented by Soft Gamma-ray Repeaters; SGRs), accreting pulsars, and Terrestrial Gamma-ray Flashes (TGFs). LEAP can also measure gamma-ray polarization in solar flares with a sensitivity that far exceeds anything that has previously been achieved.

\begin{figure}[t!]
\begin{center}
\includegraphics[width=0.32\linewidth]{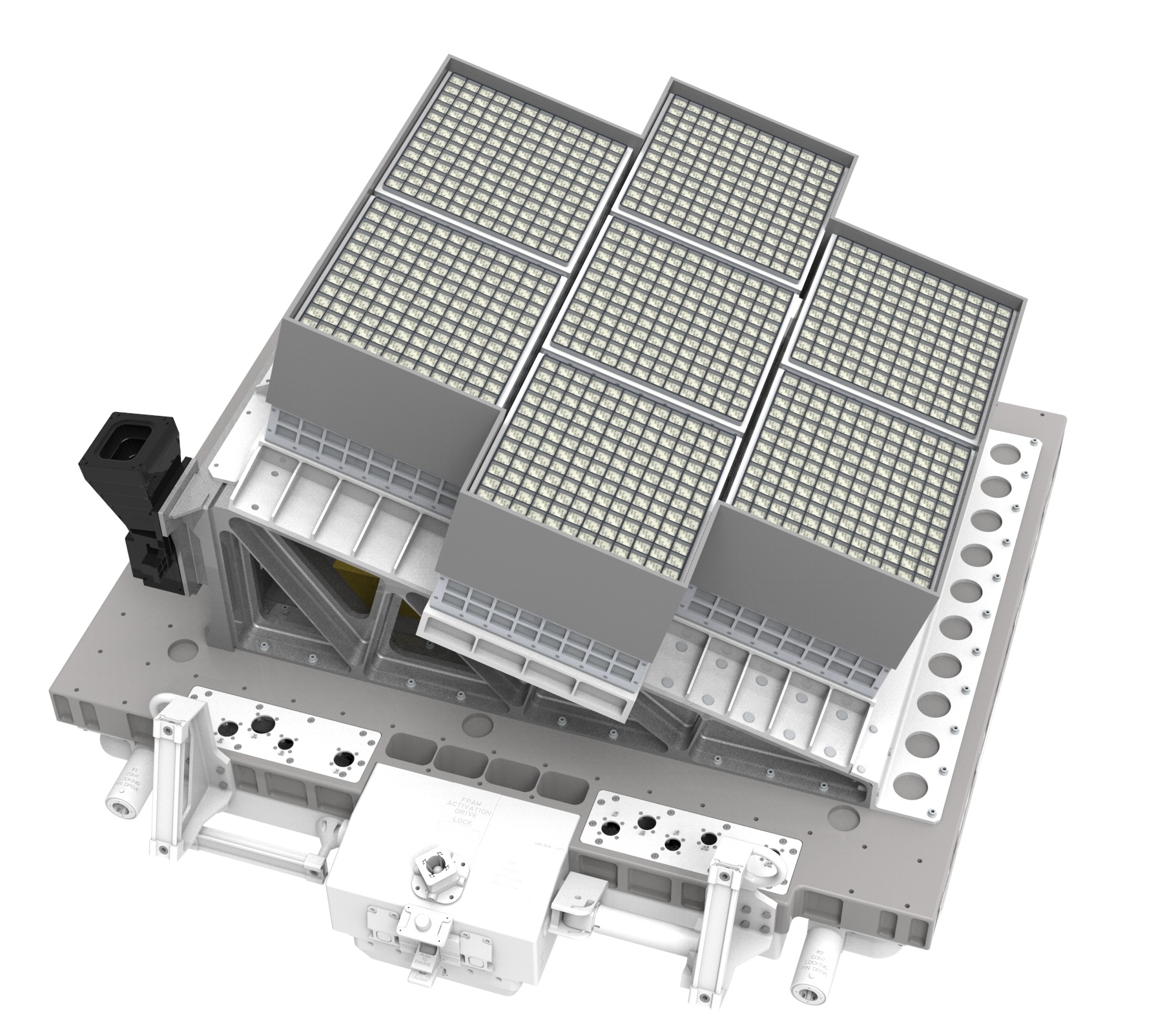}
\includegraphics[width=0.28\linewidth]{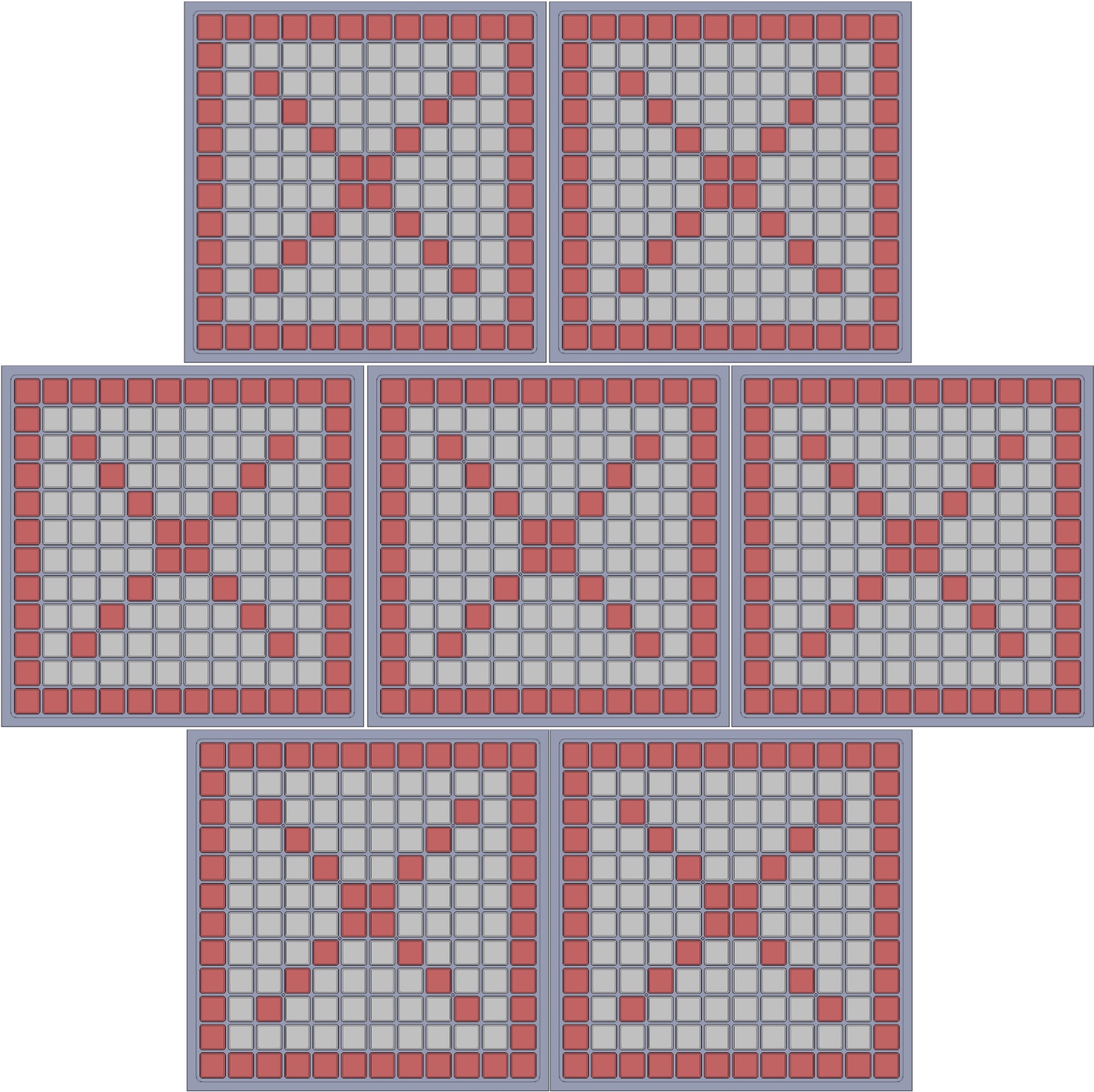} 
\includegraphics[width=0.38\textwidth]{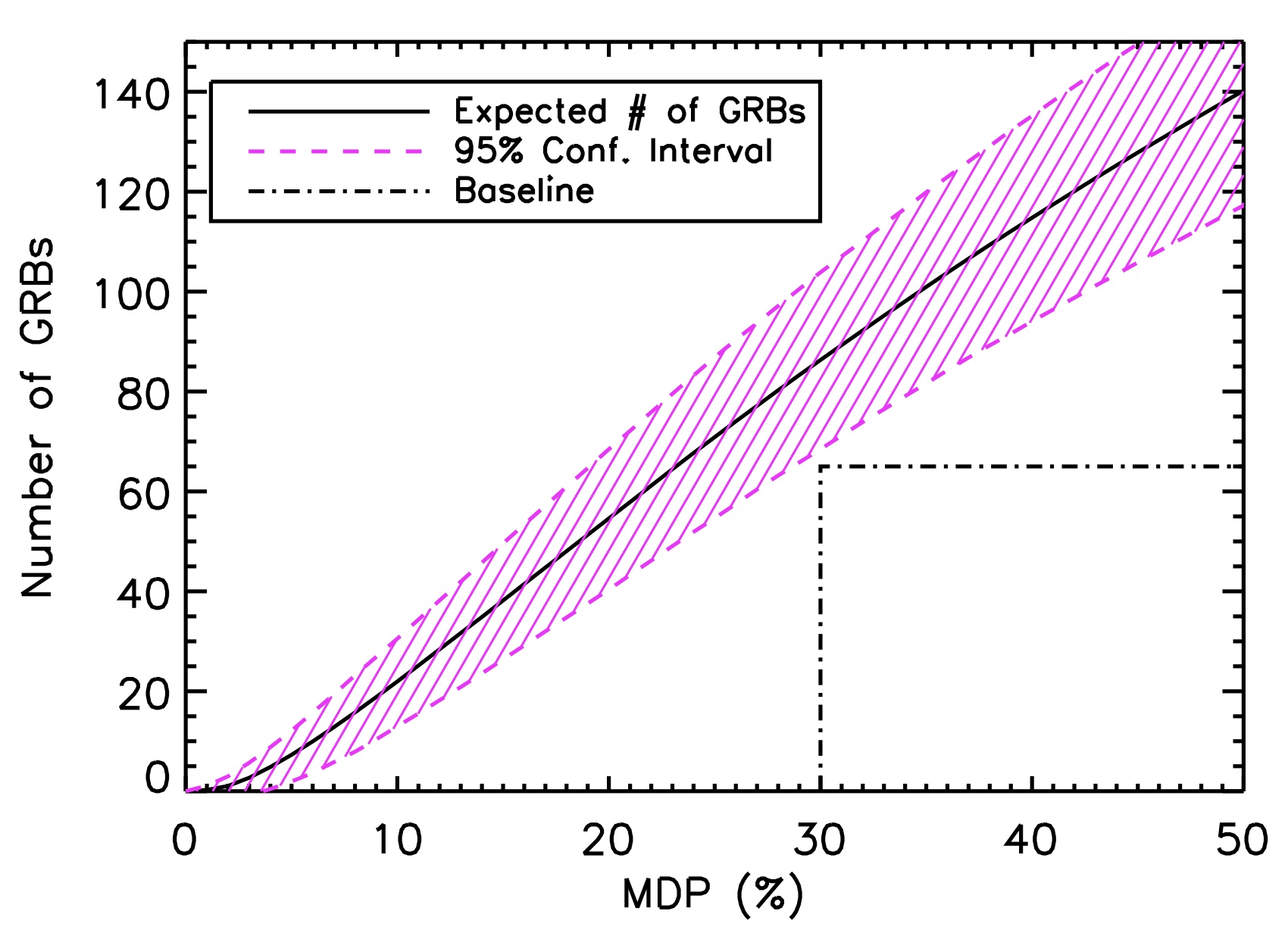}
\end{center}
\caption{ \label{fig:LEAP} \textit{Left:} The LEAP payload is designed to be mounted to one of the external attachment sites on the ISS. \textit{Middle:} The arrangement of high-Z (red) and low-Z (grey) scintillator elements is designed to optimize the polarization response. Source localization is determined from the relative response of all 420 CsI(Tl) detectors (red elements). \textit{Right:} In the baseline three-year mission, LEAP will measure 65 GRBs with a 50--300~keV MDP of 30\% or better. LEAP is also expected to measure 140 GRBs with an MDP of 50\% or better, and roughly 10 GRBs with an MDP of 5\% or better.}
\end{figure}

The LEAP payload (Figure~\ref{fig:LEAP}) consists of an array of seven independent polarimeter modules. As a Compton polarimeter, scatter events recorded by the scintillator array within each module are used to measure polarization. Within each module, the arrangement of plastic and Thallium-doped Cesium Iodide (CsI(Tl)) scintillation detectors (middle panel of Figure~\ref{fig:LEAP}) is designed to optimize the polarization response. The dominant type of scatter event is one involving only two detector elements, in which incident photons scatter from a low-Z plastic detector element into a high-Z CsI(Tl) element. The distribution of azimuthal scatter angles for these  events provides a polarization signature.  The total effective area for polarimetry is $\sim$1000~cm$^2$ at energies above 100~keV. Since the total energy deposit is a sum of the energy deposits in all triggered elements, spectroscopic information is provided by all types of events of any multiplicity (both singles event and scatter events). To characterize the GRB parameters, spectroscopic measurements (20--5000~keV) are obtained using all event types (both multiple and singles events), with a total effective area that reaches $>$3000~cm$^2$ between 50 and 500~keV.

To accurately reconstruct the source spectrum and polarization, LEAP self-sufficiently determines the source direction using singles events from all 420 CsI(Tl) elements (Figure~\ref{fig:LEAP}), whose relative response provides the source localization. Localization errors of $<$5$^{\circ}$ (1$\sigma$) are obtained at a rate of about 40 per year. This is sufficiently precise to enable rapid followup by many ground-based instruments using the rapid burst response messages that will be generated and distributed (in real time) by LEAP.

As a wide-FOV instrument, LEAP maintains some level of polarization sensitivity out to at least 75$^{\circ}$ off-axis, providing an effective FOV of $\sim1.5\pi$~sr.  Minimal obstructions within the FOV maximize sky exposure and minimize photon absorption and scattering effects. However, scattered flux from ISS structures (such as solar panels) must always be considered in the analysis of both the spectrum and the polarization.  
Response simulations spanning a range in energy, spectral shape, and incidence direction have been used in estimates of instrument performance. Figure~\ref{fig:LEAP} shows, for a three-year mission, the number of expected GRBs as a function of MDP. These estimates show that LEAP will attain its requirement of 65 GRBs with $<$30\% MDP.

%% file: COSI-Tomsick.tex
 \noindent
 \chapterauthor[]{John Tomsick}\orcidlink{0000-0001-5506-9855}
 \\
 \begin{affils}
   \chapteraffil[]{Space Sciences Laboratory, University of California, Berkeley, Berkeley, CA 94720, USA}
 \end{affils}

The Compton Spectrometer and Imager (COSI) is a wide-field telescope designed to survey the gamma-ray sky at 0.2--5~MeV~\cite{Tomsick:2021wed}. COSI has been selected as a NASA Small Explorer (SMEX) satellite mission with a planned launch in 2025. COSI is designed to have a very large field of view (FOV), covering $>$25\% of the sky instantaneously and the full sky every day. The large FOV is combined with excellent energy resolution ($<$1\%~FWHM), allowing for Galaxy-wide measurements of emission lines, including the electron-positron annihilation line at 0.511~MeV and nuclear lines at 1.157, 1.173, 1.333, and 1.809~MeV. In addition to imaging and spectroscopy, COSI will be capable of measuring the polarization of astrophysical sources such as gamma-ray bursts (GRBs) and accreting black holes.

COSI employs a novel design using a compact array of cross-strip germanium detectors (GeDs) to resolve individual gamma-ray interactions in the GeDs with high spectral and 3-dimensional spatial resolution, making COSI operate as a Compton telescope (see Section~\ref{sec:compton_tel}). The COSI array of 16 GeDs is housed in a common vacuum cryostat cooled by a mechanical cryocooler. The GeDs are read out by custom ASIC electronics integrated into the data acquisition system. An active bismuth germanate (BGO) shield encloses the cryostat on the sides and bottom to veto events from outside the FOV.

COSI addresses science discussed in Chapter~\ref{sec-fundamental} of this white paper. In particular, COSI will detect and localize short GRBs from merging neutron stars (Section~\ref{sec:GWlocalization}). For approximately ten short GRBs per year, COSI will be capable of obtaining positions with 1-degree uncertainties and reporting them $<$1 hour after they are detected. In addition, the BGO shields allow for simple detection of short GRBs and extend the FOV to well over 50\% of the sky. With arrival time determinations, this is relevant for the speed of gravity goals discussed in Section~\ref{sec:speedofgravity}. A COSI launch in 2025 would mean that it is observing when A+ generation gravitational wave detectors are planned to be in operation. In the A+ era, the joint binary neutron star (BNS)-GRB detection range is 620~Mpc (see Table~4 in Ref.~\citenum{Burns20}) for the three LIGO interferometers (HLI). A distance of 620~Mpc corresponds to z~=~0.13, and 14\% of short GRBs are within that redshift~\cite{Burns20}. 

With its polarization sensitivity, COSI will address science discussed in Section~\ref{sec:polarimetry}. COSI will study long and short GRBs to distinguish between candidate emission mechanisms and geometries~\cite{Toma09}. Another application of polarization measurements of GRBs is to test for Lorentz Invariance Violation (LIV), which is predicted to occur in some theories of quantum gravity (see Section~\ref{sec:speedofgravity}). LIV can be probed by making gamma-ray polarization measurements of high-redshift GRBs because the vacuum birefringence effect causes a wavelength-dependent rotation of the polarization angle~\cite{GleiserKozameh01, Stecker11}. Thus, with LIV, a GRB above a certain redshift should not be polarized. The highest-redshift GRB with a polarization measurement is at z~=~2.739~\cite{Gotz14}, and COSI has the potential to probe LIV by constraining polarization of high-z GRBs.

COSI will determine the distribution of 0.511~MeV emission from positron annihilation in the Galaxy. There are likely many different types of sources that contribute to the positrons in the Galaxy. We know that nucleosynthesis products such as Al-26 are positron sources and that positrons are produced in stellar flares based on solar measurements.  However, the positron totals are unclear, leaving room for the possibility that some of the 0.511~MeV emission traces exotic physics such as annihilating MeV dark matter~\cite{Ema21} and primordial black holes~\cite{Laha19}, which are discussed in Sections~\ref{sec:darkmatter} and \ref{sec:primordealBH}, respectively. In addition to 0.511~MeV emission, COSI will allow for a search for an emission line from the annihilation of candidate dark matter particles.

The cosmic gamma-ray background (CGB) at MeV energies encodes the signature of some of the most interesting source classes, but its origin is controversial. Active Galactic Nuclei (AGN) are one component, but it is unclear what fraction of the MeV emission they can explain. While another source could be gamma rays escaping from cosmological SNIa, this picture has been challenged based on updated estimates of the cosmic star formation rate density~\cite{HoriuchiBeacom10, RuizLapuente16}, leaving open the possibility that the MeV-CGB could contain a significant contribution from more exotic sources such as dark matter annihilation~\cite{Ahn05}. COSI has the capacity to detect the tail for some of the brightest Seyferts~\cite{Zdziarski00}, which will improve the determination of the AGN contribution, leading to an understanding of whether exotic physics is needed.

%% file: AMEGO-Kierans.tex
\chapterauthor[ ]{ }

 \addtocontents{toc}{
     \leftskip3cm
    \scshape\small
    \parbox{5in}{\raggedleft Carolyn Kierans, Regina Caputo, et al.}
    \upshape\normalsize
    \string\par
    \raggedright
    \vskip -0.19in
    }

\noindent
\nocontentsline\chapterauthor[]{Carolyn Kierans$^1$\orcidlink{0000-0001-6677-914X}}
\nocontentsline\chapterauthor[]{Regina Caputo$^1$\orcidlink{0000-0002-9280-836X}}
\nocontentsline\chapterauthor[]{Julie McEnery$^1$}
\nocontentsline\chapterauthor[]{Jeremy S. Perkins$^1$\orcidlink{0000-0001-9608-4023}}
\\
\begin{affils}
\chapteraffil[1]{NASA Goddard Space Flight Center, Greenbelt, MD 20771, USA}
\end{affils}

\begin{figure}[htb]
\begin{minipage}[c]{0.47\textwidth}
\centering
\includegraphics[width=0.9\textwidth]{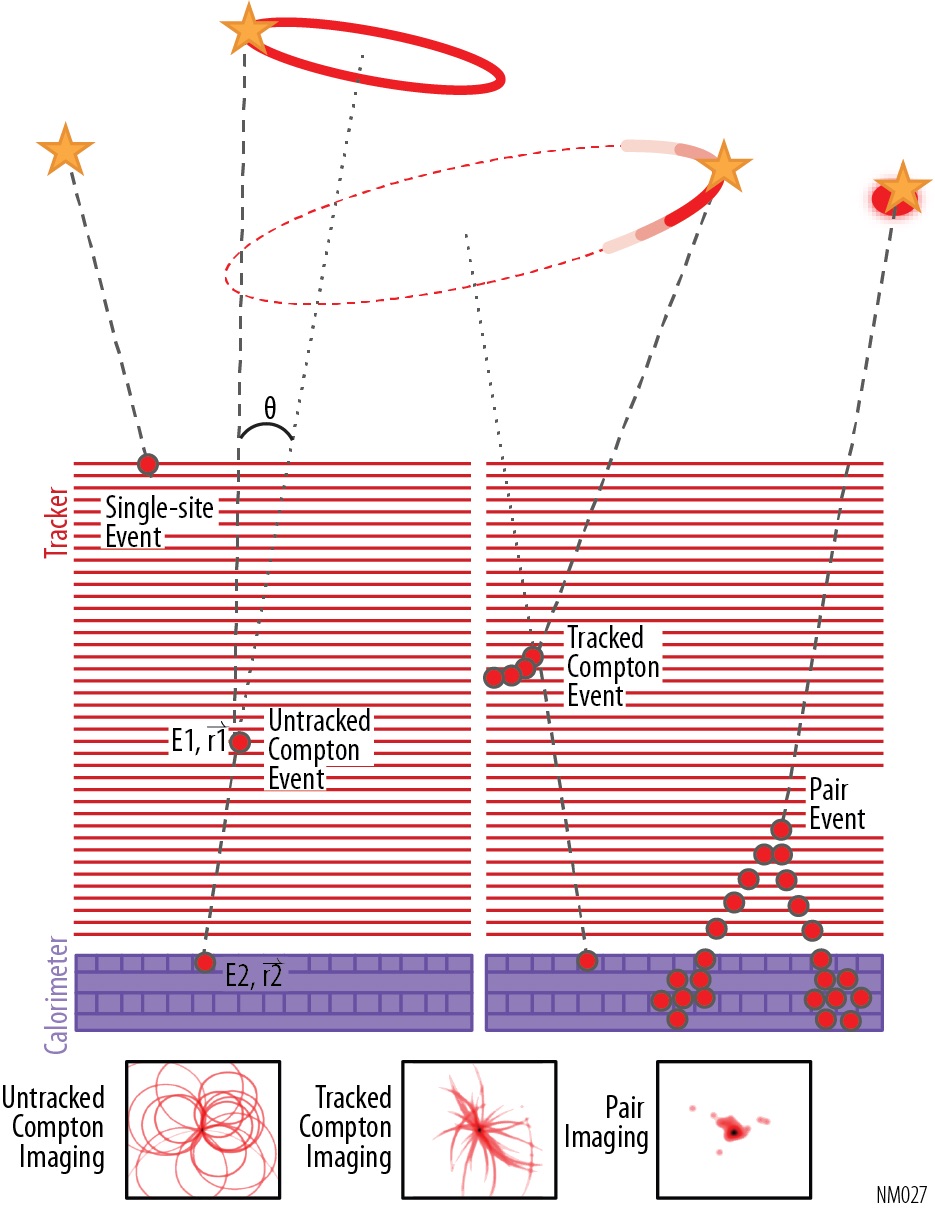}
\end{minipage}\hfill
\begin{minipage}[c]{0.52\textwidth}
\caption{AMEGO and AMEGO-X detect gamma rays through both Compton scattering and pair production. Photons with energy less than $\sim$10~MeV will Compton scatter in the Tracker and then undergo photoabsorption in the Calorimeter. The position and energy of each interaction constrain the initial direction to a circle on the sky, as shown for the Untracked Compton Events. If the direction of the Compton-scattered electron is measured, as in Tracked Compton Events, the incoming photon direction is reduced to an arc. Higher-energy photons predominately undergo electron-positron pair conversion and the measured interactions in the Tracker reconstruct the initial photon direction.}
\label{fig:amego1}
\end{minipage}
\end{figure}

A major push for high-energy astrophysics in the coming decade is the detection and study of multimessenger sources (see Chapter~\ref{sec-fundamental}). Multimessenger sources, such as merging neutrino stars or flaring super-massive black holes, are naturally gamma-ray sources due to the extreme energies and processes involved. As the last few years have shown us, detections and observations of the gamma-ray counterparts have played a critical role in the identification and study of all multimessenger sources to date~\cite{grb170817, txs0506}. While all-sky monitoring for transients is necessary, equally necessary is the long-term monitoring of the gamma-ray sky with imaging capabilities for steady-state sources similar to the capabilities of the \textit{Fermi}-Large Area Telescope (LAT) above $\sim$100~MeV~\cite{lat}. The lower MeV gamma-ray range is uniquely important since that is where the emission of known multimessenger sources peak. To enable this science, we need a telescope designed to fill the ``MeV Gap'' in sensitivity by operating as both a Compton and pair telescope (see Sections~\ref{sec:compton_tel} and \ref{sec:pair}, respectively). An instrument with unprecedented continuum sensitivity in the MeV range, with Fermi-LAT-like all-sky monitoring capabilities, will target the gamma-ray counterparts to multimessenger sources and will be a key contributor to multimessenger astrophysics. Our group at NASA Goddard Space Flight Center (GSFC) has been pursing two distinct but similar mission concepts in this vein: AMEGO and AMEGO-X. 

\begin{figure}[tb]
\centering
\includegraphics[height=6.5cm]{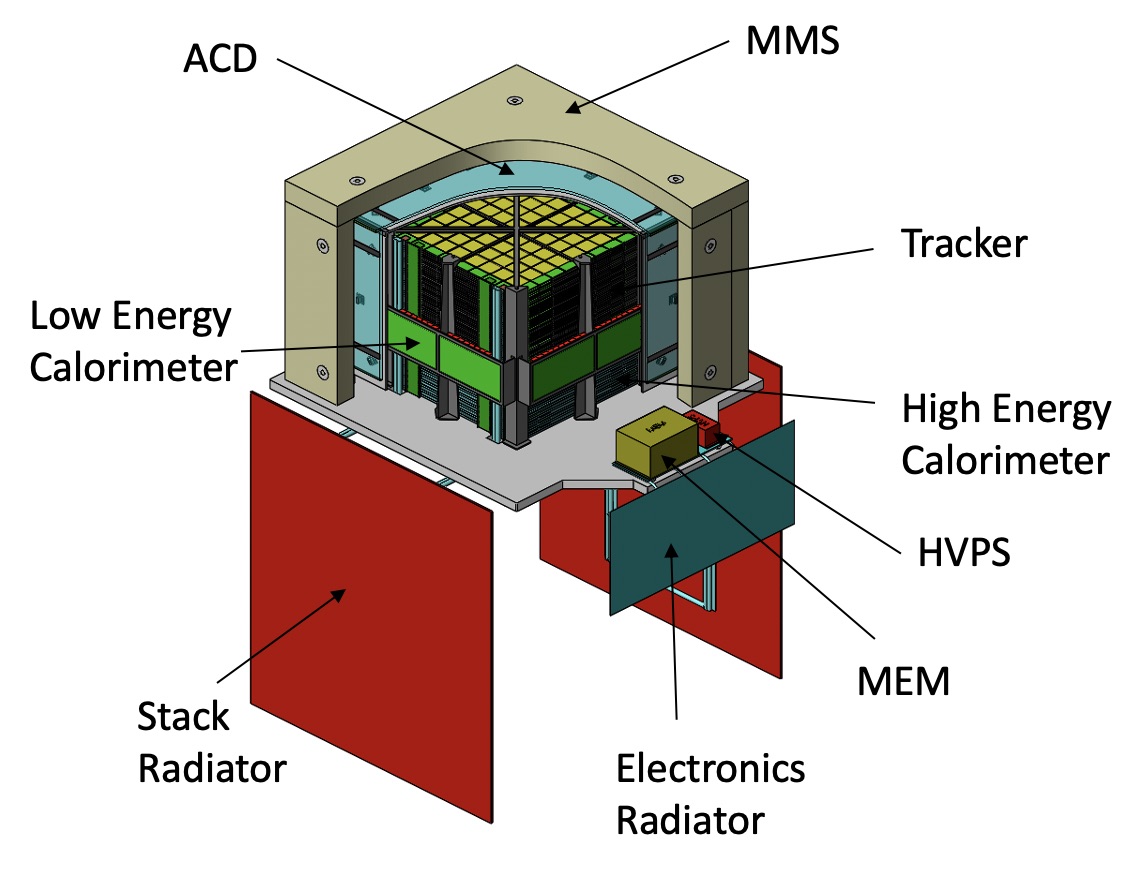}
\includegraphics[height=7.5cm]{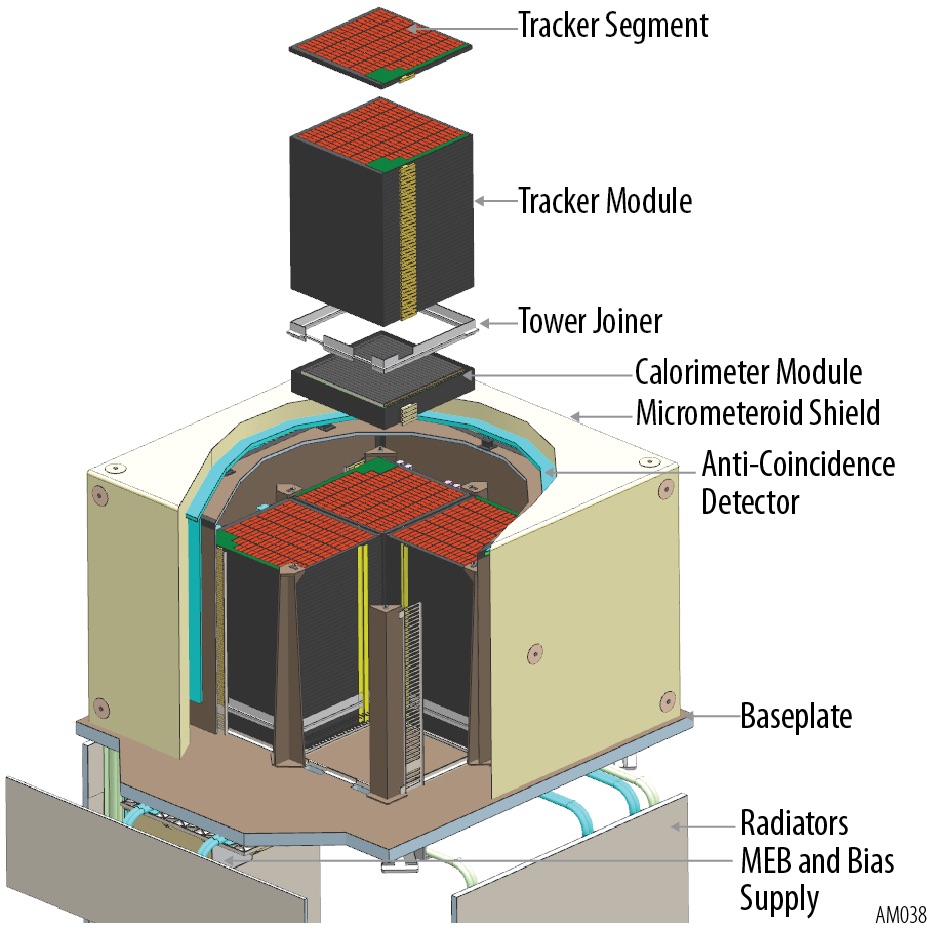}
\caption{\textit{Left:} The AMEGO Probe-class instrument consists of four main subsystems: the silicon Tracker, the CZT Low Energy Calorimeter, the CsI High-energy Calorimeter, all surrounded by the plastic ACD. A micrometeroid shield (MMS) surrounds the entire instrument. \textit{Right:} The AMEGO-X MIDEX mission consists of four towers wtih 40 layers of monolithic pixelated silicon detectors, and four layers of a CsI hodoscopic Calorimeter.}
\label{fig:amego2}
\end{figure}

The All-sky Medium Energy Gamma-ray Observatory (AMEGO) is a Probe-class concept that was submitted to the Astro2020 decadal survey~\cite{AMEGORFI, 2020SPIE11444E..31K}. The general design is similar to \textit{Fermi}-LAT, with Figure~\ref{fig:amego1} showing the detection principles of AMEGO. Four detector subsystems work together to detect and characterize photons across over four orders of magnitude in energy. Photons first interact in the silicon Tracker, which consists of 60 layers of double-sided silicon strip detectors (DSSDs; see Section~\ref{sec:dssd}). The Tracker acts as the scatterer and converter for Compton and pair events, respectively. Surrounding the bottom of the Tracker is the cadmium zinc telluride (CZT) Low-Energy Calorimeter (see Section~\ref{sec:VFG-CZT}), which has excellent spatial and spectral resolution to measure the Compton-scattered photons and low-energy pair showers. This novel subsystem also can operate as a stand-alone Compton detector, which dramatically increases the detection efficiency at low energies. The cesium iodide (CsI) High-Energy Calorimeter is at the bottom of the tower and measures the electromagnetic shower from high-energy pair events. A plastic Anti-Coincidence Detector (ACD) surrounds the other detector subsystems to reject the charge-particle background in orbit. Figure~\ref{fig:amego2}~\textit{Right} shows the full AMEGO instrument.

To fit within the Medium Explorer (MIDEX) cost box, we have developed the AMEGO eXplorer (AMEGO-X) mission concept with a similar architecture to AMEGO but slightly smaller in size. The AMEGO-X science goals are focused on the gamma-ray counterpart measurements of merging neutrons stars and flaring blazars. To enable this science, AMEGO-X utilizes emerging monolithic pixelated silicon CMOS technology (see Section~\ref{sec:astropix}) to enhance the low-energy transient response beyond even AMEGO capabilities~\cite{MartinezCastellanos2021}. The full instrument, shown in Figure~\ref{fig:amego2}~\textit{Left}, includes 40 silicon Tracker layers, and the CsI High-Energy Calorimeter similar to the AMEGO design but with two fewer layers of CsI bars. AMEGO-X does not include the expensive CZT subsystem. AMEGO-X was submitted to the NASA MIDEX call in December 2021 and, if selected, will be transformative for multimessenger astrophysics through its detection of gamma-ray counterparts~\cite{Lewis2021, Negro2021}.

AMEGO and AMEGO-X are optimized for continuum sensitivity in the MeV range in a number of ways. The design builds off the \textit{Fermi}-LAT heritage but includes advanced technology to enable Compton-event detection in the MeV range. This done in the Tracker by 1) the omission of the tungsten conversion foils that are used in the \textit{Fermi}-LAT, and 2) 2D position sensitive silicon detectors in each layer. AMEGO uses DSSDs for single-layer position measurements instead of the single-sided silicon strip detectors used in \textit{Fermi}-LAT. AMEGO-X uses monolithic silicon pixel detectors for a lower energy threshold and higher low-energy effective area than AMEGO. AMEGO also has the Low-Energy Calorimeter that enhances the response in the Compton-regime to therefore enhance the polarization and narrow-line sensitivity. For both instruments, the high-energy CsI Calorimeter is readout with silicon photomultipliers (SiPMs) which can achieve significant improvement in the energy and depth resolution for each CsI bar~\cite{woolf2018}. These enhancements allow for AMEGO and AMEGO-X to probe the full MeV Gap and bring light to this new era of multimessenger astrophysics.

%% file: GECCO-Moiseev.tex
 \noindent
 \chapterauthor[1,2]{A.A. Moiseev}
 \\
 \begin{affils}
   \chapteraffil[1]{University of Maryland, College Park, College Park, MD 20742, USA}
   \chapteraffil[2]{NASA Goddard Space Flight Center, Greenbelt, MD 20771, USA}
 \end{affils}

The MeV energy range offers great potential for astrophysics discovery in areas of nucleosynthesis, multimessenger/gravitational waves, jets, and compact objects. Among them, there are several unresolved problems connected with the dynamic structure and composition of the inner Galaxy, including the Galactic Center region and active star-forming regions that require high spatial resolution to address: the nature of unassociated \textit{Fermi} Large-Area Telescope (LAT) sources (approximately one third of all detected sources, primarily in the Galactic plane), the nature of \textit{Fermi}-eROSITA (extended ROentgen Survey with an Imaging Telescope Array) Bubbles, the origin of the 511~keV positron annihilation line, the origin of still enigmatic dark matter. High-sensitivity measurements of nuclear lines in the MeV region will also lead to resolving Galactic chemical evolution and sites of explosive element synthesis. Compton Spectrometer and Imager (COSI), recently selected for the Small Explorer medium gamma-ray mission and described in Section~\ref{sec:cosi}, is expected to provide important inputs~\cite{Tomsick, Zoglauer}.

High angular resolution and good spectral resolution are particularly valuable in these studies. Observations of the diffuse Inverse Compton component of the interstellar emission that dominates at MeV energies will allow us to obtain the spatial distribution of low-energy cosmic-ray electrons, their sources, their propagation and acceleration, and their relation to the interstellar medium, which cannot be performed by any other type of instrument due to the contamination of this component with other emission mechanisms. High-resolution observations in the MeV range can also shed light on the dynamics of Galactic winds, on the mechanisms of transport in the low-energy cosmic rays, and eventually on the role of low-energy cosmic rays on Galaxy evolution and star formation.

To date, astrophysical observations in this energy band have been limited by low cross sections and gamma-ray attenuation lengths that restrict one to indirect imaging (e.g., Compton and Coded Aperture), which when coupled with high background, limit sensitivity. Compton telescopes (Section~\ref{sec:compton_tel}) provide good, low-noise performance and allow for a wide field of view (FoV), but Doppler broadening fundamentally limits angular resolution to $\sim$1~degree. Conversely, Coded Mask telescopes (Section~\ref{sec:codedmask_tel}) can achieve high angular resolution ($\lesssim$1~arcminute) but have no inherent background rejection. Combining a coded aperture mask with an imaging detector that is also a Compton telescope, will dramatically increase the sensitivity and spatial resolution of the combined instrument. This approach has been demonstrated in simulations~\cite{Aprile, Galloway}, and tested with INTEGRAL/IBIS (the INTErnational Gamma-Ray Astrophysics Laboratory/Imager on-Board the INTEGRAL Satellite) data~\cite{Forot}, but the mature concept has never been implemented as the central motivation for a telescope design.

We are developing an Explorer-class mission concept, the Galactic Explorer with a Coded Aperture Mask Compton Telescope (GECCO) that uses the merged Coded Mask/ Compton techniques to employ benefits of two imaging modalities: superior angular resolution provided by the Coded Mask with good background rejection and wide FoV, both provided by a Compton telescope~\cite{Orlando, gecco1}. GECCO observations will extend arcminute angular resolution to high-energy images of the Galactic plane, combining the MeV spectral capabilities of INTEGRAL/IBIS and the X-ray imaging of NuSTAR (the Nuclear Spectroscopic Telescope Array~\cite{2010arXiv1008.1362H}) and eROSITA, and will make a bridge to the \textit{Fermi}-LAT observations, enabling a broad potential for discoveries in the MeV gamma-ray sky. The motivation for GECCO was intensified by the long-awaited detection of the gravitational waves and their counterpart in gamma rays, and also by the first detection of high-energy neutrino by Ice Cube, coincident with a \textit{Fermi}-LAT blazar. This was highlighted by the recent Astro2020 decadal survey~\cite{NAP26141} ``Pathways to Discovery in Astronomy and Astrophysics for the 2020s,'' which advocated for ``temporal monitoring of the sky across the electromagnetic spectrum.''

\begin{figure}[tb]
    \centering
    \includegraphics[width=\textwidth]{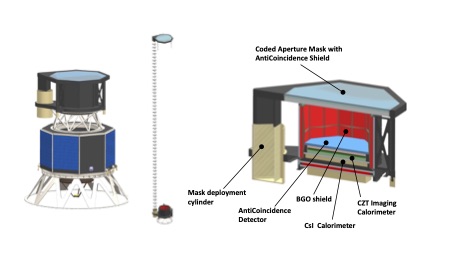}
    \caption{\textit{Left:} GECCO with Mask in stowed position, and notional SC bus; \textit{Middle:} GECCO with Mask deployed at 20m; \textit{Right:} GECCO, cutaway.}
    \label{fig:gecco}
\end{figure}

GECCO will consist of six subsystems: the CZT (cadmium, zinc, telluride) Imaging Calorimeter (see Section~\ref{sec:VFG-CZT}), a CsI (Cesium Iodide) Calorimeter, a bismuth germanate (BGO) active shield, a plastic scintillator Anticoincidence detector, a Coded Aperture Mask with plastic scintillator Anticoincidence shield, and the data-acquisition system (DAQ). In GECCO, the Imaging Calorimeter will operate as a standalone Compton telescope to detect incident photons in the energy range from 200--100~keV to $\sim$10~MeV by measuring their energy and incident direction. The CsI Calorimeter supports the Imaging Calorimeter by measuring the energy and interaction positions of radiation escaping from the Imaging Calorimeter. Monte Carlo simulations of the instrument have been performed with the MEGAlib~\cite{MEGALIB} toolkit. A similar approach is used in COSI (COmpton Spectrometer and Imager~\cite{Tomsick}), recently selected as a Small Explorer NASA mission--- see Section~\ref{sec:cosi}. To achieve arcminute angular resolution, a coded-aperture mask is placed at 20~m distance from the Imaging Calorimeter, being deployed after reaching orbit. The Imaging Calorimeter operates as a focal plane detector for the reconstruction of the coded-aperture mask image. In this concept, the instrument aperture between the Imaging Calorimeter and Coded Mask is not shielded, unlike in INTEGRAL/IBIS and other Coded Mask telescopes, and the excessive background is suppressed by selecting for analysis-only events that might have originated from the coded mask location according to their measured Compton-scattered directions. This unique feature of GECCO greatly improves its angular resolution while maintaining a high signal-to-noise ratio in coded mask imaging.

GECCO can operate in either scanning or pointed mode. In scanning mode, it will observe the Galactic Plane. It will change to pointed mode to either increase observation time for special regions of interest, (e.g., the Galactic Centre) or to observe transient events such as flares of various origins or gamma-ray bursts. The expected GECCO performance is as follows~\cite{Orlando}: energy resolution $<1$\% at 0.5--5~MeV, and angular resolution $\sim0.5$~arcmin in mask mode with $5^\circ$ FoV and $4$--$8^\circ$ in the Compton mode with $\sim80^\circ$ FoV. The effective area varies from $200~\mathrm{cm}^2$ to $\sim2000~\mathrm{cm}^2$, depending on the energy. The $3\sigma$, 1~Ms continuum (point-source) sensitivity conservative estimate is $\sim10^{-5}~\mathrm{MeV} \mathrm{cm}^{-2}\mathrm{s}^{-1}$.

%% file: GRAMS-Tsuguo.tex
 \noindent
 \chapterauthor[]{Tsuguo Aramaki}\orcidlink{0000-0002-5104-4682}
 \\
 \begin{affils}
   \chapteraffil[]{Department of Physics, Northeastern University, Boston, MA 02115, USA}
 \end{affils}

\begin{figure}[htb]
	\centering
	\includegraphics[width=\columnwidth]{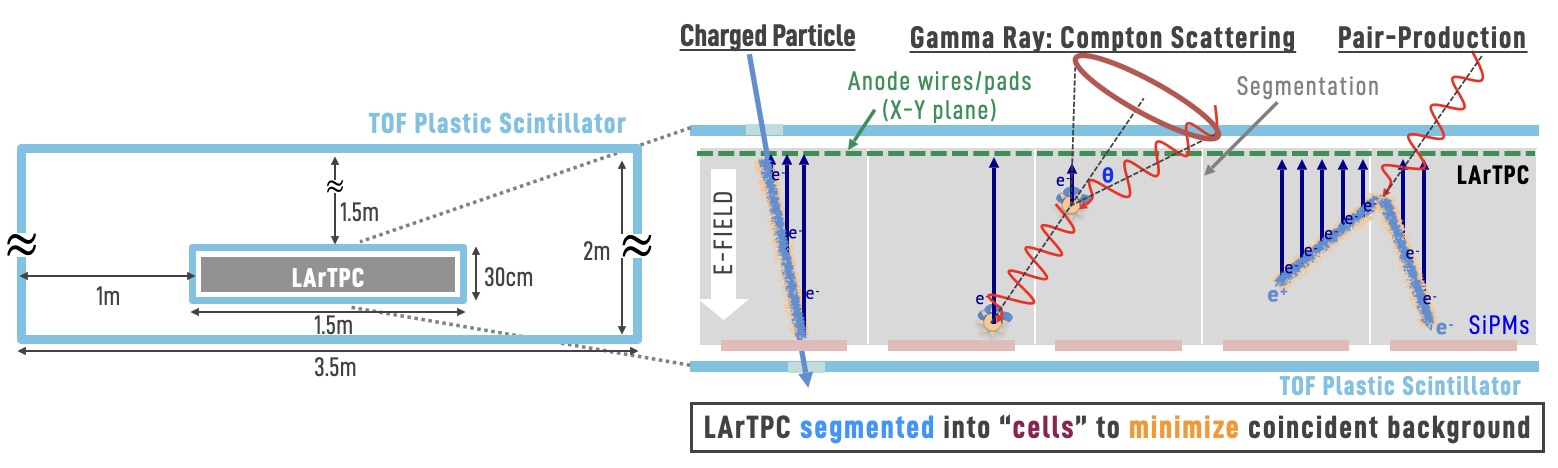}
	\caption{GRAMS instrumental design: LArTPC surrounded by plastic scintillators. Figure adapted from Ref.~\citenum{ARAMAKI2020107}.}
	\label{fig:grams}
\end{figure}

The Gamma-ray and AntiMatter Survey (GRAMS~\cite{Aramaki:2020gqm, ARAMAKI2020107}) project is a proposed next-generation balloon/satellite mission that simultaneously targets both astrophysical observations with MeV gamma rays and indirect dark matter searches with antimatter. The MeV gamma-ray energy range has long been under-explored due to the lack of large-scale detectors to reconstruct Compton scattering events efficiently. GRAMS aims to break through existing technological barriers while utilizing a cost-effective, large-scale Liquid Argon Time Projection Chamber (LArTPC) detector as a Compton camera. The LArTPC technology, successfully developed for underground dark matter and neutrino experiments over the last two decades, provides three-dimensional particle tracking capability by measuring ionization charge and scintillation light produced by particles entering or created in the argon medium. 

GRAMS will provide an affordable, scalable, and full-sky-reach solution for a Compton telescope concept with the LArTPC. The GRAMS instrumental design includes a 1.4~m~$\times$ 1.4~m~$\times$~20~cm LArTPC surrounded by two layers of plastic scintillators. The plastic scintillators veto charged particles while the LArTPC works as a Compton camera. The LArTPC volume will be segmented into small cells, localizing the signal and minimizing the coincident background events. GRAMS will be able to provide an order of magnitude improved sensitivity to MeV gamma rays with a single long-duration balloon flight, while the GRAMS satellite mission could provide another order of magnitude sensitivity. The GRAMS detector configuration also offers extensive sensitivities to antinuclei potentially produced by dark matter particles.

%% file: GammaTPC-Shutt.tex
 \noindent
 \chapterauthor[]{Thomas Shutt}
 \\
 \begin{affils}
   \chapteraffil[]{Stanford University, Stanford, CA 94305, USA}
 \end{affils}

\begin{figure}[tb]
    \centering
    \includegraphics[width=0.9\textwidth]{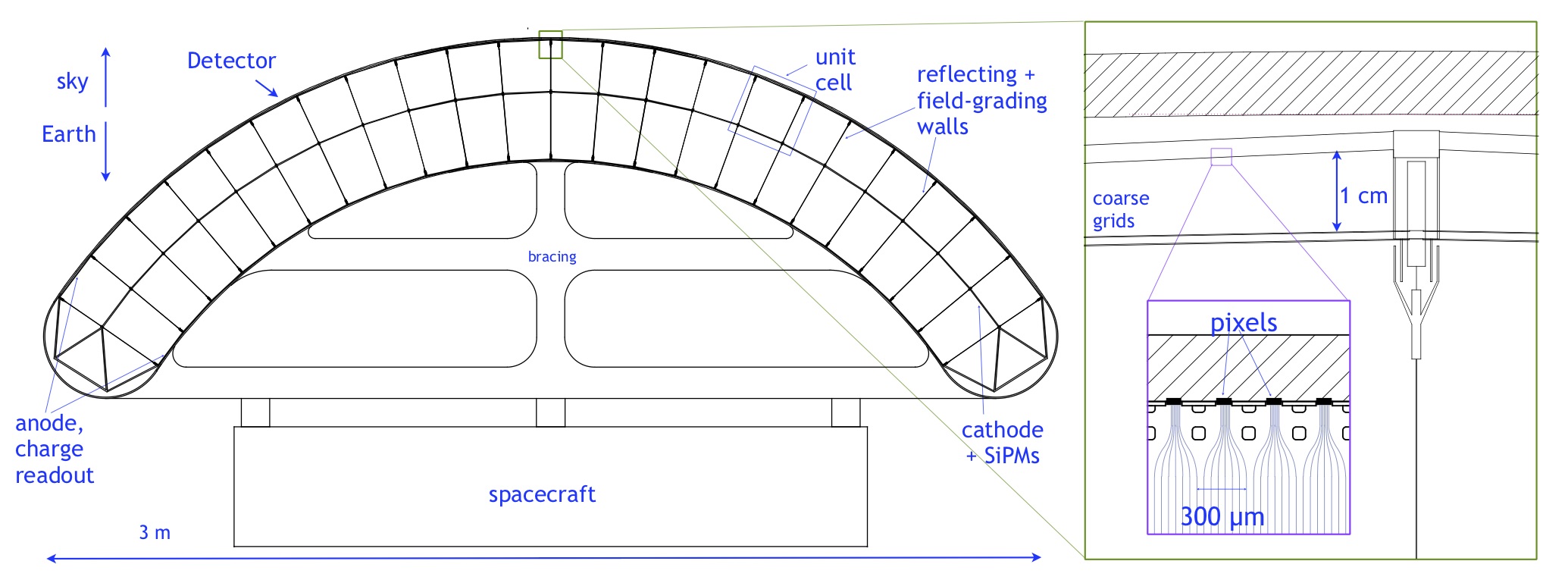}
    \caption{Cross section schematic of a 10~m$^2$, four-metric-tonne implementation of the GammaTPC concept, along with details of the dual-scale charge readout system. A 40~cm thick layer of LAr is contained in a thin-walled carbon fiber shell, along with a lightweight interior cellular readout structure. The overall geometry is a section of a sphere.}
    \label{fig:gammatpc_schematic}
\end{figure}

GammaTPC is an instrument concept for a very large, high sensitivity 0.1--10~MeV gamma-ray satellite based on liquid Argon (LAr) time-projection chamber (TPC) technology, as shown schematically in Figure~\ref{fig:gammatpc_schematic}. The underlying technology is discussed in Section~\ref{sec:LAr_tech_Shutt}. The curved geometry provides a very large field of view with relatively uniform response and also minimizes the mass of the carbon fiber pressure vessel. The LAr target is segmented as shown to reduce event pileup. An interior cathode plane is covered with silicon photomultipliers (SiPMs) to measure the scintillation signal, while the charge is read out on anode planes on the interior walls. A 10~kV voltage difference between anode and cathode establishes a 0.5~kV/cm charge drift field. The lateral segmentation is provided by thin reflecting walls. The overall detector thickness, here 40~cm, is chosen for efficiency for gamma rays up to $\sim10$~MeV.   

\begin{figure}[h!t]
    \centering
    \includegraphics[width=0.325\textwidth]{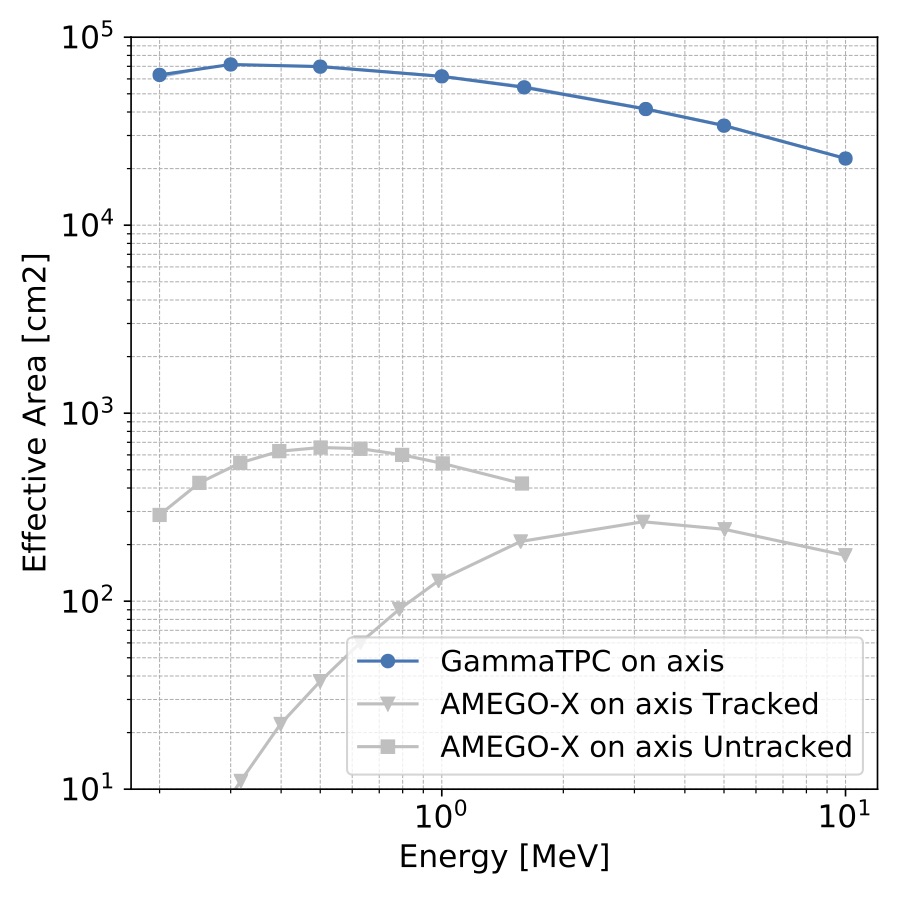} 
    \includegraphics[width=0.325\textwidth]{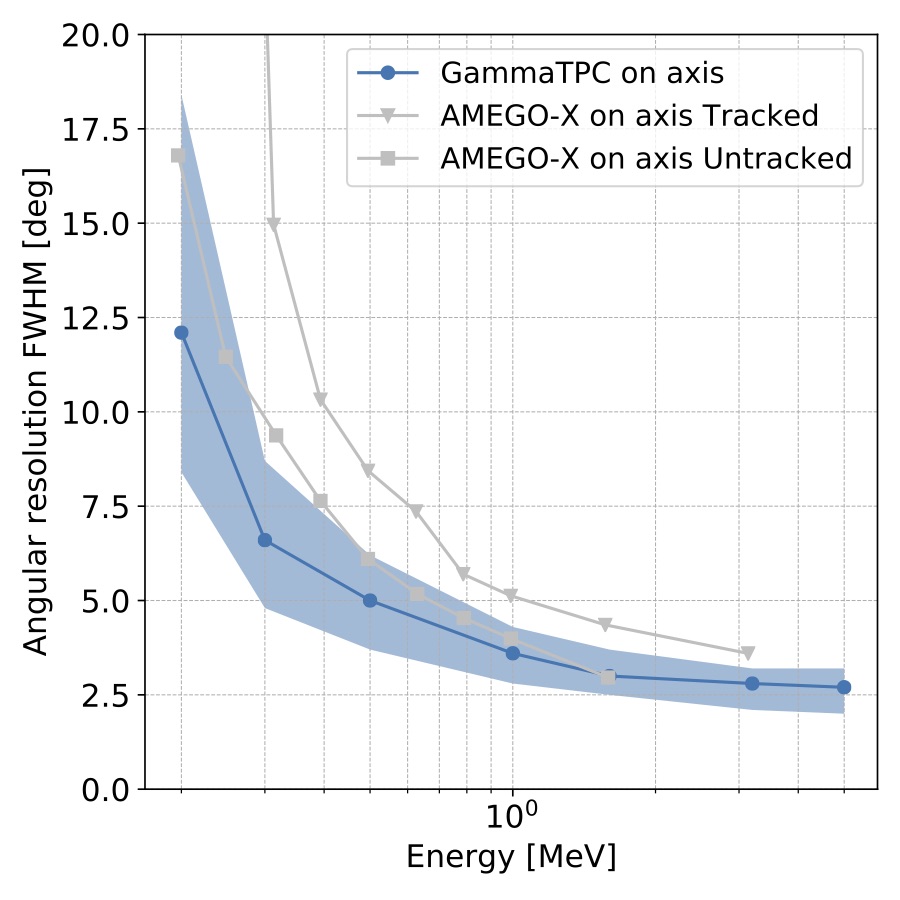}
    \includegraphics[width=0.325\textwidth]{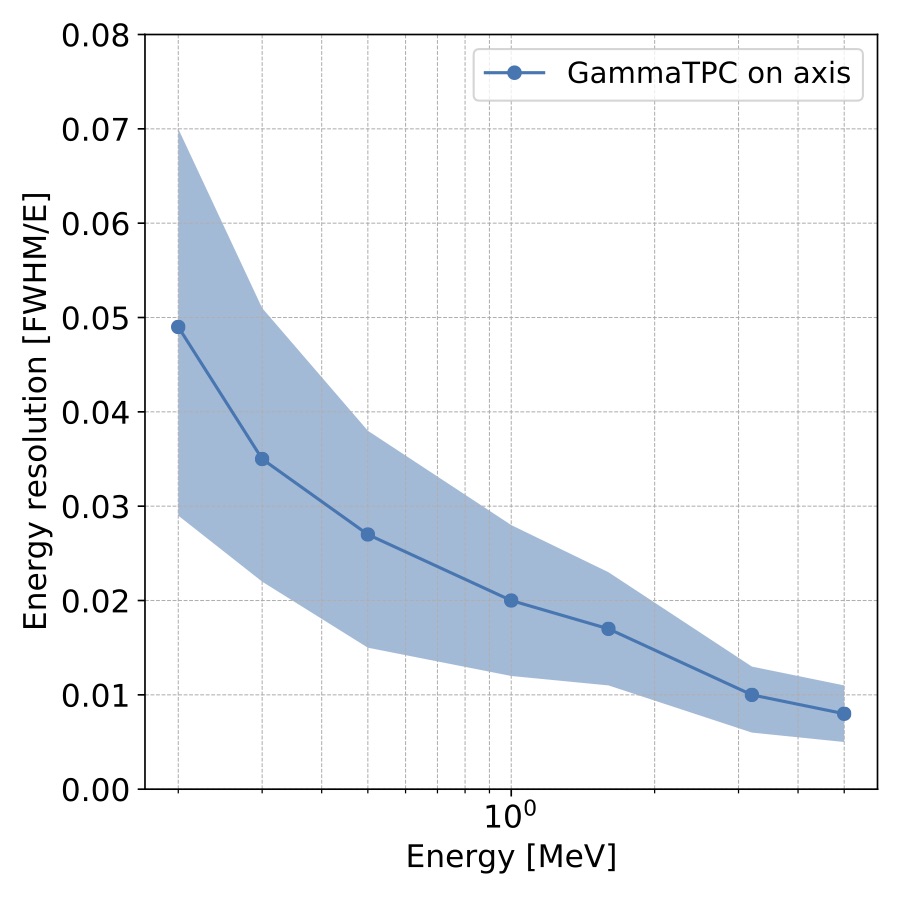}
    \caption{Effective area (left), angular resolution measure (ARM) parameter (center) and FWHM of the Energy Resolution (right). The blue bands representing the GammaTPC concept are constructed over a range of assumed readout performance parameters. The grey lines and points show the predicted performance of the AMEGO-X~\cite{Fleischhack_2021} concept for comparison. The AMEGO-X results for the effective area and angular resolution show events for which the Compton electron is tracked and those for which it is untracked separately.}
    \label{fig:irfs}
\end{figure}

We have carried out initial studies of instrument performance using MEGAlib~\cite{MEGALIB} and an idealized instrument with the same 40~cm thickness and 10~m$^2$ area as in Figure~\ref{fig:gammatpc_schematic}, but a planar shape for simplicity. An important driver of performance is use of the dual scale charge readout system described in Section~\ref{sec:LAr_tech_Shutt}. We included a treatment of the complex response of a LAr TPC, with the energy partitioned into quanta of charge and light, which are each measured with a certain efficiency and readout noise.  We used the {\tt Revan} tool from MEGALib to perform pattern recognition and Compton event reconstruction and to estimate standard instrument performance metrics for MeV telescopes, including the effective area, the angular resolution measure (ARM), and the energy resolution, shown in  Figure~\ref{fig:irfs}.  

The pointing performance and energy resolution look to be at least as good as for Si, with the pointing resolution benefiting from the $\sim$400~$\mu$m position resolution afforded by the novel fine-grained charge readout. A final estimate of instruments sensitivity will require a detailed study of the backgrounds, which has not yet been performed. Much of the background rejection comes from accurate event reconstruction, which should be good given the highly uniform target material, the good energy and spatial resolution, and the electron track directional information provided by the fine charge readout (see Figure~\ref{fig:electron_imaging}). Neutrons are very efficiently tagged by scintillation pulse shape in LAr~\cite{Ajaj:2019imk} (a fact that is central to the use of LAr for dark matter searches). From the known Ar activation cross sections for protons~\cite{Brodzinski_1955, Saldanha_2019} and neutrons~\cite{Saldanha_2019} and Low Earth Orbit (LEO) particle fluxes~\cite{Cumani_2019}, we estimate the resulting gamma rays to be a factor of $\sim$5 below the astrophysical signal, which will be further reduced by event reconstruction. LAr is also effectively immune to radiation damage.

While the final evaluation of the sensitivity of the instrument awaits these detailed background studies, the results in Figure~\ref{fig:irfs} already show that this technology promises to be transformational. The pointing performance and energy resolution look to be at least as good as for Si, but the real benefit is the effective area which is some two orders of magnitude larger than AMEGO-X~\cite{Fleischhack_2021} (see Section~\ref{sec:amego}). The combination of these will provide both superb transient sensitivity in the multi-messenger era, as well as an unprecedented all sky map. AMEGO-X is a very compelling instrument which will hopefully fly soon. Development of GammaTPC will enable a transformative leap in sensitivity beyond AMEGO-X in the future.   

This sensitivity promises to come at a comparatively low cost. The technology heavily leverages the enormous DOE investment in LAr technology for the Deep Underground Neutrino Experiment (DUNE)~\cite{Sacerdoti_2022} and advances in LAr and Liquid Xenon (LXe) for dark matter. The very low mass of the carbon fiber shell and readout structures ($\sim$150~kg each in the four tonne instrument in Figure~\ref{fig:gammatpc_schematic}) will also translate to a low cost for the instrument hardware. We are targeting development for a Medium Explorer (MIDEX) satellite mission.

%% file: CTA-Williams.tex
\chapterauthor[ ]{ }

 \addtocontents{toc}{
     \leftskip3cm
    \scshape\small
    \parbox{5in}{\raggedleft David A. Williams}
    \upshape\normalsize
    \string\par
    \raggedright
    \vskip -0.19in
    }

\noindent
\nocontentsline\chapterauthor[]{David A. Williams$^1$}\orcidlink{0000-0003-2740-9714}
\nocontentsline\chapterauthor[]{on behalf of the CTA Consortium$^2$}
 \\
 \begin{affils}
    \chapteraffil[1]{Santa Cruz Institute for Particle Physics and Department of Physics, University of California, Santa Cruz, Santa Cruz, CA 95064, USA}
    \chapteraffil[2]{\texttt{www.cta-observatory.org}}
 \end{affils}

\begin{figure} 
\centering
	\includegraphics[width=\textwidth]{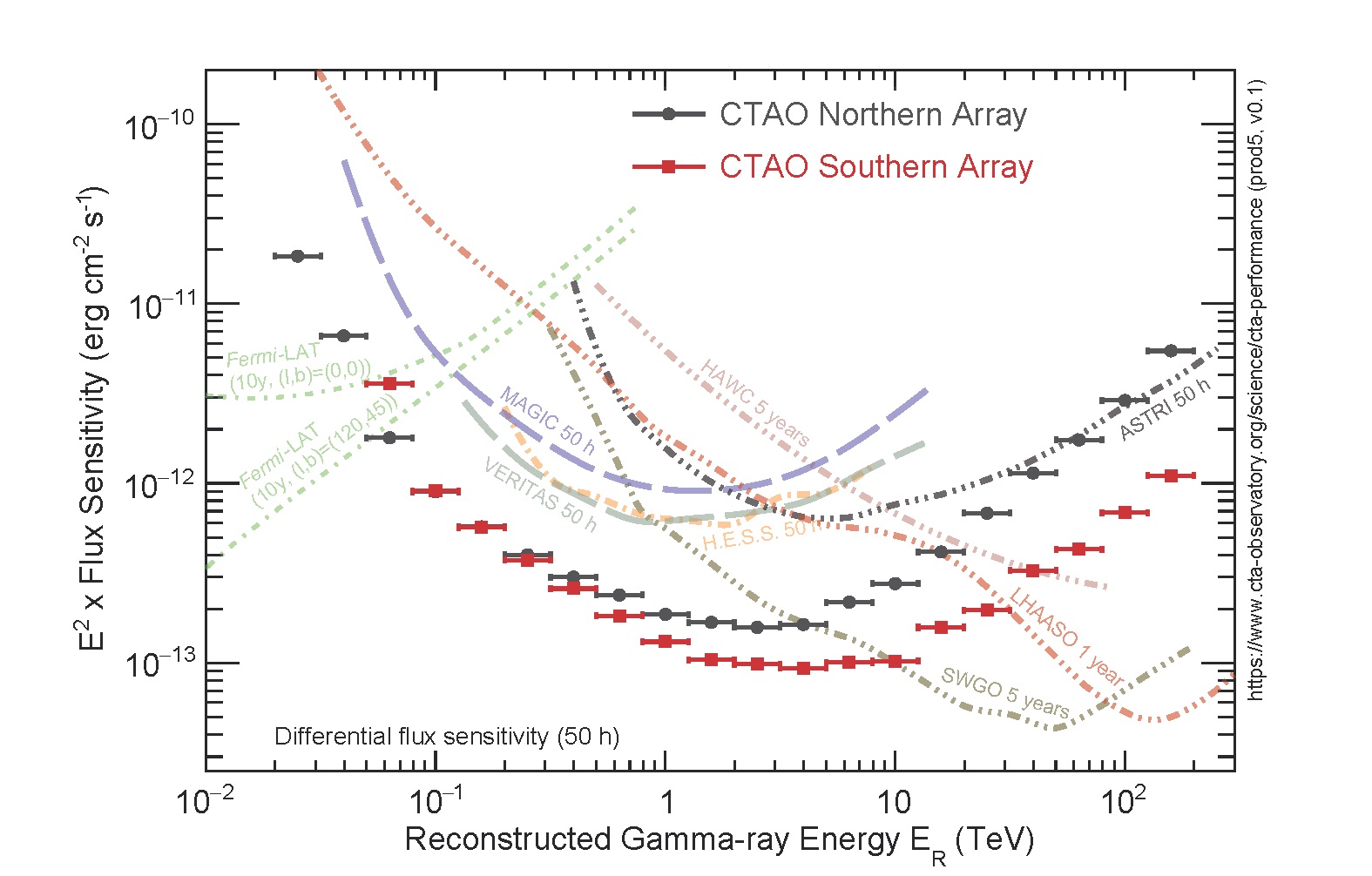}
	\caption{The differential sensitivity of the Cherenkov Telescope Array Observatory Northern and Southern Arrays in the ``Alpha Configuration,'' together those of current-generation IACT arrays MAGIC~\cite{2016APh....72...61A}, VERITAS~\cite{Holder:2006gi}, and H.E.S.S.~\cite{Aharonian:2006pe}; the planned ASTRI array~\cite{Vercellone:2015fza}; current and future Water Cherenkov Detector (WCD) arrays HAWC~\cite{Abeysekara_2019}, LHASSO~\cite{2019arXiv190502773B}, and SWGO~\cite{Albert:2019afb,Abreu:2019ahw,Hinton:2021rvp,Schoorlemmer:2019gee}; and {\it Fermi}-LAT \cite{lat} (both on and off the Galactic plane).}	
	\label{fig:ctao_alpha_sensitivity}
\end{figure}

As introduced briefly in Section~\ref{sec:atmospheric_cherenkov}, the Cherenkov Telescope Array (CTA) will be a next generation Imaging Atmospheric Cherenkov (IACT) array, using telescopes of multiple sizes to achieve state-of-the-art sensitivity in the 20~GeV--300~TeV energy range~\cite{2011ExA....32..193A,CTAConsortium:2013ofs,2019scta.book.....C}. It will be constructed and operated by the Cherenkov Telescope Array Observatory (CTAO), which is currently in the process of transitioning from a German non-profit corporation (GmbH) to a European Research Infrastructure Consortium (ERIC). Observatory installations in the Southern and Northern Hemispheres---within the grounds of the European Southern Observatory in the Atacama desert in Chile and on the Canary Island of La Palma in Spain, respectively---will provide visibility to the entire sky. Construction is about to start, with completion expected in five years. CTAO will be the first facility in this energy band operated as a open observatory rather than a principal investigator experiment: Competition for observing time will be open to scientists from all countries contributing to the costs of CTAO construction and operation.

An ``Alpha Configuration'' of CTAO will be constructed with funds committed in the ERIC application to be submitted to the European Union this year (2022). The southern installation will have 14 medium-sized telescopes (MSTs) of 12~m aperture (similar in size to the current VERITAS~\cite{Holder:2006gi} (Very Energetic Radiation Imaging Telescope Array System) and original H.E.S.S.~\cite{Aharonian:2006pe} (High-Energy Stereoscopic System) telescopes) covering the 100~GeV--10~TeV energy range and 37 small-sized telescopes (SSTs) of $\sim$4~m aperture with sensitivity from a few TeV to 300~TeV. In the north, the Alpha Configuration will have four large-sized telescopes (LSTs) with 23~m aperture to achieve a threshold of 20~GeV and nine MSTs. The projected sensitivity of the CTAO Alpha Configuration is shown in Figure~\ref{fig:ctao_alpha_sensitivity} together with other facilities in this energy range. The sensitivity of the CTAO Alpha Configuration is up to an order of magnitude better than existing IACT arrays, and significantly better at all energies. 

\begin{figure} 
\centering
	\includegraphics[width=\textwidth]{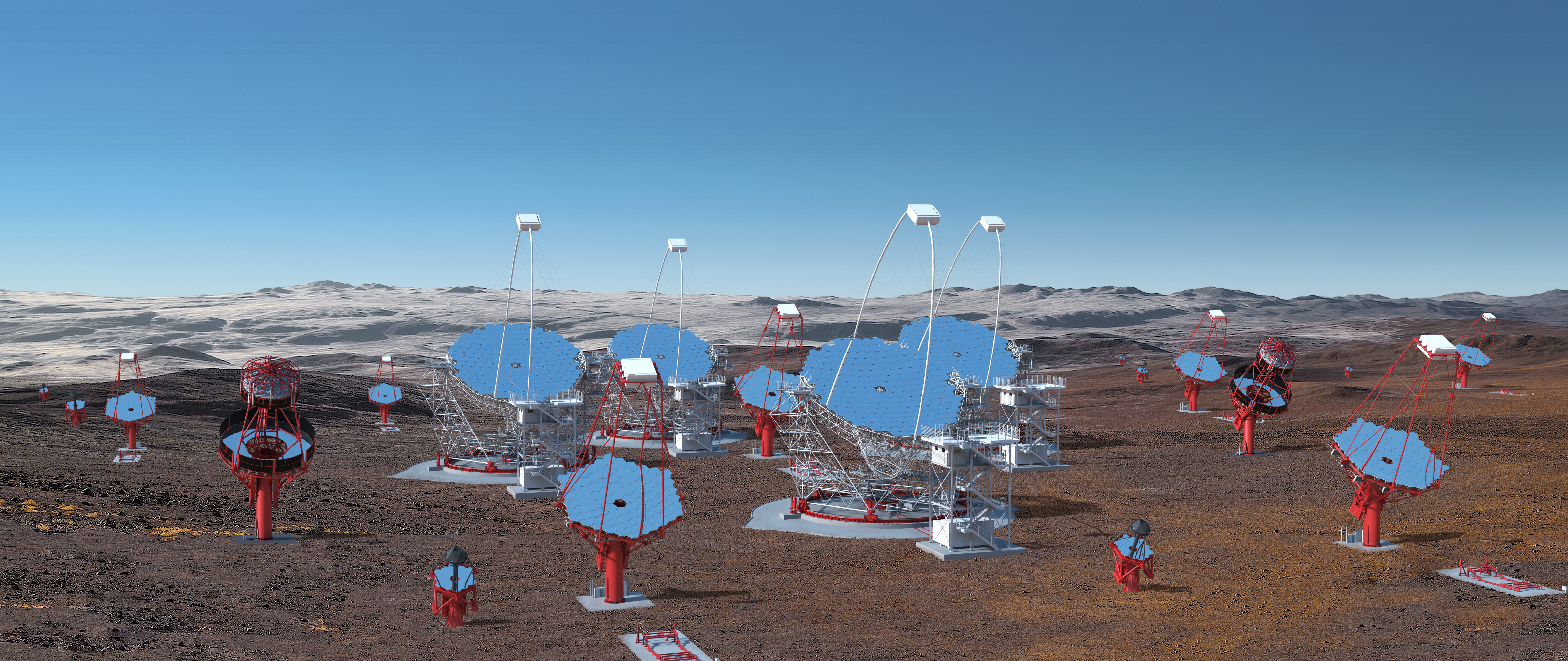}
	\caption{Artist's concept of the southern array of the Cherenkov Telescope Array Observatory, based on renderings of the actual telescope designs. In addition to the 14 medium-sized telescopes and 37 small-sized telescopes of the Alpha Configuration, the concept shows four large-sized telescopes at that center of the array and several Schwarzschild-Couder medium-sized telescopes (with a black baffle around the primary mirror) that could be added as an enhancement.}	
	\label{fig:ctao_south}
\end{figure}

It is hoped that additional funding---including from the U.S.---will allow the addition of $\sim$11 
medium-sized telescopes and four LSTs in the south (as well as several MSTs in the north), as shown in Figure~\ref{fig:ctao_south}. The additional medium-sized telescopes could use the novel Schwarzschild-Couder design, pioneered in the U.S. and prototyped at the F.L. Whipple Observatory in Arizona. This design introduces a secondary mirror for the first time in an IACT, correcting aberrations and reducing the image plate scale, yielding unprecedented optical performance across the full 8$^\circ$ field of view and enabling a high-resolution camera with 11,328 silicon photomultiplier pixels to leverage this performance. Preliminary estimates find that adding 11 Schwarzschild-Couder Telescopes (SCTs) to the southern CTAO array increases the point-source sensitivity in the 100~GeV--10~TeV energy range by as much as a factor of 1.8. U.S. support at the mid-scale level for such an enhancement to CTAO has been recommended by the Astro2020 Decadal Survey~\cite{NAP26141} as part of the Multi-messenger Program for the 2020s.

Especially if enhanced by the addition of telescopes to the southern array, CTAO will have the sensitivity to make critical advances on many of the science questions discussed in Chapter~\ref{sec-fundamental}, as recognized already in the previous Snowmass study~\cite{Snowmass2013:2013cqj} and subsequent P5 recommendations~\cite{P5Report:2014pwa}. In particular, observations of the Galactic Center region will probe dark matter annihilation for the first time at the level of the thermal relic cross section for particle masses ranging from roughly 200~GeV to 10~TeV and a variety of possible annihilation final states~\cite{CTA:2020qlo}. 
Observations of the propagation of photons over extragalactic distances will allow the extragalactic background light to be measured out to redshift $z = 2$ with a statistical uncertainty of less than 15\%, intergalactic magnetic field strengths above 0.3~pG to be probed, limits on Lorentz Invariance Violation to be improved by at least a factor of two or three, and Axion-like particles to be constrained in the parameter space where they could contribute to---or be---the dark matter~\cite{CTA:2020hii}. CTAO will also be able to search for evaporating primordial black holes and measure the very-high-energy electron-plus-positron spectrum above 5 TeV.

In summary, CTAO will have unique capabilities for exploring fundamental physics with very-high-energy gamma rays that are essential for a comprehensive approach to some of the most important questions. These capabilities will be strengthened, and U.S. scientists will have access to them, if the U.S. participates in a significant way in CTAO construction and operations, as recommended by the Astro2020 Decadal Survey.

%% file: SWGO-Engel.tex
\chapterauthor[ ]{ }

 \addtocontents{toc}{
     \leftskip3cm
    \scshape\small
    \parbox{5in}{\raggedleft Kristi Engel, J. Patrick Harding, Chad Brisbois}
    \upshape\normalsize
    \string\par
    \raggedright
    \vskip -0.19in
    }

\noindent
 \nocontentsline\chapterauthor[]{Kristi Engel$^{1,2}$}\orcidlink{0000-0001-5737-1820}
 \nocontentsline\chapterauthor[]{J. Patrick Harding$^{1,3}$}\orcidlink{0000-0001-9844-2648}
 \nocontentsline\chapterauthor[]{Chad Brisbois$^{2}$}\orcidlink{0000-0002-5493-6344}
 \nocontentsline\chapterauthor[]{on behalf of the SWGO Collaboration$^4$}
 \\
 \begin{affils}
    \chapteraffil[1]{Physics Division, Los Alamos National Laboratory, Los Alamos, NM, 87545, USA}
    \chapteraffil[2]{University of Maryland, College Park, College Park, MD, 20742, USA}
    \chapteraffil[3]{Michigan State University, East Lansing, MI, 48824, USA}
    \chapteraffil[4]{\texttt{https://www.swgo.org/SWGOWiki/doku.php?id=collaboration}}
 \end{affils}

The success of the water-Cherenkov technology (Section~\ref{sec:water_cherenkov}) implemented by the High-Altitude Water Cherenkov (HAWC) Observatory~\cite{Abeysekara_2019} (Puebla, Mexico) inspired the ambitious Chinese-lead Large High-Altitude Air-Shower Observatory (LHAASO)~\cite{2019arXiv190502773B} (Sichuan, China),  which uses a similar approach. The high duty cycle (near 100\%) of such survey instruments provides a unique opportunity to build long-term light curves for known active very-high-energy (VHE) emitters, such as Markarian 421 and 501~\cite{2017ApJ...841..100A}, and to alert Imaging Air Cherenkov Telscopes (IACTs) if bright TeV flares are detected~\cite{2017ApJ...843..116A}. Beyond their unique capabilities, Water Cherenkov Detector (WCD) arrays such as HAWC can therefore act as surveying and high-uptime monitoring instruments to provide alerts for IACTs to perform follow-ups. This monitoring capability extends that provided by GeV space telescopes, like \emph{Fermi}-LAT\cite{2009ApJ...697.1071A}, to the TeV range. The distinct geographic locations of current ground-based instruments dedicated to observations in the VHE gamma-ray band enable observations of the entire VHE gamma-ray sky with complementary capabilities
. HAWC and LHAASO will provide wide-field coverage of the Northern VHE gamma-ray sky in the coming years, but no wide-field detector yet exists to cover the Southern sky.

We propose the Southern Wide-field Gamma-ray Observatory~\cite{Albert:2019afb} (SWGO) as a next-generation WCD instrument that will provide this observational coverage of the Southern sky with nearly continuous up-time and an instantaneous field of view (FoV) of $\sim$2~sr at energies from $\gtrsim$100~GeV to above hundreds of TeV from a site in the Andes mountains of South America. The world-leading limits on fundamental physics and new TeV sources and source class discoveries of HAWC~\cite{2015ApJ...800...78A,2017ApJ...841..100A,2017A&A...607A.115I,2017ApJ...842...85A,2017ApJ...843...39A,2017ApJ...843...40A,2017ApJ...843...88A,2017ApJ...843..116A,2017ApJ...848L..12A,2017Sci...358..911A,2018ApJ...853..154A,2018JCAP...02..049A,2018JCAP...06..043A,2018Sci...361.1378I, 2018PhRvD..98l3012A,2018Natur.562...82A,2019JCAP...08..023F, 2019arXiv190512518H,2020MNRAS.497.5318F}, the excellent sensitivity expected from LHAASO at the highest energies~\cite{He:2019Qx}, and the lessons learned from both HAWC and LHAASO will inform the SWGO design, opening new opportunities for searches of gamma-ray sources in the Southern sky up to the PeV energy range.

The key science objectives for SWGO (detailed in Refs.~\citenum{Hinton:2021rvp,Schoorlemmer:2019gee,Albert:2019afb}) cover a variety of fundamental physics topics, providing new insights into several topics and performing critical all-sky continuous monitoring in the Southern Hemisphere to provide real-time alerts to pointed instruments such as CTA. The major requirements for studying each of these subjects informs the development of the design of SWGO. 

\begin{description}
\item[Probe Lorentz Invariance Violation (LIV).]
SWGO's wide FoV and sensitivity to the highest energies will make it an ideal instrument with which to search for signatures of LIV (Section~\ref{sec:speedofgravity}). Features of LIV processes can be revealed at TeV-scale energies when looked at over cosmological distances. Superluminal violations of Lorentz Invariance, for example, allow for the spontaneous decay of multi-TeV photons. To search for these violations requires an observatory that can view as many of the highest-energy astrophysical sources as possible to the highest energies, such as SWGO~\cite{2019BAAS...51c.272H,Abreu:2019ahw}.

\item[Search for Counterparts to Gravitational Wave (GW) Events.]
The foundation for GW multimessenger observation was laid with binary neutron star merger GW170817~\cite{GW170817} by the observations of the associated GRB170817A and the subsequent Kilonovae emission across the electromagnetic spectrum~\cite{GW170817_HESS,GW170817_MMA}. Since GW data analysis is inherently complex, the time needed to produce the first sky localization of GW events will exceed the minute scale throughout the foreseeable future~\cite{2019BAAS...51c.357S} (Section~\ref{sec:GWlocalization}). Even with a prompt alert, without precise localization, a pointed instrument could take less than ideal data on a counterpart emission from a GW, meaning that the wide FoV and large, unbiased duty cycle of SWGO introduces a new, ideal method for GW follow-up in the Southern Hemisphere. Should the alert be delayed, causing IACTs to miss the gamma-ray emission entirely, SWGO's ability to search archival data will be of the utmost importance. A wide-FoV ground-based observatory like SWGO will be able to record real-time, high-energy gamma-ray data for all GW events falling into its FoV without having to enforce selection criteria as IACTs must due to their limited amount of observation time~\cite{CTA_ScienceTDR}. SWGO will not only be able to ccover these high-uncertainty reegions, but also potentially locate VHE counterparts with smaller uncertainty (to the benefit of CTA and other IACTS)~\cite{2019BAAS...51c.357S}. Combined with its near-continuous duty cycle, these abilities of SWGO translate into an abundance of opportunities for scientific discovery, including currently unmodeled burst-like GW signals.

\item[Search for Neutrino VHE Counterparts.]
As evidence for the interaction of high-energy hadronic particles, high-energy neutrinos are crucial in the search for sources of high-energy cosmic-ray accelerators, but a multimessenger approach is necessary to determine the source of the astrophysical flux (Section~\ref{sec:neutrinos}). This approach, towards which SWGO's large, unbiased duty cycle will be an invaluable tool, greatly aids neutrino studies as high-energy gamma-ray observations allow precise localization of the emission region, allowing for not only the identification of their astrophysical source, but also information about overall source energetics and energy- and time-dependent power output. Such searches have already yielded a promising first result--- the detection of the flaring blazar TXS~0506+056
in coincidence with the high-energy neutrino IceCube-170922A~\cite{IceCube:2018dnn}. Ideally furthering this effort, SWGO will be the only instrument that could provide the necessary unbiased, quasi-continuous light curves for such blazars in the Southern sky, as well as monitor for sources that are spatially consistent with the neutrino direction over a wide range of time scales~\cite{Albert:2019afb}, with no limit on the number of candidates that can be observed.

\item[Search for Dark Matter (DM) Annihilation/Decay.]
Targeting the inner Galaxy/Galactic Center, SWGO will be capable of reaching the expected cross-section for DM particles thought to be thermal relics of the Big Bang for a wide range of DM particle mass, annihilation channel, and DM-halo profile shape~\cite{Viana:2021smp,Hinton:2021rvp} (Section~\ref{sec:darkmatter}). The DM flux spectra are characterized by a hard cutoff at the DM mass--- observing such a cutoff would be one of the strongest indications that an observed gamma-ray source originates from DM interactions. SWGO achieves peak sensitivity at the energy scale where these cutoffs would be apparent for multi-TeV-mass DM making its wide FoV to detect such signals and sensitivity to the highest energies ideal tools in the ongoing search for DM.

\item[Search for Primordial Black Holes (PBHs) and Axion-Like Particles (ALPs).]
Towards the ever-compelling challenge to elucidate the true nature of M, SWGO will be capable of performing beyond-the-Standard-Model physics searches for the signatures from evaporating PBHs (Section~\ref{sec:primordealBH}). Gamma-ray observatories can search for PBHs by looking for bursts of high-energy gamma rays created by the last seconds of PBH evaporation via Hawking radiation~\cite{Hawking:1974rv}. As these would appear as transient phenomena to a VHE gamma-ray observatory, SWGO's wide FoV and ability to search archival data would be crucial in the search for such signatures. The predicted sensitivity of SWGO towards the local burst rate density of PBHs is an order of magnitude deeper than with current VHE searches~\cite{Lopez-Coto:2021lxh}. Addditionally, this wide FoV and sensitivities to the highest energies also makes SWGO ideal to search for ALPs--- a generalization of standard axions and well motivated dark-matter candidates. ALPs are believed to convert to gamma rays in the presence of a magnetic field, therefore, ALPs would be found at extragalactic gamma-ray sources like Active Galactic Nuclei (AGN). SWGO's large, unbiased duty cycle lends itself ideally towards the characterization of flaring AGN, so the detection of the VHE gamma rays from extragalactic sources by SWGO would be a robust indication of ALPs~\cite{Albert:2019afb}.

\item[Search for Very Extended Emission.]
SWGO's features as a ground-based WCD instrument make it ideal for performing searches for very extended emission, including the Galactic diffuse emission, the Fermi bubbles, and extended halos around Pulsar Wind Nebula (PWN) and other accelerators~\cite{Hinton:2021rvp}. In addition to further improving our understanding of extended Gamma-Ray Halos and their evolution, blind searches for these Halos with SWGO will lead to a better understanding of the pulsar population through the discovery of new, nearby pulsars. This will, in turn, lead to a more stringent interpretation of the local positron excess and the associated implications for dark-matter interactions~\cite{Albert:2019afb}.
\end{description}

\begin{figure}[htb]
\centering
	\includegraphics[width=\textwidth]{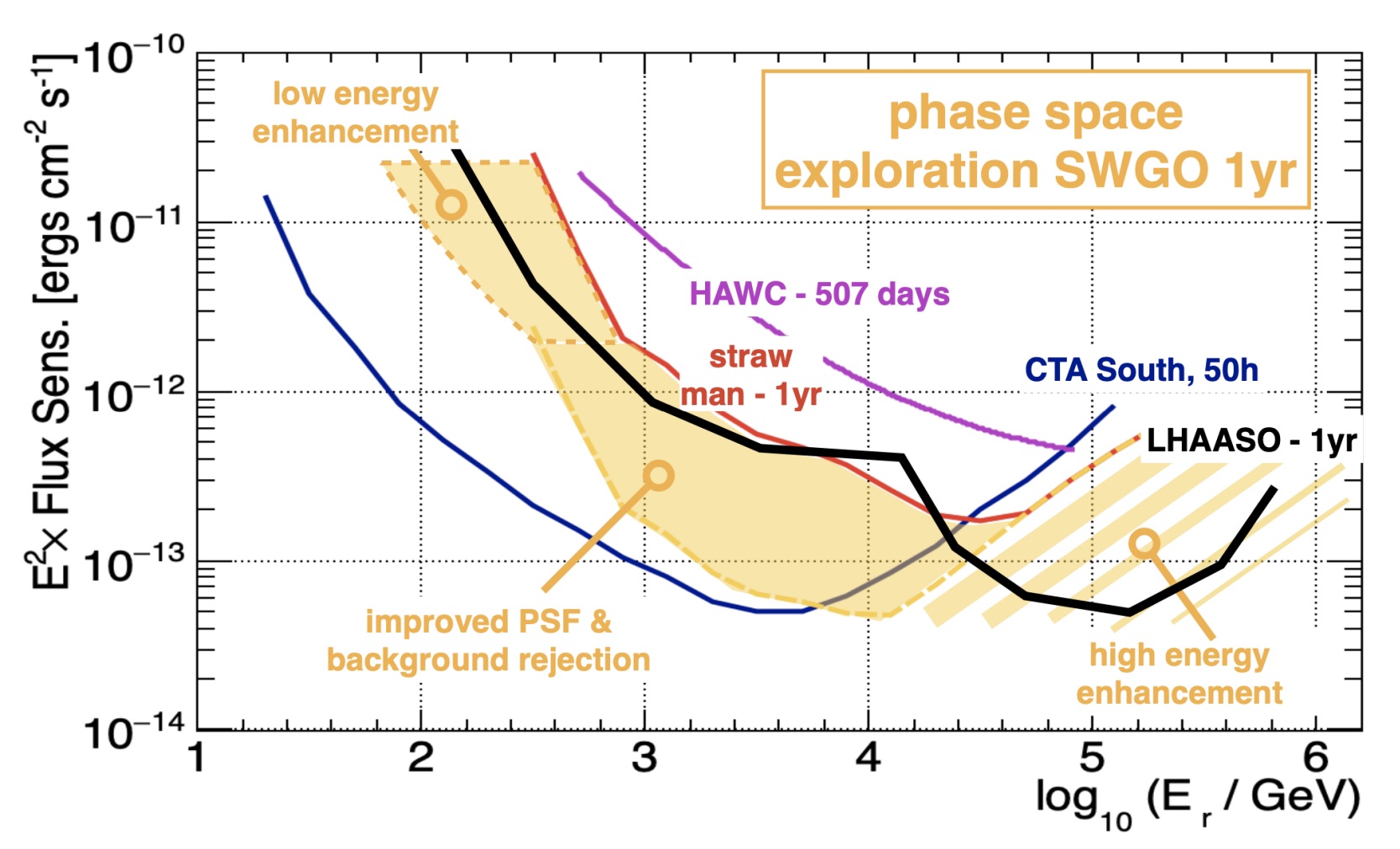}
	\caption{The differential sensitivity of the HAWC Observatory~\cite{Abeysekara_2019}, LHAASO~\cite{2019arXiv190502773B}, and the Southern portion of CTA~\cite{CTAConsortium:2018tzg} with the phase space that will be explored in the design studies for SWGO. We now consider the `strawman’ curves provided in Ref.~\citenum{Albert:2019afb} to be a lower limit to the performance that can achieved within our nominal cost envelope due to improvements in event reconstruction and background rejection. Figure from Ref.~\citenum{Hinton:2021rvp}.}	
	\label{fig:SWGO-sensi}
\end{figure}

Beyond the obvious complementarity with HAWC and LHAASO, we expect SWGO to be in simultaneous operation with the Cherenkov Telescope Array (CTA)~\cite{CTAConsortium:2018tzg} (see Section~\ref{sec:cta}), a next-generation IACT gamma-ray observatory with sensitivity in the 20~GeV to 300~TeV energy range. CTA will consist of two IACT arrays: one in the Northern Hemisphere (La Palma, Spain), and the other in the Southern Hemisphere (Atacama desert, Chile). SWGO will be complementary to the southern CTA site, therefore enabling full-sky coverage in VHE gamma rays with both IACTs and WCDs. The phase space that SWGO will occupy, showcasing ideal complementarity with existing and planned experiments, is shown in Figure~\ref{fig:SWGO-sensi}, with more details on the design of SWGO and the impact of the scientific goals of the Collaboration on that design can be found in Refs.~\citenum{Albert:2019afb,Abreu:2019ahw,Hinton:2021rvp,Schoorlemmer:2019gee}.

%% file: GRAND-Batista.tex
\noindent
\chapterauthor[ ]{Rafael {Alves Batista}}
\\
\begin{affils}
  \chapteraffil[ ]{Instituto de F\'isica Te\'orica UAM-CSIC, C/ Nicol\'as Cabrera 13-15, 28049 Madrid, Spain.} 
\end{affils}


The Giant Radio Array for Neutrino Detection (GRAND)~\cite{GRAND:2018iaj} is a proposed observatory of ultra-high-energy (UHE) particles, including photons, neutrinos, and cosmic rays, with energies above $\sim$100~PeV. The envisioned design is fully modular, comprising around twenty geographically independent sub-arrays with 10,000 radio antennas located in mountainous regions spread around the world. This staged deployment allows for incremental refinements in the design. Combined, these sub-arrays would cover an area of approximately 200,000~$\text{km}^2$, reaching sensitivities of up to $\sim$$10^{-10}~\text{GeV}~\text{cm}^{-2}~\text{s}^{-1}~\text{sr}^{-1}$ at energies of $\sim$500~PeV. 

GRAND's detection principle relies on its ability to reconstruct extensive air showers (EAS) induced by interactions between UHE particles and the atmosphere, which results in coherent radio emission. ``Bow-tie'' radio antennas---\textsc{HorizonAntenna}----were designed to detect showers with large zenith angles in the frequency range 50--200~MHz, such that the radio Cherenkov cone can be detected. This ultimately lowers the detection thresholds, improves background rejection, and consequently increases the detection efficiency~\cite{BalagopalV:2017aan}. Preliminary studies~\cite{GRAND:2018iaj} suggest an $X_\text{max}$ (nuclear composition) resolution better than 40~\text{g}~$\text{cm}^{-2}$, but improved methods could bring it down to $\simeq 20 \; \text{g}~\text{cm}^{-2}$~\cite{Corstanje:2014waa, Bezyazeekov:2015rpa, Buitink:2016nkf}. Moreover, while its design already implies sub-degree angular resolutions, the goal is to reach $\simeq 0.1^\circ$, making GRAND an exceptional facility for UHE multi-messenger studies and for tests of fundamental physics with cosmic messengers~\cite{GRAND:2018iaj}.

Currently, a smaller array with 300 antennas---GRANDProto300 (GP300)~\cite{Decoene:2019sgx}---is in the commissioning stage. The goal of this prototype is to demonstrate the feasibility of autonomous radio detection of EAS with high efficiency in the energy range between 30~PeV and 1~EeV. GP300 will feature a hybrid detection technique with an array of water-Cherenkov detectors to validate and enhance the radio-only reconstruction. 

\begin{figure}[htb]
	\centering
	\includegraphics[width=0.7\columnwidth]{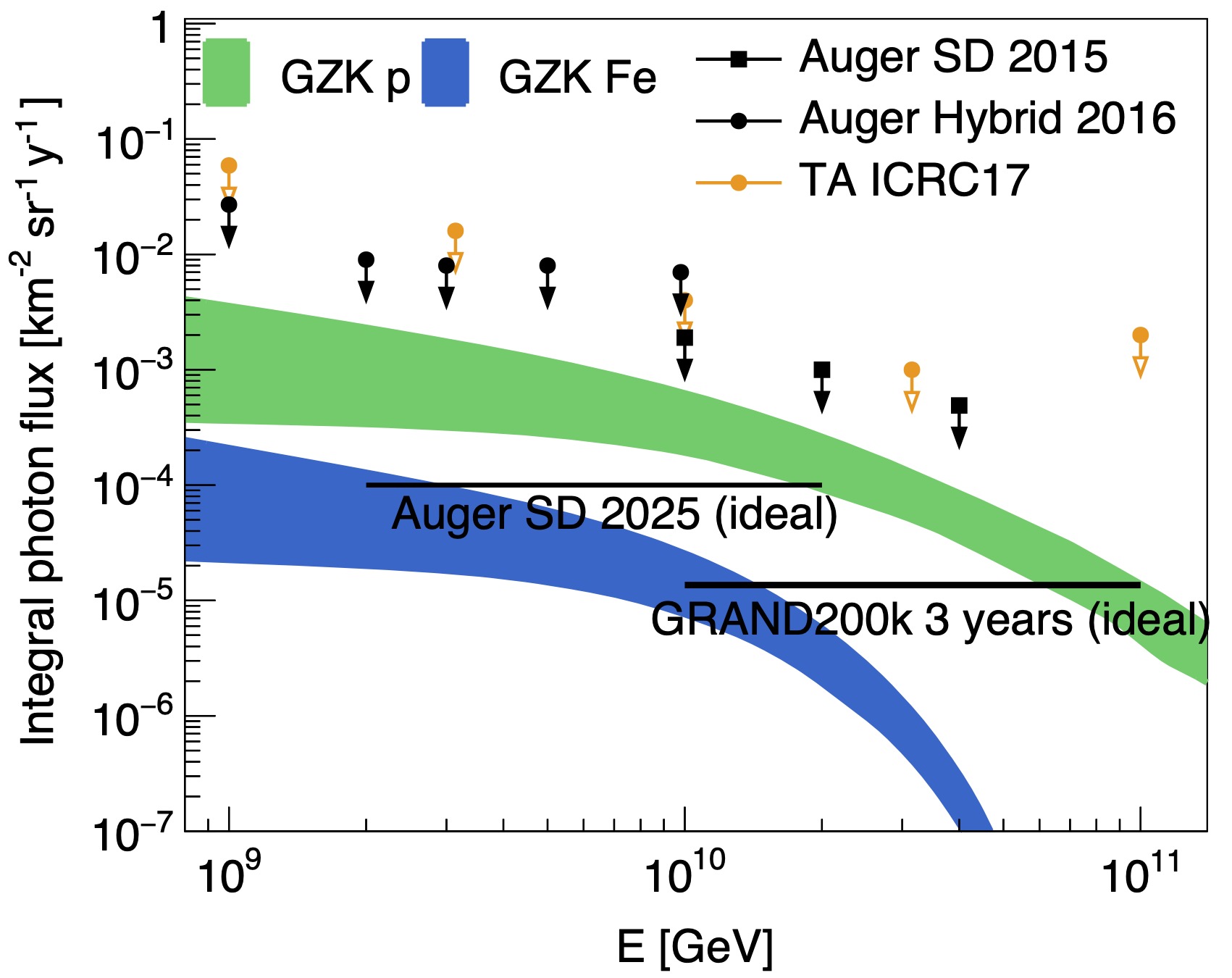}
	\caption{Cosmogenic photon fluxes expected for a pure proton or pure iron composition~\cite{Sarkar:2011hkm}. Upper limits from Auger~\cite{PierreAuger:2016kuz} and the Telescope Array (TA)~\cite{Rubtsov:2017lhs} are shown, together with the projected reach of Auger by 2025, and of GRAND after three years of operation. Figure taken from Ref.~\citenum{GRAND:2018iaj}.}
	\label{fig:uhe_grand}
\end{figure}

Preliminary studies indicate that photon-induced air showers with zenith angles between 65$^\circ$ and 85$^\circ$ can be identified with nearly 100\% efficiency~\cite{GRAND:2018iaj} with both the full array as well as with GP300. The non-detection of UHE photons with two years of measurements and a sample with at least $10^4$ events would improve present-day limits~\cite{PierreAuger:2016kuz, PierreAuger:2021mjh} on the integral UHE photon flux~\cite{Decoene:2019sgx}, as shown in Figure~\ref{fig:uhe_grand}.
The complete array would be fully efficient for detecting UHE photons above $\sim$10~EeV, with enough sensitivity to constrain cosmogenic fluxes of UHE photons stemming from UHE cosmic-ray (UHECR) interactions with background photon fields such as the Cosmic Microwave Background (CMB) and the Extragalactic Background Light (EBL)~\cite{Sarkar:2011hkm, AlvesBatista:2018zui}, as indicated in Figure~\ref{fig:uhe_grand}, in addition to a number of super-heavy dark matter models~\cite{GRAND:2018iaj}.

%% file: POEMMA-JohnK.tex
\noindent
\chapterauthor[]{John Krizmanic}
\\
 \begin{affils}
   \chapteraffil[]{Astroparticle Physics Laboratory, NASA Goddard Space Flight Center, Greenbelt, MD, USA}
\end{affils}

\begin{figure}[htb]
\begin{center}
    \includegraphics[width=0.8\textwidth]{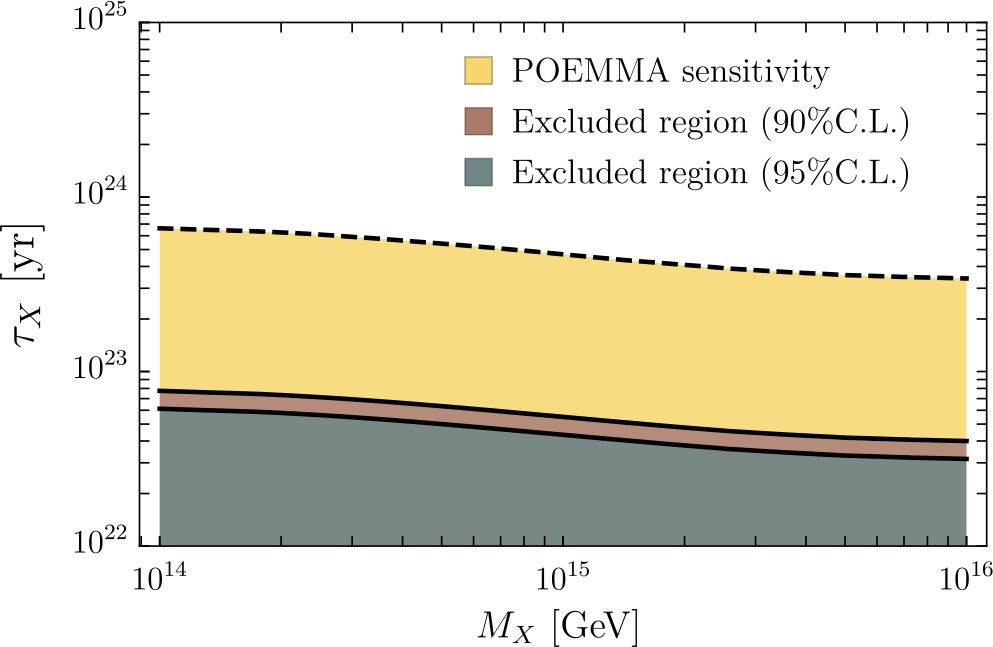}
 \end{center}
\vspace{-0.3cm}
  \caption{The lower limit on the lifetime of SHDM decays as a function of SHDM mass compared to the sensitivity (90\% \& 95\% CL) of observing one photon for $E_\gamma > 10^{11.3}$~GeV for five years of POEMMA Stereo-flourescence observations. Figure from Ref.~\citenum{Anchordoqui:2019omw}.
  }
  \label{POEMMAphotSens}
\end{figure}

Ultra-high-energy cosmic-ray (UHECR) experiments, by their nature, have considerable sensitivity to ultra-high-energy (UHE) photons by measuring the unique properties of the generated extensive air showers (EAS)~\cite{Risse:2007sd}. The combination of the characteristics of the initial electromagnetic interaction and subsequent extended EAS development, due to the Landau-Pomeranchuk-Migdal (LPM) effect~\cite{Landau:1953um,Migdal:1956tc}, leads to delayed EAS development that provides a method to separate UHE photon EAS from those initiated by UHECRs and UHE neutrinos. The distinctive UHE photon EAS signature and EAS measurement capabilities of the Probe Of Extreme Multi-Messenger Astrophysics (POEMMA) mission yields remarkable sensitivity to UHE photons~\cite{Anchordoqui:2019omw}. The POEMMA mission consists of two identical spacecraft---flying in a loose formation at 525~km altitude---that will use precision stereo air fluorescence measurements to determine the UHECR flux, nuclear composition, and full-sky distribution of the sources and to search for UHE neutrinos and photons~\cite{POEMMA:2020ykm}. Each POEMMA spacecraft contains a Schmidt telescope with $6~\mathrm{m}^{2}$ optical collecting area and a $45^\circ$ full field of view (FoV) and a hybrid focal plane array with different portions of the array tuned to finely view either the EAS fluorescence development or the beamed, optical Cherenkov signal. POEMMA will operate in a main UHECR science mode that will change into a limb-viewing mode for target-of-opportunity tau neutrino detection, triggered by an external gravitational wave or electromagnetic transient alert. In Stereo-fluorescence mode, the telescopes are oriented to co-measure the EAS air-fluorescence signal in a common volume corresponding to nearly $10^{13}$~tons of atmosphere. Due to the high accuracy of the EAS reconstruction from the stereo fluorescence technique when viewing the entire EAS development using large FoV from low Earth orbit (LEO), POEMMA can accurately reconstruct the development of the EAS with $< 20^\circ$ angular resolution, $< 20\%$ energy resolution, $< 30$~g/cm$^2$ X$_{\rm Max}$ (nuclear composition) resolution~\cite{Anchordoqui:2019omw}. This performance yields excellent sensitivity to separate EAS from UHECRs, UHE neutrinos, and UHE photons. In Limb-viewing mode, the POEMMA spacecraft will slew to allow the telescopes to tilt such that they are pointed to view slightly below the limb of the Earth to be sensitive to the beamed, optical Cherenkov signal generated by upward-moving EAS sourced from 20~PeV and above tau neutrino interactions in the Earth that lead to Earth-emergent $\tau$~leptons~\cite{Venters:2019xwi,Reno:2019jtr}. Potential sources of UHE photons include the decays or annihilation of super-heavy darkmatter (SHDM) with the contribution from the Galaxy and Galactic Halo dominating the flux~\cite{Dubovsky:1998pu,Evans:2002ry,Aloisio:2007bh,Kalashev:2017ijd}. Figure~\ref{POEMMAphotSens} presents the sensitivity to UHE photons from SHDM decays as a function of SHDM mass based an analysis presented in Ref.~\citenum{Alcantara:2019sco}. SHDM can also decay or annihilate into neutrinos and POEMMA's capability to observe neutrinos above 20~PeV will also set stringent limits on this detection channel~\cite{Guepin:2021ljb}.

%% file: DSSD-Kierans.tex
\chapterauthor[ ]{ }

 \addtocontents{toc}{
     \leftskip3cm
    \scshape\small
    \parbox{5in}{\raggedleft Carolyn Kierans}
    \upshape\normalsize
    \string\par
    \raggedright
    \vskip -0.19in
    }

 \noindent
 \nocontentsline\chapterauthor[]{Carolyn Kierans$^1$ \orcidlink{0000-0001-6677-914X}} 
 \nocontentsline\chapterauthor[]{Regina Caputo$^1$ \orcidlink{0000-0002-9280-836X}}
 \nocontentsline\chapterauthor[]{Sean Griffin$^2$ \orcidlink{0000-0002-0779-9623}}
 \nocontentsline\chapterauthor[]{Jeremy S. Perkins$^1$\orcidlink{0000-0001-9608-4023}}
\noindent
\\
\begin{affils}
\chapteraffil[1]{NASA Goddard Space Flight Center, Greenbelt, MD 20771, USA}
\chapteraffil[2]{WIPAC, University of Wisconsin--Madison, Madison, WI 53703, USA}
\end{affils}

\begin{figure}[b!]
\begin{minipage}[c]{0.45\textwidth}
\centering
\includegraphics[width=\textwidth]{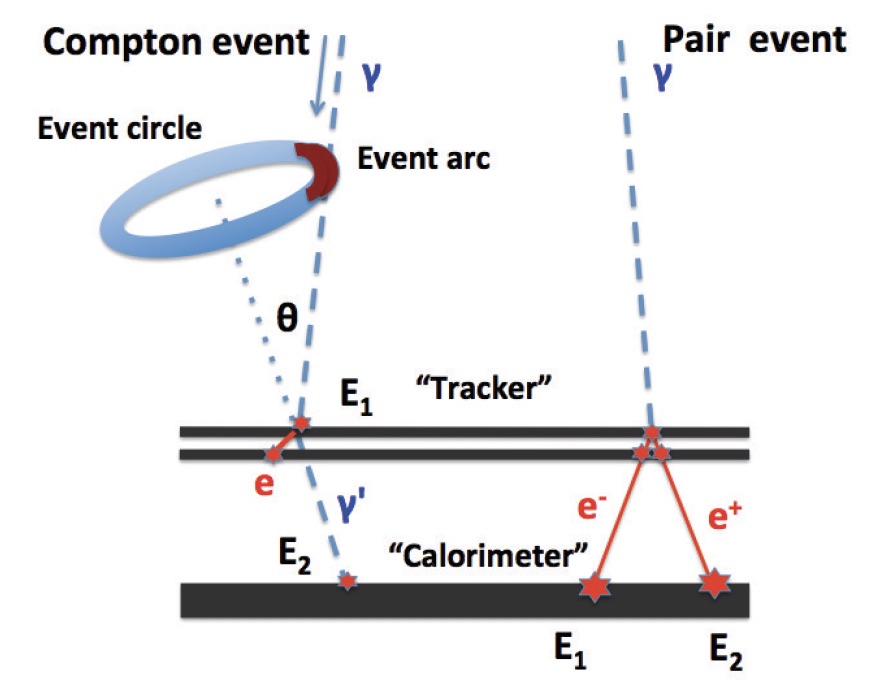}
\end{minipage}\hfill
\begin{minipage}[c]{0.55\textwidth}
\centering
\caption{To advance astroparticle physics and explore the range from $\sim$100~keV to 100~MeV, a telescope must be sensitive to Compton scattering and pair conversion interactions. 
A standard design utilizes a Tracker to measure the trajectory of charged particles coupled with a Calorimeter to contain electromagnetic showers from high-energy events.
Having many layers of DSSDs for the Tracker with high energy and spatial resolution allow for a precise measure of the Compton scattered electron and the track from electron/positron pairs.
}
\label{fig:TrackerInteractions}
\end{minipage}
\end{figure}

To achieve sensitivity in the MeV gamma-ray regime, a telescope must be sensitive to Compton scattering interactions (Section~\ref{sec:compton_tel}), in addition to pair conversion (Section~\ref{sec:pair}).
This can be achieved with a geometry similar to the \textit{Fermi} Large Area Telescope~\cite{fermi-lat}, utilizing a Tracker system to track charged particle products from gamma-ray interactions, and a Calorimeter to measure the energy; see Figure~\ref{fig:TrackerInteractions}.
However, Compton event reconstruction uses kinematic information from a sequence of scatters to determine the original direction of the photon \cite{boggs2000}, and thus the Tracker must give a precise measure of the energy ($\sim5\%$~dE/E) and position ($\sim1$~mm) of each interaction.
Since low-energy Compton-scattered electrons will not travel far before being fully absorbed, each segment of the Tracker (i.e. strips or pixels) must provide 3D position information for each interaction; therefore, single-sided silicon detectors such as those used in \textit{Fermi}-LAT do not provide sufficient information.  
The initial direction of the photon can be better constrained if the Compton-scattered electron subtends multiple Tracker segments allowing for a track to be measured.

There are four main advantages of using silicon detectors over other particle tracking detector technologies: 
1) semiconductor detectors have excellent energy resolution, 
2) silicon has good noise performance at room temperature, 
3) silicon is a relatively inexpensive and ubiquitous detector material, and
4) low Z materials, such as silicon, have a high Compton scattering cross-section and small Doppler-broadening \cite{zoglauer2003}.

The accuracy of the Compton event reconstruction, and therefore the angular resolution and the sensitivity of an MeV telescope, depends on the precision of the energy and position measurements.
The position resolution can be achieved through either double-sided silicon strip detectors (DSSDs) or pixelated silicon (see Section~\ref{sec:astropix}).
While some particle physics experiments are investing in pixelated silicon detector technology, the data rate and power constraints that are unavoidable in a space environment are more easily met with DSSDs.
Additionally, the Technology Readiness Level of pixelated silicon detectors may not be advanced enough for the development of an MeV mission in the next decade. 
DSSDs constitutes the most promising and robust technology to advance astroparticle physics in the foreseeable future.

There is a history of DSSDs being used in space-based astroparticle physics instruments, for example, PAMELA \cite{STRAULINO2004168} and AMS \cite{ZUCCON200874} have used DSSDs to study cosmic rays and dark matter.
Most recently, the Hard X-ray Imager (HXI) on Hitomi flew six layers of DSSDs to achieve sensitivity from 5 to 80~keV \cite{10.1117/1.JATIS.4.2.021410}.
In the early 2000's, there were a few efforts progressing towards an MeV telescope with DSSDs, most notably MEGA~\cite{2002NewAR..46..611B} and TIGRE~\cite{oneil2003}. However, neither of these proposed missions were sufficiently funded and development did not progress beyond the prototype stage.
Two decades later, with advancements in detector technology and electronics readout, MeV telescopes based on DSSDs remain the most compelling design and the science is more pressing than ever.

\begin{figure}[b]
\begin{minipage}[c]{0.5\textwidth}
\centering
\caption{
The AMEGO Tracker is designed around layers of DSSDs which measure the energy deposited from Compton scattering events and pair conversion interactions and is sensitive in the energy range $\sim$100~keV to $>$1~GeV.
Prototype development of the Tracker is currently underway~\cite{2019ICRC...36..565G, 2020SPIE11444E..31K}.
}
\label{fig:ComPairDSSD}
\end{minipage}\hfill
\begin{minipage}[c]{0.5\textwidth}
\centering
\includegraphics[width=0.9\textwidth]{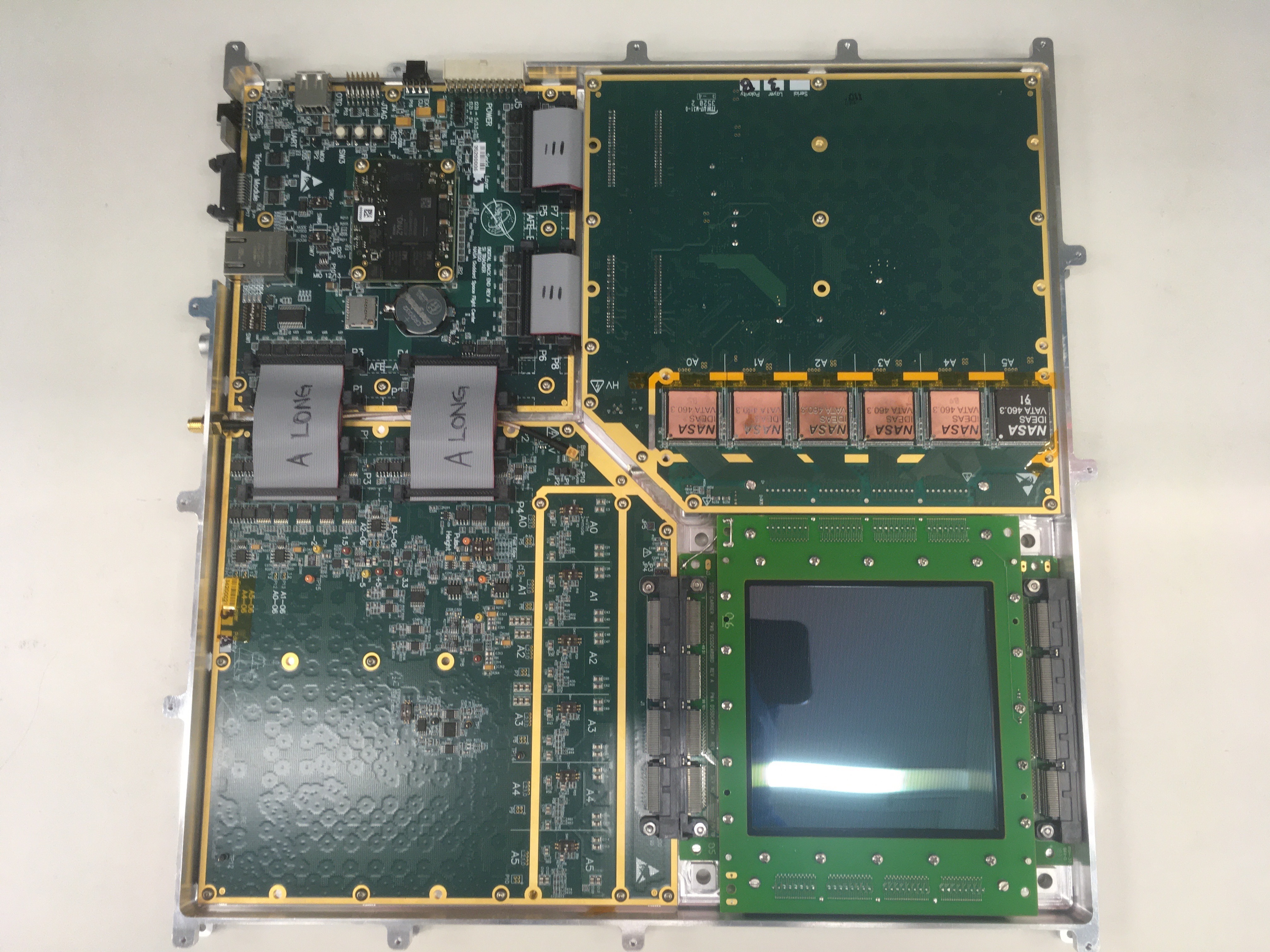}
\end{minipage}
\end{figure}

The All-sky Medium Energy Gamma-ray Observatory (AMEGO) is a NASA Probe class mission concept that was submitted to the Astro2020 Decadal Survey and will provide ground-breaking new capabilities for multimessenger astrophysics (see Section~\ref{sec:amego} \cite{2019BAAS...51g.245M}).
AMEGO consists of four subsystems that work together to operate as a Compton and pair telescope: a DSSD tracker, a 3D position sensitive virtual Frisch-grid Imaging Cadmium Zinc Telluride (CZT) calorimeter (Section~\ref{sec:VFG-CZT}), a segmented thallium-activated Cesium Iodide (CsI) calorimeter, and a plastic scintillator anti-coincidence detector.
The AMEGO Tracker consists of 60 layers of 500~$\mu$m thick DSSDs with 500~$\mu$m strip pitch.
Each layer contains four 4$\times$4 arrays of DSSDs, totaling 3840 silicon wafers each measuring 9.5~cm square.
To mature this technology, the AMEGO team is building a prototype of the four subsystems and are working towards a balloon flight to test the instrument in 2023~\cite{2019ICRC...36..565G, 2020SPIE11444E..31K}; see Figure~\ref{fig:ComPairDSSD}.

The next frontier of astroparticle physics is multimessenger astrophysics. 
With the planned advancements in neutrino and GW observatories, this science can only be enabled with a complementary advancement in gamma-ray telescope development.
Building upon their illustrious history in astroparticle physics, DSSDs are poised to play the technological lead in the unfolding drama that is multimessenger astrophysics.

%% file: FiberTracker-Nicola.tex
 \noindent
 \chapterauthor[]{M. Nicola Mazziotta}\orcidlink{0000-0001-9325-4672}
 \\
 \begin{affils}
   \chapteraffil[]{Istituto Nazionale di Fisica Nucleare, Sezione di Bari, I-70126 Bari, Italy}
 \end{affils}

The current generation of satellite-borne gamma-ray detectors for high-energy astrophysics mainly consists of multi-layer tracker-converters, based on silicon strip sensors, operating in the pair production regime~  \cite{Atwood:2007ra,Atwood:2009ez,Bulgarelli:2010zz,AGILE:2008nyq,DAMPE:2017yae,DAMPE:2019lxv}. The typical tracker layer is usually segmented in different modules, called ``ladders'', each composed of a few AC-coupled single-sided silicon micro-strip detectors (SSDs), daisy-chained via micro-wire bonding. A tracker module consists of many layers mounted on support trays, with the strips of each layer oriented perpendicularly to those of the adjacent layers, thus providing coordinate measurements along two orthogonal directions, both perpendicular to the detector pointing axis. The tracker layers are interleaved with thin tungsten layers, to enhance the probability of initiating photon conversion into electron–positron pairs. 

The length of the strips (and consequently of the ladders) is limited by the electronic noise due to the strip capacitance, which increases linearly with the strip length. The strips must be readout with a ultralow-noise front-end electronics, to cope with their large input capacitance. A typical ladder consists of four square SSDs, with a side of $\sim$ 10 cm.

With the evolution of the sensor technologies, trackers based on scintillating fiber readout with silicon photomultipliers (SiPMs) now represent a promising alternative to silicon detectors. Recent experimental results from LHCb~\cite{Joram:2015tla}, Mu3e~\cite{Bravar:2020wje} and from the balloon experiment PEBS~ \cite{vonDoetinchem:2007gw} have demonstrated that spatial resolutions below 100 $\mu$m can be achieved in large area detectors equipped with fiber trackers. 

Plastic scintillating fibers as active elements in tracking detectors have been used for more than 30 years (see for instance~ \cite{Alitti:1988za,CHORUS:1997ijc,K2K:2000kji,D0:2005cnn,Ellis:2005dx}). Their use was also proposed for gamma-ray space telescopes, such as Ref.~\cite{Antich:1990kv}, SIFTER~ \cite{Fishman:1997id}, and FiberGLAST~ \cite{Pendelton:1999ma}. Thanks to the latest development of the SiPM linear arrays for high-energy particle physics fiber tracker applications, nowadays they represent a possible option for the next generation of space-borne cosmic-ray and gamma-ray detectors.

A scintillating fiber tracker has several advantages with respect to a standard silicon tracker. First of all, the costs required to instrument a large detector area with fibers are significantly lower  than those required when using silicon detectors, as scintillating fibers are much cheaper than silicon detectors. In addition, a silicon tracker requires a high degree of segmentation, since silicon detectors usually must be assembled to form ladders, with an overall length that cannot exceed a few tens of cm, while single scintillating fibers with lengths up to 1-2 m can be used without suffering significant light attenuation. Further advantages of a scintillating fiber tracker with respect to a standard silicon tracker are given by the relative ease of the detector assembly and by the possibility of implementing different geometries with respect to the simple planar configuration. 

A fiber tracker could be an option for the new generation of instruments aimed at the detection of low-energy (MeV $-$ GeV) gamma rays, in which passive converter layers (tungsten) are removed to keep the multiple scattering angle small. These instruments will consist of many active tracking layers ($\sim 100$) to compensate the loss of the geometrical acceptance. In addition, since the typical radiation length of a plastic scintillator (40 cm) is larger than that of silicon (9.5 cm), for a given detector thickness the multiple scattering angle could be reduced. The achievable spatial resolution is correlated with the fiber diameter, and can be improved with staggered multi-layer fiber configurations.

\begin{figure*}[!t]
\centering
\includegraphics[width=0.48\textwidth, height=0.24\textheight]{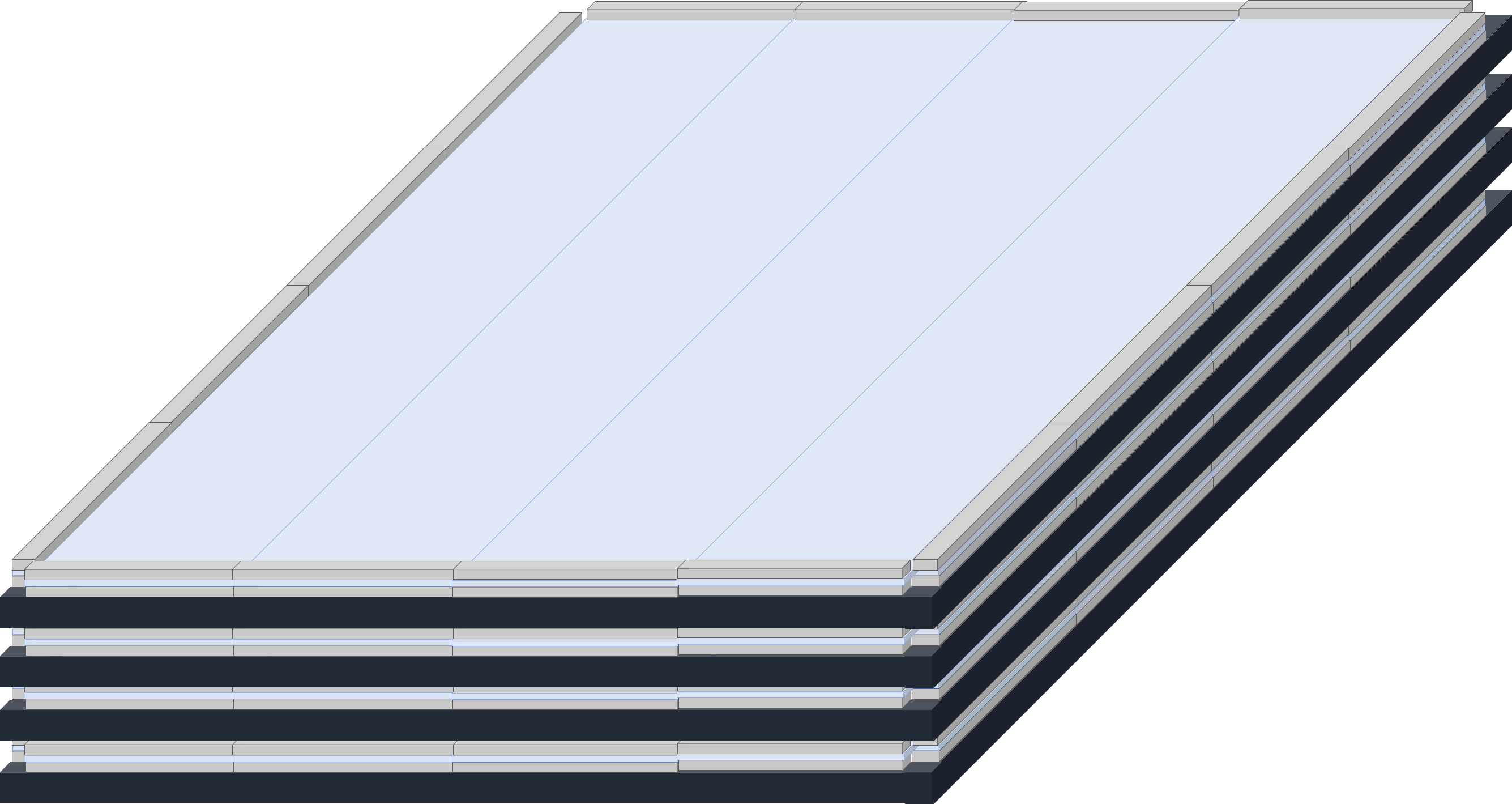}
\includegraphics[width=0.48\textwidth, height=0.24\textheight]{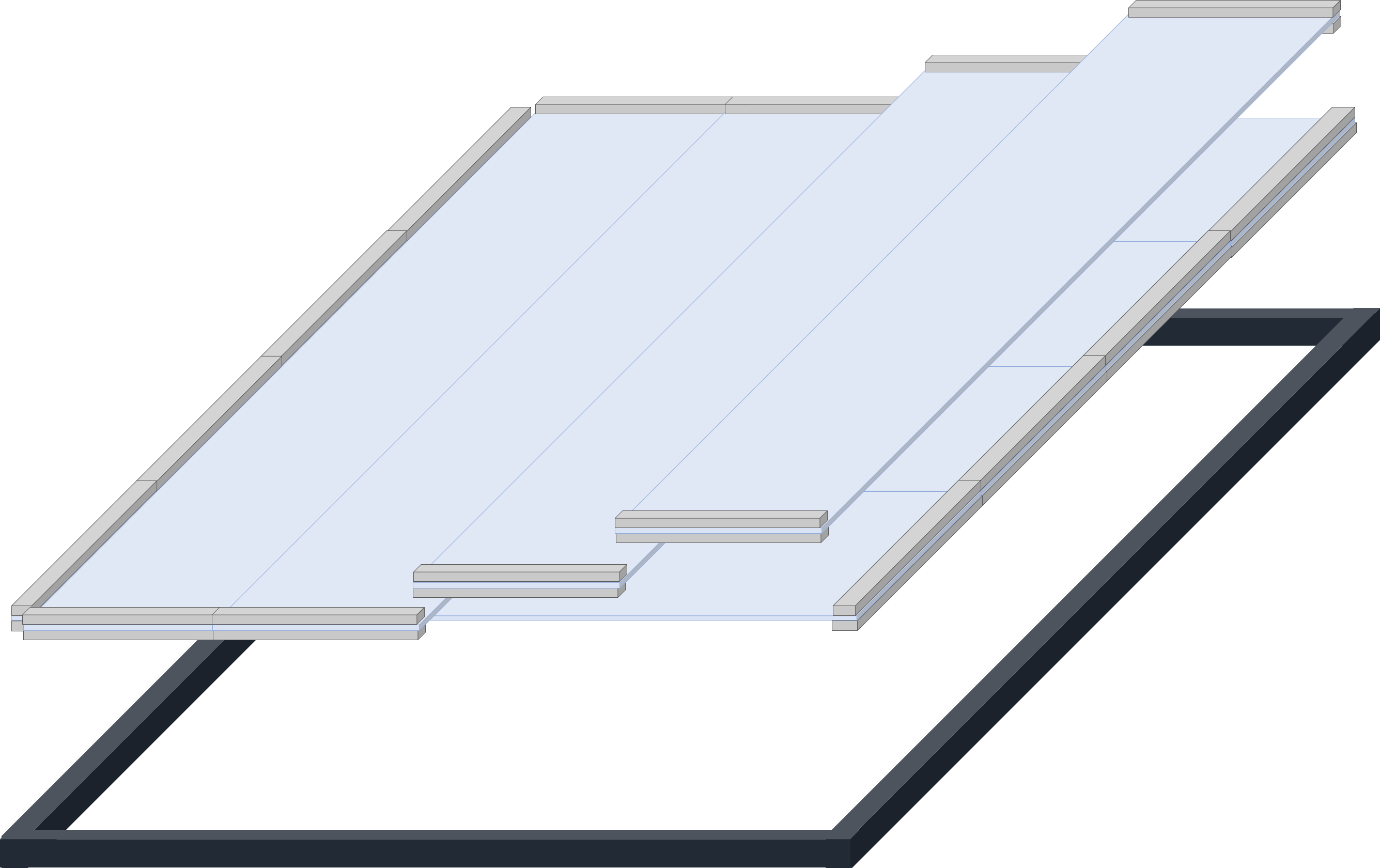}
\caption{\textit{Left:} Schematic view of the fiber tracker with a stack of four X-Y trays of scintillating fibers. \textit{Right:} Exploded view of an X-Y tray.}
\label{fig:fibmod}
\end{figure*}

A possible layout of a fiber-tracker for a low-energy gamma-ray detector could consist of a stack of several X-Y view modules (Figure~\ref{fig:fibmod}). Each view is equipped with multiple planes (ribbons) of scintillating fibers, with 250-500 $\mu$m or even larger diameter (up to 1 $-$ 2 mm). In the LHCb experiment six round 250 $\mu$m diameter fiber layers (fiber mats) are used, while in Mu3e experiment only three staggered layers of 250 $\mu$m round fiber are used. For a low-energy gamma-ray instruments two/three staggered 500 $\mu$m round fibers could ensure a good tracker detection efficiency and an adequate spatial resolution~\cite{MazSnow}. 

Scintillating fibers in a tracking detector have two functions: (i) they convert the ionization energy deposited by charged particles into optical photons with a very short decay time (2 $-$ 3 ns) and (ii) they transport these photons to the readout devices, which often are located outside the active volume. A plastic scintillating fiber consists of a core, typically made of polystyrene (n=1.59), surrounded by one or more thin cladding layers, made of polymers with lower refractive index, e.g. polymethyl methacrylate (PMMA; n=1.49) or a special fluorinated polymer (n=1.42). The thickness of each cladding layer is typically $3\%$ of the total diameter of the fiber. The photon emission spectrum is peaked in the range 400-500 nm, and the light attenuation length is $>$ 3 m. Since the typical intrinsic light yield is of about 10000 photons per MeV, a minimum ionising particle crossing a 500 $\mu$m diameter fiber will generate approximately 1000 scintillation photons. Assuming a trapping fraction of $5\%$, the average number of photons that will arrive at the photo sensor is about 50. A photo sensor with an average photon detection efficiency (PDE) of $30\%$ or more will be therefore able to detect a dozen or more of these photons (photo-electrons).

The scintillation light at one end (or both) of the fibers in each view will be collected by a SiPM array. For the LHCb experiment a 128 channels SiPM array has been developed~\cite{S13552}, with 57.5 $\mu$m $\times$ 62.5 $\mu$m pixels, arranged in columns of 4 $\times$ 26 pixels, resulting in a channel sensitive area of 230 $\mu$m $\times$ 1.625 mm, with a pitch of 250 $\mu$m. In this way a very compact layout can be achieved for a tower module, without any further wire-bonding as for the SSD front-end electronics readout. In addition the SiPM sensors are located outside the sensitive tracker area. The SiPMs could be bonded on one side of a printed wiring board (PWB) support, while the front-end electronics ASICs could be bonded on the other side of the PWB, connected to a multi-chip module which reads all the signals from a single plane.

%% file: LArTPC-Shutt.tex
 \noindent
 \chapterauthor[]{Thomas Shutt}
 \\
 \begin{affils}
   \chapteraffil[]{Stanford University, Stanford, CA 94305, USA}
 \end{affils}

Liquid argon (LAr) time projection chamber (TPC) technology holds enormous promise for measuring 0.1 - 10 MeV  gamma rays. The tremendous development of liquid noble TPCs in recent years has led to their revolutionary role in direct dark matter searches and their leading role in neutrino physics.  An early application to gamma rays was LXeGRIT~\cite{10.1117/12.962588} based on LXe, and now both GRAMS (section \ref{sec:grams}) and GammaTPC (section \ref{sec:gammatpc}) are exploring their potential with LAr.  In these detectors, particle interactions in the LAr target create scintillation light and free electrons.  An applied field drifts electrons to an anode readout plane where their X-Y locations are measured, while the light is measured by SiPMs on a cathode plane.  The depth of events (Z) is measured as the time difference between the fast ($\sim$ 10\,ns) scintillation signal and the arrival of the slower ($\sim$~170\,$\mu$s over 20\,cm) drifting electrons - hence the name TPC. The high particle rate in low earth orbit combined with the relatively slow charge drift requires segmentation at the $\sim$ 20 cm scale.

The core advantage of a TPC is that it provides 3D readout of a uniform target volume with sensors deployed only on the 2D surfaces.  This enables a large instrument with low channel count and hence low cost and power, which in turn allows a high granularity readout which directly leads to good angular resolution (see, e.g., Figure~\ref{fig:irfs}).  The minimal  interior dead material maximizes event reconstruction efficiency.  The energy resolution is optimized by efficient light collection and low noise charge readout, and is roughly comparable to silicon, though not as good as in germanium or cadmium zinc telluride (CZT). Xenon could also serve as a target, but Ar with lower atomic number gives a broader energy range over which the Compton reconstruction technique is effective (see Section~\ref{sec:compton_tel}.  A layer of LAr configured only for scintillation is a convenient option to add a calorimeter for additional stopping power, and thin outer such layers can serve as charged particle anti-coincidence detectors.
	
Much of the technology requires only modest adaptation for this application.  The light readout is much like that of LAr-based dark matter detectors \cite{Aalseth:2017fik, Ajaj:2019imk} with all possible surfaces coated with the waveshifter, and the remarkably reflective film Vikuiti$^{TM}$ \cite{vikuiti} dividing segments. The main drift field must be highly uniform, which in a curved geometry may require field shaping in the walls, such as the resistive sheet grading being developed for the DUNE near detector \cite{Berner:2019uvt}.  Purification of LAr for charge drift is a mature technology (see, e.g., \cite{Adamowski_2014}), and a modest-capacity online system consisting of a circulation pump and (likely regenerable) purifier should be adequate.  Similarly, cooling can be provided by mature mechanical space cryocoolers technology (see, e.g., \cite{cryocoolers}), but the cooling power available is a significant constraint on cold readout electronics. 

However more substantial development is required in some areas.  The particle background, in addition to presenting a dead time challenge, also gives rise to space charge of ions which have very slow drift velocity. This will alter the drift field by several percent and distort the spatial reconstruction, an effect which could possibly be corrected via a robust in situ calibration. A method to neutralize the ions by injecting electrons into the LAr would be preferable, such as a pulsed photo-cathode scheme. For a satellite (but not a balloon flight), there is need to ensure a single phase of LAr and sub-cooling to prevent boiling.  A possible method for this is  a positive displacement element such as a piston that pressurizes the liquid and ensures that the volume of liquid fully fills the vessel.   

\begin{figure}[tb]
    \centering
    \includegraphics[width=0.37\textwidth]{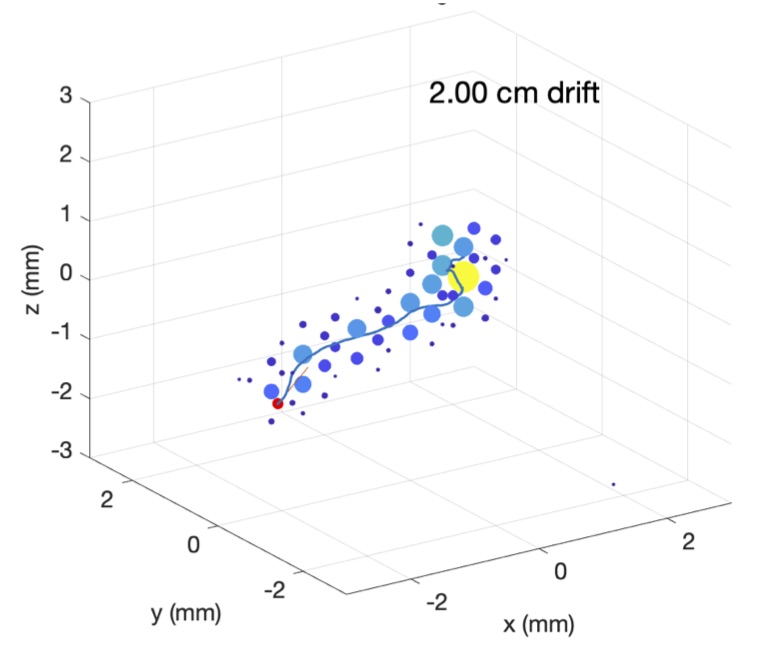}
    \includegraphics[width=0.46\textwidth]{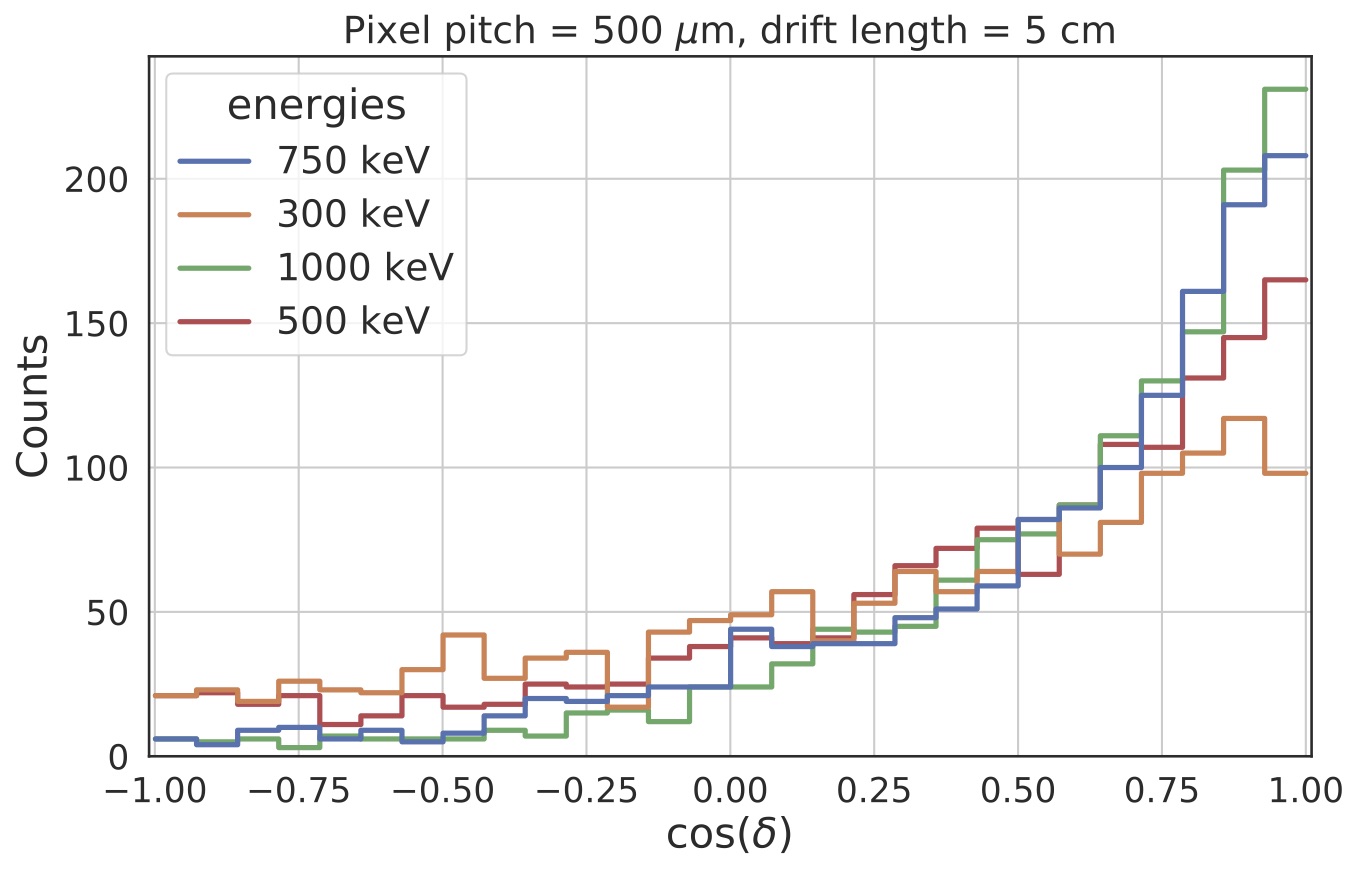}    
    \caption{\textit{Left:} Typical 1\,MeV electron recoil track (blue line) with $\sim$ 30,000\,e$^-$ electrons simulated with PENELOPE \cite{penelope}, (circles) simulated samples from a 500\,$\mu$m pixel readout with the head (red circle) and true initial recoil direction (faint red line) indicated. The tail of the track has higher charge density (yellow) from $dE/dx$.   Using ML algorithms we find better than $\sim$ 400 $\mu$m determination of the track head location. \textit{Right:} This is a powerful measurement of initial track direction, shown as the cosine of the angle between the true and measured directions.}
    \label{fig:electron_imaging}
\end{figure}

Full exploitation of the potential of this technology will also require development of a fine scale charge readout to achieve angular resolution and kinematic information competitive with silicon (see Section~\ref{sec:amego} \cite{Fleischhack_2021}) and germanium (see Section~\ref{sec:cosi} \cite{2021arXiv210910403T}) strip based detectors. The ultimate goal is to accurately image the mm-scale recoiling electron tracks to determine both their head (the location of the interaction) and their initial direction.  Recent development of cold complementary metal–oxide–semiconductor (CMOS) readout electronics for DUNE~\cite{CryoAsic1, Dwyer_2018, Adams_2020} lay a foundation for this.  However two problems must be overcome: the power resulting from high channel count, and loss of charge due to diffusion spreading the signal over many sub-threshold sensors.  Both of these are addressed by a novel dual-scale charge readout architecture being pursued at SLAC National Accelerator Laboratory which combines cm-spaced coarse grids with a switched power pixel ASIC, and shown in Figure~\ref{fig:gammatpc_schematic}. The coarse grid provides a trigger signal to power on select pixel chips, and also a charge signal unaffected by diffusion.  The power of true 3D imaging provided by the pixel readout is shown in Figure~\ref{fig:electron_imaging}.

The cost of even a very large scale instrument promises to be modest, even with a pixel readout.  Once in production, CMOS ASIC costs are $\sim$\,0.3\,\$M/m$^2$, with silicon photomultipliers (SiPMs) only somewhat greater.  The costs of a vessel and other materials should be minimal, and a cryocooler and fluid handling should also not be major costs. The mass of an instrument is by far dominated by the simple LAr target, and the cost of launching mass is greatly reduced in the current era.  The developments required to realize the potential of this technology will build on the major advances in liquid noble TPCs over the last two decades in general and the enormous DOE investment in LAr TPCs for DUNE in particular.  Demonstration of the space issues  could be accomplished via a single fully functional TPC segment flown as a CubeSat.  The payoff from such an undertaking is a transformative technology for MeV gamma rays.

%% file: SiliconPixel-Regina.tex
\chapterauthor[ ]{ }

 \addtocontents{toc}{
     \leftskip3cm
    \scshape\small
    \parbox{5in}{\raggedleft Regina Caputo, Jessica Metcalfe, Carolyn Kierans, et al.}
    \upshape\normalsize
    \string\par
    \raggedright
    \vskip -0.19in
    }
 
 \noindent
 \nocontentsline\chapterauthor[]{Regina Caputo$^1$}\orcidlink{0000-0000-0000-0000}
 \nocontentsline\chapterauthor[]{Jessica Metcalfe$^2$}
 \nocontentsline\chapterauthor[]{Carolyn Kierans$^1$\orcidlink{0000-0001-6677-914X}}
 \nocontentsline\chapterauthor[]{Jeremy S. Perkins$^1$\orcidlink{0000-0001-9608-4023}} 
 \nocontentsline\chapterauthor[]{Isabella Brewer$^1$\orcidlink{0000-0003-1908-8551}}
 \nocontentsline\chapterauthor[]{Mathieu Benoit$^3$\orcidlink{0000-0002-8623-1699}}
 \nocontentsline\chapterauthor[]{Richard Leys$^4$}
 \nocontentsline\chapterauthor[]{Ivan Peric$^4$}
 \\
 \begin{affils}
   \chapteraffil[1]{NASA Goddard Space Flight Center, Greenbelt, MD 20771, USA}
   \chapteraffil[2]{Argonne National Laboratory, Lemont, IL 60439, USA}
   \chapteraffil[3]{Brookhaven National Laboratory, Upton, NY 11973, USA}
   \chapteraffil[4]{Karlsruhe Institute of Technology, Karlsruhe, Germany}
 \end{affils}


Over the past several decades, silicon strip detectors (SSDs) have become a key detector technology in particle physics experiments both on the ground and in space. 
SSDs provide sufficient spatial and energy resolution as well as good timing capabilities with the main benefit of not requiring high voltages or pressurized gas. 
An SSD is an arrangement of strip implants on a wafer of silicon (Si) that act as charge collecting electrodes. 
The strips are patterned on a low doped fully depleted silicon wafer and form a one-dimensional array of diodes. 
By connecting each of the metalized strips to a separate charge sensitive amplifier, we can measure the position of the interaction within the bulk Si material.  
To make two dimensional detectors, we can either apply orthogonal strips on the the backside of the wafer making double sided silicon strip detectors (DSSDs; see Section~\ref{sec:dssd}), or use an additional layer of SSDs, depending on the application. 
At lower energies, in the regime where the dominant interaction within the bulk Si is via Compton scattering, it is particularly important to have multidimensional readout because the Compton scattered electron will often become absorbed within a single layer of detector material (Section~\ref{sec:compton_tel}). 
The main limitation of these DSSDs in particular is the technical complication of manufacturers to produce them and the process is both time consuming and expensive.

\begin{figure}[tb]
  \begin{minipage}[c]{0.50\textwidth}
\centering
\includegraphics[width=\textwidth]{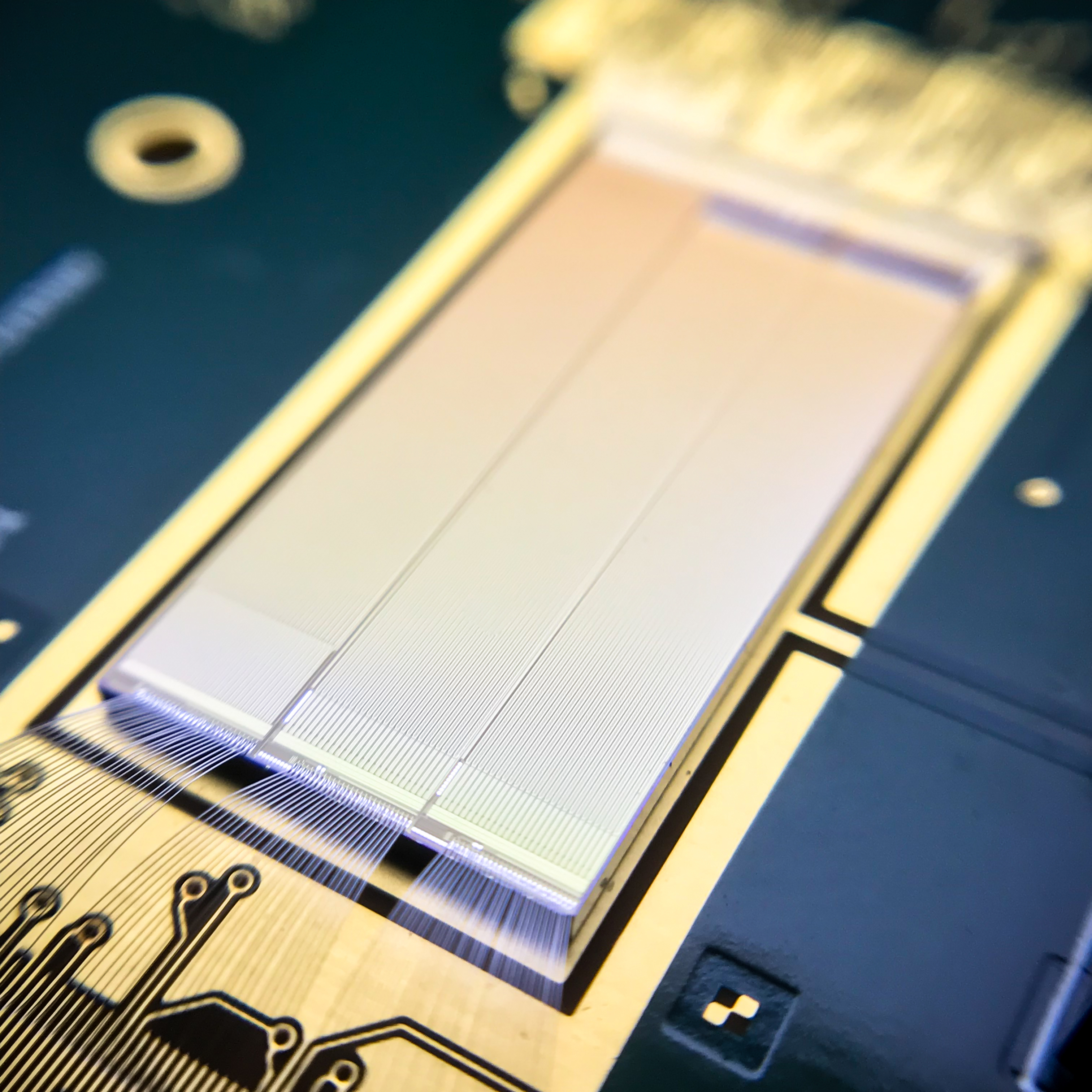}
\end{minipage}\hfill
\begin{minipage}[c]{0.45\textwidth}
\caption{ \small ATLASPix (left) has implemented CMOS-based technology for use in the ATLAS experiment. Although the current detectors are only $\sim$cm$^2$ in area (as shown), these detectors are easily scalable to 10s m$^2$ or even 100s m$^2$ by Si foundries.
Currently large-scale foundries produce $\sim$100~m$^2$ of CMOS Si-detectors per month.
Pixel detectors are currently deployed in space for dosimetry and cosmic-ray measurements \cite{pinsky_update_2016,whyntie_simulation_2014} and similar performances are to be expected after optimization of the monolithic pixel sensor design.
The process of producing CMOS detectors is fully supported by the consumer electronics industry guaranteeing the support of this technology in the foreseeable future. \label{fig:ATLASPixImage} }
\end{minipage}
\end{figure}

One promising alternative to SSDs and DSSDs currently being developed in the particle physics community is monolithic complementary metal–oxide semiconductor (CMOS) Si pixel sensors. 
Monolithic detectors do not require a separate readout application-specific integrated circuit (ASIC). Instead, they have signal amplification and readout circuits directly embedded in each pixel, reducing the pixel size, which improves spatial resolution, and limits the amount of passive material, which improves energy resolution. 
The design reduces both the overall mass of the detector and the payload size.
Integrated designs have the potential to greatly reduce power consumption due to more efficient amplification (compared to a similar detector without on-board readout).

The motivation driving the development of CMOS detectors in particle physics is to join the two main functionalities of silicon detectors: collect the deposited energy of particles interacting in the detector and amplify and discriminate that signal. 
This can be accomplished with the CMOS process. 
Combining these functions yields a tighter integration of the detector structure, fewer steps of integration, and a lower cost. 
Because of the wide-spread commercial use of CMOS sensors in industry, the CMOS detectors are mass-produced making large-scale ($\sim$100s m$^2$) Si-based detectors easily realizable.

The particle physics community, specifically the ATLAS, ALICE and Mu3e Collaborations, have invested heavily in fully monolithic silicon pixel CMOS sensors as a candidate for current and future upgrades of the Large Hadron Collider (LHC) \cite{peric_high-voltage_2018,mager_alpide_2016,berdalovic_monolithic_2018,vigani_study_2017}. 
Two important characteristics of CMOS devices are relatively low noise and low static power consumption compared with other logic families. Monolithic pixel sensors (Figure~\ref{fig:ATLASPixImage}) have amplification and readout circuits directly embedded on each pixel which enables reductions in pixel size improving spatial resolution and reduction in inactive material improving energy resolution. 
The design reduces the overall mass of the detector improving the spatial resolution and reducing the size of the payload. 
Integrated designs have the potential to greatly reduce power consumption due to more efficient amplification (compared to a similar detector without on-board readout).   
The ATLAS Collaboration's effort (ATLASPix) has focused on upgrades to their spatial resolution, radiation hardness and extremely fast timing. 

Because of their relatively lower power and mass requirements and high spacial and energy resolution capabilities CMOS silicon pixel detectors have a broad range of applications from next generation high intensity particle experiments to space-based gamma-ray telescopes.

%% file: Diamond-Bloser.tex
 \addtocontents{toc}{
     \leftskip3cm
    \scshape\small
    \parbox{5in}{\raggedleft Peter Bloser, Daniel Poulson, John Smedley, et al.}
    \upshape\normalsize
    \string\par
    \raggedright
    \vskip -0.19in
    }

\noindent
 \nocontentsline\chapterauthor[]{Peter Bloser$^1$\orcidlink{0000-0002-6664-4306}}
 \nocontentsline\chapterauthor[]{Daniel Poulson$^{1}$}
 \nocontentsline\chapterauthor[]{John Smedley$^1$\orcidlink{0000-0001-7149-9257}}
 \nocontentsline\chapterauthor[]{Jennifer Bohon$^1$\orcidlink{0000-0002-7664-9899}}
 \nocontentsline\chapterauthor[]{James Distel$^1$}
 \nocontentsline\chapterauthor[]{Dongsung Kim$^1$}
 \nocontentsline\chapterauthor[]{Mark McConnell$^2$\orcidlink{0000-0001-8186-5978}}
 \nocontentsline\chapterauthor[]{Jason Legere$^2$}
 \nocontentsline\chapterauthor[]{Keiichi Ogasawara$^3$\orcidlink{0000-0002-0429-8380}}
\\
 \begin{affils}
   \chapteraffil[1]{Los Alamos National Laboratory, Los Alamos, NM 87545, USA}
   \chapteraffil[2]{University of New Hampshire, Durham, NH 03824, USA}
   \chapteraffil[3]{Southwest Research Institute, San Antonio, TX 78238, USA}
 \end{affils}

Although currently at a relatively low technological maturity, artificial single-crystal diamond detectors (SCDDs) produced by chemical vapor deposition (CVD) show great promise as a Compton-scattering medium for soft-to-medium energy gamma rays (50 keV – 10 MeV).  
SCDDs offer the possibility of position and energy resolution comparable to those of silicon solid-state detectors (SSDs), combined with efficiency and timing resolution so-far only achievable using fast scintillators.  
When integrated with a calorimeter composed of fast inorganic scintillator~\cite{2021SPIE11821E..1DP}, SCDDs would enable a compact and efficient Compton telescope using time-of-flight (ToF) discrimination to achieve low background and high sensitivity, while potentially retaining Compton electron-tracking capabilities thus far only demonstrated using silicon strip detectors.  
The low atomic number and high density of diamond compared to silicon hold the promise of more efficient, compact instruments with very short coincidence timing windows for, e.g., Compton polarimetry.  
The inherent radiation hardness and temperature insensitivity of diamond make it attractive for space instrumentation, including use on newly available small satellite platforms.

\begin{figure}[t!]
    \centering
    \includegraphics[width=0.5\columnwidth]{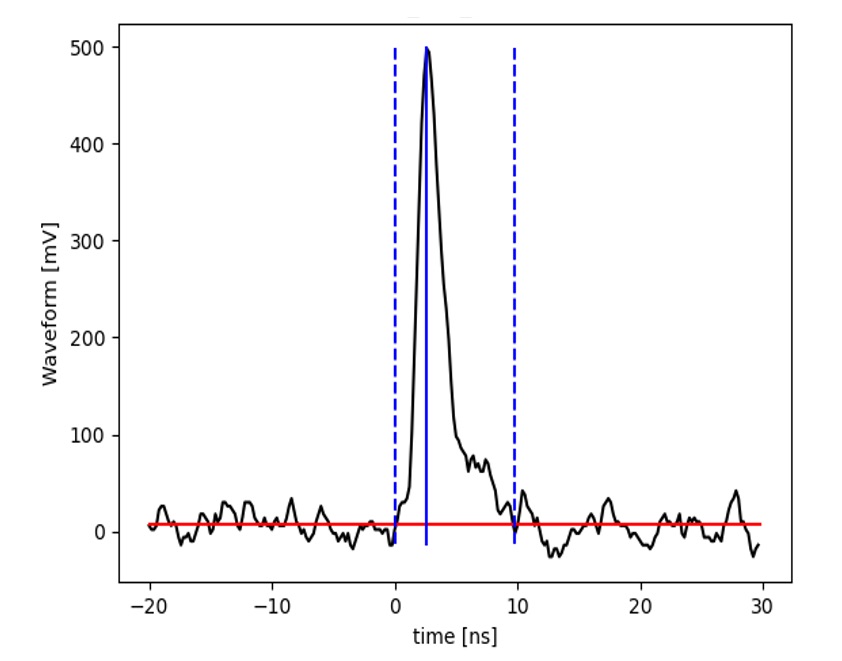}
    \caption{A $4.5 \times 4.5 \times 0.5$~mm$^3$ single-crystal diamond detector provides a pulse rise time of $\lesssim$1~ns for alpha particle interactions, showing its suitability for fast coincidence timing measurements.}
    \label{fig:diamond-bloser}
\end{figure}

SCDDs were first investigated as alternatives to Si SSDs for particle measurements, since the practical limits on the radiation hardness of silicon fall short of what is required for a variety of future space applications. 
Diamond detectors are expected to be $\sim$3 decades more tolerant of ionizing radiation doses than conventional SSDs and they reasonably operate up to $1–3 \times 10^{15}$~protons/cm$^{2}$~\cite{2002NIMPA.476..686A}.  
SCDDs also have a much faster response time due to the high mobility of electrons and holes in the detector~\cite{1369574}.
In addition, diamond detectors operate at temperatures in excess of 150~$^{\circ}$C~\cite{TANIMURA2000325} and are completely insensitive to background photons of $<5.45$~eV ($> 227.5$~nm) owing to the wide band gap~\cite{2009JPCM...21J4221B}.  
Ogasawara et al.~\cite{OGASAWARA2015131} measured an energy resolution of 7~keV (FWHM) and signal rise time of $< 0.2$~ns in a 100~$\mu$m-thick SCDD measuring protons, alpha particles, and fast electrons, and we have measured rise times of $\lesssim$1~ns in 500~$\mu$m-thick SCDDs (Fig.~\ref{fig:diamond-bloser}) for both alpha particle and gamma-ray interactions.  
SCDDs are now regularly read out with sub-mm position resolution via cross-strip electrodes for use in X-ray beam monitors~\cite{bohon2010, Zhou:pp5072}.

SCDDs remain small and expensive, with the primary commercial supplier being UK-based Element Six, Ltd.~Academic research into diamond growth continues in the U.S. however~\cite{GU2012210, LU201317}, and has resulted in the formation of domestic diamond growth companies such as Michigan-based Great Lakes Crystal Technologies (GLCT).  
Currently SCDDs up to $8 \times 8 \times 0.5$~mm$^{3}$ can be reliably produced (GLCT, private communication), and continued research and development will hopefully further increase size and reduce cost.  
SCDD elements of approximately one square inch in area and 0.5~mm thickness would permit straightforward tiling into arrays for gamma-ray collection areas competitive with current silicon strip detectors.











%% file: Scintillators-Rich.tex
\noindent
\chapterauthor[]{Richard S. Woolf}\orcidlink{0000-0003-4859-1711}
\\
\begin{affils}
\chapteraffil[]{Naval Research Laboratory, Washington, DC 20375, USA}
\end{affils}

\begin{figure}[tb]
    \begin{minipage}{0.48\textwidth}
        \centering
        \includegraphics[height = 5cm]{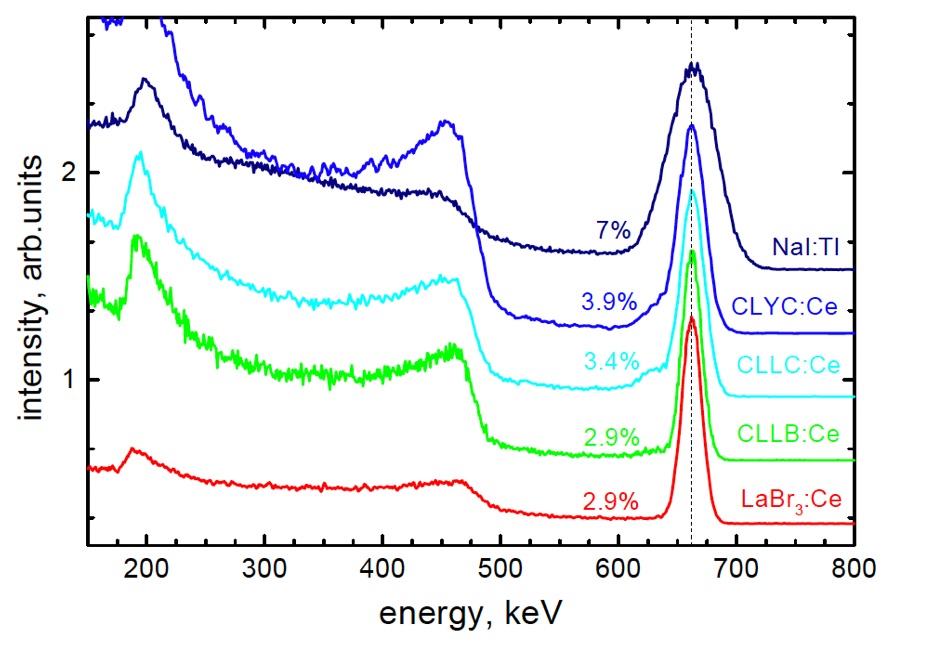}
        \caption{Comparison of energy resolution of inorganic scintillators (Shah, private communication)}
        \label{fig:Scintillators-Rich_Fig1}
    \end{minipage}%
    \hfill
    \begin{minipage}{0.48\textwidth}
        \centering
        \includegraphics[height = 5cm]{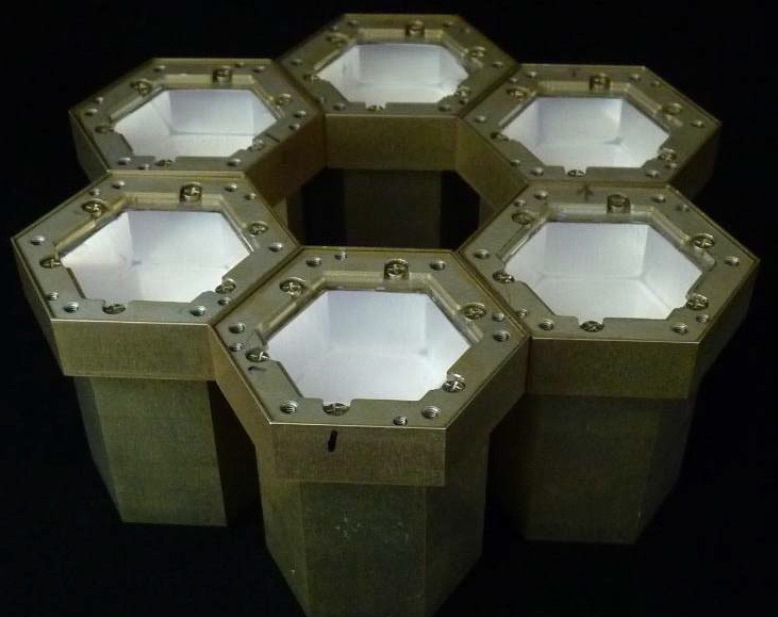}
        \caption{Partially populated SIRI-2 detectors.}
        \label{fig:Scintillators-Rich_Fig2}
    \end{minipage}
\end{figure}

Scintillation-based detectors, both organic and inorganic, have been used in the field of space-borne astrophysics for the past half century. Previous missions, such as Compton Gamma Ray Observatory (CGRO)~\cite{1993ApJS...86..657S, 1993ApJS...86..693J, 4333363}, and Fermi Gamma-Ray Space Telescope~\cite{2009ApJ...702..791M,fermi-lat}, employed inorganic scintillators, such as thallium doped sodium iodide and cesium iodide (both thallium and sodium doped), and bismuth germanate (BGO), while COMPTEL on CGRO used organic (liquid) scintillator for the detection medium. Inorganic crystal scintillators have good stopping power to MeV gamma rays and yield a typical energy resolution of 7-10\% FWHM at 662 keV. Over the past few decades, advances in crystal growing and size scaling have lead to the production of inorganic scintillators with superior energy resolution compared to their predecessors (Figure~\ref{fig:Scintillators-Rich_Fig1}). Crystals such as cerium bromide (CeBr$_3$)~\cite{QUARATI2013596}, cerium-doped lanthanum bromide (LaBr$_3$:Ce)~\cite{1239275}, and europium-doped strontium iodide (SrI$_2$:Eu)~\cite{10.1117/12.830016} all demonstrate resolutions in the 3-4\% FWHM energy range (or better). These materials have trade-offs in terms of available size, internal background, and scintillation decay time, ranging from fast (order of ns) to slow (order of $\mu$s). Several instrument payloads were recently launched to space qualify the SrI$_2$:Eu material via the SIRI-1 and SIRI-2 instrument~\cite{2019arXiv190711364M, 10.1117/12.2528073} (see Figure~\ref{fig:Scintillators-Rich_Fig2}).

\begin{figure}[tb]
    \begin{minipage}{0.48\textwidth}
        \centering
        \includegraphics[height = 5cm]{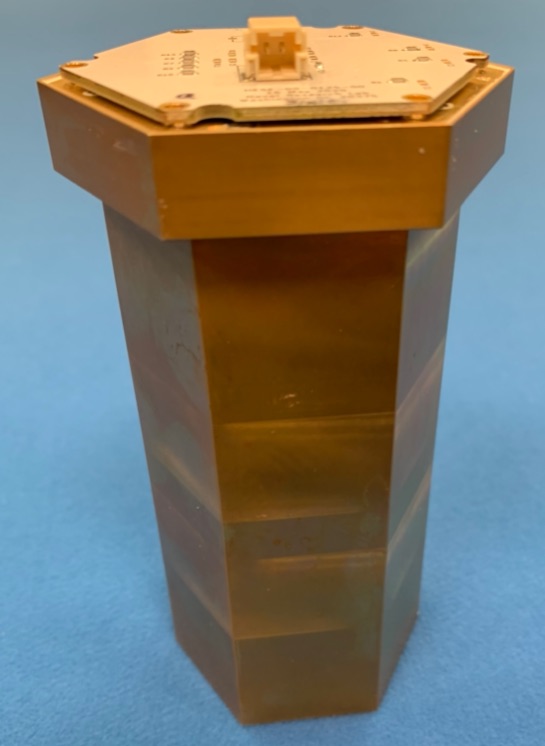}
        \caption{One of the two CLLBs detectors part of the Glowbug instrument.}
        \label{fig:scintillator_rich_3}
    \end{minipage}%
    \hfill
    \begin{minipage}{0.48\textwidth}
        \centering
        \includegraphics[height = 5cm]{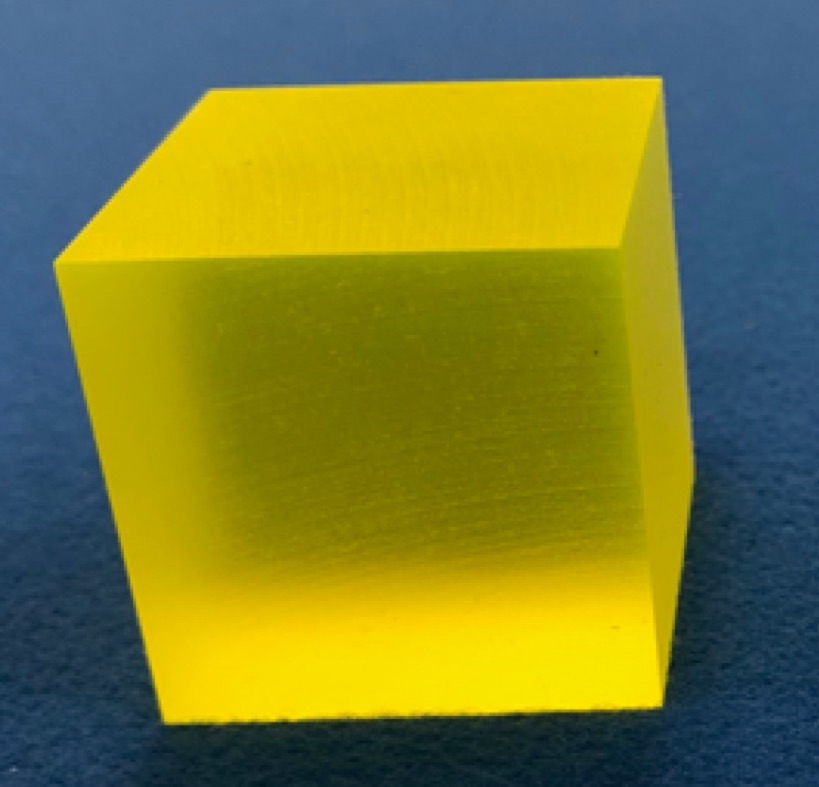}
        \caption{$30 \times 30 \times 30$~mm GAGG that comprises one of the garnet scintillators on the GARI-1 and GARI-2 payloads.}
        \label{fig:scintillator_rich_4}
    \end{minipage}
\end{figure}

Another group of inorganic scintillators have been developed over the past 20 years with crystalline properties that allow for dual mode detection of both gamma rays and neutrons. This group - known as elpasolites~\cite{osti_1096473} - are characterized by the generic formula A$_2$BLnX$_6$ where Ln: lanthanides/rare-earth metals, and X: halogens. Some examples of elpasolites are: Cs$_2$LYCl$_6$:Ce (CLYC \cite{4545124}), Cs$_2$LiLaBr$_6$:Ce (CLLB \cite{SHIRWADKAR2011268}), Cs$_2$LiLa(Br,Cl)$_6$:Ce (CLLBC \cite{10.1117/12.2060204}), and Tl$_2$LiYCl$_6$:Ce (TLYC \cite{doi:10.1021/acs.cgd.7b00583}). The intrinsic properties of these hygroscopic crystals vary but they typically have good energy resolution ($\sim$4-5\% FWHM at 662 keV), good light output ($\sim$20-40 photons/keV), density in the range of $\sim$4 g/cm$^3$, scintillation decay times in the range of ns to $\mu$s, and can detect and discriminate thermal neutrons from gamma rays due to the presence of $^6$Li~\cite{DOLYMPIA2012140}. Two elpasolites (CLLB and TLYC) will be space qualified during an upcoming mission to the International Space Station via the DOD Space Test Program (STP) H9 pallet~\cite{2020grbg.conf...57G, Hutcheson2021} (Figure~\ref{fig:scintillator_rich_3} for CLLB).

Lastly, over the past decade there has been a considerable amount of effort dedicated to crystals, primarily used as a lasing medium, for scintillation detectors. These crystals are collectively known as garnet ceramics, with the most notable examples being Gd$_3$Al$_2$Ga$_3$O$_{12}$:Ce (GAGG \cite{Kamada_2011}) and (Gd,Lu)$_3$(Ga,Al)$_5$O$_{12}$:Ce (GLuAGG \cite{GLODO2017285}). These materials have a high-density (6-7 g/cm$^3$), high-light output ($\sim$50 photons/keV) with good energy resolution (~5\% FWHM at 662 keV), and are non-hygroscopic. GAGG has higher density compared to other common scintillation materials, e.g., CsI:Tl (4.5 g/cm$^3$), NaI:Tl (3.7 g/cm$^3$), and LaBr$_3$:Ce (5.1 g/cm$^3$), which provides better photon absorption and consequently higher detection efficiency. While a material such as LaBr$_3$:Ce has higher light output ($\sim$63 photons/keV), it also has high, undesirable internal background. Additionally, GAGG is a fast scintillator with a 138 ns scintillation light decay time constant. Space qualification of this material is currently underway via the recently launched GARI-1 and GARI-2 instruments to the ISS via the DOD STP H7 and H8 pallets~\cite{10.1117/12.2598588} (Figure~\ref{fig:scintillator_rich_4}).

%% file: Phoswich-Wood.tex
\chapterauthor[ ]{ }

 \addtocontents{toc}{
     \leftskip3cm
    \scshape\small
    \parbox{5in}{\raggedleft Joshua Wood, Corinne Fletcher et al.}
    \upshape\normalsize
    \string\par
    \raggedright
    \vskip -0.19in
    }

\noindent
 \nocontentsline\chapterauthor[]{Joshua Wood$^1$\orcidlink{0000-0001-9012-2463}}
 \nocontentsline\chapterauthor[]{Corinne Fletcher$^{2}$\orcidlink{0000-0002-0186-3313}}
 \nocontentsline\chapterauthor[]{Adam Goldstein$^2$\orcidlink{0000-0002-0587-7042}}
 \nocontentsline\chapterauthor[]{Michelle Hui$^1$\orcidlink{0000-0002-0468-6025}}
\\
 \begin{affils}
   \chapteraffil[1]{NASA Marshall Space Flight Center, Huntsville, AL 35808, USA} 
   \chapteraffil[2]{Universities Space Research Association, Columbia, MD 21046, USA}
 \end{affils}
 
 Traditional gamma-ray survey instruments in the keV energy range, such as BATSE \cite{PHOS:Shaw2003} and \textit{Fermi}-GBM \cite{PHOS:Meegan2009}, have used single crystal scintillation detectors for both the detection and localization of astrophysical transients. However, this creates a fundamental limitation where off-axis detector sensitivity must be sacrificed in order to obtain a design that is thin enough to yield the asymmetric angular response \cite{PHOS:Fishman1989, PHOS:Bissaldi2009} needed to determine the location of a source flux. Phoswich detector technology can solve this limitation by combining two separate scintillation crystals into a single detector unit, allowing for a thin forward crystal with an asymmetric response as well as a second crystal which provides active area for off-axis sensitivity, enhanced high energy response, and background rejection. As a result, further development of this technology may be key to maximizing the capabilities of future gamma-ray survey instruments, especially for SmallSat missions where mass and volume requirements restrict the number of individual detector units.
 
\begin{figure}[t!]
\centering
\includegraphics[width=0.6\textwidth]{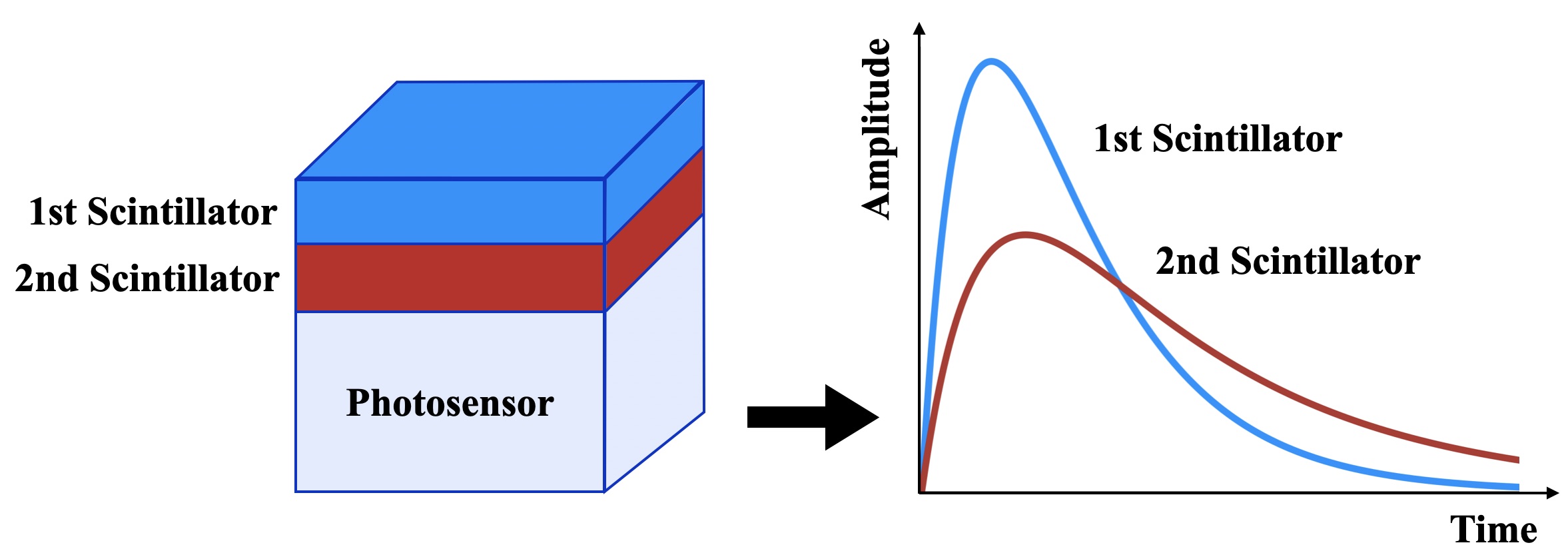}
\caption{A phoswich detector consisting of two scintillating
materials coupled to a single photosensor. Measuring the decay timescale of the pulse shape output by the photosensor allows one to identify the scintillator volume responsible for a given particle interaction.}
\label{fig:phoswich-concept}
\end{figure}

A basic phoswich detector consists of two coupled scintillating materials with distinct decay times (Figure~\ref{fig:phoswich-concept}). This has the benefit of allowing a single photosensor to distinguish between particle interactions in each scintillator, which reduces the overall detector size and weight compared to using two separate photosensors and their associated electronics. The phoswich technique was previously used by the HEAO-1 A4 experiment \cite{PHOS:Matteson1978} and RXTE HEXTE \cite{PHOS:Rothschild1998} with a NaI(Tl) / CsI(Na) scintillator combination to reduce particle background for X-ray and gamma-ray measurements. It was more recently applied by the Lomonosov \cite{PHOS:Svertilov2017} and Insight-HXMT \cite{PHOS:Zhang2020} missions to separate incident gamma-rays from rearward incident background. Here we propose efforts to modernize and optimize phoswich technology for the purpose of detecting incident gamma-ray signals, with the main application being space-based survey observations of gamma-ray bursts (GRBs) and soft gamma-ray repeaters which are typically detected at keV energies. 

\begin{figure}[tb]
\centering
\includegraphics[width=\textwidth]{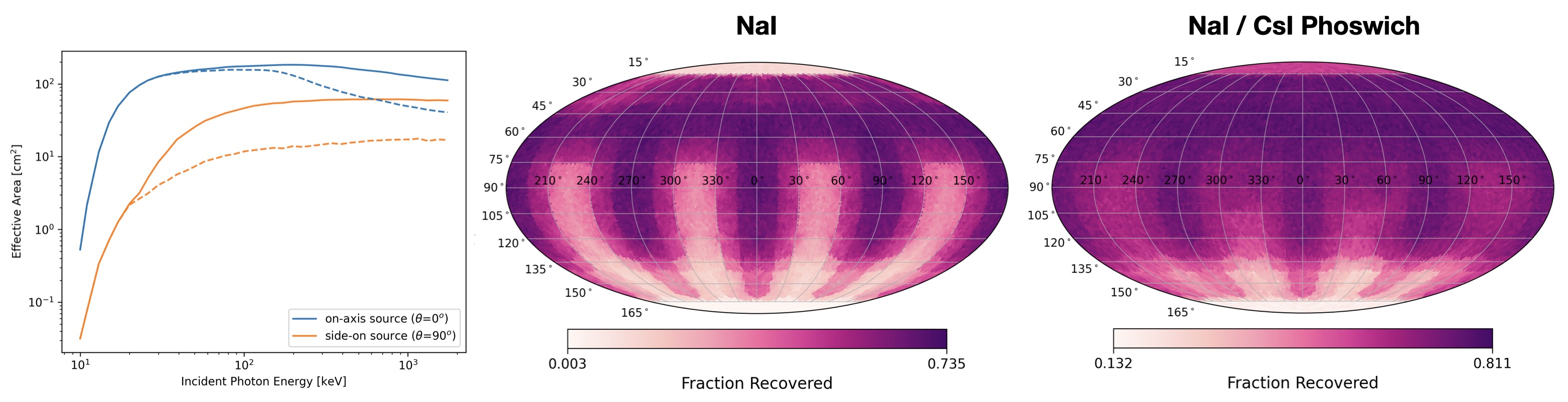}
\caption{\textit{Left}: Simulated effective areas for a NaI(Tl) / CsI(Na) detector (solid line) and a traditional NaI(Tl) detector similar in size to the \textit{Fermi}-GBM detectors (dashed line). \textit{Right:} Fraction of events detected by two different SmallSat instrument designs for the same set of simulated GRB observations. In both cases the designs have 5 detectors with the \textit{Middle} panel using a traditional 15mm thick NaI(Tl) scintillation detector and the \textit{Right} panel using a 15mm NaI(Tl) / 32mm CsI(Na) phoswich detector that provides much better sky coverage without sacrificing localization accuracy.}
\label{fig:example-smallsat}
\end{figure}

To demonstrate the merits of such a design, we consider a SmallSat concept limited in mass and volume to five detectors facing outwards from the four sides and forward face of a cube. We then simulate its response to gamma-ray signals from a population of GRBs whose distribution is representative of bursts detected by \textit{Fermi}-GBM \cite{PHOS:von_Kienlin2020, PHOS:Poolakkil2021} using two different detector configurations. The first is a traditional single crystal NaI(Tl) detector, similar to the NaI(Tl) detectors on \textit{Fermi}-GBM, with a 140 mm diameter and 15 mm thick scintillator volume. The second is a phoswich style detector with the same NaI(Tl) scintillator coupled to a 140 mm diameter and 32 mm thick volume of CsI(Na). The 32 mm CsI(Na) thickness was chosen because it rejects 90\% of rearward arriving photon backgrounds below 300 keV where GRB localizations are typically performed \cite{PHOS:Pendleton1999, PHOS:Goldstein2020}. This may not be the optimal choice for detecting off-axis signal photons but despite this Figure~\ref{fig:example-smallsat} clearly shows the benefits of the phoswich design, which has a roughly 3 times larger effective area for photons arriving at the side of the detector compared to the traditional NaI(Tl) design. This results in a notably higher number of GRB detections across the sky. Additionally, the phoswich design demonstrates better spectral response to on-axis photons above 200 keV.

Further work is needed to explore optimizations of the phoswich technique for applications in gamma-ray survey instruments but the method already shows promise. As a result, it is already being incorporated into future SmallSat missions, such as the Moon Burst Energetics All-sky Monitor (MoonBEAM; see Section~\ref{sec:moonbeam}). Additionally, we recommend research in several key areas where the existing phoswich designs with flight heritage can be improved using modern technology. This includes replacing photomultiplier photosensors with smaller devices such as Silicon Photomultipliers (see Section~\ref{sec:rad_tol_sipms}), although care must be taken to preserve the decay timescale of the scintillation pulses given the higher response times of these devices \cite{PHOS:Bloser2013}. The use of novel scintillator materials with faster decay times and higher light output like LaBr3 and CeBr3 (see Section~\ref{sec:scintillators}) can also improve energy resolution and reduce the energy threshold of event detection \cite{PHOS:Guss2010}. Lastly, modern digital pulse processing \cite{PHOS:Knoll2000} can replace much of the analog circuitry needed for pulse shape measurements, greatly reducing the complexity, package size, and power requirements of the analog electronics on-board the spacecraft.

%% file: SiliconPhotomultipliers-Perkins.tex
\noindent
\chapterauthor[]{Jeremy S. Perkins}
\\
\begin{affils}
\chapteraffil[]{Astroparticle Physics Laboratory, NASA Goddard Space Flight Center, Greenbelt, MD, USA}
\end{affils}

Scintillation detectors have a long history in high-energy astrophysics as seen in the instruments on the {\it Compton Gamma-ray Observatory} and on the \fermi~Gamma-ray Space Telescope.  Major discoveries during the last decade were enabled by scintillation detectors.  The sodium iodide (NaI) and bismuth germanium oxide (BGO) detectors of \fermi-GBM~\cite{Meegan2009} detected the first photons from the gravitational wave event GW170817A~\cite{Abbott2017}. This detection proved that short gamma-ray bursts result from neutron star mergers and expanded our understanding of jet physics and the speed of gravity.  The association of a gamma-ray flare detected by \fermi-LAT whose calorimeter~\cite{Grove2010} is comprised of logs of cesium iodide (CsI) scintilllators with a neutrino event from the active galactic nucleus TXS 0506+056~\cite{IceCube2018} showed that neutrinos are produced in the environments surrounding nature's particle accelerators, supermassive black holes. Recently, significant development has occurred to make Silicon Photomultipliers (SiPMs) a viable light collection alternative in scintillation detectors. To fully realize the benefits of using SiPMs in space, resources must be applied to fully characterize radiation damage in SiPMs and ultimately developing radiation tolerant and/or radiation hard devices.  This ensures that future missions can take advantage of the benefits of SiPMs (low voltage, mass, power, and volume).

Scintillation detectors work by coupling a scintillating material with a light collection device.  The scintillation material (such as the NaI used in \fermi-GBM) absorbs the energy of an incoming photon and emits scintillation light. The scintillators can directly detect incoming radiation in the energy range from tens of keV to about 1 MeV directly (like in \fermi-GBM). Scintillators can also be part of larger tracker systems to detect gamma rays indirectly, via secondary particles produced either by Compton scattering (primary gamma-ray energies from tens of MeV to hundreds of MeV, e.g. in the proposed AMEGO~\cite{McEnery2020} mission; see Section~\ref{sec:amego}) or electron/positron pair cascades (primary gamma-ray energies from hundreds of MeV to about 1 TeV, e.g. in \fermi-LAT). The type of scintillator and their size and shape are determined by the scientific requirements of the mission. 

Light collection devices commonly used in the past are photomultiplier tubes (PMTs) and PIN diodes.  PMTs are vacuum tubes consisting of a photocathode that emits a primary electron via the photoelectric effect and a series of dynodes that produce many secondary electrons via electron multiplication.  Thus, a single detected photon produces a large number of electrons.  In cases where high voltage or available volume are a concern, PIN diodes have been used (like the calorimeter on \fermi-LAT).  A PIN diode is a robust solid state device that produces a single electron-hole pair when a single photon is absorbed resulting in a linear relationship between the incoming flux and electrical signal.  Thus, a PIN diode does not have the large gain of a PMT and cannot be used with scintillators with low-light yields or in situations where energy resolution is a driver.  

The applications and designs of scintillation detectors in space are varied and complicated and include many creative solutions depending on the exact scientific requirements.  However, all of these devices have the basic design of a scintillation material and a light collection device.  

\begin{figure}[tb]
\includegraphics[height=1.9in]{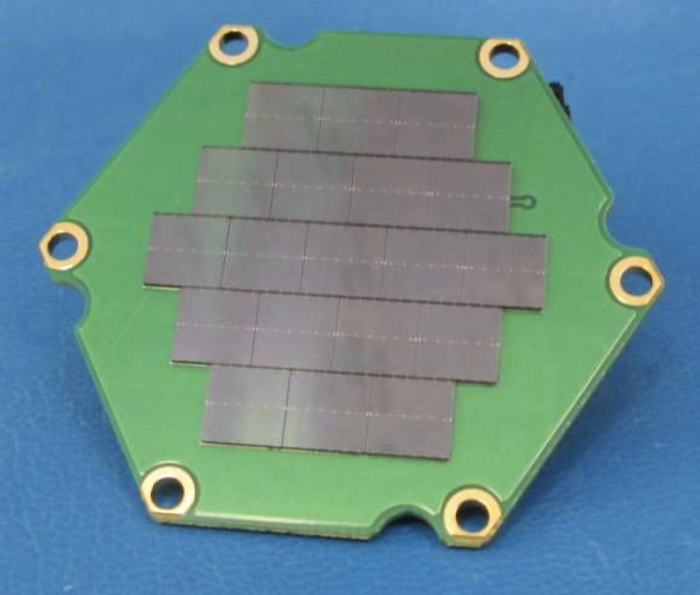}
\includegraphics[height=1.9in]{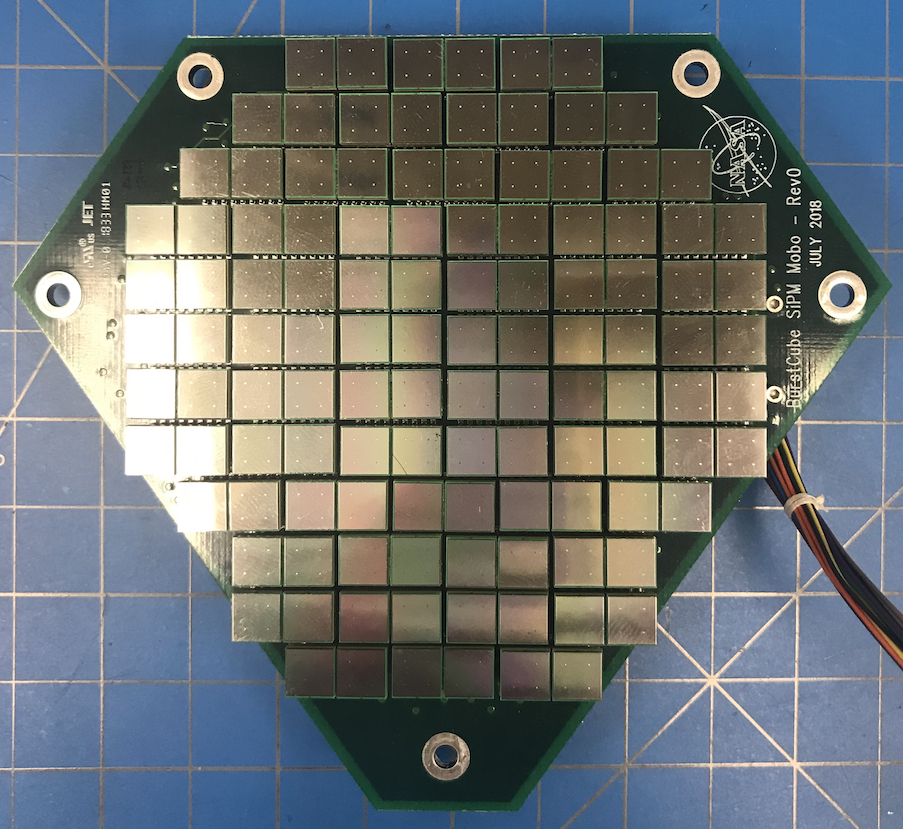}
\centering
\caption{{\it Left:} The SIRI-1 mission flew SiPMs for one year and observed a $\sim$130\% increase in the SiPM bias due to radiation damage.  SIRI-2 will use this readout board comprised of 19 6 mm SiPMs coupled to a SrI$_2$:EU scintillator~\cite{Mitchell2020}.  {\it Right:} BurstCube will fly four detectors comprised of 116 6 mm SiPMs each coupled to CsI detectors~\cite{Smith2019}.  These missions and others like them are path-finders for space qualifying SiPMs.}
\label{fig:new}
\end{figure}

Significant development has occurred to make SiPMs a viable light collection alternative (Figure \ref{fig:new}).  A SiPM is a solid-state single photon detection device based on single-photon avalanche diodes.  A single SiPM is comprised of many 10's of thousand avalanche diode cells operating in Geiger mode.  Thus, they have a high gain and a large dynamic range.  SiPMs have several advantages over traditional PMTs: they are robust, small in volume and mass, and do not require high voltage to operate (bias voltages are typically in the 10's of volts).  They have similar gains and photo-detection efficiencies as PMTs.

Several groups are space-qualifying SiPMs including SIRI-1~\cite{Mitchell2020}, GRID~\cite{Wen2019} and future missions like BurstCube~\cite{Smith2019} and MAMBO~\cite{Vestrand2019}.  During SIRI-1's one year sun-synchronous orbit, the bias current increased by $\sim$130\% due to radiation damage.  This highlights the main downside to using SiPMs in a space environment: in low earth orbit (typical for high energy missions, other orbits commonly used for heliophysics and planetary science experience higher levels of radiation), the instruments on a mission will experience radiation due to charged particles (protons cause the most damage). This radiation will cause defects in SiPMs and lead to an increase in dark current.  Several groups have recently studied this using terrestrial beam-test data~\cite{Link2019, Mitchell2020, Ulyanov2020, Tamer2020, Bartlett2020}.  It is important to note that the space environment is not as strenuous as that in most terrestrial particle physics experiments. Mitigating radiation effects for space is also beneficial to the overall particle physics community.   

Currently, radiation damage is reduced or mitigated by increasing shielding to reduce the total ionizing dose of radiation, controlling the temperature of the SiPMs to reduce the total dark current, and increasing the total bias current over the lifetime of the mission.  These solutions are not ideal since increasing shielding adds mass to usually mass-constrained instruments, controlling the temperature is difficult in a space environment and increasing the bias current does not mitigate against the increase in noise (or low-energy threshold).  A more permanent, robust solution is needed for future large-scale missions to enable the adoption of SiPMs.

To fully realize the benefits of using SiPMs as light collection devices for scintillation detectors in space, resources must be applied to fully characterize radiation damage in current SiPMs and ultimately developing radiation tolerant and/or radiation hard devices.  This ensures that future missions can take advantage of the benefits of SiPMs (low voltage, mass, power, and volume).

%% file: SiPMEAS-Krizmanic.tex
 \noindent
 \chapterauthor[]{John Krizmanic}
 \\
 \begin{affils}
   \chapteraffil[]{Astroparticle Physics Laboratory, NASA Goddard Space Flight Center, Greenbelt, MD, USA}
 \end{affils}

\begin{figure}[h]
\begin{center}
    \includegraphics[width=0.475\textwidth]{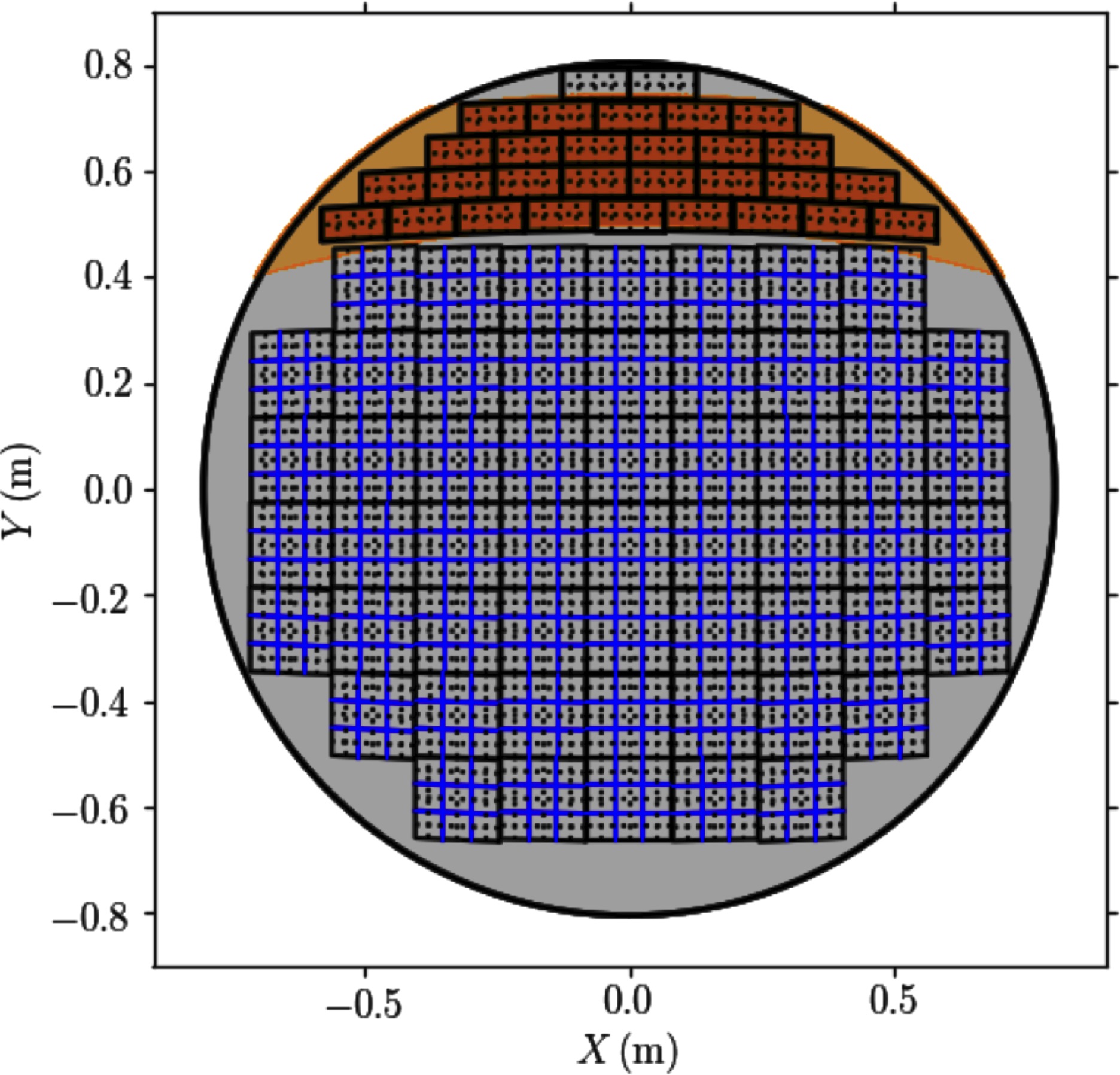}
\includegraphics[width=0.50\textwidth]{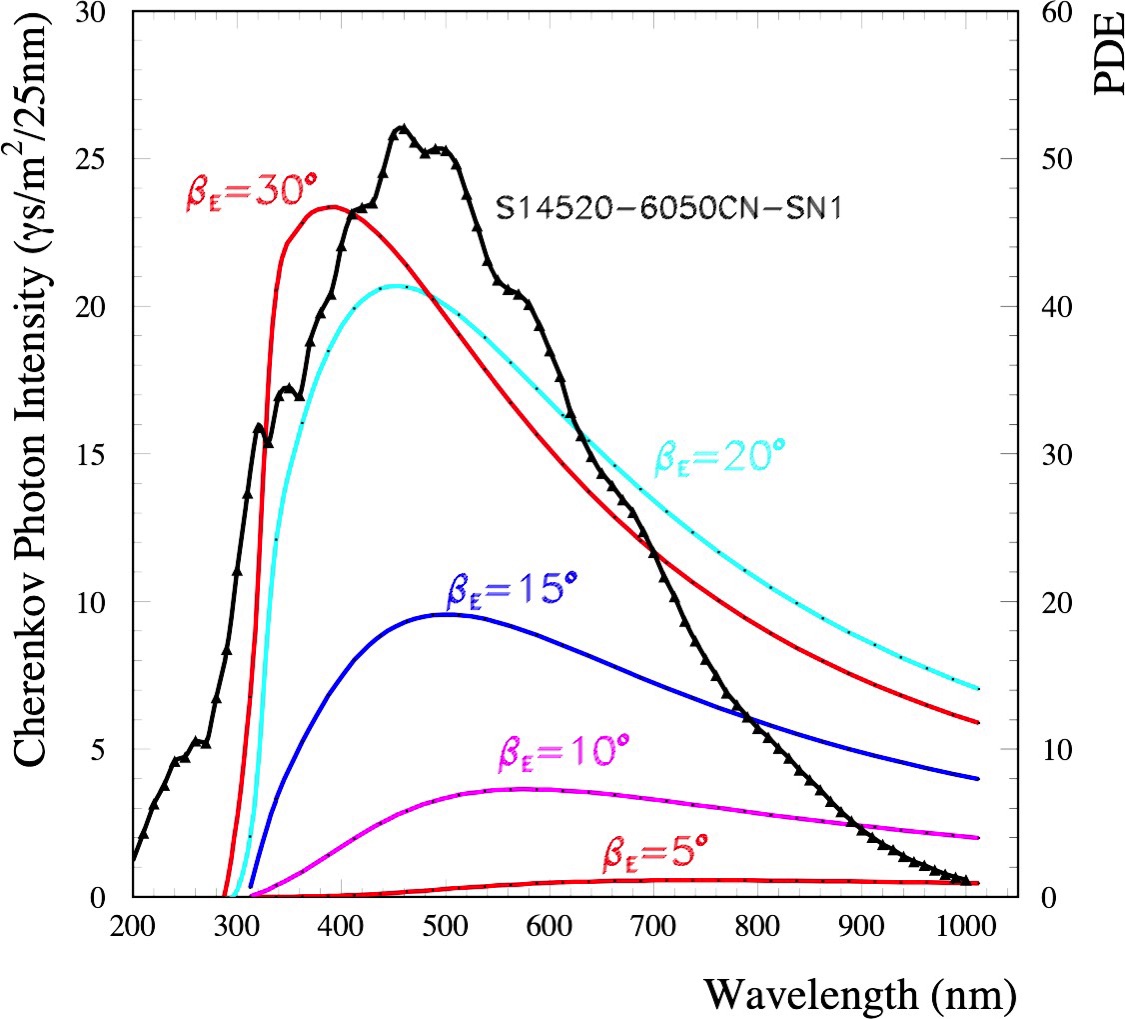}
 \end{center}
\vspace{-0.3cm}
  \caption{\textit{Left:} The layout of the POEMMA hybrid focal plane showing the MAPMT-based PFC (larger, lower blue area) and the SiPM-based PCC (upper, red area). The orange band represents the $7^\circ$ angular span, when in POEMMA-limb mode, that monitors the area below the Earth's limb for upward EAS sourced by tau neutrino interactions in the Earth (from Ref.~\cite{Krizmanic:2020shl}). \textit{Right:} The Cherenkov light intensity as function of wavelength based on simulated 100 PeV upward-moving EAS starting at sea level and as a function of $\tau$-lepton Earth emergence angle. The measured photon detection efficiency (PDE) of a Hamamatsu S14520 SiPM array \cite{Otte:2018cnv} is overlaid with the PDE scale given on the right horizontal axis (from Ref. \cite{POEMMA:2020ykm}).
  }

  \label{POEMMAphotSens}
\end{figure}

Somewhat similar to the requirements for ground-based Imaging Air shower Cherenkov Telescopes (IACTs; see Section~\ref{sec:atmospheric_cherenkov}), the wide wavelength band and spectral variability for imaging the Cherenkov light signal from upward, moving extensive air showers (EAS) are well matched to the response of SiPMs \cite{Krizmanic:2020shl}. Additionally, the ns-scale time response of SiPMs is well matched to the $\sim$10 ns temporal width (near detection threshold) of the EAS Cherenkov light signal. Fig.~\ref{POEMMAphotSens} \textit{Right} shows the simulate Cherenkov spectrum generated by 100 PeV EAS initiated at sea level and as a function of $\tau$-lepton Earth-emergence angle as measured by the Probe of Extreme Multi-Messenger Astrophysics (POEMMA), orbiting at 525 km altitude \cite{POEMMA:2020ykm,Reno:2019jtr}. The high variability of the Cherenkov spectrum in the 300$-$1000 nm band is due to the significant differences in the cumulative depth of attenuating aerosols while the reduction in the signal below $\sim$300~nm is due to atmospheric ozone absorption \cite{Krizmanic:1999gf,Reno:2019jtr,Cummings:2020ycz}. The dominant background for space- and sub-orbital based EAS optical signal detection is due to the dark-sky air glow of the atmosphere (i.e. \cite{Meier1991,Shepherd2006}) and this background becomes significant in the 300$-$1000~nm band \cite{2003A&A...407.1157H,2006JGRA..11112307C}, reaching a level of $\sim 1.5 \times 10^4$ $\gamma$m$^{-2}$nm$^{-1}$ns$^{-1}$sr$^{-1}$, compared to $\sim500~\gamma$m$^{-2}$nm$^{-1}$ns$^{-1}$ in the 300$-$500 band where the EAS air fluorescence signal dominates \cite{Krizmanic:2020shl}. The large dark-sky air glow background for Cherenkov light measurements yields a relatively high photo-electron (PE) threshold, $\gtrsim 10$ PEs, implying the SiPM instrumental effects such as dark-count rate and crosstalk become relatively reduced. Fig.~\ref{POEMMAphotSens}:Left shows the layout of a POEMMA focal surface in each of the two Schmidt telescopes \cite{POEMMA:2020ykm,Anchordoqui:2019omw}. The small upper part of the focal surface is the POEMMA Cherenkov Camera (PCC) constructed of SiPM arrays that monitor a $7^\circ$ region below the Earth's limb for the Cherenkov upward-moving EAS generated by tau neutrino interactions in the Earth \cite{Reno:2019jtr,Cummings:2020ycz}, shown in the orange band encompassing the SiPM array. The larger group of arrays in Fig.~\ref{POEMMAphotSens} \textit{Left} represent the POEMMA Fluorescence Camera (PFC), whose baseline design uses multi-anode photomultiplier tubes (MAPMTs) for the PFC components, due to the MAPMT spectral response combined with a near-UV transparent filter accepts the majority of the EAS air fluorescence signal while minimizing the dark-sky air glow background. In principle SiPMs could be used in the PFC but an appropriate filter that only has transmission in the 300$-$500~nm bandpass and no transmission above 500~nm, at least until the SiPM photo-detection efficiency becomes negligible, is required \cite{Krizmanic:2020shl}.

%% file: CZT-Moiseev.tex
 \noindent
 \chapterauthor[1,2]{Alexander A. Moiseev}
 \chapterauthor[3]{Aleksey E. Bolotnikov}
 \chapterauthor[3]{Gabriella A. Carini}
 \\
 \begin{affils}
   \chapteraffil[1]{University of Maryland, College Park, MD, USA}
   \chapteraffil[2]{CRESST, NASA Goddard Space Flight Center, Greenbelt, MD, USA}
   \chapteraffil[3]{Brookhaven National Laboratory, }
 \end{affils}

A position-sensitive Virtual Frisch-grid (VFG) CZT (cadmium zinc telluride) bar detector with a large geometrical aspect ratio, e.g., $6\times6\times20$ or $8\times8\times30$~mm$^3$, has been found to be an efficient and economically viable way for making large-area detecting planes with few Front-End Electronics (FEE) channels and large detector thickness, and, consequently, higher detection efficiency. This technique has been developed at Brookhaven National Laboratory (BNL) during last five years. The main feature of the detector is four conducting pads attached to the sides of the encapsulated CZT crystal bar. These pads, attached near the anode, act as the virtual Frisch-grid because the pads are virtually grounded through the ASIC front-ends. The detector operates as a miniature Time-Projection Chamber (TPC). The collected charge signals from the anode and the induced signals on the pads and the cathode (six signals in total per bar) are read out to provide X~\&~Y coordinates by combining their ratios, while the ratio of the cathode to the anode is used to measure Z coordinates, along with measuring the drift time to make independent Z measurements~\cite{bolNIM}. 

\begin{figure}[b!]
    \centering
    \includegraphics[height=5cm]{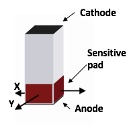}
    \hfill
    \includegraphics[height=6.5cm]{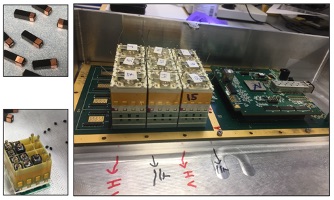}
    \caption{\textit{Left:} CZT bar artistic image. \textit{Right:} The components of the CZT Imaging Calorimeter. Upper left shows individual CZT bars with copper sensitive pads; bottom left shows the bars being inserted into the crate; and the far right shows the Calorimeter prototype $3\times3$ array, $10~\mathrm{cm}\times10~\mathrm{cm}$ footprint.}
    \label{fig:czt_moiseev}
\end{figure}

An important advantage of the position-sensitive VFG detectors is the ability to correct non-uniformity of the response caused by crystal defects that are present even inside the best quality CZT crystals. Such a correction allows one to use standard-grade crystals produced with higher acceptance yields and, thus, to reduce the overall cost of the instrument, which is critical for such a larger instrument for potential space application. Using position information, we can virtually segment a detector into voxels, equalize responses from each voxel, and apply corrections accordingly in the data analysis---we call it 3D corrections---and significantly improve the energy resolution of the detector. In tests of the individual detectors and modules, we obtained a promising energy resolution of $\sim$2\% FWHM at 200~keV and $<$1\% at $>$662~keV (after 3D corrections), with good 3D position resolution of $<$1~mm possible with these detectors.

A key feature distinguishing this design from other concepts is that the bars are placed side-by-side vertically. In this configuration, the effective detector thickness becomes equal to the bar length, whereas it is typically limited by commercial CZT detector thickness and does not exceed 15~mm. Bars with lengths up to 50~mm have been tested at BNL, and we are currently utilizing $8\times8\times32$~mm$^3$ bars, which we found to be the optimal configuration. This large effective detector thickness dramatically increases the detection efficiency and energy containment for high-energy photons, as well as signal quality.

With the support of two NASA Astrophysics Research and Analysis (APRA) grants, our team at Goddard Space Flight Center and the Department of Energy's BNL has developed a modular design, which is a prototype of what could be used in a future space instrument. The CZT bars are tightly packed inside the cells of the egg-crate structure and secured with two copper beryllium spring contacts between the anode and cathode boards, which also provide the bar cushioning. Currently, we have integrated a fully functional prototype, comprising $3\times3$ crates. The crates are plugged into a motherboard, which also carries low-voltage power regulators, analog-to-digital converters, an FPGA, and a fiber-optics communication interface.

We demonstrated the basic principles and benefits of this technology for gamma-ray space telescopes and its ability to measure with high efficiency both the photon interaction sites and the deposited energy with good accuracy: $<1$~mm for the position resolution, and $<$1\% FWHM for the energy resolution. Arrays of such detectors have been recognized as promising for use in various gamma-ray telescopes as a stand-alone Compton detector (see Section~\ref{sec:compton_tel}) and as a focal-plane detector for the instruments with a Coded-mask (see Section~\ref{sec:codedmask_tel}. Furthermore, using the crate-based modular design allows for flexibility in selecting array configurations and sizes for large-area detector systems.

%% file: ModelingSoftware-Wade.tex
 \noindent
 \chapterauthor[1]{Wade Duvall}\orcidlink{0000-0001-9322-6153}
 \chapterauthor[1]{Clio Sleator}\orcidlink{0000-0003-4732-6174}
 \\
 \begin{affils}
   \chapteraffil[1]{Naval Research Laboratory, Washington, D.C., USA}
 \end{affils}

Instrument modeling and simulations are important tools in all phases of a
gamma-ray experiments lifetime. During the proposal phase, simulations can be
used to estimate the sensitivity of an instrument before it has been
constructed, and help demonstrate that the instrument can meet the scientific
mission goals. During the design phase, simulations can be used to optimize the
experiment design by estimating the instrument performance for a variety of detector,
electronics and shielding configurations. During the operational phase,
simulations can be used to assist in instrument calibration, reconstruct
complex interactions in the instrument, and analyze the data via simulated
instrument response functions.

Here, we will discuss the radiation transport codes, and some front ends that
are commonly used for modeling gamma ray detectors. We will also discuss codes
used for modeling air showers and for ray tracing photos in Cherenkov telescopes.
While event reconstruction often uses simulation, we will not discuss it here,
see Section~\ref{sec:reconstruction}.

\subsubsection*{Radiation transport codes}

\subparagraph*{Geant4}
Geant4 \cite{geant41} is a popular  Monte Carlo simulation toolkit developed by
nuclear and particle physicists. It covers a wide range of energies with a
series of different physics lists, and includes the energies useful to space
based gamma ray missions. Geant4 has undergone extensive benchmarking and
validation \cite{geant42}.

While very powerful, Geant4 requires a lot of time and expertise to use. Models
can be created using C++ objects, or by importing a Geometry Description Markup
Language (GDML) file.  Both of these can be difficult to generate, so often
times, a Geant4 frontend is used. While geometries can be viewed with a variety
of built in rendering engines, there is not a GUI to create geometries. Geant4
is free, open source, and available from
\url{https://geant4.web.cern.ch/support/download}.

\subparagraph*{MCNP}
Monte Carlo N-Particle (MCNP) \cite{mcnp} is a general purpose Monte Carlo radiation
transport code developed by Los Alamos National Laboratory (LANL). MCNP is very
popular in the defense and medical fields, but has also been used to model
gamma ray instruments for astrophysics. MCNP has been benchmarked and validated
in the energy ranges of interest to gamma-ray astrophysics, which makes it
an attractive choice. Similar to Geant4, building models requires a lot of
experience, although several tools (including SWORD, see Section~\ref{sword}) can help
simplify the process greatly.  MCNP also features many variance reduction tools
which can reduce simulation times.  MCNP is distributed via the Radiation
Safety Information Computational Center (RSICC) at Oak Ridge National
Laboratory at \url{https://rsicc.ornl.gov/codes/ccc/ccc8/ccc-850.html}.

\subparagraph*{CORSIKA}
COsmic Ray SImulations for KAscade (\textbf{CORSIKA}) \cite{corsika} is a framework for
simulating extensive air showers from cosmic rays interacting with the
atmosphere. Air showers are difficult problems to solve using quantum electrodynamics and quantum chromodynamics calculations, so a Monte Carlo approach is currently the best option.
CORSIKA uses a number of hadronic, electromagnetic models. CORSIKA also provides
several atmospheric models \cite{corsika}, or one can be provided by the user.
CORSIKA can also model Cherenkov and Askaryan radiation and neutrinos generated
in the atmosphere. CORSIKA is avaiable from
\url{https://www.iap.kit.edu/corsika/79.php}.

\subsubsection*{Monte Carlo front ends}
\subparagraph*{SWORD}
\label{sword}
SoftWare for Optimization of Radiation Detectors (SWORD) \cite{sword} is a
vertically integrated simulation package developed at the U.S. Naval Research
Laboratory.  SWORD has a CAD-like interface for building models, interfaces to
Geant4 and MCNP Monte Carlo engines, and includes tools for analyzing the
results. SWORD uses a physics list validated under 1 GeV. SWORD can also export
models as GDML files, in case the user wishes to use the SWORD model-building
interface with a different physics list or simulation software that takes GDML
files. SWORD has been used to model several upcoming missions \cite{siri, gamma_detector}, including Starburst and Glowbug (see Sections~\ref{sec:starburst} and \ref{sec:glowbug} \cite{glowbug}). SWORD is publicly available via RSICC at
\url{https://rsicc.ornl.gov/codes/ccc/ccc7/ccc-767.html}

\subparagraph*{MEGAlib}
The Medium Energy Gamma-ray Astronomy library (MEGAlib) \cite{MEGALIB} is an
open-source toolkit designed to model gamma-ray detectors, with specialized
code for the analysis of data from Compton telescopes. MEGAlib provides a
geometry and detector description for creating models, and uses Geant4 to
conduct the radiation transport. MEGAlib also provides an extensive suite of
tools for analysis and event reconstruction which are covered in Sections~\ref{sec:analysistools} and \ref{sec:reconstruction}. MEGAlib can also apply energy
resolution, depth resolution, and trigger thresholds to simulate electronics
readouts. MEGAlib is used by many missions including NuSTAR \cite{nustar}, COSI (see Section~\ref{sec:cosi} \cite{cosi}), and AMEGO (see Section~\ref{sec:amego} \cite{amego}).  MEGAlib is open source and available on
GitHub via \url{https://github.com/zoglauer/megalib}. 

\subsubsection*{Ray tracing}
Ray tracing is an important tool for Cherenkov telescopes. These instruments
can have complex geometries, and ray tracing can help model the mirror
alignment, optical efficiency and shadowing, and effects of wavelength
dependent mirror reflectivity.

\subparagraph*{ROBAST}
ROOT-Based Simulator for Ray Tracing (ROBAST) \cite{robast} is a non-sequential
ray-tracing simulation library. Designed primarily for Cherenkov telescopes,
this project aims to be instrument agnostic. ROBAST uses the ROOT \cite{root}
geometry library and particle tracking engine and can do non-sequential ray
tracing. ROBAST also includes geometry classes for Winston cones and light
concentrators. ROBAST is used for many of the Cherenkov Telescope Array (CTA; see Section~\ref{sec:cta})
instruments. ROBAST is open source and available on github at
\url{https://github.com/ROBAST/ROBAST}.

\subparagraph*{sim\_telarray}
sim\_telarray \cite{simtel} is an extension to CORSIKA inititally designed for
the HEGRA telescope and expanded to include HESS as well as the Cherenkov
Telescope Array. Particles from the CORSIKA air shower are then ray traced
through the instrument. This package can also model the electronics of these
instruments and provide a simulated electronics readout. It is open source and
available from \url{https://www.mpi-hd.mpg.de/hfm/~bernlohr/sim\_telarray/}.

\subparagraph*{Zemax OpticStudio}
Zemax OpticStudio is a general optical design program that can be used to model
Cherenkov telescope hardware \cite{zemax}. This product has a number of
downsides including a costly licence and compatibility only with Microsoft
Windows; for these reasons ROBAST is generally favored. However, Zemax can
simulate polarization (ROBAST lists this as a coming soon feature at the time
of writing), and can import CAD files. If a project requires either of these
features, Zemax is a good alternative to ROBAST.

%% file: EventReconstruction-Andreas.tex
 \chapterauthor[]{Andreas Zoglauer}\orcidlink{0000-0001-9067-3150}
 \\
 \begin{affils}
   \chapteraffil[]{Space Sciences Laboratory, University of California at Berkeley, Berkeley, CA 94720, USA}
 \end{affils}

One of the key elements of the data-analysis pipeline of modern gamma-ray telescopes is event reconstruction. Most modern gamma-ray detectors can only measure the location and the deposited energies of the interactions of the particles in the detector, but not the sequence of interactions due to compact detector designs and the slow charge/light collection compared to the speed of light. COMPTEL, the first successful Compton telescope in space, was an exception, as it was capable of measuring the time of flight between the interactions. For non-COMPTEL-type gamma-ray telescopes the paths of the particles have to be determined from the measured locations and energies as well as the known physics the particles are following such as interaction, scatter, and absorption probabilities. When the paths are determined, the parameters of the original Compton and pair interaction can be calculated such as initial energies and scatter directions. These parameters then can then be fed into the high-level data analysis pipeline to, e.g., create sky-maps and source spectra.

Every existing Compton and pair telescope has its own specific, optimized event reconstruction toolset such as FERMI, AGILE, COMPTEL, Hitomi/SGD. However, there is one open-source toolset which can perform event reconstruction for different types of Compton and pair telescopes, and therefore is used by a wide range of current and future gamma-ray telescopes such as COSI, AMEGO, eASTROGAM, and GECCO: MEGAlib - the Medium-Energy Gamma-ray Astronomy library. MEGAlib is capable of handling the full data-analysis pipeline from simulations and calibrations, via event reconstruction to high-level data analysis such as imaging.

MEGAlib splits the event reconstruction into several independent steps:
\begin{enumerate}
\item Coincidence search: Combine hits which are within a certain detector-dependent coincidence window into events. Some detectors do this automatically as they have a built-in trigger and coincidence logic.
\item Event clustering: Some detector systems, such as ground-based Compton cameras for hadron-therapy monitoring, can measure multiple gamma rays at the same time. This step determines which hit belongs to which gamma ray.
\item Hit clustering: Some detectors are very finely pixelated so that neighboring strips or voxels are triggered by the same particle, for example, by the Compton recoil electron. This step combines those neighboring triggers into individual hits.
\item Event type determination: This step classifies the hit pattern in the detector into event types such as Compton events, pair-creation events, charged particle events, or unidentifiable events.
\item Path reconstruction:
\begin{itemize}
\item Pair events: the tracks of electron and positron are reconstructed
\item Compton events: 
\begin{itemize}
\item Recoil-electron tracking: for Compton telescopes which can determine the direction of the recoil electron, its path is reconstructed
\item Compton sequence reconstruction: this step finds the overall sequence of Compton interactions in the detector.
\end{itemize}
\end{itemize}
\item Background identification: This final step tries to identify if the event shows signatures of typical background events (e.g.~beta decays), and determines a probability that the event originated from background.
\end{enumerate}
While MEGAlib contains the full set of event reconstruction tools, the complexity of the underlying data space of modern Compton and pair telescopes still leaves significant room for future improvements. One of the most promising methods are neural networks, since all steps in the event reconstruction pipeline are well-suited for machine learning. For example, for COSI the best performing approach for Compton sequence reconstruction is a neural network, and for AMEGO-X the best performing approach for event type identification is a 3D convolutional neural network. Developing machine-learning based approaches for the other event reconstruction steps in the pipeline is an ongoing process requiring significant  resources --- especially when this is done in a way that is not optimized for a single instrument, but generally for all types of Compton and pair telescope. 
In addition, the experience of applying neural-network based approaches trained on simulations to COSI measurements has shown that it is not sufficient to just have an overall better performing approach, but it is also important to verify that a specific trained neural network works for all relevant energies, incidence direction, interaction distances, scatter angles, etc. Setting up the verification procedure is at least as much effort as setting up and optimizing the neural network approach itself.

%% file: AnalysisTools-Israel.tex
 \noindent
 \chapterauthor[1,2,3]{Israel Martinez-Castellanos}\orcidlink{0000-0002-2471-8696}
 \\
 \begin{affils}
   \chapteraffil[1]{University of Maryland, College Park, MD, USA}
   \chapteraffil[2]{Astroparticle Physics Laboratory, NASA Goddard Space Flight Center, Greenbelt, MD, USA}
   \chapteraffil[3]{Center for Research and Exploration in Space Science and Technology, NASA/GSFC, Greenbelt, MD, USA}
 \end{affils}

Analysis software has become an essential part of extracting information from gamma-ray observations. Currently the gamma-ray community expects a rigorous statistical treatment, accounting properly from detector effects, low count statistics and systematics. Sophisticated analysis methods can also increase the sensitivity of an instrument, improvements that are crucial when the observing time is limited and that can be more cost-effective than hardware upgrades. More recently there is a trend, likely to continue, of using machine-learning techniques to look for details and complex patterns that humans might miss.  

The use of machine learning methods has been motivated by its success in other fields and the desire to go beyond model-based searches. Machine learning has been explored by multiple instrument teams as a tool for background rejection. These implementations to separate gamma rays from other species have shown promising results when compared against data \cite{postnikov2019gamma, capistran2021use}. Machine learning methods have also been applied to source identification, aiding in the choice of targets for further observations \cite{Kaur_2019, Saz_Parkinson_2016, 10.1093/mnras/staa166}. The use of machine learning is poised to become a necessity to efficiently mine the increasingly richer datasets generated by modern observatories. 

Many collaborations now have the need to use ``big data'' strategies to be able to handle their large datasets. This calls for efficient algorithms that make use of high performance and high throughput computing clusters. Today, there is a tendency of moving these operations to cloud services, which can run on a variety of architectures. This has led to the increase of the practice of ``containerization'', the packaging of software including all dependencies required to run it. The community can capitalize on the use containers to increase the reproducibility of the science results, as well as making the analysis tools more accessible to institutions that lack the resources and expertise to install and manage their own data center.

Making use of all data across the electromagnetic spectrum and other astronomical messengers has been recognized as crucial to tackle the big questions of the field. There is an increasing need to perform broadband analyses using data coherently. This is a challenging task. Data formats are not always consistent, the data produced by any instrument has its own peculiarities and the systematics errors are hard to account for adequately. Nevertheless, progress has been made. Section \ref{sec:softwareinfrastructure} details some examples. The push towards open-data science is not only laudable in itself but will also promote the interoperability between the different analysis software libraries. 

Software development has become a collaborative effort in itself. Developers are moving from writing similar pieces of code for each instrument toward building libraries of general use, avoiding unnecessary duplication of code and diverting efforts towards robustness and new features. It is however important to recognize that a large fraction of the progress in open-source software is possible thanks to volunteers whose work is, on many occasions, not covered by research grants. The codebase they develop forms part of the backbone of many other analyses which can be jeopardized if long-term funding is not secured. We need to invest on the  maintenance of common software infrastructure and the training of future developers. 

Modern gamma-ray astronomy cannot be performed without analysis software. The developers that make this possible do not necessarily overlap with the end users listed as authors in publications. It is important to find avenues to recognize their work. Some progress has been achieved on the topic, with some journals encouraging the submission of publications describing relevant pieces of software and the use of citations that give credit to the authors. Without increasing the number of such initiatives that allow software developers to make a career in academia the community is at risk of losing critical talent.  

%% file: HEASARC-Smale.tex
\noindent
\chapterauthor[1]{Alan Smale}\orcidlink{0000-0001-9207-9796}
\chapterauthor[1]{Tess Jaffe}\orcidlink{0000-0003-2645-1339}
\\
 \begin{affils}
   \chapteraffil[1]{HEASARC Office, NASA Goddard Space Flight Center, Greenbelt, MD}
 \end{affils}

Since its inception in 1990, the High Energy Astrophysics Science Archive Research Center (HEASARC) has been the primary archive for NASA’s (and other space agencies’) missions studying electromagnetic radiation from extremely energetic cosmic phenomena ranging from black holes to the Big Bang (see Section \ref{sec-fundamental} for science topics). The HEASARC also includes LAMBDA (the Legacy Archive for Microwave Background Data Analysis), a thematic archive containing data from space missions, balloons, and ground-based facilities that have studied cosmic microwave background data. The HEASARC serves data from a variety of legacy gamma-ray missions, and is already working with many of the missions and facilities listed in this paper (Chapter \ref{sec-facilities}) to establish future data management plans and archiving strategies.

The astronomy community, including the HEASARC, has also established the International Virtual Observatory (VO) Alliance that provides standard API definitions so that all astronomy data worldwide can be browsed and retrieved by any VO client.  All astronomical and astroparticle research fields requiring multi-wavelength or multi-messenger data benefit from the closer collaboration of data archives using VO protocols that are now reaching maturity.  

The HEASARC, in collaboration with its partner archives MAST and IRSA, is working on a fundamental shift in how the community does research, moving away from the model where researchers use their local machines, which serves to disadvantage researchers with fewer computational resources and less technical expertise.  The HEASARC WebHera platform was one of the first sites enabling users to access and analyze high energy data via a browser, without needing to download either data or software.  Their new system is the HEASARC@SciServer platform where all standard software is prebuilt, the entire HEASARC archive available locally, and the user given powerful tools through JupyterLab (e.g., to add their own software installations) and infrastructure to manage collaborations and share proprietary results.  The next phase of this progression will be to take full advantage of cloud computing resources to provide access not only to HEASARC data but also to all astronomical data on the cloud using seamless standard interfaces.   

NASA is currently making a big push to encourage Open Science, which includes not just opening up the data but also the software generated by all NASA funded work.  The HEASARC has been releasing software to the community for decades and will continue to do so, and it is now collaborating on relevant open source projects in the community such as astropy.

%% file: SSDC-Gianluca.tex
\chapterauthor[ ]{ }

 \addtocontents{toc}{
     \leftskip3cm
    \scshape\small
    \parbox{5in}{\raggedleft Gianluca Polenta et al. on behalf of the SSDC staff}
    \upshape\normalsize
    \string\par
    \raggedright
    \vskip -0.19in
    }
 
\noindent
\nocontentsline\chapterauthor[]{Gianluca Polenta$^{1}$}\orcidlink{0000-0003-4067-9196}
\nocontentsline\chapterauthor[]{Stefano Ciprini$^{1,2}$}
\nocontentsline\chapterauthor[]{Valerio D'Elia$^{1}$}
\nocontentsline\chapterauthor[]{Dario Gasparrini$^{1,2}$}\orcidlink{0000-0002-5064-9495}
\nocontentsline\chapterauthor[]{Marco Giardino$^{1}$}
\nocontentsline\chapterauthor[]{Cristina Leto$^{1}$}
\nocontentsline\chapterauthor[]{Fabrizio Lucarelli$^{1,3}$}
\nocontentsline\chapterauthor[]{Alessandro Maselli$^{1,3}$}\orcidlink{0000-0003-3760-1910}
\nocontentsline\chapterauthor[]{Matteo Perri$^{1,3}$}
\nocontentsline\chapterauthor[]{Carlotta Pittori$^{1,3}$}\orcidlink{0000-0001-6661-9779}
\nocontentsline\chapterauthor[]{Francesco Verrecchia$^{1,3}$}
\nocontentsline\chapterauthor[]{and the SSDC staff}
\\
\begin{affils}
    \chapteraffil[1]{Space Science Data Center, Italian Space Agency, via del Politecnico snc, 00133, Roma, Italy}
    \chapteraffil[2]{INFN-Sezione di Roma Tor Vergata, 00133, Roma, Italy}
    \chapteraffil[3]{INAF-OAR, via Frascati 33, 00078 Monte Porzio Catone (RM), Italy}
\end{affils}

The Space Science Data Center\footnote{\href{https://www.ssdc.asi.it}{https://www.ssdc.asi.it}} (SSDC) is a Research Infrastructure of the Italian Space Agency designed to acquire, reduce, analyse, and distribute data from supported science missions following the open science FAIR ({\it Findable, Accessible, Interoperable, and Reusable}) principles.

The SSDC is all but a simple data repository. In a collaborative effort between ASI, National Institute for Astrophysics (INAF), and National Institute for Nuclear Physics (INFN), the SSDC develops online, user-friendly, publicly available scientific tools and services to allow researchers as well as non-expert users to effectively search, retrieve and use science data, thus removing those barriers possibly restricting the science exploitation to domain experts only.

This is an important aspect in general, but it becomes crucial when addressing science topics requiring multi-wavelength and multi-messenger analysis. Indeed, the SSDC is an intrinsic multi-wavelength, multi-messenger facility hosting data from space missions covering a broad range of frequency, from radio to gamma-rays, and different messengers (photons, cosmic rays, and neutrinos). To better illustrate this point, in the following we will highlight some of the multi-wavelength, multi-mission tools available on our web portal.

\paragraph*{The Sky Explorer:}
The SSDC {\em Sky Explorer}\footnote{\href{https://tools.ssdc.asi.it}{https://tools.ssdc.asi.it}} represents the main access gateway to SSDC services.
Users can specify the name, or corresponding coordinates, of their favourite astrophysical source to investigate in greater details using SSDC web tools.
Since 2021, a new graphic layout enables the users to perform queries also for list of sources.
The new interface, which appears after the upload of the input file, allows one to manage this list in a smart way,  simultaneously opening new, dedicated browser tabs for each source, to start further investigations with more specific SSDC web tools such as the {\em Data Explorer} and {\em SED Builder}, simplifying the quick comparison of the obtained results. 

\paragraph*{The Multi-Mission Interactive Archive:}
The SSDC {\em Multi-Mission Interactive Archive}\footnote{\href{https://www.ssdc.asi.it/mma.html}{https://www.ssdc.asi.it/mma.html}} (MMIA) is a web-based astrophysics archive and advanced database system, which provides access to extensive multi-wavelength information. 
Data from all SSDC resident archives can be obtained and visualized remotely in a simple and homogeneous way. 
Through the MMIA system, archival data from several space missions (e.g. {\it AGILE}, \fermi, {\it Swift}, {\it NuSTAR}, {\it Herschel}) can be easily explored and retrieved. 
In addition, the data can be analyzed online in an interactive way via various graphical web interfaces developed at the SSDC.
With just one click, these data can enrich the collection of those retrieved from the literature and can be readily exploited in other SSDC web tools, such as the {\em SED Builder}.

\paragraph*{The Data Explorer:}
The SSDC {\em Data Explorer} 
is a web-based tool designed to visualize and analyze the data stored in the Multi-Mission Interactive Archive. 
It allows the fast visualization of a portion of the sky centered on a specific source in all the energy bands, and the cross-matching between archival catalogs. 
It also points to external services for more extended searches as well as to internal tools to perform a more detailed analysis.

\paragraph*{The SED Builder:}

The {\em SED Builder}\footnote{\href{https://tools.ssdc.asi.it/SED/}{https://tools.ssdc.asi.it/SED/}} is a tool developed at the SSDC to build and display the Spectral Energy Distribution (SED) of astrophysical sources.

The majority of the data comes from a large and increasing number of catalogs, extracted from ground- and space-based missions and experiments covering the whole electromagnetic spectrum, from radio to TeV energies. These include resident catalogs built with major contributions from SSDC researchers and staff, but also resources imported from external archives, such as NED, CDS, etc. The collection of input catalogs is being expanded to include the contribution of multi-messenger events, such as neutrinos and cosmic rays, that will play an increasingly large role in the forthcoming years, enabling the creation of a so-called ``hybrid SED".

In addition, the {\em SED Builder} allows users to further enrich the SED for the sources of interest by adding their own data sets, as well as those resulting from recent observations of {\it Swift}, {\it NuSTAR}, {\it AGILE} and {\it Fermi} high-energy missions not included in any catalog yet. To this purpose, dedicated tools for the online analysis of these data have been developed and can be easily accessed through the SSDC web portal. Registered users are also allowed to store their data sets and SED collection.

A key point of this tool with respect to similar public ones, is that SED data available in our tool are science ready. This means that infrared, optical, and ultraviolet data from all catalogs are properly corrected for mean Galactic dust extinction, while data in the soft X-ray energy range are corrected for Galactic absorption. Concerning catalogs from high-energy missions that are generally provided in the form of count rates, the conversion to fluxes in physical units is performed taking into account the instrument response and assuming, as a first step, an absorbed power law spectral model.

The {\em SED Builder} provides several plotting options as well as features to analyze SED properties to fulfill user's requirements. In the first place, the choice of several flux units vs frequency, energy or wavelength to express the SEDs. Rest-frame luminosity is also allowed provided that a cosmological redshift is available, and to this purpose a photometric redshift calculator is also present. Then, the SED can be compared with templates taken from literature for various astrophysical sources, such as quasars, elliptical and spiral galaxies. Analytical functions as well as profiles resulting from numerical models, can be fitted to data to compare SEDs with theoretical expectations. Sensitivity curves of selected instruments can be overlapped to the plot, a particularly useful option for weak or undetected sources.

\begin{figure}
    \centering
    \includegraphics[height=10cm]{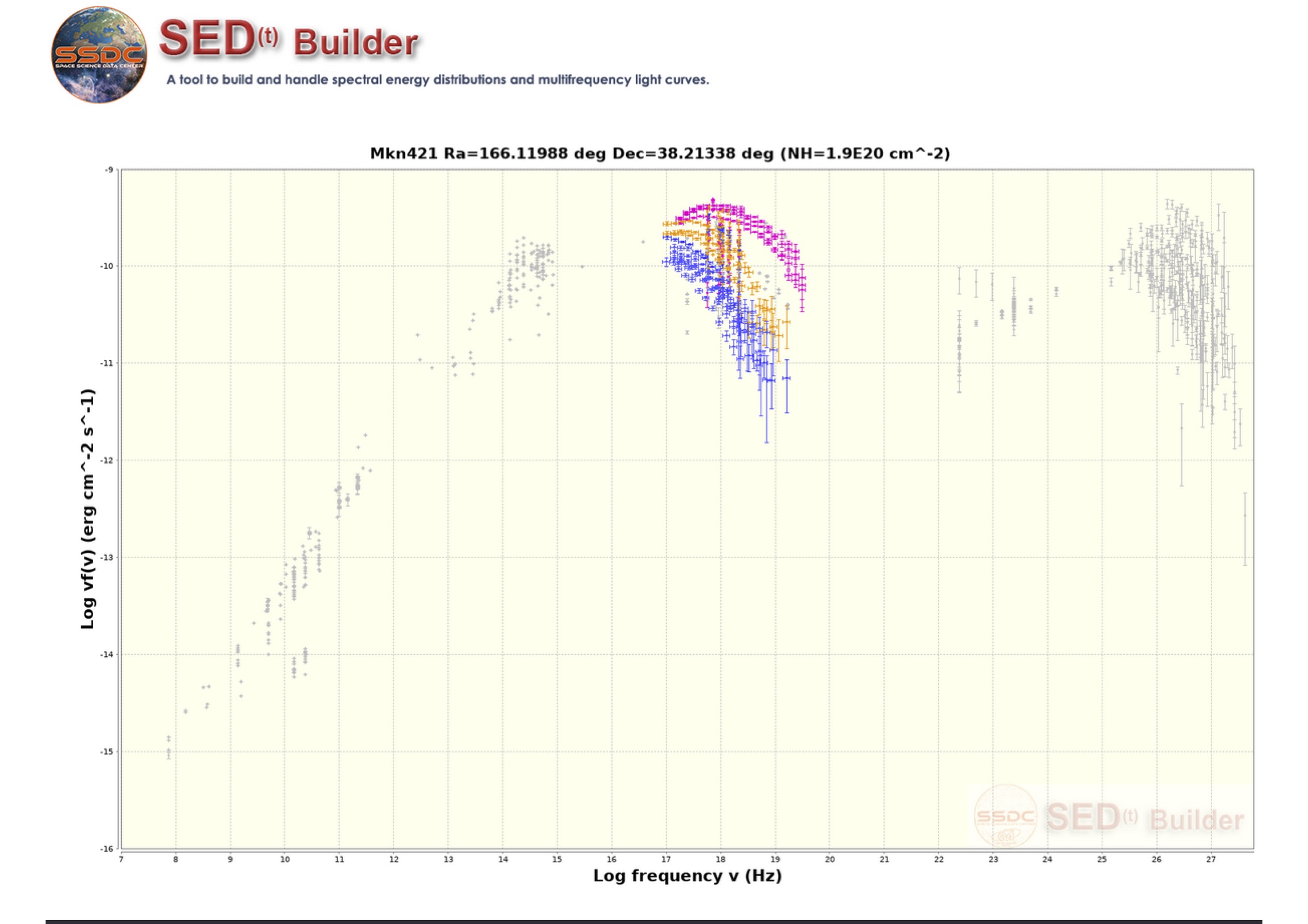}
    \caption{Time-resolved SED for MKN421 in which different colors are used to represent data collected at different epochs in order to highlight soft X-ray variability.}
    \label{fig:SEDMKN421}
\end{figure}

The growing number of telescopes operating on a very wide range of frequencies, both ground- and space-based, increases the probability for many astrophysical sources of being repeatedly observed over time. Hence, the SED Builder allows time filtering of data in order to build time-resolved SEDs, thus making variability studies far easier, also offering the possibility to display and compare light curves at different wavelengths. In Figure \ref{fig:SEDMKN421}, we show a time resolved SED for MKN421 in which we adopted different colors to distinguish observations taken at different epochs tracing high and low states.

\paragraph*{The AGILE-LV3 Tool:}
In the case of gamma-rays in the energy range above 100 MeV, as for missions such as {\it AGILE} \cite{Tavani2009} and \fermi \cite{Atwood2009}, the scientific data analysis over long time-scales may require long processing times (of the order of many hours).

\begin{figure}
    \centering
    \includegraphics[height=13cm,angle=0]{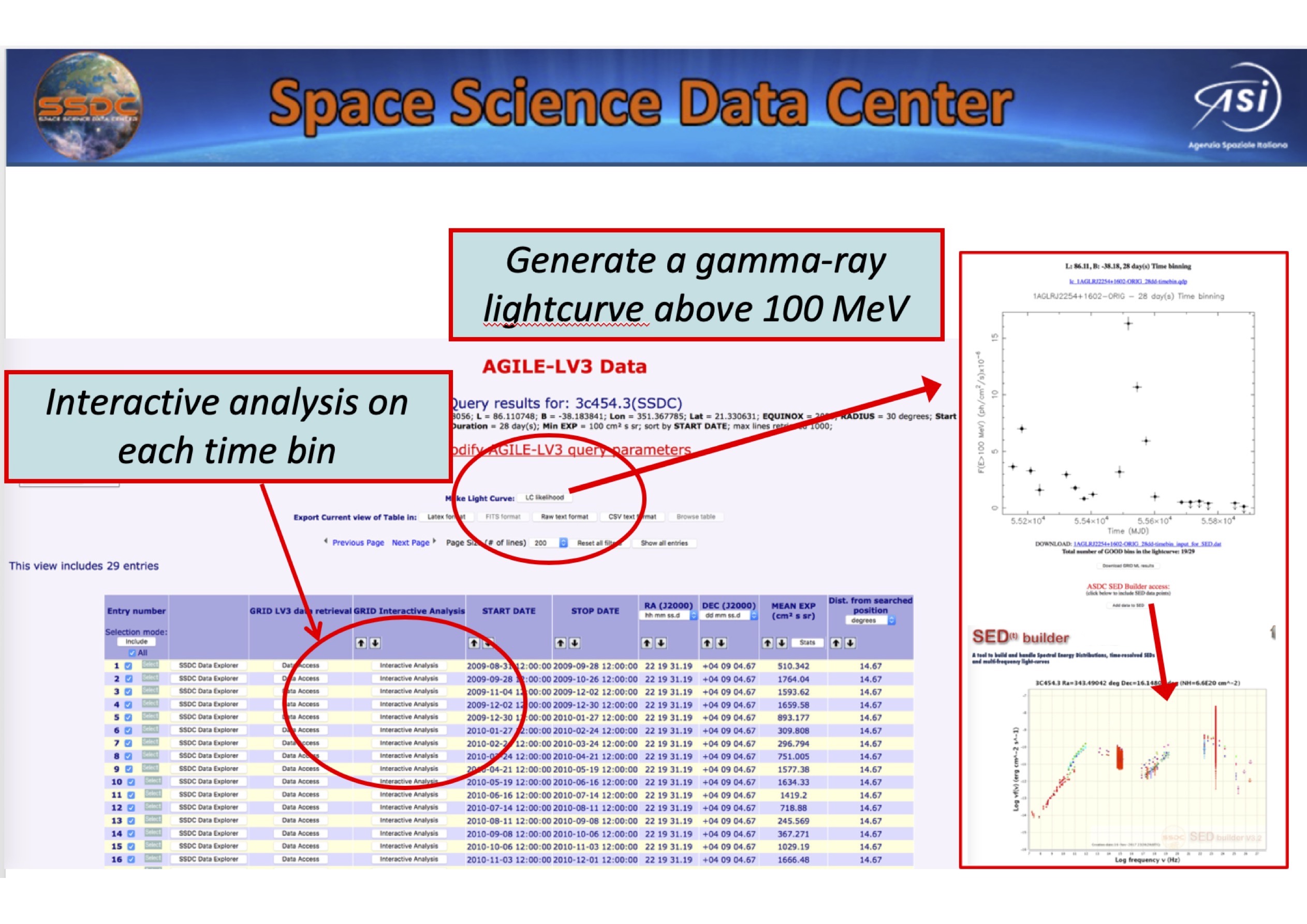}
    \caption{Combined image for illustrative purposes of hands-on data analysis with the {\it AGILE}-LV3 and SED SSDC tools.}
    \label{fig:AGILE-LV3}
\end{figure}

To speed up the {\it AGILE}-GRID scientific analysis, a complete level-3 (LV3) archive of pre-compiled exposure, counts and diffuse background maps over 1-day integration time, with fixed standard parameters was created at SSDC.
The {\it AGILE}-LV3 legacy archive can be used as basis for scientific Maximum Likelihood analysis on time scales that may vary from weeks to months, or even over the entire duration of the mission.
For an easy {\it AGILE}-GRID official data analysis, with waiting times ranging from a few seconds to a few minutes, the interested user may query the entire public LV3 archive through the {\it AGILE}-LV3 web tool\footnote{\href{https://www.ssdc.asi.it/mmia/index.php?mission=agilelv3mmia}{https://www.ssdc.asi.it/mmia/index.php?mission=agilelv3mmia}} \cite{Lucarelli2020}.
In the query page the user can enter the source name or sky coordinates of the object he/she wants to analyze, the period of interest and the duration of the LV3 maps (e.g. 1, 7 or 28 days time bins) to be used in the analysis.
The output from the query automatically selects all {\it AGILE} available observations of the source, and the gamma-ray light curve over the selected period can be directly generated, see Figure \ref{fig:AGILE-LV3}. 

The {\it AGILE}-LV3 tool is meant to be easily comprehensible, does not require any install-on-premises software or calibrations, and it has been even tested with high-school students. It is also directly interfaced with the SED builder and other SSDC tools.

\paragraph*{Fermi Online Data Analysis:}

SSDC is hosting an official mirror archive of the \fermi-LAT high level data products, thus allowing a link between \fermi-LAT data and SSDC tools.

The {\it Fermi Online Data 
Analysis}\footnote{\href{https://tools.ssdc.asi.it/fermi.jsp}{https://tools.ssdc.asi.it/fermi.jsp}} (FODA) is designed to retrieve the data in a very similar way to the NASA \fermi Science Support Center, with the advantage of allowing for a quick-look analysis of the requested data as well as the analysis of science ready data for the requested region using the standard tools of the \fermi-LAT collaboration.

Through a user-friendly web interface, the user can provide source name or sky coordinates, time interval, and energy of the interested region in order to submit the query to the archive. Given the characteristic of \fermi\ sky observations, it is not possible to use pre-computed products to speed up the computation, so we implemented an asynchronous query mechanism that notify the user by email of the successful completion and the availability of data for download.
A GTLIKE analysis can also be requested, and in that case a second email will be delivered announcing that the standard analysis is completed.

Among the results available for a given source, there are the significance of the detection during the requested periods, spectral parameters, confidence regions and high energy photons statistically associated to the requested source that could trigger TeV telescope observations. A PNG plot of the SED is also created for a quick analysis, while the resulting data point can be easily loaded into the SSDC {\em SED Builder}.

%% file: MUWCLASS-Hare.tex
 \noindent
    \chapterauthor[1,2]{Jeremy Hare}
    \chapterauthor[3]{Oleg Kargaltsev}
    \chapterauthor[3]{Hui Yang}
 \\
 \begin{affils}
    \chapteraffil[1]{Astroparticle Physics Laboratory, NASA Goddard Space Flight Center, Greenbelt, MD, USA}
    \chapteraffil[2]{NASA Postdoctoral Program Fellow}
    \chapteraffil[3]{Department of Physics, The George Washington University, 725 21st St. NW, Washington, DC 20052, USA}
 \end{affils}

Gamma-ray astrophysics is currently in an unprecedented era with observatories such as {\it Fermi}-LAT, H.E.S.S., and H.A.W.C. all viewing the sky over several decades in energy. This has led to a rapid and large increase in the number of discovered gamma-ray sources  (e.g. \cite{2022arXiv220111184F,2020ApJ...905...76A}). Consequently, many of these sources have not been classified/associated with a longer wavelength counterpart, thus their nature remains unknown. In order to extract the full scientific potential from the data produced by gamma-ray observatories, these sources must be classified. Understanding the nature of these gamma-ray sources is important in order to study their population characteristics, such as what fraction of Galactic TeV sources are pulsar wind nebulae (PWNe) or how gamma-ray PWNe evolve with
pulsar spin down age  (e.g. \cite{2012ASPC..466..167K,2018A&A...612A...2H}). Associating the TeV sources with PWNe and understanding the true extent
of relic PWNe will also shed light on the nature of the positron excess
observed by Pamela \cite{2009Natur.458..607A} and AMS
\cite{PhysRevLett.110.141102} experiments and on the nature of the
diffuse annihilation line emission in the Galactic center region
\cite{2016A&A...586A..84S}. This will constrain a possible contribution
to this excess from decaying dark matter (e.g.
\citep{2015ICRC...34...14C,PhysRevD.96.103013,2009PhRvD..80f3005M}). (See also Section \ref{sec:darkmatter}.) Additionally, classifying these sources may also lead to the discovery of new classes of gamma-ray emitting sources.

The nature of many gamma-ray sources has been understood through the use of X-ray observations to locate and classify the gamma-ray source's counterpart (e.g.  \cite{2012A&A...548A..46H,2015ApJ...803L..27B,2019ApJ...875..107H}). This is not surprising because many types of Galactic gamma-ray  sources are also bright in X-rays, such as PWNe from young energetic pulsars (e.g.  \cite{2005A&A...442L..25A,2013arXiv1305.2552K}), supernova remnants  (SNRs; e.g. \cite{2014A&A...562A..40H}), and High mass gamma-ray binaries (HMGBs; \cite{2019ApJ...884...93C}). Unfortunately, most gamma-ray observatories have a relatively poor angular resolution, leading to large positional uncertainties (from a few arcmin to tenths of a degree depending on the observatory) of the newly detected sources. The extent of these positional uncertainties often contains many potential X-ray counterparts, especially for sources in the Galactic plane. Furthermore, X-ray data alone is frequently not informative enough to confidently classify a source, making the addition of multiwavelength data (i.e., IR, NIR, optical, UV) critical for source classification. 

Traditional methods of multiwavelength source classification often rely on using various two parameter plots (e.g., X-ray hardness ratios, X-ray to optical/NIR flux) (e.g. \cite{2006ApJS..163..344K,2012ApJ...756...27L}). While this approach can be beneficial it has several drawbacks. The first is it limits exploration of relationships between multiwavelength parameters to two (or at most three) dimensions. Second, it does not provide any robust way to assign confidences of the classifications. Lastly, this process becomes cumbersome when there are many potential X-ray counterparts within the large positional uncertainties of the gamma-ray sources. To overcome these issues, we have constructed a supervised machine learning pipeline (nicknamed MUWCLASS) to classify X-ray sources based on their X-ray and multiwavelength properties. 

MUWCLASS is written in python using the Scikit learn machine learning library and relies on a Random Forest classifier \citep{2012arXiv1201.0490P}. We have two training datasets, with literature verified classes, constructed using the Chandra Source Catalog version 2 \citep{2020AAS...23515405E} and XMM-Newton 4XMM-DR11 \citep{2020A&A...641A.136W} source catalogs and containing $\sim3000$ and $\sim11000$ sources, respectively. The training dataset consists of eight classes (AGN, low and high mass stars, CVs, NSs, low and high mass X-ray binaries, and YSOs) and $\approx31$ multiwavelength features from various catalogs (see \cite{2021RNAAS...5..102Y} for details on the mutliwavelength features and catalogs used to construct the training dataset\footnote{Note that since the publication of \cite{2021RNAAS...5..102Y}, we have merged the low mass X-ray binary and Neutron Star binary (consisting of red back and black widow pulsars) classes }). MUWCLASS achieves an overall accuracy of $86\%$ increasing up to $95\%$ when only confident classifications are considered. So far MUWCLASS has been used in a variety of different fields (e.g., \cite{2016ApJ...821...54S,2017ApJ...839...59P}), including the fields of several gamma-ray sources (e.g., \cite{2016ApJ...816...52H,2017ApJ...841...81H,2020ApJ...901..157K}). 

We anticipate that MUWCLASS will continue to become more accurate as additional multiwavelength surveys release data products that can be incorporated into the pipeline (e.g., eROSITA, VLASS; \cite{2021A&A...647A...1P,2020PASP..132c5001L}). Additionally, large time domain surveys such as ZTF \citep{2019PASP..131a8002B} and VCRO \citep{2019ApJ...873..111I} have been (or soon will be) observing the sky every few days. This will allow us to include additional variability features at different wavelengths into the pipeline, which can help to further differentiate various X-ray source classes (e.g., AGN versus CVs; \cite{2021AJ....162...94S,2021AJ....161..267V}). As new gamma-ray observatories (e.g., COSI, CTA, AMEGO; see Chapter \ref{sec-facilities};  \cite{2021arXiv210910403T,2020arXiv200409213K,2019BAAS...51g.245M}) come online over the next few years the number of newly discovered gamma-ray sources will continue to climb. This coupled with the influx of high cadence multiwavelength data sets will necessitate automated classification and discovery tools such as MUWCLASS in the 2020s and beyond.

%% file: ASCL-AliceAllen.tex
 \noindent
 \chapterauthor[1,2]{Alice Allen}\orcidlink{0000-0003-3477-2845}
 \chapterauthor[2]{\& Peter Teuben}\orcidlink{0000-0003-1774-3436}
 \\
 \begin{affils}
   \chapteraffil[1]{Editor, Astrophysics Source Code Library}
   \chapteraffil[2]{Department of Astronomy, University of Maryland College Park}
 \end{affils}

\paragraph*{The Astrophysics Source Code Library's Role in Research Software:}

The Astrophysics Source Code Library (ASCL ascl.net) is a community resource for and of scientist-written software used in refereed astrophysics, astronomy, and solar system (planetary bodies) research. Its goal is to promote and strengthen the integrity, reproducibility, and falsifiability of research by making the software methods used in this research more discoverable and open for examination. 

Entries are created by ASCL editors, who inspect research papers to find suitable codes used in this research, and through submissions by software authors and community members. For each code, ASCL lists the software’s name, a description of it, who wrote it, and includes links to the code’s download site, to one or more research papers that either describe or use the software, and, when known, the preferred method of citing the software. A unique identifier, the ASCL ID, is assigned to each entry meeting the ASCL's criteria.

Started in 1999, the ASCL has grown significantly in the last decade and now contains over 2700 entries. Its entries are indexed by the \href{http://ads.harvard.edu/}{NASA/SAO Astrophysics Data System} (ADS) and Clarivate’s Web of Science Data Citation Index, and \href{https://ascl.net/wordpress/?page_id=351}{are citable}; citations to ASCL entries are tracked by ADS, Google Scholar, and Web of Science. 
\begin{figure}
\begin{center}
\includegraphics[width=0.6\textwidth]{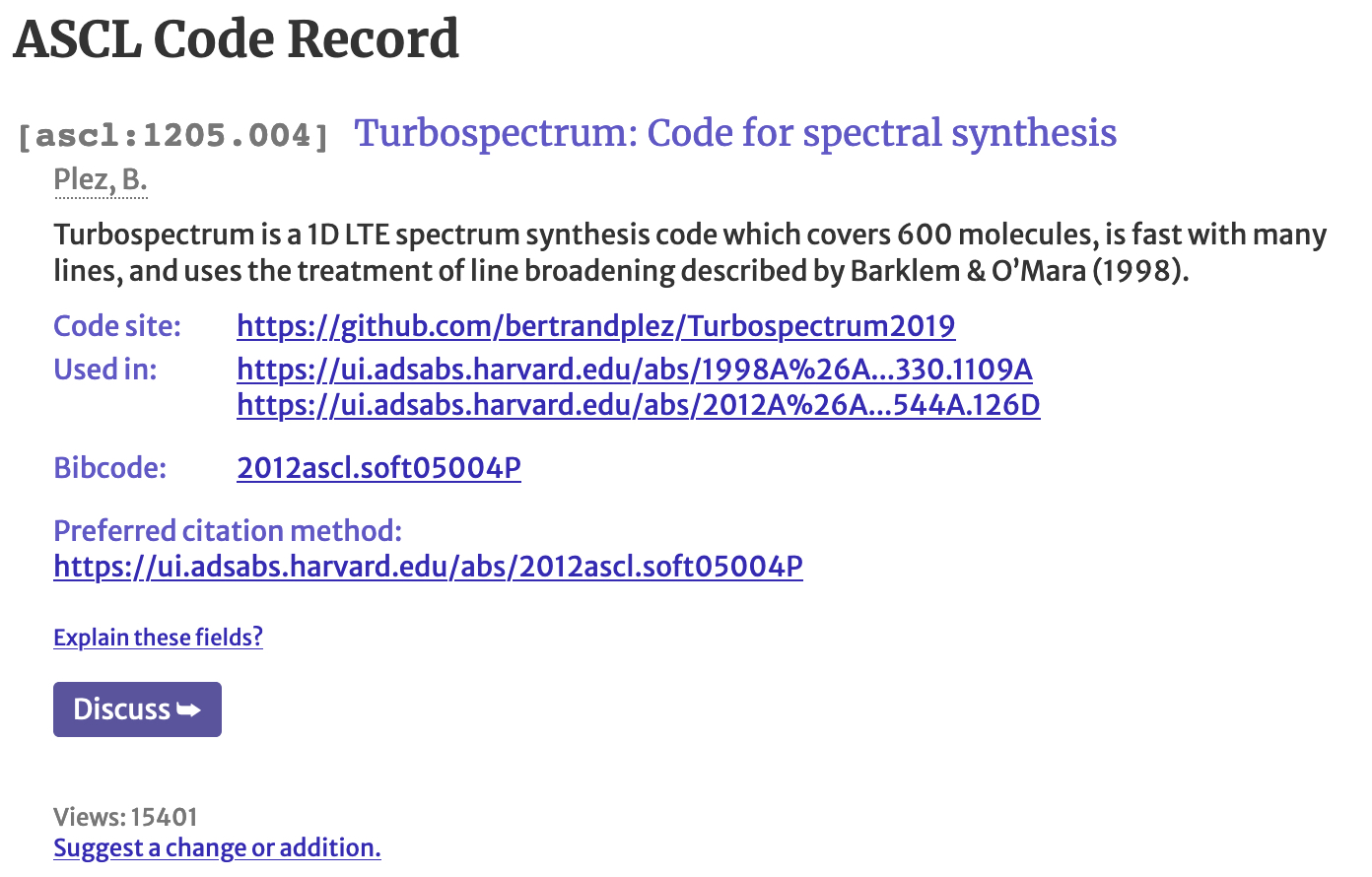}
\includegraphics[width=0.35\textwidth]{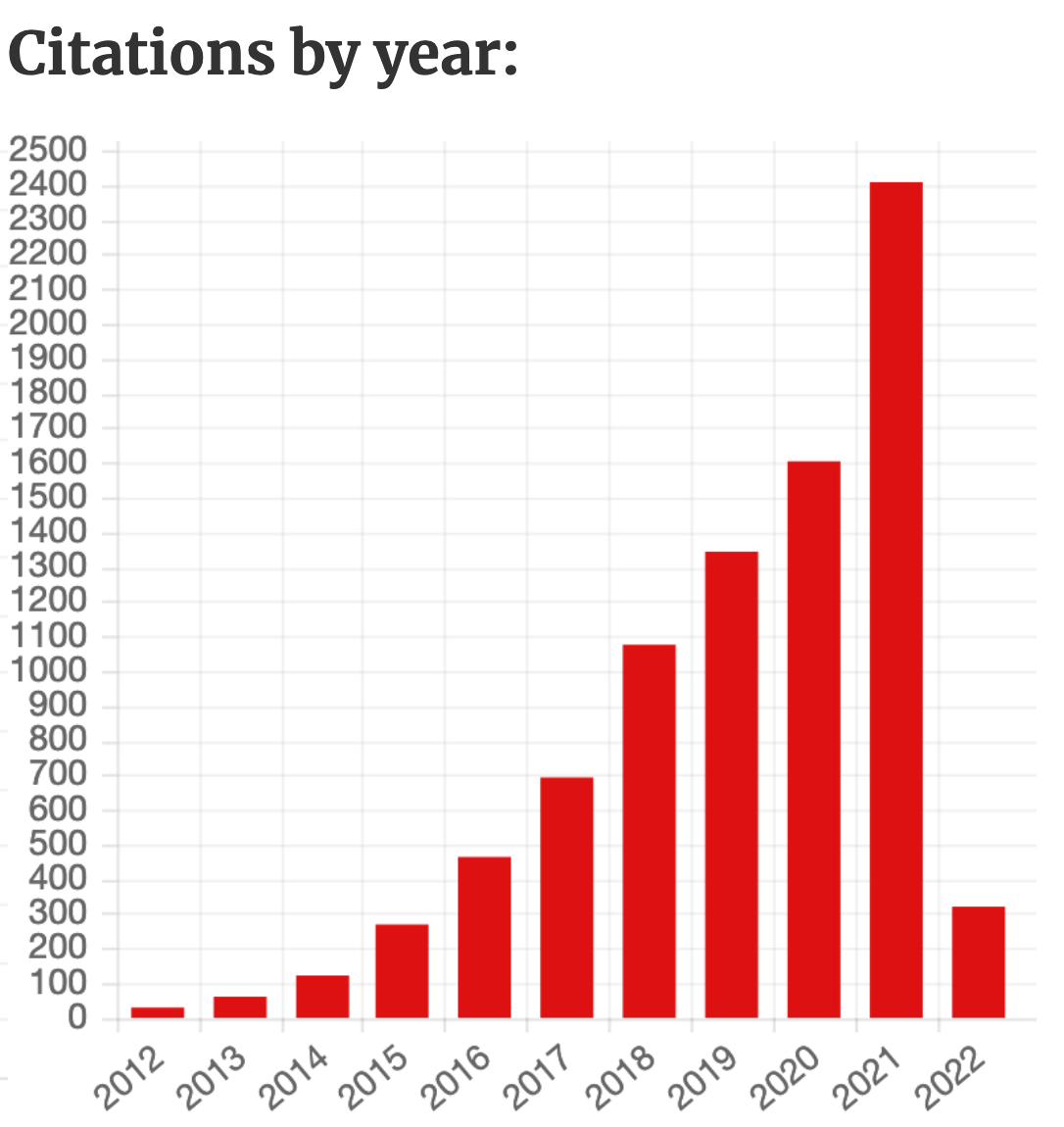}
\end{center}
\caption{Left: A typical ASCL record; right: Citations to ASCL entries by year, as tracked by ADS, as of January 30, 2022}
\end{figure}

The ASCL has worked with editors, publishers, software and article authors, indexers, and other community members and has organized sessions and given presentations to promote formal citation of software. We see software as one of the three pillars of the research record, equal to papers and data. We have long advocated for trackable software citation on par with article (and data) citation, both within the astrophysics community and in science generally, through work with organizations such as \href{https://www.force11.org}{FORCE11} (on its \href{https://force11.org/group/software-citation-working-group/}{Software Citation} and \href{https://www.force11.org/group/software-citation-implementation-working-group}{Software Citation Implementation} Working Groups), \href{https://rd-alliance.org}{Research Data Alliance}, and others. We also support, and for ASCL registration, require, that research software source code be open for examination.

\paragraph*{Software’s Role in Research and Challenges Involved:}

Carole Goble has said that ``Software is the most used instrument in science" \citep{Goble_2014}, and this is certainly true in astrophysics, as the vast majority of researchers in the discipline use it \citep{Mushotzky_2011, momcheva_software_2015}. Computational methods will continue to grow in importance with the need to reduce, manipulate, and analyze increasing amounts of data in all fields.

Infrastructure for supporting software generally exists within astrophysics and in the wider scientific community; this infrastructure includes services such as \href{https://arxiv.org/}{arXiv} and ADS for distributing information about research software through preprints, indexing, and keywords, \href{https://github.com/}{GitHub} and \href{https://about.gitlab.com/}{GitLab} for developing and versioning code, Software Heritage \citep{dicosmo:hal-01590958} and Zenodo \citep{EONR_Zenodo} for archiving software, and the ASCL for cataloging software. Still, there are numerous challenges, including:  
\begin{itemize}
    \item Research code is often not available even for examination, or if made available for download, may be on a site that proves to be ephemeral.
    \item Software often does not have a license, which because of US copyright law, we must presume to be copyrighted, thus restricting its use, or is not licensed permissively, which restricts what others may do with the code.
    \item Code is often not documented or not documented well, and it is not rigorously tested and does not offer test suites.
    \item Software methods are in danger of being lost as websites and services change content or disappear, thus threatening to make research records incomplete.
    \item It is hard to find software applicable to a current research problem when an appropriate software package was developed in another discipline.
\end{itemize}
These challenges persist in large part because of a lack of the funding necessary to solve them. First, funding for research software development should come, and increasingly does, with explicit agreement that the source code will be released with a permissive license to ensure that the software can be examined by other researchers, reused, and built upon. Further, past performance on this requirement should be a high-ranking item for evaluations of proposals for future funding; anecdotal evidence suggests this is not the case. Additionally, funding should be available to support not only the initial development of the code, but readying it for use by others by providing documentation and test suites, for maintenance if the code proves useful for additional research, for sunsetting (when and where appropriate), for storage for future access, and for discovery. 

Even when research software is made available to researchers and the public, it may be siloed by discipline, funder, or academic or research institution, making it difficult to leverage existing code applicable to new research elsewhere. Relevant to Snowmass and the particle physics community is the lack of a service that offers the services the ASCL provides for astrophysics, and the ability to link that service to the wider research community through participation in SciCodes \citep{Allen_scicodes}, a community-driven, currently unfunded consortium working to improve scientific research software libraries, such as the ASCL, DOE CODE \citep{DOECODE_2017} and Oak Ridge National Laboratory Distributed Active Archive Center \citep{ORNL_DAAC} for Department of Energy-funded software, and provide a way to search across them.


%% file: GammaPy-Axel.tex
\noindent
\chapterauthor[]{Axel Donath}\orcidlink{0000-0003-4568-7005}
\\
\begin{affils}
   \chapteraffil[]{Center for Astrophysics | Harvard \& Smithsonian, CfA, 60 Garden St., 02138 Cambridge MA, US}
\end{affils}

\paragraph*{Motivation:}
The analysis of any astronomical gamma-ray data can typically be split into multiple steps and associated data reduction levels. In a simplified scenario this first includes the reconstruction and calibration of the detector level data followed by some Poisson likelihood based higher level analysis with the goal to measure properties of astrophysical gamma-ray sources, such as flux, position, morphology and spectrum. While the detection principle of the different gamma-ray observatories and instruments is very different, the high level likelihood based analysis is often very similar. This motivated the development of \verb"Gammapy": a community developed, open source Python package for gamma-ray astronomy as well as common FITS based open data formats in an initiative called "Gamma Astro Data Formats" (GADF).

\paragraph*{Overview:}
Gammapy provides methods for the analysis of high-level data of several gamma-ray instruments including Imaging Atmospheric Cherenkov Telescopes (IACT; Section \ref{sec:atmospheric_cherenkov}), such as HESS, MAGIC and VERITAS, Water Cherenkov Observatories (Section \ref{sec:water_cherenkov}), such as HAWC as well as support for {\it Fermi}-LAT data. It is built on the scientific Python stack, using Numpy\footnote{\href{www.numpy.org}{www.numpy.org}} for N-dimensional data structures and arithmetic operations, Scipy\footnote{\href{www.scipy.org}{www.scipy.org}} for numerical algorithms such as integration and optimization and Astropy\footnote{\href{www.astropy.org}{www.astropy.org}} for astronomy specific functionality such as coordinate transforms, unit handling as well as serialisation into FITS files. Gammapy is listed as an Astropy affiliated package and thus integrates seamlessly into the Astropy ecosystem. Gammapy will be the base library for the ``Science Tools" of future Cherenkov Telescope Array (CTA; Section \ref{sec:cta})\footnote{\href{ https://www.cta-observatory.org/ctao-adopts-the-gammapy-software-package-for-science- analysis/}{https://www.cta-observatory.org/ctao-adopts-the-gammapy-software-package-for-science-analysis/}}.

\paragraph*{Analysis Methods and Validation:}
Starting from event level data and a description of the instrument response function (IRF), Gammapy allows users to reduce the data by binning gamma-like events into WCS, HEALPix and region based data structures called \verb"Map". Thereby the data structure allows for handling an arbitrary number of non-spatial extra axes, such as energy or time. In the same step the IRFs are projected on the same spatial geometry and a background model can be created. For the background model creation Gammapy features ``classical" IACT methods such as ring or reflected regions methods as well as ``modern" energy dependent templates such as those used by {\it Fermi}-LAT. Finally all the reduced data is bundled into a \verb"MapDataset". 

Together with an independent model definition the \verb"MapDataset" defines a flexible joint-likelihood interface. This allows users not only to handle multiple IACT observations with corresponding IRFs, but also to analyse multiple ``event classes" (e.g., as defined by {\it Fermi}-LAT or HAWC), combining data from multiple instruments and even combining different data types such as event based data and flux based spectra in a single likelihood analysis. Gammapy gives access to multiple optimisation backends such as \verb"scipy.optimize", iminuit\footnote{\href{https://iminuit.readthedocs.io}{https://iminuit.readthedocs.io}} or Sherpa\footnote{\href{https://sherpa.readthedocs.io}{https://sherpa.readthedocs.io}}.

Gammapy comes with a variety of built-in parametric models: including spatial, spectral and temporal models and multidimensional templates. In addition users can easily define custom models (e.g., to describe energy dependent morphology or time dependent spectra of gamma-ray sources). Gammapy also supports a selection of parametric models to describe dark matter gamma-ray emission. An overview of the available models can be found in the model gallery\footnote{\href{https://docs.gammapy.org/0.19/modeling/gallery/index.html}{https://docs.gammapy.org/0.19/modeling/gallery/index.html}}.

In addition Gammapy provides methods to estimate flux points, energy dependent light curves, flux profiles and flux and significance maps with a uniform \verb"Estimator" API. There is detailed online documentation of the package \footnote{\href{https://docs.gammapy.org/}{https://docs.gammapy.org/}} with a user guide, API reference and many detailed tutorials using different kind of data ranging from HESS, MAGIC, HAWC, simulated CTA, GeV energy {\it Fermi}-LAT data and even combinations thereof.
Gammapy is designed both for interactive use in Jupyter notebooks\footnote{\href{https://jupyter.org}{https://jupyter.org}} but also large scale offline analyses such as the upcoming CTA Galactic Plane Survey\cite{Remy2021}.

Gammapy has been validated against reference results from HESS\cite{Mohrmann2019} and HAWC\cite{OliveraNieto2021} instruments. It has also been used to demonstrate a fully reproducible Crab analysis with combined data from many gamma-ray instruments \citep{Nigro2019}.

\paragraph*{Community and Outlook:}
Gammapy is openly developed on GitHub\footnote{\href{https://github.com/gammapy/gammapy}{https://github.com/gammapy/gammapy}}. Since the beginning of the project in 2012 Gammapy has seen contributions from more than 70 developers from more than 10 different countries. Around the project an active and inclusive community of gamma-ray astronomers has formed. The community organises regular coding sprints, developer calls, beginners tutorials and user support via GitHub discussions and a dedicated Gammapy Slack channel. Currently the Gammapy development team is working towards an LST v1.0 release. Future versions of Gammapy will include support for distributed computing and scalable analyses. With an already large user and contributor base and the use of it for the future CTA Science Tools, Gammapy is expected to play a major role for the analysis of astronomical gamma-ray data in future.

%% file: FermiPy-Giacomo.tex
 \noindent
 \chapterauthor[1,2,3]{Giacomo Principe}\orcidlink{0000-0003-0406-7387}
 \\
 \begin{affils}
   \chapteraffil[1]{Dipartimento di Fisica, Universit\'a di Trieste, I-34127 Trieste, Italy}
    \chapteraffil[2]{Istituto Nazionale di Fisica Nucleare, Sezione di Trieste, I-34127 Trieste, Italy}
    \chapteraffil[3]{INAF - Istituto di Radioastronomia, I-40129 Bologna, Italy}
 \end{affils}

The \textit{Fermi} Large Area Telescope (LAT) is a pair-conversion telescope (Section \ref{sec:pair}), which collects gamma-ray photons from below 20 MeV to more than 300 GeV \citep{2009ApJ...697.1071A}. Since its launch in June 2008, the LAT has acquired more than thirteen years of data, with a remarkable 99\% uptime for the science mission \citep{2021ApJS..256...12A}.

The \textit{Fermi}-LAT events, which are classified as photon-like, are publicly available and can be found at the NASA \textit{Fermi} Science Support Center\footnote{\href{https://fermi.gsfc.nasa.gov/ssc/data/access/}{https://fermi.gsfc.nasa.gov/ssc/data/access/}} (FSSC). The FSSC also provides the \textit{Fermi} \texttt{Science Tools} (written in C++), which are used to reduce and analyse the LAT data, as well as a python interface (\texttt{pyLikelihood}) in order to simplify scripting analysis in python.

Fermipy\footnote{\href{https://fermipy.readthedocs.io/en/latest/}{https://fermipy.readthedocs.io/en/latest/}} \citep{2017ICRC...35..824W} is a python framework which provides a high-level interface for likelihood analysis of \textit{Fermi}-LAT data above 50 MeV. For a different analysis technique which can also handle lower energies with \textit{Fermi}-LAT see \citep{2017AIPC.1792g0016P, 2018A&A...618A..22P}. Fermipy lays its foundation on the \textit{Fermi} \texttt{Science Tools}, and employs the \texttt{pyLikelihood} interface.
Among the open-source python libraries, the main Fermipy dependencies consist of NumPy \citep{2013A&A...558A..33A}, Scipy \citep{2020NatMe..17..261V}, and Astropy \citep{2018AJ....156..123A}, as well as some new functionalities taken from GammaPy (Section \ref{sec:gammapy}; \citep{2015ICRC...34..789D}). In order to plot and visualise the analysis results, a further optional dependency is given by Matplotlib \citep{2007CSE.....9...90H}.

All the Fermipy functionality, such as data and model preparation, as well as the high-level analysis methods are handled by the global analysis state object (\texttt{GTAnalysis}). 
The first step of the analysis procedure consists of the definition of a configuration file, which contains the main analysis parameters, essentially the data selection, the geometry of the considered region, and the model specifications.
The configuration file is written in YAML format, the sky models are written in XML files, while the output results are provided in both FITS and NumPy formats.
The analysis state object \texttt{GTAnalysis} is initialised with the parameters reported in the configuration file. The data and model preparation are performed with the \texttt{setup} method which executes the appropriate \texttt{Science Tools}.

Once data have been reduced and the sky model of the considered region is created, it is possible to move to the high-level analysis. 
Fermipy provides a number of high-level analysis methods that automate model fine-tuning such as \texttt{optimize} for model optimisation,  \texttt{find\_sources} which iteratively searches for additional faint sources and \texttt{tsmap} to generate significance maps.
In order to spatially investigate the source the \texttt{localize} function allows a re-localisation of the sources, while \texttt{extension} is used for the source extension estimate. Finally the \texttt{sed} and \texttt{lightcurve} methods are devoted to spectral and lightcurve analyses.

\begin{figure*}
\begin{center}
\rotatebox{0}{\resizebox{!}{55mm}{\includegraphics{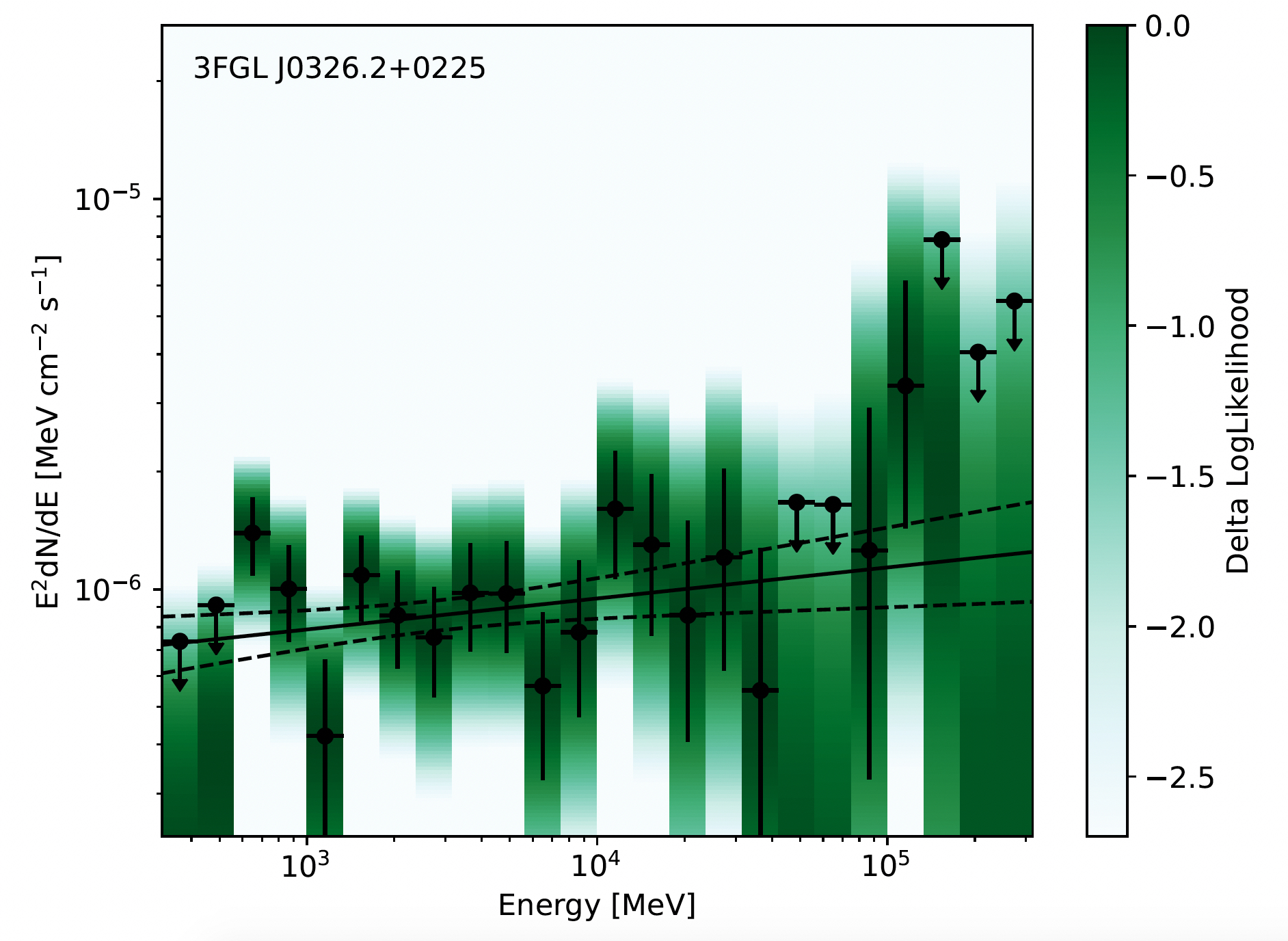}}}
\hspace{0.1cm}
\rotatebox{0}{\resizebox{!}{55mm}{\includegraphics{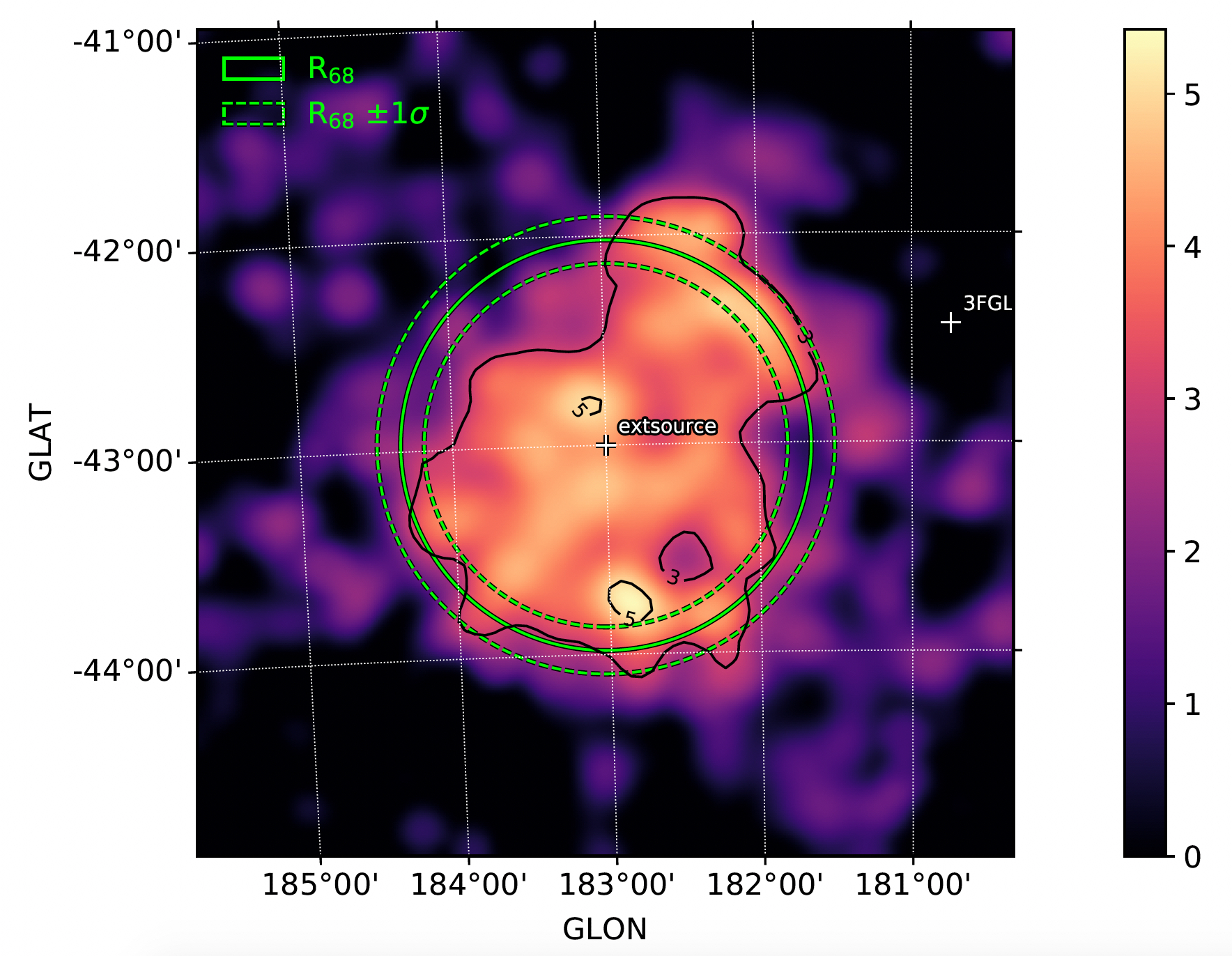}}}
\caption{\small Two examples of diagnostic plots for the high-level analysis methods. \textit{Left}: spectral plot for the \texttt{sed} method showing the comparison of the spectral points with the best-fit model. \textit{Right}: residual TS (test statistic) map of an extended source with its best-fit extension represented with green lines. The plots are taken from \citep{2017ICRC...35..824W}.}
\label{fig:fermipy_example}
\end{center}
\end{figure*}

We report here few examples of Fermipy applications. This tool has been adopted to study gamma-ray sources both Galactic, like SNRs and PWNe  \citep[][respectively]{2022arXiv220105567A,2020A&A...640A..76P} or extragalactic sources, such as for individual AGNs \citep{2020A&A...635A.185P,2021ApJ...911L..11E}, or for galaxy population studies \citep{2021MNRAS.507.4564P,2021ApJS..257...37P}. In addition it can also be used for transient events at different timescales \citep[ranging from few seconds to weeks, or even up to several years,][ respectively]{2021arXiv210903548P,2021ApJS..256...13B,2019ApJ...877...39M}, as well as for investigating multi-messenger phenomena \citep{2021NatAs...5..510S}, or for searching  dark matter in our Universe \citep{2020PhRvL.124w1101M,2020PhRvD.102j3010D}.

As it has been pointed out, Fermipy represents a remarkable example for standardising Python wrappers, for end-to-end high-level analysis of gamma-ray data, which can be used across different missions and celestial objects.

%% file: threeML-HenrikeMichael.tex
\chapterauthor[ ]{ }
\vspace{-0.2in}
 \addtocontents{toc}{
     \leftskip3cm
    \scshape\small
    \parbox{5in}{\raggedleft Henrike Fleischhack, J. Michael Burgess,  et al.}
    \upshape\normalsize
    \string\par
    \raggedright
    \vskip -0.19in
    }
 
\noindent
\nocontentsline\chapterauthor[]{Henrike Fleischhack$^{1,2,3,}$}\asteriskfootnote{H.F. acknowledges support by NASA under award number 80GSFC21M0002. Any opinions, findings, and conclusions or recommendations expressed in this material are those of the author(s) and do not necessarily reflect the views of the National Aeronautics and Space Administration.}\orcidlink{ 0000-0002-0794-8780}
\nocontentsline\chapterauthor[]{, J. Michael Burgess$^4$}\orcidlink{0000-0003-3345-9515}
\nocontentsline\chapterauthor[]{, Nicola Omodei$^5$}
\nocontentsline\chapterauthor[]{\& Niccol\`{o} Di Lalla$^5$}
\\
 \begin{affils}
   \chapteraffil[1]{Catholic University of America, Washington DC}
   \chapteraffil[2]{Astroparticle Physics Laboratory, NASA Goddard Space Flight Center, Greenbelt, MD, USA}
   \chapteraffil[3]{Center for Research and Exploration in Space Science and Technology, NASA/GSFC, Greenbelt, MD}
   \chapteraffil[4]{Max-Planck Institut f\"ur Extraterrestrische Physik, Giessenbachstrasse 1, 85740 Garching, Germany}
   \chapteraffil[5]{W. W. Hansen Experimental Physics Laboratory, Kavli Institute for Particle Astrophysics and Cosmology, Department of Physics and SLAC National Accelerator Laboratory, Stanford University, Stanford, CA 94305, USA}
\end{affils}

The multi-mission maximum likelihood framework (\texttt{3ML}, also: \texttt{threeML}, see \cite{2015arXiv150708343V}) is a python-based software package for joint likelihood analyses of  multi-wavelength astronomical data. The goal of a likelihood analysis is to obtain estimates for certain parameters of a model describing astronomical sources, by matching the model to measured data. In this case, the ``model'' is a phenomenological or physical model describing the energy spectrum, shape, and/or time evolution of one or several gamma-ray sources, with free parameters $\theta$. The data $X$ can take various formats, such as a list of gamma-ray photons with associated reconstructed energies and arrival directions, or a list of apparent magnitudes of an object as seen through various filters. The matching between model and data is done by means of the likelihood function $\mathcal{L}(\theta \mid X) = P (X\mid \theta)$, where $P(X\mid \theta)$ describes the probability of measuring $X$, given model parameters $\theta$.

\texttt{ThreeML} utilizes plugins to encapsulate data access, convolution of the models with the instrument responses, and calculation of the likelihood. Each plugin has been built to handle data and instrument response files in a specific format. Various instruments' proprietary formats (e.g. HAWC, \emph{Fermi}-LAT) as well as community standards (OGIP) are supported. Plugins can be fully implemented within \texttt{threeML} such as the OGIP plugin, implemented as wrappers around existing likelihood analysis tools (e.g. the \texttt{FermiPyLike} plugin), or provided as standalone packages (such as \texttt{hawc\_hal}). Each instance of a plugin can be configured as appropriate for the given instrument and data format, e.g. setting active data channels or defining a ``region of interest'' on the sky. 

Plugins for gamma-ray instruments such as HAWC and \emph{Fermi}-LAT use a binned forward folding likelihood approach, where $\mathcal{L}(\theta \mid X) = \prod\limits_{i} \mathrm{Poisson}\left( x_i \mid N_i \left( \theta, \alpha \right)\right)$. Here, data are binned in one or more dimensions such as energy, arrival direction, and/or time. $x_i$ is the number of measured photon candidates in bin $i$. The predicted counts $N_i$, given model parameters $\theta$ and nuisance parameters $\alpha$ (those parameters internal to a given plugin), are derived by folding the photon emission predicted by the model with the detector response, including angular resolution, energy resolution, and effective area. Nuisance parameters can contain global correction factors, e.g. the background normalization. For analyses using multiple independent data sets, the likelihoods are calculated separately for each data set (by separate plugin instances) and then multiplied to obtain the final likelihood value. 

In addition to (frequentist) maximum likelihood fits, \texttt{threeML} also supports (Bayesian) posterior probability sampling. The posterior probability is given by $p(\theta|X) = \frac{P(x|\theta)}{p(x)}p(\theta)$, where $P$ is defined as before, $p(\theta)$ is the prior probability distribution of the free parameter(s), and $p(x)$ is chosen so that the distribution is normalized appropriately.

\texttt{ThreeML} contains several options for minimizers and sampling algorithms. These are also implemented as plugins, typically as wrappers around external libraries, allowing the user to switch between minimizers without changing the rest of the analysis script.

\texttt{ThreeML} uses the \texttt{astromodels} package to model the underlying gamma-ray emission. An \texttt{astromodels} ``model'' of a given region of interest contains one or several sources. Sources are described by a gamma-ray spectrum, position, and a morphology, with various commonly used spectral and spatial functions (including point-like sources) already implemented. The user may also supply external templates for the source morphology and spectrum, or define a new function at runtime. Sources with energy-dependent morphology or time-dependent spectra are supported. The user can also select which model parameters to free or fix in the fit. Parameters can also be linked to one another, and have prior distributions associated to them (for Bayesian analyses). 

In addition to model fitting routines, \texttt{threeML} also provides convenience functions to download publicly accessible data, generate models from external catalogs such as the \emph{Fermi}-LAT's 4FGL catalog, plot fitted spectra with propagated uncertainties, and investigate the goodness-of-fit of a given model. 

The plugin approach makes \texttt{threeML} ideal for multi-wavelength analyses, as most of the code is agnostic to which and how many plugins have been configured. The user can thus easily add or remove data sets from the analysis, and a multi-instrument fit is no different than a single-instrument one except for setting up the plugins. In practice, calculation of the likelihood can be computationally expensive, depending on the complexity of the Instrument Response Functions and the number of bins. This means that the likelihood minimization may take significantly longer than e.g. first deriving a Spectral Energy Distribution (SED) or energy spectrum for each source from given data sets and then performing a $\chi^2$-fit (where $\chi^2$-fit refers generally to the act of deconvolving instrumental data with a function and then fitting these pseudo-data to a model via a Gaussian likelihood) of the model to the resulting spectrum. However, the forward-folding approach naturally accounts for correlations in the parameters due to source overlap or energy resolution, which a $\chi^2$ fit cannot easily do. Also, the forward folding approach can easily account for low-statistics situations with non-Gaussian uncertainties or null measurements, e.g. non-detections in certain energy bins, which again are difficult to include in a $\chi^2$ fit.

The code for \texttt{threeML} and \texttt{astromodels} is freely available on \texttt{github} (\url{https://github.com/threeML/}). Both packages are released via \texttt{conda} (channel ``\texttt{threeML}'') and \texttt{pip}. Documentation and worked examples can be found at  \url{https://threeml.readthedocs.io/} and \url{https://astromodels.readthedocs.io/}. The codebase is stable and has been used in dozens of publications. Future plans for the package include further optimization and improvements in resource usage, adding plugins for new and future missions/instruments, and improving some of the existing plugins.

%% file: main.bbl
\begin{thebibliography}{100}

\bibitem{LIGOScientific:2017zic}
B.~P. Abbott et~al.
\newblock {Gravitational Waves and Gamma-rays from a Binary Neutron Star
  Merger: GW170817 and GRB 170817A}.
\newblock {\em Astrophys. J. Lett.}, 848(2):L13, 2017.

\bibitem{IceCube:2018dnn}
M.~G. Aartsen et~al.
\newblock {Multimessenger observations of a flaring blazar coincident with
  high-energy neutrino IceCube-170922A}.
\newblock {\em Science}, 361(6398):eaat1378, 2018.

\bibitem{NAP26141}
{National Academies of Sciences, Engineering, and Medicine}.
\newblock {\em Pathways to Discovery in Astronomy and Astrophysics for the
  2020s}.
\newblock The National Academies Press, Washington, DC, 2021.

\bibitem{2004ApJ...611.1005G}
N.~{Gehrels}, G.~{Chincarini}, P.~{Giommi}, K.~O. {Mason}, J.~A. {Nousek},
  A.~A. {Wells}, N.~E. {White}, S.~D. {Barthelmy}, D.~N. {Burrows}, L.~R.
  {Cominsky}, K.~C. {Hurley}, F.~E. {Marshall}, P.~{M{\'e}sz{\'a}ros}, P.~W.~A.
  {Roming}, L.~{Angelini}, L.~M. {Barbier}, T.~{Belloni}, S.~{Campana}, P.~A.
  {Caraveo}, M.~M. {Chester}, O.~{Citterio}, T.~L. {Cline}, M.~S. {Cropper},
  J.~R. {Cummings}, A.~J. {Dean}, E.~D. {Feigelson}, E.~E. {Fenimore}, D.~A.
  {Frail}, A.~S. {Fruchter}, G.~P. {Garmire}, K.~{Gendreau}, G.~{Ghisellini},
  J.~{Greiner}, J.~E. {Hill}, S.~D. {Hunsberger}, H.~A. {Krimm}, S.~R.
  {Kulkarni}, P.~{Kumar}, F.~{Lebrun}, N.~M. {Lloyd-Ronning}, C.~B.
  {Markwardt}, B.~J. {Mattson}, R.~F. {Mushotzky}, J.~P. {Norris},
  J.~{Osborne}, B.~{Paczynski}, D.~M. {Palmer}, H.~S. {Park}, A.~M. {Parsons},
  J.~{Paul}, M.~J. {Rees}, C.~S. {Reynolds}, J.~E. {Rhoads}, T.~P. {Sasseen},
  B.~E. {Schaefer}, A.~T. {Short}, A.~P. {Smale}, I.~A. {Smith}, L.~{Stella},
  G.~{Tagliaferri}, T.~{Takahashi}, M.~{Tashiro}, L.~K. {Townsley},
  J.~{Tueller}, M.~J.~L. {Turner}, M.~{Vietri}, W.~{Voges}, M.~J. {Ward},
  R.~{Willingale}, F.~M. {Zerbi}, and W.~W. {Zhang}.
\newblock {The Swift Gamma-Ray Burst Mission}.
\newblock {\em \apj}, 611(2):1005--1020, August 2004.

\bibitem{lat}
W.~B. {Atwood, et al.}
\newblock {The Large Area Telescope on the Fermi Gamma-Ray Space Telescope
  Mission}.
\newblock {\em \apj}, 697(2):1071--1102, June 2009.

\bibitem{Meegan2009}
Charles {Meegan}, Giselher {Lichti}, P.~N. {Bhat}, Elisabetta {Bissaldi},
  Michael~S. {Briggs}, Valerie {Connaughton}, Roland {Diehl}, Gerald {Fishman},
  Jochen {Greiner}, Andrew~S. {Hoover}, Alexand er~J. {van der Horst}, Andreas
  {von Kienlin}, R.~Marc {Kippen}, Chryssa {Kouveliotou}, Sheila {McBreen},
  W.~S. {Paciesas}, Robert {Preece}, Helmut {Steinle}, Mark~S. {Wallace},
  Robert~B. {Wilson}, and Colleen {Wilson-Hodge}.
\newblock {The Fermi Gamma-ray Burst Monitor}.
\newblock {\em \apj}, 702(1):791--804, September 2009.

\bibitem{Abeysekara_2019}
A.~U. Abeysekara, A.~Albert, R.~Alfaro, C.~Alvarez, J.~D. {\'{A}}lvarez,
  J.~R.~Angeles Camacho, R.~Arceo, J.~C. Arteaga-Vel{\'{a}}zquez, K.~P.
  Arunbabu, D.~Avila Rojas, H.~A.~Ayala Solares, V.~Baghmanyan,
  E.~Belmont-Moreno, S.~Y. BenZvi, C.~Brisbois, K.~S. Caballero-Mora,
  T.~Capistr{\'{a}}n, A.~Carrami{\~{n}}ana, S.~Casanova, U.~Cotti, J.~Cotzomi,
  S.~Couti{\~{n}}o de~Le{\'{o}}n, E.~De la~Fuente, C.~de~Le{\'{o}}n,
  S.~Dichiara, B.~L. Dingus, M.~A. DuVernois, J.~C. D{\'{\i}}az-V{\'{e}}lez,
  R.~W. Ellsworth, K.~Engel, C.~Espinoza, B.~Fick, H.~Fleischhack, N.~Fraija,
  A.~Galv{\'{a}}n-G{\'{a}}mez, J.~A. Garc{\'{\i}}a-Gonz{\'{a}}lez, F.~Garfias,
  M.~M. Gonz{\'{a}}lez, J.~A. Goodman, J.~P. Harding, S.~Hernandez, J.~Hinton,
  B.~Hona, F.~Hueyotl-Zahuantitla, C.~M. Hui, P.~Hüntemeyer, A.~Iriarte,
  A.~Jardin-Blicq, V.~Joshi, S.~Kaufmann, D.~Kieda, A.~Lara, W.~H. Lee,
  H.~Le{\'{o}}n Vargas, J.~T. Linnemann, A.~L. Longinotti, G.~Luis-Raya,
  J.~Lundeen, K.~Malone, S.~S. Marinelli, O.~Martinez, I.~Martinez-Castellanos,
  J.~Mart{\'{\i}}nez-Castro, H.~Mart{\'{\i}}nez-Huerta, J.~A. Matthews,
  P.~Miranda-Romagnoli, J.~A. Morales-Soto, E.~Moreno, M.~Mostaf{\'{a}},
  A.~Nayerhoda, L.~Nellen, M.~Newbold, M.~U. Nisa, R.~Noriega-Papaqui,
  A.~Peisker, E.~G. P{\'{e}}rez-P{\'{e}}rez, J.~Pretz, Z.~Ren, C.~D. Rho,
  C.~Rivi{\`{e}}re, D.~Rosa-Gonz{\'{a}}lez, M.~Rosenberg, E.~Ruiz-Velasco,
  H.~Salazar, F.~Salesa Greus, A.~Sandoval, M.~Schneider, H.~Schoorlemmer,
  M.~Seglar Arroyo, G.~Sinnis, A.~J. Smith, R.~W. Springer, P.~Surajbali,
  E.~Tabachnick, M.~Tanner, O.~Tibolla, K.~Tollefson, I.~Torres, T.~Weisgarber,
  S.~Westerhoff, J.~Wood, T.~Yapici, A.~Zepeda, and H.~Zhou and.
\newblock Measurement of the crab nebula spectrum past 100 {TeV} with {HAWC}.
\newblock {\em The Astrophysical Journal}, 881(2):134, aug 2019.

\bibitem{Holder:2006gi}
Jamie Holder et~al.
\newblock {The first VERITAS telescope}.
\newblock {\em Astropart. Phys.}, 25:391--401, 2006.

\bibitem{CTAConsortium:2018tzg}
B.S. Acharya et~al.
\newblock {\em {Science with the Cherenkov Telescope Array}}.
\newblock WSP, 11 2018.

\bibitem{Albert:2019afb}
A.~Albert et~al.
\newblock {Science Case for a Wide Field-of-View Very-High-Energy Gamma-Ray
  Observatory in the Southern Hemisphere}.
\newblock 2 2019.

\bibitem{Abreu:2019ahw}
P.~Abreu et~al.
\newblock {The Southern Wide-Field Gamma-Ray Observatory (SWGO): A
  Next-Generation Ground-Based Survey Instrument for VHE Gamma-Ray Astronomy}.
\newblock 7 2019.

\bibitem{Hinton:2021rvp}
Jim Hinton.
\newblock {The Southern Wide-field Gamma-ray Observatory: Status and
  Prospects}.
\newblock {\em PoS}, ICRC2021:023, 2021.

\bibitem{Schoorlemmer:2019gee}
H.~Schoorlemmer.
\newblock {A next-generation ground-based wide field-of-view gamma-ray
  observatory in the southern hemisphere}.
\newblock {\em PoS}, ICRC2019:785, 2020.

\bibitem{Ciafaloni:2000df}
Marcello Ciafaloni, Paolo Ciafaloni, and Denis Comelli.
\newblock {Bloch-Nordsieck violating electroweak corrections to inclusive TeV
  scale hard processes}.
\newblock {\em Phys. Rev. Lett.}, 84:4810--4813, 2000.

\bibitem{ATLAS:2016jbu}
Morad Aaboud et~al.
\newblock {Measurement of $W$ boson angular distributions in events with high
  transverse momentum jets at $\sqrt{s}=$ 8 TeV using the ATLAS detector}.
\newblock {\em Phys. Lett. B}, 765:132--153, 2017.

\bibitem{CMS:2017gbl}
Albert~M Sirunyan et~al.
\newblock {Measurement of the differential cross sections for the associated
  production of a $W$ boson and jets in proton-proton collisions at
  $\sqrt{s}=13$ TeV}.
\newblock {\em Phys. Rev. D}, 96(7):072005, 2017.

\bibitem{Manohar:2014vxa}
Aneesh Manohar, Brian Shotwell, Christian Bauer, and Sascha Turczyk.
\newblock {Non-cancellation of electroweak logarithms in high-energy
  scattering}.
\newblock {\em Phys. Lett. B}, 740:179--187, 2015.

\bibitem{Chen:2016wkt}
Junmou Chen, Tao Han, and Brock Tweedie.
\newblock {Electroweak Splitting Functions and High Energy Showering}.
\newblock {\em JHEP}, 11:093, 2017.

\bibitem{Bauer:2017isx}
Christian~W. Bauer, Nicolas Ferland, and Bryan~R. Webber.
\newblock {Standard Model Parton Distributions at Very High Energies}.
\newblock {\em JHEP}, 08:036, 2017.

\bibitem{Manohar:2018kfx}
Aneesh~V. Manohar and Wouter~J. Waalewijn.
\newblock {Electroweak Logarithms in Inclusive Cross Sections}.
\newblock {\em JHEP}, 08:137, 2018.

\bibitem{Bauer:2018xag}
Christian~W. Bauer, Davide Provasoli, and Bryan~R. Webber.
\newblock {Standard Model Fragmentation Functions at Very High Energies}.
\newblock {\em JHEP}, 11:030, 2018.

\bibitem{Sjostrand:2006za}
Torbjorn Sjostrand, Stephen Mrenna, and Peter~Z. Skands.
\newblock {PYTHIA 6.4 Physics and Manual}.
\newblock {\em JHEP}, 05:026, 2006.

\bibitem{Sjostrand:2007gs}
Torbjorn Sjostrand, Stephen Mrenna, and Peter~Z. Skands.
\newblock {A Brief Introduction to PYTHIA 8.1}.
\newblock {\em Comput. Phys. Commun.}, 178:852--867, 2008.

\bibitem{Sjostrand:2014zea}
Torbj\"orn Sj\"ostrand, Stefan Ask, Jesper~R. Christiansen, Richard Corke,
  Nishita Desai, Philip Ilten, Stephen Mrenna, Stefan Prestel, Christine~O.
  Rasmussen, and Peter~Z. Skands.
\newblock {An introduction to PYTHIA 8.2}.
\newblock {\em Comput. Phys. Commun.}, 191:159--177, 2015.

\bibitem{Kleiss:2020rcg}
Ronald Kleiss and Rob Verheyen.
\newblock {Collinear electroweak radiation in antenna parton showers}.
\newblock {\em Eur. Phys. J. C}, 80(10):980, 2020.

\bibitem{Masouminia:2021kne}
M.~R. Masouminia and P.~Richardson.
\newblock {Implementation of Angularly Ordered Electroweak Parton Shower in
  Herwig 7}.
\newblock 8 2021.

\bibitem{Brooks:2021kji}
Helen Brooks, Peter Skands, and Rob Verheyen.
\newblock {Interleaved Resonance Decays and Electroweak Radiation in Vincia}.
\newblock 8 2021.

\bibitem{Ciafaloni:2010ti}
Paolo Ciafaloni, Denis Comelli, Antonio Riotto, Filippo Sala, Alessandro
  Strumia, and Alfredo Urbano.
\newblock {Weak Corrections are Relevant for Dark Matter Indirect Detection}.
\newblock {\em JCAP}, 03:019, 2011.

\bibitem{Cirelli:2010xx}
Marco Cirelli, Gennaro Corcella, Andi Hektor, Gert Hutsi, Mario Kadastik, Paolo
  Panci, Martti Raidal, Filippo Sala, and Alessandro Strumia.
\newblock {PPPC 4 DM ID: A Poor Particle Physicist Cookbook for Dark Matter
  Indirect Detection}.
\newblock {\em JCAP}, 03:051, 2011.
\newblock [Erratum: JCAP 10, E01 (2012)].

\bibitem{Bauer:2014ula}
Martin Bauer, Timothy Cohen, Richard~J. Hill, and Mikhail~P. Solon.
\newblock {Soft Collinear Effective Theory for Heavy WIMP Annihilation}.
\newblock {\em JHEP}, 01:099, 2015.

\bibitem{Ovanesyan:2014fwa}
Grigory Ovanesyan, Tracy~R. Slatyer, and Iain~W. Stewart.
\newblock {Heavy Dark Matter Annihilation from Effective Field Theory}.
\newblock {\em Phys. Rev. Lett.}, 114(21):211302, 2015.

\bibitem{Baumgart:2014saa}
Matthew Baumgart, Ira~Z. Rothstein, and Varun Vaidya.
\newblock {Constraints on Galactic Wino Densities from Gamma Ray Lines}.
\newblock {\em JHEP}, 04:106, 2015.

\bibitem{Baumgart:2015bpa}
Matthew Baumgart and Varun Vaidya.
\newblock {Semi-inclusive wino and higgsino annihilation to LL'}.
\newblock {\em JHEP}, 03:213, 2016.

\bibitem{Ovanesyan:2016vkk}
Grigory Ovanesyan, Nicholas~L. Rodd, Tracy~R. Slatyer, and Iain~W. Stewart.
\newblock {One-loop correction to heavy dark matter annihilation}.
\newblock {\em Phys. Rev. D}, 95(5):055001, 2017.
\newblock [Erratum: Phys.Rev.D 100, 119901 (2019)].

\bibitem{Baumgart:2017nsr}
Matthew Baumgart, Timothy Cohen, Ian Moult, Nicholas~L. Rodd, Tracy~R. Slatyer,
  Mikhail~P. Solon, Iain~W. Stewart, and Varun Vaidya.
\newblock {Resummed Photon Spectra for WIMP Annihilation}.
\newblock {\em JHEP}, 03:117, 2018.

\bibitem{Beneke:2018ssm}
M.~Beneke, A.~Broggio, C.~Hasner, and M.~Vollmann.
\newblock {Energetic $\gamma$-rays from TeV scale dark matter annihilation
  resummed}.
\newblock {\em Phys. Lett. B}, 786:347--354, 2018.
\newblock [Erratum: Phys.Lett.B 810, 135831 (2020)].

\bibitem{Baumgart:2018yed}
Matthew Baumgart, Timothy Cohen, Emmanuel Moulin, Ian Moult, Lucia Rinchiuso,
  Nicholas~L. Rodd, Tracy~R. Slatyer, Iain~W. Stewart, and Varun Vaidya.
\newblock {Precision Photon Spectra for Wino Annihilation}.
\newblock {\em JHEP}, 01:036, 2019.

\bibitem{Beneke:2019vhz}
M.~Beneke, A.~Broggio, C.~Hasner, K.~Urban, and M.~Vollmann.
\newblock {Resummed photon spectrum from dark matter annihilation for
  intermediate and narrow energy resolution}.
\newblock {\em JHEP}, 08:103, 2019.
\newblock [Erratum: JHEP 07, 145 (2020)].

\bibitem{Beneke:2019gtg}
Martin Beneke, Caspar Hasner, Kai Urban, and Martin Vollmann.
\newblock {Precise yield of high-energy photons from Higgsino dark matter
  annihilation}.
\newblock {\em JHEP}, 03:030, 2020.

\bibitem{Rinchiuso:2020skh}
Lucia Rinchiuso, Oscar Macias, Emmanuel Moulin, Nicholas~L. Rodd, and Tracy~R.
  Slatyer.
\newblock {Prospects for detecting heavy WIMP dark matter with the Cherenkov
  Telescope Array: The Wino and Higgsino}.
\newblock {\em Phys. Rev. D}, 103(2):023011, 2021.

\bibitem{Bauer:2020jay}
Christian~W. Bauer, Nicholas~L. Rodd, and Bryan~R. Webber.
\newblock {Dark matter spectra from the electroweak to the Planck scale}.
\newblock {\em JHEP}, 06:121, 2021.

\bibitem{SME_summary}
V~Alan Kosteleck{\`y} and Neil Russell.
\newblock {Data tables for {Lorentz} and C P T violation}.
\newblock {\em Rev. Mod. Phys.}, 83:11, 2011.

\bibitem{GR_mg_Will_1998}
Clifford~M Will.
\newblock {Bounding the mass of the graviton using gravitational-wave
  observations of inspiralling compact binaries}.
\newblock {\em \prd}, 57:2061, 1998.

\bibitem{jacob2008lorentz}
Uri Jacob and Tsvi Piran.
\newblock {{Lorentz}-violation-induced arrival delays of cosmological
  particles}.
\newblock {\em JCAP}, 2008:031, 2008.

\bibitem{nishizawa_2014_speed_of_gravity}
Atsushi Nishizawa and Takashi Nakamura.
\newblock {Measuring speed of gravitational waves by observations of photons
  and neutrinos from compact binary mergers and supernovae}.
\newblock {\em \prd}, 90:044048, 2014.

\bibitem{GW170817-GRB170817A}
B~P Abbott, R~Abbott, T~D Abbott, F~Acernese, K~Ackley, C~Adams, T~Adams,
  P~Addesso, R~X Adhikari, V~B Adya, et~al.
\newblock {Gravitational waves and gamma-rays from a binary neutron star
  merger: {GW170817} and {GRB 170817A}}.
\newblock {\em \apjl}, 848:L13, 2017.

\bibitem{cgw_GW170817_dark_matter_Boran_2018}
Sibel Boran, Shantanu Desai, E~O Kahya, and R~P Woodard.
\newblock {GW170817 falsifies dark matter emulators}.
\newblock {\em \prd}, 97:041501, 2018.

\bibitem{cgw_beyond_GR_constraints_Creminelli_2017}
Paolo Creminelli and Filippo Vernizzi.
\newblock {Dark Energy after GW170817 and GRB170817A}.
\newblock {\em \prl}, 119:251302, 2017.

\bibitem{cgw_beyond_GR_constraints_Baker_2017}
Tessa Baker, Emilio Bellini, Pedro~G Ferreira, Macarena Lagos, Johannes Noller,
  and Ignacy Sawicki.
\newblock {Strong constraints on cosmological gravity from GW170817 and GRB
  170817A}.
\newblock {\em \prl}, 119:251301, 2017.

\bibitem{cgw_beyond_GR_constraints_Ezquiaga_2017}
Jose~Mar{\'\i}a Ezquiaga and Miguel Zumalac{\'a}rregui.
\newblock {Dark energy after GW170817: dead ends and the road ahead}.
\newblock {\em \prl}, 119:251304, 2017.

\bibitem{sakstein2017implications}
Jeremy Sakstein and Bhuvnesh Jain.
\newblock {Implications of the neutron star merger GW170817 for cosmological
  scalar-tensor theories}.
\newblock {\em \prl}, 119:251303, 2017.

\bibitem{burns2020neutron}
Eric Burns.
\newblock Neutron star mergers and how to study them.
\newblock {\em Living Reviews in Relativity}, 23(1):1--177, 2020.

\bibitem{SwiftGUANO}
Aaron {Tohuvavohu}, Jamie~A. {Kennea}, James {DeLaunay}, David~M. {Palmer},
  S.~Bradley {Cenko}, and Scott {Barthelmy}.
\newblock {Gamma-Ray Urgent Archiver for Novel Opportunities (GUANO): Swift/BAT
  Event Data Dumps on Demand to Enable Sensitive Subthreshold GRB Searches}.
\newblock {\em \apj}, 900(1):35, September 2020.

\bibitem{xrt03}
S.~R. {Oates}, F.~E. {Marshall}, A.~A. {Breeveld}, N.~P.~M. {Kuin}, P.~J.
  {Brown}, M.~{De Pasquale}, P.~A. {Evans}, A.~J. {Fenney}, C.~{Gronwall},
  J.~A. {Kennea}, N.~J. {Klingler}, M.~J. {Page}, M.~H. {Siegel},
  A.~{Tohuvavohu}, E.~{Ambrosi}, S.~D. {Barthelmy}, A.~P. {Beardmore}, M.~G.
  {Bernardini}, S.~{Campana}, R.~{Caputo}, S.~B. {Cenko}, G.~{Cusumano},
  A.~{D'A{\`\i}}, P.~{D'Avanzo}, V.~{D'Elia}, P.~{Giommi}, D.~H. {Hartmann},
  H.~A. {Krimm}, S.~{Laha}, D.~B. {Malesani}, A.~{Melandri}, J.~A. {Nousek},
  P.~T. {O'Brien}, J.~P. {Osborne}, C.~{Pagani}, K.~L. {Page}, D.~M. {Palmer},
  M.~{Perri}, J.~L. {Racusin}, T.~{Sakamoto}, B.~{Sbarufatti}, J.~E.
  {Schlieder}, G.~{Tagliaferri}, and E.~{Troja}.
\newblock {Swift/UVOT follow-up of gravitational wave alerts in the O3 era}.
\newblock {\em \mnras}, 507(1):1296--1317, October 2021.

\bibitem{uvoto3}
K.~L. {Page}, P.~A. {Evans}, A.~{Tohuvavohu}, J.~A. {Kennea}, N.~J. {Klingler},
  S.~B. {Cenko}, S.~R. {Oates}, E.~{Ambrosi}, S.~D. {Barthelmy}, A.~P.
  {Beardmore}, M.~G. {Bernardini}, A.~A. {Breeveld}, P.~J. {Brown}, D.~N.
  {Burrows}, S.~{Campana}, R.~{Caputo}, G.~{Cusumano}, A.~{D'A{\`\i}},
  P.~{D'Avanzo}, V.~{D'Elia}, M.~{De Pasquale}, S.~W.~K. {Emery}, P.~{Giommi},
  C.~{Gronwall}, D.~H. {Hartmann}, H.~A. {Krimm}, N.~P.~M. {Kuin}, D.~B.
  {Malesani}, F.~E. {Marshall}, A.~{Melandri}, J.~A. {Nousek}, P.~T. {O'Brien},
  J.~P. {Osborne}, C.~{Pagani}, M.~J. {Page}, D.~M. {Palmer}, M.~{Perri}, J.~L.
  {Racusin}, T.~{Sakamoto}, B.~{Sbarufatti}, J.~E. {Schlieder}, M.~H. {Siegel},
  G.~{Tagliaferri}, and E.~{Troja}.
\newblock {Swift-XRT follow-up of gravitational wave triggers during the third
  aLIGO/Virgo observing run}.
\newblock {\em \mnras}, 499(3):3459--3480, December 2020.

\bibitem{GBMLVCO1O2}
R.~{Hamburg}, C.~{Fletcher}, E.~{Burns}, A.~{Goldstein}, E.~{Bissaldi}, M.~S.
  {Briggs}, W.~H. {Cleveland}, M.~M. {Giles}, C.~M. {Hui}, D.~{Kocevski},
  S.~{Lesage}, B.~{Mailyan}, C.~{Malacaria}, S.~{Poolakkil}, R.~{Preece}, O.~J.
  {Roberts}, P.~{Veres}, A.~{von Kienlin}, C.~A. {Wilson-Hodge}, J.~{Wood},
  {Fermi Gamma-Ray Burst Monitor}, R.~{Abbott}, T.~D. {Abbott}, S.~{Abraham},
  F.~{Acernese}, K.~{Ackley}, C.~{Adams}, R.~X. {Adhikari}, V.~B. {Adya},
  C.~{Affeldt}, M.~{Agathos}, K.~{Agatsuma}, N.~{Aggarwal}, O.~D. {Aguiar},
  A.~{Aich}, L.~{Aiello}, A.~{Ain}, P.~{Ajith}, G.~{Allen}, A.~{Allocca}, P.~A.
  {Altin}, A.~{Amato}, S.~{Anand}, A.~{Ananyeva}, S.~B. {Anderson}, W.~G.
  {Anderson}, S.~V. {Angelova}, S.~{Ansoldi}, S.~{Antier}, S.~{Appert},
  K.~{Arai}, M.~C. {Araya}, J.~S. {Areeda}, M.~{Ar{\`e}ne}, N.~{Arnaud}, S.~M.
  {Aronson}, S.~{Ascenzi}, G.~{Ashton}, S.~M. {Aston}, P.~{Astone}, F.~{Aubin},
  P.~{Aufmuth}, K.~{AultONeal}, C.~{Austin}, V.~{Avendano}, S.~{Babak},
  P.~{Bacon}, F.~{Badaracco}, M.~K.~M. {Bader}, S.~{Bae}, A.~M. {Baer},
  J.~{Baird}, F.~{Baldaccini}, G.~{Ballardin}, S.~W. {Ballmer}, A.~{Bals},
  A.~{Balsamo}, G.~{Baltus}, S.~{Banagiri}, D.~{Bankar}, R.~S. {Bankar}, J.~C.
  {Barayoga}, C.~{Barbieri}, B.~C. {Barish}, D.~{Barker}, K.~{Barkett},
  P.~{Barneo}, F.~{Barone}, B.~{Barr}, L.~{Barsotti}, M.~{Barsuglia},
  D.~{Barta}, J.~{Bartlett}, I.~{Bartos}, R.~{Bassiri}, A.~{Basti}, M.~{Bawaj},
  J.~C. {Bayley}, M.~{Bazzan}, B.~{B{\'e}csy}, M.~{Bejger}, I.~{Belahcene},
  A.~S. {Bell}, D.~{Beniwal}, M.~G. {Benjamin}, J.~D. {Bentley}, F.~{Bergamin},
  B.~K. {Berger}, G.~{Bergmann}, S.~{Bernuzzi}, C.~P.~L. {Berry},
  D.~{Bersanetti}, A.~{Bertolini}, J.~{Betzwieser}, R.~{Bhandare}, A.~V.
  {Bhandari}, J.~{Bidler}, E.~{Biggs}, I.~A. {Bilenko}, G.~{Billingsley},
  R.~{Birney}, O.~{Birnholtz}, S.~{Biscans}, M.~{Bischi}, S.~{Biscoveanu},
  A.~{Bisht}, G.~{Bissenbayeva}, M.~{Bitossi}, M.~A. {Bizouard}, J.~K.
  {Blackburn}, J.~{Blackman}, C.~D. {Blair}, D.~G. {Blair}, R.~M. {Blair},
  F.~{Bobba}, N.~{Bode}, M.~{Boer}, Y.~{Boetzel}, G.~{Bogaert}, F.~{Bondu},
  E.~{Bonilla}, R.~{Bonnand}, P.~{Booker}, B.~A. {Boom}, R.~{Bork},
  V.~{Boschi}, S.~{Bose}, V.~{Bossilkov}, J.~{Bosveld}, Y.~{Bouffanais},
  A.~{Bozzi}, C.~{Bradaschia}, P.~R. {Brady}, A.~{Bramley}, M.~{Branchesi},
  J.~E. {Brau}, M.~{Breschi}, T.~{Briant}, J.~H. {Briggs}, F.~{Brighenti},
  A.~{Brillet}, M.~{Brinkmann}, P.~{Brockill}, A.~F. {Brooks}, J.~{Brooks},
  D.~D. {Brown}, S.~{Brunett}, G.~{Bruno}, R.~{Bruntz}, A.~{Buikema},
  T.~{Bulik}, H.~J. {Bulten}, A.~{Buonanno}, D.~{Buskulic}, R.~L. {Byer},
  M.~{Cabero}, L.~{Cadonati}, G.~{Cagnoli}, C.~{Cahillane}, J.~{Calder{\'o}n
  Bustillo}, J.~D. {Callaghan}, T.~A. {Callister}, E.~{Calloni}, J.~B. {Camp},
  M.~{Canepa}, K.~C. {Cannon}, H.~{Cao}, J.~{Cao}, G.~{Carapella},
  F.~{Carbognani}, S.~{Caride}, M.~F. {Carney}, G.~{Carullo}, J.~{Casanueva
  Diaz}, C.~{Casentini}, J.~{Casta{\~n}eda}, S.~{Caudill}, M.~{Cavagli{\`a}},
  F.~{Cavalier}, R.~{Cavalieri}, G.~{Cella}, P.~{Cerd{\'a}-Dur{\'a}n},
  E.~{Cesarini}, O.~{Chaibi}, K.~{Chakravarti}, C.~{Chan}, M.~{Chan},
  S.~{Chao}, P.~{Charlton}, E.~A. {Chase}, E.~{Chassande-Mottin},
  D.~{Chatterjee}, M.~{Chaturvedi}, H.~Y. {Chen}, X.~{Chen}, Y.~{Chen}, H.~P.
  {Cheng}, C.~K. {Cheong}, H.~Y. {Chia}, F.~{Chiadini}, R.~{Chierici},
  A.~{Chincarini}, A.~{Chiummo}, G.~{Cho}, H.~S. {Cho}, M.~{Cho},
  N.~{Christensen}, Q.~{Chu}, S.~{Chua}, K.~W. {Chung}, S.~{Chung}, G.~{Ciani},
  P.~{Ciecielag}, M.~{Cie{\'s}lar}, A.~A. {Ciobanu}, R.~{Ciolfi},
  F.~{Cipriano}, A.~{Cirone}, F.~{Clara}, J.~A. {Clark}, P.~{Clearwater},
  S.~{Clesse}, F.~{Cleva}, E.~{Coccia}, P.~F. {Cohadon}, D.~{Cohen},
  M.~{Colleoni}, C.~G. {Collette}, C.~{Collins}, M.~{Colpi}, Jr. {Constancio},
  M., L.~{Conti}, S.~J. {Cooper}, P.~{Corban}, T.~R. {Corbitt},
  I.~{Cordero-Carri{\'o}n}, S.~{Corezzi}, K.~R. {Corley}, N.~{Cornish},
  D.~{Corre}, A.~{Corsi}, S.~{Cortese}, C.~A. {Costa}, R.~{Cotesta}, M.~W.
  {Coughlin}, S.~B. {Coughlin}, J.~P. {Coulon}, S.~T. {Countryman},
  P.~{Couvares}, P.~B. {Covas}, D.~M. {Coward}, M.~J. {Cowart}, D.~C. {Coyne},
  R.~{Coyne}, J.~D.~E. {Creighton}, T.~D. {Creighton}, J.~{Cripe},
  M.~{Croquette}, S.~G. {Crowder}, J.~R. {Cudell}, T.~J. {Cullen},
  A.~{Cumming}, R.~{Cummings}, L.~{Cunningham}, E.~{Cuoco}, M.~{Curylo},
  T.~{Dal Canton}, G.~{D{\'a}lya}, A.~{Dana}, L.~M. {Daneshgaran-Bajastani},
  B.~{D'Angelo}, S.~L. {Danilishin}, S.~{D'Antonio}, K.~{Danzmann},
  C.~{Darsow-Fromm}, A.~{Dasgupta}, L.~E.~H. {Datrier}, V.~{Dattilo},
  I.~{Dave}, M.~{Davier}, G.~S. {Davies}, D.~{Davis}, E.~J. {Daw}, D.~{DeBra},
  M.~{Deenadayalan}, J.~{Degallaix}, M.~{De Laurentis}, S.~{Del{\'e}glise},
  M.~{Delfavero}, N.~{De Lillo}, W.~{Del Pozzo}, L.~M. {DeMarchi},
  V.~{D'Emilio}, N.~{Demos}, T.~{Dent}, R.~{De Pietri}, R.~{De Rosa}, C.~{De
  Rossi}, R.~{DeSalvo}, O.~{de Varona}, S.~{Dhurandhar}, M.~C. {D{\'\i}az}, Jr.
  {Diaz-Ortiz}, M., T.~{Dietrich}, L.~{Di Fiore}, C.~{Di Fronzo}, C.~{Di
  Giorgio}, F.~{Di Giovanni}, M.~{Di Giovanni}, T.~{Di Girolamo}, A.~{Di
  Lieto}, B.~{Ding}, S.~{Di Pace}, I.~{Di Palma}, F.~{Di Renzo}, A.~K.
  {Divakarla}, A.~{Dmitriev}, Z.~{Doctor}, F.~{Donovan}, K.~L. {Dooley},
  S.~{Doravari}, I.~{Dorrington}, T.~P. {Downes}, M.~{Drago}, J.~C. {Driggers},
  Z.~{Du}, J.~G. {Ducoin}, P.~{Dupej}, O.~{Durante}, D.~{D'Urso}, S.~E.
  {Dwyer}, P.~J. {Easter}, G.~{Eddolls}, B.~{Edelman}, T.~B. {Edo}, O.~{Edy},
  A.~{Effler}, P.~{Ehrens}, J.~{Eichholz}, S.~S. {Eikenberry}, M.~{Eisenmann},
  R.~A. {Eisenstein}, A.~{Ejlli}, L.~{Errico}, R.~C. {Essick}, H.~{Estelles},
  D.~{Estevez}, Z.~B. {Etienne}, T.~{Etzel}, M.~{Evans}, T.~M. {Evans}, B.~E.
  {Ewing}, V.~{Fafone}, S.~{Fairhurst}, X.~{Fan}, S.~{Farinon}, B.~{Farr},
  W.~M. {Farr}, E.~J. {Fauchon-Jones}, M.~{Favata}, M.~{Fays}, M.~{Fazio},
  J.~{Feicht}, M.~M. {Fejer}, F.~{Feng}, E.~{Fenyvesi}, D.~L. {Ferguson},
  A.~{Fernandez-Galiana}, I.~{Ferrante}, E.~C. {Ferreira}, T.~A. {Ferreira},
  F.~{Fidecaro}, I.~{Fiori}, D.~{Fiorucci}, M.~{Fishbach}, R.~P. {Fisher},
  R.~{Fittipaldi}, M.~{Fitz-Axen}, V.~{Fiumara}, R.~{Flaminio}, E.~{Floden},
  E.~{Flynn}, H.~{Fong}, J.~A. {Font}, P.~W.~F. {Forsyth}, J.~D. {Fournier},
  S.~{Frasca}, F.~{Frasconi}, Z.~{Frei}, A.~{Freise}, R.~{Frey}, V.~{Frey},
  P.~{Fritschel}, V.~V. {Frolov}, G.~{Fronz{\`e}}, P.~{Fulda}, M.~{Fyffe},
  H.~A. {Gabbard}, B.~U. {Gadre}, S.~M. {Gaebel}, J.~R. {Gair}, S.~{Galaudage},
  D.~{Ganapathy}, S.~G. {Gaonkar}, C.~{Garc{\'\i}a-Quir{\'o}s}, F.~{Garufi},
  B.~{Gateley}, S.~{Gaudio}, V.~{Gayathri}, G.~{Gemme}, E.~{Genin},
  A.~{Gennai}, D.~{George}, J.~{George}, L.~{Gergely}, S.~{Ghonge}, Abhirup
  {Ghosh}, Archisman {Ghosh}, S.~{Ghosh}, B.~{Giacomazzo}, J.~A. {Giaime},
  K.~D. {Giardina}, D.~R. {Gibson}, C.~{Gier}, K.~{Gill}, J.~{Glanzer},
  J.~{Gniesmer}, P.~{Godwin}, E.~{Goetz}, R.~{Goetz}, N.~{Gohlke},
  B.~{Goncharov}, G.~{Gonz{\'a}lez}, A.~{Gopakumar}, S.~E. {Gossan},
  M.~{Gosselin}, R.~{Gouaty}, B.~{Grace}, A.~{Grado}, M.~{Granata}, A.~{Grant},
  S.~{Gras}, P.~{Grassia}, C.~{Gray}, R.~{Gray}, G.~{Greco}, A.~C. {Green},
  R.~{Green}, E.~M. {Gretarsson}, H.~L. {Griggs}, G.~{Grignani}, A.~{Grimaldi},
  S.~J. {Grimm}, H.~{Grote}, S.~{Grunewald}, P.~{Gruning}, G.~M. {Guidi}, A.~R.
  {Guimaraes}, G.~{Guix{\'e}}, H.~K. {Gulati}, Y.~{Guo}, A.~{Gupta}, Anchal
  {Gupta}, P.~{Gupta}, E.~K. {Gustafson}, R.~{Gustafson}, L.~{Haegel},
  O.~{Halim}, E.~D. {Hall}, E.~Z. {Hamilton}, G.~{Hammond}, M.~{Haney}, M.~M.
  {Hanke}, J.~{Hanks}, C.~{Hanna}, M.~D. {Hannam}, O.~A. {Hannuksela}, T.~J.
  {Hansen}, J.~{Hanson}, T.~{Harder}, T.~{Hardwick}, K.~{Haris}, J.~{Harms},
  G.~M. {Harry}, I.~W. {Harry}, R.~K. {Hasskew}, C.~J. {Haster}, K.~{Haughian},
  F.~J. {Hayes}, J.~{Healy}, A.~{Heidmann}, M.~C. {Heintze}, J.~{Heinze},
  H.~{Heitmann}, F.~{Hellman}, P.~{Hello}, G.~{Hemming}, M.~{Hendry}, I.~S.
  {Heng}, E.~{Hennes}, J.~{Hennig}, M.~{Heurs}, S.~{Hild}, T.~{Hinderer}, S.~Y.
  {Hoback}, S.~{Hochheim}, E.~{Hofgard}, D.~{Hofman}, A.~M. {Holgado}, N.~A.
  {Holland}, K.~{Holt}, D.~E. {Holz}, P.~{Hopkins}, C.~{Horst}, J.~{Hough},
  E.~J. {Howell}, C.~G. {Hoy}, Y.~{Huang}, M.~T. {H{\"u}bner}, E.~A. {Huerta},
  D.~{Huet}, B.~{Hughey}, V.~{Hui}, S.~{Husa}, S.~H. {Huttner}, R.~{Huxford},
  T.~{Huynh-Dinh}, B.~{Idzkowski}, A.~{Iess}, H.~{Inchauspe}, C.~{Ingram},
  G.~{Intini}, J.~M. {Isac}, M.~{Isi}, B.~R. {Iyer}, T.~{Jacqmin}, S.~J.
  {Jadhav}, S.~P. {Jadhav}, A.~L. {James}, K.~{Jani}, N.~N. {Janthalur},
  P.~{Jaranowski}, D.~{Jariwala}, R.~{Jaume}, A.~C. {Jenkins}, J.~{Jiang},
  G.~R. {Johns}, A.~W. {Jones}, D.~I. {Jones}, J.~D. {Jones}, P.~{Jones},
  R.~{Jones}, R.~J.~G. {Jonker}, L.~{Ju}, J.~{Junker}, C.~V. {Kalaghatgi},
  V.~{Kalogera}, B.~{Kamai}, S.~{Kandhasamy}, G.~{Kang}, J.~B. {Kanner}, S.~J.
  {Kapadia}, S.~{Karki}, R.~{Kashyap}, M.~{Kasprzack}, W.~{Kastaun},
  S.~{Katsanevas}, E.~{Katsavounidis}, W.~{Katzman}, S.~{Kaufer}, K.~{Kawabe},
  F.~{K{\'e}f{\'e}lian}, D.~{Keitel}, A.~{Keivani}, R.~{Kennedy}, J.~S. {Key},
  S.~{Khadka}, F.~Y. {Khalili}, I.~{Khan}, S.~{Khan}, Z.~A. {Khan}, E.~A.
  {Khazanov}, N.~{Khetan}, M.~{Khursheed}, N.~{Kijbunchoo}, Chunglee {Kim},
  G.~J. {Kim}, J.~C. {Kim}, K.~{Kim}, W.~{Kim}, W.~S. {Kim}, Y.~M. {Kim},
  C.~{Kimball}, P.~J. {King}, M.~{Kinley-Hanlon}, R.~{Kirchhoff}, J.~S.
  {Kissel}, L.~{Kleybolte}, S.~{Klimenko}, T.~D. {Knowles}, P.~{Koch}, S.~M.
  {Koehlenbeck}, G.~{Koekoek}, S.~{Koley}, V.~{Kondrashov}, A.~{Kontos},
  N.~{Koper}, M.~{Korobko}, W.~Z. {Korth}, M.~{Kovalam}, D.~B. {Kozak},
  V.~{Kringel}, N.~V. {Krishnendu}, A.~{Kr{\'o}lak}, N.~{Krupinski},
  G.~{Kuehn}, A.~{Kumar}, P.~{Kumar}, Rahul {Kumar}, Rakesh {Kumar},
  S.~{Kumar}, L.~{Kuo}, A.~{Kutynia}, B.~D. {Lackey}, D.~{Laghi}, E.~{Lalande},
  T.~L. {Lam}, A.~{Lamberts}, M.~{Landry}, B.~B. {Lane}, R.~N. {Lang},
  J.~{Lange}, B.~{Lantz}, R.~K. {Lanza}, I.~{La Rosa}, A.~{Lartaux-Vollard},
  P.~D. {Lasky}, M.~{Laxen}, A.~{Lazzarini}, C.~{Lazzaro}, P.~{Leaci},
  S.~{Leavey}, Y.~K. {Lecoeuche}, C.~H. {Lee}, H.~M. {Lee}, H.~W. {Lee},
  J.~{Lee}, K.~{Lee}, J.~{Lehmann}, N.~{Leroy}, N.~{Letendre}, Y.~{Levin},
  A.~K.~Y. {Li}, J.~{Li}, K.~{li}, T.~G.~F. {Li}, X.~{Li}, F.~{Linde}, S.~D.
  {Linker}, J.~N. {Linley}, T.~B. {Littenberg}, J.~{Liu}, X.~{Liu},
  M.~{Llorens-Monteagudo}, R.~K.~L. {Lo}, A.~{Lockwood}, L.~T. {London},
  A.~{Longo}, M.~{Lorenzini}, V.~{Loriette}, M.~{Lormand}, G.~{Losurdo}, J.~D.
  {Lough}, C.~O. {Lousto}, G.~{Lovelace}, H.~{L{\"u}ck}, D.~{Lumaca}, A.~P.
  {Lundgren}, Y.~{Ma}, R.~{Macas}, S.~{Macfoy}, M.~{MacInnis}, D.~M. {Macleod},
  I.~A.~O. {MacMillan}, A.~{Macquet}, I.~{Maga{\~n}a Hernandez},
  F.~{Maga{\~n}a-Sandoval}, R.~M. {Magee}, E.~{Majorana}, I.~{Maksimovic},
  A.~{Malik}, N.~{Man}, V.~{Mandic}, V.~{Mangano}, G.~L. {Mansell},
  M.~{Manske}, M.~{Mantovani}, M.~{Mapelli}, F.~{Marchesoni}, F.~{Marion},
  S.~{M{\'a}rka}, Z.~{M{\'a}rka}, C.~{Markakis}, A.~S. {Markosyan},
  A.~{Markowitz}, E.~{Maros}, A.~{Marquina}, S.~{Marsat}, F.~{Martelli}, I.~W.
  {Martin}, R.~M. {Martin}, V.~{Martinez}, D.~V. {Martynov}, H.~{Masalehdan},
  K.~{Mason}, E.~{Massera}, A.~{Masserot}, T.~J. {Massinger}, M.~{Masso-Reid},
  S.~{Mastrogiovanni}, A.~{Matas}, F.~{Matichard}, N.~{Mavalvala},
  E.~{Maynard}, J.~J. {McCann}, R.~{McCarthy}, D.~E. {McClelland},
  S.~{McCormick}, L.~{McCuller}, S.~C. {McGuire}, C.~{McIsaac}, J.~{McIver},
  D.~J. {McManus}, T.~{McRae}, S.~T. {McWilliams}, D.~{Meacher}, G.~D.
  {Meadors}, M.~{Mehmet}, A.~K. {Mehta}, E.~{Mejuto Villa}, A.~{Melatos},
  G.~{Mendell}, R.~A. {Mercer}, L.~{Mereni}, K.~{Merfeld}, E.~L. {Merilh},
  J.~D. {Merritt}, M.~{Merzougui}, S.~{Meshkov}, C.~{Messenger}, C.~{Messick},
  R.~{Metzdorff}, P.~M. {Meyers}, F.~{Meylahn}, A.~{Mhaske}, A.~{Miani},
  H.~{Miao}, I.~{Michaloliakos}, C.~{Michel}, H.~{Middleton}, L.~{Milano},
  A.~L. {Miller}, M.~{Millhouse}, J.~C. {Mills}, E.~{Milotti}, M.~C.
  {Milovich-Goff}, O.~{Minazzoli}, Y.~{Minenkov}, A.~{Mishkin}, C.~{Mishra},
  T.~{Mistry}, S.~{Mitra}, V.~P. {Mitrofanov}, G.~{Mitselmakher},
  R.~{Mittleman}, G.~{Mo}, K.~{Mogushi}, S.~R.~P. {Mohapatra}, S.~R. {Mohite},
  M.~{Molina-Ruiz}, M.~{Mondin}, M.~{Montani}, C.~J. {Moore}, D.~{Moraru},
  F.~{Morawski}, G.~{Moreno}, S.~{Morisaki}, B.~{Mours}, C.~M. {Mow-Lowry},
  S.~{Mozzon}, F.~{Muciaccia}, Arunava {Mukherjee}, D.~{Mukherjee},
  S.~{Mukherjee}, Subroto {Mukherjee}, N.~{Mukund}, A.~{Mullavey}, J.~{Munch},
  E.~A. {Mu{\~n}iz}, P.~G. {Murray}, A.~{Nagar}, I.~{Nardecchia},
  L.~{Naticchioni}, R.~K. {Nayak}, B.~F. {Neil}, J.~{Neilson}, G.~{Nelemans},
  T.~J.~N. {Nelson}, M.~{Nery}, A.~{Neunzert}, K.~Y. {Ng}, S.~{Ng},
  C.~{Nguyen}, P.~{Nguyen}, D.~{Nichols}, S.~A. {Nichols}, S.~{Nissanke},
  F.~{Nocera}, M.~{Noh}, C.~{North}, D.~{Nothard}, L.~K. {Nuttall},
  J.~{Oberling}, B.~D. {O'Brien}, G.~{Oganesyan}, G.~H. {Ogin}, J.~J. {Oh},
  S.~H. {Oh}, F.~{Ohme}, H.~{Ohta}, M.~A. {Okada}, M.~{Oliver}, C.~{Olivetto},
  P.~{Oppermann}, Richard~J. {Oram}, B.~{O'Reilly}, R.~G. {Ormiston}, L.~F.
  {Ortega}, R.~{O'Shaughnessy}, S.~{Ossokine}, C.~{Osthelder}, D.~J. {Ottaway},
  H.~{Overmier}, B.~J. {Owen}, A.~E. {Pace}, G.~{Pagano}, M.~A. {Page},
  G.~{Pagliaroli}, A.~{Pai}, S.~A. {Pai}, J.~R. {Palamos}, O.~{Palashov},
  C.~{Palomba}, H.~{Pan}, P.~K. {Panda}, P.~T.~H. {Pang}, C.~{Pankow},
  F.~{Pannarale}, B.~C. {Pant}, F.~{Paoletti}, A.~{Paoli}, A.~{Parida},
  W.~{Parker}, D.~{Pascucci}, A.~{Pasqualetti}, R.~{Passaquieti},
  D.~{Passuello}, B.~{Patricelli}, E.~{Payne}, B.~L. {Pearlstone}, T.~C.
  {Pechsiri}, A.~J. {Pedersen}, M.~{Pedraza}, A.~{Pele}, S.~{Penn},
  A.~{Perego}, C.~J. {Perez}, C.~{P{\'e}rigois}, A.~{Perreca},
  S.~{Perri{\`e}s}, J.~{Petermann}, H.~P. {Pfeiffer}, M.~{Phelps}, K.~S.
  {Phukon}, O.~J. {Piccinni}, M.~{Pichot}, M.~{Piendibene}, F.~{Piergiovanni},
  V.~{Pierro}, G.~{Pillant}, L.~{Pinard}, I.~M. {Pinto}, K.~{Piotrzkowski},
  M.~{Pirello}, M.~{Pitkin}, W.~{Plastino}, R.~{Poggiani}, D.~Y.~T. {Pong},
  S.~{Ponrathnam}, P.~{Popolizio}, E.~K. {Porter}, J.~{Powell}, A.~K.
  {Prajapati}, K.~{Prasai}, R.~{Prasanna}, G.~{Pratten}, T.~{Prestegard},
  M.~{Principe}, G.~A. {Prodi}, L.~{Prokhorov}, M.~{Punturo}, P.~{Puppo},
  M.~{P{\"u}rrer}, H.~{Qi}, V.~{Quetschke}, P.~J. {Quinonez}, F.~J. {Raab},
  G.~{Raaijmakers}, H.~{Radkins}, N.~{Radulesco}, P.~{Raffai}, H.~{Rafferty},
  S.~{Raja}, C.~{Rajan}, B.~{Rajbhandari}, M.~{Rakhmanov}, K.~E. {Ramirez},
  A.~{Ramos-Buades}, Javed {Rana}, K.~{Rao}, P.~{Rapagnani}, V.~{Raymond},
  M.~{Razzano}, J.~{Read}, T.~{Regimbau}, L.~{Rei}, S.~{Reid}, D.~H. {Reitze},
  P.~{Rettegno}, F.~{Ricci}, C.~J. {Richardson}, J.~W. {Richardson}, P.~M.
  {Ricker}, G.~{Riemenschneider}, K.~{Riles}, M.~{Rizzo}, N.~A. {Robertson},
  F.~{Robinet}, A.~{Rocchi}, R.~D. {Rodriguez-Soto}, L.~{Rolland}, J.~G.
  {Rollins}, V.~J. {Roma}, M.~{Romanelli}, R.~{Romano}, C.~L. {Romel}, I.~M.
  {Romero-Shaw}, J.~H. {Romie}, C.~A. {Rose}, D.~{Rose}, K.~{Rose},
  D.~{Rosi{\'n}ska}, S.~G. {Rosofsky}, M.~P. {Ross}, S.~{Rowan}, S.~J.
  {Rowlinson}, P.~K. {Roy}, Santosh {Roy}, Soumen {Roy}, P.~{Ruggi},
  G.~{Rutins}, K.~{Ryan}, S.~{Sachdev}, T.~{Sadecki}, M.~{Sakellariadou}, O.~S.
  {Salafia}, L.~{Salconi}, M.~{Saleem}, A.~{Samajdar}, E.~J. {Sanchez}, L.~E.
  {Sanchez}, N.~{Sanchis-Gual}, J.~R. {Sanders}, K.~A. {Santiago}, E.~{Santos},
  N.~{Sarin}, B.~{Sassolas}, B.~S. {Sathyaprakash}, O.~{Sauter}, R.~L.
  {Savage}, V.~{Savant}, D.~{Sawant}, S.~{Sayah}, D.~{Schaetzl}, P.~{Schale},
  M.~{Scheel}, J.~{Scheuer}, P.~{Schmidt}, R.~{Schnabel}, R.~M.~S. {Schofield},
  A.~{Sch{\"o}nbeck}, E.~{Schreiber}, B.~W. {Schulte}, B.~F. {Schutz},
  O.~{Schwarm}, E.~{Schwartz}, J.~{Scott}, S.~M. {Scott}, E.~{Seidel},
  D.~{Sellers}, A.~S. {Sengupta}, N.~{Sennett}, D.~{Sentenac}, V.~{Sequino},
  A.~{Sergeev}, Y.~{Setyawati}, D.~A. {Shaddock}, T.~{Shaffer}, M.~S.
  {Shahriar}, A.~{Sharma}, P.~{Sharma}, P.~{Shawhan}, H.~{Shen},
  M.~{Shikauchi}, R.~{Shink}, D.~H. {Shoemaker}, D.~M. {Shoemaker},
  K.~{Shukla}, S.~{ShyamSundar}, K.~{Siellez}, M.~{Sieniawska}, D.~{Sigg},
  L.~P. {Singer}, D.~{Singh}, N.~{Singh}, A.~{Singha}, A.~{Singhal}, A.~M.
  {Sintes}, V.~{Sipala}, V.~{Skliris}, B.~J.~J. {Slagmolen}, T.~J.
  {Slaven-Blair}, J.~{Smetana}, J.~R. {Smith}, R.~J.~E. {Smith}, S.~{Somala},
  E.~J. {Son}, S.~{Soni}, B.~{Sorazu}, V.~{Sordini}, F.~{Sorrentino},
  T.~{Souradeep}, E.~{Sowell}, A.~P. {Spencer}, M.~{Spera}, A.~K. {Srivastava},
  V.~{Srivastava}, K.~{Staats}, C.~{Stachie}, M.~{Standke}, D.~A. {Steer},
  M.~{Steinke}, J.~{Steinlechner}, S.~{Steinlechner}, D.~{Steinmeyer},
  D.~{Stocks}, D.~J. {Stops}, M.~{Stover}, K.~A. {Strain}, G.~{Stratta},
  A.~{Strunk}, R.~{Sturani}, A.~L. {Stuver}, S.~{Sudhagar}, V.~{Sudhir}, T.~Z.
  {Summerscales}, L.~{Sun}, S.~{Sunil}, A.~{Sur}, J.~{Suresh}, P.~J. {Sutton},
  B.~L. {Swinkels}, M.~J. {Szczepa{\'n}czyk}, M.~{Tacca}, S.~C. {Tait},
  C.~{Talbot}, A.~J. {Tanasijczuk}, D.~B. {Tanner}, D.~{Tao}, M.~{T{\'a}pai},
  A.~{Tapia}, E.~N. {Tapia San Martin}, J.~D. {Tasson}, R.~{Taylor},
  R.~{Tenorio}, L.~{Terkowski}, M.~P. {Thirugnanasambandam}, M.~{Thomas},
  P.~{Thomas}, J.~E. {Thompson}, S.~R. {Thondapu}, K.~A. {Thorne}, E.~{Thrane},
  C.~L. {Tinsman}, T.~R. {Saravanan}, Shubhanshu {Tiwari}, S.~{Tiwari},
  V.~{Tiwari}, K.~{Toland}, M.~{Tonelli}, Z.~{Tornasi}, A.~{Torres-Forn{\'e}},
  C.~I. {Torrie}, I.~{Tosta e Melo}, D.~{T{\"o}yr{\"a}}, E.~A. {Trail},
  F.~{Travasso}, G.~{Traylor}, M.~C. {Tringali}, A.~{Tripathee}, A.~{Trovato},
  R.~J. {Trudeau}, K.~W. {Tsang}, M.~{Tse}, R.~{Tso}, L.~{Tsukada}, D.~{Tsuna},
  T.~{Tsutsui}, M.~{Turconi}, A.~S. {Ubhi}, K.~{Ueno}, D.~{Ugolini}, C.~S.
  {Unnikrishnan}, A.~L. {Urban}, S.~A. {Usman}, A.~C. {Utina}, H.~{Vahlbruch},
  G.~{Vajente}, G.~{Valdes}, M.~{Valentini}, N.~{van Bakel}, M.~{van Beuzekom},
  J.~F.~J. {van den Brand}, C.~{Van Den Broeck}, D.~C. {Vander-Hyde}, L.~{van
  der Schaaf}, J.~V. {Van Heijningen}, A.~A. {van Veggel}, M.~{Vardaro},
  V.~{Varma}, S.~{Vass}, M.~{Vas{\'u}th}, A.~{Vecchio}, G.~{Vedovato},
  J.~{Veitch}, P.~J. {Veitch}, K.~{Venkateswara}, G.~{Venugopalan},
  D.~{Verkindt}, D.~{Veske}, F.~{Vetrano}, A.~{Vicer{\'e}}, A.~D. {Viets},
  S.~{Vinciguerra}, D.~J. {Vine}, J.~Y. {Vinet}, S.~{Vitale},
  Francisco~Hernandez {Vivanco}, T.~{Vo}, H.~{Vocca}, C.~{Vorvick}, S.~P.
  {Vyatchanin}, A.~R. {Wade}, L.~E. {Wade}, M.~{Wade}, R.~{Walet}, M.~{Walker},
  G.~S. {Wallace}, L.~{Wallace}, S.~{Walsh}, J.~Z. {Wang}, S.~{Wang}, W.~H.
  {Wang}, R.~L. {Ward}, Z.~A. {Warden}, J.~{Warner}, M.~{Was}, J.~{Watchi},
  B.~{Weaver}, L.~W. {Wei}, M.~{Weinert}, A.~J. {Weinstein}, R.~{Weiss},
  F.~{Wellmann}, L.~{Wen}, P.~{We{\ss}els}, J.~W. {Westhouse}, K.~{Wette},
  J.~T. {Whelan}, B.~F. {Whiting}, C.~{Whittle}, D.~M. {Wilken}, D.~{Williams},
  J.~L. {Willis}, B.~{Willke}, W.~{Winkler}, C.~C. {Wipf}, H.~{Wittel},
  G.~{Woan}, J.~{Woehler}, J.~K. {Wofford}, C.~{Wong}, J.~L. {Wright}, D.~S.
  {Wu}, D.~M. {Wysocki}, L.~{Xiao}, H.~{Yamamoto}, L.~{Yang}, Y.~{Yang},
  Z.~{Yang}, M.~J. {Yap}, M.~{Yazback}, D.~W. {Yeeles}, Hang {Yu}, Haocun {Yu},
  S.~H.~R. {Yuen}, A.~K. {Zadro{\.z}ny}, A.~{Zadro{\.z}ny}, M.~{Zanolin},
  T.~{Zelenova}, J.~P. {Zendri}, M.~{Zevin}, J.~{Zhang}, L.~{Zhang},
  T.~{Zhang}, C.~{Zhao}, G.~{Zhao}, M.~{Zhou}, Z.~{Zhou}, X.~J. {Zhu}, A.~B.
  {Zimmerman}, M.~E. {Zucker}, J.~{Zweizig}, {LIGO Scientific Collaboration},
  and {Virgo Collaboration}.
\newblock {A Joint Fermi-GBM and LIGO/Virgo Analysis of Compact Binary Mergers
  from the First and Second Gravitational-wave Observing Runs}.
\newblock {\em \apj}, 893(2):100, April 2020.

\bibitem{gbmtargeted}
Adam {Goldstein}, Rachel {Hamburg}, Joshua {Wood}, C.~Michelle {Hui},
  William~H. {Cleveland}, Daniel {Kocevski}, Tyson {Littenberg}, Eric {Burns},
  Tito {Dal Canton}, Peter {Veres}, Bagrat {Mailyan}, Christian {Malacaria},
  Michael~S. {Briggs}, and Colleen~A. {Wilson-Hodge}.
\newblock {Updates to the Fermi GBM Targeted Sub-threshold Search in
  Preparation for the Third Observing Run of LIGO/Virgo}.
\newblock {\em arXiv e-prints}, page arXiv:1903.12597, March 2019.

\bibitem{gbmsingleIFO}
C.~{Stachie}, T.~Dal {Canton}, E.~{Burns}, N.~{Christensen}, R.~{Hamburg},
  M.~{Briggs}, J.~{Broida}, A.~{Goldstein}, F.~{Hayes}, T.~{Littenberg},
  P.~{Shawhan}, J.~{Veitch}, P.~{Veres}, and C.~A. {Wilson-Hodge}.
\newblock {Search for advanced LIGO single interferometer compact binary
  coalescence signals in coincidence with Gamma-ray events in Fermi-GBM}.
\newblock {\em Classical and Quantum Gravity}, 37(17):175001, September 2020.

\bibitem{petrov}
Polina {Petrov}, Leo~P. {Singer}, Michael~W. {Coughlin}, Vishwesh {Kumar},
  Mouza {Almualla}, Shreya {Anand}, Mattia {Bulla}, Tim {Dietrich}, Francois
  {Foucart}, and Nidhal {Guessoum}.
\newblock {Data-driven Expectations for Electromagnetic Counterpart Searches
  Based on LIGO/Virgo Public Alerts}.
\newblock {\em \apj}, 924(2):54, January 2022.

\bibitem{GW170817MMA}
B.~P. {Abbott}, R.~{Abbott}, T.~D. {Abbott}, F.~{Acernese}, K.~{Ackley},
  C.~{Adams}, T.~{Adams}, P.~{Addesso}, R.~X. {Adhikari}, V.~B. {Adya},
  C.~{Affeldt}, M.~{Afrough}, B.~{Agarwal}, M.~{Agathos}, K.~{Agatsuma},
  N.~{Aggarwal}, O.~D. {Aguiar}, L.~{Aiello}, A.~{Ain}, P.~{Ajith}, B.~{Allen},
  G.~{Allen}, A.~{Allocca}, P.~A. {Altin}, A.~{Amato}, A.~{Ananyeva}, S.~B.
  {Anderson}, W.~G. {Anderson}, S.~V. {Angelova}, S.~{Antier}, S.~{Appert},
  K.~{Arai}, M.~C. {Araya}, J.~S. {Areeda}, N.~{Arnaud}, K.~G. {Arun},
  S.~{Ascenzi}, G.~{Ashton}, M.~{Ast}, S.~M. {Aston}, P.~{Astone}, D.~V.
  {Atallah}, P.~{Aufmuth}, C.~{Aulbert}, K.~{AultONeal}, C.~{Austin},
  A.~{Avila-Alvarez}, S.~{Babak}, P.~{Bacon}, M.~K.~M. {Bader}, S.~{Bae}, P.~T.
  {Baker}, F.~{Baldaccini}, G.~{Ballardin}, S.~W. {Ballmer}, S.~{Banagiri},
  J.~C. {Barayoga}, S.~E. {Barclay}, B.~C. {Barish}, D.~{Barker}, K.~{Barkett},
  F.~{Barone}, B.~{Barr}, L.~{Barsotti}, M.~{Barsuglia}, D.~{Barta}, S.~D.
  {Barthelmy}, J.~{Bartlett}, I.~{Bartos}, R.~{Bassiri}, A.~{Basti}, J.~C.
  {Batch}, M.~{Bawaj}, J.~C. {Bayley}, M.~{Bazzan}, B.~{B{\'e}csy}, C.~{Beer},
  M.~{Bejger}, I.~{Belahcene}, A.~S. {Bell}, B.~K. {Berger}, G.~{Bergmann},
  J.~J. {Bero}, C.~P.~L. {Berry}, D.~{Bersanetti}, A.~{Bertolini},
  J.~{Betzwieser}, S.~{Bhagwat}, R.~{Bhandare}, I.~A. {Bilenko},
  G.~{Billingsley}, C.~R. {Billman}, J.~{Birch}, R.~{Birney}, O.~{Birnholtz},
  S.~{Biscans}, S.~{Biscoveanu}, A.~{Bisht}, M.~{Bitossi}, C.~{Biwer}, M.~A.
  {Bizouard}, J.~K. {Blackburn}, J.~{Blackman}, C.~D. {Blair}, D.~G. {Blair},
  R.~M. {Blair}, S.~{Bloemen}, O.~{Bock}, N.~{Bode}, M.~{Boer}, G.~{Bogaert},
  A.~{Bohe}, F.~{Bondu}, E.~{Bonilla}, R.~{Bonnand}, B.~A. {Boom}, R.~{Bork},
  V.~{Boschi}, S.~{Bose}, K.~{Bossie}, Y.~{Bouffanais}, A.~{Bozzi},
  C.~{Bradaschia}, P.~R. {Brady}, M.~{Branchesi}, J.~E. {Brau}, T.~{Briant},
  A.~{Brillet}, M.~{Brinkmann}, V.~{Brisson}, P.~{Brockill}, J.~E. {Broida},
  A.~F. {Brooks}, D.~A. {Brown}, D.~D. {Brown}, S.~{Brunett}, C.~C. {Buchanan},
  A.~{Buikema}, T.~{Bulik}, H.~J. {Bulten}, A.~{Buonanno}, D.~{Buskulic},
  C.~{Buy}, R.~L. {Byer}, M.~{Cabero}, L.~{Cadonati}, G.~{Cagnoli},
  C.~{Cahillane}, J.~{Calder{\'o}n Bustillo}, T.~A. {Callister}, E.~{Calloni},
  J.~B. {Camp}, M.~{Canepa}, P.~{Canizares}, K.~C. {Cannon}, H.~{Cao},
  J.~{Cao}, C.~D. {Capano}, E.~{Capocasa}, F.~{Carbognani}, S.~{Caride}, M.~F.
  {Carney}, J.~{Casanueva Diaz}, C.~{Casentini}, S.~{Caudill},
  M.~{Cavagli{\`a}}, F.~{Cavalier}, R.~{Cavalieri}, G.~{Cella}, C.~B. {Cepeda},
  P.~{Cerd{\'a}-Dur{\'a}n}, G.~{Cerretani}, E.~{Cesarini}, S.~J. {Chamberlin},
  M.~{Chan}, S.~{Chao}, P.~{Charlton}, E.~{Chase}, E.~{Chassande-Mottin},
  D.~{Chatterjee}, K.~{Chatziioannou}, B.~D. {Cheeseboro}, H.~Y. {Chen},
  X.~{Chen}, Y.~{Chen}, H.~P. {Cheng}, H.~{Chia}, A.~{Chincarini},
  A.~{Chiummo}, T.~{Chmiel}, H.~S. {Cho}, M.~{Cho}, J.~H. {Chow},
  N.~{Christensen}, Q.~{Chu}, A.~J.~K. {Chua}, S.~{Chua}, A.~K.~W. {Chung},
  S.~{Chung}, G.~{Ciani}, R.~{Ciolfi}, C.~E. {Cirelli}, A.~{Cirone},
  F.~{Clara}, J.~A. {Clark}, P.~{Clearwater}, F.~{Cleva}, C.~{Cocchieri},
  E.~{Coccia}, P.~F. {Cohadon}, D.~{Cohen}, A.~{Colla}, C.~G. {Collette}, L.~R.
  {Cominsky}, Jr. {Constancio}, M., L.~{Conti}, S.~J. {Cooper}, P.~{Corban},
  T.~R. {Corbitt}, I.~{Cordero-Carri{\'o}n}, K.~R. {Corley}, N.~{Cornish},
  A.~{Corsi}, S.~{Cortese}, C.~A. {Costa}, M.~W. {Coughlin}, S.~B. {Coughlin},
  J.~P. {Coulon}, S.~T. {Countryman}, P.~{Couvares}, P.~B. {Covas}, E.~E.
  {Cowan}, D.~M. {Coward}, M.~J. {Cowart}, D.~C. {Coyne}, R.~{Coyne}, J.~D.~E.
  {Creighton}, T.~D. {Creighton}, J.~{Cripe}, S.~G. {Crowder}, T.~J. {Cullen},
  A.~{Cumming}, L.~{Cunningham}, E.~{Cuoco}, T.~{Dal Canton}, G.~{D{\'a}lya},
  S.~L. {Danilishin}, S.~{D'Antonio}, K.~{Danzmann}, A.~{Dasgupta}, C.~F. {Da
  Silva Costa}, V.~{Dattilo}, I.~{Dave}, M.~{Davier}, D.~{Davis}, E.~J. {Daw},
  B.~{Day}, S.~{De}, D.~{DeBra}, J.~{Degallaix}, M.~{De Laurentis},
  S.~{Del{\'e}glise}, W.~{Del Pozzo}, N.~{Demos}, T.~{Denker}, T.~{Dent},
  R.~{De Pietri}, V.~{Dergachev}, R.~{De Rosa}, R.~T. {DeRosa}, C.~{De Rossi},
  R.~{DeSalvo}, O.~{de Varona}, J.~{Devenson}, S.~{Dhurandhar}, M.~C.
  {D{\'\i}az}, L.~{Di Fiore}, M.~{Di Giovanni}, T.~{Di Girolamo}, A.~{Di
  Lieto}, S.~{Di Pace}, I.~{Di Palma}, F.~{Di Renzo}, Z.~{Doctor},
  V.~{Dolique}, F.~{Donovan}, K.~L. {Dooley}, S.~{Doravari}, I.~{Dorrington},
  R.~{Douglas}, M.~{Dovale {\'A}lvarez}, T.~P. {Downes}, M.~{Drago},
  C.~{Dreissigacker}, J.~C. {Driggers}, Z.~{Du}, M.~{Ducrot}, P.~{Dupej}, S.~E.
  {Dwyer}, T.~B. {Edo}, M.~C. {Edwards}, A.~{Effler}, P.~{Ehrens},
  J.~{Eichholz}, S.~S. {Eikenberry}, R.~A. {Eisenstein}, R.~C. {Essick},
  D.~{Estevez}, Z.~B. {Etienne}, T.~{Etzel}, M.~{Evans}, T.~M. {Evans},
  M.~{Factourovich}, V.~{Fafone}, H.~{Fair}, S.~{Fairhurst}, X.~{Fan},
  S.~{Farinon}, B.~{Farr}, W.~M. {Farr}, E.~J. {Fauchon-Jones}, M.~{Favata},
  M.~{Fays}, C.~{Fee}, H.~{Fehrmann}, J.~{Feicht}, M.~M. {Fejer},
  A.~{Fernandez-Galiana}, I.~{Ferrante}, E.~C. {Ferreira}, F.~{Ferrini},
  F.~{Fidecaro}, D.~{Finstad}, I.~{Fiori}, D.~{Fiorucci}, M.~{Fishbach}, R.~P.
  {Fisher}, M.~{Fitz-Axen}, R.~{Flaminio}, M.~{Fletcher}, H.~{Fong}, J.~A.
  {Font}, P.~W.~F. {Forsyth}, S.~S. {Forsyth}, J.~D. {Fournier}, S.~{Frasca},
  F.~{Frasconi}, Z.~{Frei}, A.~{Freise}, R.~{Frey}, V.~{Frey}, E.~M. {Fries},
  P.~{Fritschel}, V.~V. {Frolov}, P.~{Fulda}, M.~{Fyffe}, H.~{Gabbard}, B.~U.
  {Gadre}, S.~M. {Gaebel}, J.~R. {Gair}, L.~{Gammaitoni}, M.~R. {Ganija}, S.~G.
  {Gaonkar}, C.~{Garcia-Quiros}, F.~{Garufi}, B.~{Gateley}, S.~{Gaudio},
  G.~{Gaur}, V.~{Gayathri}, N.~{Gehrels}, G.~{Gemme}, E.~{Genin}, A.~{Gennai},
  D.~{George}, J.~{George}, L.~{Gergely}, V.~{Germain}, S.~{Ghonge}, Abhirup
  {Ghosh}, Archisman {Ghosh}, S.~{Ghosh}, J.~A. {Giaime}, K.~D. {Giardina},
  A.~{Giazotto}, K.~{Gill}, L.~{Glover}, E.~{Goetz}, R.~{Goetz}, S.~{Gomes},
  B.~{Goncharov}, G.~{Gonz{\'a}lez}, J.~M. {Gonzalez Castro}, A.~{Gopakumar},
  M.~L. {Gorodetsky}, S.~E. {Gossan}, M.~{Gosselin}, R.~{Gouaty}, A.~{Grado},
  C.~{Graef}, M.~{Granata}, A.~{Grant}, S.~{Gras}, C.~{Gray}, G.~{Greco}, A.~C.
  {Green}, E.~M. {Gretarsson}, B.~{Griswold}, P.~{Groot}, H.~{Grote},
  S.~{Grunewald}, P.~{Gruning}, G.~M. {Guidi}, X.~{Guo}, A.~{Gupta}, M.~K.
  {Gupta}, K.~E. {Gushwa}, E.~K. {Gustafson}, R.~{Gustafson}, O.~{Halim}, B.~R.
  {Hall}, E.~D. {Hall}, E.~Z. {Hamilton}, G.~{Hammond}, M.~{Haney}, M.~M.
  {Hanke}, J.~{Hanks}, C.~{Hanna}, M.~D. {Hannam}, O.~A. {Hannuksela},
  J.~{Hanson}, T.~{Hardwick}, J.~{Harms}, G.~M. {Harry}, I.~W. {Harry}, M.~J.
  {Hart}, C.~J. {Haster}, K.~{Haughian}, J.~{Healy}, A.~{Heidmann}, M.~C.
  {Heintze}, H.~{Heitmann}, P.~{Hello}, G.~{Hemming}, M.~{Hendry}, I.~S.
  {Heng}, J.~{Hennig}, A.~W. {Heptonstall}, M.~{Heurs}, S.~{Hild},
  T.~{Hinderer}, D.~{Hoak}, D.~{Hofman}, K.~{Holt}, D.~E. {Holz}, P.~{Hopkins},
  C.~{Horst}, J.~{Hough}, E.~A. {Houston}, E.~J. {Howell}, A.~{Hreibi}, Y.~M.
  {Hu}, E.~A. {Huerta}, D.~{Huet}, B.~{Hughey}, S.~{Husa}, S.~H. {Huttner},
  T.~{Huynh-Dinh}, N.~{Indik}, R.~{Inta}, G.~{Intini}, H.~N. {Isa}, J.~M.
  {Isac}, M.~{Isi}, B.~R. {Iyer}, K.~{Izumi}, T.~{Jacqmin}, K.~{Jani},
  P.~{Jaranowski}, S.~{Jawahar}, F.~{Jim{\'e}nez-Forteza}, W.~W. {Johnson},
  D.~I. {Jones}, R.~{Jones}, R.~J.~G. {Jonker}, L.~{Ju}, J.~{Junker}, C.~V.
  {Kalaghatgi}, V.~{Kalogera}, B.~{Kamai}, S.~{Kandhasamy}, G.~{Kang}, J.~B.
  {Kanner}, S.~J. {Kapadia}, S.~{Karki}, K.~S. {Karvinen}, M.~{Kasprzack},
  M.~{Katolik}, E.~{Katsavounidis}, W.~{Katzman}, S.~{Kaufer}, K.~{Kawabe},
  F.~{K{\'e}f{\'e}lian}, D.~{Keitel}, A.~J. {Kemball}, R.~{Kennedy}, C.~{Kent},
  J.~S. {Key}, F.~Y. {Khalili}, I.~{Khan}, S.~{Khan}, Z.~{Khan}, E.~A.
  {Khazanov}, N.~{Kijbunchoo}, Chunglee {Kim}, J.~C. {Kim}, K.~{Kim}, W.~{Kim},
  W.~S. {Kim}, Y.~M. {Kim}, S.~J. {Kimbrell}, E.~J. {King}, P.~J. {King},
  M.~{Kinley-Hanlon}, R.~{Kirchhoff}, J.~S. {Kissel}, L.~{Kleybolte},
  S.~{Klimenko}, T.~D. {Knowles}, P.~{Koch}, S.~M. {Koehlenbeck}, S.~{Koley},
  V.~{Kondrashov}, A.~{Kontos}, M.~{Korobko}, W.~Z. {Korth}, I.~{Kowalska},
  D.~B. {Kozak}, C.~{Kr{\"a}mer}, V.~{Kringel}, B.~{Krishnan}, A.~{Kr{\'o}lak},
  G.~{Kuehn}, P.~{Kumar}, R.~{Kumar}, S.~{Kumar}, L.~{Kuo}, A.~{Kutynia},
  S.~{Kwang}, B.~D. {Lackey}, K.~H. {Lai}, M.~{Landry}, R.~N. {Lang},
  J.~{Lange}, B.~{Lantz}, R.~K. {Lanza}, S.~L. {Larson}, A.~{Lartaux-Vollard},
  P.~D. {Lasky}, M.~{Laxen}, A.~{Lazzarini}, C.~{Lazzaro}, P.~{Leaci},
  S.~{Leavey}, C.~H. {Lee}, H.~K. {Lee}, H.~M. {Lee}, H.~W. {Lee}, K.~{Lee},
  J.~{Lehmann}, A.~{Lenon}, M.~{Leonardi}, N.~{Leroy}, N.~{Letendre},
  Y.~{Levin}, T.~G.~F. {Li}, S.~D. {Linker}, T.~B. {Littenberg}, J.~{Liu},
  R.~K.~L. {Lo}, N.~A. {Lockerbie}, L.~T. {London}, J.~E. {Lord},
  M.~{Lorenzini}, V.~{Loriette}, M.~{Lormand}, G.~{Losurdo}, J.~D. {Lough},
  C.~O. {Lousto}, G.~{Lovelace}, H.~{L{\"u}ck}, D.~{Lumaca}, A.~P. {Lundgren},
  R.~{Lynch}, Y.~{Ma}, R.~{Macas}, S.~{Macfoy}, B.~{Machenschalk},
  M.~{MacInnis}, D.~M. {Macleod}, I.~{Maga{\~n}a Hernandez},
  F.~{Maga{\~n}a-Sandoval}, L.~{Maga{\~n}a Zertuche}, R.~M. {Magee},
  E.~{Majorana}, I.~{Maksimovic}, N.~{Man}, V.~{Mandic}, V.~{Mangano}, G.~L.
  {Mansell}, M.~{Manske}, M.~{Mantovani}, F.~{Marchesoni}, F.~{Marion},
  S.~{M{\'a}rka}, Z.~{M{\'a}rka}, C.~{Markakis}, A.~S. {Markosyan},
  A.~{Markowitz}, E.~{Maros}, A.~{Marquina}, P.~{Marsh}, F.~{Martelli},
  L.~{Martellini}, I.~W. {Martin}, R.~M. {Martin}, D.~V. {Martynov},
  K.~{Mason}, E.~{Massera}, A.~{Masserot}, T.~J. {Massinger}, M.~{Masso-Reid},
  S.~{Mastrogiovanni}, A.~{Matas}, F.~{Matichard}, L.~{Matone}, N.~{Mavalvala},
  N.~{Mazumder}, R.~{McCarthy}, D.~E. {McClelland}, S.~{McCormick},
  L.~{McCuller}, S.~C. {McGuire}, G.~{McIntyre}, J.~{McIver}, D.~J. {McManus},
  L.~{McNeill}, T.~{McRae}, S.~T. {McWilliams}, D.~{Meacher}, G.~D. {Meadors},
  M.~{Mehmet}, J.~{Meidam}, E.~{Mejuto-Villa}, A.~{Melatos}, G.~{Mendell},
  R.~A. {Mercer}, E.~L. {Merilh}, M.~{Merzougui}, S.~{Meshkov}, C.~{Messenger},
  C.~{Messick}, R.~{Metzdorff}, P.~M. {Meyers}, H.~{Miao}, C.~{Michel},
  H.~{Middleton}, E.~E. {Mikhailov}, L.~{Milano}, A.~L. {Miller}, B.~B.
  {Miller}, J.~{Miller}, M.~{Millhouse}, M.~C. {Milovich-Goff}, O.~{Minazzoli},
  Y.~{Minenkov}, J.~{Ming}, C.~{Mishra}, S.~{Mitra}, V.~P. {Mitrofanov},
  G.~{Mitselmakher}, R.~{Mittleman}, D.~{Moffa}, A.~{Moggi}, K.~{Mogushi},
  M.~{Mohan}, S.~R.~P. {Mohapatra}, M.~{Montani}, C.~J. {Moore}, D.~{Moraru},
  G.~{Moreno}, S.~R. {Morriss}, B.~{Mours}, C.~M. {Mow-Lowry}, G.~{Mueller},
  A.~W. {Muir}, Arunava {Mukherjee}, D.~{Mukherjee}, S.~{Mukherjee},
  N.~{Mukund}, A.~{Mullavey}, J.~{Munch}, E.~A. {Mu{\~n}iz}, M.~{Muratore},
  P.~G. {Murray}, K.~{Napier}, I.~{Nardecchia}, L.~{Naticchioni}, R.~K.
  {Nayak}, J.~{Neilson}, G.~{Nelemans}, T.~J.~N. {Nelson}, M.~{Nery},
  A.~{Neunzert}, L.~{Nevin}, J.~M. {Newport}, G.~{Newton}, K.~K.~Y. {Ng},
  P.~{Nguyen}, T.~T. {Nguyen}, D.~{Nichols}, A.~B. {Nielsen}, S.~{Nissanke},
  A.~{Nitz}, A.~{Noack}, F.~{Nocera}, D.~{Nolting}, C.~{North}, L.~K.
  {Nuttall}, J.~{Oberling}, G.~D. {O'Dea}, G.~H. {Ogin}, J.~J. {Oh}, S.~H.
  {Oh}, F.~{Ohme}, M.~A. {Okada}, M.~{Oliver}, P.~{Oppermann}, Richard~J.
  {Oram}, B.~{O'Reilly}, R.~{Ormiston}, L.~F. {Ortega}, R.~{O'Shaughnessy},
  S.~{Ossokine}, D.~J. {Ottaway}, H.~{Overmier}, B.~J. {Owen}, A.~E. {Pace},
  J.~{Page}, M.~A. {Page}, A.~{Pai}, S.~A. {Pai}, J.~R. {Palamos},
  O.~{Palashov}, C.~{Palomba}, A.~{Pal-Singh}, Howard {Pan}, Huang-Wei {Pan},
  B.~{Pang}, P.~T.~H. {Pang}, C.~{Pankow}, F.~{Pannarale}, B.~C. {Pant},
  F.~{Paoletti}, A.~{Paoli}, M.~A. {Papa}, A.~{Parida}, W.~{Parker},
  D.~{Pascucci}, A.~{Pasqualetti}, R.~{Passaquieti}, D.~{Passuello},
  M.~{Patil}, B.~{Patricelli}, B.~L. {Pearlstone}, M.~{Pedraza}, R.~{Pedurand},
  L.~{Pekowsky}, A.~{Pele}, S.~{Penn}, C.~J. {Perez}, A.~{Perreca}, L.~M.
  {Perri}, H.~P. {Pfeiffer}, M.~{Phelps}, O.~J. {Piccinni}, M.~{Pichot},
  F.~{Piergiovanni}, V.~{Pierro}, G.~{Pillant}, L.~{Pinard}, I.~M. {Pinto},
  M.~{Pirello}, M.~{Pitkin}, M.~{Poe}, R.~{Poggiani}, P.~{Popolizio}, E.~K.
  {Porter}, A.~{Post}, J.~{Powell}, J.~{Prasad}, J.~W.~W. {Pratt},
  G.~{Pratten}, V.~{Predoi}, T.~{Prestegard}, L.~R. {Price}, M.~{Prijatelj},
  M.~{Principe}, S.~{Privitera}, G.~A. {Prodi}, L.~G. {Prokhorov},
  O.~{Puncken}, M.~{Punturo}, P.~{Puppo}, M.~{P{\"u}rrer}, H.~{Qi},
  V.~{Quetschke}, E.~A. {Quintero}, R.~{Quitzow-James}, F.~J. {Raab}, D.~S.
  {Rabeling}, H.~{Radkins}, P.~{Raffai}, S.~{Raja}, C.~{Rajan},
  B.~{Rajbhandari}, M.~{Rakhmanov}, K.~E. {Ramirez}, A.~{Ramos-Buades},
  P.~{Rapagnani}, V.~{Raymond}, M.~{Razzano}, J.~{Read}, T.~{Regimbau},
  L.~{Rei}, S.~{Reid}, D.~H. {Reitze}, W.~{Ren}, S.~D. {Reyes}, F.~{Ricci},
  P.~M. {Ricker}, S.~{Rieger}, K.~{Riles}, M.~{Rizzo}, N.~A. {Robertson},
  R.~{Robie}, F.~{Robinet}, A.~{Rocchi}, L.~{Rolland}, J.~G. {Rollins}, V.~J.
  {Roma}, R.~{Romano}, C.~L. {Romel}, J.~H. {Romie}, D.~{Rosi{\'n}ska}, M.~P.
  {Ross}, S.~{Rowan}, A.~{R{\"u}diger}, P.~{Ruggi}, G.~{Rutins}, K.~{Ryan},
  S.~{Sachdev}, T.~{Sadecki}, L.~{Sadeghian}, M.~{Sakellariadou}, L.~{Salconi},
  M.~{Saleem}, F.~{Salemi}, A.~{Samajdar}, L.~{Sammut}, L.~M. {Sampson}, E.~J.
  {Sanchez}, L.~E. {Sanchez}, N.~{Sanchis-Gual}, V.~{Sandberg}, J.~R.
  {Sanders}, B.~{Sassolas}, B.~S. {Sathyaprakash}, P.~R. {Saulson},
  O.~{Sauter}, R.~L. {Savage}, A.~{Sawadsky}, P.~{Schale}, M.~{Scheel},
  J.~{Scheuer}, J.~{Schmidt}, P.~{Schmidt}, R.~{Schnabel}, R.~M.~S.
  {Schofield}, A.~{Sch{\"o}nbeck}, E.~{Schreiber}, D.~{Schuette}, B.~W.
  {Schulte}, B.~F. {Schutz}, S.~G. {Schwalbe}, J.~{Scott}, S.~M. {Scott},
  E.~{Seidel}, D.~{Sellers}, A.~S. {Sengupta}, D.~{Sentenac}, V.~{Sequino},
  A.~{Sergeev}, D.~A. {Shaddock}, T.~J. {Shaffer}, A.~A. {Shah}, M.~S.
  {Shahriar}, M.~B. {Shaner}, L.~{Shao}, B.~{Shapiro}, P.~{Shawhan},
  A.~{Sheperd}, D.~H. {Shoemaker}, D.~M. {Shoemaker}, K.~{Siellez},
  X.~{Siemens}, M.~{Sieniawska}, D.~{Sigg}, A.~D. {Silva}, L.~P. {Singer},
  A.~{Singh}, A.~{Singhal}, A.~M. {Sintes}, B.~J.~J. {Slagmolen}, B.~{Smith},
  J.~R. {Smith}, R.~J.~E. {Smith}, S.~{Somala}, E.~J. {Son}, J.~A.
  {Sonnenberg}, B.~{Sorazu}, F.~{Sorrentino}, T.~{Souradeep}, A.~P. {Spencer},
  A.~K. {Srivastava}, K.~{Staats}, A.~{Staley}, M.~{Steinke},
  J.~{Steinlechner}, S.~{Steinlechner}, D.~{Steinmeyer}, S.~P. {Stevenson},
  R.~{Stone}, D.~J. {Stops}, K.~A. {Strain}, G.~{Stratta}, S.~E. {Strigin},
  A.~{Strunk}, R.~{Sturani}, A.~L. {Stuver}, T.~Z. {Summerscales}, L.~{Sun},
  S.~{Sunil}, J.~{Suresh}, P.~J. {Sutton}, B.~L. {Swinkels}, M.~J.
  {Szczepa{\'n}czyk}, M.~{Tacca}, S.~C. {Tait}, C.~{Talbot}, D.~{Talukder},
  D.~B. {Tanner}, M.~{T{\'a}pai}, A.~{Taracchini}, J.~D. {Tasson}, J.~A.
  {Taylor}, R.~{Taylor}, S.~V. {Tewari}, T.~{Theeg}, F.~{Thies}, E.~G.
  {Thomas}, M.~{Thomas}, P.~{Thomas}, K.~A. {Thorne}, K.~S. {Thorne},
  E.~{Thrane}, S.~{Tiwari}, V.~{Tiwari}, K.~V. {Tokmakov}, K.~{Toland},
  M.~{Tonelli}, Z.~{Tornasi}, A.~{Torres-Forn{\'e}}, C.~I. {Torrie},
  D.~{T{\"o}yr{\"a}}, F.~{Travasso}, G.~{Traylor}, J.~{Trinastic}, M.~C.
  {Tringali}, L.~{Trozzo}, K.~W. {Tsang}, M.~{Tse}, R.~{Tso}, L.~{Tsukada},
  D.~{Tsuna}, D.~{Tuyenbayev}, K.~{Ueno}, D.~{Ugolini}, C.~S. {Unnikrishnan},
  A.~L. {Urban}, S.~A. {Usman}, H.~{Vahlbruch}, G.~{Vajente}, G.~{Valdes},
  N.~{van Bakel}, M.~{van Beuzekom}, J.~F.~J. {van den Brand}, C.~{Van Den
  Broeck}, D.~C. {Vander-Hyde}, L.~{van der Schaaf}, J.~V. {van Heijningen},
  A.~A. {van Veggel}, M.~{Vardaro}, V.~{Varma}, S.~{Vass}, M.~{Vas{\'u}th},
  A.~{Vecchio}, G.~{Vedovato}, J.~{Veitch}, P.~J. {Veitch}, K.~{Venkateswara},
  G.~{Venugopalan}, D.~{Verkindt}, F.~{Vetrano}, A.~{Vicer{\'e}}, A.~D.
  {Viets}, S.~{Vinciguerra}, D.~J. {Vine}, J.~Y. {Vinet}, S.~{Vitale}, T.~{Vo},
  H.~{Vocca}, C.~{Vorvick}, S.~P. {Vyatchanin}, A.~R. {Wade}, L.~E. {Wade},
  M.~{Wade}, R.~{Walet}, M.~{Walker}, L.~{Wallace}, S.~{Walsh}, G.~{Wang},
  H.~{Wang}, J.~Z. {Wang}, W.~H. {Wang}, Y.~F. {Wang}, R.~L. {Ward},
  J.~{Warner}, M.~{Was}, J.~{Watchi}, B.~{Weaver}, L.~W. {Wei}, M.~{Weinert},
  A.~J. {Weinstein}, R.~{Weiss}, L.~{Wen}, E.~K. {Wessel}, P.~{Wessels},
  J.~{Westerweck}, T.~{Westphal}, K.~{Wette}, J.~T. {Whelan}, S.~E. {Whitcomb},
  B.~F. {Whiting}, C.~{Whittle}, D.~{Wilken}, D.~{Williams}, R.~D. {Williams},
  A.~R. {Williamson}, J.~L. {Willis}, B.~{Willke}, M.~H. {Wimmer},
  W.~{Winkler}, C.~C. {Wipf}, H.~{Wittel}, G.~{Woan}, J.~{Woehler},
  J.~{Wofford}, K.~W.~K. {Wong}, J.~{Worden}, J.~L. {Wright}, D.~S. {Wu}, D.~M.
  {Wysocki}, S.~{Xiao}, H.~{Yamamoto}, C.~C. {Yancey}, L.~{Yang}, M.~J. {Yap},
  M.~{Yazback}, Hang {Yu}, Haocun {Yu}, M.~{Yvert}, A.~{Zadro{\.z}ny},
  M.~{Zanolin}, T.~{Zelenova}, J.~P. {Zendri}, M.~{Zevin}, L.~{Zhang},
  M.~{Zhang}, T.~{Zhang}, Y.~H. {Zhang}, C.~{Zhao}, M.~{Zhou}, Z.~{Zhou}, S.~J.
  {Zhu}, X.~J. {Zhu}, A.~B. {Zimmerman}, M.~E. {Zucker}, J.~{Zweizig}, {LIGO
  Scientific Collaboration}, {Virgo Collaboration}, C.~A. {Wilson-Hodge},
  E.~{Bissaldi}, L.~{Blackburn}, M.~S. {Briggs}, E.~{Burns}, W.~H. {Cleveland},
  V.~{Connaughton}, M.~H. {Gibby}, M.~M. {Giles}, A.~{Goldstein}, R.~{Hamburg},
  P.~{Jenke}, C.~M. {Hui}, R.~M. {Kippen}, D.~{Kocevski}, S.~{McBreen}, C.~A.
  {Meegan}, W.~S. {Paciesas}, S.~{Poolakkil}, R.~D. {Preece}, J.~{Racusin},
  O.~J. {Roberts}, M.~{Stanbro}, P.~{Veres}, A.~{von Kienlin}, Fermi {GBM},
  V.~{Savchenko}, C.~{Ferrigno}, E.~{Kuulkers}, A.~{Bazzano}, E.~{Bozzo},
  S.~{Brandt}, J.~{Chenevez}, T.~J.~L. {Courvoisier}, R.~{Diehl}, A.~{Domingo},
  L.~{Hanlon}, E.~{Jourdain}, P.~{Laurent}, F.~{Lebrun}, A.~{Lutovinov},
  A.~{Martin-Carrillo}, S.~{Mereghetti}, L.~{Natalucci}, J.~{Rodi}, J.~P.
  {Roques}, R.~{Sunyaev}, P.~{Ubertini}, {INTEGRAL}, M.~G. {Aartsen},
  M.~{Ackermann}, J.~{Adams}, J.~A. {Aguilar}, M.~{Ahlers}, M.~{Ahrens}, I.~Al
  {Samarai}, D.~{Altmann}, K.~{Andeen}, T.~{Anderson}, I.~{Ansseau},
  G.~{Anton}, C.~{Arg{\"u}elles}, J.~{Auffenberg}, S.~{Axani}, H.~{Bagherpour},
  X.~{Bai}, J.~P. {Barron}, S.~W. {Barwick}, V.~{Baum}, R.~{Bay}, J.~J.
  {Beatty}, J.~{Becker Tjus}, E.~{Bernardini}, D.~Z. {Besson}, G.~{Binder},
  D.~{Bindig}, E.~{Blaufuss}, S.~{Blot}, C.~{Bohm}, M.~{B{\"o}rner}, F.~{Bos},
  D.~{Bose}, S.~{B{\"o}ser}, O.~{Botner}, E.~{Bourbeau}, J.~{Bourbeau},
  F.~{Bradascio}, J.~{Braun}, L.~{Brayeur}, M.~{Brenzke}, H.~P. {Bretz},
  S.~{Bron}, J.~{Brostean-Kaiser}, A.~{Burgman}, T.~{Carver}, J.~{Casey},
  M.~{Casier}, E.~{Cheung}, D.~{Chirkin}, A.~{Christov}, K.~{Clark},
  L.~{Classen}, S.~{Coenders}, G.~H. {Collin}, J.~M. {Conrad}, D.~F. {Cowen},
  R.~{Cross}, M.~{Day}, J.~P.~A.~M. {de Andr{\'e}}, C.~{De Clercq}, J.~J.
  {DeLaunay}, H.~{Dembinski}, S.~{De Ridder}, P.~{Desiati}, K.~D. {de Vries},
  G.~{de Wasseige}, M.~{de With}, T.~{DeYoung}, J.~C. {D{\'\i}az-V{\'e}lez},
  V.~{di Lorenzo}, H.~{Dujmovic}, J.~P. {Dumm}, M.~{Dunkman}, E.~{Dvorak},
  B.~{Eberhardt}, T.~{Ehrhardt}, B.~{Eichmann}, P.~{Eller}, P.~A. {Evenson},
  S.~{Fahey}, A.~R. {Fazely}, J.~{Felde}, K.~{Filimonov}, C.~{Finley},
  S.~{Flis}, A.~{Franckowiak}, E.~{Friedman}, T.~{Fuchs}, T.~K. {Gaisser},
  J.~{Gallagher}, L.~{Gerhardt}, K.~{Ghorbani}, W.~{Giang}, T.~{Glauch},
  T.~{Gl{\"u}senkamp}, A.~{Goldschmidt}, J.~G. {Gonzalez}, D.~{Grant},
  Z.~{Griffith}, C.~{Haack}, A.~{Hallgren}, F.~{Halzen}, K.~{Hanson},
  D.~{Hebecker}, D.~{Heereman}, K.~{Helbing}, R.~{Hellauer}, S.~{Hickford},
  J.~{Hignight}, G.~C. {Hill}, K.~D. {Hoffman}, R.~{Hoffmann},
  B.~{Hokanson-Fasig}, K.~{Hoshina}, F.~{Huang}, M.~{Huber}, K.~{Hultqvist},
  M.~{H{\"u}nnefeld}, S.~{In}, A.~{Ishihara}, E.~{Jacobi}, G.~S. {Japaridze},
  M.~{Jeong}, K.~{Jero}, B.~J.~P. {Jones}, P.~{Kalaczynski}, W.~{Kang},
  A.~{Kappes}, T.~{Karg}, A.~{Karle}, M.~{Kauer}, A.~{Keivani}, J.~L. {Kelley},
  A.~{Kheirandish}, J.~{Kim}, M.~{Kim}, T.~{Kintscher}, J.~{Kiryluk},
  T.~{Kittler}, S.~R. {Klein}, G.~{Kohnen}, R.~{Koirala}, H.~{Kolanoski},
  L.~{K{\"o}pke}, C.~{Kopper}, S.~{Kopper}, J.~P. {Koschinsky}, D.~J.
  {Koskinen}, M.~{Kowalski}, K.~{Krings}, M.~{Kroll}, G.~{Kr{\"u}ckl},
  J.~{Kunnen}, S.~{Kunwar}, N.~{Kurahashi}, T.~{Kuwabara}, A.~{Kyriacou},
  M.~{Labare}, J.~L. {Lanfranchi}, M.~J. {Larson}, F.~{Lauber},
  M.~{Lesiak-Bzdak}, M.~{Leuermann}, Q.~R. {Liu}, L.~{Lu}, J.~{L{\"u}nemann},
  W.~{Luszczak}, J.~{Madsen}, G.~{Maggi}, K.~B.~M. {Mahn}, S.~{Mancina},
  R.~{Maruyama}, K.~{Mase}, R.~{Maunu}, F.~{McNally}, K.~{Meagher},
  M.~{Medici}, M.~{Meier}, T.~{Menne}, G.~{Merino}, T.~{Meures}, S.~{Miarecki},
  J.~{Micallef}, G.~{Moment{\'e}}, T.~{Montaruli}, R.~W. {Moore}, M.~{Moulai},
  R.~{Nahnhauer}, P.~{Nakarmi}, U.~{Naumann}, G.~{Neer}, H.~{Niederhausen},
  S.~C. {Nowicki}, D.~R. {Nygren}, A.~{Obertacke Pollmann}, A.~{Olivas},
  A.~{O'Murchadha}, T.~{Palczewski}, H.~{Pandya}, D.~V. {Pankova},
  P.~{Peiffer}, J.~A. {Pepper}, C.~{P{\'e}rez de los Heros}, D.~{Pieloth},
  E.~{Pinat}, P.~B. {Price}, G.~T. {Przybylski}, C.~{Raab}, L.~{R{\"a}del},
  M.~{Rameez}, K.~{Rawlins}, I.~C. {Rea}, R.~{Reimann}, B.~{Relethford},
  M.~{Relich}, E.~{Resconi}, W.~{Rhode}, M.~{Richman}, S.~{Robertson},
  M.~{Rongen}, C.~{Rott}, T.~{Ruhe}, D.~{Ryckbosch}, D.~{Rysewyk},
  T.~{S{\"a}lzer}, S.~E. {Sanchez Herrera}, A.~{Sandrock}, J.~{Sandroos},
  M.~{Santander}, S.~{Sarkar}, S.~{Sarkar}, K.~{Satalecka}, P.~{Schlunder},
  T.~{Schmidt}, A.~{Schneider}, S.~{Schoenen}, S.~{Sch{\"o}neberg},
  L.~{Schumacher}, D.~{Seckel}, S.~{Seunarine}, J.~{Soedingrekso}, D.~{Soldin},
  M.~{Song}, G.~M. {Spiczak}, C.~{Spiering}, J.~{Stachurska}, M.~{Stamatikos},
  T.~{Stanev}, A.~{Stasik}, J.~{Stettner}, A.~{Steuer}, T.~{Stezelberger},
  R.~G. {Stokstad}, A.~{St{\"o}ssl}, N.~L. {Strotjohann}, T.~{Stuttard}, G.~W.
  {Sullivan}, M.~{Sutherland}, I.~{Taboada}, J.~{Tatar}, F.~{Tenholt},
  S.~{Ter-Antonyan}, A.~{Terliuk}, G.~{Te{\v{s}}i{\'c}}, S.~{Tilav}, P.~A.
  {Toale}, M.~N. {Tobin}, S.~{Toscano}, D.~{Tosi}, M.~{Tselengidou}, C.~F.
  {Tung}, A.~{Turcati}, C.~F. {Turley}, B.~{Ty}, E.~{Unger}, M.~{Usner},
  J.~{Vandenbroucke}, W.~{Van Driessche}, N.~{van Eijndhoven}, S.~{Vanheule},
  J.~{van Santen}, M.~{Vehring}, E.~{Vogel}, M.~{Vraeghe}, C.~{Walck},
  A.~{Wallace}, M.~{Wallraff}, F.~D. {Wandler}, N.~{Wandkowsky}, A.~{Waza},
  C.~{Weaver}, M.~J. {Weiss}, C.~{Wendt}, J.~{Werthebach}, B.~J. {Whelan},
  K.~{Wiebe}, C.~H. {Wiebusch}, L.~{Wille}, D.~R. {Williams}, L.~{Wills},
  M.~{Wolf}, T.~R. {Wood}, E.~{Woolsey}, K.~{Woschnagg}, D.~L. {Xu}, X.~W.
  {Xu}, Y.~{Xu}, J.~P. {Yanez}, G.~{Yodh}, S.~{Yoshida}, T.~{Yuan}, M.~{Zoll},
  {IceCube Collaboration}, A.~{Balasubramanian}, S.~{Mate}, V.~{Bhalerao},
  D.~{Bhattacharya}, A.~{Vibhute}, G.~C. {Dewangan}, A.~R. {Rao}, S.~V.
  {Vadawale}, {AstroSat Cadmium Zinc Telluride Imager Team}, D.~S. {Svinkin},
  K.~{Hurley}, R.~L. {Aptekar}, D.~D. {Frederiks}, S.~V. {Golenetskii}, A.~V.
  {Kozlova}, A.~L. {Lysenko}, Ph.~P. {Oleynik}, A.~E. {Tsvetkova}, M.~V.
  {Ulanov}, T.~{Cline}, {IPN Collaboration}, T.~P. {Li}, S.~L. {Xiong}, S.~N.
  {Zhang}, F.~J. {Lu}, L.~M. {Song}, X.~L. {Cao}, Z.~{Chang}, G.~{Chen},
  L.~{Chen}, T.~X. {Chen}, Y.~{Chen}, Y.~B. {Chen}, Y.~P. {Chen}, W.~{Cui},
  W.~W. {Cui}, J.~K. {Deng}, Y.~W. {Dong}, Y.~Y. {Du}, M.~X. {Fu}, G.~H. {Gao},
  H.~{Gao}, M.~{Gao}, M.~Y. {Ge}, Y.~D. {Gu}, J.~{Guan}, C.~C. {Guo}, D.~W.
  {Han}, W.~{Hu}, Y.~{Huang}, J.~{Huo}, S.~M. {Jia}, L.~H. {Jiang}, W.~C.
  {Jiang}, J.~{Jin}, Y.~J. {Jin}, B.~{Li}, C.~K. {Li}, G.~{Li}, M.~S. {Li},
  W.~{Li}, X.~{Li}, X.~B. {Li}, X.~F. {Li}, Y.~G. {Li}, Z.~J. {Li}, Z.~W. {Li},
  X.~H. {Liang}, J.~Y. {Liao}, C.~Z. {Liu}, G.~Q. {Liu}, H.~W. {Liu}, S.~Z.
  {Liu}, X.~J. {Liu}, Y.~{Liu}, Y.~N. {Liu}, B.~{Lu}, X.~F. {Lu}, T.~{Luo},
  X.~{Ma}, B.~{Meng}, Y.~{Nang}, J.~Y. {Nie}, G.~{Ou}, J.~L. {Qu}, N.~{Sai},
  L.~{Sun}, Y.~{Tan}, L.~{Tao}, W.~H. {Tao}, Y.~L. {Tuo}, G.~F. {Wang}, H.~Y.
  {Wang}, J.~{Wang}, W.~S. {Wang}, Y.~S. {Wang}, X.~Y. {Wen}, B.~B. {Wu},
  M.~{Wu}, G.~C. {Xiao}, H.~{Xu}, Y.~P. {Xu}, L.~L. {Yan}, J.~W. {Yang},
  S.~{Yang}, Y.~J. {Yang}, A.~M. {Zhang}, C.~L. {Zhang}, C.~M. {Zhang},
  F.~{Zhang}, H.~M. {Zhang}, J.~{Zhang}, Q.~{Zhang}, S.~{Zhang}, T.~{Zhang},
  W.~{Zhang}, W.~C. {Zhang}, W.~Z. {Zhang}, Y.~{Zhang}, Y.~{Zhang}, Y.~F.
  {Zhang}, Y.~J. {Zhang}, Z.~{Zhang}, Z.~L. {Zhang}, H.~S. {Zhao}, J.~L.
  {Zhao}, X.~F. {Zhao}, S.~J. {Zheng}, Y.~{Zhu}, Y.~X. {Zhu}, C.~L. {Zou},
  {Insight-HXMT Collaboration}, A.~{Albert}, M.~{Andr{\'e}}, M.~{Anghinolfi},
  M.~{Ardid}, J.~J. {Aubert}, J.~{Aublin}, T.~{Avgitas}, B.~{Baret},
  J.~{Barrios-Mart{\'\i}}, S.~{Basa}, B.~{Belhorma}, V.~{Bertin}, S.~{Biagi},
  R.~{Bormuth}, S.~{Bourret}, M.~C. {Bouwhuis}, H.~{Br{\^a}nza{\c{s}}},
  R.~{Bruijn}, J.~{Brunner}, J.~{Busto}, A.~{Capone}, L.~{Caramete}, J.~{Carr},
  S.~{Celli}, R.~{Cherkaoui El Moursli}, T.~{Chiarusi}, M.~{Circella}, J.~A.~B.
  {Coelho}, A.~{Coleiro}, R.~{Coniglione}, H.~{Costantini}, P.~{Coyle},
  A.~{Creusot}, A.~F. {D{\'\i}az}, A.~{Deschamps}, G.~{De Bonis},
  C.~{Distefano}, I.~{Di Palma}, A.~{Domi}, C.~{Donzaud}, D.~{Dornic},
  D.~{Drouhin}, T.~{Eberl}, I.~{El Bojaddaini}, N.~{El Khayati},
  D.~{Els{\"a}sser}, A.~{Enzenh{\"o}fer}, A.~{Ettahiri}, F.~{Fassi},
  I.~{Felis}, L.~A. {Fusco}, P.~{Gay}, V.~{Giordano}, H.~{Glotin},
  T.~{Gr{\'e}goire}, R.~Gracia {Ruiz}, K.~{Graf}, S.~{Hallmann}, H.~{van
  Haren}, A.~J. {Heijboer}, Y.~{Hello}, J.~J. {Hern{\'a}ndez-Rey},
  J.~{H{\"o}ssl}, J.~{Hofest{\"a}dt}, C.~{Hugon}, G.~{Illuminati}, C.~W.
  {James}, M.~{de Jong}, M.~{Jongen}, M.~{Kadler}, O.~{Kalekin}, U.~{Katz},
  D.~{Kiessling}, A.~{Kouchner}, M.~{Kreter}, I.~{Kreykenbohm},
  V.~{Kulikovskiy}, C.~{Lachaud}, R.~{Lahmann}, D.~{Lef{\`e}vre}, E.~{Leonora},
  M.~{Lotze}, S.~{Loucatos}, M.~{Marcelin}, A.~{Margiotta}, A.~{Marinelli},
  J.~A. {Mart{\'\i}nez-Mora}, R.~{Mele}, K.~{Melis}, T.~{Michael},
  P.~{Migliozzi}, A.~{Moussa}, S.~{Navas}, E.~{Nezri}, M.~{Organokov}, G.~E.
  {P{\u{a}}v{\u{a}}la{\c{s}}}, C.~{Pellegrino}, C.~{Perrina}, P.~{Piattelli},
  V.~{Popa}, T.~{Pradier}, L.~{Quinn}, C.~{Racca}, G.~{Riccobene},
  A.~{S{\'a}nchez-Losa}, M.~{Salda{\~n}a}, I.~{Salvadori}, D.~F.~E.
  {Samtleben}, M.~{Sanguineti}, P.~{Sapienza}, C.~{Sieger}, M.~{Spurio}, Th.
  {Stolarczyk}, M.~{Taiuti}, Y.~{Tayalati}, A.~{Trovato}, D.~{Turpin},
  C.~{T{\"o}nnis}, B.~{Vallage}, V.~{Van Elewyck}, F.~{Versari}, D.~{Vivolo},
  A.~{Vizzoca}, J.~{Wilms}, J.~D. {Zornoza}, J.~{Z{\'u}{\~n}iga}, {ANTARES
  Collaboration}, A.~P. {Beardmore}, A.~A. {Breeveld}, D.~N. {Burrows}, S.~B.
  {Cenko}, G.~{Cusumano}, A.~{D'A{\`\i}}, M.~{de Pasquale}, S.~W.~K. {Emery},
  P.~A. {Evans}, P.~{Giommi}, C.~{Gronwall}, J.~A. {Kennea}, H.~A. {Krimm},
  N.~P.~M. {Kuin}, A.~{Lien}, F.~E. {Marshall}, A.~{Melandri}, J.~A. {Nousek},
  S.~R. {Oates}, J.~P. {Osborne}, C.~{Pagani}, K.~L. {Page}, D.~M. {Palmer},
  M.~{Perri}, M.~H. {Siegel}, B.~{Sbarufatti}, G.~{Tagliaferri},
  A.~{Tohuvavohu}, {Swift Collaboration}, M.~{Tavani}, F.~{Verrecchia},
  A.~{Bulgarelli}, Y.~{Evangelista}, L.~{Pacciani}, M.~{Feroci}, C.~{Pittori},
  A.~{Giuliani}, E.~{Del Monte}, I.~{Donnarumma}, A.~{Argan}, A.~{Trois},
  A.~{Ursi}, M.~{Cardillo}, G.~{Piano}, F.~{Longo}, F.~{Lucarelli},
  P.~{Munar-Adrover}, F.~{Fuschino}, C.~{Labanti}, M.~{Marisaldi},
  G.~{Minervini}, V.~{Fioretti}, N.~{Parmiggiani}, F.~{Gianotti},
  M.~{Trifoglio}, G.~{Di Persio}, L.~A. {Antonelli}, G.~{Barbiellini},
  P.~{Caraveo}, P.~W. {Cattaneo}, E.~{Costa}, S.~{Colafrancesco}, F.~{D'Amico},
  A.~{Ferrari}, A.~{Morselli}, F.~{Paoletti}, P.~{Picozza}, M.~{Pilia},
  A.~{Rappoldi}, P.~{Soffitta}, S.~{Vercellone}, {AGILE Team}, R.~J. {Foley},
  D.~A. {Coulter}, C.~D. {Kilpatrick}, M.~R. {Drout}, A.~L. {Piro}, B.~J.
  {Shappee}, M.~R. {Siebert}, J.~D. {Simon}, N.~{Ulloa}, D.~{Kasen}, B.~F.
  {Madore}, A.~{Murguia-Berthier}, Y.~C. {Pan}, J.~X. {Prochaska},
  E.~{Ramirez-Ruiz}, A.~{Rest}, C.~{Rojas-Bravo}, {1M2H Team}, E.~{Berger},
  M.~{Soares-Santos}, J.~{Annis}, K.~D. {Alexander}, S.~{Allam}, E.~{Balbinot},
  P.~{Blanchard}, D.~{Brout}, R.~E. {Butler}, R.~{Chornock}, E.~R. {Cook},
  P.~{Cowperthwaite}, H.~T. {Diehl}, A.~{Drlica-Wagner}, M.~R. {Drout},
  F.~{Durret}, T.~{Eftekhari}, D.~A. {Finley}, W.~{Fong}, J.~A. {Frieman},
  C.~L. {Fryer}, J.~{Garc{\'\i}a-Bellido}, R.~A. {Gruendl}, W.~{Hartley},
  K.~{Herner}, R.~{Kessler}, H.~{Lin}, P.~A.~A. {Lopes}, A.~C.~C.
  {Louren{\c{c}}o}, R.~{Margutti}, J.~L. {Marshall}, T.~{Matheson}, G.~E.
  {Medina}, B.~D. {Metzger}, R.~R. {Mu{\~n}oz}, J.~{Muir}, M.~{Nicholl},
  P.~{Nugent}, A.~{Palmese}, F.~{Paz-Chinch{\'o}n}, E.~{Quataert}, M.~{Sako},
  M.~{Sauseda}, D.~J. {Schlegel}, D.~{Scolnic}, L.~F. {Secco}, N.~{Smith},
  F.~{Sobreira}, V.~A. {Villar}, A.~K. {Vivas}, W.~{Wester}, P.~K.~G.
  {Williams}, B.~{Yanny}, A.~{Zenteno}, Y.~{Zhang}, T.~M.~C. {Abbott},
  M.~{Banerji}, K.~{Bechtol}, A.~{Benoit-L{\'e}vy}, E.~{Bertin}, D.~{Brooks},
  E.~{Buckley-Geer}, D.~L. {Burke}, D.~{Capozzi}, A.~{Carnero Rosell},
  M.~{Carrasco Kind}, F.~J. {Castander}, M.~{Crocce}, C.~E. {Cunha}, C.~B.
  {D'Andrea}, L.~N. {da Costa}, C.~{Davis}, D.~L. {DePoy}, S.~{Desai}, J.~P.
  {Dietrich}, T.~F. {Eifler}, E.~{Fernandez}, B.~{Flaugher}, P.~{Fosalba},
  E.~{Gaztanaga}, D.~W. {Gerdes}, T.~{Giannantonio}, D.~A. {Goldstein},
  D.~{Gruen}, J.~{Gschwend}, G.~{Gutierrez}, K.~{Honscheid}, D.~J. {James},
  T.~{Jeltema}, M.~W.~G. {Johnson}, M.~D. {Johnson}, S.~{Kent}, E.~{Krause},
  R.~{Kron}, K.~{Kuehn}, O.~{Lahav}, M.~{Lima}, M.~A.~G. {Maia}, M.~{March},
  P.~{Martini}, R.~G. {McMahon}, F.~{Menanteau}, C.~J. {Miller}, R.~{Miquel},
  J.~J. {Mohr}, R.~C. {Nichol}, R.~L.~C. {Ogando}, A.~A. {Plazas}, A.~K.
  {Romer}, A.~{Roodman}, E.~S. {Rykoff}, E.~{Sanchez}, V.~{Scarpine},
  R.~{Schindler}, M.~{Schubnell}, I.~{Sevilla-Noarbe}, E.~{Sheldon},
  M.~{Smith}, R.~C. {Smith}, A.~{Stebbins}, E.~{Suchyta}, M.~E.~C. {Swanson},
  G.~{Tarle}, R.~C. {Thomas}, M.~A. {Troxel}, D.~L. {Tucker}, V.~{Vikram},
  A.~R. {Walker}, R.~H. {Wechsler}, J.~{Weller}, J.~L. {Carlin}, M.~S.~S.
  {Gill}, T.~S. {Li}, J.~{Marriner}, E.~{Neilsen}, {Dark Energy Camera GW-EM
  Collaboration}, {DES Collaboration}, J.~B. {Haislip}, V.~V. {Kouprianov},
  D.~E. {Reichart}, D.~J. {Sand}, L.~{Tartaglia}, S.~{Valenti}, S.~{Yang},
  {DLT40 Collaboration}, S.~{Benetti}, E.~{Brocato}, S.~{Campana},
  E.~{Cappellaro}, S.~{Covino}, P.~{D'Avanzo}, V.~{D'Elia}, F.~{Getman},
  G.~{Ghirlanda}, G.~{Ghisellini}, L.~{Limatola}, L.~{Nicastro}, E.~{Palazzi},
  E.~{Pian}, S.~{Piranomonte}, A.~{Possenti}, A.~{Rossi}, O.~S. {Salafia},
  L.~{Tomasella}, L.~{Amati}, L.~A. {Antonelli}, M.~G. {Bernardini},
  F.~{Bufano}, M.~{Capaccioli}, P.~{Casella}, M.~{Dadina}, G.~{De Cesare},
  A.~{Di Paola}, G.~{Giuffrida}, A.~{Giunta}, G.~L. {Israel}, M.~{Lisi},
  E.~{Maiorano}, M.~{Mapelli}, N.~{Masetti}, A.~{Pescalli}, L.~{Pulone},
  R.~{Salvaterra}, P.~{Schipani}, M.~{Spera}, A.~{Stamerra}, L.~{Stella},
  V.~{Testa}, M.~{Turatto}, D.~{Vergani}, G.~{Aresu}, M.~{Bachetti},
  F.~{Buffa}, M.~{Burgay}, M.~{Buttu}, T.~{Caria}, E.~{Carretti},
  V.~{Casasola}, P.~{Castangia}, G.~{Carboni}, S.~{Casu}, R.~{Concu},
  A.~{Corongiu}, G.~L. {Deiana}, E.~{Egron}, A.~{Fara}, F.~{Gaudiomonte},
  V.~{Gusai}, A.~{Ladu}, S.~{Loru}, S.~{Leurini}, L.~{Marongiu}, A.~{Melis},
  G.~{Melis}, Carlo {Migoni}, Sabrina {Milia}, Alessandro {Navarrini},
  A.~{Orlati}, P.~{Ortu}, S.~{Palmas}, A.~{Pellizzoni}, D.~{Perrodin},
  T.~{Pisanu}, S.~{Poppi}, S.~{Righini}, A.~{Saba}, G.~{Serra}, M.~{Serrau},
  M.~{Stagni}, G.~{Surcis}, V.~{Vacca}, G.~P. {Vargiu}, L.~K. {Hunt}, Z.~P.
  {Jin}, S.~{Klose}, C.~{Kouveliotou}, P.~A. {Mazzali}, P.~{M{\o}ller},
  L.~{Nava}, T.~{Piran}, J.~{Selsing}, S.~D. {Vergani}, K.~{Wiersema},
  K.~{Toma}, A.~B. {Higgins}, C.~G. {Mundell}, S.~{di Serego Alighieri},
  D.~{G{\'o}tz}, W.~{Gao}, A.~{Gomboc}, L.~{Kaper}, S.~{Kobayashi}, D.~{Kopac},
  J.~{Mao}, R.~L.~C. {Starling}, I.~{Steele}, A.~J. {van der Horst}, {GRAWITA:
  GRAvitational Wave Inaf TeAm}, F.~{Acero}, W.~B. {Atwood}, L.~{Baldini},
  G.~{Barbiellini}, D.~{Bastieri}, B.~{Berenji}, R.~{Bellazzini},
  E.~{Bissaldi}, R.~D. {Blandford}, E.~D. {Bloom}, R.~{Bonino}, E.~{Bottacini},
  J.~{Bregeon}, R.~{Buehler}, S.~{Buson}, R.~A. {Cameron}, R.~{Caputo}, P.~A.
  {Caraveo}, E.~{Cavazzuti}, A.~{Chekhtman}, C.~C. {Cheung}, J.~{Chiang},
  S.~{Ciprini}, J.~{Cohen-Tanugi}, L.~R. {Cominsky}, D.~{Costantin},
  A.~{Cuoco}, F.~{D'Ammando}, F.~{de Palma}, S.~W. {Digel}, N.~{Di Lalla},
  M.~{Di Mauro}, L.~{Di Venere}, R.~{Dubois}, S.~J. {Fegan}, W.~B. {Focke},
  A.~{Franckowiak}, Y.~{Fukazawa}, S.~{Funk}, P.~{Fusco}, F.~{Gargano},
  D.~{Gasparrini}, N.~{Giglietto}, F.~{Giordano}, M.~{Giroletti},
  T.~{Glanzman}, D.~{Green}, M.~H. {Grondin}, L.~{Guillemot}, S.~{Guiriec},
  A.~K. {Harding}, D.~{Horan}, G.~{J{\'o}hannesson}, T.~{Kamae}, S.~{Kensei},
  M.~{Kuss}, G.~{La Mura}, L.~{Latronico}, M.~{Lemoine-Goumard}, F.~{Longo},
  F.~{Loparco}, M.~N. {Lovellette}, P.~{Lubrano}, J.~D. {Magill}, S.~{Maldera},
  A.~{Manfreda}, M.~N. {Mazziotta}, J.~E. {McEnery}, M.~{Meyer}, P.~F.
  {Michelson}, N.~{Mirabal}, M.~E. {Monzani}, E.~{Moretti}, A.~{Morselli},
  I.~V. {Moskalenko}, M.~{Negro}, E.~{Nuss}, R.~{Ojha}, N.~{Omodei},
  M.~{Orienti}, E.~{Orlando}, M.~{Palatiello}, V.~S. {Paliya}, D.~{Paneque},
  M.~{Pesce-Rollins}, F.~{Piron}, T.~A. {Porter}, G.~{Principe},
  S.~{Rain{\`o}}, R.~{Rando}, M.~{Razzano}, S.~{Razzaque}, A.~{Reimer},
  O.~{Reimer}, T.~{Reposeur}, L.~S. {Rochester}, P.~M. {Saz Parkinson},
  C.~{Sgr{\`o}}, E.~J. {Siskind}, F.~{Spada}, G.~{Spandre}, D.~J. {Suson},
  M.~{Takahashi}, Y.~{Tanaka}, J.~G. {Thayer}, J.~B. {Thayer}, D.~J.
  {Thompson}, L.~{Tibaldo}, D.~F. {Torres}, E.~{Torresi}, E.~{Troja}, T.~M.
  {Venters}, G.~{Vianello}, G.~{Zaharijas}, {Fermi Large Area Telescope
  Collaboration}, J.~R. {Allison}, K.~W. {Bannister}, D.~{Dobie}, D.~L.
  {Kaplan}, E.~{Lenc}, C.~{Lynch}, T.~{Murphy}, E.~M. {Sadler}, ATCA:
  {Australia Telescope Compact Array}, A.~{Hotan}, C.~W. {James},
  S.~{Oslowski}, W.~{Raja}, R.~M. {Shannon}, M.~{Whiting}, ASKAP: {Australian
  SKA Pathfinder}, I.~{Arcavi}, D.~A. {Howell}, C.~{McCully},
  G.~{Hosseinzadeh}, D.~{Hiramatsu}, D.~{Poznanski}, J.~{Barnes},
  M.~{Zaltzman}, S.~{Vasylyev}, D.~{Maoz}, {Las Cumbres Observatory Group},
  J.~{Cooke}, M.~{Bailes}, C.~{Wolf}, A.~T. {Deller}, C.~{Lidman}, L.~{Wang},
  B.~{Gendre}, I.~{Andreoni}, K.~{Ackley}, T.~A. {Pritchard}, M.~S. {Bessell},
  S.~W. {Chang}, A.~{M{\"o}ller}, C.~A. {Onken}, R.~A. {Scalzo},
  R.~{Ridden-Harper}, R.~G. {Sharp}, B.~E. {Tucker}, T.~J. {Farrell},
  E.~{Elmer}, S.~{Johnston}, V.~{Venkatraman Krishnan}, E.~F. {Keane}, J.~A.
  {Green}, A.~{Jameson}, L.~{Hu}, B.~{Ma}, T.~{Sun}, X.~{Wu}, X.~{Wang},
  Z.~{Shang}, Y.~{Hu}, M.~C.~B. {Ashley}, X.~{Yuan}, X.~{Li}, C.~{Tao},
  Z.~{Zhu}, H.~{Zhang}, N.~B. {Suntzeff}, J.~{Zhou}, J.~{Yang}, B.~{Orange},
  D.~{Morris}, A.~{Cucchiara}, T.~{Giblin}, A.~{Klotz}, J.~{Staff},
  P.~{Thierry}, B.~P. {Schmidt}, {OzGrav}, DWF {(Deeper}, {Wider}, Faster
  {program}, {AST3}, {CAASTRO Collaborations}, N.~R. {Tanvir}, A.~J. {Levan},
  Z.~{Cano}, A.~{de Ugarte-Postigo}, C.~{Gonz{\'a}lez-Fern{\'a}ndez},
  J.~{Greiner}, J.~{Hjorth}, M.~{Irwin}, T.~{Kr{\"u}hler}, I.~{Mandel},
  B.~{Milvang-Jensen}, P.~{O'Brien}, E.~{Rol}, S.~{Rosetti}, S.~{Rosswog},
  A.~{Rowlinson}, D.~T.~H. {Steeghs}, C.~C. {Th{\"o}ne}, K.~{Ulaczyk},
  D.~{Watson}, S.~H. {Bruun}, R.~{Cutter}, R.~{Figuera Jaimes}, Y.~I. {Fujii},
  A.~S. {Fruchter}, B.~{Gompertz}, P.~{Jakobsson}, G.~{Hodosan}, U.~G.
  {J{\`e}rgensen}, T.~{Kangas}, D.~A. {Kann}, M.~{Rabus}, S.~L. {Schr{\o}der},
  E.~R. {Stanway}, R.~A.~M.~J. {Wijers}, {VINROUGE Collaboration}, V.~M.
  {Lipunov}, E.~S. {Gorbovskoy}, V.~G. {Kornilov}, N.~V. {Tyurina}, P.~V.
  {Balanutsa}, A.~S. {Kuznetsov}, D.~M. {Vlasenko}, R.~C. {Podesta},
  C.~{Lopez}, F.~{Podesta}, H.~O. {Levato}, C.~{Saffe}, C.~C. {Mallamaci},
  N.~M. {Budnev}, O.~A. {Gress}, D.~A. {Kuvshinov}, I.~A. {Gorbunov}, V.~V.
  {Vladimirov}, D.~S. {Zimnukhov}, A.~V. {Gabovich}, V.~V. {Yurkov}, Yu.~P.
  {Sergienko}, R.~{Rebolo}, M.~{Serra-Ricart}, A.~G. {Tlatov}, Yu.~V.
  {Ishmuhametova}, {MASTER Collaboration}, F.~{Abe}, K.~{Aoki}, W.~{Aoki},
  Y.~{Asakura}, S.~{Baar}, S.~{Barway}, I.~A. {Bond}, M.~{Doi}, F.~{Finet},
  T.~{Fujiyoshi}, H.~{Furusawa}, S.~{Honda}, R.~{Itoh}, N.~{Kanda}, K.~S.
  {Kawabata}, M.~{Kawabata}, J.~H. {Kim}, S.~{Koshida}, D.~{Kuroda}, C.~H.
  {Lee}, W.~{Liu}, K.~{Matsubayashi}, S.~{Miyazaki}, K.~{Morihana},
  T.~{Morokuma}, K.~{Motohara}, K.~L. {Murata}, H.~{Nagai}, H.~{Nagashima},
  T.~{Nagayama}, T.~{Nakaoka}, F.~{Nakata}, R.~{Ohsawa}, T.~{Ohshima},
  K.~{Ohta}, H.~{Okita}, T.~{Saito}, Y.~{Saito}, S.~{Sako}, Y.~{Sekiguchi},
  T.~{Sumi}, A.~{Tajitsu}, J.~{Takahashi}, M.~{Takayama}, Y.~{Tamura},
  I.~{Tanaka}, M.~{Tanaka}, T.~{Terai}, N.~{Tominaga}, P.~J. {Tristram},
  M.~{Uemura}, Y.~{Utsumi}, M.~S. {Yamaguchi}, N.~{Yasuda}, M.~{Yoshida},
  T.~{Zenko}, {J-GEM}, S.~M. {Adams}, G.~C. {Anupama}, J.~{Bally}, S.~{Barway},
  E.~{Bellm}, N.~{Blagorodnova}, C.~{Cannella}, P.~{Chandra}, D.~{Chatterjee},
  T.~E. {Clarke}, B.~E. {Cobb}, D.~O. {Cook}, C.~{Copperwheat}, K.~{De},
  S.~W.~K. {Emery}, U.~{Feindt}, K.~{Foster}, O.~D. {Fox}, D.~A. {Frail},
  C.~{Fremling}, C.~{Frohmaier}, J.~A. {Garcia}, S.~{Ghosh}, S.~{Giacintucci},
  A.~{Goobar}, O.~{Gottlieb}, B.~W. {Grefenstette}, G.~{Hallinan},
  F.~{Harrison}, M.~{Heida}, G.~{Helou}, A.~Y.~Q. {Ho}, A.~{Horesh},
  K.~{Hotokezaka}, W.~H. {Ip}, R.~{Itoh}, Bob {Jacobs}, J.~E. {Jencson},
  D.~{Kasen}, M.~M. {Kasliwal}, N.~E. {Kassim}, H.~{Kim}, B.~S. {Kiran},
  N.~P.~M. {Kuin}, S.~R. {Kulkarni}, T.~{Kupfer}, R.~M. {Lau}, K.~{Madsen},
  P.~A. {Mazzali}, A.~A. {Miller}, H.~{Miyasaka}, K.~{Mooley}, S.~T. {Myers},
  E.~{Nakar}, C.~C. {Ngeow}, P.~{Nugent}, E.~O. {Ofek}, N.~{Palliyaguru},
  M.~{Pavana}, D.~A. {Perley}, W.~M. {Peters}, S.~{Pike}, T.~{Piran}, H.~{Qi},
  R.~M. {Quimby}, J.~{Rana}, S.~{Rosswog}, F.~{Rusu}, E.~M. {Sadler}, A.~{Van
  Sistine}, J.~{Sollerman}, Y.~{Xu}, L.~{Yan}, Y.~{Yatsu}, P.~C. {Yu},
  C.~{Zhang}, W.~{Zhao}, {GROWTH}, {JAGWAR}, {Caltech-NRAO}, {TTU-NRAO},
  {NuSTAR Collaborations}, K.~C. {Chambers}, M.~E. {Huber}, A.~S.~B. {Schultz},
  J.~{Bulger}, H.~{Flewelling}, E.~A. {Magnier}, T.~B. {Lowe}, R.~J.
  {Wainscoat}, C.~{Waters}, M.~{Willman}, {Pan-STARRS}, K.~{Ebisawa},
  C.~{Hanyu}, S.~{Harita}, T.~{Hashimoto}, K.~{Hidaka}, T.~{Hori},
  M.~{Ishikawa}, N.~{Isobe}, W.~{Iwakiri}, H.~{Kawai}, N.~{Kawai},
  T.~{Kawamuro}, T.~{Kawase}, Y.~{Kitaoka}, K.~{Makishima}, M.~{Matsuoka},
  T.~{Mihara}, T.~{Morita}, K.~{Morita}, S.~{Nakahira}, M.~{Nakajima},
  Y.~{Nakamura}, H.~{Negoro}, S.~{Oda}, A.~{Sakamaki}, R.~{Sasaki},
  M.~{Serino}, M.~{Shidatsu}, R.~{Shimomukai}, Y.~{Sugawara}, S.~{Sugita},
  M.~{Sugizaki}, Y.~{Tachibana}, Y.~{Takao}, A.~{Tanimoto}, H.~{Tomida},
  Y.~{Tsuboi}, H.~{Tsunemi}, Y.~{Ueda}, S.~{Ueno}, S.~{Yamada}, K.~{Yamaoka},
  M.~{Yamauchi}, F.~{Yatabe}, T.~{Yoneyama}, T.~{Yoshii}, {MAXI Team}, D.~M.
  {Coward}, H.~{Crisp}, D.~{Macpherson}, I.~{Andreoni}, R.~{Laugier},
  K.~{Noysena}, A.~{Klotz}, B.~{Gendre}, P.~{Thierry}, D.~{Turpin}, TZAC
  {Consortium}, M.~{Im}, C.~{Choi}, J.~{Kim}, Y.~{Yoon}, G.~{Lim}, S.~K. {Lee},
  C.~U. {Lee}, S.~L. {Kim}, S.~W. {Ko}, J.~{Joe}, M.~K. {Kwon}, P.~J. {Kim},
  S.~K. {Lim}, J.~S. {Choi}, {KU Collaboration}, J.~P.~U. {Fynbo},
  D.~{Malesani}, D.~{Xu}, Nordic {Optical Telescope}, S.~J. {Smartt},
  A.~{Jerkstrand}, E.~{Kankare}, S.~A. {Sim}, M.~{Fraser}, C.~{Inserra},
  K.~{Maguire}, G.~{Leloudas}, M.~{Magee}, L.~J. {Shingles}, K.~W. {Smith},
  D.~R. {Young}, R.~{Kotak}, A.~{Gal-Yam}, J.~D. {Lyman}, D.~S. {Homan},
  C.~{Agliozzo}, J.~P. {Anderson}, C.~R. {Angus}, C.~{Ashall}, C.~{Barbarino},
  F.~E. {Bauer}, M.~{Berton}, M.~T. {Botticella}, M.~{Bulla}, G.~{Cannizzaro},
  R.~{Cartier}, A.~{Cikota}, P.~{Clark}, A.~{De Cia}, M.~{Della Valle},
  M.~{Dennefeld}, L.~{Dessart}, G.~{Dimitriadis}, N.~{Elias-Rosa}, R.~E.
  {Firth}, A.~{Fl{\"o}rs}, C.~{Frohmaier}, L.~{Galbany},
  S.~{Gonz{\'a}lez-Gait{\'a}n}, M.~{Gromadzki}, C.~P. {Guti{\'e}rrez},
  A.~{Hamanowicz}, J.~{Harmanen}, K.~E. {Heintz}, M.~S. {Hernandez}, S.~T.
  {Hodgkin}, I.~M. {Hook}, L.~{Izzo}, P.~A. {James}, P.~G. {Jonker}, W.~E.
  {Kerzendorf}, Z.~{Kostrzewa-Rutkowska}, M.~{Kromer}, H.~{Kuncarayakti},
  A.~{Lawrence}, I.~{Manulis}, S.~{Mattila}, O.~{McBrien}, A.~{M{\"u}ller},
  J.~{Nordin}, D.~{O'Neill}, F.~{Onori}, J.~T. {Palmerio}, A.~{Pastorello},
  F.~{Patat}, G.~{Pignata}, P.~{Podsiadlowski}, A.~{Razza}, T.~{Reynolds},
  R.~{Roy}, A.~J. {Ruiter}, K.~A. {Rybicki}, L.~{Salmon}, M.~L. {Pumo}, S.~J.
  {Prentice}, I.~R. {Seitenzahl}, M.~{Smith}, J.~{Sollerman}, M.~{Sullivan},
  H.~{Szegedi}, F.~{Taddia}, S.~{Taubenberger}, G.~{Terreran}, B.~{Van Soelen},
  J.~{Vos}, N.~A. {Walton}, D.~E. {Wright}, {\L}.~{Wyrzykowski}, O.~{Yaron},
  <author {pre=''(''>ePESSTO}, T.~W. {Chen}, T.~{Kr{\"u}hler}, P.~{Schady},
  P.~{Wiseman}, J.~{Greiner}, A.~{Rau}, T.~{Schweyer}, S.~{Klose}, A.~{Nicuesa
  Guelbenzu}, {GROND}, N.~T. {Palliyaguru}, Texas {Tech University}, M.~M.
  {Shara}, T.~{Williams}, P.~{Vaisanen}, S.~B. {Potter}, E.~{Romero Colmenero},
  S.~{Crawford}, D.~A.~H. {Buckley}, J.~{Mao}, {SALT Group}, M.~C. {D{\'\i}az},
  L.~M. {Macri}, D.~{Garc{\'\i}a Lambas}, C.~{Mendes de Oliveira}, J.~L. {Nilo
  Castell{\'o}n}, T.~{Ribeiro}, B.~{S{\'a}nchez}, W.~{Schoenell}, L.~R.
  {Abramo}, S.~{Akras}, J.~S. {Alcaniz}, R.~{Artola}, M.~{Beroiz}, S.~{Bonoli},
  J.~{Cabral}, R.~{Camuccio}, V.~{Chavushyan}, P.~{Coelho}, C.~{Colazo}, M.~V.
  {Costa-Duarte}, H.~{Cuevas Larenas}, M.~{Dom{\'\i}nguez Romero},
  D.~{Dultzin}, D.~{Fern{\'a}ndez}, J.~{Garc{\'\i}a}, C.~{Girardini}, D.~R.
  {Gon{\c{c}}alves}, T.~S. {Gon{\c{c}}alves}, S.~{Gurovich},
  Y.~{Jim{\'e}nez-Teja}, A.~{Kanaan}, M.~{Lares}, R.~{Lopes de Oliveira},
  O.~{L{\'o}pez-Cruz}, R.~{Melia}, A.~{Molino}, N.~{Padilla}, T.~{Pe{\~n}uela},
  V.~M. {Placco}, C.~{Qui{\~n}ones}, A.~{Ram{\'\i}rez Rivera}, V.~{Renzi},
  L.~{Riguccini}, E.~{R{\'\i}os-L{\'o}pez}, H.~{Rodriguez}, L.~{Sampedro},
  M.~{Schneiter}, L.~{Sodr{\'e}}, M.~{Starck}, S.~{Torres-Flores},
  M.~{Tornatore}, A.~{Zadro{\.z}ny}, M.~{Castillo}, {TOROS: Transient Robotic
  Observatory of South Collaboration}, A.~J. {Castro-Tirado}, J.~C. {Tello},
  Y.~D. {Hu}, B.~B. {Zhang}, R.~{Cunniffe}, A.~{Castell{\'o}n}, D.~{Hiriart},
  M.~D. {Caballero-Garc{\'\i}a}, M.~{Jel{\'\i}nek}, P.~{Kub{\'a}nek},
  C.~{P{\'e}rez del Pulgar}, I.~H. {Park}, S.~{Jeong}, J.~M. {Castro
  Cer{\'o}n}, S.~B. {Pandey}, P.~C. {Yock}, R.~{Querel}, Y.~{Fan}, C.~{Wang},
  {BOOTES Collaboration}, A.~{Beardsley}, I.~S. {Brown}, B.~{Crosse},
  D.~{Emrich}, T.~{Franzen}, B.~M. {Gaensler}, L.~{Horsley},
  M.~{Johnston-Hollitt}, D.~{Kenney}, M.~F. {Morales}, D.~{Pallot},
  M.~{Sokolowski}, K.~{Steele}, S.~J. {Tingay}, C.~M. {Trott}, M.~{Walker},
  R.~{Wayth}, A.~{Williams}, C.~{Wu}, MWA: {Murchison Widefield Array},
  A.~{Yoshida}, T.~{Sakamoto}, Y.~{Kawakubo}, K.~{Yamaoka}, I.~{Takahashi},
  Y.~{Asaoka}, S.~{Ozawa}, S.~{Torii}, Y.~{Shimizu}, T.~{Tamura},
  W.~{Ishizaki}, M.~L. {Cherry}, S.~{Ricciarini}, A.~V. {Penacchioni}, P.~S.
  {Marrocchesi}, {CALET Collaboration}, A.~S. {Pozanenko}, A.~A. {Volnova},
  E.~D. {Mazaeva}, P.~Yu. {Minaev}, M.~A. {Krugov}, A.~V. {Kusakin}, I.~V.
  {Reva}, A.~S. {Moskvitin}, V.~V. {Rumyantsev}, R.~{Inasaridze}, E.~V.
  {Klunko}, N.~{Tungalag}, S.~E. {Schmalz}, O.~{Burhonov}, {IKI-GW Follow-up
  Collaboration}, H.~{Abdalla}, A.~{Abramowski}, F.~{Aharonian}, F.~{Ait
  Benkhali}, E.~O. {Ang{\"u}ner}, M.~{Arakawa}, M.~{Arrieta}, P.~{Aubert},
  M.~{Backes}, A.~{Balzer}, M.~{Barnard}, Y.~{Becherini}, J.~{Becker Tjus},
  D.~{Berge}, S.~{Bernhard}, K.~{Bernl{\"o}hr}, R.~{Blackwell},
  M.~{B{\"o}ttcher}, C.~{Boisson}, J.~{Bolmont}, S.~{Bonnefoy}, P.~{Bordas},
  J.~{Bregeon}, F.~{Brun}, P.~{Brun}, M.~{Bryan}, M.~{B{\"u}chele}, T.~{Bulik},
  M.~{Capasso}, S.~{Caroff}, A.~{Carosi}, S.~{Casanova}, M.~{Cerruti},
  N.~{Chakraborty}, R.~C.~G. {Chaves}, A.~{Chen}, J.~{Chevalier},
  S.~{Colafrancesco}, B.~{Condon}, J.~{Conrad}, I.~D. {Davids}, J.~{Decock},
  C.~{Deil}, J.~{Devin}, P.~{deWilt}, L.~{Dirson}, A.~{Djannati-Ata{\"\i}},
  A.~{Donath}, L.~{O'C. Drury}, K.~{Dutson}, J.~{Dyks}, T.~{Edwards},
  K.~{Egberts}, G.~{Emery}, J.~P. {Ernenwein}, S.~{Eschbach}, C.~{Farnier},
  S.~{Fegan}, M.~V. {Fernandes}, A.~{Fiasson}, G.~{Fontaine}, S.~{Funk},
  M.~{F{\"u}ssling}, S.~{Gabici}, Y.~A. {Gallant}, T.~{Garrigoux},
  F.~{Gat{\'e}}, G.~{Giavitto}, B.~{Giebels}, D.~{Glawion}, J.~F.
  {Glicenstein}, D.~{Gottschall}, M.~H. {Grondin}, J.~{Hahn}, M.~{Haupt},
  J.~{Hawkes}, G.~{Heinzelmann}, G.~{Henri}, G.~{Hermann}, J.~A. {Hinton},
  W.~{Hofmann}, C.~{Hoischen}, T.~L. {Holch}, M.~{Holler}, D.~{Horns},
  A.~{Ivascenko}, H.~{Iwasaki}, A.~{Jacholkowska}, M.~{Jamrozy},
  D.~{Jankowsky}, F.~{Jankowsky}, M.~{Jingo}, L.~{Jouvin}, I.~{Jung-Richardt},
  M.~A. {Kastendieck}, K.~{Katarzy{\'n}ski}, M.~{Katsuragawa}, D.~{Kerszberg},
  D.~{Khangulyan}, B.~{Kh{\'e}lifi}, J.~{King}, S.~{Klepser}, D.~{Klochkov},
  W.~{Klu{\'z}niak}, Nu. {Komin}, K.~{Kosack}, S.~{Krakau}, M.~{Kraus}, P.~P.
  {Kr{\"u}ger}, H.~{Laffon}, G.~{Lamanna}, J.~{Lau}, J.~P. {Lees},
  J.~{Lefaucheur}, A.~{Lemi{\`e}re}, M.~{Lemoine-Goumard}, J.~P. {Lenain},
  E.~{Leser}, T.~{Lohse}, M.~{Lorentz}, R.~{Liu}, I.~{Lypova}, D.~{Malyshev},
  V.~{Marandon}, A.~{Marcowith}, C.~{Mariaud}, R.~{Marx}, G.~{Maurin},
  N.~{Maxted}, M.~{Mayer}, P.~J. {Meintjes}, M.~{Meyer}, A.~M.~W. {Mitchell},
  R.~{Moderski}, M.~{Mohamed}, L.~{Mohrmann}, K.~{Mor{\r{a}}}, E.~{Moulin},
  T.~{Murach}, S.~{Nakashima}, M.~{de Naurois}, H.~{Ndiyavala},
  F.~{Niederwanger}, J.~{Niemiec}, L.~{Oakes}, P.~{O'Brien}, H.~{Odaka},
  S.~{Ohm}, M.~{Ostrowski}, I.~{Oya}, M.~{Padovani}, M.~{Panter}, R.~D.
  {Parsons}, N.~W. {Pekeur}, G.~{Pelletier}, C.~{Perennes}, P.~O. {Petrucci},
  B.~{Peyaud}, Q.~{Piel}, S.~{Pita}, V.~{Poireau}, H.~{Poon}, D.~{Prokhorov},
  H.~{Prokoph}, G.~{P{\"u}hlhofer}, M.~{Punch}, A.~{Quirrenbach}, S.~{Raab},
  R.~{Rauth}, A.~{Reimer}, O.~{Reimer}, M.~{Renaud}, R.~{de los Reyes},
  F.~{Rieger}, L.~{Rinchiuso}, C.~{Romoli}, G.~{Rowell}, B.~{Rudak}, C.~B.
  {Rulten}, V.~{Sahakian}, S.~{Saito}, D.~A. {Sanchez}, A.~{Santangelo},
  M.~{Sasaki}, R.~{Schlickeiser}, F.~{Sch{\"u}ssler}, A.~{Schulz},
  U.~{Schwanke}, S.~{Schwemmer}, M.~{Seglar-Arroyo}, M.~{Settimo}, A.~S.
  {Seyffert}, N.~{Shafi}, I.~{Shilon}, K.~{Shiningayamwe}, R.~{Simoni},
  H.~{Sol}, F.~{Spanier}, M.~{Spir-Jacob}, {\L}.~{Stawarz}, R.~{Steenkamp},
  C.~{Stegmann}, C.~{Steppa}, I.~{Sushch}, T.~{Takahashi}, J.~P. {Tavernet},
  T.~{Tavernier}, A.~M. {Taylor}, R.~{Terrier}, L.~{Tibaldo}, D.~{Tiziani},
  M.~{Tluczykont}, C.~{Trichard}, M.~{Tsirou}, N.~{Tsuji}, R.~{Tuffs},
  Y.~{Uchiyama}, D.~J. {van der Walt}, C.~{van Eldik}, C.~{van Rensburg},
  B.~{van Soelen}, G.~{Vasileiadis}, J.~{Veh}, C.~{Venter}, A.~{Viana},
  P.~{Vincent}, J.~{Vink}, F.~{Voisin}, H.~J. {V{\"o}lk}, T.~{Vuillaume},
  Z.~{Wadiasingh}, S.~J. {Wagner}, P.~{Wagner}, R.~M. {Wagner}, R.~{White},
  A.~{Wierzcholska}, P.~{Willmann}, A.~{W{\"o}rnlein}, D.~{Wouters}, R.~{Yang},
  D.~{Zaborov}, M.~{Zacharias}, R.~{Zanin}, A.~A. {Zdziarski}, A.~{Zech},
  F.~{Zefi}, A.~{Ziegler}, J.~{Zorn}, N.~{{\.Z}ywucka}, {H.~E.~S.~S.
  Collaboration}, R.~P. {Fender}, J.~W. {Broderick}, A.~{Rowlinson},
  R.~A.~M.~J. {Wijers}, A.~J. {Stewart}, S.~{ter Veen}, A.~{Shulevski}, {LOFAR
  Collaboration}, M.~{Kavic}, J.~H. {Simonetti}, C.~{League}, J.~{Tsai}, K.~S.
  {Obenberger}, K.~{Nathaniel}, G.~B. {Taylor}, J.~D. {Dowell}, S.~L.
  {Liebling}, J.~A. {Estes}, M.~{Lippert}, I.~{Sharma}, P.~{Vincent},
  B.~{Farella}, LWA:~Long {Wavelength Array}, A.~U. {Abeysekara}, A.~{Albert},
  R.~{Alfaro}, C.~{Alvarez}, R.~{Arceo}, J.~C. {Arteaga-Vel{\'a}zquez},
  D.~{Avila Rojas}, H.~A. {Ayala Solares}, A.~S. {Barber}, J.~{Becerra
  Gonzalez}, A.~{Becerril}, E.~{Belmont-Moreno}, S.~Y. {BenZvi}, D.~{Berley},
  A.~{Bernal}, J.~{Braun}, C.~{Brisbois}, K.~S. {Caballero-Mora},
  T.~{Capistr{\'a}n}, A.~{Carrami{\~n}ana}, S.~{Casanova}, M.~{Castillo},
  U.~{Cotti}, J.~{Cotzomi}, S.~{Couti{\~n}o de Le{\'o}n}, C.~{De Le{\'o}n},
  E.~{De la Fuente}, R.~{Diaz Hernandez}, S.~{Dichiara}, B.~L. {Dingus}, M.~A.
  {DuVernois}, J.~C. {D{\'\i}az-V{\'e}lez}, R.~W. {Ellsworth}, K.~{Engel},
  O.~{Enr{\'\i}quez-Rivera}, D.~W. {Fiorino}, H.~{Fleischhack}, N.~{Fraija},
  J.~A. {Garc{\'\i}a-Gonz{\'a}lez}, F.~{Garfias}, M.~{Gerhardt},
  A.~{Gonz{\~o}lez Mu{\~n}oz}, M.~M. {Gonz{\'a}lez}, J.~A. {Goodman},
  Z.~{Hampel-Arias}, J.~P. {Harding}, S.~{Hernandez}, A.~{Hernandez-Almada},
  B.~{Hona}, P.~{H{\"u}ntemeyer}, A.~{Iriarte}, A.~{Jardin-Blicq}, V.~{Joshi},
  S.~{Kaufmann}, D.~{Kieda}, A.~{Lara}, R.~J. {Lauer}, D.~{Lennarz},
  H.~{Le{\'o}n Vargas}, J.~T. {Linnemann}, A.~L. {Longinotti}, G.~Luis {Raya},
  R.~{Luna-Garc{\'\i}a}, R.~{L{\'o}pez-Coto}, K.~{Malone}, S.~S. {Marinelli},
  O.~{Martinez}, I.~{Martinez-Castellanos}, J.~{Mart{\'\i}nez-Castro},
  H.~{Mart{\'\i}nez-Huerta}, J.~A. {Matthews}, P.~{Miranda-Romagnoli},
  E.~{Moreno}, M.~{Mostaf{\'a}}, L.~{Nellen}, M.~{Newbold}, M.~U. {Nisa},
  R.~{Noriega-Papaqui}, R.~{Pelayo}, J.~{Pretz}, E.~G. {P{\'e}rez-P{\'e}rez},
  Z.~{Ren}, C.~D. {Rho}, C.~{Rivi{\`e}re}, D.~{Rosa-Gonz{\'a}lez},
  M.~{Rosenberg}, E.~{Ruiz-Velasco}, H.~{Salazar}, F.~{Salesa Greus},
  A.~{Sandoval}, M.~{Schneider}, H.~{Schoorlemmer}, G.~{Sinnis}, A.~J. {Smith},
  R.~W. {Springer}, P.~{Surajbali}, O.~{Tibolla}, K.~{Tollefson}, I.~{Torres},
  T.~N. {Ukwatta}, T.~{Weisgarber}, S.~{Westerhoff}, I.~G. {Wisher}, J.~{Wood},
  T.~{Yapici}, G.~B. {Yodh}, P.~W. {Younk}, H.~{Zhou}, J.~D. {{\'A}lvarez},
  {HAWC Collaboration}, A.~{Aab}, P.~{Abreu}, M.~{Aglietta}, I.~F.~M.
  {Albuquerque}, J.~M. {Albury}, I.~{Allekotte}, A.~{Almela}, J.~{Alvarez
  Castillo}, J.~{Alvarez-Mu{\~n}iz}, G.~A. {Anastasi}, L.~{Anchordoqui},
  B.~{Andrada}, S.~{Andringa}, C.~{Aramo}, N.~{Arsene}, H.~{Asorey},
  P.~{Assis}, G.~{Avila}, A.~M. {Badescu}, A.~{Balaceanu}, F.~{Barbato}, R.~J.
  {Barreira Luz}, K.~H. {Becker}, J.~A. {Bellido}, C.~{Berat}, M.~E.
  {Bertaina}, X.~{Bertou}, P.~L. {Biermann}, J.~{Biteau}, S.~G. {Blaess},
  A.~{Blanco}, J.~{Blazek}, C.~{Bleve}, M.~{Boh{\'a}{\v{c}}ov{\'a}},
  C.~{Bonifazi}, N.~{Borodai}, A.~M. {Botti}, J.~{Brack}, I.~{Brancus},
  T.~{Bretz}, A.~{Bridgeman}, F.~L. {Briechle}, P.~{Buchholz}, A.~{Bueno},
  S.~{Buitink}, M.~{Buscemi}, K.~S. {Caballero-Mora}, L.~{Caccianiga},
  A.~{Cancio}, F.~{Canfora}, R.~{Caruso}, A.~{Castellina}, F.~{Catalani},
  G.~{Cataldi}, L.~{Cazon}, A.~G. {Chavez}, J.~A. {Chinellato}, J.~{Chudoba},
  R.~W. {Clay}, A.~C. {Cobos Cerutti}, R.~{Colalillo}, A.~{Coleman},
  L.~{Collica}, M.~R. {Coluccia}, R.~{Concei{\c{c}}{\~a}o}, G.~{Consolati},
  F.~{Contreras}, M.~J. {Cooper}, S.~{Coutu}, C.~E. {Covault}, J.~{Cronin},
  S.~{D'Amico}, B.~{Daniel}, S.~{Dasso}, K.~{Daumiller}, B.~R. {Dawson}, J.~A.
  {Day}, R.~M. {de Almeida}, S.~J. {de Jong}, G.~{De Mauro}, J.~R.~T. {de Mello
  Neto}, I.~{De Mitri}, J.~{de Oliveira}, V.~{de Souza}, J.~{Debatin},
  O.~{Deligny}, M.~L. {D{\'\i}az Castro}, F.~{Diogo}, C.~{Dobrigkeit}, J.~C.
  {D'Olivo}, Q.~{Dorosti}, R.~C. {Dos Anjos}, M.~T. {Dova}, A.~{Dundovic},
  J.~{Ebr}, R.~{Engel}, M.~{Erdmann}, M.~{Erfani}, C.~O. {Escobar},
  J.~{Espadanal}, A.~{Etchegoyen}, H.~{Falcke}, J.~{Farmer}, G.~{Farrar}, A.~C.
  {Fauth}, N.~{Fazzini}, F.~{Feldbusch}, F.~{Fenu}, B.~{Fick}, J.~M.
  {Figueira}, A.~{Filip{\v{c}}i{\v{c}}}, M.~M. {Freire}, T.~{Fujii},
  A.~{Fuster}, R.~{Ga{\"\i}or}, B.~{Garc{\'\i}a}, F.~{Gat{\'e}}, H.~{Gemmeke},
  A.~{Gherghel-Lascu}, P.~L. {Ghia}, U.~{Giaccari}, M.~{Giammarchi},
  M.~{Giller}, D.~{G{\l}as}, C.~{Glaser}, G.~{Golup}, M.~{G{\'o}mez Berisso},
  P.~F. {G{\'o}mez Vitale}, N.~{Gonz{\'a}lez}, A.~{Gorgi}, M.~{Gottowik}, A.~F.
  {Grillo}, T.~D. {Grubb}, F.~{Guarino}, G.~P. {Guedes}, R.~{Halliday}, M.~R.
  {Hampel}, P.~{Hansen}, D.~{Harari}, T.~A. {Harrison}, V.~M. {Harvey},
  A.~{Haungs}, T.~{Hebbeker}, D.~{Heck}, P.~{Heimann}, A.~E. {Herve}, G.~C.
  {Hill}, C.~{Hojvat}, E.~{Holt}, P.~{Homola}, J.~R. {H{\"o}randel},
  P.~{Horvath}, M.~{Hrabovsk{\'y}}, T.~{Huege}, J.~{Hulsman}, A.~{Insolia},
  P.~G. {Isar}, I.~{Jandt}, J.~A. {Johnsen}, M.~{Josebachuili}, J.~{Jurysek},
  A.~{K{\"a}{\"a}p{\"a}}, K.~H. {Kampert}, B.~{Keilhauer}, N.~{Kemmerich},
  J.~{Kemp}, R.~M. {Kieckhafer}, H.~O. {Klages}, M.~{Kleifges},
  J.~{Kleinfeller}, R.~{Krause}, N.~{Krohm}, D.~{Kuempel}, G.~{Kukec Mezek},
  N.~{Kunka}, A.~{Kuotb Awad}, B.~L. {Lago}, D.~{LaHurd}, R.~G. {Lang},
  M.~{Lauscher}, R.~{Legumina}, M.~A. {Leigui de Oliveira},
  A.~{Letessier-Selvon}, I.~{Lhenry-Yvon}, K.~{Link}, D.~{Lo Presti},
  L.~{Lopes}, R.~{L{\'o}pez}, A.~{L{\'o}pez Casado}, R.~{Lorek}, Q.~{Luce},
  A.~{Lucero}, M.~{Malacari}, M.~{Mallamaci}, D.~{Mandat}, P.~{Mantsch}, A.~G.
  {Mariazzi}, I.~C. {Maris}, G.~{Marsella}, D.~{Martello}, H.~{Martinez},
  O.~{Mart{\'\i}nez Bravo}, J.~J. {Mas{\'\i}as Meza}, H.~J. {Mathes},
  S.~{Mathys}, J.~{Matthews}, G.~{Matthiae}, E.~{Mayotte}, P.~O. {Mazur},
  C.~{Medina}, G.~{Medina-Tanco}, D.~{Melo}, A.~{Menshikov}, K.~D. {Merenda},
  S.~{Michal}, M.~I. {Micheletti}, L.~{Middendorf}, L.~{Miramonti},
  B.~{Mitrica}, D.~{Mockler}, S.~{Mollerach}, F.~{Montanet}, C.~{Morello},
  G.~{Morlino}, A.~L. {M{\"u}ller}, G.~{M{\"u}ller}, M.~A. {Muller},
  S.~{M{\"u}ller}, R.~{Mussa}, I.~{Naranjo}, P.~H. {Nguyen},
  M.~{Niculescu-Oglinzanu}, M.~{Niechciol}, L.~{Niemietz}, T.~{Niggemann},
  D.~{Nitz}, D.~{Nosek}, V.~{Novotny}, L.~{No{\v{z}}ka}, L.~A. {N{\'u}{\~n}ez},
  F.~{Oikonomou}, A.~{Olinto}, M.~{Palatka}, J.~{Pallotta}, P.~{Papenbreer},
  G.~{Parente}, A.~{Parra}, T.~{Paul}, M.~{Pech}, F.~{Pedreira},
  J.~{P{\c{e}}kala}, J.~{Pe{\~n}a-Rodriguez}, L.~A.~S. {Pereira}, M.~{Perlin},
  L.~{Perrone}, C.~{Peters}, S.~{Petrera}, J.~{Phuntsok}, T.~{Pierog},
  M.~{Pimenta}, V.~{Pirronello}, M.~{Platino}, M.~{Plum}, J.~{Poh},
  C.~{Porowski}, R.~R. {Prado}, P.~{Privitera}, M.~{Prouza}, E.~J. {Quel},
  S.~{Querchfeld}, S.~{Quinn}, R.~{Ramos-Pollan}, J.~{Rautenberg},
  D.~{Ravignani}, J.~{Ridky}, F.~{Riehn}, M.~{Risse}, P.~{Ristori}, V.~{Rizi},
  W.~{Rodrigues de Carvalho}, G.~{Rodriguez Fernandez}, J.~{Rodriguez Rojo},
  M.~J. {Roncoroni}, M.~{Roth}, E.~{Roulet}, A.~C. {Rovero}, P.~{Ruehl}, S.~J.
  {Saffi}, A.~{Saftoiu}, F.~{Salamida}, H.~{Salazar}, A.~{Saleh}, G.~{Salina},
  F.~{S{\'a}nchez}, P.~{Sanchez-Lucas}, E.~M. {Santos}, E.~{Santos},
  F.~{Sarazin}, R.~{Sarmento}, C.~{Sarmiento-Cano}, R.~{Sato}, M.~{Schauer},
  V.~{Scherini}, H.~{Schieler}, M.~{Schimp}, D.~{Schmidt}, O.~{Scholten},
  P.~{Schov{\'a}nek}, F.~G. {Schr{\"o}der}, S.~{Schr{\"o}der}, A.~{Schulz},
  J.~{Schumacher}, S.~J. {Sciutto}, A.~{Segreto}, A.~{Shadkam}, R.~C.
  {Shellard}, G.~{Sigl}, G.~{Silli}, R.~{{\v{S}}m{\'\i}da}, G.~R. {Snow},
  P.~{Sommers}, S.~{Sonntag}, J.~F. {Soriano}, R.~{Squartini}, D.~{Stanca},
  S.~{Stani{\v{c}}}, J.~{Stasielak}, P.~{Stassi}, M.~{Stolpovskiy},
  F.~{Strafella}, A.~{Streich}, F.~{Suarez}, M.~{Suarez-Dur{\'a}n},
  T.~{Sudholz}, T.~{Suomij{\"a}rvi}, A.~D. {Supanitsky}, J.~{{\v{S}}up{\'\i}k},
  J.~{Swain}, Z.~{Szadkowski}, A.~{Taboada}, O.~A. {Taborda}, C.~{Timmermans},
  C.~J. {Todero Peixoto}, L.~{Tomankova}, B.~{Tom{\'e}}, G.~{Torralba Elipe},
  P.~{Travnicek}, M.~{Trini}, M.~{Tueros}, R.~{Ulrich}, M.~{Unger}, M.~{Urban},
  J.~F. {Vald{\'e}s Galicia}, I.~{Vali{\~n}o}, L.~{Valore}, G.~{van Aar},
  P.~{van Bodegom}, A.~M. {van den Berg}, A.~{van Vliet}, E.~{Varela},
  B.~{Vargas C{\'a}rdenas}, R.~A. {V{\'a}zquez}, D.~{Veberi{\v{c}}},
  C.~{Ventura}, I.~D. {Vergara Quispe}, V.~{Verzi}, J.~{Vicha},
  L.~{Villase{\~n}or}, S.~{Vorobiov}, H.~{Wahlberg}, O.~{Wainberg}, D.~{Walz},
  A.~A. {Watson}, M.~{Weber}, A.~{Weindl}, M.~{Wiede{\'n}ski}, L.~{Wiencke},
  H.~{Wilczy{\'n}ski}, M.~{Wirtz}, D.~{Wittkowski}, B.~{Wundheiler}, L.~{Yang},
  A.~{Yushkov}, E.~{Zas}, D.~{Zavrtanik}, M.~{Zavrtanik}, A.~{Zepeda},
  B.~{Zimmermann}, M.~{Ziolkowski}, Z.~{Zong}, F.~{Zuccarello}, {Pierre Auger
  Collaboration}, S.~{Kim}, S.~{Schulze}, F.~E. {Bauer}, J.~M.
  {Corral-Santana}, I.~{de Gregorio-Monsalvo}, J.~{Gonz{\'a}lez-L{\'o}pez},
  D.~H. {Hartmann}, C.~H. {Ishwara-Chandra}, S.~{Mart{\'\i}n}, A.~{Mehner},
  K.~{Misra}, M.~J. {Micha{\l}owski}, L.~{Resmi}, {ALMA Collaboration},
  Z.~{Paragi}, I.~{Agudo}, T.~{An}, R.~{Beswick}, C.~{Casadio}, S.~{Frey},
  P.~{Jonker}, M.~{Kettenis}, B.~{Marcote}, J.~{Moldon}, A.~{Szomoru}, H.~J.
  {van Langevelde}, J.~{Yang}, {Euro VLBI Team}, A.~{Cwiek}, M.~{Cwiok},
  H.~{Czyrkowski}, R.~{Dabrowski}, G.~{Kasprowicz}, L.~{Mankiewicz},
  K.~{Nawrocki}, R.~{Opiela}, L.~W. {Piotrowski}, G.~{Wrochna}, M.~{Zaremba},
  A.~F. {{\.Z}arnecki}, {Pi of Sky Collaboration}, D.~{Haggard}, M.~{Nynka},
  J.~J. {Ruan}, {Chandra Team at McGill University}, P.~A. {Bland},
  T.~{Booler}, H.~A.~R. {Devillepoix}, J.~S. {de Gois}, P.~J. {Hancock}, R.~M.
  {Howie}, J.~{Paxman}, E.~K. {Sansom}, M.~C. {Towner}, DFN: {Desert Fireball
  Network}, J.~{Tonry}, M.~{Coughlin}, C.~W. {Stubbs}, L.~{Denneau},
  A.~{Heinze}, B.~{Stalder}, H.~{Weiland}, {ATLAS}, R.~P. {Eatough},
  M.~{Kramer}, A.~{Kraus}, High {Time Resolution Universe Survey}, E.~{Troja},
  L.~{Piro}, J.~{Becerra Gonz{\'a}lez}, N.~R. {Butler}, O.~D. {Fox}, H.~G.
  {Khandrika}, A.~{Kutyrev}, W.~H. {Lee}, R.~{Ricci}, Jr. {Ryan}, R.~E.,
  R.~{S{\'a}nchez-Ram{\'\i}rez}, S.~{Veilleux}, A.~M. {Watson}, M.~H.
  {Wieringa}, J.~M. {Burgess}, H.~{van Eerten}, C.~J. {Fontes}, C.~L. {Fryer},
  O.~{Korobkin}, R.~T. {Wollaeger}, {RIMAS}, {RATIR}, F.~{Camilo}, A.~R.
  {Foley}, S.~{Goedhart}, S.~{Makhathini}, N.~{Oozeer}, O.~M. {Smirnov}, R.~P.
  {Fender}, P.~A. {Woudt}, and SKA {South Africa/MeerKAT}.
\newblock {Multi-messenger Observations of a Binary Neutron Star Merger}.
\newblock {\em \apjl}, 848(2):L12, October 2017.

\bibitem{Hillas:1985is}
A.M. Hillas.
\newblock {The Origin of Ultrahigh-Energy Cosmic Rays}.
\newblock {\em Ann. Rev. Astron. Astrophys.}, 22:425--444, 1984.

\bibitem{Fang:2017zjf}
K.~Fang and K.~Murase.
\newblock {Linking High-Energy Cosmic Particles by Black Hole Jets Embedded in
  Large-Scale Structures}.
\newblock {\em Nature Phys.}, 14(4):396--398, 2018.

\bibitem{Bustamante:2019sdb}
M.~Bustamante and M.~Ahlers.
\newblock {Inferring the Flavor of High-Energy Astrophysical Neutrinos at Their
  Sources}.
\newblock {\em Phys. Rev. Lett.}, 122(24):241101, 2019.

\bibitem{Ackermann:2019cxh}
M.~Ackermann et~al.
\newblock {Fundamental Physics with High-Energy Cosmic Neutrinos}.
\newblock {\em Bull. Am. Astron. Soc.}, 51:215, 2019.

\bibitem{Aartsen:2013jdh}
M.G. Aartsen et~al.
\newblock {Evidence for High-Energy Extraterrestrial Neutrinos at the IceCube
  Detector}.
\newblock {\em Science}, 342:1242856, 2013.

\bibitem{Aartsen:2014gkd}
M.G. Aartsen et~al.
\newblock {Observation of High-Energy Astrophysical Neutrinos in Three Years of
  IceCube Data}.
\newblock {\em Phys. Rev. Lett.}, 113:101101, 2014.

\bibitem{Aartsen:2015wto}
M.G. Aartsen et~al.
\newblock {Searches for Time Dependent Neutrino Sources with IceCube Data from
  2008 to 2012}.
\newblock {\em Astrophys. J.}, 807(1):46, 2015.

\bibitem{Aartsen:2016oji}
M.G. Aartsen et~al.
\newblock {All-sky Search for Time-integrated Neutrino Emission from
  Astrophysical Sources with 7 yr of IceCube Data}.
\newblock {\em Astrophys. J.}, 835(2):151, 2017.

\bibitem{IceCube:2018cha}
M.G. Aartsen et~al.
\newblock {Neutrino Emission from the Direction of the Blazar TXS 0506+056
  Prior to the IceCube-170922A Alert}.
\newblock {\em Science}, 361(6398):147--151, 2018.

\bibitem{Murase:2018iyl}
K.~Murase, F.~Oikonomou, and M.~Petropoulou.
\newblock {Blazar Flares as an Origin of High-Energy Cosmic Neutrinos?}
\newblock {\em Astrophys. J.}, 865(2):124, 2018.

\bibitem{2019ApJ...881...46R}
A.~{Reimer}, M.~{B{\"o}ttcher}, and S.~{Buson}.
\newblock {Cascading Constraints from Neutrino-emitting Blazars: The Case of
  TXS 0506+056}.
\newblock {\em Astrophys. J.}, 881(1):46, 2019.

\bibitem{2019ApJ...874L..29R}
X.~{Rodrigues}, S.~{Gao}, A.~{Fedynitch}, A.~{Palladino}, and W.~{Winter}.
\newblock {Leptohadronic Blazar Models Applied to the 2014-2015 Flare of TXS
  0506+056}.
\newblock {\em Astrophys. J. Lett.}, 874(2):L29, 2019.

\bibitem{2019PhRvD..99f3008L}
R.~Y. {Liu}, K.~{Wang}, R.~{Xue}, A.~M. {Taylor}, X.~Y. {Wang}, Z.~{Li}, and
  H.~{Yan}.
\newblock {Hadronuclear Interpretation of a High-Energy Neutrino Event
  Coincident with a Blazar Flare}.
\newblock {\em Phys. Rev. D}, 99(6):063008, 2019.

\bibitem{2019ApJ...886...23X}
R.~{Xue}, R.~Y. {Liu}, M.~{Petropoulou}, F.~{Oikonomou}, Z.~R. {Wang},
  K.~{Wang}, and X.~Y. {Wang}.
\newblock {A Two-zone Model for Blazar Emission: Implications for TXS 0506+056
  and the Neutrino Event IceCube-170922A}.
\newblock {\em Astrophys. J.}, 886(1):23, 2019.

\bibitem{2020ApJ...889..118Z}
B.~T. {Zhang}, M.~{Petropoulou}, K.~{Murase}, and F.~{Oikonomou}.
\newblock {A Neutral Beam Model for High-energy Neutrino Emission from the
  Blazar TXS 0506+056}.
\newblock {\em Astrophys. J.}, 889(2):118, 2020.

\bibitem{2021ApJ...906...51X}
R.~{Xue}, R.~Y. {Liu}, Z.~R. {Wang}, N.~{Ding}, and X.~Y. {Wang}.
\newblock {A Two-zone Blazar Radiation Model for ``Orphan'' Neutrino Flares}.
\newblock {\em Astrophys. J.}, 906(1):51, 2021.

\bibitem{2019ApJ...880...40I}
Y.~{Inoue}, D.~{Khangulyan}, S.~{Inoue}, and A.~{Doi}.
\newblock {On High-energy Particles in Accretion Disk Coronae of Supermassive
  Black Holes: Implications for MeV Gamma-rays and High-energy Neutrinos from
  AGN Cores}.
\newblock {\em Astrophys. J.}, 880(1):40, 2019.

\bibitem{2020PhRvL.125a1101M}
K.~{Murase}, S.~S. {Kimura}, and P.~{M{\'e}sz{\'a}ros}.
\newblock {Hidden Cores of Active Galactic Nuclei as the Origin of
  Medium-Energy Neutrinos: Critical Tests with the MeV Gamma-Ray Connection}.
\newblock {\em Phys. Rev. Lett.}, 125(1):011101, 2020.

\bibitem{Murase:2015ndr}
K.~{Murase}.
\newblock {Active Galactic Nuclei as High-Energy Neutrino Sources}.
\newblock In {\em Neutrino Astronomy: Current Status, Future Prospects}, pages
  15--31. 2017.

\bibitem{Rani:2019ber}
B.~Rani et~al.
\newblock {High-Energy Polarimetry - A New Window to Probe Extreme Physics in
  AGN Jets}.
\newblock {\em arXiv 1903.04607}, 2019.

\bibitem{Rani:2019cba}
B.~Rani et~al.
\newblock {Multi-Physics of AGN Jets in the Multi-Messenger Era}.
\newblock {\em arXiv 1903.04504}, 2019.

\bibitem{Ojha:2019xan}
R.~Ojha et~al.
\newblock {Neutrinos, Cosmic Rays and the MeV Band}.
\newblock {\em arXiv 1903.05765}, 2019.

\bibitem{2021arXiv210802860F}
H.~Fleischhack.
\newblock {AMEGO-X: MeV Gamma-Ray Astronomy in the Multi-messenger Era}.
\newblock {\em PoS}, ICRC2021:649, 2021.

\bibitem{Lewis:2021roc}
T.~R. Lewis, C.~M. Karwin, T.~M. Venters, H.~Fleischhack, Y.~Sheng, C.~A.
  Kierans, R.~Caputo, and J.~McEnery.
\newblock {Modeling and Simulations of TXS 0506+056 Neutrino Events in the MeV
  Band}.
\newblock {\em arXiv 2111.10600}, 2021.

\bibitem{Marscher2008}
Alan~P. {Marscher}, Svetlana~G. {Jorstad}, Francesca~D. {D'Arcangelo}, Paul~S.
  {Smith}, G.~Grant {Williams}, Valeri~M. {Larionov}, Haruki {Oh}, Alice~R.
  {Olmstead}, Margo~F. {Aller}, Hugh~D. {Aller}, Ian~M. {McHardy}, Anne
  {L{\"a}hteenm{\"a}ki}, Merja {Tornikoski}, Esko {Valtaoja}, Vladimir~A.
  {Hagen-Thorn}, Eugenia~N. {Kopatskaya}, Walter~K. {Gear}, Gino {Tosti}, Omar
  {Kurtanidze}, Maria {Nikolashvili}, Lorand {Sigua}, H.~Richard {Miller}, and
  Wesley~T. {Ryle}.
\newblock {The inner jet of an active galactic nucleus as revealed by a
  radio-to-{\ensuremath{\gamma}}-ray outburst}.
\newblock {\em \nat}, 452(7190):966--969, April 2008.

\bibitem{Mundell2007}
Carole~G. {Mundell}, Iain~A. {Steele}, Robert~J. {Smith}, Shiho {Kobayashi},
  Andrea {Melandri}, Cristiano {Guidorzi}, Andreja {Gomboc}, Chris~J.
  {Mottram}, David {Clarke}, Alessandro {Monfardini}, David {Carter}, and David
  {Bersier}.
\newblock {Early Optical Polarization of a Gamma-Ray Burst Afterglow}.
\newblock {\em Science}, 315(5820):1822, March 2007.

\bibitem{Slowikowska2009}
A.~{S{\l}owikowska}, G.~{Kanbach}, M.~{Kramer}, and A.~{Stefanescu}.
\newblock {Optical polarization of the Crab pulsar: precision measurements and
  comparison to the radio emission}.
\newblock {\em \mnras}, 397(1):103--123, July 2009.

\bibitem{Chauvin2017}
M.~{Chauvin}, H.~G. {Flor{\'e}n}, M.~{Friis}, M.~{Jackson}, T.~{Kamae},
  J.~{Kataoka}, T.~{Kawano}, M.~{Kiss}, V.~{Mikhalev}, T.~{Mizuno},
  N.~{Ohashi}, T.~{Stana}, H.~{Tajima}, H.~{Takahashi}, N.~{Uchida}, and
  M.~{Pearce}.
\newblock {Shedding new light on the Crab with polarized X-rays}.
\newblock {\em Scientific Reports}, 7:7816, August 2017.

\bibitem{ZhangSN2019}
Shuang-Nan {Zhang}, Merlin {Kole}, Tian-Wei {Bao}, Tadeusz {Batsch}, Tancredi
  {Bernasconi}, Franck {Cadoux}, Jun-Ying {Chai}, Zi-Gao {Dai}, Yong-Wei
  {Dong}, Neal {Gauvin}, Wojtek {Hajdas}, Mi-Xiang {Lan}, Han-Cheng {Li},
  Lu~{Li}, Zheng-Heng {Li}, Jiang-Tao {Liu}, Xin {Liu}, Radoslaw
  {Marcinkowski}, Nicolas {Produit}, Silvio {Orsi}, Martin {Pohl}, Dominik
  {Rybka}, Hao-Li {Shi}, Li-Ming {Song}, Jian-Chao {Sun}, Jacek {Szabelski},
  Teresa {Tymieniecka}, Rui-Jie {Wang}, Yuan-Hao {Wang}, Xing {Wen}, Bo-Bing
  {Wu}, Xin {Wu}, Xue-Feng {Wu}, Hua-Lin {Xiao}, Shao-Lin {Xiong}, Lai-Yu
  {Zhang}, Li~{Zhang}, Xiao-Feng {Zhang}, Yong-Jie {Zhang}, and Anna
  {Zwolinska}.
\newblock {Detailed polarization measurements of the prompt emission of five
  gamma-ray bursts}.
\newblock {\em Nature Astronomy}, 3:258--264, January 2019.

\bibitem{Weisskopf2016}
Martin~C. {Weisskopf}, Brian {Ramsey}, Stephen {O'Dell}, Allyn {Tennant},
  Ronald {Elsner}, Paolo {Soffitta}, Ronaldo {Bellazzini}, Enrico {Costa},
  Jeffrey {Kolodziejczak}, Victoria {Kaspi}, Fabio {Muleri}, Herman {Marshall},
  Giorgio {Matt}, and Roger {Romani}.
\newblock {The Imaging X-ray Polarimetry Explorer (IXPE)}.
\newblock In Jan-Willem~A. {den Herder}, Tadayuki {Takahashi}, and Marshall
  {Bautz}, editors, {\em Space Telescopes and Instrumentation 2016: Ultraviolet
  to Gamma Ray}, volume 9905 of {\em Society of Photo-Optical Instrumentation
  Engineers (SPIE) Conference Series}, page 990517, July 2016.

\bibitem{marscher1985}
A.~P. {Marscher} and W.~K. {Gear}.
\newblock {Models for high-frequency radio outbursts in extragalactic sources,
  with application to the early 1983 millimeter-to-infrared flare of 3C 273}.
\newblock {\em \apj}, 298:114--127, November 1985.

\bibitem{maraschi1992}
L.~{Maraschi}, G.~{Ghisellini}, and A.~{Celotti}.
\newblock {A jet model for the gamma-ray emitting blazar 3C 279}.
\newblock {\em \apjl}, 397:L5--L9, September 1992.

\bibitem{dermer1992}
C.~D. {Dermer}, R.~{Schlickeiser}, and A.~{Mastichiadis}.
\newblock {High-energy gamma radiation from extragalactic radio sources}.
\newblock {\em \aap}, 256:L27--L30, March 1992.

\bibitem{sikora1994}
M.~{Sikora}, M.~C. {Begelman}, and M.~J. {Rees}.
\newblock {Comptonization of diffuse ambient radiation by a relativistic jet:
  The source of gamma rays from blazars?}
\newblock {\em \apj}, 421:153--162, January 1994.

\bibitem{Mannheim1993}
K.~{Mannheim}.
\newblock {The proton blazar.}
\newblock {\em \aap}, 269:67--76, March 1993.

\bibitem{mucke2001}
A.~{M{\"u}cke} and R.~J. {Protheroe}.
\newblock {A proton synchrotron blazar model for flaring in Markarian 501}.
\newblock {\em Astroparticle Physics}, 15:121--136, March 2001.

\bibitem{Boettcher2013}
M.~{B{\"o}ttcher}, A.~{Reimer}, K.~{Sweeney}, and A.~{Prakash}.
\newblock {Leptonic and Hadronic Modeling of Fermi-detected Blazars}.
\newblock {\em \apj}, 768(1):54, May 2013.

\bibitem{Petropoulou2015}
M.~{Petropoulou}, S.~{Dimitrakoudis}, P.~{Padovani}, A.~{Mastichiadis}, and
  E.~{Resconi}.
\newblock {Photohadronic origin of {$\gamma$} -ray BL Lac emission:
  implications for IceCube neutrinos}.
\newblock {\em \mnras}, 448:2412--2429, April 2015.

\bibitem{IceCube2018}
{IceCube Collaboration}, M.~G. {Aartsen}, M.~{Ackermann}, J.~{Adams}, J.~A.
  {Aguilar}, M.~{Ahlers}, M.~{Ahrens}, I.~{Al Samarai}, D.~{Altmann},
  K.~{Andeen}, and et~al.
\newblock {Multimessenger observations of a flaring blazar coincident with
  high-energy neutrino IceCube-170922A}.
\newblock {\em Science}, 361(6398):eaat1378, July 2018.

\bibitem{Rees1994}
M.~J. {Rees} and P.~{Meszaros}.
\newblock {Unsteady Outflow Models for Cosmological Gamma-Ray Bursts}.
\newblock {\em \apjl}, 430:L93, August 1994.

\bibitem{Shaviv1995}
Nir~J. {Shaviv} and Arnon {Dar}.
\newblock {Gamma-Ray Bursts from Minijets}.
\newblock {\em \apj}, 447:863, July 1995.

\bibitem{Ioka2007}
Kunihito {Ioka}, Kohta {Murase}, Kenji {Toma}, Shigehiro {Nagataki}, and
  Takashi {Nakamura}.
\newblock {Unstable GRB Photospheres and e$^{+/-}$ Annihilation Lines}.
\newblock {\em \apjl}, 670(2):L77--L80, December 2007.

\bibitem{Panitescu2000}
A.~{Panaitescu} and P.~{M{\'e}sz{\'a}ros}.
\newblock {Gamma-Ray Bursts from Upscattered Self-absorbed Synchrotron
  Emission}.
\newblock {\em \apjl}, 544(1):L17--L21, November 2000.

\bibitem{Krawczynski2012}
H.~{Krawczynski}.
\newblock {The Polarization Properties of Inverse Compton Emission and
  Implications for Blazar Observations with the GEMS X-Ray Polarimeter}.
\newblock {\em \apj}, 744(1):30, January 2012.

\bibitem{Zhang2013}
H.~{Zhang} and M.~{B{\"o}ttcher}.
\newblock {X-Ray and Gamma-Ray Polarization in Leptonic and Hadronic Jet Models
  of Blazars}.
\newblock {\em \apj}, 774(1):18, September 2013.

\bibitem{Paliya2018}
Vaidehi~S. {Paliya}, Haocheng {Zhang}, Markus {B{\"o}ttcher}, M.~{Ajello},
  A.~{Dom{\'\i}nguez}, M.~{Joshi}, D.~{Hartmann}, and C.~S. {Stalin}.
\newblock {Leptonic and Hadronic Modeling of Fermi-LAT Hard Spectrum Quasars
  and Predictions for High-energy Polarization}.
\newblock {\em \apj}, 863(1):98, August 2018.

\bibitem{Zhang2019}
Haocheng {Zhang}, Ke~{Fang}, Hui {Li}, Dimitrios {Giannios}, Markus
  {B{\"o}ttcher}, and Sara {Buson}.
\newblock {Probing the Emission Mechanism and Magnetic Field of Neutrino
  Blazars with Multiwavelength Polarization Signatures}.
\newblock {\em \apj}, 876(2):109, May 2019.

\bibitem{Toma2009}
Kenji {Toma}, Takanori {Sakamoto}, Bing {Zhang}, Joanne~E. {Hill}, Mark~L.
  {McConnell}, Peter~F. {Bloser}, Ryo {Yamazaki}, Kunihito {Ioka}, and Takashi
  {Nakamura}.
\newblock {Statistical Properties of Gamma-Ray Burst Polarization}.
\newblock {\em \apj}, 698(2):1042--1053, June 2009.

\bibitem{Gill2020}
Ramandeep {Gill}, Jonathan {Granot}, and Pawan {Kumar}.
\newblock {Linear polarization in gamma-ray burst prompt emission}.
\newblock {\em \mnras}, 491(3):3343--3373, January 2020.

\bibitem{Zhang2016}
Haocheng {Zhang}, Chris {Diltz}, and Markus {B{\"o}ttcher}.
\newblock {Radiation and Polarization Signatures of the 3D Multizone
  Time-dependent Hadronic Blazar Model}.
\newblock {\em \apj}, 829(2):69, October 2016.

\bibitem{Deng2016}
W.~{Deng}, H.~{Zhang}, B.~{Zhang}, and H.~{Li}.
\newblock {Collision-induced Magnetic Reconnection and a Unified Interpretation
  of Polarization Properties of GRBs and Blazars}.
\newblock {\em \apjl}, 821:L12, April 2016.

\bibitem{Rani2019}
Bindu {Rani}, H.~{Zhang}, S.~D. {Hunter}, F.~{Kislat}, M.~{B{\"o}ttcher}, J.~E.
  {McEnery}, D.~J. {Thompson}, D.~{Giannios}, F.~{Guo}, H.~{Li}, M.~{Baring},
  I.~{Agudo}, S.~{Buson}, M.~{Petropoulou}, V.~{Pavlidou}, E.~{Angelakis},
  I.~{Myserlis}, Z.~{Wadiasingh}, R.~M.~Curado {da Silva}, P.~{Kilian},
  S.~{Guiriec}, V.~V. {Bozhilov}, J.~{Hodgson}, S.~{Ant{\'o}n}, D.~{Kazanas},
  P.~{Coppi}, T.~{Venters}, F.~{Longo}, E.~{Bottacini}, R.~{Ojha}, B.~{Zhang},
  S.~{Ciprini}, A.~{Moiseev}, and C.~{Shrader}.
\newblock {High-Energy Polarimetry - a new window to probe extreme physics in
  AGN jets}.
\newblock {\em \baas}, 51(3):348, May 2019.

\bibitem{McEnery2019}
Julie {McEnery}, Alexander {van der Horst}, Alberto {Dominguez}, Alexander
  {Moiseev}, Alexandre {Marcowith}, Alice {Harding}, Amy {Lien}, Andrea
  {Giuliani}, Andrew {Inglis}, Stefano {Ansoldi}, Antonio {Stamerra}, Antonios
  {Manousakis}, Andy {Strong}, Cosimo {Bambi}, Barbara {Patricelli}, Matthew
  {Baring}, Juan~Abel {Barrio}, Denis {Bastieri}, Brian {Fields}, John
  {Beacom}, Volker {Beckmann}, Wlodek {Bednarek}, Bindu {Rani}, Steven {Boggs},
  Aleksey {Bolotnikov}, S.~Brad {Cenko}, Jim {Buckley}, Brian {Grefenstette},
  Michelle {Hui}, Carlotta {Pittori}, Chanda {Prescod-Weinstein}, Chris
  {Shrader}, Christian {Gouiffes}, Carolyn {Kierans}, Colleen {Wilson-Hodge},
  Filippo {D'Ammando}, Daniel {Castro}, Daniel {Kocveski}, Dario {Gasparrini},
  David {Thompson}, David {Williams}, Alessandro {De Angelis}, Denis {Bernard},
  Seth {Digel}, Daniel {Morcuende}, Eric {Charles}, Elisabetta {Bissaldi},
  Elizabeth {Hays}, Elizabeth {Ferrara}, Enrico {Bozzo}, Eric {Grove}, Eric
  {Wulf}, Eugenio {Bottacini}, Ezio {Caroli}, Fabian {Kislat}, Foteini
  {Oikonomou}, Francesco {Giordano}, Francesco {Longo}, Chris {Fryer}, Yasushi
  {Fukazawa}, Markos {Georganopoulos}, Georgia {De Nolfo}, Giacomo {Vianello},
  Gottfried {Kanbach}, George {Younes}, Harsha {Blumer}, Dieter {Hartmann},
  Margarita {Hernanz}, Hiromitsu {Takahashi}, Hui {Li}, Ivan {Agudo}, Igor
  {Moskalenko}, Inga {Stumke}, Isabelle {Grenier}, Jacob {Smith}, James {Rodi},
  Jeremy {Perkins}, Joseph {Gelfand}, Jamie {Holder}, Jurgen {Knodlseder},
  Joachim {Kopp}, Jean-Philippe {Lenain}, Jos{\'e}-Manuel {{\'A}lvarez},
  Jessica {Metcalfe}, John {Krizmanic}, John~B. {Stephen}, Jack {Hewitt}, John
  {Mitchell}, Pat {Harding}, John {Tomsick}, Judith {Racusin}, Justin {Finke},
  Oleg {Kargaltsev}, Alexei~V. {Klimenko}, Henric {Krawczynski}, Karl {Smith},
  Hidetoshi {Kubo}, Leonardo {Di Venere}, Lea {Marcotulli}, Jan {Lommler},
  Lucas {Parker}, Luca {Baldini}, Luca {Foffano}, Luca {Zampieri}, Luigi
  {Tibaldo}, Maria {Petropoulou}, Marco {Ajello}, Manuel {Meyer}, Marcos
  {L{\'o}pez}, Marc {McConnell}, Markus {Boettcher}, Martina {Cardillo}, Manel
  {Martinez}, Matthew {Kerr}, M.~Nicola {Mazziotta}, Julie {McEnery}, Mattia
  {Di Mauro}, Matthew {Wood}, Eileen {Meyer}, Michael {Briggs}, Micha{\"e}l {De
  Becker}, Michael {Lovellette}, Michele {Doro}, Miguel~A. {Sanchez-Conde},
  Michael {Moss}, Tsunefumi {Mizuno}, Marc {Rib{\'o}}, Kazuhiro {Nakazawa},
  Naoko~Kurahashi {Neilson}, Natalia {Auricchio}, Nicola {Omodei}, Uwe
  {Oberlack}, Masanori {Ohno}, Elena {Orlando}, Nepomuk {Otte}, Paolo {Coppi},
  Peter {Bloser}, Haocheng {Zhang}, Philippe {Laurent}, Martin {Pohl}, Elisa
  {Prandini}, Peter {Shawhan}, Regina {Caputo}, Riccardo {Campana}, Riccardo
  {Rando}, Richard {Woolf}, Robert {Johnson}, Roberto {Mignani}, Roland
  {Walter}, Roopesh {Ojha}, Rui~Curado {da Silva}, Stefano {Dietrich}, Stefan
  {Funk}, Silvia {Zane}, Sonia {Anton}, Sara {Buson}, Sara {Cutini}, Pablo {Saz
  Parkinson}, Richard {Schirato}, Sean {Griffin}, S.~{Kaufmann}, Lukasz
  {Stawarz}, Stefano {Ciprini}, Stefano {Del Sordo}, Sam {Jones}, Sylvain
  {Guiriec}, Hiro {Tajima}, Teddy {Cheung}, Lih-Sin {The}, Tonia {Venters},
  Troy {Porter}, Tim {Linden}, Ulisses {Barres}, Vaidehi~S. {Paliya}, Vladimir
  {Bozhilov}, Tom {Vestrand}, Vincent {Tatischeff}, Wenlei {Chen}, Xilu {Wang},
  Yasuyuki {Tanaka}, Lucas {Uhm}, Bing {Zhang}, Stephan {Zimmer}, Andreas
  {Zoglauer}, and Zorawar {Wadiasingh}.
\newblock {All-sky Medium Energy Gamma-ray Observatory: Exploring the Extreme
  Multimessenger Universe}.
\newblock In {\em Bulletin of the American Astronomical Society}, volume~51,
  page 245, September 2019.

\bibitem{Spitkovsky2008}
A.~{Spitkovsky}.
\newblock {On the Structure of Relativistic Collisionless Shocks in
  Electron-Ion Plasmas}.
\newblock {\em \apjl}, 673:L39, January 2008.

\bibitem{Sironi2009}
L.~{Sironi} and A.~{Spitkovsky}.
\newblock {Synthetic Spectra from Particle-In-Cell Simulations of Relativistic
  Collisionless Shocks}.
\newblock {\em \apjl}, 707:L92--L96, December 2009.

\bibitem{Sironi2014}
L.~{Sironi} and A.~{Spitkovsky}.
\newblock {Relativistic Reconnection: An Efficient Source of Non-thermal
  Particles}.
\newblock {\em \apjl}, 783:L21, March 2014.

\bibitem{Guo2016}
F.~{Guo}, X.~{Li}, H.~{Li}, W.~{Daughton}, B.~{Zhang}, N.~{Lloyd-Ronning},
  Y.-H. {Liu}, H.~{Zhang}, and W.~{Deng}.
\newblock {Efficient Production of High-energy Nonthermal Particles during
  Magnetic Reconnection in a Magnetically Dominated Ion-Electron Plasma}.
\newblock {\em \apjl}, 818:L9, February 2016.

\bibitem{Zhdankin2017}
Vladimir {Zhdankin}, Gregory~R. {Werner}, Dmitri~A. {Uzdensky}, and Mitchell~C.
  {Begelman}.
\newblock {Kinetic Turbulence in Relativistic Plasma: From Thermal Bath to
  Nonthermal Continuum}.
\newblock {\em \prl}, 118(5):055103, February 2017.

\bibitem{Comisso2018}
Luca {Comisso} and Lorenzo {Sironi}.
\newblock {Particle Acceleration in Relativistic Plasma Turbulence}.
\newblock {\em \prl}, 121(25):255101, December 2018.

\bibitem{Achterberg2001}
A.~{Achterberg}, Y.~A. {Gallant}, J.~G. {Kirk}, and A.~W. {Guthmann}.
\newblock {Particle acceleration by ultrarelativistic shocks: theory and
  simulations}.
\newblock {\em \mnras}, 328:393--408, December 2001.

\bibitem{Drake2019}
J.~F. {Drake}, H.~{Arnold}, M.~{Swisdak}, and J.~T. {Dahlin}.
\newblock {A computational model for exploring particle acceleration during
  reconnection in macroscale systems}.
\newblock {\em Physics of Plasmas}, 26(1):012901, January 2019.

\bibitem{Li2018}
X.~{Li}, F.~{Guo}, H.~{Li}, and S.~{Li}.
\newblock {Large-scale Compression Acceleration during Magnetic Reconnection in
  a Low-{$\beta$} Plasma}.
\newblock {\em \apj}, 866:4, October 2018.

\bibitem{Bai2015}
X.-N. {Bai}, D.~{Caprioli}, L.~{Sironi}, and A.~{Spitkovsky}.
\newblock {Magnetohydrodynamic-particle-in-cell Method for Coupling Cosmic Rays
  with a Thermal Plasma: Application to Non-relativistic Shocks}.
\newblock {\em \apj}, 809:55, August 2015.

\bibitem{Zhang2018}
Haocheng {Zhang}, Xiaocan {Li}, Fan {Guo}, and Dimitrios {Giannios}.
\newblock {Large-amplitude Blazar Polarization Angle Swing as a Signature of
  Magnetic Reconnection}.
\newblock {\em \apjl}, 862(2):L25, August 2018.

\bibitem{1995MNRAS.275..255T}
C.~{Thompson} and R.~C. {Duncan}.
\newblock {The soft gamma repeaters as very strongly magnetized neutron stars -
  I. Radiative mechanism for outbursts}.
\newblock {\em \mnras}, 275:255--300, July 1995.

\bibitem{2008A&ARv..15..225M}
S.~{Mereghetti}.
\newblock {The strongest cosmic magnets: soft gamma-ray repeaters and anomalous
  X-ray pulsars}.
\newblock {\em \aapr}, 15:225--287, July 2008.

\bibitem{turolla15:mag}
R.~{Turolla}, S.~{Zane}, and A.~L. {Watts}.
\newblock {Magnetars: the physics behind observations. A review}.
\newblock {\em Reports on Progress in Physics}, 78(11):116901, November 2015.

\bibitem{1966RvMP...38..626E}
T.~{Erber}.
\newblock {High-Energy Electromagnetic Conversion Processes in Intense Magnetic
  Fields}.
\newblock {\em Reviews of Modern Physics}, 38:626--659, October 1966.

\bibitem{2006RPPh...69.2631H}
A.~K. {Harding} and D.~{Lai}.
\newblock {Physics of strongly magnetized neutron stars}.
\newblock {\em Reports on Progress in Physics}, 69:2631--2708, September 2006.

\bibitem{2013LNP...854.....M}
Donald {Melrose}.
\newblock {\em {Quantum Plasmadynamics}}, volume 854.
\newblock 2013.

\bibitem{PhysRevD.37.1237}
Georg Raffelt and Leo Stodolsky.
\newblock Mixing of the photon with low-mass particles.
\newblock {\em Phys. Rev. D}, 37:1237--1249, Mar 1988.

\bibitem{Kuiper-2004-ApJ}
L.~{Kuiper}, W.~{Hermsen}, and M.~{Mendez}.
\newblock {Discovery of Hard Nonthermal Pulsed X-Ray Emission from the
  Anomalous X-Ray Pulsar 1E 1841-045}.
\newblock {\em \apj}, 613:1173--1178, October 2004.

\bibitem{Kuiper-2006-ApJ}
L.~{Kuiper}, W.~{Hermsen}, P.~R. {den Hartog}, and W.~{Collmar}.
\newblock {Discovery of Luminous Pulsed Hard X-Ray Emission from Anomalous
  X-Ray Pulsars 1RXS J1708-4009, 4U 0142+61, and 1E 2259+586 by INTEGRAL and
  RXTE}.
\newblock {\em \apj}, 645:556--575, July 2006.

\bibitem{Enoto-2017-ApJS}
T.~{Enoto}, S.~{Shibata}, T.~{Kitaguchi}, Y.~{Suwa}, T.~{Uchide},
  H.~{Nishioka}, S.~{Kisaka}, T.~{Nakano}, H.~{Murakami}, and K.~{Makishima}.
\newblock {Magnetar Broadband X-Ray Spectra Correlated with Magnetic Fields:
  Suzaku Archive of SGRs and AXPs Combined with NuSTAR, Swift, and RXTE}.
\newblock {\em \apjs}, 231:8, July 2017.

\bibitem{2019BAAS...51c.292W}
Zorawar {Wadiasingh}, George {Younes}, Matthew~G. {Baring}, Alice~K. {Harding},
  Peter~L. {Gonthier}, Kun {Hu}, Alexander {van der Horst}, Silvia {Zane},
  Chryssa {Kouveliotou}, Andrei~M. {Beloborodov}, Chanda {Prescod-Weinstein},
  Tanmoy {Chattopadhyay}, Sunil {Chand ra}, Constantinos {Kalapotharakos}, Kyle
  {Parfrey}, and Demos {Kazanas}.
\newblock {Magnetars as Astrophysical Laboratories of Extreme Quantum
  Electrodynamics: The Case for a Compton Telescope}.
\newblock {\em \baas}, 51(3):292, May 2019.

\bibitem{2019BAAS...51g.245M}
Julie {McEnery}, Alexander {van der Horst}, Alberto {Dominguez}, Alexander
  {Moiseev}, Alexandre {Marcowith}, Alice {Harding}, Amy {Lien}, Andrea
  {Giuliani}, Andrew {Inglis}, Stefano {Ansoldi}, Antonio {Stamerra}, Antonios
  {Manousakis}, Andy {Strong}, Cosimo {Bambi}, Barbara {Patricelli}, Matthew
  {Baring}, Juan~Abel {Barrio}, Denis {Bastieri}, Brian {Fields}, John
  {Beacom}, Volker {Beckmann}, Wlodek {Bednarek}, Bindu {Rani}, Steven {Boggs},
  Aleksey {Bolotnikov}, S.~Brad {Cenko}, Jim {Buckley}, Brian {Grefenstette},
  Michelle {Hui}, Carlotta {Pittori}, Chanda {Prescod-Weinstein}, Chris
  {Shrader}, Christian {Gouiffes}, Carolyn {Kierans}, Colleen {Wilson-Hodge},
  Filippo {D'Ammando}, Daniel {Castro}, Daniel {Kocveski}, Dario {Gasparrini},
  David {Thompson}, David {Williams}, Alessandro {De Angelis}, Denis {Bernard},
  Seth {Digel}, Daniel {Morcuende}, Eric {Charles}, Elisabetta {Bissaldi},
  Elizabeth {Hays}, Elizabeth {Ferrara}, Enrico {Bozzo}, Eric {Grove}, Eric
  {Wulf}, Eugenio {Bottacini}, Ezio {Caroli}, Fabian {Kislat}, Foteini
  {Oikonomou}, Francesco {Giordano}, Francesco {Longo}, Chris {Fryer}, Yasushi
  {Fukazawa}, Markos {Georganopoulos}, Georgia {De Nolfo}, Giacomo {Vianello},
  Gottfried {Kanbach}, George {Younes}, Harsha {Blumer}, Dieter {Hartmann},
  Margarita {Hernanz}, Hiromitsu {Takahashi}, Hui {Li}, Ivan {Agudo}, Igor
  {Moskalenko}, Inga {Stumke}, Isabelle {Grenier}, Jacob {Smith}, James {Rodi},
  Jeremy {Perkins}, Joseph {Gelfand}, Jamie {Holder}, Jurgen {Knodlseder},
  Joachim {Kopp}, Jean-Philippe {Lenain}, Jos{\'e}-Manuel {{\'A}lvarez},
  Jessica {Metcalfe}, John {Krizmanic}, John~B. {Stephen}, Jack {Hewitt}, John
  {Mitchell}, Pat {Harding}, John {Tomsick}, Judith {Racusin}, Justin {Finke},
  Oleg {Kargaltsev}, Alexei~V. {Klimenko}, Henric {Krawczynski}, Karl {Smith},
  Hidetoshi {Kubo}, Leonardo {Di Venere}, Lea {Marcotulli}, Jan {Lommler},
  Lucas {Parker}, Luca {Baldini}, Luca {Foffano}, Luca {Zampieri}, Luigi
  {Tibaldo}, Maria {Petropoulou}, Marco {Ajello}, Manuel {Meyer}, Marcos
  {L{\'o}pez}, Marc {McConnell}, Markus {Boettcher}, Martina {Cardillo}, Manel
  {Martinez}, Matthew {Kerr}, M.~Nicola {Mazziotta}, Julie {McEnery}, Mattia
  {Di Mauro}, Matthew {Wood}, Eileen {Meyer}, Michael {Briggs}, Micha{\"e}l {De
  Becker}, Michael {Lovellette}, Michele {Doro}, Miguel~A. {Sanchez-Conde},
  Michael {Moss}, Tsunefumi {Mizuno}, Marc {Rib{\'o}}, Kazuhiro {Nakazawa},
  Naoko~Kurahashi {Neilson}, Natalia {Auricchio}, Nicola {Omodei}, Uwe
  {Oberlack}, Masanori {Ohno}, Elena {Orlando}, Nepomuk {Otte}, Paolo {Coppi},
  Peter {Bloser}, Haocheng {Zhang}, Philippe {Laurent}, Martin {Pohl}, Elisa
  {Prandini}, Peter {Shawhan}, Regina {Caputo}, Riccardo {Campana}, Riccardo
  {Rando}, Richard {Woolf}, Robert {Johnson}, Roberto {Mignani}, Roland
  {Walter}, Roopesh {Ojha}, Rui~Curado {da Silva}, Stefano {Dietrich}, Stefan
  {Funk}, Silvia {Zane}, Sonia {Anton}, Sara {Buson}, Sara {Cutini}, Pablo {Saz
  Parkinson}, Richard {Schirato}, Sean {Griffin}, S.~{Kaufmann}, Lukasz
  {Stawarz}, Stefano {Ciprini}, Stefano {Del Sordo}, Sam {Jones}, Sylvain
  {Guiriec}, Hiro {Tajima}, Teddy {Cheung}, Lih-Sin {The}, Tonia {Venters},
  Troy {Porter}, Tim {Linden}, Ulisses {Barres}, Vaidehi~S. {Paliya}, Vladimir
  {Bozhilov}, Tom {Vestrand}, Vincent {Tatischeff}, Wenlei {Chen}, Xilu {Wang},
  Yasuyuki {Tanaka}, Lucas {Uhm}, Bing {Zhang}, Stephan {Zimmer}, Andreas
  {Zoglauer}, and Zorawar {Wadiasingh}.
\newblock {All-sky Medium Energy Gamma-ray Observatory: Exploring the Extreme
  Multimessenger Universe}.
\newblock In {\em Bulletin of the American Astronomical Society}, volume~51,
  page 245, September 2019.

\bibitem{1986ApJ...309..362D}
J.~K. {Daugherty} and A.~K. {Harding}.
\newblock {Compton scattering in strong magnetic fields}.
\newblock {\em \apj}, 309:362--371, October 1986.

\bibitem{2005ApJ...630..430B}
M.~G. {Baring}, P.~L. {Gonthier}, and A.~K. {Harding}.
\newblock {Spin-dependent Cyclotron Decay Rates in Strong Magnetic Fields}.
\newblock {\em \apj}, 630:430--440, September 2005.

\bibitem{2007Ap&SS.308..109B}
M.~G. {Baring} and A.~K. {Harding}.
\newblock {Resonant Compton upscattering in anomalous X-ray pulsars}.
\newblock {\em \apss}, 308:109--118, April 2007.

\bibitem{2011ApJ...733...61B}
M.~G. {Baring}, Z.~{Wadiasingh}, and P.~L. {Gonthier}.
\newblock {Cooling Rates for Relativistic Electrons Undergoing Compton
  Scattering in Strong Magnetic Fields}.
\newblock {\em \apj}, 733:61, May 2011.

\bibitem{2014PhRvD..90d3014G}
P.~L. {Gonthier}, M.~G. {Baring}, M.~T. {Eiles}, Z.~{Wadiasingh}, C.~A.
  {Taylor}, and C.~J. {Fitch}.
\newblock {Compton scattering in strong magnetic fields: Spin-dependent
  influences at the cyclotron resonance}.
\newblock {\em \prd}, 90(4):043014, August 2014.

\bibitem{2016PhRvD..93j5003M}
Alexander~A. {Mushtukov}, Dmitrij~I. {Nagirner}, and Juri {Poutanen}.
\newblock {Compton scattering S matrix and cross section in strong magnetic
  field}.
\newblock {\em \prd}, 93(10):105003, May 2016.

\bibitem{2018ApJ...854...98W}
Z.~{Wadiasingh}, M.~G. {Baring}, P.~L. {Gonthier}, and A.~K. {Harding}.
\newblock {Resonant Inverse Compton Scattering Spectra from Highly Magnetized
  Neutron Stars}.
\newblock {\em \apj}, 854:98, February 2018.

\bibitem{1983ApJ...273..761D}
J.~K. {Daugherty} and A.~K. {Harding}.
\newblock {Pair production in superstrong magnetic fields.}
\newblock {\em \apj}, 273:761--773, Oct 1983.

\bibitem{2001ApJ...547..929B}
M.~G. {Baring} and A.~K. {Harding}.
\newblock {Photon Splitting and Pair Creation in Highly Magnetized Pulsars}.
\newblock {\em \apj}, 547:929--948, February 2001.

\bibitem{2019MNRAS.486.3327H}
Kun {Hu}, Matthew~G. {Baring}, Zorawar {Wadiasingh}, and Alice~K. {Harding}.
\newblock {Opacities for photon splitting and pair creation in neutron star
  magnetospheres}.
\newblock {\em \mnras}, 486(3):3327--3349, July 2019.

\bibitem{1975PhRvD..12.1132T}
Wu-Yang {Tsai} and Thomas {Erber}.
\newblock {Propagation of photons in homogeneous magnetic fields: Index of
  refraction}.
\newblock {\em \prd}, 12(4):1132--1137, August 1975.

\bibitem{1971AnPhy..67..599A}
S.~L. {Adler}.
\newblock {Photon splitting and photon dispersion in a strong magnetic field.}
\newblock {\em Annals of Physics}, 67:599--647, Jan 1971.

\bibitem{1997ApJ...482..372B}
M.~G. {Baring} and A.~K. {Harding}.
\newblock {Magnetic Photon Splitting: Computations of Proper-Time Rates and
  Spectra}.
\newblock {\em \apj}, 482:372--376, June 1997.

\bibitem{2019JHEP...01..163F}
Jean-Fran{\c{c}}ois {Fortin} and Kuver {Sinha}.
\newblock {X-ray polarization signals from magnetars with
  axion-like-particles}.
\newblock {\em Journal of High Energy Physics}, 2019(1):163, January 2019.

\bibitem{2018JHEP...06..048F}
Jean-Fran{\c{c}}ois {Fortin} and Kuver {Sinha}.
\newblock {Constraining axion-like-particles with hard X-ray emission from
  magnetars}.
\newblock {\em Journal of High Energy Physics}, 2018(6):48, June 2018.

\bibitem{2019PhRvD.100f3005L}
Sheridan~J. {Lloyd}, Paula~M. {Chadwick}, and Anthony~M. {Brown}.
\newblock {Constraining the axion mass through gamma-ray observations of
  pulsars}.
\newblock {\em \prd}, 100(6):063005, September 2019.

\bibitem{2020arXiv200110849L}
Sheridan~J. {Lloyd}, Paula~M. {Chadwick}, Anthony~M. {Brown}, Huai-ke {Guo},
  and Kuver {Sinha}.
\newblock {Axion Constraints from Quiescent Soft Gamma-ray Emission from
  Magnetars}.
\newblock {\em arXiv e-prints}, page arXiv:2001.10849, January 2020.

\bibitem{Fortin-2021-arxiv}
Jean-Fran{\c{c}}ois {Fortin}, Huai-Ke {Guo}, Steven~P. {Harris}, Elijah
  {Sheridan}, and Kuver {Sinha}.
\newblock {Magnetars and Axion-like Particles: Probes with the Hard X-ray
  Spectrum}.
\newblock {\em arXiv e-prints}, page arXiv:2101.05302, January 2021.

\bibitem{Fortin-2021-review}
Jean-Fran{\c{c}}ois {Fortin}, Huai-Ke {Guo}, Steven~P. {Harris}, Doojin {Kim},
  Kuver {Sinha}, and Chen {Sun}.
\newblock {Axions: From Magnetars and Neutron Star Mergers to Beam Dumps and
  BECs}.
\newblock {\em arXiv e-prints}, page arXiv:2102.12503, February 2021.

\bibitem{Zwicky1933}
F~Zwicky.
\newblock {Die Rotverschiebung von extragalaktischen Nebeln}.
\newblock {\em Helvetica Physica Acta}, 6:110--127, 1933.

\bibitem{Rubin1983}
Vera~C. Rubin.
\newblock {The rotation of spiral galaxies}.
\newblock {\em Science}, 220(4604), 1983.

\bibitem{Bosma1981a}
A.~Bosma.
\newblock {21-cm line studies of spiral galaxies. I - Observations of the
  galaxies NGC 5033, 3198, 5055, 2841, and 7331.}
\newblock {\em AJ}, 86, 1981.

\bibitem{Bosma1981b}
A.~Bosma.
\newblock {21-cm line studies of spiral galaxies. II. The distribution and
  kinematics of neutral hydrogen in spiral galaxies of various morphological
  types.}
\newblock {\em The Astronomical Journal}, 86, 1981.

\bibitem{Bertone2005}
Gianfranco Bertone, Dan Hooper, and Joseph Silk.
\newblock {Particle dark matter: Evidence, candidates and constraints}, 2005.

\bibitem{Arbey2021}
A.~Arbey and F.~Mahmoudi.
\newblock {Dark matter and the early Universe: A review}, 2021.

\bibitem{Boyarsky2019}
A.~Boyarsky, M.~Drewes, T.~Lasserre, S.~Mertens, and O.~Ruchayskiy.
\newblock {Sterile neutrino Dark Matter}, 2019.

\bibitem{Bertone2007}
Gianfranco Bertone, Wilfried Buchm{\"{u}}ller, Laura Covi, and Alejandro
  Ibarra.
\newblock {Gamma-rays from decaying dark matter}.
\newblock {\em Journal of Cosmology and Astroparticle Physics}, (11), 2007.

\bibitem{Poulin2016}
Vivian Poulin, Pasquale~D. Serpico, and Julien Lesgourgues.
\newblock {A fresh look at linear cosmological constraints on a decaying Dark
  Matter component}.
\newblock {\em Journal of Cosmology and Astroparticle Physics}, 2016(8), 2016.

\bibitem{Nygaard2021}
Andreas Nygaard, Thomas Tram, and Steen Hannestad.
\newblock {Updated constraints on decaying cold dark matter}.
\newblock {\em Journal of Cosmology and Astroparticle Physics}, 2021(5), 2021.

\bibitem{Griest1990}
Kim Griest and Marc Kamionkowski.
\newblock {Unitarity limits on the mass and radius of dark-matter particles}.
\newblock {\em Physical Review Letters}, 64(6), 1990.

\bibitem{Chung1998}
Daniel~J.H. Chung, Edward~W. Kolb, and Antonio Riotto.
\newblock {Nonthermal supermassive dark matter}.
\newblock {\em Physical Review Letters}, 81(19), 1998.

\bibitem{Kolb2017}
Edward~W. Kolb and Andrew~J. Long.
\newblock {Superheavy dark matter through Higgs portal operators}.
\newblock {\em Physical Review D}, 96(10), 2017.

\bibitem{Stecker1978}
F.~W. Stecker.
\newblock {The cosmic gamma-ray background from the annihilation of primordial
  stable neutral heavy leptons}.
\newblock {\em The Astrophysical Journal}, 223, 1978.

\bibitem{Ibarra2013}
Alejandro Ibarra, David Tran, and Christoph Weniger.
\newblock {Indirect searches for decaying dark matter}, 2013.

\bibitem{Ando2015}
Shin'Ichiro Ando and Koji Ishiwata.
\newblock {Constraints on decaying dark matter from the extragalactic gamma-ray
  background}.
\newblock {\em Journal of Cosmology and Astroparticle Physics}, 2015(5), 2015.

\bibitem{Fichtel1993}
C.~E. Fichtel, D.~L. Bertsch, B.~Dingus, R.~C. Hartman, S.~D. Hunter,
  G.~Kanbach, D.~A. Kniffen, P.~W. Kwok, Y.~C. Lin, J.~R. Mattox, H.~A.
  Mayer-Hasselwander, P.~F. Michelson, C.~von Montigny, P.~L. Nolan, K.~Pinkau,
  H.~Rothermel, E.~J. Schneid, M.~Sommer, P.~Sreekumar, and D.~J. Thompson.
\newblock {Results from the Energetic Gamma-Ray Experiment Telescope (EGRET) on
  the Compton Observatory}.
\newblock {\em Advances in Space Research}, 13(12), 1993.

\bibitem{Viana2019}
Aion Viana, Harm Schoorlemmer, Andrea Albert, Vitor {De Souza}, J.~{Patrick
  Harding}, and Jim Hinton.
\newblock {Searching for dark matter in the Galactic halo with a wide field of
  view TeV gamma-ray observatory in the Southern Hemisphere}.
\newblock {\em Journal of Cosmology and Astroparticle Physics}, 2019(12), 2019.

\bibitem{Anderhub2013}
H.~Anderhub, M.~Backes, A.~Biland, V.~Boccone, I.~Braun, T.~Bretz, J.~Bu{\ss},
  F.~Cadoux, V.~Commichau, L.~Djambazov, D.~Dorner, S.~Einecke, D.~Eisenacher,
  A.~Gendotti, O.~Grimm, H.~{Von Gunten}, C.~Haller, D.~Hildebrand,
  U.~Horisberger, B.~Huber, K.~S. Kim, M.~L. Knoetig, J.~H. K{\"{o}}hne,
  T.~Kr{\"{a}}henb{\"{u}}hl, B.~Krumm, M.~Lee, E.~Lorenz, W.~Lustermann,
  E.~Lyard, K.~Mannheim, M.~Meharga, K.~Meier, T.~Montaruli, D.~Neise,
  F.~Nessi-Tedaldi, A.~K. Overkemping, A.~Paravac, F.~Pauss, D.~Renker,
  W.~Rhode, M.~Ribordy, U.~R{\"{o}}ser, J.~P. Stucki, J.~Schneider,
  T.~Steinbring, F.~Temme, J.~Thaele, S.~Tobler, G.~Viertel, P.~Vogler,
  R.~Walter, K.~Warda, Q.~Weitzel, and M.~Z{\"{a}}nglein.
\newblock {Design and operation of FACT-the first G-APD Cherenkov telescope}.
\newblock {\em Journal of Instrumentation}, 8(6), 2013.

\bibitem{Knodlseder2011}
J{\"{u}}rgen Kn{\"{o}}dlseder.
\newblock {The CTA observatory}.
\newblock In {\em AIP Conference Proceedings}, volume 1379, 2011.

\bibitem{Sivertsson2012}
Sofia Sivertsson and Joakim Edsj{\"{o}}.
\newblock {WIMP diffusion in the Solar System including solar WIMP-nucleon
  scattering}.
\newblock {\em Physical Review D - Particles, Fields, Gravitation and
  Cosmology}, 85(12), 2012.

\bibitem{Bell2011}
Nicole~F. Bell and Kalliopi Petraki.
\newblock {Enhanced neutrino signals from dark matter annihilation in the Sun
  via metastable mediators}.
\newblock {\em Journal of Cosmology and Astroparticle Physics}, 2011(4), 2011.

\bibitem{Albert2018}
A.~Albert, R.~Alfaro, C.~Alvarez, R.~Arceo, J.~C. Arteaga-Vel{\'{a}}zquez,
  D.~{Avila Rojas}, H.~A. {Ayala Solares}, E.~Belmont-Moreno, S.~Y. Benzvi,
  C.~Brisbois, K.~S. Caballero-Mora, T.~Capistr{\'{a}}n, A.~Carrami{\~{n}}ana,
  S.~Casanova, M.~Castillo, J.~Cotzomi, S.~{Couti{\~{n}}o De Le{\'{o}}n},
  C.~{De Le{\'{o}}n}, E.~{De La Fuente}, S.~Dichiara, B.~L. Dingus, M.~A.
  Duvernois, J.~C. D{\'{i}}az-V{\'{e}}lez, K.~Engel, O.~Enr{\'{i}}quez-Rivera,
  C.~Espinoza, H.~Fleischhack, N.~Fraija, J.~A. Garc{\'{i}}a-Gonz{\'{a}}lez,
  F.~Garfias, M.~M. Gonz{\'{a}}lez, J.~A. Goodman, Z.~Hampel-Arias, J.~P.
  Harding, S.~Hernandez, B.~Hona, F.~Hueyotl-Zahuantitla, P.~H{\"{u}}ntemeyer,
  A.~Iriarte, A.~Jardin-Blicq, V.~Joshi, S.~Kaufmann, H.~{Le{\'{o}}n Vargas},
  G.~Luis-Raya, J.~Lundeen, R.~L{\'{o}}pez-Coto, K.~Malone, S.~S. Marinelli,
  O.~Martinez, I.~Martinez-Castellanos, J.~Mart{\'{i}}nez-Castro,
  P.~Miranda-Romagnoli, E.~Moreno, M.~Mostaf{\'{a}}, A.~Nayerhoda, L.~Nellen,
  M.~Newbold, M.~U. Nisa, R.~Noriega-Papaqui, J.~Pretz, E.~G.
  P{\'{e}}rez-P{\'{e}}rez, Z.~Ren, C.~D. Rho, C.~Rivi{\`{e}}re,
  D.~Rosa-Gonz{\'{a}}lez, M.~Rosenberg, E.~Ruiz-Velasco, H.~Salazar, F.~{Salesa
  Greus}, A.~Sandoval, M.~Schneider, M.~{Seglar Arroyo}, G.~Sinnis, A.~J.
  Smith, R.~W. Springer, P.~Surajbali, I.~Taboada, O.~Tibolla, K.~Tollefson,
  I.~Torres, L.~Villase{\~{n}}or, T.~Weisgarber, S.~Westerhoff, I.~G. Wisher,
  J.~Wood, T.~Yapici, A.~Zepeda, H.~Zhou, J.~D. {\'{A}}lvarez, J.~F. Beacom,
  R.~K. Leane, T.~Linden, K.~C.Y. Ng, A.~H.G. Peter, and B.~Zhou.
\newblock {Constraints on spin-dependent dark matter scattering with long-lived
  mediators from TeV observations of the Sun with HAWC}.
\newblock {\em Physical Review D}, 98(12), 2018.

\bibitem{Ng2016}
Kenny~C.Y. Ng, John~F. Beacom, Annika~H.G. Peter, and Carsten Rott.
\newblock {First observation of time variation in the solar-disk gamma-ray flux
  with Fermi}.
\newblock {\em Physical Review D}, 94(2), 2016.

\bibitem{Ackermann2015}
A.~A. Abdo and for the Fermi-LAT Collaboration.
\newblock {The First Catalog of Active Galactic Nuclei Detected by the Fermi
  Large Area Telescope}.
\newblock {\em The Astrophysical Journal}, 715(1):429--457, 2010.

\bibitem{Blanco2019}
Carlos Blanco and Dan Hooper.
\newblock {Constraints on decaying dark matter from the isotropic gamma-ray
  background}.
\newblock {\em Journal of Cosmology and Astroparticle Physics}, 2019(3), 2019.

\bibitem{Ando2021}
Shin'ichiro Ando, Suvendu~K. Barik, Zhuoran Feng, Marco Finetti, Andreas~Guerra
  Chaves, Sahaja Kanuri, Jorinde Kleverlaan, Yixuan Ma, Nicolo~Maresca {Di
  Serracapriola}, Matthew~S.P. Meinema, Imanol~Navarro Martinez, Kenny~C.Y. Ng,
  Ebo Peerbooms, Casper~A. {Van Veen}, and Fabian Zimmer.
\newblock {Decaying dark matter in dwarf spheroidal galaxies: Prospects for
  x-ray and gamma-ray telescopes}.
\newblock {\em Physical Review D}, 104(2), 2021.

\bibitem{Ishiwata2020}
Koji Ishiwata, Oscar MacIas, Shin'Ichiro Ando, and Makoto Arimoto.
\newblock {Probing heavy dark matter decays with multi-messenger astrophysical
  data}.
\newblock {\em Journal of Cosmology and Astroparticle Physics}, 2020(1), 2020.

\bibitem{Hooper2011a}
Dan Hooper and Lisa Goodenough.
\newblock {Dark matter annihilation in the Galactic Center as seen by the Fermi
  Gamma Ray Space Telescope}.
\newblock {\em Physics Letters, Section B: Nuclear, Elementary Particle and
  High-Energy Physics}, 697(5):412--428, mar 2011.

\bibitem{Hooper2011b}
Dan Hooper and Tim Linden.
\newblock {Origin of the gamma rays from the Galactic Center}.
\newblock {\em Physical Review D - Particles, Fields, Gravitation and
  Cosmology}, 84(12):123005, dec 2011.

\bibitem{Daylan2016}
Tansu Daylan, Douglas~P. Finkbeiner, Dan Hooper, Tim Linden, Stephen K~N
  Portillo, Nicholas~L. Rodd, and Tracy~R. Slatyer.
\newblock {The characterization of the gamma-ray signal from the central Milky
  Way: A case for annihilating dark matter}.
\newblock {\em Physics of the Dark Universe}, 12:1--23, 2016.

\bibitem{Hoof2020}
Sebastian Hoof, Alex Geringer-Sameth, and Roberto Trotta.
\newblock {A global analysis of dark matter signals from 27 dwarf spheroidal
  galaxies using 11 years of Fermi-LAT observations}.
\newblock {\em Journal of Cosmology and Astroparticle Physics}, 2020(2), 2020.

\bibitem{1966AZh....43..758Z}
Y.~B. {Zel'dovich} and I.~D. {Novikov}.
\newblock {The Hypothesis of Cores Retarded during Expansion and the Hot
  Cosmological Model}.
\newblock {\em Astronomicheskii Zhurnal}, 43:758, 1966.

\bibitem{Hawking:1971ei}
Stephen Hawking.
\newblock {Gravitationally collapsed objects of very low mass}.
\newblock {\em Mon. Not. Roy. Astron. Soc.}, 152:75, 1971.

\bibitem{Carr:1974nx}
Bernard~J. Carr and S.~W. Hawking.
\newblock {Black holes in the early Universe}.
\newblock {\em Mon. Not. Roy. Astron. Soc.}, 168:399--415, 1974.

\bibitem{Chapline:1975ojl}
George~F. Chapline.
\newblock {Cosmological effects of primordial black holes}.
\newblock {\em Nature}, 253(5489):251--252, 1975.

\bibitem{Page:1976df}
Don~N. Page.
\newblock {Particle Emission Rates from a Black Hole: Massless Particles from
  an Uncharged, Nonrotating Hole}.
\newblock {\em Phys. Rev.}, D13:198--206, 1976.

\bibitem{Page:1976ki}
Don~N. Page.
\newblock {Particle Emission Rates from a Black Hole. 2. Massless Particles
  from a Rotating Hole}.
\newblock {\em Phys. Rev.}, D14:3260--3273, 1976.

\bibitem{MacGibbon:2007yq}
Jane~H. MacGibbon, Bernard~J. Carr, and Don~N. Page.
\newblock {Do Evaporating Black Holes Form Photospheres?}
\newblock {\em Phys. Rev.}, D78:064043, 2008.

\bibitem{Arbey:2019jmj}
Alexandre Arbey, J\'er\'emy Auffinger, and Joseph Silk.
\newblock {Evolution of primordial black hole spin due to Hawking radiation}.
\newblock {\em Mon. Not. Roy. Astron. Soc.}, 494(1):1257--1262, 2020.

\bibitem{Bai:2019zcd}
Yang Bai and Nicholas Orlofsky.
\newblock {Primordial Extremal Black Holes as Dark Matter}.
\newblock {\em Phys. Rev. D}, 101(5):055006, 2020.

\bibitem{Chongchitnan:2021ehn}
Siri Chongchitnan and Joseph Silk.
\newblock {Extreme-value statistics of the spin of primordial black holes}.
\newblock {\em Phys. Rev. D}, 104(8):083018, 2021.

\bibitem{MacGibbon:1990zk}
J.~H. MacGibbon and B.~R. Webber.
\newblock {Quark and gluon jet emission from primordial black holes: The
  instantaneous spectra}.
\newblock {\em Phys. Rev.}, D41:3052--3079, 1990.

\bibitem{MacGibbon:1991tj}
Jane~H. MacGibbon.
\newblock {Quark and gluon jet emission from primordial black holes. 2. The
  Lifetime emission}.
\newblock {\em Phys. Rev.}, D44:376--392, 1991.

\bibitem{Arbey:2019mbc}
Alexandre Arbey and J\'er\'emy Auffinger.
\newblock {BlackHawk: A public code for calculating the Hawking evaporation
  spectra of any black hole distribution}.
\newblock {\em Eur. Phys. J. C}, 79(8):693, 2019.

\bibitem{Arbey:2021mbl}
Alexandre Arbey and J\'er\'emy Auffinger.
\newblock {Physics Beyond the Standard Model with BlackHawk v2.0}.
\newblock {\em Eur. Phys. J. C}, 81:910, 2021.

\bibitem{Auffinger:2022dic}
J\'er\'emy Auffinger.
\newblock {Limits on primordial black holes detectability with Isatis: A
  BlackHawk tool}.
\newblock 1 2022.

\bibitem{Poulin:2016anj}
Vivian Poulin, Julien Lesgourgues, and Pasquale~D. Serpico.
\newblock {Cosmological constraints on exotic injection of electromagnetic
  energy}.
\newblock {\em JCAP}, 03:043, 2017.

\bibitem{Boudaud:2018hqb}
Mathieu Boudaud and Marco Cirelli.
\newblock {Voyager 1 e$^\pm$ Further Constrain Primordial Black Holes as Dark
  Matter}.
\newblock {\em Phys. Rev. Lett.}, 122(4):041104, 2019.

\bibitem{Arbey:2019vqx}
Alexandre Arbey, J\'er\'emy Auffinger, and Joseph Silk.
\newblock {Constraining primordial black hole masses with the isotropic gamma
  ray background}.
\newblock {\em Phys. Rev. D}, 101(2):023010, 2020.

\bibitem{Ballesteros:2019exr}
Guillermo Ballesteros, Javier Coronado-Bl\'azquez, and Daniele Gaggero.
\newblock {X-ray and gamma-ray limits on the primordial black hole abundance
  from Hawking radiation}.
\newblock {\em Phys. Lett. B}, 808:135624, 2020.

\bibitem{Iguaz:2021irx}
J.~Iguaz, P.~D. Serpico, and T.~Siegert.
\newblock {Isotropic X-ray bound on Primordial Black Hole Dark Matter}.
\newblock {\em Phys. Rev. D}, 103(10):103025, 2021.

\bibitem{Chen:2021ngo}
Siyu Chen, Hong-Hao Zhang, and Guangbo Long.
\newblock {Revisiting the constraints on primordial black hole abundance with
  the isotropic gamma ray background}.
\newblock 12 2021.

\bibitem{DeRocco:2019fjq}
William DeRocco and Peter~W. Graham.
\newblock {Constraining Primordial Black Hole Abundance with the Galactic 511
  keV Line}.
\newblock {\em Phys. Rev. Lett.}, 123(25):251102, 2019.

\bibitem{Laha:2019ssq}
Ranjan Laha.
\newblock {Primordial Black Holes as a Dark Matter Candidate Are Severely
  Constrained by the Galactic Center 511 keV gamma -ray Line}.
\newblock {\em Phys. Rev. Lett.}, 123(25):251101, 2019.

\bibitem{Dasgupta:2019cae}
Basudeb Dasgupta, Ranjan Laha, and Anupam Ray.
\newblock {Neutrino and positron constraints on spinning primordial black hole
  dark matter}.
\newblock {\em Phys. Rev. Lett.}, 125(10):101101, 2020.

\bibitem{Laha:2020ivk}
Ranjan Laha, Julian~B. Munoz, and Tracy~R. Slatyer.
\newblock {INTEGRAL constraints on primordial black holes and particle dark
  matter}.
\newblock {\em Phys. Rev. D}, 101(12):123514, 2020.

\bibitem{Coogan:2020tuf}
Adam Coogan, Logan Morrison, and Stefano Profumo.
\newblock {Direct Detection of Hawking Radiation from Asteroid-Mass Primordial
  Black Holes}.
\newblock {\em Phys. Rev. Lett.}, 126(17):171101, 2021.

\bibitem{Laha:2020vhg}
Ranjan Laha, Philip Lu, and Volodymyr Takhistov.
\newblock {Gas heating from spinning and non-spinning evaporating primordial
  black holes}.
\newblock {\em Phys. Lett. B}, 820:136459, 2021.

\bibitem{Kim:2020ngi}
Hyungjin Kim.
\newblock {A constraint on light primordial black holes from the interstellar
  medium temperature}.
\newblock 7 2020.

\bibitem{Siegert:2021upf}
Thomas Siegert, Celine Boehm, Francesca Calore, Roland Diehl, Martin G.~H.
  Krause, Pasquale~D. Serpico, and Aaron~C. Vincent.
\newblock {Reticulum II: Particle Dark Matter and Primordial Black Holes
  Limits}.
\newblock 9 2021.

\bibitem{Lee:2021qhe}
Chak~Man Lee and Man Ho~Chan.
\newblock {The Evaporating Primordial Black Hole Fraction in Cool-core Galaxy
  Clusters}.
\newblock {\em Astrophys. J.}, 912(1):24, 2021.

\bibitem{Clark:2018ghm}
Steven Clark, Bhaskar Dutta, Yu~Gao, Yin-Zhe Ma, and Louis~E. Strigari.
\newblock {21 cm limits on decaying dark matter and primordial black holes}.
\newblock {\em Phys. Rev. D}, 98(4):043006, 2018.

\bibitem{Mittal:2021egv}
Shikhar Mittal, Anupam Ray, Girish Kulkarni, and Basudeb Dasgupta.
\newblock {Constraining primordial black holes as dark matter using the global
  21-cm signal with X-ray heating and excess radio background}.
\newblock 7 2021.

\bibitem{Natwariya:2021xki}
Pravin~Kumar Natwariya, Alekha~C. Nayak, and Tripurari Srivastava.
\newblock {Constraining spinning primordial black holes with global 21-cm
  signal}.
\newblock {\em Mon. Not. Roy. Astron. Soc.}, 510:4236, 2021.

\bibitem{Cang:2021owu}
Junsong Cang, Yu~Gao, and Yin-Zhe Ma.
\newblock {21-cm constraints on spinning primordial black holes}.
\newblock 8 2021.

\bibitem{Halder:2021jiv}
Ashadul Halder and Madhurima Pandey.
\newblock {Probing the effects of primordial black holes on 21-cm EDGES signal
  along with interacting dark energy and dark matter-baryon scattering}.
\newblock {\em Mon. Not. Roy. Astron. Soc.}, 508(3):3446--3454, 2021.

\bibitem{Saha:2021pqf}
Akash~Kumar Saha and Ranjan Laha.
\newblock {Sensitivities on non-spinning and spinning primordial black hole
  dark matter with global 21 cm troughs}.
\newblock 12 2021.

\bibitem{Berteaud:2022tws}
J.~Berteaud, F.~Calore, J.~Iguaz, P.~D. Serpico, and T.~Siegert.
\newblock {Strong constraints on primordial black hole dark matter from 16
  years of INTEGRAL/SPI observations}.
\newblock 2 2022.

\bibitem{1975Ap&SS..32L...1F}
Y.~{Fukada}, S.~{Hayakawa}, M.~{Ikeda}, I.~{Kasahara}, F.~{Makino}, and
  Y.~{Tanaka}.
\newblock {Rocket Observation of Energy Spectrum of Diffuse Hard X-Rays}.
\newblock {\em Astrophysics and Space Science}, 32:L1, January 1975.

\bibitem{1997AIPC..410.1223W}
K.~{Watanabe}, D.~H. {Hartmann}, M.~D. {Leising}, L.~S. {The}, G.~H. {Share},
  and R.~L. {Kinzer}.
\newblock {The Cosmic gamma-ray Background from supernovae}.
\newblock In Charles~D. {Dermer}, Mark~S. {Strickman}, and James~D. {Kurfess},
  editors, {\em Proceedings of the Fourth Compton Symposium}, volume 410 of
  {\em American Institute of Physics Conference Series}, pages 1223--1227, May
  1997.

\bibitem{1995AAS...187.5801K}
S.~C. {Kappadath}, J.~{Ryan}, D.~{Forrest}, R.~M. {Kippen}, M.~{McConnell},
  K.~{Bennett}, C.~{Winkler}, H.~{Bloemen}, W.~{Hermsen}, R.~{van Dijk},
  R.~{Diehl}, V.~{Schonfelder}, M.~{Varendorff}, and G.~{Weidenspointer}.
\newblock {The Preliminary Cosmic Diffuse gamma -Ray Spectrum from 800 keV to
  30 MeV Measured with COMPTEL}.
\newblock In {\em American Astronomical Society Meeting Abstracts}, volume 187
  of {\em American Astronomical Society Meeting Abstracts}, page 58.01,
  December 1995.

\bibitem{1997ApJ...481..205H}
S.~D. {Hunter}, D.~L. {Bertsch}, J.~R. {Catelli}, T.~M. {Dame}, S.~W. {Digel},
  B.~L. {Dingus}, J.~A. {Esposito}, C.~E. {Fichtel}, R.~C. {Hartman},
  G.~{Kanbach}, D.~A. {Kniffen}, Y.~C. {Lin}, H.~A. {Mayer-Hasselwander}, P.~F.
  {Michelson}, C.~{von Montigny}, R.~{Mukherjee}, P.~L. {Nolan}, E.~{Schneid},
  P.~{Sreekumar}, P.~{Thaddeus}, and D.~J. {Thompson}.
\newblock {EGRET Observations of the Diffuse Gamma-Ray Emission from the
  Galactic Plane}.
\newblock {\em Astrophysical Journal}, 481(1):205--240, May 1997.

\bibitem{2015ApJ...799...86A}
M.~{Ackermann}, M.~{Ajello}, A.~{Albert}, W.~B. {Atwood}, L.~{Baldini},
  J.~{Ballet}, G.~{Barbiellini}, D.~{Bastieri}, K.~{Bechtol}, R.~{Bellazzini},
  E.~{Bissaldi}, R.~D. {Blandford}, E.~D. {Bloom}, E.~{Bottacini}, T.~J.
  {Brandt}, J.~{Bregeon}, P.~{Bruel}, R.~{Buehler}, S.~{Buson}, G.~A.
  {Caliandro}, R.~A. {Cameron}, M.~{Caragiulo}, P.~A. {Caraveo},
  E.~{Cavazzuti}, C.~{Cecchi}, E.~{Charles}, A.~{Chekhtman}, J.~{Chiang},
  G.~{Chiaro}, S.~{Ciprini}, R.~{Claus}, J.~{Cohen-Tanugi}, J.~{Conrad},
  A.~{Cuoco}, S.~{Cutini}, F.~{D'Ammando}, A.~{de Angelis}, F.~{de Palma},
  C.~D. {Dermer}, S.~W. {Digel}, E.~do Couto~e. {Silva}, P.~S. {Drell},
  C.~{Favuzzi}, E.~C. {Ferrara}, W.~B. {Focke}, A.~{Franckowiak},
  Y.~{Fukazawa}, S.~{Funk}, P.~{Fusco}, F.~{Gargano}, D.~{Gasparrini},
  S.~{Germani}, N.~{Giglietto}, P.~{Giommi}, F.~{Giordano}, M.~{Giroletti},
  G.~{Godfrey}, G.~A. {Gomez-Vargas}, I.~A. {Grenier}, S.~{Guiriec},
  M.~{Gustafsson}, D.~{Hadasch}, K.~{Hayashi}, E.~{Hays}, J.~W. {Hewitt},
  P.~{Ippoliti}, T.~{Jogler}, G.~{J{\'o}hannesson}, A.~S. {Johnson}, W.~N.
  {Johnson}, T.~{Kamae}, J.~{Kataoka}, J.~{Kn{\"o}dlseder}, M.~{Kuss},
  S.~{Larsson}, L.~{Latronico}, J.~{Li}, L.~{Li}, F.~{Longo}, F.~{Loparco},
  B.~{Lott}, M.~N. {Lovellette}, P.~{Lubrano}, G.~M. {Madejski}, A.~{Manfreda},
  F.~{Massaro}, M.~{Mayer}, M.~N. {Mazziotta}, J.~E. {McEnery}, P.~F.
  {Michelson}, W.~{Mitthumsiri}, T.~{Mizuno}, A.~A. {Moiseev}, M.~E. {Monzani},
  A.~{Morselli}, I.~V. {Moskalenko}, S.~{Murgia}, R.~{Nemmen}, E.~{Nuss},
  T.~{Ohsugi}, N.~{Omodei}, E.~{Orlando}, J.~F. {Ormes}, D.~{Paneque}, J.~H.
  {Panetta}, J.~S. {Perkins}, M.~{Pesce-Rollins}, F.~{Piron}, G.~{Pivato},
  T.~A. {Porter}, S.~{Rain{\`o}}, R.~{Rando}, M.~{Razzano}, S.~{Razzaque},
  A.~{Reimer}, O.~{Reimer}, T.~{Reposeur}, S.~{Ritz}, R.~W. {Romani},
  M.~{S{\'a}nchez-Conde}, M.~{Schaal}, A.~{Schulz}, C.~{Sgr{\`o}}, E.~J.
  {Siskind}, G.~{Spandre}, P.~{Spinelli}, A.~W. {Strong}, D.~J. {Suson},
  H.~{Takahashi}, J.~G. {Thayer}, J.~B. {Thayer}, L.~{Tibaldo}, M.~{Tinivella},
  D.~F. {Torres}, G.~{Tosti}, E.~{Troja}, Y.~{Uchiyama}, G.~{Vianello},
  M.~{Werner}, B.~L. {Winer}, K.~S. {Wood}, M.~{Wood}, G.~{Zaharijas}, and
  S.~{Zimmer}.
\newblock {The Spectrum of Isotropic Diffuse Gamma-Ray Emission between 100 MeV
  and 820 GeV}.
\newblock {\em Astrophysical Journal}, 799(1):86, January 2015.

\bibitem{Prantzos:2010wi}
N.~Prantzos et~al.
\newblock {The 511 keV emission from positron annihilation in the Galaxy}.
\newblock {\em Rev. Mod. Phys.}, 83:1001--1056, 2011.

\bibitem{Kierans:2019pkh}
Carolyn~A. Kierans et~al.
\newblock {Positron Annihilation in the Galaxy}.
\newblock 3 2019.

\bibitem{Siegert:2019tus}
Thomas Siegert, Roland~M. Crocker, Roland Diehl, Martin G.~H. Krause, Fiona~H.
  Panther, Moritz M.~M. Pleintinger, and Christoph Weinberger.
\newblock {Constraints on positron annihilation kinematics in the inner
  Galaxy}.
\newblock {\em Astron. Astrophys.}, 627:A126, 2019.

\bibitem{Kierans:2019aqz}
C.~A. Kierans et~al.
\newblock {Detection of the 511keV Galactic Positron Annihilation Line with
  COSI}.
\newblock {\em Astrophys. J.}, 895(1):44, 2020.

\bibitem{1980A&A....81..263O}
P.~N. {Okele} and M.~J. {Rees}.
\newblock {Observational consequences of positron production by evaporating
  black holes}.
\newblock {\em Astronomy and astrophysics}, 81(1-2):263, January 1980.

\bibitem{1991ApJ...371..447M}
Jane~H. {MacGibbon} and B.~J. {Carr}.
\newblock {Cosmic Rays from Primordial Black Holes}.
\newblock {\em Astrophysical Journal}, 371:447, April 1991.

\bibitem{Bambi:2008kx}
Cosimo Bambi, Alexander~D. Dolgov, and Alexey~A. Petrov.
\newblock {Primordial black holes and the observed Galactic 511-keV line}.
\newblock {\em Phys. Lett. B}, 670:174--178, 2008.
\newblock [Erratum: Phys.Lett.B 681, 504--504 (2009)].

\bibitem{Keith:2021guq}
Celeste Keith and Dan Hooper.
\newblock {511~keV excess and primordial black holes}.
\newblock {\em Phys. Rev. D}, 104(6):063033, 2021.

\bibitem{AMEGO:2019gny}
Regina Caputo et~al.
\newblock {All-sky Medium Energy Gamma-ray Observatory: Exploring the Extreme
  Multimessenger Universe}.
\newblock 7 2019.

\bibitem{2021arXiv211207190O}
Elena {Orlando}, Eugenio {Bottacini}, Alexander {Moiseev}, Arash {Bodaghee},
  Werner {Collmar}, Torsten {Ensslin}, Igor~V. {Moskalenko}, Michela {Negro},
  Stefano {Profumo}, Matthew~G. {Baring}, Aleksey {Bolotnikov}, Nicholas
  {Cannady}, Gabriella~A. {Carini}, Seth {Digel}, Isabelle~A. {Grenier},
  Alice~K. {Harding}, Dieter {Hartmann}, Sven {Herrmann}, Matthew {Kerr}, Roman
  {Krivonos}, Philippe {Laurent}, Francesco {Longo}, Aldo {Morselli}, Makoto
  {Sasaki}, Peter {Shawhan}, Gerry {Skinner}, Lucas~D. {Smith}, Floyd~W.
  {Stecker}, Andrew {Strong}, Steven {Sturner}, David~J. {Thompson}, John~A.
  {Tomsick}, Zorawar {Wadiasingh}, Richard~S. {Woolf}, Eric {Yates}, and
  Andreas {Zoglauer}.
\newblock {Exploring the MeV Sky with a Combined Coded Mask and Compton
  Telescope: The Galactic Explorer with a Coded Aperture Mask Compton Telescope
  (GECCO)}.
\newblock {\em arXiv e-prints}, page arXiv:2112.07190, December 2021.

\bibitem{2010SPIE.7732E..21H}
Stanley~D. {Hunter}, Peter~F. {Bloser}, Michael~P. {Dion}, Mark~L. {McConnell},
  Georgia~A. {de Nolfo}, Seunghee {Son}, James~M. {Ryan}, and Floyd~W.
  {Stecker}.
\newblock {Development of the Advance Energetic Pair Telescope (AdEPT) for
  medium-energy gamma-ray astronomy}.
\newblock In Monique {Arnaud}, Stephen~S. {Murray}, and Tadayuki {Takahashi},
  editors, {\em Space Telescopes and Instrumentation 2010: Ultraviolet to Gamma
  Ray}, volume 7732 of {\em Society of Photo-Optical Instrumentation Engineers
  (SPIE) Conference Series}, page 773221, July 2010.

\bibitem{2019APh...112....1D}
Timur {Dzhatdoev} and Egor {Podlesnyi}.
\newblock {Massive Argon Space Telescope (MAST): A concept of heavy time
  projection chamber for {\ensuremath{\gamma}}-ray astronomy in the 100 MeV-1
  TeV energy range}.
\newblock {\em Astroparticle Physics}, 112:1--7, November 2019.

\bibitem{2016SPIE.9905E..6EW}
X.~{Wu}, R.~{Walter}, M.~{Su}, G.~{Ambrosi}, P.~{Azzarello}, M.~{B{\"o}ttcher},
  J.~{Chang}, M.~{Chernyakova}, Y.~{Fan}, C.~{Farnier}, F.~{Gargano},
  I.~{Grenier}, W.~{Hajdas}, M.~N. {Mazziotta}, M.~{Pearce}, M.~{Pohl}, and
  A.~{Zdziarski}.
\newblock {PANGU: a wide field gamma-ray imager and polarimeter}.
\newblock In Jan-Willem~A. {den Herder}, Tadayuki {Takahashi}, and Marshall
  {Bautz}, editors, {\em Space Telescopes and Instrumentation 2016: Ultraviolet
  to Gamma Ray}, volume 9905 of {\em Society of Photo-Optical Instrumentation
  Engineers (SPIE) Conference Series}, page 99056E, July 2016.

\bibitem{2020APh...114..107A}
Tsuguo {Aramaki}, Per Ola~Hansson {Adrian}, Georgia {Karagiorgi}, and Hirokazu
  {Odaka}.
\newblock {Dual MeV gamma-ray and dark matter observatory - GRAMS Project}.
\newblock {\em Astroparticle Physics}, 114:107--114, January 2020.

\bibitem{2021arXiv210208701L}
Claudio {Labanti}, Lorenzo {Amati}, Filippo {Frontera}, Sandro {Mereghetti},
  Jos{\'e}~Luis {Gasent-Blesa}, Christoph {Tenzer}, Piotr {Orleanski}, Irfan
  {Kuvvetli}, Riccardo {Campana}, Fabio {Fuschino}, Luca {Terenzi}, Enrico
  {Virgilli}, Gianluca {Morgante}, Mauro {Orlandini}, Reginald~C. {Butler},
  John~B. {Stephen}, Natalia {Auricchio}, Adriano {De Rosa}, Vanni {Da Ronco},
  Federico {Evangelisti}, Michele {Melchiorri}, Stefano {Squerzanti}, Mauro
  {Fiorini}, Giuseppe {Bertuccio}, Filippo {Mele}, Massimo {Gandola}, Piero
  {Malcovati}, Marco {Grassi}, Pierluigi {Bellutti}, Giacomo {Borghi},
  Francesco {Ficorella}, Antonino {Picciotto}, Vittorio {Zanini}, Nicola
  {Zorzi}, Evgeny {Demenev}, Irina {Rashevskaya}, Alexander {Rachevski},
  Gianluigi {Zampa}, Andrea {Vacchi}, Nicola {Zampa}, Giuseppe {Baldazzi},
  Giovanni {La Rosa}, Giuseppe {Sottile}, Angela {Volpe}, Marek {Winkler},
  Victor {Reglero}, Paul~H. {Connell}, Benjamin {Pinazo-Herrero}, Javier
  {Navarro-Gonz{\'a}lez}, Pedro {Rodr{\'\i}guez-Mart{\'\i}nez}, Alberto~J.
  {Castro-Tirado}, Andrea {Santangelo}, Paul {Hedderman}, Paolo {Lorenzi},
  Paolo {Sarra}, S{\o}ren~M. {Pedersen}, Denis {Tcherniak}, Cristiano
  {Guidorzi}, Piero {Rosati}, Alessio {Trois}, and Raffaele {Piazzolla}.
\newblock {The X/Gamma-ray Imaging Spectrometer (XGIS) on-board THESEUS:
  design, main characteristics, and concept of operation}.
\newblock {\em arXiv e-prints}, page arXiv:2102.08701, February 2021.

\bibitem{2021ExA...tmp..137A}
L.~{Amati}, P.~T. {O'Brien}, D.~{G{\"o}tz}, E.~{Bozzo}, A.~{Santangelo},
  N.~{Tanvir}, F.~{Frontera}, S.~{Mereghetti}, J.~P. {Osborne}, A.~{Blain},
  S.~{Basa}, M.~{Branchesi}, L.~{Burderi}, M.~{Caballero-Garc{\'\i}a}, A.~J.
  {Castro-Tirado}, L.~{Christensen}, R.~{Ciolfi}, A.~{De Rosa},
  V.~{Doroshenko}, A.~{Ferrara}, G.~{Ghirlanda}, L.~{Hanlon}, P.~{Heddermann},
  I.~{Hutchinson}, C.~{Labanti}, E.~{Le Floch}, H.~{Lerman}, S.~{Paltani},
  V.~{Reglero}, L.~{Rezzolla}, P.~{Rosati}, R.~{Salvaterra}, G.~{Stratta},
  C.~{Tenzer}, and {Theseus Consortium}.
\newblock {The THESEUS space mission: science goals, requirements and mission
  concept}.
\newblock {\em Experimental Astronomy}, November 2021.

\bibitem{Ray:2021mxu}
Anupam Ray, Ranjan Laha, Julian~B. Mu\~noz, and Regina Caputo.
\newblock {Near future MeV telescopes can discover asteroid-mass primordial
  black hole dark matter}.
\newblock {\em Phys. Rev. D}, 104(2):023516, 2021.

\bibitem{Ghosh:2021gfa}
Diptimoy Ghosh, Divya Sachdeva, and Praniti Singh.
\newblock {Future Constraints on Primordial Black Holes from XGIS-THESEUS}.
\newblock 10 2021.

\bibitem{2018ApSpe..72..663H}
Ryley {Hill}, Kiyoshi~W. {Masui}, and Douglas {Scott}.
\newblock {The Spectrum of the Universe}.
\newblock {\em Applied Spectroscopy}, 72(5):663--688, may 2018.

\bibitem{2001ARA&A..39..249H}
Michael~G. {Hauser} and Eli {Dwek}.
\newblock {The Cosmic Infrared Background: Measurements and Implications}.
\newblock {\em \araa}, 39:249--307, jan 2001.

\bibitem{2021arXiv210212089D}
Simon~P. {Driver}.
\newblock {Measuring energy production in the Universe over all wavelengths and
  all time}.
\newblock {\em arXiv e-prints}, page arXiv:2102.12089, feb 2021.

\bibitem{2019ConPh..60...23M}
Kalevi {Mattila} and Petri {V{\"a}is{\"a}nen}.
\newblock {Extragalactic background light: inventory of light throughout the
  cosmic history}.
\newblock {\em Contemporary Physics}, 60(1):23--44, jan 2019.

\bibitem{Nikishov1962}
A.~I. {Nikishov}.
\newblock {Absorption of High-Enegy Photons in the Universe}.
\newblock {\em Soviet Physics JETP}, 14(2):393--394, feb 1962.

\bibitem{1967PhRv..155.1408G}
Robert~J. {Gould} and G{\'e}rard~P. {Schr{\'e}der}.
\newblock {Opacity of the Universe to High-Energy Photons}.
\newblock {\em Physical Review}, 155(5):1408--1411, mar 1967.

\bibitem{1967PhRv..155.1404G}
Robert~J. {Gould} and G{\'e}rard~P. {Schr{\'e}der}.
\newblock {Pair Production in Photon-Photon Collisions}.
\newblock {\em Physical Review}, 155(5):1404--1407, mar 1967.

\bibitem{2012Sci...338.1190A}
{Fermi-LAT Collaboration}, M.~{Ackermann}, and \textit{et al}.
\newblock {The Imprint of the Extragalactic Background Light in the Gamma-Ray
  Spectra of Blazars}.
\newblock {\em Science}, 338(6111):1190, nov 2012.

\bibitem{2018Sci...362.1031F}
{Fermi-LAT Collaboration}, S.~{Abdollahi}, and \textit{et al}.
\newblock {A gamma-ray determination of the Universe's star formation history}.
\newblock {\em Science}, 362(6418):1031--1034, nov 2018.

\bibitem{2013A&A...550A...4H}
{H.E.S.S. Collaboration}, A.~{Abramowski}, and \textit{et al}.
\newblock {Measurement of the extragalactic background light imprint on the
  spectra of the brightest blazars observed with H.E.S.S.}
\newblock {\em \aap}, 550:A4, feb 2013.

\bibitem{2015ApJ...812...60B}
J.~{Biteau} and D.~A. {Williams}.
\newblock {The Extragalactic Background Light, the Hubble Constant, and
  Anomalies: Conclusions from 20 Years of TeV Gamma-ray Observations}.
\newblock {\em \apj}, 812(1):60, oct 2015.

\bibitem{2017A&A...606A..59H}
{H.E.S.S. Collaboration}, H.~{Abdalla}, and \textit{et al}.
\newblock {Measurement of the EBL spectral energy distribution using the VHE
  {\ensuremath{\gamma}}-ray spectra of H.E.S.S. blazars}.
\newblock {\em \aap}, 606:A59, oct 2017.

\bibitem{2019ApJ...885..150A}
{VERITAS Collaboration}, A.~U. {Abeysekara}, and \textit{et al}.
\newblock {Measurement of the Extragalactic Background Light Spectral Energy
  Distribution with VERITAS}.
\newblock {\em \apj}, 885(2):150, nov 2019.

\bibitem{2019MNRAS.486.4233A}
{MAGIC Collaboration}, V.~A. {Acciari}, and \textit{et al}.
\newblock {Measurement of the extragalactic background light using MAGIC and
  Fermi-LAT gamma-ray observations of blazars up to z = 1}.
\newblock {\em \mnras}, 486(3):4233--4251, jul 2019.

\bibitem{2019ApJ...874L...7D}
A.~{Desai}, K.~{Helgason}, M.~{Ajello}, and \textit{et al}.
\newblock {A GeV-TeV Measurement of the Extragalactic Background Light}.
\newblock {\em \apjl}, 874(1):L7, mar 2019.

\bibitem{BiteauMeyer2022}
J.~{Biteau} and M.~{Meyer}.
\newblock {Gamma-Ray Cosmology and Fundamental Physics}.
\newblock {\em submitted to \textit{Galaxies}}, 2022.

\bibitem{2021ApJ...906...77L}
Tod~R. {Lauer}, Marc {Postman}, Harold~A. {Weaver}, and \textit{et al}.
\newblock {New Horizons Observations of the Cosmic Optical Background}.
\newblock {\em \apj}, 906(2):77, jan 2021.

\bibitem{2021MNRAS.503.2033K}
Soheil {Koushan}, Simon~P. {Driver}, Sabine {Bellstedt}, and \textit{et al}.
\newblock {GAMA/DEVILS: constraining the cosmic star formation history from
  improved measurements of the 0.3-2.2 {\ensuremath{\mu}}m extragalactic
  background light}.
\newblock {\em \mnras}, 503(2):2033--2052, may 2021.

\bibitem{2019ApJ...885..137D}
A.~{Dom{\'\i}nguez}, R.~{Wojtak}, J.~{Finke}, and \textit{et al}.
\newblock {A New Measurement of the Hubble Constant and Matter Content of the
  Universe Using Extragalactic Background Light {\ensuremath{\gamma}}-Ray
  Attenuation}.
\newblock {\em \apj}, 885(2):137, nov 2019.

\bibitem{2012MNRAS.420..800G}
Rudy~C. {Gilmore}.
\newblock {Constraining the near-infrared background light from Population III
  stars using high-redshift gamma-ray sources}.
\newblock {\em \mnras}, 420(1):800--809, February 2012.

\bibitem{2012ApJ...745..166M}
A.~{Maurer}, M.~{Raue}, T.~{Kneiske}, D.~{Horns}, D.~{Els{\"a}sser}, and P.~H.
  {Hauschildt}.
\newblock {Dark Matter Powered Stars: Constraints from the Extragalactic
  Background Light}.
\newblock {\em \apj}, 745(2):166, February 2012.

\bibitem{2012JCAP...02..032C}
Davide {Cadamuro} and Javier {Redondo}.
\newblock {Cosmological bounds on pseudo Nambu-Goldstone bosons}.
\newblock {\em \jcap}, 2012(2):032, feb 2012.

\bibitem{2019PhRvD..99b3002K}
Oleg~E. {Kalashev}, Alexander {Kusenko}, and Edoardo {Vitagliano}.
\newblock {Cosmic infrared background excess from axionlike particles and
  implications for multimessenger observations of blazars}.
\newblock {\em \prd}, 99(2):023002, jan 2019.

\bibitem{2020JCAP...03..064K}
A.~{Korochkin}, A.~{Neronov}, and D.~{Semikoz}.
\newblock {Search for decaying eV-mass axion-like particles using gamma-ray
  signal from blazars}.
\newblock {\em \jcap}, 2020(3):064, mar 2020.

\bibitem{2020A&A...633A..74K}
A.~{Korochkin}, A.~{Neronov}, and D.~{Semikoz}.
\newblock {Search for spectral features in extragalactic background light with
  gamma-ray telescopes}.
\newblock {\em \aap}, 633:A74, jan 2020.

\bibitem{2007PhRvD..76b3001M}
Alessandro {Mirizzi}, Georg~G. {Raffelt}, and Pasquale~D. {Serpico}.
\newblock {Signatures of axionlike particles in the spectra of TeV gamma-ray
  sources}.
\newblock {\em \prd}, 76(2):023001, jul 2007.

\bibitem{2007PhRvL..99w1102H}
Dan {Hooper} and Pasquale~D. {Serpico}.
\newblock {Detecting Axionlike Particles with Gamma Ray Telescopes}.
\newblock {\em \prl}, 99(23):231102, dec 2007.

\bibitem{2007PhRvD..76l1301D}
Alessandro {de Angelis}, Marco {Roncadelli}, and Oriana {Mansutti}.
\newblock {Evidence for a new light spin-zero boson from cosmological gamma-ray
  propagation?}
\newblock {\em \prd}, 76(12):121301, dec 2007.

\bibitem{2009PhRvD..79l3511S}
M.~A. {S{\'a}nchez-Conde}, D.~{Paneque}, E.~{Bloom}, and \textit{et al}.
\newblock {Hints of the existence of axionlike particles from the gamma-ray
  spectra of cosmological sources}.
\newblock {\em \prd}, 79(12):123511, jun 2009.

\bibitem{2011PhRvD..84j5030D}
Alessandro {de Angelis}, Giorgio {Galanti}, and Marco {Roncadelli}.
\newblock {Relevance of axionlike particles for very-high-energy astrophysics}.
\newblock {\em \prd}, 84(10):105030, nov 2011.

\bibitem{2012JCAP...02..033H}
D.~{Horns} and M.~{Meyer}.
\newblock {Indications for a pair-production anomaly from the propagation of
  VHE gamma-rays}.
\newblock {\em \jcap}, 2012(2):033, feb 2012.

\bibitem{2011JCAP...11..020D}
A.~{Dom{\'\i}nguez}, M.~A. {S{\'a}nchez-Conde}, and F.~{Prada}.
\newblock {Axion-like particle imprint in cosmological very-high-energy
  sources}.
\newblock {\em \jcap}, 2011(11):020, nov 2011.

\bibitem{2014JETPL.100..355R}
G.~I. {Rubtsov} and S.~V. {Troitsky}.
\newblock {Breaks in gamma-ray spectra of distant blazars and transparency of
  the Universe}.
\newblock {\em Soviet Journal of Experimental and Theoretical Physics Letters},
  100(6):355--359, nov 2014.

\bibitem{2017PhRvD..96e1701K}
Kazunori {Kohri} and Hideo {Kodama}.
\newblock {Axion-like particles and recent observations of the cosmic infrared
  background radiation}.
\newblock {\em \prd}, 96(5):051701, sep 2017.

\bibitem{2020MNRAS.493.1553G}
G.~{Galanti}, M.~{Roncadelli}, A.~{De Angelis}, and G.~F. {Bignami}.
\newblock {Hint at an axion-like particle from the redshift dependence of
  blazar spectra}.
\newblock {\em \mnras}, 493(2):1553--1564, apr 2020.

\bibitem{2015ApJ...813L..34D}
Alberto {Dom{\'\i}nguez} and Marco {Ajello}.
\newblock {Spectral Analysis of Fermi-LAT Blazars above 50 GeV}.
\newblock {\em \apjl}, 813(2):L34, nov 2015.

\bibitem{1998Natur.393..763A}
G.~{Amelino-Camelia}, John {Ellis}, N.~E. {Mavromatos}, and \textit{et al}.
\newblock {Tests of quantum gravity from observations of
  {\ensuremath{\gamma}}-ray bursts}.
\newblock {\em \nat}, 393(6687):763--765, jun 1998.

\bibitem{2009NJPh...11h5003S}
Floyd~W. {Stecker} and Sean~T. {Scully}.
\newblock {Searching for new physics with ultrahigh energy cosmic rays}.
\newblock {\em New Journal of Physics}, 11(8):085003, aug 2009.

\bibitem{2009ApJ...691L..91S}
Floyd~W. {Stecker} and Sean~T. {Scully}.
\newblock {Is the Universe More Transparent to Gamma Rays Than Previously
  Thought?}
\newblock {\em \apjl}, 691(2):L91--L94, feb 2009.

\bibitem{2019ApJ...870...93A}
{H.E.S.S. Collaboration}, H.~{Abdalla}, and \textit{et al}.
\newblock {The 2014 TeV {\ensuremath{\gamma}}-Ray Flare of Mrk 501 Seen with
  H.E.S.S.: Temporal and Spectral Constraints on Lorentz Invariance Violation}.
\newblock {\em \apj}, 870(2):93, jan 2019.

\bibitem{2019PhRvD..99d3015L}
Rodrigo~Guedes {Lang}, Humberto {Mart{\'\i}nez-Huerta}, and Vitor {de Souza}.
\newblock {Improved limits on Lorentz invariance violation from astrophysical
  gamma-ray sources}.
\newblock {\em \prd}, 99(4):043015, feb 2019.

\bibitem{2019scta.book.....C}
{Cherenkov Telescope Array Consortium}, B.~S. {Acharya}, and \textit{et al}.
\newblock {\em {Science with the Cherenkov Telescope Array}}.
\newblock 2019.

\bibitem{2021JCAP...02..048A}
{Cherenkov Telescope Array Consortium}, H.~Abdalla, and \textit{et al}.
\newblock {Sensitivity of the Cherenkov Telescope Array for probing cosmology
  and fundamental physics with gamma-ray propagation}.
\newblock {\em \jcap}, 2021(2):048, feb 2021.

\bibitem{2019Natur.575..455M}
{MAGIC Collaboration}, V.~A. {Acciari}, and \textit{et al}.
\newblock {Teraelectronvolt emission from the {\ensuremath{\gamma}}-ray burst
  GRB 190114C}.
\newblock {\em \nat}, 575(7783):455--458, nov 2019.

\bibitem{2021Sci...372.1081H}
{H.E.S.S. Collaboration}, H.~{Abdalla}, and \textit{et al}.
\newblock {Revealing x-ray and gamma ray temporal and spectral similarities in
  the GRB 190829A afterglow}.
\newblock {\em Science}, 372(6546):1081--1085, jun 2021.

\bibitem{2021ApJ...908...90A}
{MAGIC Collaboration}, V.~A. {Acciari}, and \textit{et al}.
\newblock {MAGIC Observations of the Nearby Short Gamma-Ray Burst GRB 160821B}.
\newblock {\em \apj}, 908(1):90, feb 2021.

\bibitem{2013ExA....35..413G}
Rudy~C. {Gilmore}, Aurelien {Bouvier}, Valerie {Connaughton}, and \textit{et
  al}.
\newblock {IACT observations of gamma-ray bursts: prospects for the Cherenkov
  Telescope Array}.
\newblock {\em Experimental Astronomy}, 35(3):413--457, apr 2013.

\bibitem{2014JCAP...12..016M}
Manuel {Meyer} and Jan {Conrad}.
\newblock {Sensitivity of the Cherenkov Telescope Array to the detection of
  axion-like particles at high gamma-ray opacities}.
\newblock {\em \jcap}, 2014(12):016, dec 2014.

\bibitem{2017JCAP...01..024K}
A.~{Kartavtsev}, G.~{Raffelt}, and H.~{Vogel}.
\newblock {Extragalactic photon-ALP conversion at CTA energies}.
\newblock {\em \jcap}, 2017(1):024, jan 2017.

\bibitem{2019arXiv190502773B}
{LHAASO Collaboration}, X.~{Bai}, and \textit{et al}.
\newblock {The Large High Altitude Air Shower Observatory (LHAASO) Science
  White Paper}.
\newblock {\em arXiv e-prints}, page arXiv:1905.02773, may 2019.

\bibitem{MillenniumProject}
V.~{Springel} et~al.
\newblock {Simulations of the formation, evolution and clustering of galaxies
  and quasars}.
\newblock {\em Nature}, 435(7042):629--636, June 2005.

\bibitem{Taoso_2008}
Marco Taoso, Gianfranco Bertone, and Antonio Masiero.
\newblock Dark matter candidates: a ten-point test.
\newblock {\em Journal of Cosmology and Astroparticle Physics}, 2008(03):022,
  mar 2008.

\bibitem{xia11}
J.-Q. Xia et~al.
\newblock {A cross-correlation study of the Fermi-LAT $\gamma$-ray diffuse
  extragalactic signal}.
\newblock {\em Mon.Not.Roy.Astron.Soc.}, 416:2247--2264, 2011.

\bibitem{Cuoco:2017bpv}
A.~Cuoco et~al.
\newblock {Tomographic imaging of the Fermi-LAT gamma-ray sky through
  cross-correlations: A wider and deeper look}.
\newblock {\em Astrophys. J. Suppl.}, 232(10), 2017.

\bibitem{Ammazzalorso:2018evf}
S.~Ammazzalorso et~al.
\newblock {Characterizing the local gamma-ray Universe via angular
  cross-correlations}.
\newblock {\em Phys. Rev.}, D98(10):103007, 2018.

\bibitem{Branchini:2016glc}
E.~Branchini et~al.
\newblock {Cross-correlating the $\gamma$-ray sky with Catalogs of Galaxy
  Clusters}.
\newblock {\em Astrophys. J. Suppl.}, 228(1):8, 2017.

\bibitem{2017arXiv170809385L}
Mariangela {Lisanti}, Siddharth {Mishra-Sharma}, Nicholas~L. {Rodd}, and
  Benjamin~R. {Safdi}.
\newblock {Search for Dark Matter Annihilation in Galaxy Groups}.
\newblock {\em \prl}, 120(10):101101, March 2018.

\bibitem{2017arXiv170900416L}
Mariangela {Lisanti}, Siddharth {Mishra-Sharma}, Nicholas~L. {Rodd},
  Benjamin~R. {Safdi}, and Risa~H. {Wechsler}.
\newblock {Mapping extragalactic dark matter annihilation with galaxy surveys:
  A systematic study of stacked group searches}.
\newblock {\em \prd}, 97(6):063005, March 2018.

\bibitem{2018PASJ...70S..25M}
R.~{Mandelbaum} et~al.
\newblock {The first-year shear catalog of the Subaru Hyper Suprime-Cam Subaru
  Strategic Program Survey}.
\newblock {\em pasj}, 70:S25, January 2018.

\bibitem{Camera:2012cj}
S.~Camera et~al.
\newblock {A Novel Approach in the Weakly Interacting Massive Particle Quest:
  Cross-correlation of Gamma-Ray Anisotropies and Cosmic Shear}.
\newblock {\em Astrophys.J.}, 771:L5, 2013.

\bibitem{Camera:2014rja}
S.~{Camera} et~al.
\newblock {Tomographic-spectral approach for dark matter detection in the
  cross-correlation between cosmic shear and diffuse {$\gamma$}-ray emission}.
\newblock {\em JCAP}, 6:029, June 2015.

\bibitem{Shirasaki:2014noa}
M.~Shirasaki et~al.
\newblock {Cross-Correlation of Cosmic Shear and Extragalactic Gamma-ray
  Background: Constraints on the Dark Matter Annihilation Cross-Section}.
\newblock {\em Phys.Rev.}, D90:063502, 2014.

\bibitem{AmmazzalorsoDES}
S.~{Ammazzalorso}, D.~{Gruen}, M.~{Regis}, et~al.
\newblock {Detection of Cross-Correlation between Gravitational Lensing and
  {\ensuremath{\gamma}} Rays}.
\newblock {\em \prl}, 124(10):101102, March 2020.

\bibitem{Fornengo:2014cya}
N.~Fornengo et~al.
\newblock {Evidence of Cross-correlation between the CMB Lensing and the
  $\Gamma$-ray sky}.
\newblock {\em Astrophys. J.}, 802(1):L1, 2015.

\bibitem{Fornasa:2015qua}
M.~Fornasa and M.~A. Sanchez-Conde.
\newblock {The nature of the Diffuse Gamma-Ray Background}.
\newblock {\em Phys. Rept.}, 598:1--58, 2015.

\bibitem{1974ApJ...187..425P}
William~H. {Press} and Paul {Schechter}.
\newblock {Formation of Galaxies and Clusters of Galaxies by Self-Similar
  Gravitational Condensation}.
\newblock {\em \apj}, 187:425--438, February 1974.

\bibitem{2002PhR...372....1C}
A.~{Cooray} and R.~{Sheth}.
\newblock {Halo models of large scale structure}.
\newblock {\em Phys.Rep}, 372:1--129, December 2002.

\bibitem{Serpico2012}
P.~D. Serpico, E.~Sefusatti, M.~Gustafsson, and G.~Zaharijas.
\newblock Extragalactic gamma-ray signal from dark matter annihilation: a power
  spectrum based computation.
\newblock {\em Monthly Notices of the Royal Astronomical Society: Letters},
  421(1):L87–L91, Feb 2012.

\bibitem{Sefusatti2014}
E.~Sefusatti, G.~Zaharijas, P.~D. Serpico, D.~Theurel, and M.~Gustafsson.
\newblock Extragalactic gamma-ray signal from dark matter annihilation: an
  appraisal.
\newblock {\em Monthly Notices of the Royal Astronomical Society},
  441(3):1861–1878, May 2014.

\bibitem{PhysRevLett.115.231301}
M.~Ackermann and other.
\newblock Searching for dark matter annihilation from milky way dwarf
  spheroidal galaxies with six years of fermi large area telescope data.
\newblock {\em Phys. Rev. Lett.}, 115:231301, Nov 2015.

\bibitem{2017ApJ...834..110A}
A.~{Albert}, others {Fermi-LAT Collaboration}, and {DES Collaboration}.
\newblock {Searching for Dark Matter Annihilation in Recently Discovered Milky
  Way Satellites with Fermi-Lat}.
\newblock {\em \apj}, 834(2):110, January 2017.

\bibitem{Calore2019}
Francesca Calore, Moritz Hütten, and Martin Stref.
\newblock Gamma-ray sensitivity to dark matter subhalo modelling at high
  latitudes.
\newblock {\em Galaxies}, 7(4):90, Nov 2019.

\bibitem{10.1117/12.2308289}
N.~Barrière, P.~von Ballmoos, L.~Natalucci, G.~Roudil, P.~Bastie, P.~Courtois,
  M.~Jentschel, N.~V. Abrosimov, and J.~Rousselle.
\newblock {Laue lens: the challenge of focusing gamma rays}.
\newblock In Josiane Costeraste, Errico Armandillo, and Nikos Karafolas,
  editors, {\em International Conference on Space Optics — ICSO 2008}, volume
  10566, pages 16 -- 22. International Society for Optics and Photonics, SPIE,
  2017.

\bibitem{refId0}
{Hurley, K.}, {Mitrofanov, I.G.}, {Golovin, D.}, {Litvak, M.L.}, {Sanin, A.B.},
  {Boynton, W.}, {Fellows, C.}, {Harshman, K.}, {Starr, R.}, {Golenetskii, S.},
  {Aptekar, R.}, {Mazets, E.}, {Pal\'{}shin, V.}, {Frederiks, D.}, {Svinkin,
  D.}, {Smith, D.M.}, {Hajdas, W.}, {von Kienlin, A.}, {Zhang, X.}, {Rau, A.},
  {Yamaoka, K.}, {Takahashi, T.}, {Ohno, M.}, {Hanabata, Y.}, {Fukazawa, Y.},
  {Tashiro, M.}, {Terada, Y.}, {Murakami, T.}, {Makishima, K.}, {Cline, T.},
  {Barthelmy, S.}, {Cummings, J.}, {Gehrels, N.}, {Krimm, H.}, {Palmer, D.},
  {Goldsten, J.}, {Del Monte, E.}, {Feroci, M.}, {Marisaldi, M.}, {Connaughton,
  V.}, {Briggs, M.S.}, and {Meegan, C.}
\newblock The interplanetary network.
\newblock {\em EAS Publications Series}, 61:459--464, 2013.

\bibitem{pal2020grbalpha}
Andr{\'a}s P{\'a}l, Masanori Ohno, L{\'a}szl{\'o} M{\'e}sz{\'a}ros, Norbert
  Werner, Jakub Ripa, Marcel Frajt, Naoyoshi Hirade, J{\'a}n Hudec, Jakub
  Kapu{\v{s}}, Martin Koleda, et~al.
\newblock Grbalpha: A 1u cubesat mission for validating timing-based gamma-ray
  burst localization.
\newblock In {\em Space Telescopes and Instrumentation 2020: Ultraviolet to
  Gamma Ray}, volume 11444, pages 825--833. SPIE, 2020.

\bibitem{kieransthesis}
Carolyn Kierans.
\newblock {\em Detection of the 511 keV positron annihilation line with the
  Compton Spectrometer and Imager}.
\newblock PhD thesis, University of California, Berkeley, 2018.

\bibitem{2000A&AS..143..145S}
V.~{Sch{\"o}nfelder}, K.~{Bennett}, J.~J. {Blom}, H.~{Bloemen}, W.~{Collmar},
  A.~{Connors}, R.~{Diehl}, W.~{Hermsen}, A.~{Iyudin}, R.~M. {Kippen},
  J.~{Kn{\"o}dlseder}, L.~{Kuiper}, G.~G. {Lichti}, M.~{McConnell},
  D.~{Morris}, R.~{Much}, U.~{Oberlack}, J.~{Ryan}, G.~{Stacy}, H.~{Steinle},
  A.~{Strong}, R.~{Suleiman}, R.~{van Dijk}, M.~{Varendorff}, C.~{Winkler}, and
  O.~R. {Williams}.
\newblock {The first COMPTEL source catalogue}.
\newblock {\em \aaps}, 143:145--179, April 2000.

\bibitem{KAMAE1987254}
T.~Kamae, R.~Enomoto, and N.~Hanada.
\newblock A new method to measure energy, direction, and polarization of gamma
  rays.
\newblock {\em Nuclear Instruments and Methods in Physics Research Section A:
  Accelerators, Spectrometers, Detectors and Associated Equipment},
  260(1):254--257, 1987.

\bibitem{DOGAN1990501}
N.~Dogan, D.K. Wehe, and G.F. Knoll.
\newblock Multiple compton scattering gamma ray imaging camera.
\newblock {\em Nuclear Instruments and Methods in Physics Research Section A:
  Accelerators, Spectrometers, Detectors and Associated Equipment},
  299(1):501--506, 1990.

\bibitem{2021arXiv210910403T}
John~A. {Tomsick}, Steven~E. {Boggs}, Andreas {Zoglauer}, Eric {Wulf}, Lee
  {Mitchell}, Bernard {Phlips}, Clio {Sleator}, Terri {Brandt}, Albert {Shih},
  Jarred {Roberts}, Pierre {Jean}, Peter {von Ballmoos}, Juan {Martinez
  Oliveros}, Alan {Smale}, Carolyn {Kierans}, Dieter {Hartmann}, Mark
  {Leising}, Marco {Ajello}, Eric {Burns}, Chris {Fryer}, Pascal
  {Saint-Hilaire}, Julien {Malzac}, Fabrizio {Tavecchio}, Valentina {Fioretti},
  Andrea {Bulgarelli}, Giancarlo {Ghirlanda}, Hsiang-Kuang {Chang}, Tadayuki
  {Takahashi}, Kazuhiro {Nakazawa}, Shigeki {Matsumoto}, Tom {Melia}, Thomas
  {Siegert}, Alexander {Lowell}, Hadar {Lazar}, Jacqueline {Beechert}, and
  Hannah {Gulick}.
\newblock {The Compton Spectrometer and Imager Project for MeV Astronomy}.
\newblock {\em arXiv e-prints}, page arXiv:2109.10403, September 2021.

\bibitem{cosinasa}
{NASA Selects Gamma-ray Telescope to Chart Milky Way Evolution}.
\newblock \newline \url
  https://www.nasa.gov/press-release/nasa-selects-gamma-ray-telescope-to-chart-milky-way-evolution.

\bibitem{10.1117/12.962588}
Elena Aprile, Reshmi Mukherjee, and Masayo Suzuki.
\newblock {A Liquid Xenon Imaging Telescope For 1-30 MeV Gamma-Ray
  Astrophysics}.
\newblock In Charles~J. Hailey and Oswald H.~W. Siegmund, editors, {\em EUV,
  X-Ray, and Gamma-Ray Instrumentation for Astronomy and Atomic Physics},
  volume 1159, pages 295 -- 305. International Society for Optics and
  Photonics, SPIE, 1989.

\bibitem{gammatpc_loi}
T.~Shutt, D.S. Akerib, S.~Breur, M.~Buuck, A.~Dragone, S.W. Digel, G.~Haller,
  O.A. Hitchcock, R.~Linehan, S.~Luitz, G.M. Madejski, M.E. Monzani,
  G.~Petrillo, M.J.~Pivovaroff andH.A. Tanaka, L.~Tompkins, and Y.-T. Tsa.
\newblock {A next-generation LAr TPC-based MeV Gamma ray instrument}.
\newblock \newline \url
  https://www.snowmass21.org/docs/files/summaries/CF/SNOWMASS21-CF7\_CF1-NF7\_NF10-IF8\_IF0\_Shutt-224.pdf.

\bibitem{Weisskopf:1999hm}
Martin~C. Weisskopf.
\newblock {The chandra x-ray observatory (cxo): an overview}.
\newblock 12 1999.

\bibitem{2010arXiv1008.1362H}
Fiona~A. {Harrison}, Steven {Boggs}, Finn {Christensen}, William {Craig},
  Charles {Hailey}, Daniel {Stern}, William {Zhang}, Lorella {Angelini},
  HongJun {An}, Varun {Bhalereo}, Nicolai {Brejnholt}, Lynn {Cominsky}, W.~Rick
  {Cook}, Melania {Doll}, Paolo {Giommi}, Brian {Grefenstette}, Allan
  {Hornstrup}, Victoria~M. {Kaspi}, Yunjin {Kim}, Takao {Kitaguchi}, Jason
  {Koglin}, Carl~Christian {Liebe}, Greg {Madejski}, Kristin {Kruse Madsen},
  Peter {Mao}, David {Meier}, Hiromasa {Miyasaka}, Kaya {Mori}, Matteo {Perri},
  Michael {Pivovaroff}, Simonetta {Puccetti}, Vikram {Rana}, and Andreas
  {Zoglauer}.
\newblock {The Nuclear Spectroscopic Telescope Array (NuSTAR)}.
\newblock {\em arXiv e-prints}, page arXiv:1008.1362, August 2010.

\bibitem{2005SSRv..120..143B}
S.~D. {Barthelmy} et~al.
\newblock {The Burst Alert Telescope (BAT) on the SWIFT Midex Mission}.
\newblock {\em Space Science Reviews}, 120:143--164, October 2005.

\bibitem{Caroli}
E.~{Caroli}, J.~B. {Stephen}, G.~{Di Cocco}, L.~{Natalucci}, and
  A.~{Spizzichino}.
\newblock {Coded Aperture Imaging in X-Ray and Gamma-Ray Astronomy}.
\newblock {\em \ssr}, 45(3-4):349--403, September 1987.

\bibitem{Skinner}
Gerald~K. {Skinner}.
\newblock {Sensitivity of coded mask telescopes}.
\newblock {\em \ao}, 47(15):2739--2749, May 2008.

\bibitem{IBIS}
P.~{Ubertini}, F.~{Lebrun}, G.~{Di Cocco}, A.~{Bazzano}, A.~J. {Bird},
  K.~{Broenstad}, A.~{Goldwurm}, G.~{La Rosa}, C.~{Labanti}, P.~{Laurent},
  I.~F. {Mirabel}, E.~M. {Quadrini}, B.~{Ramsey}, V.~{Reglero}, L.~{Sabau},
  B.~{Sacco}, R.~{Staubert}, L.~{Vigroux}, M.~C. {Weisskopf}, and A.~A.
  {Zdziarski}.
\newblock {IBIS: The Imager on-board INTEGRAL}.
\newblock {\em Astronomy and Astrophysics}, 411:L131--L139, November 2003.

\bibitem{INTEGRAL}
Christoph Winkler, Roland Diehl, Pietro Ubertini, and Jorn Wilms.
\newblock {INTEGRAL: science highlights and future prospects}.
\newblock {\em Space Sci. Rev.}, 161:149--177, 2011.

\bibitem{SPI}
G.~{Vedrenne}, J.~P. {Roques}, V.~{Sch{\"o}nfelder}, P.~{Mandrou}, G.~G.
  {Lichti}, A.~{von Kienlin}, B.~{Cordier}, S.~{Schanne}, J.~{Kn{\"o}dlseder},
  G.~{Skinner}, P.~{Jean}, F.~{Sanchez}, P.~{Caraveo}, B.~{Teegarden}, P.~{von
  Ballmoos}, L.~{Bouchet}, P.~{Paul}, J.~{Matteson}, S.~{Boggs}, C.~{Wunderer},
  P.~{Leleux}, G.~{Weidenspointner}, Ph. {Durouchoux}, R.~{Diehl}, A.~{Strong},
  M.~{Cass{\'e}}, M.~A. {Clair}, and Y.~{Andr{\'e}}.
\newblock {SPI: The spectrometer aboard INTEGRAL}.
\newblock {\em Astronomy and Astrophysics}, 411:L63--L70, November 2003.

\bibitem{bolNIM}
A.~E. {Bolotnikov}, G.~S. {Camarda}, G.~De {Geronimo}, J.~{Fried}, D.~{Hodges},
  A.~{Hossain}, K.~{Kim}, G.~{Mahler}, L.~Ocampo {Giraldo}, E.~{Vernon},
  G.~{Yang}, and R.~B. {James}.
\newblock {A 4 x 4 array module of position-sensitive virtual Frisch-grid
  CdZnTe detectors for gamma-ray imaging spectrometers}.
\newblock {\em Nuclear Instruments and Methods in Physics Research A},
  954:161036, February 2020.

\bibitem{Aprile}
E.~{Aprile}, A.~{Bolotnikov}, D.~{Chen}, and R.~{Mukherjee}.
\newblock A monte carlo analysis of the liquid xenon tpc as gamma-ray
  telescope.
\newblock {\em Nuclear Instruments and Methods in Physics Research},
  A327:216--221, 1993.

\bibitem{Galloway}
M.~Galloway et~al.
\newblock {A Combined Compton and Coded-aperture Telescope for Medium- energy
  Gamma-ray Astrophysics}.
\newblock 01 2018.

\bibitem{Forot}
M.~Forot et~al.
\newblock {Compton Telescope with a coded aperture mask: imaging with the
  INTEGRAL/IBIS Compton mode}.
\newblock {\em Astrophysical Journal}, 668:1259, 2007.

\bibitem{gecco1}
A.~Moiseev et~al.
\newblock {New Mission Concept: Galactic Explorer with a Coded Aperture Mask
  Compton Telescope (GECCO)}.
\newblock {\em PoS(ICRC2021)648, 2021 International Cosmic Ray Conference},
  Berlin, 2021.

\bibitem{2012ApJS..203....4A}
M.~{Ackermann} et~al.
\newblock {The Fermi Large Area Telescope on Orbit: Event Classification,
  Instrument Response Functions, and Calibration}.
\newblock {\em Astrophysical Journal Supplmental Series}, 203:4, November 2012.

\bibitem{PhysRevLett.8.106}
W.~L. Kraushaar and G.~W. Clark.
\newblock Search for primary cosmic gamma rays with the satellite explorer xi.
\newblock {\em Phys. Rev. Lett.}, 8:106--109, Feb 1962.

\bibitem{1975ApJ...198..163F}
C.~E. {Fichtel}, R.~C. {Hartman}, D.~A. {Kniffen}, D.~J. {Thompson}, G.~F.
  {Bignami}, H.~{{\"O}gelman}, M.~E. {{\"O}zel}, and T.~{T{\"u}mer}.
\newblock {High-energy gamma-ray results from the second Small Astronomy
  Satellite.}
\newblock {\em ApJ}, 198:163--182, May 1975.

\bibitem{CHANG20176}
J.~Chang, G.~Ambrosi, Q.~An, R.~Asfandiyarov, P.~Azzarello, P.~Bernardini,
  B.~Bertucci, M.S. Cai, M.~Caragiulo, D.Y. Chen, H.F. Chen, J.L. Chen,
  W.~Chen, M.Y. Cui, T.S. Cui, A.~D’Amone, A.~{De Benedittis}, I.~{De Mitri},
  M.~{Di Santo}, J.N. Dong, T.K. Dong, Y.F. Dong, Z.X. Dong, G.~Donvito,
  D.~Droz, K.K. Duan, J.L. Duan, M.~Duranti, D.~D’Urso, R.R. Fan, Y.Z. Fan,
  F.~Fang, C.Q. Feng, L.~Feng, P.~Fusco, V.~Gallo, F.J. Gan, W.Q. Gan, M.~Gao,
  S.S. Gao, F.~Gargano, K.~Gong, Y.Z. Gong, J.H. Guo, Y.M. Hu, G.S. Huang, Y.Y.
  Huang, M.~Ionica, D.~Jiang, W.~Jiang, X.~Jin, J.~Kong, S.J. Lei, S.~Li,
  X.~Li, W.L. Li, Y.~Li, Y.F. Liang, Y.M. Liang, N.H. Liao, Q.Z. Liu, H.~Liu,
  J.~Liu, S.B. Liu, Q.Z. Liu, W.Q. Liu, Y.~Liu, F.~Loparco, J.~Lü, M.~Ma, P.X.
  Ma, S.Y. Ma, T.~Ma, X.Q. Ma, X.Y. Ma, G.~Marsella, M.N. Mazziotta, D.~Mo,
  T.T. Miao, X.Y. Niu, M.~Pohl, X.Y. Peng, W.X. Peng, R.~Qiao, J.N. Rao, M.M.
  Salinas, G.Z. Shang, W.H. Shen, Z.Q. Shen, Z.T. Shen, J.X. Song, H.~Su,
  M.~Su, Z.Y. Sun, A.~Surdo, X.J. Teng, X.B. Tian, A.~Tykhonov, V.~Vagelli,
  S.~Vitillo, C.~Wang, Chi Wang, H.~Wang, H.Y. Wang, J.Z. Wang, L.G. Wang,
  Q.~Wang, S.~Wang, X.H. Wang, X.L. Wang, Y.F. Wang, Y.P. Wang, Y.Z. Wang, S.C.
  Wen, Z.M. Wang, D.M. Wei, J.J. Wei, Y.F. Wei, D.~Wu, J.~Wu, S.S. Wu, X.~Wu,
  K.~Xi, Z.Q. Xia, Y.L. Xin, H.T. Xu, Z.L. Xu, Z.Z. Xu, G.F. Xue, H.B. Yang,
  J.~Yang, P.~Yang, Y.Q. Yang, Z.L. Yang, H.J. Yao, Y.H. Yu, Q.~Yuan, C.~Yue,
  J.J. Zang, C.~Zhang, D.L. Zhang, F.~Zhang, J.B. Zhang, J.Y. Zhang, J.Z.
  Zhang, L.~Zhang, P.F. Zhang, S.X. Zhang, W.Z. Zhang, Y.~Zhang, Y.J. Zhang,
  Y.Q. Zhang, Y.L. Zhang, Y.P. Zhang, Z.~Zhang, Z.Y. Zhang, H.~Zhao, H.Y. Zhao,
  X.F. Zhao, C.Y. Zhou, Y.~Zhou, X.~Zhu, Y.~Zhu, and S.~Zimmer.
\newblock The dark matter particle explorer mission.
\newblock {\em Astroparticle Physics}, 95:6--24, 2017.

\bibitem{2019A&A...627A..13B}
A.~{Bulgarelli}, V.~{Fioretti}, N.~{Parmiggiani}, F.~{Verrecchia},
  C.~{Pittori}, F.~{Lucarelli}, M.~{Tavani}, A.~{Aboudan}, M.~{Cardillo},
  A.~{Giuliani}, P.~W. {Cattaneo}, A.~W. {Chen}, G.~{Piano}, A.~{Rappoldi},
  L.~{Baroncelli}, A.~{Argan}, L.~A. {Antonelli}, I.~{Donnarumma},
  F.~{Gianotti}, P.~{Giommi}, M.~{Giusti}, F.~{Longo}, A.~{Pellizzoni},
  M.~{Pilia}, M.~{Trifoglio}, A.~{Trois}, S.~{Vercellone}, and A.~{Zoli}.
\newblock {Second AGILE catalogue of gamma-ray sources}.
\newblock {\em A\& A}, 627:A13, July 2019.

\bibitem{gaisser_engel_resconi_2016}
Thomas~K. Gaisser, Ralph Engel, and Elisa Resconi.
\newblock {\em Extensive air showers}, page 313–340.
\newblock Cambridge University Press, 2 edition, 2016.

\bibitem{CYGNUS:1986dze}
B.~L. Dingus et~al.
\newblock {CYGNUS Experiment at Los Alamos}.
\newblock {\em AIP Conf. Proc.}, 150:1078--1082, 1986.

\bibitem{2004ApJ...608..680A}
R.~{Atkins}, W.~{Benbow}, D.~{Berley}, E.~{Blaufuss}, J.~{Bussons}, D.~G.
  {Coyne}, T.~{De Young}, B.~L. {Dingus}, D.~E. {Dorfan}, R.~W. {Ellsworth},
  L.~{Fleysher}, R.~{Fleysher}, G.~{Gisler}, M.~M. {Gonzalez}, J.~A. {Goodman},
  T.~J. {Haines}, E.~{Hays}, C.~M. {Hoffman}, L.~A. {Kelley}, C.~P. {Lansdell},
  J.~T. {Linnemann}, J.~E. {McEnery}, R.~S. {Miller}, A.~I. {Mincer}, M.~F.
  {Morales}, P.~{Nemethy}, D.~{Noyes}, J.~M. {Ryan}, F.~W. {Samuelson},
  A.~{Shoup}, G.~{Sinnis}, A.~J. {Smith}, G.~W. {Sullivan}, D.~A. {Williams},
  S.~{Westerhoff}, M.~E. {Wilson}, X.~W. {Xu}, and G.~B. {Yodh}.
\newblock {TeV Gamma-Ray Survey of the Northern Hemisphere Sky Using the
  Milagro Observatory}.
\newblock {\em ApJ}, 608(2):680--685, June 2004.

\bibitem{3HWC_2020}
A.~Albert, R.~Alfaro, C.~Alvarez, J.~R.~Angeles Camacho, J.~C.
  Arteaga-Velázquez, K.~P. Arunbabu, D.~Avila Rojas, H.~A.~Ayala Solares,
  V.~Baghmanyan, E.~Belmont-Moreno, S.~Y. BenZvi, C.~Brisbois, K.~S.
  Caballero-Mora, T.~Capistrán, A.~Carramiñana, S.~Casanova, U.~Cotti,
  S.~Coutiño de~León, E.~De~la Fuente, R.~Diaz Hernandez, L.~Diaz-Cruz, B.~L.
  Dingus, M.~A. DuVernois, M.~Durocher, J.~C. Díaz-Vélez, R.~W. Ellsworth,
  K.~Engel, C.~Espinoza, K.~L. Fan, K.~Fang, M.~Fernández Alonso,
  H.~Fleischhack, N.~Fraija, A.~Galván-Gámez, D.~Garcia, J.~A.
  García-González, F.~Garfias, G.~Giacinti, M.~M. González, J.~A. Goodman,
  J.~P. Harding, S.~Hernandez, J.~Hinton, B.~Hona, D.~Huang,
  F.~Hueyotl-Zahuantitla, P.~Hüntemeyer, A.~Iriarte, A.~Jardin-Blicq,
  V.~Joshi, D.~Kieda, A.~Lara, W.~H. Lee, H.~León Vargas, J.~T. Linnemann,
  A.~L. Longinotti, G.~Luis-Raya, J.~Lundeen, R.~López-Coto, K.~Malone,
  V.~Marandon, O.~Martinez, I.~Martinez-Castellanos, J.~Martínez-Castro, J.~A.
  Matthews, P.~Miranda-Romagnoli, J.~A. Morales-Soto, E.~Moreno, M.~Mostafá,
  A.~Nayerhoda, L.~Nellen, M.~Newbold, M.~U. Nisa, R.~Noriega-Papaqui,
  L.~Olivera-Nieto, N.~Omodei, A.~Peisker, Y.~Pérez Araujo, E.~G.
  Pérez-Pérez, Z.~Ren, C.~D. Rho, C.~Rivière, D.~Rosa-González,
  E.~Ruiz-Velasco, H.~Salazar, F.~Salesa Greus, A.~Sandoval, M.~Schneider,
  H.~Schoorlemmer, F.~Serna, G.~Sinnis, A.~J. Smith, R.~W. Springer,
  P.~Surajbali, K.~Tollefson, I.~Torres, R.~Torres-Escobedo, T.~N. Ukwatta,
  F.~Ureña-Mena, T.~Weisgarber, F.~Werner, E.~Willox, A.~Zepeda, H.~Zhou,
  C.~de León, and J.~D. Álvarez.
\newblock 3hwc: The third hawc catalog of very-high-energy gamma-ray sources.
\newblock {\em The Astrophysical Journal}, 905(1):76, Dec 2020.

\bibitem{Vernetto_2016}
S~Vernetto and.
\newblock Gamma ray astronomy with {LHAASO}.
\newblock {\em Journal of Physics: Conference Series}, 718:052043, may 2016.

\bibitem{Amenomori:2019rjd}
M.~Amenomori et~al.
\newblock {First Detection of Photons with Energy Beyond 100 TeV from an
  Astrophysical Source}.
\newblock {\em Phys. Rev. Lett.}, 123(5):051101, 2019.

\bibitem{Sako:2021fyf}
Takashi Sako.
\newblock {Current status of ALPACA for exploring sub-PeV gamma-ray sky in
  Bolivia}.
\newblock {\em PoS}, ICRC2021:733, 2021.

\bibitem{Adams:2021hiq}
C.~B. Adams et~al.
\newblock {Prototype Schwarzschild-Couder Telescope for the Cherenkov Telescope
  Array: Commissioning the Optical System}.
\newblock {\em PoS}, ICRC2021:717, 2021.

\bibitem{2017ICRC...35..855M}
M.~C. {Maccarone} and Cta {Astri Project}.
\newblock {ASTRI for the Cherenkov Telescope Array}.
\newblock In {\em 35th International Cosmic Ray Conference (ICRC2017)}, volume
  301 of {\em International Cosmic Ray Conference}, page 855, January 2017.

\bibitem{Hofmann1999ComparisonOT}
W.~Hofmann, Ira Jung, Alexander~K. Konopelko, Henric Krawczynski, Hubert
  Lampeitl, and Gerd Puehlhofer.
\newblock {Comparison of techniques to reconstruct VHE gamma-ray showers from
  multiple stereoscopic Cherenkov images}.
\newblock {\em Astroparticle Physics}, 12:135--143, 1999.

\bibitem{Impact2014}
R.D. Parsons and J.A. Hinton.
\newblock {A Monte Carlo template based analysis for air-Cherenkov arrays}.
\newblock {\em Astroparticle Physics}, 56:26–34, Apr 2014.

\bibitem{1989ApJ...342..379W}
T.~C. {Weekes}, M.~F. {Cawley}, D.~J. {Fegan}, K.~G. {Gibbs}, A.~M. {Hillas},
  P.~W. {Kowk}, R.~C. {Lamb}, D.~A. {Lewis}, D.~{Macomb}, N.~A. {Porter}, P.~T.
  {Reynolds}, and G.~{Vacanti}.
\newblock {Observation of TeV Gamma Rays from the Crab Nebula Using the
  Atmospheric Cerenkov Imaging Technique}.
\newblock {\em Astrophysical Journal}, 342:379, July 1989.

\bibitem{Aharonian:2006pe}
F.~Aharonian et~al.
\newblock {Observations of the Crab Nebula with H.E.S.S}.
\newblock {\em Astron. Astrophys.}, 457:899--915, 2006.

\bibitem{2016APh....72...61A}
J.~{Aleksi{\'c}}, S.~{Ansoldi}, L.~A. {Antonelli}, P.~{Antoranz}, A.~{Babic},
  P.~{Bangale}, M.~{Barcel{\'o}}, J.~A. {Barrio}, J.~{Becerra Gonz{\'a}lez},
  W.~{Bednarek}, E.~{Bernardini}, B.~{Biasuzzi}, A.~{Biland}, M.~{Bitossi},
  O.~{Blanch}, S.~{Bonnefoy}, G.~{Bonnoli}, F.~{Borracci}, T.~{Bretz},
  E.~{Carmona}, A.~{Carosi}, R.~{Cecchi}, P.~{Colin}, E.~{Colombo}, J.~L.
  {Contreras}, D.~{Corti}, J.~{Cortina}, S.~{Covino}, P.~{Da Vela}, F.~{Dazzi},
  A.~{DeAngelis}, G.~{De Caneva}, B.~{De Lotto}, E.~{de O{\~n}a Wilhelmi},
  C.~{Delgado Mendez}, A.~{Dettlaff}, D.~{Dominis Prester}, D.~{Dorner},
  M.~{Doro}, S.~{Einecke}, D.~{Eisenacher}, D.~{Elsaesser}, D.~{Fidalgo},
  D.~{Fink}, M.~V. {Fonseca}, L.~{Font}, K.~{Frantzen}, C.~{Fruck},
  D.~{Galindo}, R.~J. {Garc{\'\i}a L{\'o}pez}, M.~{Garczarczyk}, D.~{Garrido
  Terrats}, M.~{Gaug}, G.~{Giavitto}, N.~{Godinovi{\'c}}, A.~{Gonz{\'a}lez
  Mu{\~n}oz}, S.~R. {Gozzini}, W.~{Haberer}, D.~{Hadasch}, Y.~{Hanabata},
  M.~{Hayashida}, J.~{Herrera}, D.~{Hildebrand }, J.~{Hose}, D.~{Hrupec},
  W.~{Idec}, J.~M. {Illa}, V.~{Kadenius}, H.~{Kellermann}, M.~L. {Knoetig},
  K.~{Kodani}, Y.~{Konno}, J.~{Krause}, H.~{Kubo}, J.~{Kushida}, A.~{La
  Barbera}, D.~{Lelas}, J.~L. {Lemus}, N.~{Lewandowska}, E.~{Lindfors},
  S.~{Lombardi}, F.~{Longo}, M.~{L{\'o}pez}, R.~{L{\'o}pez-Coto},
  A.~{L{\'o}pez-Oramas}, A.~{Lorca}, E.~{Lorenz}, I.~{Lozano}, M.~{Makariev},
  K.~{Mallot}, G.~{Maneva}, N.~{Mankuzhiyil}, K.~{Mannheim}, L.~{Maraschi},
  B.~{Marcote}, M.~{Mariotti}, M.~{Mart{\'\i}nez}, D.~{Mazin}, U.~{Menzel},
  J.~M. {Miranda}, R.~{Mirzoyan}, A.~{Moralejo}, P.~{Munar-Adrover},
  D.~{Nakajima}, M.~{Negrello}, V.~{Neustroev}, A.~{Niedzwiecki}, K.~{Nilsson},
  K.~{Nishijima}, K.~{Noda}, R.~{Orito}, A.~{Overkemping}, S.~{Paiano},
  M.~{Palatiello}, D.~{Paneque}, R.~{Paoletti}, J.~M. {Paredes},
  X.~{Paredes-Fortuny}, M.~{Persic}, J.~{Poutanen}, P.~G. {Prada Moroni},
  E.~{Prandini}, I.~{Puljak}, R.~{Reinthal}, W.~{Rhode}, M.~{Rib{\'o}},
  J.~{Rico}, J.~{Rodriguez Garcia}, S.~{R{\"u}gamer}, T.~{Saito}, K.~{Saito},
  K.~{Satalecka}, V.~{Scalzotto}, V.~{Scapin}, C.~{Schultz}, J.~{Schlammer},
  S.~{Schmidl}, T.~{Schweizer}, A.~{Sillanp{\"a}{\"a}}, J.~{Sitarek},
  I.~{Snidaric}, D.~{Sobczynska}, F.~{Spanier}, A.~{Stamerra}, T.~{Steinbring},
  J.~{Storz}, M.~{Strzys}, L.~{Takalo}, H.~{Takami}, F.~{Tavecchio}, L.~A.
  {Tejedor}, P.~{Temnikov}, T.~{Terzi{\'c}}, D.~{Tescaro}, M.~{Teshima},
  J.~{Thaele}, O.~{Tibolla}, D.~F. {Torres}, T.~{Toyama}, A.~{Treves},
  P.~{Vogler}, H.~{Wetteskind}, M.~{Will}, and R.~{Zanin}.
\newblock {The major upgrade of the MAGIC telescopes, Part I: The hardware
  improvements and the commissioning of the system}.
\newblock {\em Astroparticle Physics}, 72:61--75, January 2016.

\bibitem{Anderhub:2013cqa}
H.~Anderhub et~al.
\newblock {Design and Operation of FACT -- The First G-APD Cherenkov
  Telescope}.
\newblock {\em JINST}, 8:P06008, 2013.

\bibitem{spiacs}
V.~{Savchenko}, C.~{Ferrigno}, E.~{Kuulkers}, A.~{Bazzano}, E.~{Bozzo},
  S.~{Brandt}, J.~{Chenevez}, T.~J.~L. {Courvoisier}, R.~{Diehl}, A.~{Domingo},
  L.~{Hanlon}, E.~{Jourdain}, A.~{von Kienlin}, P.~{Laurent}, F.~{Lebrun},
  A.~{Lutovinov}, A.~{Martin-Carrillo}, S.~{Mereghetti}, L.~{Natalucci},
  J.~{Rodi}, J.~P. {Roques}, R.~{Sunyaev}, and P.~{Ubertini}.
\newblock {INTEGRAL Detection of the First Prompt Gamma-Ray Signal Coincident
  with the Gravitational-wave Event GW170817}.
\newblock {\em \apjl}, 848(2):L15, October 2017.

\bibitem{Abbott2017}
B.~P. Abbott, R.~Abbott, T.~D. Abbott, F.~Acernese, K.~Ackley, C.~Adams,
  T.~Adams, P.~Addesso, R.~X. Adhikari, V.~B. Adya, C.~Affeldt, M.~Afrough,
  B.~Agarwal, M.~Agathos, K.~Agatsuma, N.~Aggarwal, O.~D. Aguiar, L.~Aiello,
  A.~Ain, P.~Ajith, B.~Allen, G.~Allen, A.~Allocca, P.~A. Altin, A.~Amato,
  A.~Ananyeva, S.~B. Anderson, W.~G. Anderson, S.~V. Angelova, S.~Antier,
  S.~Appert, K.~Arai, M.~C. Araya, J.~S. Areeda, N.~Arnaud, K.~G. Arun,
  S.~Ascenzi, G.~Ashton, M.~Ast, S.~M. Aston, P.~Astone, D.~V. Atallah,
  P.~Aufmuth, C.~Aulbert, K.~AultONeal, C.~Austin, A.~Avila-Alvarez, S.~Babak,
  P.~Bacon, M.~K.~M. Bader, S.~Bae, M.~Bailes, P.~T. Baker, F.~Baldaccini,
  G.~Ballardin, S.~W. Ballmer, S.~Banagiri, J.~C. Barayoga, S.~E. Barclay,
  B.~C. Barish, D.~Barker, K.~Barkett, F.~Barone, B.~Barr, L.~Barsotti,
  M.~Barsuglia, D.~Barta, S.~D. Barthelmy, J.~Bartlett, I.~Bartos, R.~Bassiri,
  A.~Basti, J.~C. Batch, M.~Bawaj, J.~C. Bayley, M.~Bazzan, B.~B\'ecsy,
  C.~Beer, M.~Bejger, I.~Belahcene, A.~S. Bell, B.~K. Berger, G.~Bergmann,
  S.~Bernuzzi, J.~J. Bero, C.~P.~L. Berry, D.~Bersanetti, A.~Bertolini,
  J.~Betzwieser, S.~Bhagwat, R.~Bhandare, I.~A. Bilenko, G.~Billingsley, C.~R.
  Billman, J.~Birch, R.~Birney, O.~Birnholtz, S.~Biscans, S.~Biscoveanu,
  A.~Bisht, M.~Bitossi, C.~Biwer, M.~A. Bizouard, J.~K. Blackburn, J.~Blackman,
  C.~D. Blair, D.~G. Blair, R.~M. Blair, S.~Bloemen, O.~Bock, N.~Bode, M.~Boer,
  G.~Bogaert, A.~Bohe, F.~Bondu, E.~Bonilla, R.~Bonnand, B.~A. Boom, R.~Bork,
  V.~Boschi, S.~Bose, K.~Bossie, Y.~Bouffanais, A.~Bozzi, C.~Bradaschia, P.~R.
  Brady, M.~Branchesi, J.~E. Brau, T.~Briant, A.~Brillet, M.~Brinkmann,
  V.~Brisson, P.~Brockill, J.~E. Broida, A.~F. Brooks, D.~A. Brown, D.~D.
  Brown, S.~Brunett, C.~C. Buchanan, A.~Buikema, T.~Bulik, H.~J. Bulten,
  A.~Buonanno, D.~Buskulic, C.~Buy, R.~L. Byer, M.~Cabero, L.~Cadonati,
  G.~Cagnoli, C.~Cahillane, J.~Calder\'on~Bustillo, T.~A. Callister,
  E.~Calloni, J.~B. Camp, M.~Canepa, P.~Canizares, K.~C. Cannon, H.~Cao,
  J.~Cao, C.~D. Capano, E.~Capocasa, F.~Carbognani, S.~Caride, M.~F. Carney,
  G.~Carullo, J.~Casanueva~Diaz, C.~Casentini, S.~Caudill, M.~Cavagli\`a,
  F.~Cavalier, R.~Cavalieri, G.~Cella, C.~B. Cepeda, P.~Cerd\'a-Dur\'an,
  G.~Cerretani, E.~Cesarini, S.~J. Chamberlin, M.~Chan, S.~Chao, P.~Charlton,
  E.~Chase, E.~Chassande-Mottin, D.~Chatterjee, K.~Chatziioannou, B.~D.
  Cheeseboro, H.~Y. Chen, X.~Chen, Y.~Chen, H.-P. Cheng, H.~Chia,
  A.~Chincarini, A.~Chiummo, T.~Chmiel, H.~S. Cho, M.~Cho, J.~H. Chow,
  N.~Christensen, Q.~Chu, A.~J.~K. Chua, S.~Chua, A.~K.~W. Chung, S.~Chung,
  G.~Ciani, R.~Ciolfi, C.~E. Cirelli, A.~Cirone, F.~Clara, J.~A. Clark,
  P.~Clearwater, F.~Cleva, C.~Cocchieri, E.~Coccia, P.-F. Cohadon, D.~Cohen,
  A.~Colla, C.~G. Collette, L.~R. Cominsky, M.~Constancio, L.~Conti, S.~J.
  Cooper, P.~Corban, T.~R. Corbitt, I.~Cordero-Carri\'on, K.~R. Corley,
  N.~Cornish, A.~Corsi, S.~Cortese, C.~A. Costa, M.~W. Coughlin, S.~B.
  Coughlin, J.-P. Coulon, S.~T. Countryman, P.~Couvares, P.~B. Covas, E.~E.
  Cowan, D.~M. Coward, M.~J. Cowart, D.~C. Coyne, R.~Coyne, J.~D.~E. Creighton,
  T.~D. Creighton, J.~Cripe, S.~G. Crowder, T.~J. Cullen, A.~Cumming,
  L.~Cunningham, E.~Cuoco, T.~Dal~Canton, G.~D\'alya, S.~L. Danilishin,
  S.~D'Antonio, K.~Danzmann, A.~Dasgupta, C.~F. Da~Silva~Costa, V.~Dattilo,
  I.~Dave, M.~Davier, D.~Davis, E.~J. Daw, B.~Day, S.~De, D.~DeBra,
  J.~Degallaix, M.~De~Laurentis, S.~Del\'eglise, W.~Del~Pozzo, N.~Demos,
  T.~Denker, T.~Dent, R.~De~Pietri, V.~Dergachev, R.~De~Rosa, R.~T. DeRosa,
  C.~De~Rossi, R.~DeSalvo, O.~de~Varona, J.~Devenson, S.~Dhurandhar, M.~C.
  D\'{\i}az, T.~Dietrich, L.~Di~Fiore, M.~Di~Giovanni, T.~Di~Girolamo,
  A.~Di~Lieto, S.~Di~Pace, I.~Di~Palma, F.~Di~Renzo, Z.~Doctor, V.~Dolique,
  F.~Donovan, K.~L. Dooley, S.~Doravari, I.~Dorrington, R.~Douglas,
  M.~Dovale~\'Alvarez, T.~P. Downes, M.~Drago, C.~Dreissigacker, J.~C.
  Driggers, Z.~Du, M.~Ducrot, R.~Dudi, P.~Dupej, S.~E. Dwyer, T.~B. Edo, M.~C.
  Edwards, A.~Effler, H.-B. Eggenstein, P.~Ehrens, J.~Eichholz, S.~S.
  Eikenberry, R.~A. Eisenstein, R.~C. Essick, D.~Estevez, Z.~B. Etienne,
  T.~Etzel, M.~Evans, T.~M. Evans, M.~Factourovich, V.~Fafone, H.~Fair,
  S.~Fairhurst, X.~Fan, S.~Farinon, B.~Farr, W.~M. Farr, E.~J. Fauchon-Jones,
  M.~Favata, M.~Fays, C.~Fee, H.~Fehrmann, J.~Feicht, M.~M. Fejer,
  A.~Fernandez-Galiana, I.~Ferrante, E.~C. Ferreira, F.~Ferrini, F.~Fidecaro,
  D.~Finstad, I.~Fiori, D.~Fiorucci, M.~Fishbach, R.~P. Fisher, M.~Fitz-Axen,
  R.~Flaminio, M.~Fletcher, H.~Fong, J.~A. Font, P.~W.~F. Forsyth, S.~S.
  Forsyth, J.-D. Fournier, S.~Frasca, F.~Frasconi, Z.~Frei, A.~Freise, R.~Frey,
  V.~Frey, E.~M. Fries, P.~Fritschel, V.~V. Frolov, P.~Fulda, M.~Fyffe,
  H.~Gabbard, B.~U. Gadre, S.~M. Gaebel, J.~R. Gair, L.~Gammaitoni, M.~R.
  Ganija, S.~G. Gaonkar, C.~Garcia-Quiros, F.~Garufi, B.~Gateley, S.~Gaudio,
  G.~Gaur, V.~Gayathri, N.~Gehrels, G.~Gemme, E.~Genin, A.~Gennai, D.~George,
  J.~George, L.~Gergely, V.~Germain, S.~Ghonge, Abhirup Ghosh, Archisman Ghosh,
  S.~Ghosh, J.~A. Giaime, K.~D. Giardina, A.~Giazotto, K.~Gill, L.~Glover,
  E.~Goetz, R.~Goetz, S.~Gomes, B.~Goncharov, G.~Gonz\'alez, J.~M.
  Gonzalez~Castro, A.~Gopakumar, M.~L. Gorodetsky, S.~E. Gossan, M.~Gosselin,
  R.~Gouaty, A.~Grado, C.~Graef, M.~Granata, A.~Grant, S.~Gras, C.~Gray,
  G.~Greco, A.~C. Green, E.~M. Gretarsson, P.~Groot, H.~Grote, S.~Grunewald,
  P.~Gruning, G.~M. Guidi, X.~Guo, A.~Gupta, M.~K. Gupta, K.~E. Gushwa, E.~K.
  Gustafson, R.~Gustafson, O.~Halim, B.~R. Hall, E.~D. Hall, E.~Z. Hamilton,
  G.~Hammond, M.~Haney, M.~M. Hanke, J.~Hanks, C.~Hanna, M.~D. Hannam, O.~A.
  Hannuksela, J.~Hanson, T.~Hardwick, J.~Harms, G.~M. Harry, I.~W. Harry, M.~J.
  Hart, C.-J. Haster, K.~Haughian, J.~Healy, A.~Heidmann, M.~C. Heintze,
  H.~Heitmann, P.~Hello, G.~Hemming, M.~Hendry, I.~S. Heng, J.~Hennig, A.~W.
  Heptonstall, M.~Heurs, S.~Hild, T.~Hinderer, W.~C.~G. Ho, D.~Hoak, D.~Hofman,
  K.~Holt, D.~E. Holz, P.~Hopkins, C.~Horst, J.~Hough, E.~A. Houston, E.~J.
  Howell, A.~Hreibi, Y.~M. Hu, E.~A. Huerta, D.~Huet, B.~Hughey, S.~Husa, S.~H.
  Huttner, T.~Huynh-Dinh, N.~Indik, R.~Inta, G.~Intini, H.~N. Isa, J.-M. Isac,
  M.~Isi, B.~R. Iyer, K.~Izumi, T.~Jacqmin, K.~Jani, P.~Jaranowski, S.~Jawahar,
  F.~Jim\'enez-Forteza, W.~W. Johnson, N.~K. Johnson-McDaniel, D.~I. Jones,
  R.~Jones, R.~J.~G. Jonker, L.~Ju, J.~Junker, C.~V. Kalaghatgi, V.~Kalogera,
  B.~Kamai, S.~Kandhasamy, G.~Kang, J.~B. Kanner, S.~J. Kapadia, S.~Karki,
  K.~S. Karvinen, M.~Kasprzack, W.~Kastaun, M.~Katolik, E.~Katsavounidis,
  W.~Katzman, S.~Kaufer, K.~Kawabe, F.~K\'ef\'elian, D.~Keitel, A.~J. Kemball,
  R.~Kennedy, C.~Kent, J.~S. Key, F.~Y. Khalili, I.~Khan, S.~Khan, Z.~Khan,
  E.~A. Khazanov, N.~Kijbunchoo, Chunglee Kim, J.~C. Kim, K.~Kim, W.~Kim, W.~S.
  Kim, Y.-M. Kim, S.~J. Kimbrell, E.~J. King, P.~J. King, M.~Kinley-Hanlon,
  R.~Kirchhoff, J.~S. Kissel, L.~Kleybolte, S.~Klimenko, T.~D. Knowles,
  P.~Koch, S.~M. Koehlenbeck, S.~Koley, V.~Kondrashov, A.~Kontos, M.~Korobko,
  W.~Z. Korth, I.~Kowalska, D.~B. Kozak, C.~Kr\"amer, V.~Kringel, B.~Krishnan,
  A.~Kr\'olak, G.~Kuehn, P.~Kumar, R.~Kumar, S.~Kumar, L.~Kuo, A.~Kutynia,
  S.~Kwang, B.~D. Lackey, K.~H. Lai, M.~Landry, R.~N. Lang, J.~Lange, B.~Lantz,
  R.~K. Lanza, S.~L. Larson, A.~Lartaux-Vollard, P.~D. Lasky, M.~Laxen,
  A.~Lazzarini, C.~Lazzaro, P.~Leaci, S.~Leavey, C.~H. Lee, H.~K. Lee, H.~M.
  Lee, H.~W. Lee, K.~Lee, J.~Lehmann, A.~Lenon, E.~Leon, M.~Leonardi, N.~Leroy,
  N.~Letendre, Y.~Levin, T.~G.~F. Li, S.~D. Linker, T.~B. Littenberg, J.~Liu,
  X.~Liu, R.~K.~L. Lo, N.~A. Lockerbie, L.~T. London, J.~E. Lord, M.~Lorenzini,
  V.~Loriette, M.~Lormand, G.~Losurdo, J.~D. Lough, C.~O. Lousto, G.~Lovelace,
  H.~L\"uck, D.~Lumaca, A.~P. Lundgren, R.~Lynch, Y.~Ma, R.~Macas, S.~Macfoy,
  B.~Machenschalk, M.~MacInnis, D.~M. Macleod, I.~Maga\~na Hernandez,
  F.~Maga\~na Sandoval, L.~Maga\~na Zertuche, R.~M. Magee, E.~Majorana,
  I.~Maksimovic, N.~Man, V.~Mandic, V.~Mangano, G.~L. Mansell, M.~Manske,
  M.~Mantovani, F.~Marchesoni, F.~Marion, S.~M\'arka, Z.~M\'arka, C.~Markakis,
  A.~S. Markosyan, A.~Markowitz, E.~Maros, A.~Marquina, P.~Marsh, F.~Martelli,
  L.~Martellini, I.~W. Martin, R.~M. Martin, D.~V. Martynov, J.~N. Marx,
  K.~Mason, E.~Massera, A.~Masserot, T.~J. Massinger, M.~Masso-Reid,
  S.~Mastrogiovanni, A.~Matas, F.~Matichard, L.~Matone, N.~Mavalvala,
  N.~Mazumder, R.~McCarthy, D.~E. McClelland, S.~McCormick, L.~McCuller, S.~C.
  McGuire, G.~McIntyre, J.~McIver, D.~J. McManus, L.~McNeill, T.~McRae, S.~T.
  McWilliams, D.~Meacher, G.~D. Meadors, M.~Mehmet, J.~Meidam, E.~Mejuto-Villa,
  A.~Melatos, G.~Mendell, R.~A. Mercer, E.~L. Merilh, M.~Merzougui, S.~Meshkov,
  C.~Messenger, C.~Messick, R.~Metzdorff, P.~M. Meyers, H.~Miao, C.~Michel,
  H.~Middleton, E.~E. Mikhailov, L.~Milano, A.~L. Miller, B.~B. Miller,
  J.~Miller, M.~Millhouse, M.~C. Milovich-Goff, O.~Minazzoli, Y.~Minenkov,
  J.~Ming, C.~Mishra, S.~Mitra, V.~P. Mitrofanov, G.~Mitselmakher,
  R.~Mittleman, D.~Moffa, A.~Moggi, K.~Mogushi, M.~Mohan, S.~R.~P. Mohapatra,
  I.~Molina, M.~Montani, C.~J. Moore, D.~Moraru, G.~Moreno, S.~Morisaki, S.~R.
  Morriss, B.~Mours, C.~M. Mow-Lowry, G.~Mueller, A.~W. Muir, Arunava
  Mukherjee, D.~Mukherjee, S.~Mukherjee, N.~Mukund, A.~Mullavey, J.~Munch,
  E.~A. Mu\~niz, M.~Muratore, P.~G. Murray, A.~Nagar, K.~Napier, I.~Nardecchia,
  L.~Naticchioni, R.~K. Nayak, J.~Neilson, G.~Nelemans, T.~J.~N. Nelson,
  M.~Nery, A.~Neunzert, L.~Nevin, J.~M. Newport, G.~Newton, K.~K.~Y. Ng,
  P.~Nguyen, T.~T. Nguyen, D.~Nichols, A.~B. Nielsen, S.~Nissanke, A.~Nitz,
  A.~Noack, F.~Nocera, D.~Nolting, C.~North, L.~K. Nuttall, J.~Oberling, G.~D.
  O'Dea, G.~H. Ogin, J.~J. Oh, S.~H. Oh, F.~Ohme, M.~A. Okada, M.~Oliver,
  P.~Oppermann, Richard~J. Oram, B.~O'Reilly, R.~Ormiston, L.~F. Ortega,
  R.~O'Shaughnessy, S.~Ossokine, D.~J. Ottaway, H.~Overmier, B.~J. Owen, A.~E.
  Pace, J.~Page, M.~A. Page, A.~Pai, S.~A. Pai, J.~R. Palamos, O.~Palashov,
  C.~Palomba, A.~Pal-Singh, Howard Pan, Huang-Wei Pan, B.~Pang, P.~T.~H. Pang,
  C.~Pankow, F.~Pannarale, B.~C. Pant, F.~Paoletti, A.~Paoli, M.~A. Papa,
  A.~Parida, W.~Parker, D.~Pascucci, A.~Pasqualetti, R.~Passaquieti,
  D.~Passuello, M.~Patil, B.~Patricelli, B.~L. Pearlstone, M.~Pedraza,
  R.~Pedurand, L.~Pekowsky, A.~Pele, S.~Penn, C.~J. Perez, A.~Perreca, L.~M.
  Perri, H.~P. Pfeiffer, M.~Phelps, O.~J. Piccinni, M.~Pichot, F.~Piergiovanni,
  V.~Pierro, G.~Pillant, L.~Pinard, I.~M. Pinto, M.~Pirello, M.~Pitkin, M.~Poe,
  R.~Poggiani, P.~Popolizio, E.~K. Porter, A.~Post, J.~Powell, J.~Prasad,
  J.~W.~W. Pratt, G.~Pratten, V.~Predoi, T.~Prestegard, M.~Prijatelj,
  M.~Principe, S.~Privitera, R.~Prix, G.~A. Prodi, L.~G. Prokhorov, O.~Puncken,
  M.~Punturo, P.~Puppo, M.~P\"urrer, H.~Qi, V.~Quetschke, E.~A. Quintero,
  R.~Quitzow-James, F.~J. Raab, D.~S. Rabeling, H.~Radkins, P.~Raffai, S.~Raja,
  C.~Rajan, B.~Rajbhandari, M.~Rakhmanov, K.~E. Ramirez, A.~Ramos-Buades,
  P.~Rapagnani, V.~Raymond, M.~Razzano, J.~Read, T.~Regimbau, L.~Rei, S.~Reid,
  D.~H. Reitze, W.~Ren, S.~D. Reyes, F.~Ricci, P.~M. Ricker, S.~Rieger,
  K.~Riles, M.~Rizzo, N.~A. Robertson, R.~Robie, F.~Robinet, A.~Rocchi,
  L.~Rolland, J.~G. Rollins, V.~J. Roma, J.~D. Romano, R.~Romano, C.~L. Romel,
  J.~H. Romie, D.~Rosi\ifmmode~\acute{n}\else \'{n}\fi{}ska, M.~P. Ross,
  S.~Rowan, A.~R\"udiger, P.~Ruggi, G.~Rutins, K.~Ryan, S.~Sachdev, T.~Sadecki,
  L.~Sadeghian, M.~Sakellariadou, L.~Salconi, M.~Saleem, F.~Salemi,
  A.~Samajdar, L.~Sammut, L.~M. Sampson, E.~J. Sanchez, L.~E. Sanchez,
  N.~Sanchis-Gual, V.~Sandberg, J.~R. Sanders, B.~Sassolas, B.~S.
  Sathyaprakash, P.~R. Saulson, O.~Sauter, R.~L. Savage, A.~Sawadsky,
  P.~Schale, M.~Scheel, J.~Scheuer, J.~Schmidt, P.~Schmidt, R.~Schnabel,
  R.~M.~S. Schofield, A.~Sch\"onbeck, E.~Schreiber, D.~Schuette, B.~W. Schulte,
  B.~F. Schutz, S.~G. Schwalbe, J.~Scott, S.~M. Scott, E.~Seidel, D.~Sellers,
  A.~S. Sengupta, D.~Sentenac, V.~Sequino, A.~Sergeev, D.~A. Shaddock, T.~J.
  Shaffer, A.~A. Shah, M.~S. Shahriar, M.~B. Shaner, L.~Shao, B.~Shapiro,
  P.~Shawhan, A.~Sheperd, D.~H. Shoemaker, D.~M. Shoemaker, K.~Siellez,
  X.~Siemens, M.~Sieniawska, D.~Sigg, A.~D. Silva, L.~P. Singer, A.~Singh,
  A.~Singhal, A.~M. Sintes, B.~J.~J. Slagmolen, B.~Smith, J.~R. Smith, R.~J.~E.
  Smith, S.~Somala, E.~J. Son, J.~A. Sonnenberg, B.~Sorazu, F.~Sorrentino,
  T.~Souradeep, A.~P. Spencer, A.~K. Srivastava, K.~Staats, A.~Staley,
  M.~Steinke, J.~Steinlechner, S.~Steinlechner, D.~Steinmeyer, S.~P. Stevenson,
  R.~Stone, D.~J. Stops, K.~A. Strain, G.~Stratta, S.~E. Strigin, A.~Strunk,
  R.~Sturani, A.~L. Stuver, T.~Z. Summerscales, L.~Sun, S.~Sunil, J.~Suresh,
  P.~J. Sutton, B.~L. Swinkels, M.~J. Szczepa\ifmmode~\acute{n}\else
  \'{n}\fi{}czyk, M.~Tacca, S.~C. Tait, C.~Talbot, D.~Talukder, D.~B. Tanner,
  M.~T\'apai, A.~Taracchini, J.~D. Tasson, J.~A. Taylor, R.~Taylor, S.~V.
  Tewari, T.~Theeg, F.~Thies, E.~G. Thomas, M.~Thomas, P.~Thomas, K.~A. Thorne,
  K.~S. Thorne, E.~Thrane, S.~Tiwari, V.~Tiwari, K.~V. Tokmakov, K.~Toland,
  M.~Tonelli, Z.~Tornasi, A.~Torres-Forn\'e, C.~I. Torrie, D.~T\"oyr\"a,
  F.~Travasso, G.~Traylor, J.~Trinastic, M.~C. Tringali, L.~Trozzo, K.~W.
  Tsang, M.~Tse, R.~Tso, L.~Tsukada, D.~Tsuna, D.~Tuyenbayev, K.~Ueno,
  D.~Ugolini, C.~S. Unnikrishnan, A.~L. Urban, S.~A. Usman, H.~Vahlbruch,
  G.~Vajente, G.~Valdes, M.~Vallisneri, N.~van Bakel, M.~van Beuzekom, J.~F.~J.
  van~den Brand, C.~Van Den~Broeck, D.~C. Vander-Hyde, L.~van~der Schaaf, J.~V.
  van Heijningen, A.~A. van Veggel, M.~Vardaro, V.~Varma, S.~Vass, M.~Vas\'uth,
  A.~Vecchio, G.~Vedovato, J.~Veitch, P.~J. Veitch, K.~Venkateswara,
  G.~Venugopalan, D.~Verkindt, F.~Vetrano, A.~Vicer\'e, A.~D. Viets,
  S.~Vinciguerra, D.~J. Vine, J.-Y. Vinet, S.~Vitale, T.~Vo, H.~Vocca,
  C.~Vorvick, S.~P. Vyatchanin, A.~R. Wade, L.~E. Wade, M.~Wade, R.~Walet,
  M.~Walker, L.~Wallace, S.~Walsh, G.~Wang, H.~Wang, J.~Z. Wang, W.~H. Wang,
  Y.~F. Wang, R.~L. Ward, J.~Warner, M.~Was, J.~Watchi, B.~Weaver, L.-W. Wei,
  M.~Weinert, A.~J. Weinstein, R.~Weiss, L.~Wen, E.~K. Wessel, P.~We\ss{}els,
  J.~Westerweck, T.~Westphal, K.~Wette, J.~T. Whelan, S.~E. Whitcomb, B.~F.
  Whiting, C.~Whittle, D.~Wilken, D.~Williams, R.~D. Williams, A.~R.
  Williamson, J.~L. Willis, B.~Willke, M.~H. Wimmer, W.~Winkler, C.~C. Wipf,
  H.~Wittel, G.~Woan, J.~Woehler, J.~Wofford, K.~W.~K. Wong, J.~Worden, J.~L.
  Wright, D.~S. Wu, D.~M. Wysocki, S.~Xiao, H.~Yamamoto, C.~C. Yancey, L.~Yang,
  M.~J. Yap, M.~Yazback, Hang Yu, Haocun Yu, M.~Yvert,
  A.~Zadro\ifmmode~\dot{z}\else \.{z}\fi{}ny, M.~Zanolin, T.~Zelenova, J.-P.
  Zendri, M.~Zevin, L.~Zhang, M.~Zhang, T.~Zhang, Y.-H. Zhang, C.~Zhao,
  M.~Zhou, Z.~Zhou, S.~J. Zhu, X.~J. Zhu, A.~B. Zimmerman, M.~E. Zucker, and
  J.~Zweizig.
\newblock Gw170817: Observation of gravitational waves from a binary neutron
  star inspiral.
\newblock {\em Phys. Rev. Lett.}, 119:161101, Oct 2017.

\bibitem{Abbott2020}
R.~{Abbott}, T.~D. {Abbott}, S.~{Abraham}, F.~{Acernese}, et~al.
\newblock {GWTC-2: Compact Binary Coalescences Observed by LIGO and Virgo
  During the First Half of the Third Observing Run}.
\newblock {\em arXiv e-prints}, 0:arXiv:2010.14527, October 2020.

\bibitem{Abbott2018}
B.~P. {Abbott}, R.~{Abbott}, T.~D. {Abbott}, M.~R. {Abernathy}, et~al.
\newblock {Prospects for observing and localizing gravitational-wave transients
  with Advanced LIGO, Advanced Virgo and KAGRA}.
\newblock {\em Living Reviews in Relativity}, 21(1):3, April 2018.

\bibitem{perkins2020}
Jeremy~S. Perkins, Isabella Brewer, Michael~S. Briggs, Alessandro Bruno, et~al.
\newblock {BurstCube: a CubeSat for gravitational wave counterparts}.
\newblock In Jan-Willem~A. den Herder, Shouleh Nikzad, and Kazuhiro Nakazawa,
  editors, {\em Space Telescopes and Instrumentation 2020: Ultraviolet to Gamma
  Ray}, volume 11444, pages 277 -- 285. International Society for Optics and
  Photonics, SPIE, 2020.

\bibitem{Martinez2021}
Israel Martinez, Isabella Brewer, Michael~S. Briggs, Alessandro Bruno, et~al.
\newblock {BurstCube: status and public alerts}.
\newblock In {\em Proceedings of 37th International Cosmic Ray Conference
  {\textemdash} PoS(ICRC2021)}, volume 395, page 656, 2021.

\bibitem{StarBurst}
Elizabeth Landau.
\newblock Nasa selects 4 concepts for small missions to study universe’s
  secrets.
\newblock Technical Report Jan2021, NASA, 2021.
\newblock
  \url{https://www.nasa.gov/feature/nasa-selects-4-concepts-for-small-missions-to-study-universe-s-secret}.

\bibitem{Grove2020}
J.~E. {Grove}, C.~C. {Cheung}, M.~{Kerr}, L.~J. {Mitchell}, B.~F. {Phlips},
  R.~S. {Woolf}, C.~{Wilson-Hodge}, D.~{Kocevski}, and M.~S. {Briggs}.
\newblock {Glowbug, a Gamma-Ray Telescope for Bursts and Other Transients
  (including TGFs!)}.
\newblock In {\em AGU Fall Meeting Abstracts}, volume 2020, pages AE005--07,
  December 2020.

\bibitem{2020ascl.soft10002F}
{Fermi Science Support Development Team}.
\newblock {GSpec: Gamma-ray Burst Monitor analyzer}, October 2020.
\newblock
  \url{https://fermi.gsfc.nasa.gov/ssc/data/analysis/gbm/gbm_data_tools/gdt-docs/}.

\bibitem{2020grbg.conf...57G}
J.~E. {Grove}, C.~C. {Cheung}, M.~{Kerr}, L.~J. {Mitchell}, B.~F. {Phlips},
  R.~S. {Woolf}, E.~A. {Wulf}, M.~S. {Briggs}, C.~A. {Wilson-Hodge},
  D.~{Kocevski}, and J.~{Perkins}.
\newblock {Glowbug, a Low-Cost, High-Sensitivity Gamma-Ray Burst Telescope}.
\newblock In {\em Gamma-ray Bursts in the Gravitational Wave Era 2019}, pages
  57--59, May 2020.

\bibitem{PhysRevLett.119.161101}
B.~P. Abbott, R.~Abbott, T.~D. Abbott, F.~Acernese, K.~Ackley, C.~Adams,
  T.~Adams, P.~Addesso, R.~X. Adhikari, V.~B. Adya, C.~Affeldt, M.~Afrough,
  B.~Agarwal, M.~Agathos, K.~Agatsuma, N.~Aggarwal, O.~D. Aguiar, L.~Aiello,
  A.~Ain, P.~Ajith, B.~Allen, G.~Allen, A.~Allocca, P.~A. Altin, A.~Amato,
  A.~Ananyeva, S.~B. Anderson, W.~G. Anderson, S.~V. Angelova, S.~Antier,
  S.~Appert, K.~Arai, M.~C. Araya, J.~S. Areeda, N.~Arnaud, K.~G. Arun,
  S.~Ascenzi, G.~Ashton, M.~Ast, S.~M. Aston, P.~Astone, D.~V. Atallah,
  P.~Aufmuth, C.~Aulbert, K.~AultONeal, C.~Austin, A.~Avila-Alvarez, S.~Babak,
  P.~Bacon, M.~K.~M. Bader, S.~Bae, M.~Bailes, P.~T. Baker, F.~Baldaccini,
  G.~Ballardin, S.~W. Ballmer, S.~Banagiri, J.~C. Barayoga, S.~E. Barclay,
  B.~C. Barish, D.~Barker, K.~Barkett, F.~Barone, B.~Barr, L.~Barsotti,
  M.~Barsuglia, D.~Barta, S.~D. Barthelmy, J.~Bartlett, I.~Bartos, R.~Bassiri,
  A.~Basti, J.~C. Batch, M.~Bawaj, J.~C. Bayley, M.~Bazzan, B.~B\'ecsy,
  C.~Beer, M.~Bejger, I.~Belahcene, A.~S. Bell, B.~K. Berger, G.~Bergmann,
  S.~Bernuzzi, J.~J. Bero, C.~P.~L. Berry, D.~Bersanetti, A.~Bertolini,
  J.~Betzwieser, S.~Bhagwat, R.~Bhandare, I.~A. Bilenko, G.~Billingsley, C.~R.
  Billman, J.~Birch, R.~Birney, O.~Birnholtz, S.~Biscans, S.~Biscoveanu,
  A.~Bisht, M.~Bitossi, C.~Biwer, M.~A. Bizouard, J.~K. Blackburn, J.~Blackman,
  C.~D. Blair, D.~G. Blair, R.~M. Blair, S.~Bloemen, O.~Bock, N.~Bode, M.~Boer,
  G.~Bogaert, A.~Bohe, F.~Bondu, E.~Bonilla, R.~Bonnand, B.~A. Boom, R.~Bork,
  V.~Boschi, S.~Bose, K.~Bossie, Y.~Bouffanais, A.~Bozzi, C.~Bradaschia, P.~R.
  Brady, M.~Branchesi, J.~E. Brau, T.~Briant, A.~Brillet, M.~Brinkmann,
  V.~Brisson, P.~Brockill, J.~E. Broida, A.~F. Brooks, D.~A. Brown, D.~D.
  Brown, S.~Brunett, C.~C. Buchanan, A.~Buikema, T.~Bulik, H.~J. Bulten,
  A.~Buonanno, D.~Buskulic, C.~Buy, R.~L. Byer, M.~Cabero, L.~Cadonati,
  G.~Cagnoli, C.~Cahillane, J.~Calder\'on~Bustillo, T.~A. Callister,
  E.~Calloni, J.~B. Camp, M.~Canepa, P.~Canizares, K.~C. Cannon, H.~Cao,
  J.~Cao, C.~D. Capano, E.~Capocasa, F.~Carbognani, S.~Caride, M.~F. Carney,
  G.~Carullo, J.~Casanueva~Diaz, C.~Casentini, S.~Caudill, M.~Cavagli\`a,
  F.~Cavalier, R.~Cavalieri, G.~Cella, C.~B. Cepeda, P.~Cerd\'a-Dur\'an,
  G.~Cerretani, E.~Cesarini, S.~J. Chamberlin, M.~Chan, S.~Chao, P.~Charlton,
  E.~Chase, E.~Chassande-Mottin, D.~Chatterjee, K.~Chatziioannou, B.~D.
  Cheeseboro, H.~Y. Chen, X.~Chen, Y.~Chen, H.-P. Cheng, H.~Chia,
  A.~Chincarini, A.~Chiummo, T.~Chmiel, H.~S. Cho, M.~Cho, J.~H. Chow,
  N.~Christensen, Q.~Chu, A.~J.~K. Chua, S.~Chua, A.~K.~W. Chung, S.~Chung,
  G.~Ciani, R.~Ciolfi, C.~E. Cirelli, A.~Cirone, F.~Clara, J.~A. Clark,
  P.~Clearwater, F.~Cleva, C.~Cocchieri, E.~Coccia, P.-F. Cohadon, D.~Cohen,
  A.~Colla, C.~G. Collette, L.~R. Cominsky, M.~Constancio, L.~Conti, S.~J.
  Cooper, P.~Corban, T.~R. Corbitt, I.~Cordero-Carri\'on, K.~R. Corley,
  N.~Cornish, A.~Corsi, S.~Cortese, C.~A. Costa, M.~W. Coughlin, S.~B.
  Coughlin, J.-P. Coulon, S.~T. Countryman, P.~Couvares, P.~B. Covas, E.~E.
  Cowan, D.~M. Coward, M.~J. Cowart, D.~C. Coyne, R.~Coyne, J.~D.~E. Creighton,
  T.~D. Creighton, J.~Cripe, S.~G. Crowder, T.~J. Cullen, A.~Cumming,
  L.~Cunningham, E.~Cuoco, T.~Dal~Canton, G.~D\'alya, S.~L. Danilishin,
  S.~D'Antonio, K.~Danzmann, A.~Dasgupta, C.~F. Da~Silva~Costa, V.~Dattilo,
  I.~Dave, M.~Davier, D.~Davis, E.~J. Daw, B.~Day, S.~De, D.~DeBra,
  J.~Degallaix, M.~De~Laurentis, S.~Del\'eglise, W.~Del~Pozzo, N.~Demos,
  T.~Denker, T.~Dent, R.~De~Pietri, V.~Dergachev, R.~De~Rosa, R.~T. DeRosa,
  C.~De~Rossi, R.~DeSalvo, O.~de~Varona, J.~Devenson, S.~Dhurandhar, M.~C.
  D\'{\i}az, T.~Dietrich, L.~Di~Fiore, M.~Di~Giovanni, T.~Di~Girolamo,
  A.~Di~Lieto, S.~Di~Pace, I.~Di~Palma, F.~Di~Renzo, Z.~Doctor, V.~Dolique,
  F.~Donovan, K.~L. Dooley, S.~Doravari, I.~Dorrington, R.~Douglas,
  M.~Dovale~\'Alvarez, T.~P. Downes, M.~Drago, C.~Dreissigacker, J.~C.
  Driggers, Z.~Du, M.~Ducrot, R.~Dudi, P.~Dupej, S.~E. Dwyer, T.~B. Edo, M.~C.
  Edwards, A.~Effler, H.-B. Eggenstein, P.~Ehrens, J.~Eichholz, S.~S.
  Eikenberry, R.~A. Eisenstein, R.~C. Essick, D.~Estevez, Z.~B. Etienne,
  T.~Etzel, M.~Evans, T.~M. Evans, M.~Factourovich, V.~Fafone, H.~Fair,
  S.~Fairhurst, X.~Fan, S.~Farinon, B.~Farr, W.~M. Farr, E.~J. Fauchon-Jones,
  M.~Favata, M.~Fays, C.~Fee, H.~Fehrmann, J.~Feicht, M.~M. Fejer,
  A.~Fernandez-Galiana, I.~Ferrante, E.~C. Ferreira, F.~Ferrini, F.~Fidecaro,
  D.~Finstad, I.~Fiori, D.~Fiorucci, M.~Fishbach, R.~P. Fisher, M.~Fitz-Axen,
  R.~Flaminio, M.~Fletcher, H.~Fong, J.~A. Font, P.~W.~F. Forsyth, S.~S.
  Forsyth, J.-D. Fournier, S.~Frasca, F.~Frasconi, Z.~Frei, A.~Freise, R.~Frey,
  V.~Frey, E.~M. Fries, P.~Fritschel, V.~V. Frolov, P.~Fulda, M.~Fyffe,
  H.~Gabbard, B.~U. Gadre, S.~M. Gaebel, J.~R. Gair, L.~Gammaitoni, M.~R.
  Ganija, S.~G. Gaonkar, C.~Garcia-Quiros, F.~Garufi, B.~Gateley, S.~Gaudio,
  G.~Gaur, V.~Gayathri, N.~Gehrels, G.~Gemme, E.~Genin, A.~Gennai, D.~George,
  J.~George, L.~Gergely, V.~Germain, S.~Ghonge, Abhirup Ghosh, Archisman Ghosh,
  S.~Ghosh, J.~A. Giaime, K.~D. Giardina, A.~Giazotto, K.~Gill, L.~Glover,
  E.~Goetz, R.~Goetz, S.~Gomes, B.~Goncharov, G.~Gonz\'alez, J.~M.
  Gonzalez~Castro, A.~Gopakumar, M.~L. Gorodetsky, S.~E. Gossan, M.~Gosselin,
  R.~Gouaty, A.~Grado, C.~Graef, M.~Granata, A.~Grant, S.~Gras, C.~Gray,
  G.~Greco, A.~C. Green, E.~M. Gretarsson, P.~Groot, H.~Grote, S.~Grunewald,
  P.~Gruning, G.~M. Guidi, X.~Guo, A.~Gupta, M.~K. Gupta, K.~E. Gushwa, E.~K.
  Gustafson, R.~Gustafson, O.~Halim, B.~R. Hall, E.~D. Hall, E.~Z. Hamilton,
  G.~Hammond, M.~Haney, M.~M. Hanke, J.~Hanks, C.~Hanna, M.~D. Hannam, O.~A.
  Hannuksela, J.~Hanson, T.~Hardwick, J.~Harms, G.~M. Harry, I.~W. Harry, M.~J.
  Hart, C.-J. Haster, K.~Haughian, J.~Healy, A.~Heidmann, M.~C. Heintze,
  H.~Heitmann, P.~Hello, G.~Hemming, M.~Hendry, I.~S. Heng, J.~Hennig, A.~W.
  Heptonstall, M.~Heurs, S.~Hild, T.~Hinderer, W.~C.~G. Ho, D.~Hoak, D.~Hofman,
  K.~Holt, D.~E. Holz, P.~Hopkins, C.~Horst, J.~Hough, E.~A. Houston, E.~J.
  Howell, A.~Hreibi, Y.~M. Hu, E.~A. Huerta, D.~Huet, B.~Hughey, S.~Husa, S.~H.
  Huttner, T.~Huynh-Dinh, N.~Indik, R.~Inta, G.~Intini, H.~N. Isa, J.-M. Isac,
  M.~Isi, B.~R. Iyer, K.~Izumi, T.~Jacqmin, K.~Jani, P.~Jaranowski, S.~Jawahar,
  F.~Jim\'enez-Forteza, W.~W. Johnson, N.~K. Johnson-McDaniel, D.~I. Jones,
  R.~Jones, R.~J.~G. Jonker, L.~Ju, J.~Junker, C.~V. Kalaghatgi, V.~Kalogera,
  B.~Kamai, S.~Kandhasamy, G.~Kang, J.~B. Kanner, S.~J. Kapadia, S.~Karki,
  K.~S. Karvinen, M.~Kasprzack, W.~Kastaun, M.~Katolik, E.~Katsavounidis,
  W.~Katzman, S.~Kaufer, K.~Kawabe, F.~K\'ef\'elian, D.~Keitel, A.~J. Kemball,
  R.~Kennedy, C.~Kent, J.~S. Key, F.~Y. Khalili, I.~Khan, S.~Khan, Z.~Khan,
  E.~A. Khazanov, N.~Kijbunchoo, Chunglee Kim, J.~C. Kim, K.~Kim, W.~Kim, W.~S.
  Kim, Y.-M. Kim, S.~J. Kimbrell, E.~J. King, P.~J. King, M.~Kinley-Hanlon,
  R.~Kirchhoff, J.~S. Kissel, L.~Kleybolte, S.~Klimenko, T.~D. Knowles,
  P.~Koch, S.~M. Koehlenbeck, S.~Koley, V.~Kondrashov, A.~Kontos, M.~Korobko,
  W.~Z. Korth, I.~Kowalska, D.~B. Kozak, C.~Kr\"amer, V.~Kringel, B.~Krishnan,
  A.~Kr\'olak, G.~Kuehn, P.~Kumar, R.~Kumar, S.~Kumar, L.~Kuo, A.~Kutynia,
  S.~Kwang, B.~D. Lackey, K.~H. Lai, M.~Landry, R.~N. Lang, J.~Lange, B.~Lantz,
  R.~K. Lanza, S.~L. Larson, A.~Lartaux-Vollard, P.~D. Lasky, M.~Laxen,
  A.~Lazzarini, C.~Lazzaro, P.~Leaci, S.~Leavey, C.~H. Lee, H.~K. Lee, H.~M.
  Lee, H.~W. Lee, K.~Lee, J.~Lehmann, A.~Lenon, E.~Leon, M.~Leonardi, N.~Leroy,
  N.~Letendre, Y.~Levin, T.~G.~F. Li, S.~D. Linker, T.~B. Littenberg, J.~Liu,
  X.~Liu, R.~K.~L. Lo, N.~A. Lockerbie, L.~T. London, J.~E. Lord, M.~Lorenzini,
  V.~Loriette, M.~Lormand, G.~Losurdo, J.~D. Lough, C.~O. Lousto, G.~Lovelace,
  H.~L\"uck, D.~Lumaca, A.~P. Lundgren, R.~Lynch, Y.~Ma, R.~Macas, S.~Macfoy,
  B.~Machenschalk, M.~MacInnis, D.~M. Macleod, I.~Maga\~na Hernandez,
  F.~Maga\~na Sandoval, L.~Maga\~na Zertuche, R.~M. Magee, E.~Majorana,
  I.~Maksimovic, N.~Man, V.~Mandic, V.~Mangano, G.~L. Mansell, M.~Manske,
  M.~Mantovani, F.~Marchesoni, F.~Marion, S.~M\'arka, Z.~M\'arka, C.~Markakis,
  A.~S. Markosyan, A.~Markowitz, E.~Maros, A.~Marquina, P.~Marsh, F.~Martelli,
  L.~Martellini, I.~W. Martin, R.~M. Martin, D.~V. Martynov, J.~N. Marx,
  K.~Mason, E.~Massera, A.~Masserot, T.~J. Massinger, M.~Masso-Reid,
  S.~Mastrogiovanni, A.~Matas, F.~Matichard, L.~Matone, N.~Mavalvala,
  N.~Mazumder, R.~McCarthy, D.~E. McClelland, S.~McCormick, L.~McCuller, S.~C.
  McGuire, G.~McIntyre, J.~McIver, D.~J. McManus, L.~McNeill, T.~McRae, S.~T.
  McWilliams, D.~Meacher, G.~D. Meadors, M.~Mehmet, J.~Meidam, E.~Mejuto-Villa,
  A.~Melatos, G.~Mendell, R.~A. Mercer, E.~L. Merilh, M.~Merzougui, S.~Meshkov,
  C.~Messenger, C.~Messick, R.~Metzdorff, P.~M. Meyers, H.~Miao, C.~Michel,
  H.~Middleton, E.~E. Mikhailov, L.~Milano, A.~L. Miller, B.~B. Miller,
  J.~Miller, M.~Millhouse, M.~C. Milovich-Goff, O.~Minazzoli, Y.~Minenkov,
  J.~Ming, C.~Mishra, S.~Mitra, V.~P. Mitrofanov, G.~Mitselmakher,
  R.~Mittleman, D.~Moffa, A.~Moggi, K.~Mogushi, M.~Mohan, S.~R.~P. Mohapatra,
  I.~Molina, M.~Montani, C.~J. Moore, D.~Moraru, G.~Moreno, S.~Morisaki, S.~R.
  Morriss, B.~Mours, C.~M. Mow-Lowry, G.~Mueller, A.~W. Muir, Arunava
  Mukherjee, D.~Mukherjee, S.~Mukherjee, N.~Mukund, A.~Mullavey, J.~Munch,
  E.~A. Mu\~niz, M.~Muratore, P.~G. Murray, A.~Nagar, K.~Napier, I.~Nardecchia,
  L.~Naticchioni, R.~K. Nayak, J.~Neilson, G.~Nelemans, T.~J.~N. Nelson,
  M.~Nery, A.~Neunzert, L.~Nevin, J.~M. Newport, G.~Newton, K.~K.~Y. Ng,
  P.~Nguyen, T.~T. Nguyen, D.~Nichols, A.~B. Nielsen, S.~Nissanke, A.~Nitz,
  A.~Noack, F.~Nocera, D.~Nolting, C.~North, L.~K. Nuttall, J.~Oberling, G.~D.
  O'Dea, G.~H. Ogin, J.~J. Oh, S.~H. Oh, F.~Ohme, M.~A. Okada, M.~Oliver,
  P.~Oppermann, Richard~J. Oram, B.~O'Reilly, R.~Ormiston, L.~F. Ortega,
  R.~O'Shaughnessy, S.~Ossokine, D.~J. Ottaway, H.~Overmier, B.~J. Owen, A.~E.
  Pace, J.~Page, M.~A. Page, A.~Pai, S.~A. Pai, J.~R. Palamos, O.~Palashov,
  C.~Palomba, A.~Pal-Singh, Howard Pan, Huang-Wei Pan, B.~Pang, P.~T.~H. Pang,
  C.~Pankow, F.~Pannarale, B.~C. Pant, F.~Paoletti, A.~Paoli, M.~A. Papa,
  A.~Parida, W.~Parker, D.~Pascucci, A.~Pasqualetti, R.~Passaquieti,
  D.~Passuello, M.~Patil, B.~Patricelli, B.~L. Pearlstone, M.~Pedraza,
  R.~Pedurand, L.~Pekowsky, A.~Pele, S.~Penn, C.~J. Perez, A.~Perreca, L.~M.
  Perri, H.~P. Pfeiffer, M.~Phelps, O.~J. Piccinni, M.~Pichot, F.~Piergiovanni,
  V.~Pierro, G.~Pillant, L.~Pinard, I.~M. Pinto, M.~Pirello, M.~Pitkin, M.~Poe,
  R.~Poggiani, P.~Popolizio, E.~K. Porter, A.~Post, J.~Powell, J.~Prasad,
  J.~W.~W. Pratt, G.~Pratten, V.~Predoi, T.~Prestegard, M.~Prijatelj,
  M.~Principe, S.~Privitera, R.~Prix, G.~A. Prodi, L.~G. Prokhorov, O.~Puncken,
  M.~Punturo, P.~Puppo, M.~P\"urrer, H.~Qi, V.~Quetschke, E.~A. Quintero,
  R.~Quitzow-James, F.~J. Raab, D.~S. Rabeling, H.~Radkins, P.~Raffai, S.~Raja,
  C.~Rajan, B.~Rajbhandari, M.~Rakhmanov, K.~E. Ramirez, A.~Ramos-Buades,
  P.~Rapagnani, V.~Raymond, M.~Razzano, J.~Read, T.~Regimbau, L.~Rei, S.~Reid,
  D.~H. Reitze, W.~Ren, S.~D. Reyes, F.~Ricci, P.~M. Ricker, S.~Rieger,
  K.~Riles, M.~Rizzo, N.~A. Robertson, R.~Robie, F.~Robinet, A.~Rocchi,
  L.~Rolland, J.~G. Rollins, V.~J. Roma, J.~D. Romano, R.~Romano, C.~L. Romel,
  J.~H. Romie, D.~Rosi\ifmmode~\acute{n}\else \'{n}\fi{}ska, M.~P. Ross,
  S.~Rowan, A.~R\"udiger, P.~Ruggi, G.~Rutins, K.~Ryan, S.~Sachdev, T.~Sadecki,
  L.~Sadeghian, M.~Sakellariadou, L.~Salconi, M.~Saleem, F.~Salemi,
  A.~Samajdar, L.~Sammut, L.~M. Sampson, E.~J. Sanchez, L.~E. Sanchez,
  N.~Sanchis-Gual, V.~Sandberg, J.~R. Sanders, B.~Sassolas, B.~S.
  Sathyaprakash, P.~R. Saulson, O.~Sauter, R.~L. Savage, A.~Sawadsky,
  P.~Schale, M.~Scheel, J.~Scheuer, J.~Schmidt, P.~Schmidt, R.~Schnabel,
  R.~M.~S. Schofield, A.~Sch\"onbeck, E.~Schreiber, D.~Schuette, B.~W. Schulte,
  B.~F. Schutz, S.~G. Schwalbe, J.~Scott, S.~M. Scott, E.~Seidel, D.~Sellers,
  A.~S. Sengupta, D.~Sentenac, V.~Sequino, A.~Sergeev, D.~A. Shaddock, T.~J.
  Shaffer, A.~A. Shah, M.~S. Shahriar, M.~B. Shaner, L.~Shao, B.~Shapiro,
  P.~Shawhan, A.~Sheperd, D.~H. Shoemaker, D.~M. Shoemaker, K.~Siellez,
  X.~Siemens, M.~Sieniawska, D.~Sigg, A.~D. Silva, L.~P. Singer, A.~Singh,
  A.~Singhal, A.~M. Sintes, B.~J.~J. Slagmolen, B.~Smith, J.~R. Smith, R.~J.~E.
  Smith, S.~Somala, E.~J. Son, J.~A. Sonnenberg, B.~Sorazu, F.~Sorrentino,
  T.~Souradeep, A.~P. Spencer, A.~K. Srivastava, K.~Staats, A.~Staley,
  M.~Steinke, J.~Steinlechner, S.~Steinlechner, D.~Steinmeyer, S.~P. Stevenson,
  R.~Stone, D.~J. Stops, K.~A. Strain, G.~Stratta, S.~E. Strigin, A.~Strunk,
  R.~Sturani, A.~L. Stuver, T.~Z. Summerscales, L.~Sun, S.~Sunil, J.~Suresh,
  P.~J. Sutton, B.~L. Swinkels, M.~J. Szczepa\ifmmode~\acute{n}\else
  \'{n}\fi{}czyk, M.~Tacca, S.~C. Tait, C.~Talbot, D.~Talukder, D.~B. Tanner,
  M.~T\'apai, A.~Taracchini, J.~D. Tasson, J.~A. Taylor, R.~Taylor, S.~V.
  Tewari, T.~Theeg, F.~Thies, E.~G. Thomas, M.~Thomas, P.~Thomas, K.~A. Thorne,
  K.~S. Thorne, E.~Thrane, S.~Tiwari, V.~Tiwari, K.~V. Tokmakov, K.~Toland,
  M.~Tonelli, Z.~Tornasi, A.~Torres-Forn\'e, C.~I. Torrie, D.~T\"oyr\"a,
  F.~Travasso, G.~Traylor, J.~Trinastic, M.~C. Tringali, L.~Trozzo, K.~W.
  Tsang, M.~Tse, R.~Tso, L.~Tsukada, D.~Tsuna, D.~Tuyenbayev, K.~Ueno,
  D.~Ugolini, C.~S. Unnikrishnan, A.~L. Urban, S.~A. Usman, H.~Vahlbruch,
  G.~Vajente, G.~Valdes, M.~Vallisneri, N.~van Bakel, M.~van Beuzekom, J.~F.~J.
  van~den Brand, C.~Van Den~Broeck, D.~C. Vander-Hyde, L.~van~der Schaaf, J.~V.
  van Heijningen, A.~A. van Veggel, M.~Vardaro, V.~Varma, S.~Vass, M.~Vas\'uth,
  A.~Vecchio, G.~Vedovato, J.~Veitch, P.~J. Veitch, K.~Venkateswara,
  G.~Venugopalan, D.~Verkindt, F.~Vetrano, A.~Vicer\'e, A.~D. Viets,
  S.~Vinciguerra, D.~J. Vine, J.-Y. Vinet, S.~Vitale, T.~Vo, H.~Vocca,
  C.~Vorvick, S.~P. Vyatchanin, A.~R. Wade, L.~E. Wade, M.~Wade, R.~Walet,
  M.~Walker, L.~Wallace, S.~Walsh, G.~Wang, H.~Wang, J.~Z. Wang, W.~H. Wang,
  Y.~F. Wang, R.~L. Ward, J.~Warner, M.~Was, J.~Watchi, B.~Weaver, L.-W. Wei,
  M.~Weinert, A.~J. Weinstein, R.~Weiss, L.~Wen, E.~K. Wessel, P.~We\ss{}els,
  J.~Westerweck, T.~Westphal, K.~Wette, J.~T. Whelan, S.~E. Whitcomb, B.~F.
  Whiting, C.~Whittle, D.~Wilken, D.~Williams, R.~D. Williams, A.~R.
  Williamson, J.~L. Willis, B.~Willke, M.~H. Wimmer, W.~Winkler, C.~C. Wipf,
  H.~Wittel, G.~Woan, J.~Woehler, J.~Wofford, K.~W.~K. Wong, J.~Worden, J.~L.
  Wright, D.~S. Wu, D.~M. Wysocki, S.~Xiao, H.~Yamamoto, C.~C. Yancey, L.~Yang,
  M.~J. Yap, M.~Yazback, Hang Yu, Haocun Yu, M.~Yvert,
  A.~Zadro\ifmmode~\dot{z}\else \.{z}\fi{}ny, M.~Zanolin, T.~Zelenova, J.-P.
  Zendri, M.~Zevin, L.~Zhang, M.~Zhang, T.~Zhang, Y.-H. Zhang, C.~Zhao,
  M.~Zhou, Z.~Zhou, S.~J. Zhu, X.~J. Zhu, A.~B. Zimmerman, M.~E. Zucker, and
  J.~Zweizig.
\newblock Gw170817: Observation of gravitational waves from a binary neutron
  star inspiral.
\newblock {\em Phys. Rev. Lett.}, 119:161101, Oct 2017.

\bibitem{Abbott_2017}
B.~P. Abbott, R.~Abbott, T.~D. Abbott, F.~Acernese, K.~Ackley, C.~Adams,
  T.~Adams, P.~Addesso, R.~X. Adhikari, V.~B. Adya, C.~Affeldt, M.~Afrough,
  B.~Agarwal, M.~Agathos, K.~Agatsuma, N.~Aggarwal, O.~D. Aguiar, L.~Aiello,
  A.~Ain, P.~Ajith, B.~Allen, G.~Allen, A.~Allocca, M.~A. Aloy, P.~A. Altin,
  A.~Amato, A.~Ananyeva, S.~B. Anderson, W.~G. Anderson, S.~V. Angelova,
  S.~Antier, S.~Appert, K.~Arai, M.~C. Araya, J.~S. Areeda, N.~Arnaud, K.~G.
  Arun, S.~Ascenzi, G.~Ashton, M.~Ast, S.~M. Aston, P.~Astone, D.~V. Atallah,
  P.~Aufmuth, C.~Aulbert, K.~AultONeal, C.~Austin, A.~Avila-Alvarez, S.~Babak,
  P.~Bacon, M.~K.~M. Bader, S.~Bae, P.~T. Baker, F.~Baldaccini, G.~Ballardin,
  S.~W. Ballmer, S.~Banagiri, J.~C. Barayoga, S.~E. Barclay, B.~C. Barish,
  D.~Barker, K.~Barkett, F.~Barone, B.~Barr, L.~Barsotti, M.~Barsuglia,
  D.~Barta, J.~Bartlett, I.~Bartos, R.~Bassiri, A.~Basti, J.~C. Batch,
  M.~Bawaj, J.~C. Bayley, M.~Bazzan, B.~B{\'{e}}csy, C.~Beer, M.~Bejger,
  I.~Belahcene, A.~S. Bell, B.~K. Berger, G.~Bergmann, J.~J. Bero, C.~P.~L.
  Berry, D.~Bersanetti, A.~Bertolini, J.~Betzwieser, S.~Bhagwat, R.~Bhandare,
  I.~A. Bilenko, G.~Billingsley, C.~R. Billman, J.~Birch, R.~Birney,
  O.~Birnholtz, S.~Biscans, S.~Biscoveanu, A.~Bisht, M.~Bitossi, C.~Biwer,
  M.~A. Bizouard, J.~K. Blackburn, J.~Blackman, C.~D. Blair, D.~G. Blair, R.~M.
  Blair, S.~Bloemen, O.~Bock, N.~Bode, M.~Boer, G.~Bogaert, A.~Bohe, F.~Bondu,
  E.~Bonilla, R.~Bonnand, B.~A. Boom, R.~Bork, V.~Boschi, S.~Bose, K.~Bossie,
  Y.~Bouffanais, A.~Bozzi, C.~Bradaschia, P.~R. Brady, M.~Branchesi, J.~E.
  Brau, T.~Briant, A.~Brillet, M.~Brinkmann, V.~Brisson, P.~Brockill, J.~E.
  Broida, A.~F. Brooks, D.~A. Brown, D.~D. Brown, S.~Brunett, C.~C. Buchanan,
  A.~Buikema, T.~Bulik, H.~J. Bulten, A.~Buonanno, D.~Buskulic, C.~Buy, R.~L.
  Byer, M.~Cabero, L.~Cadonati, G.~Cagnoli, C.~Cahillane, J.~Calder{\'{o}}n
  Bustillo, T.~A. Callister, E.~Calloni, J.~B. Camp, M.~Canepa, P.~Canizares,
  K.~C. Cannon, H.~Cao, J.~Cao, C.~D. Capano, E.~Capocasa, F.~Carbognani,
  S.~Caride, M.~F. Carney, J.~Casanueva Diaz, C.~Casentini, S.~Caudill,
  M.~Cavagli{\`{a}}, F.~Cavalier, R.~Cavalieri, G.~Cella, C.~B. Cepeda,
  P.~Cerd{\'{a}}-Dur{\'{a}}n, G.~Cerretani, E.~Cesarini, S.~J. Chamberlin,
  M.~Chan, S.~Chao, P.~Charlton, E.~Chase, E.~Chassande-Mottin, D.~Chatterjee,
  K.~Chatziioannou, B.~D. Cheeseboro, H.~Y. Chen, X.~Chen, Y.~Chen, H.-P.
  Cheng, H.~Chia, A.~Chincarini, A.~Chiummo, T.~Chmiel, H.~S. Cho, M.~Cho,
  J.~H. Chow, N.~Christensen, Q.~Chu, A.~J.~K. Chua, S.~Chua, A.~K.~W. Chung,
  S.~Chung, G.~Ciani, R.~Ciolfi, C.~E. Cirelli, A.~Cirone, F.~Clara, J.~A.
  Clark, P.~Clearwater, F.~Cleva, C.~Cocchieri, E.~Coccia, P.-F. Cohadon,
  D.~Cohen, A.~Colla, C.~G. Collette, L.~R. Cominsky, M.~Constancio Jr.,
  L.~Conti, S.~J. Cooper, P.~Corban, T.~R. Corbitt, I.~Cordero-Carri{\'{o}}n,
  K.~R. Corley, N.~Cornish, A.~Corsi, S.~Cortese, C.~A. Costa, M.~W. Coughlin,
  S.~B. Coughlin, J.-P. Coulon, S.~T. Countryman, P.~Couvares, P.~B. Covas,
  E.~E. Cowan, D.~M. Coward, M.~J. Cowart, D.~C. Coyne, R.~Coyne, J.~D.~E.
  Creighton, T.~D. Creighton, J.~Cripe, S.~G. Crowder, T.~J. Cullen,
  A.~Cumming, L.~Cunningham, E.~Cuoco, T.~Dal Canton, G.~D{\'{a}}lya, S.~L.
  Danilishin, S.~D'Antonio, K.~Danzmann, A.~Dasgupta, C.~F. Da~Silva Costa,
  V.~Dattilo, I.~Dave, M.~Davier, D.~Davis, E.~J. Daw, B.~Day, S.~De, D.~DeBra,
  J.~Degallaix, M.~De Laurentis, S.~Del{\'{e}}glise, W.~Del Pozzo, N.~Demos,
  T.~Denker, T.~Dent, R.~De Pietri, V.~Dergachev, R.~De Rosa, R.~T. DeRosa,
  C.~De Rossi, R.~DeSalvo, O.~de~Varona, J.~Devenson, S.~Dhurandhar, M.~C.
  D{\'{\i}}az, L.~Di Fiore, M.~Di Giovanni, T.~Di Girolamo, A.~Di Lieto, S.~Di
  Pace, I.~Di Palma, F.~Di Renzo, Z.~Doctor, V.~Dolique, F.~Donovan, K.~L.
  Dooley, S.~Doravari, I.~Dorrington, R.~Douglas, M.~Dovale {\'{A}}lvarez,
  T.~P. Downes, M.~Drago, C.~Dreissigacker, J.~C. Driggers, Z.~Du, M.~Ducrot,
  P.~Dupej, S.~E. Dwyer, T.~B. Edo, M.~C. Edwards, A.~Effler, H.-B. Eggenstein,
  P.~Ehrens, J.~Eichholz, S.~S. Eikenberry, R.~A. Eisenstein, R.~C. Essick,
  D.~Estevez, Z.~B. Etienne, T.~Etzel, M.~Evans, T.~M. Evans, M.~Factourovich,
  V.~Fafone, H.~Fair, S.~Fairhurst, X.~Fan, S.~Farinon, B.~Farr, W.~M. Farr,
  E.~J. Fauchon-Jones, M.~Favata, M.~Fays, C.~Fee, H.~Fehrmann, J.~Feicht,
  M.~M. Fejer, A.~Fernandez-Galiana, I.~Ferrante, E.~C. Ferreira, F.~Ferrini,
  F.~Fidecaro, D.~Finstad, I.~Fiori, D.~Fiorucci, M.~Fishbach, R.~P. Fisher,
  M.~Fitz-Axen, R.~Flaminio, M.~Fletcher, H.~Fong, J.~A. Font, P.~W.~F.
  Forsyth, S.~S. Forsyth, J.-D. Fournier, S.~Frasca, F.~Frasconi, Z.~Frei,
  A.~Freise, R.~Frey, V.~Frey, E.~M. Fries, P.~Fritschel, V.~V. Frolov,
  P.~Fulda, M.~Fyffe, H.~Gabbard, B.~U. Gadre, S.~M. Gaebel, J.~R. Gair,
  L.~Gammaitoni, M.~R. Ganija, S.~G. Gaonkar, C.~Garcia-Quiros, F.~Garufi,
  B.~Gateley, S.~Gaudio, G.~Gaur, V.~Gayathri, N.~Gehrels, G.~Gemme, E.~Genin,
  A.~Gennai, D.~George, J.~George, L.~Gergely, V.~Germain, S.~Ghonge, Abhirup
  Ghosh, Archisman Ghosh, S.~Ghosh, J.~A. Giaime, K.~D. Giardina, A.~Giazotto,
  K.~Gill, L.~Glover, E.~Goetz, R.~Goetz, S.~Gomes, B.~Goncharov,
  G.~Gonz{\'{a}}lez, J.~M.~Gonzalez Castro, A.~Gopakumar, M.~L. Gorodetsky,
  S.~E. Gossan, M.~Gosselin, R.~Gouaty, A.~Grado, C.~Graef, M.~Granata,
  A.~Grant, S.~Gras, C.~Gray, G.~Greco, A.~C. Green, E.~M. Gretarsson,
  P.~Groot, H.~Grote, S.~Grunewald, P.~Gruning, G.~M. Guidi, X.~Guo, A.~Gupta,
  M.~K. Gupta, K.~E. Gushwa, E.~K. Gustafson, R.~Gustafson, O.~Halim, B.~R.
  Hall, E.~D. Hall, E.~Z. Hamilton, G.~Hammond, M.~Haney, M.~M. Hanke,
  J.~Hanks, C.~Hanna, M.~D. Hannam, O.~A. Hannuksela, J.~Hanson, T.~Hardwick,
  J.~Harms, G.~M. Harry, I.~W. Harry, M.~J. Hart, C.-J. Haster, K.~Haughian,
  J.~Healy, A.~Heidmann, M.~C. Heintze, H.~Heitmann, P.~Hello, G.~Hemming,
  M.~Hendry, I.~S. Heng, J.~Hennig, A.~W. Heptonstall, M.~Heurs, S.~Hild,
  T.~Hinderer, D.~Hoak, D.~Hofman, K.~Holt, D.~E. Holz, P.~Hopkins, C.~Horst,
  J.~Hough, E.~A. Houston, E.~J. Howell, A.~Hreibi, Y.~M. Hu, E.~A. Huerta,
  D.~Huet, B.~Hughey, S.~Husa, S.~H. Huttner, T.~Huynh-Dinh, N.~Indik, R.~Inta,
  G.~Intini, H.~N. Isa, J.-M. Isac, M.~Isi, B.~R. Iyer, K.~Izumi, T.~Jacqmin,
  K.~Jani, P.~Jaranowski, S.~Jawahar, F.~Jim{\'{e}}nez-Forteza, W.~W. Johnson,
  N.~K. Johnson-McDaniel, D.~I. Jones, R.~Jones, R.~J.~G. Jonker, L.~Ju,
  J.~Junker, C.~V. Kalaghatgi, V.~Kalogera, B.~Kamai, S.~Kandhasamy, G.~Kang,
  J.~B. Kanner, S.~J. Kapadia, S.~Karki, K.~S. Karvinen, M.~Kasprzack,
  W.~Kastaun, M.~Katolik, E.~Katsavounidis, W.~Katzman, S.~Kaufer, K.~Kawabe,
  F.~K{\'{e}}f{\'{e}}lian, D.~Keitel, A.~J. Kemball, R.~Kennedy, C.~Kent, J.~S.
  Key, F.~Y. Khalili, I.~Khan, S.~Khan, Z.~Khan, E.~A. Khazanov, N.~Kijbunchoo,
  Chunglee Kim, J.~C. Kim, K.~Kim, W.~Kim, W.~S. Kim, Y.-M. Kim, S.~J.
  Kimbrell, E.~J. King, P.~J. King, M.~Kinley-Hanlon, R.~Kirchhoff, J.~S.
  Kissel, L.~Kleybolte, S.~Klimenko, T.~D. Knowles, P.~Koch, S.~M. Koehlenbeck,
  S.~Koley, V.~Kondrashov, A.~Kontos, M.~Korobko, W.~Z. Korth, I.~Kowalska,
  D.~B. Kozak, C.~Krämer, V.~Kringel, B.~Krishnan, A.~Kr{\'{o}}lak, G.~Kuehn,
  P.~Kumar, R.~Kumar, S.~Kumar, L.~Kuo, A.~Kutynia, S.~Kwang, B.~D. Lackey,
  K.~H. Lai, M.~Landry, R.~N. Lang, J.~Lange, B.~Lantz, R.~K. Lanza,
  A.~Lartaux-Vollard, P.~D. Lasky, M.~Laxen, A.~Lazzarini, C.~Lazzaro,
  P.~Leaci, S.~Leavey, C.~H. Lee, H.~K. Lee, H.~M. Lee, H.~W. Lee, K.~Lee,
  J.~Lehmann, A.~Lenon, M.~Leonardi, N.~Leroy, N.~Letendre, Y.~Levin, T.~G.~F.
  Li, S.~D. Linker, T.~B. Littenberg, J.~Liu, R.~K.~L. Lo, N.~A. Lockerbie,
  L.~T. London, J.~E. Lord, M.~Lorenzini, V.~Loriette, M.~Lormand, G.~Losurdo,
  J.~D. Lough, C.~O. Lousto, G.~Lovelace, H.~Lück, D.~Lumaca, A.~P. Lundgren,
  R.~Lynch, Y.~Ma, R.~Macas, S.~Macfoy, B.~Machenschalk, M.~MacInnis, D.~M.
  Macleod, I.~Maga{\~{n}}a Hernandez, F.~Maga{\~{n}}a-Sandoval, L.~Maga{\~{n}}a
  Zertuche, R.~M. Magee, E.~Majorana, I.~Maksimovic, N.~Man, V.~Mandic,
  V.~Mangano, G.~L. Mansell, M.~Manske, M.~Mantovani, F.~Marchesoni, F.~Marion,
  S.~M{\'{a}}rka, Z.~M{\'{a}}rka, C.~Markakis, A.~S. Markosyan, A.~Markowitz,
  E.~Maros, A.~Marquina, F.~Martelli, L.~Martellini, I.~W. Martin, R.~M.
  Martin, D.~V. Martynov, K.~Mason, E.~Massera, A.~Masserot, T.~J. Massinger,
  M.~Masso-Reid, S.~Mastrogiovanni, A.~Matas, F.~Matichard, L.~Matone,
  N.~Mavalvala, N.~Mazumder, R.~McCarthy, D.~E. McClelland, S.~McCormick,
  L.~McCuller, S.~C. McGuire, G.~McIntyre, J.~McIver, D.~J. McManus,
  L.~McNeill, T.~McRae, S.~T. McWilliams, D.~Meacher, G.~D. Meadors, M.~Mehmet,
  J.~Meidam, E.~Mejuto-Villa, A.~Melatos, G.~Mendell, R.~A. Mercer, E.~L.
  Merilh, M.~Merzougui, S.~Meshkov, C.~Messenger, C.~Messick, R.~Metzdorff,
  P.~M. Meyers, H.~Miao, C.~Michel, H.~Middleton, E.~E. Mikhailov, L.~Milano,
  A.~L. Miller, B.~B. Miller, J.~Miller, M.~Millhouse, M.~C. Milovich-Goff,
  O.~Minazzoli, Y.~Minenkov, J.~Ming, C.~Mishra, S.~Mitra, V.~P. Mitrofanov,
  G.~Mitselmakher, R.~Mittleman, D.~Moffa, A.~Moggi, K.~Mogushi, M.~Mohan,
  S.~R.~P. Mohapatra, M.~Montani, C.~J. Moore, D.~Moraru, G.~Moreno, S.~R.
  Morriss, B.~Mours, C.~M. Mow-Lowry, G.~Mueller, A.~W. Muir, Arunava
  Mukherjee, D.~Mukherjee, S.~Mukherjee, N.~Mukund, A.~Mullavey, J.~Munch,
  E.~A. Mu{\~{n}}iz, M.~Muratore, P.~G. Murray, K.~Napier, I.~Nardecchia,
  L.~Naticchioni, R.~K. Nayak, J.~Neilson, G.~Nelemans, T.~J.~N. Nelson,
  M.~Nery, A.~Neunzert, L.~Nevin, J.~M. Newport, G.~Newton, K.~K.~Y. Ng, T.~T.
  Nguyen, D.~Nichols, A.~B. Nielsen, S.~Nissanke, A.~Nitz, A.~Noack, F.~Nocera,
  D.~Nolting, C.~North, L.~K. Nuttall, J.~Oberling, G.~D. O'Dea, G.~H. Ogin,
  J.~J. Oh, S.~H. Oh, F.~Ohme, M.~A. Okada, M.~Oliver, P.~Oppermann, Richard~J.
  Oram, B.~O'Reilly, R.~Ormiston, L.~F. Ortega, R.~O'Shaughnessy, S.~Ossokine,
  D.~J. Ottaway, H.~Overmier, B.~J. Owen, A.~E. Pace, J.~Page, M.~A. Page,
  A.~Pai, S.~A. Pai, J.~R. Palamos, O.~Palashov, C.~Palomba, A.~Pal-Singh,
  Howard Pan, Huang-Wei Pan, B.~Pang, P.~T.~H. Pang, C.~Pankow, F.~Pannarale,
  B.~C. Pant, F.~Paoletti, A.~Paoli, M.~A. Papa, A.~Parida, W.~Parker,
  D.~Pascucci, A.~Pasqualetti, R.~Passaquieti, D.~Passuello, M.~Patil,
  B.~Patricelli, B.~L. Pearlstone, M.~Pedraza, R.~Pedurand, L.~Pekowsky,
  A.~Pele, S.~Penn, C.~J. Perez, A.~Perreca, L.~M. Perri, H.~P. Pfeiffer,
  M.~Phelps, O.~J. Piccinni, M.~Pichot, F.~Piergiovanni, V.~Pierro, G.~Pillant,
  L.~Pinard, I.~M. Pinto, M.~Pirello, M.~Pitkin, M.~Poe, R.~Poggiani,
  P.~Popolizio, E.~K. Porter, A.~Post, J.~Powell, J.~Prasad, J.~W.~W. Pratt,
  G.~Pratten, V.~Predoi, T.~Prestegard, M.~Prijatelj, M.~Principe,
  S.~Privitera, G.~A. Prodi, L.~G. Prokhorov, O.~Puncken, M.~Punturo, P.~Puppo,
  M.~Pürrer, H.~Qi, V.~Quetschke, E.~A. Quintero, R.~Quitzow-James, F.~J.
  Raab, D.~S. Rabeling, H.~Radkins, P.~Raffai, S.~Raja, C.~Rajan,
  B.~Rajbhandari, M.~Rakhmanov, K.~E. Ramirez, A.~Ramos-Buades, P.~Rapagnani,
  V.~Raymond, M.~Razzano, J.~Read, T.~Regimbau, L.~Rei, S.~Reid, D.~H. Reitze,
  W.~Ren, S.~D. Reyes, F.~Ricci, P.~M. Ricker, S.~Rieger, K.~Riles, M.~Rizzo,
  N.~A. Robertson, R.~Robie, F.~Robinet, A.~Rocchi, L.~Rolland, J.~G. Rollins,
  V.~J. Roma, R.~Romano, C.~L. Romel, J.~H. Romie, D.~Rosi{\'{n}}ska, M.~P.
  Ross, S.~Rowan, A.~Rüdiger, P.~Ruggi, G.~Rutins, K.~Ryan, S.~Sachdev,
  T.~Sadecki, L.~Sadeghian, M.~Sakellariadou, L.~Salconi, M.~Saleem, F.~Salemi,
  A.~Samajdar, L.~Sammut, L.~M. Sampson, E.~J. Sanchez, L.~E. Sanchez,
  N.~Sanchis-Gual, V.~Sandberg, J.~R. Sanders, B.~Sassolas, B.~S.
  Sathyaprakash, P.~R. Saulson, O.~Sauter, R.~L. Savage, A.~Sawadsky,
  P.~Schale, M.~Scheel, J.~Scheuer, J.~Schmidt, P.~Schmidt, R.~Schnabel,
  R.~M.~S. Schofield, A.~Schönbeck, E.~Schreiber, D.~Schuette, B.~W. Schulte,
  B.~F. Schutz, S.~G. Schwalbe, J.~Scott, S.~M. Scott, E.~Seidel, D.~Sellers,
  A.~S. Sengupta, D.~Sentenac, V.~Sequino, A.~Sergeev, D.~A. Shaddock, T.~J.
  Shaffer, A.~A. Shah, M.~S. Shahriar, M.~B. Shaner, L.~Shao, B.~Shapiro,
  P.~Shawhan, A.~Sheperd, D.~H. Shoemaker, D.~M. Shoemaker, K.~Siellez,
  X.~Siemens, M.~Sieniawska, D.~Sigg, A.~D. Silva, L.~P. Singer, A.~Singh,
  A.~Singhal, A.~M. Sintes, B.~J.~J. Slagmolen, B.~Smith, J.~R. Smith, R.~J.~E.
  Smith, S.~Somala, E.~J. Son, J.~A. Sonnenberg, B.~Sorazu, F.~Sorrentino,
  T.~Souradeep, A.~P. Spencer, A.~K. Srivastava, K.~Staats, A.~Staley,
  M.~Steinke, J.~Steinlechner, S.~Steinlechner, D.~Steinmeyer, S.~P. Stevenson,
  R.~Stone, D.~J. Stops, K.~A. Strain, G.~Stratta, S.~E. Strigin, A.~Strunk,
  R.~Sturani, A.~L. Stuver, T.~Z. Summerscales, L.~Sun, S.~Sunil, J.~Suresh,
  P.~J. Sutton, B.~L. Swinkels, M.~J. Szczepa{\'{n}}czyk, M.~Tacca, S.~C. Tait,
  C.~Talbot, D.~Talukder, D.~B. Tanner, M.~T{\'{a}}pai, A.~Taracchini, J.~D.
  Tasson, J.~A. Taylor, R.~Taylor, S.~V. Tewari, T.~Theeg, F.~Thies, E.~G.
  Thomas, M.~Thomas, P.~Thomas, K.~A. Thorne, K.~S. Thorne, E.~Thrane,
  S.~Tiwari, V.~Tiwari, K.~V. Tokmakov, K.~Toland, M.~Tonelli, Z.~Tornasi,
  A.~Torres-Forn{\'{e}}, C.~I. Torrie, D.~Töyrä, F.~Travasso, G.~Traylor,
  J.~Trinastic, M.~C. Tringali, L.~Trozzo, K.~W. Tsang, M.~Tse, R.~Tso,
  L.~Tsukada, D.~Tsuna, D.~Tuyenbayev, K.~Ueno, D.~Ugolini, C.~S. Unnikrishnan,
  A.~L. Urban, S.~A. Usman, H.~Vahlbruch, G.~Vajente, G.~Valdes, N.~van Bakel,
  M.~van Beuzekom, J.~F.~J. van~den Brand, C.~Van~Den Broeck, D.~C.
  Vander-Hyde, L.~van~der Schaaf, J.~V. van Heijningen, A.~A. van Veggel,
  M.~Vardaro, V.~Varma, S.~Vass, M.~Vas{\'{u}}th, A.~Vecchio, G.~Vedovato,
  J.~Veitch, P.~J. Veitch, K.~Venkateswara, G.~Venugopalan, D.~Verkindt,
  F.~Vetrano, A.~Vicer{\'{e}}, A.~D. Viets, S.~Vinciguerra, D.~J. Vine, J.-Y.
  Vinet, S.~Vitale, T.~Vo, H.~Vocca, C.~Vorvick, S.~P. Vyatchanin, A.~R. Wade,
  L.~E. Wade, M.~Wade, R.~Walet, M.~Walker, L.~Wallace, S.~Walsh, G.~Wang,
  H.~Wang, J.~Z. Wang, W.~H. Wang, Y.~F. Wang, R.~L. Ward, J.~Warner, M.~Was,
  J.~Watchi, B.~Weaver, L.-W. Wei, M.~Weinert, A.~J. Weinstein, R.~Weiss,
  L.~Wen, E.~K. Wessel, P.~We{\ss}els, J.~Westerweck, T.~Westphal, K.~Wette,
  J.~T. Whelan, S.~E. Whitcomb, B.~F. Whiting, C.~Whittle, D.~Wilken,
  D.~Williams, R.~D. Williams, A.~R. Williamson, J.~L. Willis, B.~Willke, M.~H.
  Wimmer, W.~Winkler, C.~C. Wipf, H.~Wittel, G.~Woan, J.~Woehler, J.~Wofford,
  K.~W.~K. Wong, J.~Worden, J.~L. Wright, D.~S. Wu, D.~M. Wysocki, S.~Xiao,
  H.~Yamamoto, C.~C. Yancey, L.~Yang, M.~J. Yap, M.~Yazback, Hang Yu, Haocun
  Yu, M.~Yvert, A.~Zadro{\.{z}}ny, M.~Zanolin, T.~Zelenova, J.-P. Zendri,
  M.~Zevin, L.~Zhang, M.~Zhang, T.~Zhang, Y.-H. Zhang, C.~Zhao, M.~Zhou,
  Z.~Zhou, S.~J. Zhu, X.~J. Zhu, A.~B. Zimmerman, M.~E. Zucker, J.~Zweizig,
  E.~Burns, P.~Veres, D.~Kocevski, J.~Racusin, A.~Goldstein, V.~Connaughton,
  M.~S. Briggs, L.~Blackburn, R.~Hamburg, C.~M. Hui, A.~von Kienlin,
  J.~McEnery, R.~D. Preece, C.~A. Wilson-Hodge, E.~Bissaldi, W.~H. Cleveland,
  M.~H. Gibby, M.~M. Giles, R.~M. Kippen, S.~McBreen, C.~A. Meegan, W.~S.
  Paciesas, S.~Poolakkil, O.~J. Roberts, M.~Stanbro, V.~Savchenko, C.~Ferrigno,
  E.~Kuulkers, A.~Bazzano, E.~Bozzo, S.~Brandt, J.~Chenevez, T.~J.-L.
  Courvoisier, R.~Diehl, A.~Domingo, L.~Hanlon, E.~Jourdain, P.~Laurent,
  F.~Lebrun, A.~Lutovinov, S.~Mereghetti, L.~Natalucci, J.~Rodi, J.-P. Roques,
  R.~Sunyaev, P.~Ubertini, , and and.
\newblock Gravitational waves and gamma-rays from a binary neutron star merger:
  {GW}170817 and {GRB} 170817a.
\newblock {\em The Astrophysical Journal}, 848(2):L13, oct 2017.

\bibitem{2020LRR....23....3A}
B.~P. {Abbott}, R.~{Abbott}, T.~D. {Abbott}, S.~{Abraham}, F.~{Acernese},
  K.~{Ackley}, C.~{Adams}, V.~B. {Adya}, C.~{Affeldt}, M.~{Agathos},
  K.~{Agatsuma}, N.~{Aggarwal}, O.~D. {Aguiar}, L.~{Aiello}, A.~{Ain},
  P.~{Ajith}, T.~{Akutsu}, G.~{Allen}, A.~{Allocca}, M.~A. {Aloy}, P.~A.
  {Altin}, A.~{Amato}, A.~{Ananyeva}, S.~B. {Anderson}, W.~G. {Anderson},
  M.~{Ando}, S.~V. {Angelova}, S.~{Antier}, S.~{Appert}, K.~{Arai}, Koya
  {Arai}, Y.~{Arai}, S.~{Araki}, A.~{Araya}, M.~C. {Araya}, J.~S. {Areeda},
  M.~{Ar{\`e}ne}, N.~{Aritomi}, N.~{Arnaud}, K.~G. {Arun}, S.~{Ascenzi},
  G.~{Ashton}, Y.~{Aso}, S.~M. {Aston}, P.~{Astone}, F.~{Aubin}, P.~{Aufmuth},
  K.~{Aultoneal}, C.~{Austin}, V.~{Avendano}, A.~{Avila-Alvarez}, S.~{Babak},
  P.~{Bacon}, F.~{Badaracco}, M.~K.~M. {Bader}, S.~W. {Bae}, Y.~B. {Bae},
  L.~{Baiotti}, R.~{Bajpai}, P.~T. {Baker}, F.~{Baldaccini}, G.~{Ballardin},
  S.~W. {Ballmer}, S.~{Banagiri}, J.~C. {Barayoga}, S.~E. {Barclay}, B.~C.
  {Barish}, D.~{Barker}, K.~{Barkett}, S.~{Barnum}, F.~{Barone}, B.~{Barr},
  L.~{Barsotti}, M.~{Barsuglia}, D.~{Barta}, J.~{Bartlett}, M.~A. {Barton},
  I.~{Bartos}, R.~{Bassiri}, A.~{Basti}, M.~{Bawaj}, J.~C. {Bayley},
  M.~{Bazzan}, B.~{B{\'e}csy}, M.~{Bejger}, I.~{Belahcene}, A.~S. {Bell},
  D.~{Beniwal}, B.~K. {Berger}, G.~{Bergmann}, S.~{Bernuzzi}, J.~J. {Bero},
  C.~P.~L. {Berry}, D.~{Bersanetti}, A.~{Bertolini}, J.~{Betzwieser},
  R.~{Bhandare}, J.~{Bidler}, I.~A. {Bilenko}, S.~A. {Bilgili},
  G.~{Billingsley}, J.~{Birch}, R.~{Birney}, O.~{Birnholtz}, S.~{Biscans},
  S.~{Biscoveanu}, A.~{Bisht}, M.~{Bitossi}, M.~A. {Bizouard}, J.~K.
  {Blackburn}, C.~D. {Blair}, D.~G. {Blair}, R.~M. {Blair}, S.~{Bloemen},
  N.~{Bode}, M.~{Boer}, Y.~{Boetzel}, G.~{Bogaert}, F.~{Bondu}, E.~{Bonilla},
  R.~{Bonnand}, P.~{Booker}, B.~A. {Boom}, C.~D. {Booth}, R.~{Bork},
  V.~{Boschi}, S.~{Bose}, K.~{Bossie}, V.~{Bossilkov}, J.~{Bosveld},
  Y.~{Bouffanais}, A.~{Bozzi}, C.~{Bradaschia}, P.~R. {Brady}, A.~{Bramley},
  M.~{Branchesi}, J.~E. {Brau}, T.~{Briant}, J.~H. {Briggs}, F.~{Brighenti},
  A.~{Brillet}, M.~{Brinkmann}, V.~{Brisson}, P.~{Brockill}, A.~F. {Brooks},
  D.~A. {Brown}, D.~D. {Brown}, S.~{Brunett}, A.~{Buikema}, T.~{Bulik}, H.~J.
  {Bulten}, A.~{Buonanno}, D.~{Buskulic}, C.~{Buy}, R.~L. {Byer}, M.~{Cabero},
  L.~{Cadonati}, G.~{Cagnoli}, C.~{Cahillane}, J.~Calder{\'o}n {Bustillo},
  T.~A. {Callister}, E.~{Calloni}, J.~B. {Camp}, W.~A. {Campbell}, M.~{Canepa},
  K.~{Cannon}, K.~C. {Cannon}, H.~{Cao}, J.~{Cao}, E.~{Capocasa},
  F.~{Carbognani}, S.~{Caride}, M.~F. {Carney}, G.~{Carullo}, J.~{Casanueva
  Diaz}, C.~{Casentini}, S.~{Caudill}, M.~{Cavagli{\`a}}, F.~{Cavalier},
  R.~{Cavalieri}, G.~{Cella}, P.~{Cerd{\'a}-Dur{\'a}n}, G.~{Cerretani},
  E.~{Cesarini}, O.~{Chaibi}, K.~{Chakravarti}, S.~J. {Chamberlin}, M.~{Chan},
  M.~L. {Chan}, S.~{Chao}, P.~{Charlton}, E.~A. {Chase}, E.~{Chassande-Mottin},
  D.~{Chatterjee}, M.~{Chaturvedi}, K.~{Chatziioannou}, B.~D. {Cheeseboro},
  C.~S. {Chen}, H.~Y. {Chen}, K.~H. {Chen}, X.~{Chen}, Y.~{Chen}, Y.~R. {Chen},
  H.~P. {Cheng}, C.~K. {Cheong}, H.~Y. {Chia}, A.~{Chincarini}, A.~{Chiummo},
  G.~{Cho}, H.~S. {Cho}, M.~{Cho}, N.~{Christensen}, H.~Y. {Chu}, Q.~{Chu},
  Y.~K. {Chu}, S.~{Chua}, K.~W. {Chung}, S.~{Chung}, G.~{Ciani}, A.~A.
  {Ciobanu}, R.~{Ciolfi}, F.~{Cipriano}, A.~{Cirone}, F.~{Clara}, J.~A.
  {Clark}, P.~{Clearwater}, F.~{Cleva}, C.~{Cocchieri}, E.~{Coccia}, P.~F.
  {Cohadon}, D.~{Cohen}, R.~{Colgan}, M.~{Colleoni}, C.~G. {Collette},
  C.~{Collins}, L.~R. {Cominsky}, M.~{Constancio}, L.~{Conti}, S.~J. {Cooper},
  P.~{Corban}, T.~R. {Corbitt}, I.~{Cordero-Carri{\'o}n}, K.~R. {Corley},
  N.~{Cornish}, A.~{Corsi}, S.~{Cortese}, C.~A. {Costa}, R.~{Cotesta}, M.~W.
  {Coughlin}, S.~B. {Coughlin}, J.~P. {Coulon}, S.~T. {Countryman},
  P.~{Couvares}, P.~B. {Covas}, E.~E. {Cowan}, D.~M. {Coward}, M.~J. {Cowart},
  D.~C. {Coyne}, R.~{Coyne}, J.~D.~E. {Creighton}, T.~D. {Creighton},
  J.~{Cripe}, M.~{Croquette}, S.~G. {Crowder}, T.~J. {Cullen}, A.~{Cumming},
  L.~{Cunningham}, E.~{Cuoco}, T.~{Dal Canton}, G.~{D{\'a}lya}, S.~L.
  {Danilishin}, S.~{D'Antonio}, K.~{Danzmann}, A.~{Dasgupta}, C.~F. {da Silva
  Costa}, L.~E.~H. {Datrier}, V.~{Dattilo}, I.~{Dave}, M.~{Davier}, D.~{Davis},
  E.~J. {Daw}, D.~{Debra}, M.~{Deenadayalan}, J.~{Degallaix}, M.~{de
  Laurentis}, S.~{Del{\'e}glise}, W.~Del {Pozzo}, L.~M. {Demarchi}, N.~{Demos},
  T.~{Dent}, R.~{de Pietri}, J.~{Derby}, R.~{De Rosa}, C.~{de Rossi},
  R.~{Desalvo}, O.~{de Varona}, S.~{Dhurandhar}, M.~C. {D{\'\i}az},
  T.~{Dietrich}, L.~{di Fiore}, M.~{di Giovanni}, T.~{di Girolamo}, A.~{di
  Lieto}, B.~{Ding}, S.~{di Pace}, I.~{di Palma}, F.~{di Renzo}, A.~{Dmitriev},
  Z.~{Doctor}, K.~{Doi}, F.~{Donovan}, K.~L. {Dooley}, S.~{Doravari},
  I.~{Dorrington}, T.~P. {Downes}, M.~{Drago}, J.~C. {Driggers}, Z.~{Du}, J.~G.
  {Ducoin}, P.~{Dupej}, S.~E. {Dwyer}, P.~J. {Easter}, T.~B. {Edo}, M.~C.
  {Edwards}, A.~{Effler}, S.~{Eguchi}, P.~{Ehrens}, J.~{Eichholz}, S.~S.
  {Eikenberry}, M.~{Eisenmann}, R.~A. {Eisenstein}, Y.~{Enomoto}, R.~C.
  {Essick}, H.~{Estelles}, D.~{Estevez}, Z.~B. {Etienne}, T.~{Etzel},
  M.~{Evans}, T.~M. {Evans}, V.~{Fafone}, H.~{Fair}, S.~{Fairhurst}, X.~{Fan},
  S.~{Farinon}, B.~{Farr}, W.~M. {Farr}, E.~J. {Fauchon-Jones}, M.~{Favata},
  M.~{Fays}, M.~{Fazio}, C.~{Fee}, J.~{Feicht}, M.~M. {Fejer}, F.~{Feng},
  A.~{Fernandez-Galiana}, I.~{Ferrante}, E.~C. {Ferreira}, T.~A. {Ferreira},
  F.~{Ferrini}, F.~{Fidecaro}, I.~{Fiori}, D.~{Fiorucci}, M.~{Fishbach}, R.~P.
  {Fisher}, J.~M. {Fishner}, M.~{Fitz-Axen}, R.~{Flaminio}, M.~{Fletcher},
  E.~{Flynn}, H.~{Fong}, J.~A. {Font}, P.~W.~F. {Forsyth}, J.~D. {Fournier},
  S.~{Frasca}, F.~{Frasconi}, Z.~{Frei}, A.~{Freise}, R.~{Frey}, V.~{Frey},
  P.~{Fritschel}, V.~V. {Frolov}, Y.~{Fujii}, M.~{Fukunaga}, M.~{Fukushima},
  P.~{Fulda}, M.~{Fyffe}, H.~A. {Gabbard}, B.~U. {Gadre}, S.~M. {Gaebel}, J.~R.
  {Gair}, L.~{Gammaitoni}, M.~R. {Ganija}, S.~G. {Gaonkar}, A.~{Garcia},
  C.~{Garc{\'\i}a-Quir{\'o}s}, F.~{Garufi}, B.~{Gateley}, S.~{Gaudio},
  G.~{Gaur}, V.~{Gayathri}, G.~G. {Ge}, G.~{Gemme}, E.~{Genin}, A.~{Gennai},
  D.~{George}, J.~{George}, L.~{Gergely}, V.~{Germain}, S.~{Ghonge}, Abhirup
  {Ghosh}, Archisman {Ghosh}, S.~{Ghosh}, B.~{Giacomazzo}, J.~A. {Giaime},
  K.~D. {Giardina}, A.~{Giazotto}, K.~{Gill}, G.~{Giordano}, L.~{Glover},
  P.~{Godwin}, E.~{Goetz}, R.~{Goetz}, B.~{Goncharov}, G.~{Gonz{\'a}lez}, J.~M.
  {Gonzalez Castro}, A.~{Gopakumar}, M.~L. {Gorodetsky}, S.~E. {Gossan},
  M.~{Gosselin}, R.~{Gouaty}, A.~{Grado}, C.~{Graef}, M.~{Granata}, A.~{Grant},
  S.~{Gras}, P.~{Grassia}, C.~{Gray}, R.~{Gray}, G.~{Greco}, A.~C. {Green},
  R.~{Green}, E.~M. {Gretarsson}, P.~{Groot}, H.~{Grote}, S.~{Grunewald},
  P.~{Gruning}, G.~M. {Guidi}, H.~K. {Gulati}, Y.~{Guo}, A.~{Gupta}, M.~K.
  {Gupta}, E.~K. {Gustafson}, R.~{Gustafson}, L.~{Haegel}, A.~{Hagiwara},
  S.~{Haino}, O.~{Halim}, B.~R. {Hall}, E.~D. {Hall}, E.~Z. {Hamilton},
  G.~{Hammond}, M.~{Haney}, M.~M. {Hanke}, J.~{Hanks}, C.~{Hanna}, M.~D.
  {Hannam}, O.~A. {Hannuksela}, J.~{Hanson}, T.~{Hardwick}, K.~{Haris},
  J.~{Harms}, G.~M. {Harry}, I.~W. {Harry}, K.~{Hasegawa}, C.~J. {Haster},
  K.~{Haughian}, H.~{Hayakawa}, K.~{Hayama}, F.~J. {Hayes}, J.~{Healy},
  A.~{Heidmann}, M.~C. {Heintze}, H.~{Heitmann}, P.~{Hello}, G.~{Hemming},
  M.~{Hendry}, I.~S. {Heng}, J.~{Hennig}, A.~W. {Heptonstall}, M.~{Heurs},
  S.~{Hild}, Y.~{Himemoto}, T.~{Hinderer}, Y.~{Hiranuma}, N.~{Hirata},
  E.~{Hirose}, D.~{Hoak}, S.~{Hochheim}, D.~{Hofman}, A.~M. {Holgado}, N.~A.
  {Holland}, K.~{Holt}, D.~E. {Holz}, Z.~{Hong}, P.~{Hopkins}, C.~{Horst},
  J.~{Hough}, E.~J. {Howell}, C.~G. {Hoy}, A.~{Hreibi}, B.~H. {Hsieh}, G.~Z.
  {Huang}, P.~W. {Huang}, Y.~J. {Huang}, E.~A. {Huerta}, D.~{Huet},
  B.~{Hughey}, M.~{Hulko}, S.~{Husa}, S.~H. {Huttner}, T.~{Huynh-Dinh},
  B.~{Idzkowski}, A.~{Iess}, B.~{Ikenoue}, S.~{Imam}, K.~{Inayoshi},
  C.~{Ingram}, Y.~{Inoue}, R.~{Inta}, G.~{Intini}, K.~{Ioka}, B.~{Irwin}, H.~N.
  {Isa}, J.~M. {Isac}, M.~{Isi}, Y.~{Itoh}, B.~R. {Iyer}, K.~{Izumi},
  T.~{Jacqmin}, S.~J. {Jadhav}, K.~{Jani}, N.~N. {Janthalur}, P.~{Jaranowski},
  A.~C. {Jenkins}, J.~{Jiang}, D.~S. {Johnson}, A.~W. {Jones}, D.~I. {Jones},
  R.~{Jones}, R.~J.~G. {Jonker}, L.~{Ju}, K.~{Jung}, P.~{Jung}, J.~{Junker},
  T.~{Kajita}, C.~V. {Kalaghatgi}, V.~{Kalogera}, B.~{Kamai}, M.~{Kamiizumi},
  N.~{Kanda}, S.~{Kandhasamy}, G.~W. {Kang}, J.~B. {Kanner}, S.~J. {Kapadia},
  S.~{Karki}, K.~S. {Karvinen}, R.~{Kashyap}, M.~{Kasprzack}, S.~{Katsanevas},
  E.~{Katsavounidis}, W.~{Katzman}, S.~{Kaufer}, K.~{Kawabe}, K.~{Kawaguchi},
  N.~{Kawai}, T.~{Kawasaki}, N.~V. {Keerthana}, F.~{K{\'e}f{\'e}lian},
  D.~{Keitel}, R.~{Kennedy}, J.~S. {Key}, F.~Y. {Khalili}, H.~{Khan},
  I.~{Khan}, S.~{Khan}, Z.~{Khan}, E.~A. {Khazanov}, M.~{Khursheed},
  N.~{Kijbunchoo}, Chunglee {Kim}, C.~{Kim}, J.~C. {Kim}, J.~{Kim}, K.~{Kim},
  W.~{Kim}, W.~S. {Kim}, Y.~M. {Kim}, C.~{Kimball}, N.~{Kimura}, E.~J. {King},
  P.~J. {King}, M.~{Kinley-Hanlon}, R.~{Kirchhoff}, J.~S. {Kissel}, N.~{Kita},
  H.~{Kitazawa}, L.~{Kleybolte}, J.~H. {Klika}, S.~{Klimenko}, T.~D. {Knowles},
  E.~{Knyazev}, P.~{Koch}, S.~M. {Koehlenbeck}, G.~{Koekoek}, Y.~{Kojima},
  K.~{Kokeyama}, S.~{Koley}, K.~{Komori}, V.~{Kondrashov}, A.~K.~H. {Kong},
  A.~{Kontos}, N.~{Koper}, M.~{Korobko}, W.~Z. {Korth}, K.~{Kotake},
  I.~{Kowalska}, D.~B. {Kozak}, C.~{Kozakai}, R.~{Kozu}, V.~{Kringel},
  N.~{Krishnendu}, A.~{Kr{\'o}lak}, G.~{Kuehn}, A.~{Kumar}, P.~{Kumar}, Rahul
  {Kumar}, R.~{Kumar}, S.~{Kumar}, J.~{Kume}, C.~M. {Kuo}, H.~S. {Kuo},
  L.~{Kuo}, S.~{Kuroyanagi}, K.~{Kusayanagi}, A.~{Kutynia}, K.~{Kwak},
  S.~{Kwang}, B.~D. {Lackey}, K.~H. {Lai}, T.~L. {Lam}, M.~{Landry}, B.~B.
  {Lane}, R.~N. {Lang}, J.~{Lange}, B.~{Lantz}, R.~K. {Lanza},
  A.~{Lartaux-Vollard}, P.~D. {Lasky}, M.~{Laxen}, A.~{Lazzarini},
  C.~{Lazzaro}, P.~{Leaci}, S.~{Leavey}, Y.~K. {Lecoeuche}, C.~H. {Lee}, H.~K.
  {Lee}, H.~M. {Lee}, H.~W. {Lee}, J.~{Lee}, K.~{Lee}, R.~K. {Lee},
  J.~{Lehmann}, A.~{Lenon}, M.~{Leonardi}, N.~{Leroy}, N.~{Letendre},
  Y.~{Levin}, J.~{Li}, K.~J.~L. {Li}, T.~G.~F. {Li}, X.~{Li}, C.~Y. {Lin},
  F.~{Lin}, F.~L. {Lin}, L.~C.~C. {Lin}, F.~{Linde}, S.~D. {Linker}, T.~B.
  {Littenberg}, G.~C. {Liu}, J.~{Liu}, X.~{Liu}, R.~K.~L. {Lo}, N.~A.
  {Lockerbie}, L.~T. {London}, A.~{Longo}, M.~{Lorenzini}, V.~{Loriette},
  M.~{Lormand}, G.~{Losurdo}, J.~D. {Lough}, C.~O. {Lousto}, G.~{Lovelace},
  M.~E. {Lower}, H.~{L{\"u}ck}, D.~{Lumaca}, A.~P. {Lundgren}, L.~W. {Luo},
  R.~{Lynch}, Y.~{Ma}, R.~{Macas}, S.~{Macfoy}, M.~{Macinnis}, D.~M. {MacLeod},
  A.~{Macquet}, F.~{Maga{\~n}a-Sandoval}, L.~Maga{\~n}a {Zertuche}, R.~M.
  {Magee}, E.~{Majorana}, I.~{Maksimovic}, A.~{Malik}, N.~{Man}, V.~{Mandic},
  V.~{Mangano}, G.~L. {Mansell}, M.~{Manske}, M.~{Mantovani}, F.~{Marchesoni},
  M.~{Marchio}, F.~{Marion}, S.~{M{\'a}rka}, Z.~{M{\'a}rka}, C.~{Markakis},
  A.~S. {Markosyan}, A.~{Markowitz}, E.~{Maros}, A.~{Marquina}, S.~{Marsat},
  F.~{Martelli}, I.~W. {Martin}, R.~M. {Martin}, D.~V. {Martynov}, K.~{Mason},
  E.~{Massera}, A.~{Masserot}, T.~J. {Massinger}, M.~{Masso-Reid},
  S.~{Mastrogiovanni}, A.~{Matas}, F.~{Matichard}, L.~{Matone}, N.~{Mavalvala},
  N.~{Mazumder}, J.~J. {McCann}, R.~{McCarthy}, D.~E. {McClelland},
  S.~{McCormick}, L.~{McCuller}, S.~C. {McGuire}, J.~{McIver}, D.~J. {McManus},
  T.~{McRae}, S.~T. {McWilliams}, D.~{Meacher}, G.~D. {Meadors}, M.~{Mehmet},
  A.~K. {Mehta}, J.~{Meidam}, A.~{Melatos}, G.~{Mendell}, R.~A. {Mercer},
  L.~{Mereni}, E.~L. {Merilh}, M.~{Merzougui}, S.~{Meshkov}, C.~{Messenger},
  C.~{Messick}, R.~{Metzdorff}, P.~M. {Meyers}, H.~{Miao}, C.~{Michel},
  Y.~{Michimura}, H.~{Middleton}, E.~E. {Mikhailov}, L.~{Milano}, A.~L.
  {Miller}, A.~{Miller}, M.~{Millhouse}, J.~C. {Mills}, M.~C. {Milovich-Goff},
  O.~{Minazzoli}, Y.~{Minenkov}, N.~{Mio}, A.~{Mishkin}, C.~{Mishra},
  T.~{Mistry}, S.~{Mitra}, V.~P. {Mitrofanov}, G.~{Mitselmakher},
  R.~{Mittleman}, O.~{Miyakawa}, A.~{Miyamoto}, Y.~{Miyazaki}, K.~{Miyo},
  S.~{Miyoki}, G.~{Mo}, D.~{Moffa}, K.~{Mogushi}, S.~R.~P. {Mohapatra},
  M.~{Montani}, C.~J. {Moore}, D.~{Moraru}, G.~{Moreno}, S.~{Morisaki},
  Y.~{Moriwaki}, B.~{Mours}, C.~M. {Mow-Lowry}, Arunava {Mukherjee},
  D.~{Mukherjee}, S.~{Mukherjee}, N.~{Mukund}, A.~{Mullavey}, J.~{Munch}, E.~A.
  {Mu{\~n}iz}, M.~{Muratore}, P.~G. {Murray}, K.~{Nagano}, S.~{Nagano},
  A.~{Nagar}, K.~{Nakamura}, H.~{Nakano}, M.~{Nakano}, R.~{Nakashima},
  I.~{Nardecchia}, T.~{Narikawa}, L.~{Naticchioni}, R.~K. {Nayak},
  R.~{Negishi}, J.~{Neilson}, G.~{Nelemans}, T.~J.~N. {Nelson}, M.~{Nery},
  A.~{Neunzert}, K.~Y. {Ng}, S.~{Ng}, P.~{Nguyen}, W.~T. {Ni}, D.~{Nichols},
  A.~{Nishizawa}, S.~{Nissanke}, F.~{Nocera}, C.~{North}, L.~K. {Nuttall},
  M.~{Obergaulinger}, J.~{Oberling}, B.~D. {O'Brien}, Y.~{Obuchi}, G.~D.
  {O'Dea}, W.~{Ogaki}, G.~H. {Ogin}, J.~J. {Oh}, S.~H. {Oh}, M.~{Ohashi},
  N.~{Ohishi}, M.~{Ohkawa}, F.~{Ohme}, H.~{Ohta}, M.~A. {Okada}, K.~{Okutomi},
  M.~{Oliver}, K.~{Oohara}, C.~P. {Ooi}, P.~{Oppermann}, Richard~J. {Oram},
  B.~{O'Reilly}, R.~G. {Ormiston}, L.~F. {Ortega}, R.~{O'Shaughnessy},
  S.~{Oshino}, S.~{Ossokine}, D.~J. {Ottaway}, H.~{Overmier}, B.~J. {Owen},
  A.~E. {Pace}, G.~{Pagano}, M.~A. {Page}, A.~{Pai}, S.~A. {Pai}, J.~R.
  {Palamos}, O.~{Palashov}, C.~{Palomba}, A.~{Pal-Singh}, Huang-Wei {Pan},
  K.~C. {Pan}, B.~{Pang}, H.~F. {Pang}, P.~T.~H. {Pang}, C.~{Pankow},
  F.~{Pannarale}, B.~C. {Pant}, F.~{Paoletti}, A.~{Paoli}, M.~A. {Papa},
  A.~{Parida}, J.~{Park}, W.~{Parker}, D.~{Pascucci}, A.~{Pasqualetti},
  R.~{Passaquieti}, D.~{Passuello}, M.~{Patil}, B.~{Patricelli}, B.~L.
  {Pearlstone}, C.~{Pedersen}, M.~{Pedraza}, R.~{Pedurand}, A.~{Pele},
  F.~E.~Pe{\~n}a {Arellano}, S.~{Penn}, C.~J. {Perez}, A.~{Perreca}, H.~P.
  {Pfeiffer}, M.~{Phelps}, K.~S. {Phukon}, O.~J. {Piccinni}, M.~{Pichot},
  F.~{Piergiovanni}, G.~{Pillant}, L.~{Pinard}, I.~{Pinto}, M.~{Pirello},
  M.~{Pitkin}, R.~{Poggiani}, D.~Y.~T. {Pong}, S.~{Ponrathnam}, P.~{Popolizio},
  E.~K. {Porter}, J.~{Powell}, A.~K. {Prajapati}, J.~{Prasad}, K.~{Prasai},
  R.~{Prasanna}, G.~{Pratten}, T.~{Prestegard}, S.~{Privitera}, G.~A. {Prodi},
  L.~G. {Prokhorov}, O.~{Puncken}, M.~{Punturo}, P.~{Puppo}, M.~{P{\"u}rrer},
  H.~{Qi}, V.~{Quetschke}, P.~J. {Quinonez}, E.~A. {Quintero},
  R.~{Quitzow-James}, F.~J. {Raab}, H.~{Radkins}, N.~{Radulescu}, P.~{Raffai},
  S.~{Raja}, C.~{Rajan}, B.~{Rajbhandari}, M.~{Rakhmanov}, K.~E. {Ramirez},
  A.~{Ramos-Buades}, Javed {Rana}, K.~{Rao}, P.~{Rapagnani}, V.~{Raymond},
  M.~{Razzano}, J.~{Read}, T.~{Regimbau}, L.~{Rei}, S.~{Reid}, D.~H. {Reitze},
  W.~{Ren}, F.~{Ricci}, C.~J. {Richardson}, J.~W. {Richardson}, P.~M. {Ricker},
  K.~{Riles}, M.~{Rizzo}, N.~A. {Robertson}, R.~{Robie}, F.~{Robinet},
  A.~{Rocchi}, L.~{Rolland}, J.~G. {Rollins}, V.~J. {Roma}, M.~{Romanelli},
  R.~{Romano}, C.~L. {Romel}, J.~H. {Romie}, K.~{Rose}, D.~{Rosi{\'n}ska},
  S.~G. {Rosofsky}, M.~P. {Ross}, S.~{Rowan}, A.~{R{\"u}diger}, P.~{Ruggi},
  G.~{Rutins}, K.~{Ryan}, S.~{Sachdev}, T.~{Sadecki}, N.~{Sago}, S.~{Saito},
  Y.~{Saito}, K.~{Sakai}, Y.~{Sakai}, H.~{Sakamoto}, M.~{Sakellariadou},
  Y.~{Sakuno}, L.~{Salconi}, M.~{Saleem}, A.~{Samajdar}, L.~{Sammut}, E.~J.
  {Sanchez}, L.~E. {Sanchez}, N.~{Sanchis-Gual}, V.~{Sandberg}, J.~R.
  {Sanders}, K.~A. {Santiago}, N.~{Sarin}, B.~{Sassolas}, B.~S.
  {Sathyaprakash}, S.~{Sato}, T.~{Sato}, O.~{Sauter}, R.~L. {Savage},
  T.~{Sawada}, P.~{Schale}, M.~{Scheel}, J.~{Scheuer}, P.~{Schmidt},
  R.~{Schnabel}, R.~M.~S. {Schofield}, A.~{Sch{\"o}nbeck}, E.~{Schreiber},
  B.~W. {Schulte}, B.~F. {Schutz}, S.~G. {Schwalbe}, J.~{Scott}, S.~M. {Scott},
  E.~{Seidel}, T.~{Sekiguchi}, Y.~{Sekiguchi}, D.~{Sellers}, A.~S. {Sengupta},
  N.~{Sennett}, D.~{Sentenac}, V.~{Sequino}, A.~{Sergeev}, Y.~{Setyawati},
  D.~A. {Shaddock}, T.~{Shaffer}, M.~S. {Shahriar}, M.~B. {Shaner}, L.~{Shao},
  P.~{Sharma}, P.~{Shawhan}, H.~{Shen}, S.~{Shibagaki}, R.~{Shimizu},
  T.~{Shimoda}, K.~{Shimode}, R.~{Shink}, H.~{Shinkai}, T.~{Shishido},
  A.~{Shoda}, D.~H. {Shoemaker}, D.~M. {Shoemaker}, S.~{Shyamsundar},
  K.~{Siellez}, M.~{Sieniawska}, D.~{Sigg}, A.~D. {Silva}, L.~P. {Singer},
  N.~{Singh}, A.~{Singhal}, A.~M. {Sintes}, S.~{Sitmukhambetov}, V.~{Skliris},
  B.~J.~J. {Slagmolen}, T.~J. {Slaven-Blair}, J.~R. {Smith}, R.~J.~E. {Smith},
  S.~{Somala}, K.~{Somiya}, E.~J. {Son}, B.~{Sorazu}, F.~{Sorrentino},
  H.~{Sotani}, T.~{Souradeep}, E.~{Sowell}, A.~P. {Spencer}, A.~K.
  {Srivastava}, V.~{Srivastava}, K.~{Staats}, C.~{Stachie}, M.~{Standke}, D.~A.
  {Steer}, M.~{Steinke}, J.~{Steinlechner}, S.~{Steinlechner}, D.~{Steinmeyer},
  S.~P. {Stevenson}, D.~{Stocks}, R.~{Stone}, D.~J. {Stops}, K.~A. {Strain},
  G.~{Stratta}, S.~E. {Strigin}, A.~{Strunk}, R.~{Sturani}, A.~L. {Stuver},
  V.~{Sudhir}, R.~{Sugimoto}, T.~Z. {Summerscales}, L.~{Sun}, S.~{Sunil},
  J.~{Suresh}, P.~J. {Sutton}, Takamasa {Suzuki}, Toshikazu {Suzuki}, B.~L.
  {Swinkels}, M.~J. {Szczepa{\'n}czyk}, M.~{Tacca}, H.~{Tagoshi}, S.~C. {Tait},
  H.~{Takahashi}, R.~{Takahashi}, A.~{Takamori}, S.~{Takano}, H.~{Takeda},
  M.~{Takeda}, C.~{Talbot}, D.~{Talukder}, H.~{Tanaka}, Kazuyuki {Tanaka},
  Kenta {Tanaka}, Taiki {Tanaka}, Takahiro {Tanaka}, S.~{Tanioka}, D.~B.
  {Tanner}, M.~{T{\'a}pai}, E.~N. {Tapia San Martin}, A.~{Taracchini}, J.~D.
  {Tasson}, R.~{Taylor}, S.~{Telada}, F.~{Thies}, M.~{Thomas}, P.~{Thomas},
  S.~R. {Thondapu}, K.~A. {Thorne}, E.~{Thrane}, Shubhanshu {Tiwari}, Srishti
  {Tiwari}, V.~{Tiwari}, K.~{Toland}, T.~{Tomaru}, Y.~{Tomigami}, T.~{Tomura},
  M.~{Tonelli}, Z.~{Tornasi}, A.~{Torres-Forn{\'e}}, C.~I. {Torrie},
  D.~{T{\"o}yr{\"a}}, F.~{Travasso}, G.~{Traylor}, M.~C. {Tringali},
  A.~{Trovato}, L.~{Trozzo}, R.~{Trudeau}, K.~W. {Tsang}, T.~T.~L. {Tsang},
  M.~{Tse}, R.~{Tso}, K.~{Tsubono}, S.~{Tsuchida}, L.~{Tsukada}, D.~{Tsuna},
  T.~{Tsuzuki}, D.~{Tuyenbayev}, N.~{Uchikata}, T.~{Uchiyama}, A.~{Ueda},
  T.~{Uehara}, K.~{Ueno}, G.~{Ueshima}, D.~{Ugolini}, C.~S. {Unnikrishnan},
  F.~{Uraguchi}, A.~L. {Urban}, T.~{Ushiba}, S.~A. {Usman}, H.~{Vahlbruch},
  G.~{Vajente}, G.~{Valdes}, N.~{van Bakel}, M.~{van Beuzekom}, J.~F.~J. {van
  den Brand}, C.~{van den Broeck}, D.~C. {Vander-Hyde}, L.~{van der Schaaf},
  J.~V. {van Heijningen}, M.~H.~P.~M. {van Putten}, A.~A. {van Veggel},
  M.~{Vardaro}, V.~{Varma}, S.~{Vass}, M.~{Vas{\'u}th}, A.~{Vecchio},
  G.~{Vedovato}, J.~{Veitch}, P.~J. {Veitch}, K.~{Venkateswara},
  G.~{Venugopalan}, D.~{Verkindt}, F.~{Vetrano}, A.~{Vicer{\'e}}, A.~D.
  {Viets}, D.~J. {Vine}, J.~Y. {Vinet}, S.~{Vitale}, Francisco~Hernandez
  {Vivanco}, T.~{Vo}, H.~{Vocca}, C.~{Vorvick}, S.~P. {Vyatchanin}, A.~R.
  {Wade}, L.~E. {Wade}, M.~{Wade}, R.~{Walet}, M.~{Walker}, L.~{Wallace},
  S.~{Walsh}, G.~{Wang}, H.~{Wang}, J.~{Wang}, J.~Z. {Wang}, W.~H. {Wang},
  Y.~F. {Wang}, R.~L. {Ward}, Z.~A. {Warden}, J.~{Warner}, M.~{Was},
  J.~{Watchi}, B.~{Weaver}, L.~W. {Wei}, M.~{Weinert}, A.~J. {Weinstein},
  R.~{Weiss}, F.~{Wellmann}, L.~{Wen}, E.~K. {Wessel}, P.~{We{\ss}els}, J.~W.
  {Westhouse}, K.~{Wette}, J.~T. {Whelan}, B.~F. {Whiting}, C.~{Whittle}, D.~M.
  {Wilken}, D.~{Williams}, A.~R. {Williamson}, J.~L. {Willis}, B.~{Willke},
  M.~H. {Wimmer}, W.~{Winkler}, C.~C. {Wipf}, H.~{Wittel}, G.~{Woan},
  J.~{Woehler}, J.~K. {Wofford}, J.~{Worden}, J.~L. {Wright}, C.~M. {Wu}, D.~S.
  {Wu}, H.~C. {Wu}, S.~R. {Wu}, D.~M. {Wysocki}, L.~{Xiao}, W.~R. {Xu},
  T.~{Yamada}, H.~{Yamamoto}, Kazuhiro {Yamamoto}, Kohei {Yamamoto},
  T.~{Yamamoto}, C.~C. {Yancey}, L.~{Yang}, M.~J. {Yap}, M.~{Yazback}, D.~W.
  {Yeeles}, K.~{Yokogawa}, J.~{Yokoyama}, T.~{Yokozawa}, T.~{Yoshioka}, Hang
  {Yu}, Haocun {Yu}, S.~H.~R. {Yuen}, H.~{Yuzurihara}, M.~{Yvert}, A.~K.
  {Zadro{\.z}ny}, M.~{Zanolin}, S.~{Zeidler}, T.~{Zelenova}, J.~P. {Zendri},
  M.~{Zevin}, J.~{Zhang}, L.~{Zhang}, T.~{Zhang}, C.~{Zhao}, Y.~{Zhao},
  M.~{Zhou}, Z.~{Zhou}, X.~J. {Zhu}, Z.~H. {Zhu}, A.~B. {Zimmerman}, M.~E.
  {Zucker}, J.~{Zweizig}, Ligo Scientific~Collaboration {Kagra Collaboration},
  and {VIRGO Collaboration}.
\newblock {Prospects for observing and localizing gravitational-wave transients
  with Advanced LIGO, Advanced Virgo and KAGRA}.
\newblock {\em Living Reviews in Relativity}, 23(1):3, September 2020.

\bibitem{2019arXiv190711364M}
Lee~J. {Mitchell}, Bernard~F. {Phlips}, J.~Eric {Grove}, Theodore {Finne}, Mary
  {Johnson-Rambert}, and W.~Neil {Johnson}.
\newblock {Strontium Iodide Radiation Instrument (SIRI) -- Early On-Orbit
  Results}.
\newblock {\em arXiv e-prints}, page arXiv:1907.11364, July 2019.

\bibitem{Smith:2019zra}
Jacob~R. Smith et~al.
\newblock {BurstCube: Mission Concept, Performance, and Status}.
\newblock {\em PoS}, ICRC2019:604, 2021.

\bibitem{MB:Hurley2013}
K.~{Hurley}, V.~D. {Pal'shin}, R.~L. {Aptekar}, S.~V. {Golenetskii}, D.~D.
  {Frederiks}, E.~P. {Mazets}, D.~S. {Svinkin}, M.~S. {Briggs},
  V.~{Connaughton}, C.~{Meegan}, J.~{Goldsten}, W.~{Boynton}, C.~{Fellows},
  K.~{Harshman}, I.~G. {Mitrofanov}, D.~V. {Golovin}, A.~S. {Kozyrev}, M.~L.
  {Litvak}, A.~B. {Sanin}, A.~{Rau}, A.~{von Kienlin}, X.~{Zhang},
  K.~{Yamaoka}, Y.~{Fukazawa}, Y.~{Hanabata}, M.~{Ohno}, T.~{Takahashi},
  M.~{Tashiro}, Y.~{Terada}, T.~{Murakami}, K.~{Makishima}, S.~{Barthelmy},
  T.~{Cline}, N.~{Gehrels}, J.~{Cummings}, H.~A. {Krimm}, D.~M. {Smith},
  E.~{Del Monte}, M.~{Feroci}, and M.~{Marisaldi}.
\newblock {The Interplanetary Network Supplement to the Fermi GBM Catalog of
  Cosmic Gamma-Ray Bursts}.
\newblock {\em ApJS}, 207(2):39, August 2013.

\bibitem{MB:Petrov2022}
Polina {Petrov}, Leo~P. {Singer}, Michael~W. {Coughlin}, Vishwesh {Kumar},
  Mouza {Almualla}, Shreya {Anand}, Mattia {Bulla}, Tim {Dietrich}, Francois
  {Foucart}, and Nidhal {Guessoum}.
\newblock {Data-driven Expectations for Electromagnetic Counterpart Searches
  Based on LIGO/Virgo Public Alerts}.
\newblock {\em ApJ}, 924(2):54, January 2022.

\bibitem{MB:Feindt2019}
Ulrich {Feindt}, Jakob {Nordin}, Mickael {Rigault}, Val{\'e}ry {Brinnel},
  Suhail {Dhawan}, Ariel {Goobar}, and Marek {Kowalski}.
\newblock {simsurvey: estimating transient discovery rates for the Zwicky
  transient facility}.
\newblock {\em JCAP}, 2019(10):005, October 2019.

\bibitem{MB:Burns2021}
E.~{Burns}, D.~{Svinkin}, K.~{Hurley}, Z.~{Wadiasingh}, M.~{Negro},
  G.~{Younes}, R.~{Hamburg}, A.~{Ridnaia}, D.~{Cook}, S.~B. {Cenko},
  R.~{Aloisi}, G.~{Ashton}, M.~{Baring}, M.~S. {Briggs}, N.~{Christensen},
  D.~{Frederiks}, A.~{Goldstein}, C.~M. {Hui}, D.~L. {Kaplan}, M.~M.
  {Kasliwal}, D.~{Kocevski}, O.~J. {Roberts}, V.~{Savchenko}, A.~{Tohuvavohu},
  P.~{Veres}, and C.~A. {Wilson-Hodge}.
\newblock {Identification of a Local Sample of Gamma-Ray Bursts Consistent with
  a Magnetar Giant Flare Origin}.
\newblock {\em ApJL}, 907(2):L28, February 2021.

\bibitem{McConnell.2021}
Mark~L McConnell, Matthew Baring, Peter~F Bloser, Michael Briggs, Camden
  Ertley, Gregory Fletcher, Jessica Gaskin, Karen Gelmis, Adam Goldstein,
  J~Eric Grove, Dieter Hartmann, Michelle Hui, Peter Jenke, R~Marc Kippen,
  Fabian Kislat, Daniel Kocevski, Merlin Kole, John~F Krizmanic, Jason Legere,
  Tyson Littenberg, Neil Martin, Sheila McBreen, Donald McQueen, Charles
  Meegan, Karla Oñate-Melecio, Mark Pearce, Rob Preece, Nicolas Produit, James
  Ryan, Steven~J Sturner, Peter Veres, W~Thomas Vestrand, Colleen Wilson-Hodge,
  and Bing Zhang.
\newblock {The LargE Area burst Polarimeter (LEAP) – A NASA mission of
  opportunity for the ISS}.
\newblock {\em UV, X-Ray, and Gamma-Ray Space Instrumentation for Astronomy
  XXII}, page~35, 2021.

\bibitem{Onate-Melecio.2021}
Karla Oñate-Melecio, Camden Ertley, Mark~L McConnell, Jason Legere, Peter~F
  Bloser, Michael Briggs, Jessica Gaskin, Adam Goldstein, J~Eric Grove,
  Michelle Hui, Peter Jenke, R~Marc Kippen, Fabian Kislat, Daniel Kocevski,
  John~F Krizmanic, Charles Meegan, Rob Preece, James Ryan, Steven~J Sturner,
  Peter Veres, W~Thomas Vestrand, and Colleen Wilson-Hodge.
\newblock {Evaluation of a prototype detector for the LargE Area burst
  Polarimeter (LEAP)}.
\newblock {\em UV, X-Ray, and Gamma-Ray Space Instrumentation for Astronomy
  XXII}, page~36, 2021.

\bibitem{Toma.20097gfh}
Kenji Toma, Takanori Sakamoto, Bing Zhang, Joanne~E Hill, M~L McConnell,
  Peter~F Bloser, Ryo Yamazaki, Kunihito Ioka, and Takashi Nakamura.
\newblock {Statistical Properties of Gamma-Ray Burst Polarization}.
\newblock {\em Astrophysical Journal}, 698(2):1042 -- 1053, 05 2009.
\newblock 12 pages, 12 figures, 1 table, submitted to ApJ.

\bibitem{McConnell.2017}
M~L McConnell.
\newblock {High energy polarimetry of prompt GRB emission}.
\newblock {\em New Astronomy Reviews}, 76:1 -- 21, 2017.

\bibitem{Tatischeff.2019}
Vincent Tatischeff, Mark~L. McConnell, and Philippe Laurent.
\newblock {Astronomical Polarisation from the Infrared to Gamma Rays}.
\newblock Astrophysics and Space Science Library, pages 109--146. 2019.

\bibitem{Chattopadhyay.2021ny}
Tanmoy Chattopadhyay.
\newblock {Hard X-ray polarimetry—an overview of the method, science drivers,
  and recent findings}.
\newblock {\em Journal of Astrophysics and Astronomy}, 42(2):106, 2021.

\bibitem{Tomsick:2021wed}
John~A. Tomsick.
\newblock {The Compton Spectrometer and Imager Project for MeV Astronomy}.
\newblock {\em PoS}, ICRC2021:652, 2021.

\bibitem{Burns20}
Eric {Burns}.
\newblock {Neutron star mergers and how to study them}.
\newblock {\em Living Reviews in Relativity}, 23(1):4, December 2020.

\bibitem{Toma09}
Kenji {Toma}, Takanori {Sakamoto}, Bing {Zhang}, Joanne~E. {Hill}, Mark~L.
  {McConnell}, Peter~F. {Bloser}, Ryo {Yamazaki}, Kunihito {Ioka}, and Takashi
  {Nakamura}.
\newblock {Statistical Properties of Gamma-Ray Burst Polarization}.
\newblock {\em \apj}, 698(2):1042--1053, June 2009.

\bibitem{GleiserKozameh01}
Reinaldo~J. {Gleiser} and Carlos~N. {Kozameh}.
\newblock {Astrophysical limits on quantum gravity motivated birefringence}.
\newblock {\em \prd}, 64(8):083007, October 2001.

\bibitem{Stecker11}
Floyd~W. {Stecker}.
\newblock {A new limit on Planck scale Lorentz violation from
  {\ensuremath{\gamma}}-ray burst polarization}.
\newblock {\em Astroparticle Physics}, 35(2):95--97, September 2011.

\bibitem{Gotz14}
D.~{G{\"o}tz}, P.~{Laurent}, S.~{Antier}, S.~{Covino}, P.~{D'Avanzo},
  V.~{D'Elia}, and A.~{Melandri}.
\newblock {GRB 140206A: the most distant polarized gamma-ray burst}.
\newblock {\em \mnras}, 444(3):2776--2782, November 2014.

\bibitem{Ema21}
Yohei {Ema}, Filippo {Sala}, and Ryosuke {Sato}.
\newblock {Dark matter models for the 511 keV galactic line predict keV
  electron recoils on Earth}.
\newblock {\em European Physical Journal C}, 81(2):129, February 2021.

\bibitem{Laha19}
Ranjan {Laha}.
\newblock {Primordial Black Holes as a Dark Matter Candidate Are Severely
  Constrained by the Galactic Center 511 keV {\ensuremath{\gamma}} -Ray Line}.
\newblock {\em \prl}, 123(25):251101, December 2019.

\bibitem{HoriuchiBeacom10}
Shunsaku {Horiuchi} and John~F. {Beacom}.
\newblock {Revealing Type Ia Supernova Physics with Cosmic Rates and Nuclear
  Gamma Rays}.
\newblock {\em \apj}, 723(1):329--341, November 2010.

\bibitem{RuizLapuente16}
Pilar {Ruiz-Lapuente}, Lih-Sin {The}, Dieter~H. {Hartmann}, Marco {Ajello},
  Ramon {Canal}, Friedrich~K. {R{\"o}pke}, Sebastian~T. {Ohlmann}, and Wolfgang
  {Hillebrandt}.
\newblock {The Origin of the Cosmic Gamma-ray Background in the MeV Range}.
\newblock {\em \apj}, 820(2):142, April 2016.

\bibitem{Ahn05}
Kyungjin {Ahn}, Eiichiro {Komatsu}, and Peter {H{\"o}flich}.
\newblock {Cosmic gamma-ray background from type Ia supernovae reexamined:
  Evidence for missing gamma rays at MeV energy}.
\newblock {\em \prd}, 71(12):121301, June 2005.

\bibitem{Zdziarski00}
Andrzej~A. {Zdziarski}, Juri {Poutanen}, and W.~Neil {Johnson}.
\newblock {Observations of Seyfert Galaxies by OSSE and Parameters of Their
  X-Ray/Gamma-Ray Sources}.
\newblock {\em \apj}, 542(2):703--709, October 2000.

\bibitem{grb170817}
A.~{Goldstein}, P.~{Veres}, E.~{Burns}, M.~S. {Briggs}, R.~{Hamburg},
  D.~{Kocevski}, C.~A. {Wilson-Hodge}, R.~D. {Preece}, S.~{Poolakkil}, O.~J.
  {Roberts}, C.~M. {Hui}, V.~{Connaughton}, J.~{Racusin}, A.~{von Kienlin},
  T.~{Dal Canton}, N.~{Christensen}, T.~{Littenberg}, K.~{Siellez},
  L.~{Blackburn}, J.~{Broida}, E.~{Bissaldi}, W.~H. {Cleveland}, M.~H. {Gibby},
  M.~M. {Giles}, R.~M. {Kippen}, S.~{McBreen}, J.~{McEnery}, C.~A. {Meegan},
  W.~S. {Paciesas}, and M.~{Stanbro}.
\newblock {An Ordinary Short Gamma-Ray Burst with Extraordinary Implications:
  Fermi-GBM Detection of GRB 170817A}.
\newblock {\em \apjl}, 848(2):L14, October 2017.

\bibitem{txs0506}
{IceCube Collaboration, et al.}
\newblock {Multimessenger observations of a flaring blazar coincident with
  high-energy neutrino IceCube-170922A}.
\newblock {\em Science}, 361(6398):eaat1378, July 2018.

\bibitem{AMEGORFI}
J.~McEnery et~al.
\newblock Amego: A multimessenger mission for the extreme universe.
\newblock \url{https://asd.gsfc.nasa.gov/amego/files/AMEGO_Decadal_RFI.pdf}.

\bibitem{2020SPIE11444E..31K}
Carolyn~A. {Kierans}.
\newblock {AMEGO: exploring the extreme multimessenger universe}.
\newblock In {\em Society of Photo-Optical Instrumentation Engineers (SPIE)
  Conference Series}, volume 11444 of {\em Society of Photo-Optical
  Instrumentation Engineers (SPIE) Conference Series}, page 1144431, December
  2020.

\bibitem{MartinezCastellanos2021}
I.~{Martinez-Castellanos}, H.~{Fleischhack}, C.~{Karwin}, M.~{Negro}, D.~{Tak},
  Amy {Lien}, C.~A. {Kierans}, Zorawar {Wadiasingh}, Yasushi {Fukazawa}, Marco
  {Ajello}, Matthew~G. {Baring}, E.~{Burns}, R.~{Caputo}, Jeremy~S. {Perkins},
  Judith~L. {Racusin}, and Yong {Sheng}.
\newblock {Improving the low-energy transient sensitivity of AMEGO-X using
  single-site events}.
\newblock {\em arXiv e-prints}, page arXiv:2111.09209, November 2021.

\bibitem{Lewis2021}
Tiffany~R. {Lewis}, Christopher~M. {Karwin}, Tonia~M. {Venters}, Henrike
  {Fleischhack}, Yong {Sheng}, Carolyn~A. {Kierans}, Regina {Caputo}, and Julie
  {McEnery}.
\newblock {Modeling and Simulations of TXS 0506+056 Neutrino Events in the MeV
  Band}.
\newblock {\em arXiv e-prints}, page arXiv:2111.10600, November 2021.

\bibitem{Negro2021}
Michela {Negro}, Henrike {Fleischhack}, Andreas {Zoglauer}, Seth {Digel}, and
  Marco {Ajello}.
\newblock {Unveiling the Fermi Bubbles origin with MeV photon telescopes}.
\newblock {\em arXiv e-prints}, page arXiv:2111.10362, November 2021.

\bibitem{woolf2018}
R.~S. {Woolf}, J.~E. {Grove}, B.~F. {Phlips}, and E.~A. {Wulf}.
\newblock {Development of a CsI:Tl Calorimeter Subsystem for the All-Sky
  Medium-Energy Gamma-Ray Observatory (AMEGO)}.
\newblock In {\em The Space Astrophysics Landscape for the 2020s and Beyond},
  volume 2135 of {\em LPI Contributions}, page 5016, April 2019.

\bibitem{Tomsick}
J.A. Tomsick and {\em et al.}
\newblock {The Compton Spectrometer and Imager Project for MeV Astronomy}.
\newblock {\em PoS(ICRC2021)652, 2021 International Cosmic Ray Conference},
  Berlin, 2021.

\bibitem{Zoglauer}
A.~Zoglauer et~al.
\newblock {COSI: From Calibrations and Observations to All-sky Images}.
\newblock 02 2021.

\bibitem{Orlando}
Elena Orlando et~al.
\newblock {Exploring the MeV Sky with a Combined Coded Mask and Compton
  Telescope: The Galactic Explorer with a Coded Aperture Mask Compton Telescope
  (GECCO)}.
\newblock 12 2021.

\bibitem{MEGALIB}
A.~{Zoglauer}, R.~{Andritschke}, and F.~{Schopper}.
\newblock {MEGAlib The Medium Energy Gamma-ray Astronomy Library}.
\newblock 50:629--632, October 2006.

\bibitem{ARAMAKI2020107}
Tsuguo Aramaki, Per Ola~Hansson Adrian, Georgia Karagiorgi, and Hirokazu Odaka.
\newblock Dual mev gamma-ray and dark matter observatory - grams project.
\newblock {\em Astroparticle Physics}, 114:107--114, 2020.

\bibitem{Aramaki:2020gqm}
Tsuguo Aramaki et~al.
\newblock {Snowmass 2021 Letter of Interest: The GRAMS Project: MeV Gamma-Ray
  Observations and Antimatter-Based Dark Matter Searches}.
\newblock 9 2020.

\bibitem{Ajaj:2019imk}
R.~Ajaj et~al.
\newblock {Search for dark matter with a 231-day exposure of liquid argon using
  DEAP-3600 at SNOLAB}.
\newblock {\em Phys. Rev.}, D100(2):022004, 2019.

\bibitem{Brodzinski_1955}
R.~L. Brodzinski and N.~A. Wogman.
\newblock High-energy proton spallation of argon.
\newblock {\em Phys. Rev. C}, 1:1955--1959, Jun 1970.

\bibitem{Saldanha_2019}
R.~Saldanha, H.~O. Back, R.~H.~M. Tsang, T.~Alexander, S.~R. Elliott,
  S.~Ferrara, E.~Mace, C.~Overman, and M.~Zalavadia.
\newblock Cosmogenic production of $^{39}\mathrm{Ar}$ and $^{37}\mathrm{Ar}$ in
  argon.
\newblock {\em Phys. Rev. C}, 100:024608, Aug 2019.

\bibitem{Cumani_2019}
P.~Cumani, M.~Hernanz, J.~Kiener, V.~Tatischeff, and A.~Zoglauer.
\newblock Background for a gamma-ray satellite on a low-earth orbit.
\newblock {\em Experimental Astronomy}, 47(3):273–302, Mar 2019.

\bibitem{Fleischhack_2021}
Henrike {Fleischhack}.
\newblock {AMEGO-X: MeV gamma-ray Astronomy in the Multimessenger Era}.
\newblock {\em arXiv e-prints}, page arXiv:2108.02860, August 2021.

\bibitem{Sacerdoti_2022}
S.~Sacerdoti.
\newblock Development of analog signal transmission in {LAr} for {DUNE}.
\newblock {\em Journal of Instrumentation}, 17(01):C01069, jan 2022.

\bibitem{Vercellone:2015fza}
Stefano Vercellone.
\newblock {The ASTRI mini-array within the future Cherenkov Telescope Array}.
\newblock {\em EPJ Web Conf.}, 121:04006, 2016.

\bibitem{2011ExA....32..193A}
M.~{Actis}, G.~{Agnetta}, F.~{Aharonian}, A.~{Akhperjanian}, J.~{Aleksi{\'c}},
  E.~{Aliu}, D.~{Allan}, I.~{Allekotte}, F.~{Antico}, L.~A. {Antonelli},
  P.~{Antoranz}, A.~{Aravantinos}, T.~{Arlen}, H.~{Arnaldi}, S.~{Artmann},
  K.~{Asano}, H.~{Asorey}, J.~{B{\"a}hr}, A.~{Bais}, C.~{Baixeras},
  S.~{Bajtlik}, D.~{Balis}, A.~{Bamba}, C.~{Barbier}, M.~{Barcel{\'o}},
  A.~{Barnacka}, J.~{Barnstedt}, U.~{Barres de Almeida}, J.~A. {Barrio},
  S.~{Basso}, D.~{Bastieri}, C.~{Bauer}, J.~{Becerra}, Y.~{Becherini},
  K.~{Bechtol}, J.~{Becker}, V.~{Beckmann}, W.~{Bednarek}, B.~{Behera},
  M.~{Beilicke}, M.~{Belluso}, M.~{Benallou}, W.~{Benbow}, J.~{Berdugo},
  K.~{Berger}, T.~{Bernardino}, K.~{Bernl{\"o}hr}, A.~{Biland}, S.~{Billotta},
  T.~{Bird}, E.~{Birsin}, E.~{Bissaldi}, S.~{Blake}, O.~{Blanch}, A.~A.
  {Bobkov}, L.~{Bogacz}, M.~{Bogdan}, C.~{Boisson}, J.~{Boix}, J.~{Bolmont},
  G.~{Bonanno}, A.~{Bonardi}, T.~{Bonev}, J.~{Borkowski}, O.~{Botner},
  A.~{Bottani}, M.~{Bourgeat}, C.~{Boutonnet}, A.~{Bouvier},
  S.~{Brau-Nogu{\'e}}, I.~{Braun}, T.~{Bretz}, M.~S. {Briggs}, P.~{Brun},
  L.~{Brunetti}, J.~H. {Buckley}, V.~{Bugaev}, R.~{B{\"u}hler}, T.~{Bulik},
  G.~{Busetto}, S.~{Buson}, K.~{Byrum}, M.~{Cailles}, R.~{Cameron},
  R.~{Canestrari}, S.~{Cantu}, E.~{Carmona}, A.~{Carosi}, J.~{Carr}, P.~H.
  {Carton}, M.~{Casiraghi}, H.~{Castarede}, O.~{Catalano}, S.~{Cavazzani},
  S.~{Cazaux}, B.~{Cerruti}, M.~{Cerruti}, P.~M. {Chadwick}, J.~{Chiang},
  M.~{Chikawa}, M.~{Cie{\'s}lar}, M.~{Ciesielska}, A.~{Cillis}, C.~{Clerc},
  P.~{Colin}, J.~{Colom{\'e}}, M.~{Compin}, P.~{Conconi}, V.~{Connaughton},
  J.~{Conrad}, J.~L. {Contreras}, P.~{Coppi}, M.~{Corlier}, P.~{Corona},
  O.~{Corpace}, D.~{Corti}, J.~{Cortina}, H.~{Costantini}, G.~{Cotter},
  B.~{Courty}, S.~{Couturier}, S.~{Covino}, J.~{Croston}, G.~{Cusumano}, M.~K.
  {Daniel}, F.~{Dazzi}, A.~{de Angelis}, E.~{de Cea Del Pozo}, E.~M. {de
  Gouveia Dal Pino}, O.~{de Jager}, I.~{de La Calle P{\'e}rez}, G.~{de La
  Vega}, B.~{de Lotto}, M.~{de Naurois}, E.~{de O{\~n}a Wilhelmi}, V.~{de
  Souza}, B.~{Decerprit}, C.~{Deil}, E.~{Delagnes}, G.~{Deleglise},
  C.~{Delgado}, T.~{Dettlaff}, A.~{di Paolo}, F.~{di Pierro}, C.~{D{\'\i}az},
  J.~{Dick}, H.~{Dickinson}, S.~W. {Digel}, D.~{Dimitrov}, G.~{Disset},
  A.~{Djannati-Ata{\"\i}}, M.~{Doert}, W.~{Domainko}, D.~{Dorner}, M.~{Doro},
  J.~L. {Dournaux}, D.~{Dravins}, L.~{Drury}, F.~{Dubois}, R.~{Dubois},
  G.~{Dubus}, C.~{Dufour}, D.~{Durand}, J.~{Dyks}, M.~{Dyrda}, E.~{Edy},
  K.~{Egberts}, C.~{Eleftheriadis}, S.~{Elles}, D.~{Emmanoulopoulos},
  R.~{Enomoto}, J.~P. {Ernenwein}, M.~{Errando}, A.~{Etchegoyen}, A.~D.
  {Falcone}, K.~{Farakos}, C.~{Farnier}, S.~{Federici}, F.~{Feinstein},
  D.~{Ferenc}, E.~{Fillin-Martino}, D.~{Fink}, C.~{Finley}, J.~P. {Finley},
  R.~{Firpo}, D.~{Florin}, C.~{F{\"o}hr}, E.~{Fokitis}, Ll. {Font},
  G.~{Fontaine}, A.~{Fontana}, A.~{F{\"o}rster}, L.~{Fortson}, N.~{Fouque},
  C.~{Fransson}, G.~W. {Fraser}, L.~{Fresnillo}, C.~{Fruck}, Y.~{Fujita},
  Y.~{Fukazawa}, S.~{Funk}, W.~{G{\"a}bele}, S.~{Gabici}, A.~{Gadola},
  N.~{Galante}, Y.~{Gallant}, B.~{Garc{\'\i}a}, R.~J. {Garc{\'\i}a L{\'o}pez},
  D.~{Garrido}, L.~{Garrido}, D.~{Gasc{\'o}n}, C.~{Gasq}, M.~{Gaug},
  J.~{Gaweda}, N.~{Geffroy}, C.~{Ghag}, A.~{Ghedina}, M.~{Ghigo},
  E.~{Gianakaki}, S.~{Giarrusso}, G.~{Giavitto}, B.~{Giebels}, E.~{Giro},
  P.~{Giubilato}, T.~{Glanzman}, J.~F. {Glicenstein}, M.~{Gochna}, V.~{Golev},
  M.~{G{\'o}mez Berisso}, A.~{Gonz{\'a}lez}, F.~{Gonz{\'a}lez},
  F.~{Gra{\~n}ena}, R.~{Graciani}, J.~{Granot}, R.~{Gredig}, A.~{Green},
  T.~{Greenshaw}, O.~{Grimm}, J.~{Grube}, M.~{Grudzi{\'n}ska}, J.~{Grygorczuk},
  V.~{Guarino}, L.~{Guglielmi}, F.~{Guilloux}, S.~{Gunji}, G.~{Gyuk},
  D.~{Hadasch}, D.~{Haefner}, R.~{Hagiwara}, J.~{Hahn}, A.~{Hallgren},
  S.~{Hara}, M.~J. {Hardcastle}, T.~{Hassan}, T.~{Haubold}, M.~{Hauser},
  M.~{Hayashida}, R.~{Heller}, G.~{Henri}, G.~{Hermann}, A.~{Herrero}, J.~A.
  {Hinton}, D.~{Hoffmann}, W.~{Hofmann}, P.~{Hofverberg}, D.~{Horns},
  D.~{Hrupec}, H.~{Huan}, B.~{Huber}, J.~M. {Huet}, G.~{Hughes},
  K.~{Hultquist}, T.~B. {Humensky}, J.~F. {Huppert}, A.~{Ibarra}, J.~M. {Illa},
  J.~{Ingjald}, Y.~{Inoue}, S.~{Inoue}, K.~{Ioka}, C.~{Jablonski},
  A.~{Jacholkowska}, M.~{Janiak}, P.~{Jean}, H.~{Jensen}, T.~{Jogler},
  I.~{Jung}, P.~{Kaaret}, S.~{Kabuki}, J.~{Kakuwa}, C.~{Kalkuhl},
  R.~{Kankanyan}, M.~{Kapala}, A.~{Karastergiou}, M.~{Karczewski}, S.~{Karkar},
  N.~{Karlsson}, J.~{Kasperek}, H.~{Katagiri}, K.~{Katarzy{\'n}ski},
  N.~{Kawanaka}, B.~{K{\c{e}}dziora}, E.~{Kendziorra}, B.~{Kh{\'e}lifi},
  D.~{Kieda}, T.~{Kifune}, T.~{Kihm}, S.~{Klepser}, W.~{Klu{\'z}niak},
  J.~{Knapp}, A.~R. {Knappy}, T.~{Kneiske}, J.~{Kn{\"o}dlseder}, F.~{K{\"o}ck},
  K.~{Kodani}, K.~{Kohri}, K.~{Kokkotas}, N.~{Komin}, A.~{Konopelko},
  K.~{Kosack}, R.~{Kossakowski}, P.~{Kostka}, J.~{Kotu{\l}a}, G.~{Kowal},
  J.~{Kozio{\l}}, T.~{Kr{\"a}henb{\"u}hl}, J.~{Krause}, H.~{Krawczynski},
  F.~{Krennrich}, A.~{Kretzschmann}, H.~{Kubo}, V.~A. {Kudryavtsev},
  J.~{Kushida}, N.~{La Barbera}, V.~{La Parola}, G.~{La Rosa}, A.~{L{\'o}pez},
  G.~{Lamanna}, P.~{Laporte}, C.~{Lavalley}, T.~{Le Flour}, A.~{Le Padellec},
  J.~P. {Lenain}, L.~{Lessio}, B.~{Lieunard}, E.~{Lindfors}, A.~{Liolios},
  T.~{Lohse}, S.~{Lombardi}, A.~{Lopatin}, E.~{Lorenz}, P.~{Lubi{\'n}ski},
  O.~{Luz}, E.~{Lyard}, M.~C. {Maccarone}, T.~{Maccarone}, G.~{Maier},
  P.~{Majumdar}, S.~{Maltezos}, P.~{Ma{\l}kiewicz}, C.~{Ma{\~n}{\'a}},
  A.~{Manalaysay}, G.~{Maneva}, A.~{Mangano}, P.~{Manigot}, J.~{Mar{\'\i}n},
  M.~{Mariotti}, S.~{Markoff}, G.~{Mart{\'\i}nez}, M.~{Mart{\'\i}nez},
  A.~{Mastichiadis}, H.~{Matsumoto}, S.~{Mattiazzo}, D.~{Mazin}, T.~J.~L.
  {McComb}, N.~{McCubbin}, I.~{McHardy}, C.~{Medina}, D.~{Melkumyan},
  A.~{Mendes}, P.~{Mertsch}, M.~{Meucci}, J.~{Micha{\l}owski}, P.~{Micolon},
  T.~{Mineo}, N.~{Mirabal}, F.~{Mirabel}, J.~M. {Miranda}, R.~{Mirzoyan},
  T.~{Mizuno}, B.~{Moal}, R.~{Moderski}, E.~{Molinari}, I.~{Monteiro},
  A.~{Moralejo}, C.~{Morello}, K.~{Mori}, G.~{Motta}, F.~{Mottez}, E.~{Moulin},
  R.~{Mukherjee}, P.~{Munar}, H.~{Muraishi}, K.~{Murase}, A.~Stj. {Murphy},
  S.~{Nagataki}, T.~{Naito}, T.~{Nakamori}, K.~{Nakayama}, C.~{Naumann},
  D.~{Naumann}, P.~{Nayman}, D.~{Nedbal}, A.~{Nied{\'z}wiecki}, J.~{Niemiec},
  A.~{Nikolaidis}, K.~{Nishijima}, S.~J. {Nolan}, N.~{Nowak}, P.~T. {O'Brien},
  I.~{Ochoa}, Y.~{Ohira}, M.~{Ohishi}, H.~{Ohka}, A.~{Okumura}, C.~{Olivetto},
  R.~A. {Ong}, R.~{Orito}, M.~{Orr}, J.~P. {Osborne}, M.~{Ostrowski},
  L.~{Otero}, A.~N. {Otte}, E.~{Ovcharov}, I.~{Oya}, A.~{Ozi{\c{e}}b{\l}o},
  S.~{Paiano}, J.~{Pallota}, J.~L. {Panazol}, D.~{Paneque}, M.~{Panter},
  R.~{Paoletti}, G.~{Papyan}, J.~M. {Paredes}, G.~{Pareschi}, R.~D. {Parsons},
  M.~{Paz Arribas}, G.~{Pedaletti}, A.~{Pepato}, M.~{Persic}, P.~O. {Petrucci},
  B.~{Peyaud}, W.~{Piechocki}, S.~{Pita}, G.~{Pivato}, {\L}.~{P{\l}atos},
  R.~{Platzer}, L.~{Pogosyan}, M.~{Pohl}, G.~{Pojma{\'n}ski}, J.~D. {Ponz},
  W.~{Potter}, E.~{Prandini}, R.~{Preece}, H.~{Prokoph}, G.~{P{\"u}hlhofer},
  M.~{Punch}, E.~{Quel}, A.~{Quirrenbach}, P.~{Rajda}, R.~{Rando}, M.~{Rataj},
  M.~{Raue}, C.~{Reimann}, O.~{Reimann}, A.~{Reimer}, O.~{Reimer}, M.~{Renaud},
  S.~{Renner}, J.~M. {Reymond}, W.~{Rhode}, M.~{Rib{\'o}}, M.~{Ribordy},
  J.~{Rico}, F.~{Rieger}, P.~{Ringegni}, J.~{Ripken}, P.~{Ristori},
  S.~{Rivoire}, L.~{Rob}, S.~{Rodriguez}, U.~{Roeser}, P.~{Romano}, G.~E.
  {Romero}, S.~{Rosier-Lees}, A.~C. {Rovero}, F.~{Roy}, S.~{Royer}, B.~{Rudak},
  C.~B. {Rulten}, J.~{Ruppel}, F.~{Russo}, F.~{Ryde}, B.~{Sacco}, A.~{Saggion},
  V.~{Sahakian}, K.~{Saito}, T.~{Saito}, N.~{Sakaki}, E.~{Salazar},
  A.~{Salini}, F.~{S{\'a}nchez}, M.~{\'A}. {S{\'a}nchez Conde},
  A.~{Santangelo}, E.~M. {Santos}, A.~{Sanuy}, L.~{Sapozhnikov}, S.~{Sarkar},
  V.~{Scalzotto}, V.~{Scapin}, M.~{Scarcioffolo}, T.~{Schanz},
  S.~{Schlenstedt}, R.~{Schlickeiser}, T.~{Schmidt}, J.~{Schmoll},
  M.~{Schroedter}, C.~{Schultz}, J.~{Schultze}, A.~{Schulz}, U.~{Schwanke},
  S.~{Schwarzburg}, T.~{Schweizer}, J.~{Seiradakis}, S.~{Selmane},
  K.~{Seweryn}, M.~{Shayduk}, R.~C. {Shellard}, T.~{Shibata}, M.~{Sikora},
  J.~{Silk}, A.~{Sillanp{\"a}{\"a}}, J.~{Sitarek}, C.~{Skole}, N.~{Smith},
  D.~{Sobczy{\'n}ska}, M.~{Sofo Haro}, H.~{Sol}, F.~{Spanier}, D.~{Spiga},
  S.~{Spyrou}, V.~{Stamatescu}, A.~{Stamerra}, R.~L.~C. {Starling},
  {\L}.~{Stawarz}, R.~{Steenkamp}, C.~{Stegmann}, S.~{Steiner},
  N.~{Stergioulas}, R.~{Sternberger}, F.~{Stinzing}, M.~{Stodulski},
  U.~{Straumann}, A.~{Su{\'a}rez}, M.~{Suchenek}, R.~{Sugawara}, K.~H.
  {Sulanke}, S.~{Sun}, A.~D. {Supanitsky}, P.~{Sutcliffe}, M.~{Szanecki},
  T.~{Szepieniec}, A.~{Szostek}, A.~{Szymkowiak}, G.~{Tagliaferri},
  H.~{Tajima}, H.~{Takahashi}, K.~{Takahashi}, L.~{Takalo}, H.~{Takami}, R.~G.
  {Talbot}, P.~H. {Tam}, M.~{Tanaka}, T.~{Tanimori}, M.~{Tavani}, J.~P.
  {Tavernet}, C.~{Tchernin}, L.~A. {Tejedor}, I.~{Telezhinsky}, P.~{Temnikov},
  C.~{Tenzer}, Y.~{Terada}, R.~{Terrier}, M.~{Teshima}, V.~{Testa},
  L.~{Tibaldo}, O.~{Tibolla}, M.~{Tluczykont}, C.~J. {Todero Peixoto},
  F.~{Tokanai}, M.~{Tokarz}, K.~{Toma}, D.~F. {Torres}, G.~{Tosti},
  T.~{Totani}, F.~{Toussenel}, P.~{Vallania}, G.~{Vallejo}, J.~{van der Walt},
  C.~{van Eldik}, J.~{Vandenbroucke}, H.~{Vankov}, G.~{Vasileiadis}, V.~V.
  {Vassiliev}, I.~{Vegas}, L.~{Venter}, S.~{Vercellone}, C.~{Veyssiere}, J.~P.
  {Vialle}, M.~{Videla}, P.~{Vincent}, J.~{Vink}, N.~{Vlahakis}, L.~{Vlahos},
  P.~{Vogler}, A.~{Vollhardt}, F.~{Volpe}, H.~P. {von Gunten}, S.~{Vorobiov},
  S.~{Wagner}, R.~M. {Wagner}, B.~{Wagner}, S.~P. {Wakely}, P.~{Walter},
  R.~{Walter}, R.~{Warwick}, P.~{Wawer}, R.~{Wawrzaszek}, N.~{Webb},
  P.~{Wegner}, A.~{Weinstein}, Q.~{Weitzel}, R.~{Welsing}, H.~{Wetteskind},
  R.~{White}, A.~{Wierzcholska}, M.~I. {Wilkinson}, D.~A. {Williams},
  M.~{Winde}, R.~{Wischnewski}, {\L}.~{Wi{\'s}niewski}, A.~{Wolczko},
  M.~{Wood}, Q.~{Xiong}, T.~{Yamamoto}, K.~{Yamaoka}, R.~{Yamazaki},
  S.~{Yanagita}, B.~{Yoffo}, M.~{Yonetani}, A.~{Yoshida}, T.~{Yoshida},
  T.~{Yoshikoshi}, V.~{Zabalza}, A.~{Zagda{\'n}ski}, A.~{Zajczyk},
  A.~{Zdziarski}, A.~{Zech}, K.~{Zi{\c{e}}tara}, P.~{Zi{\'o}{\l}kowski},
  V.~{Zitelli}, and P.~{Zychowski}.
\newblock {Design concepts for the Cherenkov Telescope Array CTA: an advanced
  facility for ground-based high-energy gamma-ray astronomy}.
\newblock {\em Experimental Astronomy}, 32(3):193--316, December 2011.

\bibitem{CTAConsortium:2013ofs}
B.~S. Acharya et~al.
\newblock {Introducing the CTA concept}.
\newblock {\em Astroparticle Physics}, 43:3--18, 2013.

\bibitem{Snowmass2013:2013cqj}
Norman~A. Graf, Michael~E. Peskin, and Jonathan~L. Rosner, editors.
\newblock {\em {Proceedings, 2013 Community Summer Study on the Future of U.S.
  Particle Physics: Snowmass on the Mississippi (CSS2013)}: {Minneapolis, MN,
  USA, July 29-August 6, 2013}}, 2013.

\bibitem{P5Report:2014pwa}
Steve Ritz et~al.
\newblock {Building for Discovery: Strategic Plan for U.S. Particle Physics in
  the Global Context}.
\newblock 5 2014.

\bibitem{CTA:2020qlo}
A.~Acharyya et~al.
\newblock {Sensitivity of the Cherenkov Telescope Array to a dark matter signal
  from the Galactic centre}.
\newblock {\em Journal of Cosmology and Astroparticle Physics}, 01:057, 2021.

\bibitem{CTA:2020hii}
H.~Abdalla et~al.
\newblock {Sensitivity of the Cherenkov Telescope Array for probing cosmology
  and fundamental physics with gamma-ray propagation}.
\newblock {\em JCAP}, 02:048, 2021.

\bibitem{2017ApJ...841..100A}
A.~U. {Abeysekara}, A.~{Albert}, R.~{Alfaro}, C.~{Alvarez}, J.~D.
  {{\'A}lvarez}, R.~{Arceo}, J.~C. {Arteaga-Vel{\'a}zquez}, D.~{Avila Rojas},
  H.~A. {Ayala Solares}, and A.~S. {Barber}.
\newblock {Daily Monitoring of TeV Gamma-Ray Emission from Mrk 421, Mrk 501,
  and the Crab Nebula with HAWC}.
\newblock {\em \apj}, 841(2):100, Jun 2017.

\bibitem{2017ApJ...843..116A}
A.~U. {Abeysekara}, R.~{Alfaro}, C.~{Alvarez}, J.~D. {{\'A}lvarez}, R.~{Arceo},
  J.~C. {Arteaga-Vel{\'a}zquez}, D.~{Avila Rojas}, H.~A. {Ayala Solares}, A.~S.
  {Barber}, and N.~{Bautista-Elivar}.
\newblock {The HAWC Real-time Flare Monitor for Rapid Detection of Transient
  Events}.
\newblock {\em \apj}, 843(2):116, Jul 2017.

\bibitem{2009ApJ...697.1071A}
W.~B. {Atwood}, A.~A. {Abdo}, M.~{Ackermann}, W.~{Althouse}, B.~{Anderson},
  M.~{Axelsson}, L.~{Baldini}, J.~{Ballet}, D.~L. {Band}, G.~{Barbiellini},
  J.~{Bartelt}, D.~{Bastieri}, B.~M. {Baughman}, K.~{Bechtol},
  D.~{B{\'e}d{\'e}r{\`e}de}, F.~{Bellardi}, R.~{Bellazzini}, B.~{Berenji},
  G.~F. {Bignami}, D.~{Bisello}, E.~{Bissaldi}, R.~D. {Blandford}, E.~D.
  {Bloom}, J.~R. {Bogart}, E.~{Bonamente}, J.~{Bonnell}, A.~W. {Borgland},
  A.~{Bouvier}, J.~{Bregeon}, A.~{Brez}, M.~{Brigida}, P.~{Bruel}, T.~H.
  {Burnett}, G.~{Busetto}, G.~A. {Caliandro}, R.~A. {Cameron}, P.~A. {Caraveo},
  S.~{Carius}, P.~{Carlson}, J.~M. {Casandjian}, E.~{Cavazzuti}, M.~{Ceccanti},
  C.~{Cecchi}, E.~{Charles}, A.~{Chekhtman}, C.~C. {Cheung}, J.~{Chiang},
  R.~{Chipaux}, A.~N. {Cillis}, S.~{Ciprini}, R.~{Claus}, J.~{Cohen-Tanugi},
  S.~{Condamoor}, J.~{Conrad}, R.~{Corbet}, L.~{Corucci}, L.~{Costamante},
  S.~{Cutini}, D.~S. {Davis}, D.~{Decotigny}, M.~{DeKlotz}, C.~D. {Dermer},
  A.~{de Angelis}, S.~W. {Digel}, E.~{do Couto e Silva}, P.~S. {Drell},
  R.~{Dubois}, D.~{Dumora}, Y.~{Edmonds}, D.~{Fabiani}, C.~{Farnier},
  C.~{Favuzzi}, D.~L. {Flath}, P.~{Fleury}, W.~B. {Focke}, S.~{Funk},
  P.~{Fusco}, F.~{Gargano}, D.~{Gasparrini}, N.~{Gehrels}, F.~X. {Gentit},
  S.~{Germani}, B.~{Giebels}, N.~{Giglietto}, P.~{Giommi}, F.~{Giordano},
  T.~{Glanzman}, G.~{Godfrey}, I.~A. {Grenier}, M.~H. {Grondin}, J.~E. {Grove},
  L.~{Guillemot}, S.~{Guiriec}, G.~{Haller}, A.~K. {Harding}, P.~A. {Hart},
  E.~{Hays}, S.~E. {Healey}, M.~{Hirayama}, L.~{Hjalmarsdotter}, R.~{Horn},
  R.~E. {Hughes}, G.~{J{\'o}hannesson}, G.~{Johansson}, A.~S. {Johnson}, R.~P.
  {Johnson}, T.~J. {Johnson}, W.~N. {Johnson}, T.~{Kamae}, H.~{Katagiri},
  J.~{Kataoka}, A.~{Kavelaars}, N.~{Kawai}, H.~{Kelly}, M.~{Kerr}, W.~{Klamra},
  J.~{Kn{\"o}dlseder}, M.~L. {Kocian}, N.~{Komin}, F.~{Kuehn}, M.~{Kuss},
  D.~{Landriu}, L.~{Latronico}, B.~{Lee}, S.~H. {Lee}, M.~{Lemoine-Goumard},
  A.~M. {Lionetto}, F.~{Longo}, F.~{Loparco}, B.~{Lott}, M.~N. {Lovellette},
  P.~{Lubrano}, G.~M. {Madejski}, A.~{Makeev}, B.~{Marangelli}, M.~M. {Massai},
  M.~N. {Mazziotta}, J.~E. {McEnery}, N.~{Menon}, C.~{Meurer}, P.~F.
  {Michelson}, M.~{Minuti}, N.~{Mirizzi}, W.~{Mitthumsiri}, T.~{Mizuno}, A.~A.
  {Moiseev}, C.~{Monte}, M.~E. {Monzani}, E.~{Moretti}, A.~{Morselli}, I.~V.
  {Moskalenko}, S.~{Murgia}, T.~{Nakamori}, S.~{Nishino}, P.~L. {Nolan}, J.~P.
  {Norris}, E.~{Nuss}, M.~{Ohno}, T.~{Ohsugi}, N.~{Omodei}, E.~{Orlando}, J.~F.
  {Ormes}, A.~{Paccagnella}, D.~{Paneque}, J.~H. {Panetta}, D.~{Parent},
  M.~{Pearce}, M.~{Pepe}, A.~{Perazzo}, M.~{Pesce-Rollins}, P.~{Picozza},
  L.~{Pieri}, M.~{Pinchera}, F.~{Piron}, T.~A. {Porter}, L.~{Poupard},
  S.~{Rain{\`o}}, R.~{Rando}, E.~{Rapposelli}, M.~{Razzano}, A.~{Reimer},
  O.~{Reimer}, T.~{Reposeur}, L.~C. {Reyes}, S.~{Ritz}, L.~S. {Rochester},
  A.~Y. {Rodriguez}, R.~W. {Romani}, M.~{Roth}, J.~J. {Russell}, F.~{Ryde},
  S.~{Sabatini}, H.~F.~W. {Sadrozinski}, D.~{Sanchez}, A.~{Sander},
  L.~{Sapozhnikov}, P.~M.~Saz {Parkinson}, J.~D. {Scargle}, T.~L. {Schalk},
  G.~{Scolieri}, C.~{Sgr{\`o}}, G.~H. {Share}, M.~{Shaw}, T.~{Shimokawabe},
  C.~{Shrader}, A.~{Sierpowska-Bartosik}, E.~J. {Siskind}, D.~A. {Smith}, P.~D.
  {Smith}, G.~{Spandre}, P.~{Spinelli}, J.~L. {Starck}, T.~E. {Stephens}, M.~S.
  {Strickman}, A.~W. {Strong}, D.~J. {Suson}, H.~{Tajima}, H.~{Takahashi},
  T.~{Takahashi}, T.~{Tanaka}, A.~{Tenze}, S.~{Tether}, J.~B. {Thayer}, J.~G.
  {Thayer}, D.~J. {Thompson}, L.~{Tibaldo}, O.~{Tibolla}, D.~F. {Torres},
  G.~{Tosti}, A.~{Tramacere}, M.~{Turri}, T.~L. {Usher}, N.~{Vilchez},
  V.~{Vitale}, P.~{Wang}, K.~{Watters}, B.~L. {Winer}, K.~S. {Wood},
  T.~{Ylinen}, and M.~{Ziegler}.
\newblock {The Large Area Telescope on the Fermi Gamma-Ray Space Telescope
  Mission}.
\newblock {\em \apj}, 697(2):1071--1102, June 2009.

\bibitem{2015ApJ...800...78A}
A.~U. {Abeysekara}, R.~{Alfaro}, C.~{Alvarez}, J.~D. {{\'A}lvarez}, R.~{Arceo},
  J.~C. {Arteaga-Vel{\'a}zquez}, H.~A. {Ayala Solares}, A.~S. {Barber}, B.~M.
  {Baughman}, and N.~{Bautista-Elivar}.
\newblock {Search for Gamma-Rays from the Unusually Bright GRB 130427A with the
  HAWC Gamma-Ray Observatory}.
\newblock {\em \apj}, 800(2):78, Feb 2015.

\bibitem{2017A&A...607A.115I}
{Icecube Collaboration}, M.~G. {Aartsen}, M.~{Ackermann}, J.~{Adams}, J.~A.
  {Aguilar}, M.~{Ahlers}, M.~{Ahrens}, I.~{Al Samarai}, D.~{Altmann}, and
  K.~{Andeen}.
\newblock {Multiwavelength follow-up of a rare IceCube neutrino multiplet}.
\newblock {\em Astronomy \& Astrophysics}, 607:A115, Nov 2017.

\bibitem{2017ApJ...842...85A}
A.~U. {Abeysekara}, A.~{Albert}, R.~{Alfaro}, C.~{Alvarez}, J.~D.
  {{\'A}lvarez}, R.~{Arceo}, J.~C. {Arteaga-Vel{\'a}zquez}, H.~A. {Ayala
  Solares}, A.~S. {Barber}, and N.~{Bautista-Elivar}.
\newblock {Search for Very High-energy Gamma Rays from the Northern Fermi
  Bubble Region with HAWC}.
\newblock {\em \apj}, 842(2):85, Jun 2017.

\bibitem{2017ApJ...843...39A}
A.~U. {Abeysekara}, A.~{Albert}, R.~{Alfaro}, C.~{Alvarez}, J.~D.
  {{\'A}lvarez}, R.~{Arceo}, J.~C. {Arteaga-Vel{\'a}zquez}, H.~A. {Ayala
  Solares}, A.~S. {Barber}, N.~{Bautista-Elivar}, A.~{Becerril},
  E.~{Belmont-Moreno}, S.~Y. {BenZvi}, D.~{Berley}, J.~{Braun}, C.~{Brisbois},
  K.~S. {Caballero-Mora}, T.~{Capistr{\'a}n}, A.~{Carrami{\~n}ana},
  S.~{Casanova}, M.~{Castillo}, U.~{Cotti}, J.~{Cotzomi}, S.~{Couti{\~n}o de
  Le{\'o}n}, E.~{de la Fuente}, C.~{De Le{\'o}n}, T.~{DeYoung}, B.~L. {Dingus},
  M.~A. {DuVernois}, J.~C. {D{\'{\i}}az-V{\'e}lez}, R.~W. {Ellsworth}, D.~W.
  {Fiorino}, N.~{Fraija}, J.~A. {Garc{\'{\i}}a-Gonz{\'a}lez}, M.~{Gerhardt},
  A.~{Gonz{\'a}lez Mun{\"o}z}, M.~M. {Gonz{\'a}lez}, J.~A. {Goodman},
  Z.~{Hampel-Arias}, J.~P. {Harding}, S.~{Hernandez}, A.~{Hernandez-Almada},
  J.~{Hinton}, C.~M. {Hui}, P.~{H{\"u}ntemeyer}, A.~{Iriarte},
  A.~{Jardin-Blicq}, V.~{Joshi}, S.~{Kaufmann}, D.~{Kieda}, A.~{Lara}, R.~J.
  {Lauer}, W.~H. {Lee}, D.~{Lennarz}, H.~{Le{\'o}n Vargas}, J.~T. {Linnemann},
  A.~L. {Longinotti}, G.~L. {Raya}, R.~{Luna-Garc{\'{\i}}a},
  R.~{L{\'o}pez-Coto}, K.~{Malone}, S.~S. {Marinelli}, O.~{Martinez},
  I.~{Martinez-Castellanos}, J.~{Mart{\'{\i}}nez-Castro},
  H.~{Mart{\'{\i}}nez-Huerta}, J.~A. {Matthews}, P.~{Miranda-Romagnoli},
  E.~{Moreno}, M.~{Mostaf{\'a}}, L.~{Nellen}, M.~{Newbold}, M.~U. {Nisa},
  R.~{Noriega-Papaqui}, R.~{Pelayo}, J.~{Pretz}, E.~G. {P{\'e}rez-P{\'e}rez},
  Z.~{Ren}, C.~D. {Rho}, C.~{Rivi{\`e}re}, D.~{Rosa-Gonz{\'a}lez},
  M.~{Rosenberg}, E.~{Ruiz-Velasco}, H.~{Salazar}, F.~{Salesa Greus},
  A.~{Sandoval}, M.~{Schneider}, H.~{Schoorlemmer}, G.~{Sinnis}, A.~J. {Smith},
  R.~W. {Springer}, P.~{Surajbali}, I.~{Taboada}, O.~{Tibolla}, K.~{Tollefson},
  I.~{Torres}, T.~N. {Ukwatta}, L.~{Villase{\~n}or}, T.~{Weisgarber},
  S.~{Westerhoff}, I.~G. {Wisher}, J.~{Wood}, T.~{Yapici}, G.~B. {Yodh}, P.~W.
  {Younk}, A.~{Zepeda}, and H.~{Zhou}.
\newblock {Observation of the Crab Nebula with the HAWC Gamma-Ray Observatory}.
\newblock {\em \apj}, 843:39, July 2017.

\bibitem{2017ApJ...843...40A}
A.~U. {Abeysekara}, A.~{Albert}, R.~{Alfaro}, C.~{Alvarez}, J.~D.
  {{\'A}lvarez}, R.~{Arceo}, J.~C. {Arteaga-Vel{\'a}zquez}, H.~A. {Ayala
  Solares}, A.~S. {Barber}, and B.~{Baughman}.
\newblock {The 2HWC HAWC Observatory Gamma-Ray Catalog}.
\newblock {\em \apj}, 843(1):40, Jul 2017.

\bibitem{2017ApJ...843...88A}
R.~{Alfaro}, C.~{Alvarez}, J.~D. {{\'A}lvarez}, R.~{Arceo}, J.~C.
  {Arteaga-Vel{\'a}zquez}, D.~{Avila Rojas}, H.~A. {Ayala Solares}, A.~S.
  {Barber}, N.~{Bautista-Elivar}, and A.~{Becerril}.
\newblock {Search for Very-high-energy Emission from Gamma-Ray Bursts Using the
  First 18 Months of Data from the HAWC Gamma-Ray Observatory}.
\newblock {\em \apj}, 843(2):88, Jul 2017.

\bibitem{2017ApJ...848L..12A}
B.~P. {Abbott}, R.~{Abbott}, T.~D. {Abbott}, F.~{Acernese}, K.~{Ackley},
  C.~{Adams}, T.~{Adams}, P.~{Addesso}, R.~X. {Adhikari}, V.~B. {Adya},
  C.~{Affeldt}, M.~{Afrough}, B.~{Agarwal}, M.~{Agathos}, K.~{Agatsuma},
  N.~{Aggarwal}, O.~D. {Aguiar}, L.~{Aiello}, A.~{Ain}, P.~{Ajith}, B.~{Allen},
  G.~{Allen}, A.~{Allocca}, P.~A. {Altin}, A.~{Amato}, A.~{Ananyeva}, S.~B.
  {Anderson}, W.~G. {Anderson}, S.~V. {Angelova}, S.~{Antier}, S.~{Appert},
  K.~{Arai}, M.~C. {Araya}, J.~S. {Areeda}, N.~{Arnaud}, K.~G. {Arun}, and
  S.~and {Ascenzi}.
\newblock {Multi-messenger Observations of a Binary Neutron Star Merger}.
\newblock {\em \apjl}, 848(2):L12, October 2017.

\bibitem{2017Sci...358..911A}
A.~U. {Abeysekara}, A.~{Albert}, R.~{Alfaro}, C.~{Alvarez}, J.~D.
  {{\'A}lvarez}, R.~{Arceo}, J.~C. {Arteaga-Vel{\'a}zquez}, D.~{Avila Rojas},
  H.~A. {Ayala Solares}, A.~S. {Barber}, N.~{Bautista-Elivar}, A.~{Becerril},
  E.~{Belmont-Moreno}, S.~Y. {BenZvi}, D.~{Berley}, A.~{Bernal}, J.~{Braun},
  C.~{Brisbois}, K.~S. {Caballero-Mora}, T.~{Capistr{\'a}n},
  A.~{Carrami{\~n}ana}, S.~{Casanova}, M.~{Castillo}, U.~{Cotti}, J.~{Cotzomi},
  S.~{Couti{\~n}o de Le{\'o}n}, C.~{De Le{\'o}n}, E.~{De la Fuente}, B.~L.
  {Dingus}, M.~A. {DuVernois}, J.~C. {D{\'{\i}}az-V{\'e}lez}, R.~W.
  {Ellsworth}, K.~{Engel}, O.~{Enr{\'{\i}}quez-Rivera}, D.~W. {Fiorino},
  N.~{Fraija}, J.~A. {Garc{\'{\i}}a-Gonz{\'a}lez}, F.~{Garfias}, M.~{Gerhardt},
  A.~{Gonz{\'a}lez Mu{\~n}oz}, M.~M. {Gonz{\'a}lez}, J.~A. {Goodman},
  Z.~{Hampel-Arias}, J.~P. {Harding}, S.~{Hern{\'a}ndez},
  A.~{Hern{\'a}ndez-Almada}, J.~{Hinton}, B.~{Hona}, C.~M. {Hui},
  P.~{H{\"u}ntemeyer}, A.~{Iriarte}, A.~{Jardin-Blicq}, V.~{Joshi},
  S.~{Kaufmann}, D.~{Kieda}, A.~{Lara}, R.~J. {Lauer}, W.~H. {Lee},
  D.~{Lennarz}, H.~L. {Vargas}, J.~T. {Linnemann}, A.~L. {Longinotti}, G.~{Luis
  Raya}, R.~{Luna-Garc{\'{\i}}a}, R.~{L{\'o}pez-Coto}, K.~{Malone}, S.~S.
  {Marinelli}, O.~{Martinez}, I.~{Martinez-Castellanos},
  J.~{Mart{\'{\i}}nez-Castro}, H.~{Mart{\'{\i}}nez-Huerta}, J.~A. {Matthews},
  P.~{Miranda-Romagnoli}, E.~{Moreno}, M.~{Mostaf{\'a}}, L.~{Nellen},
  M.~{Newbold}, M.~U. {Nisa}, R.~{Noriega-Papaqui}, R.~{Pelayo}, J.~{Pretz},
  E.~G. {P{\'e}rez-P{\'e}rez}, Z.~{Ren}, C.~D. {Rho}, C.~{Rivi{\`e}re},
  D.~{Rosa-Gonz{\'a}lez}, M.~{Rosenberg}, E.~{Ruiz-Velasco}, H.~{Salazar},
  F.~{Salesa Greus}, A.~{Sandoval}, M.~{Schneider}, H.~{Schoorlemmer},
  G.~{Sinnis}, A.~J. {Smith}, R.~W. {Springer}, P.~{Surajbali}, I.~{Taboada},
  O.~{Tibolla}, K.~{Tollefson}, I.~{Torres}, T.~N. {Ukwatta}, G.~{Vianello},
  T.~{Weisgarber}, S.~{Westerhoff}, I.~G. {Wisher}, J.~{Wood}, T.~{Yapici},
  G.~{Yodh}, P.~W. {Younk}, A.~{Zepeda}, H.~{Zhou}, F.~{Guo}, J.~{Hahn},
  H.~{Li}, and H.~{Zhang}.
\newblock {Extended gamma-ray sources around pulsars constrain the origin of
  the positron flux at Earth}.
\newblock {\em Science}, 358:911--914, November 2017.

\bibitem{2018ApJ...853..154A}
A.~{Albert}, R.~{Alfaro}, C.~{Alvarez}, J.~D. {{\'A}lvarez}, R.~{Arceo}, J.~C.
  {Arteaga-Vel{\'a}zquez}, D.~{Avila Rojas}, H.~A. {Ayala Solares},
  N.~{Bautista-Elivar}, and A.~{Becerril}.
\newblock {Dark Matter Limits from Dwarf Spheroidal Galaxies with the HAWC
  Gamma-Ray Observatory}.
\newblock {\em \apj}, 853(2):154, Feb 2018.

\bibitem{2018JCAP...02..049A}
A.~U. {Abeysekara}, A.~{Albert}, R.~{Alfaro}, C.~{Alvarez}, R.~{Arceo}, J.~C.
  {Arteaga-Vel{\'a}zquez}, D.~{Avila Rojas}, H.~A. {Ayala Solares},
  A.~{Becerril}, and E.~{Belmont-Moreno}.
\newblock {A search for dark matter in the Galactic halo with HAWC}.
\newblock {\em \jcap}, 2018(2):049, Feb 2018.

\bibitem{2018JCAP...06..043A}
A.~{Albert}, R.~{Alfaro}, C.~{Alvarez}, J.~D. {{\'A}lvarez}, R.~{Arceo}, J.~C.
  {Arteaga-Vel{\'a}zquez}, D.~{Avila Rojas}, H.~A. {Ayala Solares},
  A.~{Becerril}, and E.~{Belmont-Moreno}.
\newblock {Search for dark matter gamma-ray emission from the Andromeda Galaxy
  with the High-Altitude Water Cherenkov Observatory}.
\newblock {\em \jcap}, 2018(6):043, Jun 2018.

\bibitem{2018Sci...361.1378I}
{IceCube Collaboration}, M.~G. {Aartsen}, M.~{Ackermann}, J.~{Adams}, J.~A.
  {Aguilar}, M.~{Ahlers}, M.~{Ahrens}, I.~{Al Samarai}, D.~{Altmann},
  K.~{Andeen}, T.~{Anderson}, I.~{Ansseau}, G.~{Anton}, C.~{Arg{\"u}elles},
  J.~{Auffenberg}, S.~{Axani}, H.~{Bagherpour}, X.~{Bai}, J.~P. {Barron}, S.~W.
  {Barwick}, V.~{Baum}, R.~{Bay}, J.~J. {Beatty}, J.~{Becker Tjus}, K.~H.
  {Becker}, S.~{BenZvi}, D.~{Berley}, E.~{Bernardini}, D.~Z. {Besson},
  G.~{Binder}, D.~{Bindig}, E.~{Blaufuss}, S.~{Blot}, C.~{Bohm},
  M.~{B{\"o}rner}, F.~{Bos}, S.~{B{\"o}ser}, O.~{Botner}, E.~{Bourbeau},
  J.~{Bourbeau}, F.~{Bradascio}, J.~{Braun}, M.~{Brenzke}, H.~P. {Bretz},
  S.~{Bron}, J.~{Brostean-Kaiser}, A.~{Burgman}, R.~S. {Busse}, T.~{Carver},
  E.~{Cheung}, D.~{Chirkin}, A.~{Christov}, K.~{Clark}, L.~{Classen},
  S.~{Coenders}, G.~H. {Collin}, J.~M. {Conrad}, P.~{Coppin}, P.~{Correa},
  D.~F. {Cowen}, R.~{Cross}, P.~{Dave}, M.~{Day}, J.~P.~A.~M. {de Andr{\'e}},
  C.~{De Clercq}, J.~J. {DeLaunay}, H.~{Dembinski}, S.~{De Ridder},
  P.~{Desiati}, K.~D. {de Vries}, G.~{de Wasseige}, M.~{de With}, T.~{DeYoung},
  J.~C. {D{\'\i}az-V{\'e}lez}, V.~{di Lorenzo}, H.~{Dujmovic}, J.~P. {Dumm},
  M.~{Dunkman}, E.~{Dvorak}, B.~{Eberhardt}, T.~{Ehrhardt}, B.~{Eichmann},
  P.~{Eller}, P.~A. {Evenson}, S.~{Fahey}, A.~R. {Fazely}, J.~{Felde},
  K.~{Filimonov}, C.~{Finley}, S.~{Flis}, A.~{Franckowiak}, E.~{Friedman},
  A.~{Fritz}, T.~K. {Gaisser}, J.~{Gallagher}, L.~{Gerhardt}, K.~{Ghorbani},
  T.~{Glauch}, T.~{Gl{\"u}senkamp}, A.~{Goldschmidt}, J.~G. {Gonzalez},
  D.~{Grant}, Z.~{Griffith}, C.~{Haack}, A.~{Hallgren}, F.~{Halzen},
  K.~{Hanson}, D.~{Hebecker}, D.~{Heereman}, K.~{Helbing}, R.~{Hellauer},
  S.~{Hickford}, J.~{Hignight}, G.~C. {Hill}, K.~D. {Hoffman}, R.~{Hoffmann},
  T.~{Hoinka}, B.~{Hokanson-Fasig}, K.~{Hoshina}, F.~{Huang}, M.~{Huber},
  K.~{Hultqvist}, M.~{H{\"u}nnefeld}, R.~{Hussain}, S.~{In}, N.~{Iovine},
  A.~{Ishihara}, E.~{Jacobi}, G.~S. {Japaridze}, M.~{Jeong}, K.~{Jero},
  B.~J.~P. {Jones}, P.~{Kalaczynski}, W.~{Kang}, A.~{Kappes}, D.~{Kappesser},
  T.~{Karg}, A.~{Karle}, U.~{Katz}, M.~{Kauer}, A.~{Keivani}, J.~L. {Kelley},
  A.~{Kheirandish}, J.~{Kim}, M.~{Kim}, T.~{Kintscher}, J.~{Kiryluk},
  T.~{Kittler}, S.~R. {Klein}, R.~{Koirala}, H.~{Kolanoski}, L.~{K{\"o}pke},
  C.~{Kopper}, S.~{Kopper}, J.~P. {Koschinsky}, D.~J. {Koskinen},
  M.~{Kowalski}, K.~{Krings}, M.~{Kroll}, G.~{Kr{\"u}ckl}, S.~{Kunwar},
  N.~{Kurahashi}, T.~{Kuwabara}, A.~{Kyriacou}, M.~{Labare}, J.~L.
  {Lanfranchi}, M.~J. {Larson}, F.~{Lauber}, K.~{Leonard}, M.~{Lesiak-Bzdak},
  M.~{Leuermann}, Q.~R. {Liu}, C.~J. {Lozano Mariscal}, L.~{Lu},
  J.~{L{\"u}nemann}, W.~{Luszczak}, J.~{Madsen}, G.~{Maggi}, K.~B.~M. {Mahn},
  S.~{Mancina}, R.~{Maruyama}, K.~{Mase}, R.~{Maunu}, K.~{Meagher},
  M.~{Medici}, M.~{Meier}, T.~{Menne}, G.~{Merino}, T.~{Meures}, S.~{Miarecki},
  J.~{Micallef}, G.~{Moment{\'e}}, T.~{Montaruli}, R.~W. {Moore}, {S},
  R.~{Morse}, M.~{Moulai}, R.~{Nahnhauer}, P.~{Nakarmi}, U.~{Naumann},
  G.~{Neer}, H.~{Niederhausen}, S.~C. {Nowicki}, D.~R. {Nygren}, A.~{Obertacke
  Pollmann}, A.~{Olivas}, A.~{O'Murchadha}, E.~{O'Sullivan}, T.~{Palczewski},
  H.~{Pandya}, D.~V. {Pankova}, P.~{Peiffer}, J.~A. {Pepper}, C.~{P{\'e}rez de
  los Heros}, D.~{Pieloth}, E.~{Pinat}, M.~{Plum}, P.~B. {Price}, G.~T.
  {Przybylski}, C.~{Raab}, L.~{R{\"a}del}, M.~{Rameez}, L.~{Rauch},
  K.~{Rawlins}, I.~C. {Rea}, R.~{Reimann}, B.~{Relethford}, M.~{Relich},
  E.~{Resconi}, W.~{Rhode}, M.~{Richman}, S.~{Robertson}, M.~{Rongen},
  C.~{Rott}, T.~{Ruhe}, D.~{Ryckbosch}, D.~{Rysewyk}, I.~{Safa},
  T.~{S{\"a}lzer}, S.~E. {Sanchez Herrera}, A.~{Sandrock}, J.~{Sandroos},
  M.~{Santander}, S.~{Sarkar}, S.~{Sarkar}, K.~{Satalecka}, P.~{Schlunder},
  T.~{Schmidt}, A.~{Schneider}, S.~{Schoenen}, S.~{Sch{\"o}neberg},
  L.~{Schumacher}, S.~{Sclafani}, D.~{Seckel}, S.~{Seunarine},
  J.~{Soedingrekso}, D.~{Soldin}, M.~{Song}, G.~M. {Spiczak}, C.~{Spiering},
  J.~{Stachurska}, M.~{Stamatikos}, T.~{Stanev}, A.~{Stasik}, R.~{Stein},
  J.~{Stettner}, A.~{Steuer}, T.~{Stezelberger}, R.~G. {Stokstad},
  A.~{St{\"o}{\ss}l}, N.~L. {Strotjohann}, T.~{Stuttard}, G.~W. {Sullivan},
  M.~{Sutherland}, I.~{Taboada}, J.~{Tatar}, F.~{Tenholt}, S.~{Ter-Antonyan},
  A.~{Terliuk}, S.~{Tilav}, P.~A. {Toale}, M.~N. {Tobin}, C.~{Toennis},
  S.~{Toscano}, D.~{Tosi}, M.~{Tselengidou}, C.~F. {Tung}, A.~{Turcati}, C.~F.
  {Turley}, B.~{Ty}, E.~{Unger}, M.~{Usner}, J.~{Vandenbroucke}, W.~{Van
  Driessche}, D.~{van Eijk}, N.~{van Eijndhoven}, S.~{Vanheule}, J.~{van
  Santen}, E.~{Vogel}, M.~{Vraeghe}, C.~{Walck}, A.~{Wallace}, M.~{Wallraff},
  F.~D. {Wandler}, N.~{Wandkowsky}, A.~{Waza}, C.~{Weaver}, M.~J. {Weiss},
  C.~{Wendt}, J.~{Werthebach}, S.~{Westerhoff}, B.~J. {Whelan}, N.~{Whitehorn},
  K.~{Wiebe}, C.~H. {Wiebusch}, L.~{Wille}, D.~R. {Williams}, L.~{Wills},
  M.~{Wolf}, J.~{Wood}, T.~R. {Wood}, K.~{Woschnagg}, D.~L. {Xu}, X.~W. {Xu},
  Y.~{Xu}, J.~P. {Yanez}, G.~{Yodh}, S.~{Yoshida}, T.~{Yuan}, {Fermi-LAT
  Collaboration}, S.~{Abdollahi}, M.~{Ajello}, R.~{Angioni}, L.~{Baldini},
  J.~{Ballet}, G.~{Barbiellini}, D.~{Bastieri}, K.~{Bechtol}, R.~{Bellazzini},
  B.~{Berenji}, E.~{Bissaldi}, R.~D. {Blandford}, R.~{Bonino}, E.~{Bottacini},
  J.~{Bregeon}, P.~{Bruel}, R.~{Buehler}, T.~H. {Burnett}, E.~{Burns},
  S.~{Buson}, R.~A. {Cameron}, R.~{Caputo}, P.~A. {Caraveo}, E.~{Cavazzuti},
  E.~{Charles}, S.~{Chen}, C.~C. {Cheung}, J.~{Chiang}, G.~{Chiaro},
  S.~{Ciprini}, J.~{Cohen-Tanugi}, J.~{Conrad}, D.~{Costantin}, S.~{Cutini},
  F.~{D'Ammando}, F.~{de Palma}, S.~W. {Digel}, N.~{Di Lalla}, M.~{Di Mauro},
  L.~{Di Venere}, A.~{Dom{\'\i}nguez}, C.~{Favuzzi}, A.~{Franckowiak},
  Y.~{Fukazawa}, S.~{Funk}, P.~{Fusco}, F.~{Gargano}, D.~{Gasparrini},
  N.~{Giglietto}, M.~{Giomi}, P.~{Giommi}, F.~{Giordano}, M.~{Giroletti},
  T.~{Glanzman}, D.~{Green}, I.~A. {Grenier}, M.~H. {Grondin}, S.~{Guiriec},
  A.~K. {Harding}, M.~{Hayashida}, E.~{Hays}, J.~W. {Hewitt}, D.~{Horan},
  G.~{J{\'o}hannesson}, M.~{Kadler}, S.~{Kensei}, D.~{Kocevski}, F.~{Krauss},
  M.~{Kreter}, M.~{Kuss}, G.~{La Mura}, S.~{Larsson}, L.~{Latronico},
  M.~{Lemoine-Goumard}, J.~{Li}, F.~{Longo}, F.~{Loparco}, M.~N. {Lovellette},
  P.~{Lubrano}, J.~D. {Magill}, S.~{Maldera}, D.~{Malyshev}, A.~{Manfreda},
  M.~N. {Mazziotta}, J.~E. {McEnery}, M.~{Meyer}, P.~F. {Michelson},
  T.~{Mizuno}, M.~E. {Monzani}, A.~{Morselli}, I.~V. {Moskalenko}, M.~{Negro},
  E.~{Nuss}, R.~{Ojha}, N.~{Omodei}, M.~{Orienti}, E.~{Orlando},
  M.~{Palatiello}, V.~S. {Paliya}, J.~S. {Perkins}, M.~{Persic},
  M.~{Pesce-Rollins}, F.~{Piron}, T.~A. {Porter}, G.~{Principe},
  S.~{Rain{\`o}}, R.~{Rando}, B.~{Rani}, M.~{Razzano}, S.~{Razzaque},
  A.~{Reimer}, O.~{Reimer}, N.~{Renault-Tinacci}, S.~{Ritz}, L.~S. {Rochester},
  P.~M. {Saz Parkinson}, C.~{Sgr{\`o}}, E.~J. {Siskind}, G.~{Spandre},
  P.~{Spinelli}, D.~J. {Suson}, H.~{Tajima}, M.~{Takahashi}, Y.~{Tanaka}, J.~B.
  {Thayer}, D.~J. {Thompson}, L.~{Tibaldo}, D.~F. {Torres}, E.~{Torresi},
  G.~{Tosti}, E.~{Troja}, J.~{Valverde}, G.~{Vianello}, M.~{Vogel}, K.~{Wood},
  M.~{Wood}, G.~{Zaharijas}, {MAGIC Collaboration}, M.~L. {Ahnen},
  S.~{Ansoldi}, L.~A. {Antonelli}, C.~{Arcaro}, D.~{Baack}, A.~{Babi{\'c}},
  B.~{Banerjee}, P.~{Bangale}, U.~{Barres de Almeida}, J.~A. {Barrio},
  J.~{Becerra Gonz{\'a}lez}, W.~{Bednarek}, E.~{Bernardini}, A.~{Berti},
  W.~{Bhattacharyya}, A.~{Biland}, O.~{Blanch}, G.~{Bonnoli}, A.~{Carosi},
  R.~{Carosi}, G.~{Ceribella}, A.~{Chatterjee}, S.~M. {Colak}, P.~{Colin},
  E.~{Colombo}, J.~L. {Contreras}, J.~{Cortina}, S.~{Covino}, P.~{Cumani},
  P.~{Da Vela}, F.~{Dazzi}, A.~{De Angelis}, B.~{De Lotto}, M.~{Delfino},
  J.~{Delgado}, F.~{Di Pierro}, A.~{Dom{\'\i}nguez}, D.~{Dominis Prester},
  D.~{Dorner}, M.~{Doro}, S.~{Einecke}, D.~{Elsaesser}, V.~{Fallah Ramazani},
  A.~{Fern{\'a}ndez-Barral}, D.~{Fidalgo}, L.~{Foffano}, K.~{Pfrang}, M.~V.
  {Fonseca}, L.~{Font}, A.~{Franceschini}, C.~{Fruck}, D.~{Galindo},
  S.~{Gallozzi}, R.~J. {Garc{\'\i}a L{\'o}pez}, M.~{Garczarczyk}, M.~{Gaug},
  P.~{Giammaria}, N.~{Godinovi{\'c}}, D.~{Gora}, D.~{Guberman}, D.~{Hadasch},
  A.~{Hahn}, T.~{Hassan}, M.~{Hayashida}, J.~{Herrera}, J.~{Hose}, D.~{Hrupec},
  S.~{Inoue}, K.~{Ishio}, Y.~{Konno}, H.~{Kubo}, J.~{Kushida}, D.~{Lelas},
  E.~{Lindfors}, S.~{Lombardi}, F.~{Longo}, M.~{L{\'o}pez}, C.~{Maggio},
  P.~{Majumdar}, M.~{Makariev}, G.~{Maneva}, M.~{Manganaro}, K.~{Mannheim},
  L.~{Maraschi}, M.~{Mariotti}, M.~{Mart{\'\i}nez}, S.~{Masuda}, D.~{Mazin},
  M.~{Minev}, J.~M. {M}, R.~{Mirzoyan}, A.~{Moralejo}, V.~{Moreno},
  E.~{Moretti}, T.~{Nagayoshi}, V.~{Neustroev}, A.~{Niedzwiecki}, M.~{Nievas
  Rosillo}, C.~{Nigro}, K.~{Nilsson}, D.~{Ninci}, K.~{Nishijima}, K.~{Noda},
  L.~{Nogu{\'e}s}, S.~{Paiano}, J.~{Palacio}, D.~{Paneque}, R.~{Paoletti},
  J.~M. {Paredes}, G.~{Pedaletti}, M.~{Peresano}, M.~{Persic}, P.~G. {Prada
  Moroni}, E.~{Prandini}, I.~{Puljak}, J.~{Rodriguez Garcia}, I.~{Reichardt},
  W.~{Rhode}, M.~{Rib{\'o}}, J.~{Rico}, C.~{Righi}, A.~{Rugliancich},
  T.~{Saito}, K.~{Satalecka}, T.~{Schweizer}, J.~{Sitarek},
  I.~{{\v{S}}nidaric}, D.~{Sobczynska}, A.~{Stamerra}, M.~{Strzys},
  T.~{Suri{\'c}}, M.~{Takahashi}, F.~{Tavecchio}, P.~{Temnikov},
  T.~{Terzi{\'c}}, M.~{Teshima}, N.~{Torres-Alb{\`a}}, A.~{Treves},
  S.~{Tsujimoto}, G.~{Vanzo}, M.~{Vazquez Acosta}, I.~{Vovk}, J.~E. {Ward},
  M.~{Will}, {S}, D.~{Zaric}, {AGILE Team}, F.~{Lucarelli}, M.~{Tavani},
  G.~{Piano}, I.~{Donnarumma}, C.~{Pittori}, F.~{Verrecchia}, G.~{Barbiellini},
  A.~{Bulgarelli}, P.~{Caraveo}, P.~W. {Cattaneo}, S.~{Colafrancesco},
  E.~{Costa}, G.~{Di Cocco}, A.~{Ferrari}, F.~{Gianotti}, A.~{Giuliani},
  P.~{Lipari}, S.~{Mereghetti}, A.~{Morselli}, L.~{Pacciani}, F.~{Paoletti},
  N.~{Parmiggiani}, A.~{Pellizzoni}, P.~{Picozza}, M.~{Pilia}, A.~{Rappoldi},
  A.~{Trois}, S.~{Vercellone}, V.~{Vittorini}, {ASAS-SN Team}, A.~Franckowiak
  K.~Z. {Stanek}, C.~S. {Kochanek}, J.~F. {Beacom}, T.~A. {Thompson}, T.~W.~S.
  {Holoien}, S.~{Dong}, J.~L. {Prieto}, B.~J. {Shappee}, S.~{Holmbo}, {HAWC
  Collaboration}, A.~U. {Abeysekara}, A.~{Albert}, R.~{Alfaro}, C.~{Alvarez},
  R.~{Arceo}, J.~C. {Arteaga-Vel{\'a}zquez}, D.~{Avila Rojas}, H.~A. {Ayala
  Solares}, A.~{Becerril}, E.~{Belmont-Moreno}, A.~{Bernal}, K.~S.
  {Caballero-Mora}, T.~{Capistr{\'a}n}, A.~{Carrami{\~n}ana}, S.~{Casanova},
  M.~{Castillo}, U.~{Cotti}, J.~{Cotzomi}, S.~{Couti{\~n}o de Le{\'o}n}, C.~{De
  Le{\'o}n}, E.~{De la Fuente}, R.~{Diaz Hernandez}, S.~{Dichiara}, B.~L.
  {Dingus}, M.~A. {DuVernois}, J.~C. {D{\'\i}az-V{\'e}lez}, R.~W. {Ellsworth},
  K.~{Engel}, D.~W. {Fiorino}, H.~{Fleischhack}, N.~{Fraija}, J.~A.
  {Garc{\'\i}a-Gonz{\'a}lez}, F.~{Garfias}, A.~{Gonz{\'a}lez Mu{\~n}oz}, M.~M.
  {Gonz{\'a}lez}, J.~A. {Goodman}, Z.~{Hampel-Arias}, J.~P. {Harding},
  S.~{Hernand ez}, B.~{Hona}, F.~{Hueyotl-Zahuantitla}, C.~M. {Hui},
  P.~{H{\"u}ntemeyer}, A.~{Iriarte}, A.~{Jardin-Blicq}, V.~{Joshi},
  S.~{Kaufmann}, G.~J. {Kunde}, A.~{Lara}, R.~J. {Lauer}, W.~H. {Lee},
  D.~{Lennarz}, H.~{Le{\'o}n Vargas}, J.~T. {Linnemann}, A.~L. {Longinotti},
  G.~{Luis-Raya}, R.~{Luna-Garc{\'\i}a}, K.~{Malone}, S.~S. {Marinelli},
  O.~{Martinez}, I.~{Martinez-Castellanos}, J.~{Mart{\'\i}nez-Castro},
  H.~{Mart{\'\i}nez-Huerta}, J.~A. {Matthews}, P.~{Miranda-Romagnoli},
  E.~{Moreno}, M.~{Mostaf{\'a}}, A.~{Nayerhoda}, L.~{Nellen}, M.~{Newbold},
  M.~U. {Nisa}, R.~{Noriega-Papaqui}, R.~{Pelayo}, J.~{Pretz}, E.~G.
  {P{\'e}rez-P{\'e}rez}, Z.~{Ren}, C.~D. {Rho}, C.~{Rivi{\`e}re},
  D.~{Rosa-Gonz{\'a}lez}, M.~{Rosenberg}, E.~{Ruiz-Velasco}, E.~{Ruiz-Velasco},
  F.~{Salesa Greus}, A.~{Sandoval}, M.~{Schneider}, H.~{Schoorlemmer},
  G.~{Sinnis}, A.~J. {Smith}, R.~W. {Springer}, P.~{Surajbali}, O.~{Tibolla},
  K.~{Tollefson}, I.~{Torres}, L.~{Villase{\~n}or}, T.~{Weisgarber},
  F.~{Werner}, T.~{Yapici}, Y.~{Gaurang}, A.~{Zepeda}, H.~{Zhou}, J.~D.
  {{\'A}lvarez}, {H.~E.~S.~S. Collaboration}, H.~{Abdalla}, E.~O.
  {Ang{\"u}ner}, C.~{Armand}, M.~{Backes}, Y.~{Becherini}, D.~{Berge},
  M.~{B{\"o}ttcher}, C.~{Boisson}, J.~{Bolmont}, S.~{Bonnefoy}, P.~{Bordas},
  F.~{Brun}, M.~{B{\"u}chele}, T.~{Bulik}, S.~{Caroff}, A.~{Carosi},
  S.~{Casanova}, M.~{Cerruti}, N.~{Chakraborty}, S.~{Chandra}, A.~{Chen},
  S.~{Colafrancesco}, I.~D. {Davids}, C.~{Deil}, J.~{Devin},
  A.~{Djannati-Ata{\"\i}}, K.~{Egberts}, G.~{Emery}, S.~{Eschbach},
  A.~{Fiasson}, G.~{Fontaine}, S.~{Funk}, M.~{F{\"u}{\ss}ling}, Y.~A.
  {Gallant}, F.~{Gat{\'e}}, G.~{Giavitto}, D.~{Glawion}, J.~F. {Glicenstein},
  D.~{Gottschall}, M.~H. {Grondin}, M.~{Haupt}, G.~{Henri}, J.~A. {Hinton},
  C.~{Hoischen}, T.~L. {Holch}, D.~{Huber}, M.~{Jamrozy}, D.~{Jankowsky},
  F.~{Jankowsky}, L.~{Jouvin}, I.~{Jung-Richardt}, D.~{Kerszberg},
  B.~{Kh{\'e}lifi}, J.~{King}, S.~{Klepser}, W.~{Kluzniak}, Nu. {Komin},
  M.~{Kraus}, J.~{Lefaucheur}, A.~{Lemi{\`e}re}, M.~{Lemoine-Goumard}, J.~P.
  {Lenain}, E.~{Leser}, T.~{Lohse}, R.~{L{\'o}pez-Coto}, M.~{Lorentz},
  I.~{Lypova}, V.~{Marandon}, G.~{Guillem Mart{\'\i}-Devesa}, G.~{Maurin},
  A.~M.~W. {Mitchell}, R.~{Moderski}, M.~{Mohamed}, L.~{Mohrmann}, E.~{Moulin},
  T.~{Murach}, M.~{de Naurois}, F.~{Niederwanger}, J.~{Niemiec}, L.~{Oakes},
  P.~{O'Brien}, S.~{Ohm}, M.~{Ostrowski}, I.~{Oya}, M.~{Panter}, R.~D.
  {Parsons}, C.~{Perennes}, Q.~{Piel}, S.~{Pita}, V.~{Poireau}, A.~{Priyana
  Noel}, H.~{Prokoph}, G.~{P{\"u}hlhofer}, A.~{Quirrenbach}, S.~{Raab},
  R.~{Rauth}, M.~{Renaud}, F.~{Rieger}, L.~{Rinchiuso}, C.~{Romoli},
  G.~{Rowell}, B.~{Rudak}, D.~A. {Sasaki}, M.~{Sanchez}, R.~{Schlickeiser},
  F.~{Sch{\"u}ssler}, A.~{Schulz}, U.~{Schwanke}, M.~{Seglar-Arroyo},
  N.~{Shafi}, R.~{Simoni}, H.~{Sol}, C.~{Stegmann}, C.~{Steppa},
  T.~{Tavernier}, A.~M. {Taylor}, D.~{Tiziani}, C.~{Trichard}, M.~{Tsirou},
  C.~{van Eldik}, C.~{van Rensburg}, B.~{van Soelen}, J.~{Veh}, P.~{Vincent},
  F.~{Voisin}, S.~J. {Wagner}, R.~M. {Wagner}, A.~{Wierzcholska}, R.~{Zanin},
  A.~A. {Zdziarski}, A.~{Zech}, A.~{Ziegler}, J.~{Zorn}, N.~{{\.Z}ywucka},
  {INTEGRAL Team}, V.~{Savchenko}, C.~{Ferrigno}, A.~{Bazzano}, R.~{Diehl},
  E.~{Kuulkers}, P.~{Laurent}, S.~{Mereghetti}, L.~{Natalucci}, F.~{Panessa},
  J.~{Rodi}, P.~{Ubertini}, Kiso {Kanata}, Subaru~Observing Teams,
  T.~{Morokuma}, K.~{Ohta}, Y.~T. {Tanaka}, H.~{Mori}, M.~{Yamanaka}, K.~S.
  {Kawabata}, Y.~{Utsumi}, T.~{Nakaoka}, M.~{Kawabata}, H.~{Nagashima},
  M.~{Yoshida}, Y.~{Matsuoka}, R.~{Itoh}, {Kapteyn Team}, W.~{Keel}, {Liverpool
  Telescope Team}, C.~{Copperwheat}, I.~{Steele}, {Swift/NuSTAR Team}, S.~B.
  {Cenko}, D.~F. {Cowen}, J.~J. {DeLaunay}, P.~A. {Evans}, D.~B. {Fox},
  A.~{Keivani}, J.~A. {Kennea}, F.~E. {Marshall}, J.~P. {Osborne},
  M.~{Santander}, A.~{Tohuvavohu}, C.~F. {Turley}, {VERITAS Collaboration},
  A.~U. {Abeysekara}, A.~{Archer}, W.~{Benbow}, R.~{Bird}, A.~{Brill},
  R.~{Brose}, M.~{Buchovecky}, J.~H. {Buckley}, V.~{Bugaev}, J.~L.
  {Christiansen}, M.~P. {Connolly}, W.~{Cui}, M.~K. {Daniel}, M.~{Errando},
  A.~{Falcone}, Q.~{Feng}, J.~P. {Finley}, L.~{Fortson}, A.~{Furniss},
  O.~{Gueta}, M.~{H{\"u}tten}, O.~{Hervet}, G.~{Hughes}, T.~B. {Humensky},
  C.~A. {Johnson}, P.~{Kaaret}, P.~{Kar}, N.~{Kelley-Hoskins}, M.~{Kertzman},
  D.~{Kieda}, M.~{Krause}, F.~{Krennrich}, S.~{Kumar}, M.~J. {Lang}, T.~T.~Y.
  {Lin}, G.~{Maier}, S.~{McArthur}, P.~{Moriarty}, R.~{Mukherjee}, D.~{Nieto},
  S.~{O'Brien}, R.~A. {Ong}, A.~N. {Otte}, N.~{Park}, A.~{Petrashyk},
  M.~{Pohl}, A.~{Popkow}, S.~E. {Pueschel}, J.~{Quinn}, K.~{Ragan}, P.~T.
  {Reynolds}, G.~T. {Richards}, E.~{Roache}, C.~{Rulten}, I.~{Sadeh},
  M.~{Santander}, S.~S. {Scott}, G.~H. {Sembroski}, K.~{Shahinyan},
  I.~{Sushch}, S.~{Tr{\'e}panier}, J.~{Tyler}, V.~V. {Vassiliev}, S.~P.
  {Wakely}, A.~{Weinstein}, R.~M. {Wells}, P.~{Wilcox}, A.~{Wilhelm}, D.~A.
  {Williams}, B.~{Zitzer}, {VLA/B Team}, A.~J. {Tetarenko}, A.~E. {Kimball},
  J.~C.~A. {Miller-Jones}, and G.~R. {Sivakoff}.
\newblock {Multimessenger observations of a flaring blazar coincident with
  high-energy neutrino IceCube-170922A}.
\newblock {\em Science}, 361(6398):eaat1378, July 2018.

\bibitem{2018PhRvD..98l3012A}
A.~{Albert}, R.~{Alfaro}, C.~{Alvarez}, R.~{Arceo}, J.~C.
  {Arteaga-Vel{\'a}zquez}, D.~{Avila Rojas}, H.~A. {Ayala Solares},
  E.~{Belmont-Moreno}, S.~Y. {BenZvi}, and C.~{Brisbois}.
\newblock {Constraints on spin-dependent dark matter scattering with long-lived
  mediators from TeV observations of the Sun with HAWC}.
\newblock {\em \prd}, 98(12):123012, Dec 2018.

\bibitem{2018Natur.562...82A}
A.~U. {Abeysekara}, A.~{Albert}, R.~{Alfaro}, C.~{Alvarez}, J.~D.
  {{\'A}lvarez}, R.~{Arceo}, J.~C. {Arteaga-Vel{\'a}zquez}, D.~{Avila Rojas},
  H.~A. {Ayala Solares}, E.~{Belmont-Moreno}, S.~Y. {BenZvi}, C.~{Brisbois},
  K.~S. {Caballero-Mora}, T.~{Capistr{\'a}n}, A.~{Carrami{\~n}ana},
  S.~{Casanova}, M.~{Castillo}, U.~{Cotti}, J.~{Cotzomi}, S.~{Couti{\~n}o de
  Le{\'o}n}, C.~{De Le{\'o}n}, E.~{De la Fuente}, J.~C.
  {D{\'{\i}}az-V{\'e}lez}, S.~{Dichiara}, B.~L. {Dingus}, M.~A. {DuVernois},
  R.~W. {Ellsworth}, K.~{Engel}, C.~{Espinoza}, K.~{Fang}, H.~{Fleischhack},
  N.~{Fraija}, A.~{Galv{\'a}n-G{\'a}mez}, J.~A. {Garc{\'{\i}}a-Gonz{\'a}lez},
  F.~{Garfias}, A.~{Gonz{\'a}lez-Mu{\~n}oz}, M.~M. {Gonz{\'a}lez}, J.~A.
  {Goodman}, Z.~{Hampel-Arias}, J.~P. {Harding}, S.~{Hernandez}, J.~{Hinton},
  B.~{Hona}, F.~{Hueyotl-Zahuantitla}, C.~M. {Hui}, P.~{H{\"u}ntemeyer},
  A.~{Iriarte}, A.~{Jardin-Blicq}, V.~{Joshi}, S.~{Kaufmann}, P.~{Kar}, G.~J.
  {Kunde}, R.~J. {Lauer}, W.~H. {Lee}, H.~{Le{\'o}n Vargas}, H.~{Li}, J.~T.
  {Linnemann}, A.~L. {Longinotti}, G.~{Luis-Raya}, R.~{L{\'o}pez-Coto},
  K.~{Malone}, S.~S. {Marinelli}, O.~{Martinez}, I.~{Martinez-Castellanos},
  J.~{Mart{\'{\i}}nez-Castro}, J.~A. {Matthews}, P.~{Miranda-Romagnoli},
  E.~{Moreno}, M.~{Mostaf{\'a}}, A.~{Nayerhoda}, L.~{Nellen}, M.~{Newbold},
  M.~U. {Nisa}, R.~{Noriega-Papaqui}, J.~{Pretz}, E.~G. {P{\'e}rez-P{\'e}rez},
  Z.~{Ren}, C.~D. {Rho}, C.~{Rivi{\`e}re}, D.~{Rosa-Gonz{\'a}lez},
  M.~{Rosenberg}, E.~{Ruiz-Velasco}, F.~{Salesa Greus}, A.~{Sandoval},
  M.~{Schneider}, H.~{Schoorlemmer}, M.~{Seglar Arroyo}, G.~{Sinnis}, A.~J.
  {Smith}, R.~W. {Springer}, P.~{Surajbali}, I.~{Taboada}, O.~{Tibolla},
  K.~{Tollefson}, I.~{Torres}, G.~{Vianello}, L.~{Villase{\~n}or},
  T.~{Weisgarber}, F.~{Werner}, S.~{Westerhoff}, J.~{Wood}, T.~{Yapici},
  G.~{Yodh}, A.~{Zepeda}, H.~{Zhang}, and H.~{Zhou}.
\newblock {Very-high-energy particle acceleration powered by the jets of the
  microquasar SS 433}.
\newblock {\em \nat}, 562:82--85, October 2018.

\bibitem{2019JCAP...08..023F}
N.~{Fraija}, M.~{Araya}, A.~{Galv{\'a}n-G{\'a}mez}, and J.~A. {de Diego}.
\newblock {Analysis of Fermi-LAT observations, UHECRs and neutrinos from the
  radio galaxy Centaurus B}.
\newblock {\em \jcap}, 2019(8):023, August 2019.

\bibitem{2019arXiv190512518H}
{HAWC Collaboration}, A.~U. {Abeysekara}, A.~{Albert}, R.~{Alfaro},
  C.~{Alvarez}, J.~D. {{\'A}lvarez}, J.~R.~Angeles {Camacho}, R.~{Acero}, J.~C.
  {Arteaga-Vel{\'a}zquez}, and K.~P. {Arunbabu}.
\newblock {Measurement of the Crab Nebula at the Highest Energies with HAWC}.
\newblock {\em arXiv e-prints (accepted for pub.~in \apj)}, page
  arXiv:1905.12518, May 2019.

\bibitem{2020MNRAS.497.5318F}
N.~{Fraija}, E.~{Aguilar-Ruiz}, and A.~{Galv{\'a}n-G{\'a}mez}.
\newblock {Electron-positron pair plasma in TXS 0506+056 and the 'neutrino
  flare' in 2014-2015}.
\newblock {\em \mnras}, 497(4):5318--5325, August 2020.

\bibitem{He:2019Qx}
Huihai He.
\newblock {Status and First Results of the LHAASO Experiment}.
\newblock {\em PoS}, ICRC2019:693, 2019.

\bibitem{2019BAAS...51c.272H}
J.~Patrick {Harding}, A.~{Albert}, C.~{Alvarez}, R.~{Arceo}, K.~S.
  {Caballero-Mora}, G.~{Cotter}, B.~L. {Dingus}, K.~L. {Engel},
  H.~{Fleischhack}, J.~A. {Goodman}, T.~{Greenshaw}, B.~{Hona},
  P.~{Huentemeyer}, J.~S. {Lapington}, J.~{Lundeen}, J.~{Martinez-Castro},
  H.~{Martinez-Huerta}, K.~{Murase}, M.~U. {Nisa}, H.~{Schoorlemmer},
  K.~{Tollefson}, A.~{Viana}, and A.~{Zepeda}.
\newblock {Exploring Beyond-the-Standard-Model Physics with TeV Gamma-rays}.
\newblock {\em \baas}, 51(3):272, May 2019.

\bibitem{GW170817}
B.~P. {Abbott}, R.~{Abbott}, T.~D. {Abbott}, F.~{Acernese}, K.~{Ackley},
  C.~{Adams}, T.~{Adams}, P.~{Addesso}, R.~X. {Adhikari}, V.~B. {Adya}, and
  et~al.
\newblock {GW170817: Observation of Gravitational Waves from a Binary Neutron
  Star Inspiral}.
\newblock {\em Physical Review Letters}, 119(16):161101, October 2017.

\bibitem{GW170817_HESS}
H.~{Abdalla}, A.~{Abramowski}, F.~{Aharonian}, F.~{Ait Benkhali}, E.~O.
  {Ang{\"u}ner}, M.~{Arakawa}, M.~{Arrieta}, P.~{Aubert}, M.~{Backes},
  A.~{Balzer}, and et~al.
\newblock {TeV Gamma-Ray Observations of the Binary Neutron Star Merger
  GW170817 with H.E.S.S.}
\newblock {\em \apjl}, 850:L22, December 2017.

\bibitem{GW170817_MMA}
B.~P. {Abbott}, R.~{Abbott}, T.~D. {Abbott}, F.~{Acernese}, K.~{Ackley},
  C.~{Adams}, T.~{Adams}, P.~{Addesso}, R.~X. {Adhikari}, V.~B. {Adya}, and
  et~al.
\newblock {Multi-messenger Observations of a Binary Neutron Star Merger}.
\newblock {\em APJL}, 848:L12, October 2017.

\bibitem{2019BAAS...51c.357S}
Fabian {Sch\"ussler} and Konstancja {Satalecka}.
\newblock {All-Sky time domain astrophysics with Very High Energy Gamma rays}.
\newblock {\em \baas}, 51(3):357, May 2019.

\bibitem{CTA_ScienceTDR}
{The CTA Consortium}.
\newblock {Science with the Cherenkov Telescope Array}.
\newblock {\em World Scientific}, {DOI 10.1142/10986}, 2019.

\bibitem{Viana:2021smp}
Aion Viana, Andrea Albert, J.~Patrick Harding, Jim Hinton, Harm Schoorlemmer,
  and Vitor de~Souza.
\newblock {Searching for Dark Matter with the Southern Wide-field Gamma-ray
  Observatory (SWGO)}.
\newblock {\em PoS}, ICRC2021:555, 2021.

\bibitem{Hawking:1974rv}
S.~W. Hawking.
\newblock {Black hole explosions}.
\newblock {\em Nature}, 248:30--31, 1974.

\bibitem{Lopez-Coto:2021lxh}
R.~L\'opez-Coto, M.~Doro, A.~de~Angelis, M.~Mariotti, and J.~P. Harding.
\newblock {Prospects for the observation of Primordial Black Hole evaporation
  with the Southern Wide field of view Gamma-ray Observatory}.
\newblock {\em JCAP}, 08:040, 2021.

\bibitem{GRAND:2018iaj}
Jaime \'Alvarez-Mu\~niz et~al.
\newblock {The Giant Radio Array for Neutrino Detection (GRAND): Science and
  Design}.
\newblock {\em Sci. China Phys. Mech. Astron.}, 63(1):219501, 2020.

\bibitem{BalagopalV:2017aan}
A.~Balagopal~V., A.~Haungs, T.~Huege, and F.~G. Schroeder.
\newblock {Search for PeVatrons at the Galactic Center using a radio air-shower
  array at the South Pole}.
\newblock {\em Eur. Phys. J. C}, 78(2):111, 2018.
\newblock [Erratum: Eur.Phys.J.C 78, 1017 (2018), Erratum: Eur.Phys.J.C 81, 483
  (2021)].

\bibitem{Corstanje:2014waa}
A.~Corstanje et~al.
\newblock {The shape of the radio wavefront of extensive air showers as
  measured with LOFAR}.
\newblock {\em Astropart. Phys.}, 61:22--31, 2015.

\bibitem{Bezyazeekov:2015rpa}
P.~A. Bezyazeekov et~al.
\newblock {Measurement of cosmic-ray air showers with the Tunka Radio Extension
  (Tunka-Rex)}.
\newblock {\em Nucl. Instrum. Meth. A}, 802:89--96, 2015.

\bibitem{Buitink:2016nkf}
S.~Buitink et~al.
\newblock {A large light-mass component of cosmic rays at
  10\textasciicircum{}{17} - 10\textasciicircum{}{17.5} eV from radio
  observations}.
\newblock {\em Nature}, 531:70, 2016.

\bibitem{Decoene:2019sgx}
Valentin Decoene.
\newblock {GRANDProto300 experiment: a pathfinder with rich astroparticle and
  radio-astronomy science case}.
\newblock {\em PoS}, ICRC2019:233, 2020.

\bibitem{PierreAuger:2016kuz}
Alexander Aab et~al.
\newblock {Search for photons with energies above 10$^{18}$ eV using the hybrid
  detector of the Pierre Auger Observatory}.
\newblock {\em JCAP}, 04:009, 2017.
\newblock [Erratum: JCAP 09, E02 (2020)].

\bibitem{PierreAuger:2021mjh}
Pedro Abreu et~al.
\newblock {A search for ultra-high-energy photons at the Pierre Auger
  Observatory exploiting air-shower universality}.
\newblock {\em PoS}, ICRC2021:373, 2021.

\bibitem{Sarkar:2011hkm}
Biswajit Sarkar, Karl-Heinz Kampert, and Joerg Kulbartz.
\newblock {Ultra-High Energy Photon and Neutrino Fluxes in Realistic
  Astrophysical Scenarios}.
\newblock In {\em {32nd International Cosmic Ray Conference}}, volume~2, page
  198, 2011.

\bibitem{AlvesBatista:2018zui}
Rafael Alves~Batista, Rogerio~M. de~Almeida, Bruno Lago, and Kumiko Kotera.
\newblock {Cosmogenic photon and neutrino fluxes in the Auger era}.
\newblock {\em JCAP}, 01:002, 2019.

\bibitem{Rubtsov:2017lhs}
Grigory Rubtsov, Masaki Fukushima, Dmitri Ivanov, Mikhail Kuznetsov, Maxim
  Piskunov, Gordon Thomson, Sergey Troitsky, and Yana Zhezher.
\newblock {Telescope Array search for EeV photons and neutrinos}.
\newblock {\em PoS}, ICRC2017:551, 2018.

\bibitem{Risse:2007sd}
Markus Risse and Piotr Homola.
\newblock {Search for ultrahigh energy photons using air showers}.
\newblock {\em Mod. Phys. Lett. A}, 22:749--766, 2007.

\bibitem{Landau:1953um}
L.~D. Landau and I.~Pomeranchuk.
\newblock {Limits of applicability of the theory of bremsstrahlung electrons
  and pair production at high-energies}.
\newblock {\em Dokl. Akad. Nauk Ser. Fiz.}, 92:535--536, 1953.

\bibitem{Migdal:1956tc}
A.~B. Migdal.
\newblock {Bremsstrahlung and pair production in condensed media at
  high-energies}.
\newblock {\em Phys. Rev.}, 103:1811--1820, 1956.

\bibitem{Anchordoqui:2019omw}
Luis~A. Anchordoqui et~al.
\newblock {Performance and science reach of the Probe of Extreme Multimessenger
  Astrophysics for ultrahigh-energy particles}.
\newblock {\em Phys. Rev. D}, 101(2):023012, 2020.

\bibitem{POEMMA:2020ykm}
A.~V. Olinto et~al.
\newblock {The POEMMA (Probe of Extreme Multi-Messenger Astrophysics)
  observatory}.
\newblock {\em JCAP}, 06:007, 2021.

\bibitem{Venters:2019xwi}
Tonia~M. Venters, Mary~Hall Reno, John~F. Krizmanic, Luis~A. Anchordoqui,
  Claire Gu\'epin, and Angela~V. Olinto.
\newblock {POEMMA's Target of Opportunity Sensitivity to Cosmic Neutrino
  Transient Sources}.
\newblock {\em Phys. Rev. D}, 102:123013, 2020.

\bibitem{Reno:2019jtr}
Mary~Hall Reno, John~F. Krizmanic, and Tonia~M. Venters.
\newblock {Cosmic tau neutrino detection via Cherenkov signals from air showers
  from Earth-emerging taus}.
\newblock {\em Phys. Rev. D}, 100(6):063010, 2019.

\bibitem{Dubovsky:1998pu}
S.~L. Dubovsky and P.~G. Tinyakov.
\newblock {Galactic anisotropy as signature of CDM related ultrahigh-energy
  cosmic rays}.
\newblock {\em JETP Lett.}, 68:107--111, 1998.

\bibitem{Evans:2002ry}
N.~W. Evans, F.~Ferrer, and Subir Sarkar.
\newblock {Clustering of ultrahigh-energy cosmic rays and their sources}.
\newblock {\em Phys. Rev. D}, 67:103005, 2003.

\bibitem{Aloisio:2007bh}
R.~Aloisio and F.~Tortorici.
\newblock {Super Heavy Dark Matter and UHECR Anisotropy at Low Energy}.
\newblock {\em Astropart. Phys.}, 29:307--316, 2008.

\bibitem{Kalashev:2017ijd}
O.~E. Kalashev and M.~Yu Kuznetsov.
\newblock {Heavy decaying dark matter and large-scale anisotropy of high-energy
  cosmic rays}.
\newblock {\em JETP Lett.}, 106(2):73--80, 2017.

\bibitem{Alcantara:2019sco}
Esteban Alcantara, Luis~A. Anchordoqui, and Jorge~F. Soriano.
\newblock {Hunting for superheavy dark matter with the highest-energy cosmic
  rays}.
\newblock {\em Phys. Rev. D}, 99(10):103016, 2019.

\bibitem{Guepin:2021ljb}
Claire Gu\'epin, Roberto Aloisio, Luis~A. Anchordoqui, Austin Cummings, John~F.
  Krizmanic, Angela~V. Olinto, Mary~Hall Reno, and Tonia~M. Venters.
\newblock {Indirect dark matter searches at ultrahigh energy neutrino
  detectors}.
\newblock {\em Phys. Rev. D}, 104(8):083002, 2021.

\bibitem{fermi-lat}
W.~B. {Atwood}, A.~A. {Abdo}, M.~{Ackermann}, W.~{Althouse}, B.~{Anderson},
  M.~{Axelsson}, L.~{Baldini}, J.~{Ballet}, D.~L. {Band}, G.~{Barbiellini},
  J.~{Bartelt}, D.~{Bastieri}, B.~M. {Baughman}, K.~{Bechtol},
  D.~{B{\'e}d{\'e}r{\`e}de}, F.~{Bellardi}, R.~{Bellazzini}, B.~{Berenji},
  G.~F. {Bignami}, D.~{Bisello}, E.~{Bissaldi}, R.~D. {Blandford}, E.~D.
  {Bloom}, J.~R. {Bogart}, E.~{Bonamente}, J.~{Bonnell}, A.~W. {Borgland },
  A.~{Bouvier}, J.~{Bregeon}, A.~{Brez}, M.~{Brigida}, P.~{Bruel}, T.~H.
  {Burnett}, G.~{Busetto}, G.~A. {Caliandro}, R.~A. {Cameron}, P.~A. {Caraveo},
  S.~{Carius}, P.~{Carlson}, J.~M. {Casandjian}, E.~{Cavazzuti}, M.~{Ceccanti},
  C.~{Cecchi}, E.~{Charles}, A.~{Chekhtman}, C.~C. {Cheung}, J.~{Chiang},
  R.~{Chipaux}, A.~N. {Cillis}, S.~{Ciprini}, R.~{Claus}, J.~{Cohen-Tanugi},
  S.~{Condamoor}, J.~{Conrad}, R.~{Corbet}, L.~{Corucci}, L.~{Costamante},
  S.~{Cutini}, D.~S. {Davis}, D.~{Decotigny}, M.~{DeKlotz}, C.~D. {Dermer},
  A.~{de Angelis}, S.~W. {Digel}, E.~{do Couto e Silva}, P.~S. {Drell},
  R.~{Dubois}, D.~{Dumora}, Y.~{Edmonds}, D.~{Fabiani}, C.~{Farnier},
  C.~{Favuzzi}, D.~L. {Flath}, P.~{Fleury}, W.~B. {Focke}, S.~{Funk},
  P.~{Fusco}, F.~{Gargano}, D.~{Gasparrini}, N.~{Gehrels}, F.~X. {Gentit},
  S.~{Germani}, B.~{Giebels}, N.~{Giglietto}, P.~{Giommi}, F.~{Giordano},
  T.~{Glanzman}, G.~{Godfrey}, I.~A. {Grenier}, M.~H. {Grondin}, J.~E. {Grove},
  L.~{Guillemot}, S.~{Guiriec}, G.~{Haller}, A.~K. {Harding}, P.~A. {Hart},
  E.~{Hays}, S.~E. {Healey}, M.~{Hirayama}, L.~{Hjalmarsdotter}, R.~{Horn},
  R.~E. {Hughes}, G.~{J{\'o}hannesson}, G.~{Johansson}, A.~S. {Johnson}, R.~P.
  {Johnson}, T.~J. {Johnson}, W.~N. {Johnson}, T.~{Kamae}, H.~{Katagiri},
  J.~{Kataoka}, A.~{Kavelaars}, N.~{Kawai}, H.~{Kelly}, M.~{Kerr}, W.~{Klamra},
  J.~{Kn{\"o}dlseder}, M.~L. {Kocian}, N.~{Komin}, F.~{Kuehn}, M.~{Kuss},
  D.~{Landriu}, L.~{Latronico}, B.~{Lee}, S.~H. {Lee}, M.~{Lemoine-Goumard},
  A.~M. {Lionetto}, F.~{Longo}, F.~{Loparco}, B.~{Lott}, M.~N. {Lovellette},
  P.~{Lubrano}, G.~M. {Madejski}, A.~{Makeev}, B.~{Marangelli}, M.~M. {Massai},
  M.~N. {Mazziotta}, J.~E. {McEnery}, N.~{Menon}, C.~{Meurer}, P.~F.
  {Michelson}, M.~{Minuti}, N.~{Mirizzi}, W.~{Mitthumsiri}, T.~{Mizuno}, A.~A.
  {Moiseev}, C.~{Monte}, M.~E. {Monzani}, E.~{Moretti}, A.~{Morselli}, I.~V.
  {Moskalenko}, S.~{Murgia}, T.~{Nakamori}, S.~{Nishino}, P.~L. {Nolan}, J.~P.
  {Norris}, E.~{Nuss}, M.~{Ohno}, T.~{Ohsugi}, N.~{Omodei}, E.~{Orlando}, J.~F.
  {Ormes}, A.~{Paccagnella}, D.~{Paneque}, J.~H. {Panetta}, D.~{Parent},
  M.~{Pearce}, M.~{Pepe}, A.~{Perazzo}, M.~{Pesce-Rollins}, P.~{Picozza},
  L.~{Pieri}, M.~{Pinchera}, F.~{Piron}, T.~A. {Porter}, L.~{Poupard},
  S.~{Rain{\`o}}, R.~{Rando}, E.~{Rapposelli}, M.~{Razzano}, A.~{Reimer},
  O.~{Reimer}, T.~{Reposeur}, L.~C. {Reyes}, S.~{Ritz}, L.~S. {Rochester},
  A.~Y. {Rodriguez}, R.~W. {Romani}, M.~{Roth}, J.~J. {Russell}, F.~{Ryde},
  S.~{Sabatini}, H.~F.~W. {Sadrozinski}, D.~{Sanchez}, A.~{Sand er},
  L.~{Sapozhnikov}, P.~M.~Saz {Parkinson}, J.~D. {Scargle}, T.~L. {Schalk},
  G.~{Scolieri}, C.~{Sgr{\`o}}, G.~H. {Share}, M.~{Shaw}, T.~{Shimokawabe},
  C.~{Shrader}, A.~{Sierpowska-Bartosik}, E.~J. {Siskind}, D.~A. {Smith}, P.~D.
  {Smith}, G.~{Spandre}, P.~{Spinelli}, J.~L. {Starck}, T.~E. {Stephens}, M.~S.
  {Strickman}, A.~W. {Strong}, D.~J. {Suson}, H.~{Tajima}, H.~{Takahashi},
  T.~{Takahashi}, T.~{Tanaka}, A.~{Tenze}, S.~{Tether}, J.~B. {Thayer}, J.~G.
  {Thayer}, D.~J. {Thompson}, L.~{Tibaldo}, O.~{Tibolla}, D.~F. {Torres},
  G.~{Tosti}, A.~{Tramacere}, M.~{Turri}, T.~L. {Usher}, N.~{Vilchez},
  V.~{Vitale}, P.~{Wang}, K.~{Watters}, B.~L. {Winer}, K.~S. {Wood},
  T.~{Ylinen}, and M.~{Ziegler}.
\newblock {The Large Area Telescope on the Fermi Gamma-Ray Space Telescope
  Mission}.
\newblock {\em \apj}, 697(2):1071--1102, June 2009.

\bibitem{2020ApJS..247...33A}
S.~{Abdollahi}, F.~{Acero}, M.~{Ackermann}, M.~{Ajello}, W.~B. {Atwood},
  M.~{Axelsson}, L.~{Baldini}, J.~{Ballet}, G.~{Barbiellini}, D.~{Bastieri},
  J.~{Becerra Gonzalez}, R.~{Bellazzini}, A.~{Berretta}, E.~{Bissaldi}, R.~D.
  {Bland ford}, E.~D. {Bloom}, R.~{Bonino}, E.~{Bottacini}, T.~J. {Brandt},
  J.~{Bregeon}, P.~{Bruel}, R.~{Buehler}, T.~H. {Burnett}, S.~{Buson}, R.~A.
  {Cameron}, R.~{Caputo}, P.~A. {Caraveo}, J.~M. {Casandjian}, D.~{Castro},
  E.~{Cavazzuti}, E.~{Charles}, S.~{Chaty}, S.~{Chen}, C.~C. {Cheung},
  G.~{Chiaro}, S.~{Ciprini}, J.~{Cohen-Tanugi}, L.~R. {Cominsky},
  J.~{Coronado-Bl{\'a}zquez}, D.~{Costantin}, A.~{Cuoco}, S.~{Cutini},
  F.~{D'Ammando}, M.~{DeKlotz}, P.~{de la Torre Luque}, F.~{de Palma},
  A.~{Desai}, S.~W. {Digel}, N.~{Di Lalla}, M.~{Di Mauro}, L.~{Di Venere},
  A.~{Dom{\'\i}nguez}, D.~{Dumora}, F.~{Fana Dirirsa}, S.~J. {Fegan}, E.~C.
  {Ferrara}, A.~{Franckowiak}, Y.~{Fukazawa}, S.~{Funk}, P.~{Fusco},
  F.~{Gargano}, D.~{Gasparrini}, N.~{Giglietto}, P.~{Giommi}, F.~{Giordano},
  M.~{Giroletti}, T.~{Glanzman}, D.~{Green}, I.~A. {Grenier}, S.~{Griffin},
  M.~H. {Grondin}, J.~E. {Grove}, S.~{Guiriec}, A.~K. {Harding}, K.~{Hayashi},
  E.~{Hays}, J.~W. {Hewitt}, D.~{Horan}, G.~{J{\'o}hannesson}, T.~J. {Johnson},
  T.~{Kamae}, M.~{Kerr}, D.~{Kocevski}, M.~{Kovac'evic'}, M.~{Kuss},
  D.~{Landriu}, S.~{Larsson}, L.~{Latronico}, M.~{Lemoine-Goumard}, J.~{Li},
  I.~{Liodakis}, F.~{Longo}, F.~{Loparco}, B.~{Lott}, M.~N. {Lovellette},
  P.~{Lubrano}, G.~M. {Madejski}, S.~{Maldera}, D.~{Malyshev}, A.~{Manfreda},
  E.~J. {Marchesini}, L.~{Marcotulli}, G.~{Mart{\'\i}-Devesa}, P.~{Martin},
  F.~{Massaro}, M.~N. {Mazziotta}, J.~E. {McEnery}, I.~{Mereu}, M.~{Meyer},
  P.~F. {Michelson}, N.~{Mirabal}, T.~{Mizuno}, M.~E. {Monzani}, A.~{Morselli},
  I.~V. {Moskalenko}, M.~{Negro}, E.~{Nuss}, R.~{Ojha}, N.~{Omodei},
  M.~{Orienti}, E.~{Orlando}, J.~F. {Ormes}, M.~{Palatiello}, V.~S. {Paliya},
  D.~{Paneque}, Z.~{Pei}, H.~{Pe{\~n}a-Herazo}, J.~S. {Perkins}, M.~{Persic},
  M.~{Pesce-Rollins}, V.~{Petrosian}, L.~{Petrov}, F.~{Piron}, H.~{Poon}, T.~A.
  {Porter}, G.~{Principe}, S.~{Rain{\`o}}, R.~{Rando}, M.~{Razzano},
  S.~{Razzaque}, A.~{Reimer}, O.~{Reimer}, Q.~{Remy}, T.~{Reposeur}, R.~W.
  {Romani}, P.~M. {Saz Parkinson}, F.~K. {Schinzel}, D.~{Serini},
  C.~{Sgr{\`o}}, E.~J. {Siskind}, D.~A. {Smith}, G.~{Spandre}, P.~{Spinelli},
  A.~W. {Strong}, D.~J. {Suson}, H.~{Tajima}, M.~N. {Takahashi}, D.~{Tak},
  J.~B. {Thayer}, D.~J. {Thompson}, L.~{Tibaldo}, D.~F. {Torres}, E.~{Torresi},
  J.~{Valverde}, B.~{Van Klaveren}, P.~{van Zyl}, K.~{Wood}, M.~{Yassine}, and
  G.~{Zaharijas}.
\newblock {Fermi Large Area Telescope Fourth Source Catalog}.
\newblock {\em \apjs}, 247(1):33, March 2020.

\bibitem{2016PhR...636....1C}
E.~{Charles}, M.~{S{\'a}nchez-Conde}, B.~{Anderson}, R.~{Caputo}, A.~{Cuoco},
  M.~{Di Mauro}, A.~{Drlica-Wagner}, G.~A. {Gomez-Vargas}, M.~{Meyer},
  L.~{Tibaldo}, M.~{Wood}, G.~{Zaharijas}, S.~{Zimmer}, M.~{Ajello},
  A.~{Albert}, L.~{Baldini}, K.~{Bechtol}, E.~D. {Bloom}, F.~{Ceraudo},
  J.~{Cohen-Tanugi}, S.~W. {Digel}, J.~{Gaskins}, M.~{Gustafsson},
  N.~{Mirabal}, and M.~{Razzano}.
\newblock {Sensitivity projections for dark matter searches with the Fermi
  large area telescope}.
\newblock {\em \physrep}, 636:1--46, June 2016.

\bibitem{boggs2000}
S.~E. Boggs and P.~Jean.
\newblock Event reconstruction in high resolution compton telescopes.
\newblock {\em Astronomy and Astrophysics Supplement Series}, 145(2):311--321,
  2000.

\bibitem{zoglauer2003}
A.~Zoglauer and G.~Kanbach.
\newblock Doppler broadening as a lower limit to the angular resolution of next
  generation compon telescopes.
\newblock {\em Proceedings of SPIE}, 4851, 2003.

\bibitem{STRAULINO2004168}
S.~Straulino, O.~Adriani, L.~Bonechi, M.~Bongi, G.~Castellini, R.~D'Alessandro,
  A.~Gabbanini, M.~Grandi, P.~Papini, S.B. Ricciarini, P.~Spillantini,
  F.~Taccetti, M.~Tesi, and E.~Vannuccini.
\newblock The pamela silicon tracker.
\newblock {\em Nuclear Instruments and Methods in Physics Research Section A},
  530(1):168 -- 172, 2004.
\newblock Proceedings of the 6th International Conference on Large Scale
  Applications and Radiation Hardness of Semiconductor Detectors.

\bibitem{ZUCCON200874}
P.~Zuccon.
\newblock The ams silicon tracker: Construction and performance.
\newblock {\em Nuclear Instruments and Methods in Physics Research Section A},
  596(1):74 -- 78, 2008.
\newblock Proceedings of the 8th International Conference on Large Scale
  Applications and Radiation Hardness of Semiconductor Detectors.

\bibitem{10.1117/1.JATIS.4.2.021410}
Kazuhiro Nakazawa, Goro Sato, Motohide Kokubun, Teruaki Enoto, Yasushi
  Fukazawa, Kouichi Hagino, Katsuhiro Hayashi, Jun Kataoka, Junichiro Katsuta,
  Shogo~B. Kobayashi, Philippe Laurent, Fran{\c c}ois Lebrun, Olivier Limousin,
  Daniel Maier, Kazuo Makishima, Tsunefumi Mizuno, Kunishiro Mori, Takeshi
  Nakamori, Toshio Nakano, Hirofumi Noda, Hirokazu Odaka, Masanori Ohno,
  Masayuki Ohta, Shinya Saito, Rie Sato, Hiroyasu Tajima, Hiromitsu Takahashi,
  Tadayuki Takahashi, Shin'ichiro Takeda, Takaaki Tanaka, Yukikatsu Terada,
  Hideki Uchiyama, Yasunobu Uchiyama, Shin Watanabe, Kazutaka Yamaoka, Yoichi
  Yatsu, and Takayuki Yuasa.
\newblock {Hard x-ray imager onboard Hitomi (ASTRO-H)}.
\newblock {\em Journal of Astronomical Telescopes, Instruments, and Systems},
  4(2):1 -- 12, 2018.

\bibitem{2002NewAR..46..611B}
P.~F. {Bloser}, R.~{Andritschke}, G.~{Kanbach}, V.~{Sch{\"o}nfelder},
  F.~{Schopper}, and A.~{Zoglauer}.
\newblock {The MEGA advanced Compton telescope project}.
\newblock {\em \nar}, 46(8-10):611--616, July 2002.

\bibitem{oneil2003}
T.~J. O'Neill, D.~Bhattacharya, M.~Polsen, A.~D. Zych, J.~Samimi, and A.~Akyuz.
\newblock Development of the tigre compton telescope for intermediate-energy
  gamma-ray astronomy.
\newblock {\em IEEE Transactions on Nuclear Science}, 50(2):251--257, 2003.

\bibitem{2019ICRC...36..565G}
S.~{Griffin}.
\newblock {Subsystem Development for the All-Sky Medium Energy Gamma-ray
  Observatory (AMEGO) prototype}.
\newblock In {\em 36th International Cosmic Ray Conference (ICRC2019)},
  volume~36 of {\em International Cosmic Ray Conference}, page 565, July 2019.

\bibitem{Atwood:2007ra}
W.~B. Atwood et~al.
\newblock {Design and Initial Tests of the Tracker-Converter of the Gamma-ray
  Large Area Space Telescope}.
\newblock {\em Astropart. Phys.}, 28:422--434, 2007.

\bibitem{Atwood:2009ez}
W.~B. Atwood et~al.
\newblock {The Large Area Telescope on the Fermi Gamma-ray Space Telescope
  Mission}.
\newblock {\em Astrophys. J.}, 697:1071--1102, 2009.

\bibitem{Bulgarelli:2010zz}
A.~Bulgarelli et~al.
\newblock {The AGILE silicon tracker: Pre-launch and in-flight configuration}.
\newblock {\em Nucl. Instrum. Meth. A}, 614:213--226, 2010.

\bibitem{AGILE:2008nyq}
M.~Tavani et~al.
\newblock {The AGILE Mission}.
\newblock {\em Astron. Astrophys.}, 502:995--1013, 2009.

\bibitem{DAMPE:2017yae}
A.~Tykhonov et~al.
\newblock {Internal alignment and position resolution of the silicon tracker of
  DAMPE determined with orbit data}.
\newblock {\em Nucl. Instrum. Meth. A}, 893:43--56, 2018.

\bibitem{DAMPE:2019lxv}
G.~Ambrosi et~al.
\newblock {The on-orbit calibration of DArk Matter Particle Explorer}.
\newblock {\em Astropart. Phys.}, 106:18--34, 2019.

\bibitem{Joram:2015tla}
Christian Joram, G.~Haefeli, and B.~Leverington.
\newblock {Scintillating Fibre Tracking at High Luminosity Colliders}.
\newblock {\em JINST}, 10(08):C08005, 2015.

\bibitem{Bravar:2020wje}
A.~Bravar, K.~Briggl, S.~Corrodi, A.~Damyanova, L.~Gerritzen, C.~Grab,
  M.~Hildebrandt, A.~Papa, and G.~Rutar.
\newblock {The Mu3e scintillating fiber timing detector}.
\newblock {\em Nucl. Instrum. Meth. A}, 958:162564, 2020.

\bibitem{vonDoetinchem:2007gw}
P.~von Doetinchem, Henning Gast, T.~Kirn, G.~Roper Yearwood, and S.~Schael.
\newblock {PEBS: Positron Electron Balloon Spectrometer}.
\newblock {\em Nucl. Instrum. Meth. A}, 581:151--155, 2007.

\bibitem{Alitti:1988za}
J.~Alitti et~al.
\newblock {Performance of the Scintillating Fiber Detector in the Upgraded Ua2
  Detector}.
\newblock {\em Nucl. Instrum. Meth. A}, 279:364--375, 1989.

\bibitem{CHORUS:1997ijc}
P.~Annis et~al.
\newblock {The CHORUS scintillating fiber tracker and optoelectronic readout
  system}.
\newblock {\em Nucl. Instrum. Meth. A}, 412:19--37, 1998.

\bibitem{K2K:2000kji}
A.~Suzuki et~al.
\newblock {Design, construction, and operation of SciFi tracking detector for
  K2K experiment}.
\newblock {\em Nucl. Instrum. Meth. A}, 453:165--176, 2000.

\bibitem{D0:2005cnn}
V.~M. Abazov et~al.
\newblock {The Upgraded D0 detector}.
\newblock {\em Nucl. Instrum. Meth. A}, 565:463--537, 2006.

\bibitem{Ellis:2005dx}
Malcolm Ellis.
\newblock {A scintillating fibre tracker for MICE}.
\newblock {\em Int. J. Mod. Phys. A}, 20:3815--3819, 2005.

\bibitem{Antich:1990kv}
P.~Antich, M.~Atac, R.~Chaney, D.~Chrisman, D.~Cline, E.~Fenyves, and J.~Park.
\newblock {Development of a High Resolution Scintillating Fiber $\gamma^-$ ray
  Telescope}.
\newblock {\em Nucl. Instrum. Meth. A}, 297:514--520, 1990.

\bibitem{Fishman:1997id}
G.~J. Fishman.
\newblock {GLAST: Using scintillation fibers for both the tracker and the
  calorimeter}.
\newblock {\em AIP Conf. Proc.}, 450(1):571--577, 1998.

\bibitem{Pendelton:1999ma}
Geffrey~N. Pendelton et~al.
\newblock {FiberGLAST: A Scintillating fiber approach to the GLAST mission}.
\newblock {\em Proc. SPIE Int. Soc. Opt. Eng.}, 3765:12--21, 1999.

\bibitem{MazSnow}
\url{https://www.snowmass21.org/docs/files/summaries/IF/SNOWMASS21-IF3_IF2_Mazziotta-100.pdf}.

\bibitem{S13552}
\url{https://www.hamamatsu.com/eu/en/product/type/S13552/index.html}.

\bibitem{Aalseth:2017fik}
C.~E. Aalseth et~al.
\newblock {DarkSide-20k: A 20 tonne two-phase LAr TPC for direct dark matter
  detection at LNGS}.
\newblock {\em Eur. Phys. J. Plus}, 133:131, 2018.

\bibitem{vikuiti}
Vikuiti$^{TM}$ Enhanced Specular Reflector Film (ESR),
  \url{https://www.3.com/displayfilms}.

\bibitem{Berner:2019uvt}
Roman Berner et~al.
\newblock {First Operation of a Resistive Shell Liquid Argon Time Projection
  Chamber: A New Approach to Electric-Field Shaping}.
\newblock {\em Instruments}, 3(2):28, 2019.

\bibitem{Adamowski_2014}
M~Adamowski, B~Carls, E~Dvorak, A~Hahn, W~Jaskierny, C~Johnson, H~Jostlein,
  C~Kendziora, S~Lockwitz, B~Pahlka, R~Plunkett, S~Pordes, B~Rebel, R~Schmitt,
  M~Stancari, T~Tope, E~Voirin, and T~Yang.
\newblock The liquid argon purity demonstrator.
\newblock {\em Journal of Instrumentation}, 9(07):P07005--P07005, jul 2014.

\bibitem{cryocoolers}
\url{https://www.northropgrumman.com/high-efficiency-cryocoolers/}.

\bibitem{penelope}
Nuclear~Energy Agency.
\newblock {\em PENELOPE 2018: A code system for Monte Carlo simulation of
  electron and photon transport}.
\newblock 2019.

\bibitem{CryoAsic1}
A.~{Pena-Perez}, D.~{Doering}, A.~{Gupta}, C.~{Tamma}, B.~{Markovic},
  P.~{Caragiulo} H.~{Ali}, L.~{Rota}, U.~{Kamath}, S.~{Petrignani}, X.~{Xu},
  F.~{Abu-Nimeh}, P.~A. {(Sander) Breur}, P.~{Tsang}, M.~{Convery}, and
  A.~{Dragone}.
\newblock {CRYO: A System-On-Chip ASIC for Noble Liquid TPC Experiments}.
\newblock In {\em 2020 IEEE Nuclear Science Symposium and Medical Imaging
  Conference Record (NSS/MIC)}, pages 1--2, 2020.

\bibitem{Dwyer_2018}
D.A. Dwyer, M.~Garcia-Sciveres, D.~Gnani, C.~Grace, S.~Kohn, M.~Kramer,
  A.~Krieger, C.J. Lin, K.B. Luk, P.~Madigan, C.~Marshall, H.~Steiner, and
  T.~Stezelberger.
\newblock {LArPix}: demonstration of low-power 3d pixelated charge readout for
  liquid argon time projection chambers.
\newblock {\em Journal of Instrumentation}, 13(10):P10007--P10007, oct 2018.

\bibitem{Adams_2020}
D.~Adams, M.~Bass, M.~Bishai, C.~Bromberg, J.~Calcutt, H.~Chen, J.~Fried,
  I.~Furic, S.~Gao, D.~Gastler, J.~Hugon, J.~Joshi, B.~Kirby, F.~Liu, K.~Mahn,
  M.~Mooney, C.~Morris, C.~Pereyra, X.~Pons, V.~Radeka, E.~Raguzin, D.~Shooltz,
  M.~Spanu, A.~Timilsina, S.~Tufanli, M.~Tzanov, B.~Viren, W.~Gu, Z.~Williams,
  K.~Wood, E.~Worcester, M.~Worcester, G.~Yang, and J.~Zhang.
\newblock The {ProtoDUNE}-{SP} {LArTPC} electronics production, commissioning,
  and performance.
\newblock {\em Journal of Instrumentation}, 15(06):P06017--P06017, jun 2020.

\bibitem{pinsky_update_2016}
L~Pinsky, T~Campbell-Ricketts, A~Empl, S~George, L~Tlustos, D~Turecek, D~Fry,
  M~Kroupa, R~Rios, E~Semones, N~Stoffle, S~Wheeler, C~Zeitlin, H~Kitamura, and
  S~Kodaira.
\newblock An {Update} on {Medipix} in {Space} and {Future} {Plans} ({Medipix}2,
  3 \& 4).
\newblock {\em AMICSA \& DSP 2016 Prceedings}, page~4, 2016.

\bibitem{whyntie_simulation_2014}
T~Whyntie and M~A Harrison.
\newblock Simulation and analysis of the {LUCID} experiment in the {Low}
  {Earth} {Orbit} radiation environment.
\newblock {\em Journal of Physics: Conference Series}, 513(2):022038, June
  2014.

\bibitem{peric_high-voltage_2018}
Ivan Peric, Mridula Prathapan, Heiko Augustin, Mathieu Benoit, Raimon~Casanova
  Mohr, Dominik Dannheim, Felix Ehrler, Fadoua~Guezzi Messaoud, Moritz Kiehn,
  Andreas Nürnberg, Rudolf Schimassek, Mateus~Vicente Barreto, Eva~Vilella
  Figueras, Alena Weber, Winnie Wong, and Hui Zhang.
\newblock A high-voltage pixel sensor for the {ATLAS} upgrade.
\newblock {\em Nuclear Instruments and Methods in Physics Research Section A:
  Accelerators, Spectrometers, Detectors and Associated Equipment}, June 2018.

\bibitem{mager_alpide_2016}
M.~Mager.
\newblock {ALPIDE}, the {Monolithic} {Active} {Pixel} {Sensor} for the {ALICE}
  {ITS} upgrade.
\newblock {\em Nuclear Instruments and Methods in Physics Research Section A:
  Accelerators, Spectrometers, Detectors and Associated Equipment},
  824:434--438, July 2016.

\bibitem{berdalovic_monolithic_2018}
I.~Berdalovic, R.~Bates, C.~Buttar, R.~Cardella, N.~Egidos Plaja, T.~Hemperek,
  B.~Hiti, J.W.~van Hoorne, T.~Kugathasan, I.~Mandic, D.~Maneuski, C.A.~Marin
  Tobon, K.~Moustakas, L.~Musa, H.~Pernegger, P.~Riedler, C.~Riegel,
  D.~Schaefer, E.J. Schioppa, A.~Sharma, W.~Snoeys, C.~Solans Sanchez, T.~Wang,
  and N.~Wermes.
\newblock Monolithic pixel development in {TowerJazz} 180 nm {CMOS} for the
  outer pixel layers in the {ATLAS} experiment.
\newblock {\em Journal of Instrumentation}, 13(01):C01023--C01023, January
  2018.

\bibitem{vigani_study_2017}
L.~Vigani, D.~Bortoletto, L.~Ambroz, R.~Plackett, T.~Hemperek, P.~Rymaszewski,
  T.~Wang, H.~Krueger, T.~Hirono, I.~Caicedo Sierra, N.~Wermes, M.~Barbero,
  S.~Bhat, P.~Breugnon, Z.~Chen, S.~Godiot, P.~Pangaud, and A.~Rozanov.
\newblock Study of prototypes of {LFoundry} active and monolithic {CMOS} pixels
  sensors for the {ATLAS} detector.
\newblock {\em Proceeding for PSD11}, October 2017.

\bibitem{2021SPIE11821E..1DP}
D.~{Poulson}, P.~F. {Bloser}, K.~{Ogasawara}, J.~A. {Trevino}, J.~S. {Legere},
  J.~M. {Ryan}, and M.~L. {McConnell}.
\newblock {Development of a Compton telescope based on single-crystal diamond
  detectors and fast scintillators}.
\newblock In {\em Society of Photo-Optical Instrumentation Engineers (SPIE)
  Conference Series}, volume 11821 of {\em Society of Photo-Optical
  Instrumentation Engineers (SPIE) Conference Series}, page 118211D, August
  2021.

\bibitem{2002NIMPA.476..686A}
W.~{Adam}, E.~{Berdermann}, P.~{Bergonzo}, G.~{Bertuccio}, F.~{Bogani},
  E.~{Borchi}, A.~{Brambilla}, M.~{Bruzzi}, C.~{Colledani}, J.~{Conway},
  P.~{D'Angelo}, W.~{Dabrowski}, P.~{Delpierre}, A.~{Deneuville},
  W.~{Dulinski}, B.~{van Eijk}, A.~{Fallou}, F.~{Fizzotti}, F.~{Foulon},
  M.~{Friedl}, K.~K. {Gan}, E.~{Gheeraert}, G.~{Hallewell}, S.~{Han},
  F.~{Hartjes}, J.~{Hrubec}, D.~{Husson}, H.~{Kagan}, D.~{Kania}, J.~{Kaplon},
  R.~{Kass}, T.~{Koeth}, M.~{Krammer}, A.~{Logiudice}, R.~{Lu}, L.~{mac Lynne},
  C.~{Manfredotti}, D.~{Meier}, M.~{Mishina}, L.~{Moroni}, J.~{Noomen},
  A.~{Oh}, L.~S. {Pan}, M.~{Pernicka}, A.~{Peitz}, L.~{Perera}, S.~{Pirollo},
  M.~{Procario}, J.~L. {Riester}, S.~{Roe}, L.~{Rousseau}, A.~{Rudge},
  J.~{Russ}, S.~{Sala}, M.~{Sampietro}, S.~{Schnetzer}, S.~{Sciortino},
  H.~{Stelzer}, R.~{Stone}, B.~{Suter}, R.~J. {Tapper}, R.~{Tesarek},
  W.~{Trischuk}, D.~{Tromson}, E.~{Vittone}, A.~M. {Walsh}, R.~{Wedenig},
  P.~{Weilhammer}, M.~{Wetstein}, C.~{White}, W.~{Zeuner}, and M.~{Zoeller}.
\newblock {Radiation tolerance of CVD diamond detectors for pions and protons}.
\newblock {\em Nuclear Instruments and Methods in Physics Research A},
  476(3):686--693, January 2002.

\bibitem{1369574}
H.~Frais-Kolbl, E.~Griesmayer, H.~Kagan, and H.~Pernegger.
\newblock A fast low-noise charged-particle cvd diamond detector.
\newblock {\em IEEE Transactions on Nuclear Science}, 51(6):3833--3837, 2004.

\bibitem{TANIMURA2000325}
Yoshihiko Tanimura, Junichi Kaneko, Masaki Katagiri, {Yujiro Ikeda}, Takeo
  Nishitani, Hiroshi Takeuchi, and Toshiyuki Iida.
\newblock High-temperature operation of a radiation detector made of a type iia
  diamond single crystal synthesized by a hp/ht method.
\newblock {\em Nuclear Instruments and Methods in Physics Research Section A:
  Accelerators, Spectrometers, Detectors and Associated Equipment},
  443(2):325--330, 2000.

\bibitem{2009JPCM...21J4221B}
R.~S. {Balmer}, J.~R. {Brandon}, S.~L. {Clewes}, H.~K. {Dhillon}, J.~M.
  {Dodson}, I.~{Friel}, P.~N. {Inglis}, T.~D. {Madgwick}, M.~L. {Markham},
  T.~P. {Mollart}, N.~{Perkins}, G.~A. {Scarsbrook}, D.~J. {Twitchen}, A.~J.
  {Whitehead}, J.~J. {Wilman}, and S.~M. {Woollard}.
\newblock {Chemical vapour deposition synthetic diamond: materials, technology
  and applications}.
\newblock {\em Journal of Physics Condensed Matter}, 21(36):364221, September
  2009.

\bibitem{OGASAWARA2015131}
K.~Ogasawara, T.W. Broiles, K.E. Coulter, M.A. Dayeh, M.I. Desai, S.A. Livi,
  D.J. McComas, and B.C. Walther.
\newblock Single crystal chemical vapor deposit diamond detector for energetic
  plasma measurement in space.
\newblock {\em Nuclear Instruments and Methods in Physics Research Section A:
  Accelerators, Spectrometers, Detectors and Associated Equipment},
  777:131--137, 2015.

\bibitem{bohon2010}
J.~{Bohon}, E.~{Muller}, and J.~{Smedley}.
\newblock {Development of diamond-based X-ray detection for high-flux beamline
  diagnostics}.
\newblock {\em J. Synchrotron Rad.}, 17:711, 2010.

\bibitem{Zhou:pp5072}
Tianyi Zhou, Wenxiang Ding, Mengjia Gaowei, Gianluigi De~Geronimo, Jen Bohon,
  John Smedley, and Erik Muller.
\newblock {Pixelated transmission-mode diamond X-ray detector}.
\newblock {\em Journal of Synchrotron Radiation}, 22(6):1396--1402, Nov 2015.

\bibitem{GU2012210}
Yajun Gu, J.~Lu, T.~Grotjohn, T.~Schuelke, and J.~Asmussen.
\newblock Microwave plasma reactor design for high pressure and high power
  density diamond synthesis.
\newblock {\em Diamond and Related Materials}, 24:210--214, 2012.

\bibitem{LU201317}
J.~Lu, Y.~Gu, T.A. Grotjohn, T.~Schuelke, and J.~Asmussen.
\newblock Experimentally defining the safe and efficient, high pressure
  microwave plasma assisted cvd operating regime for single crystal diamond
  synthesis.
\newblock {\em Diamond and Related Materials}, 37:17--28, 2013.

\bibitem{1993ApJS...86..657S}
V.~{Schoenfelder}, H.~{Aarts}, K.~{Bennett}, H.~{de Boer}, J.~{Clear},
  W.~{Collmar}, A.~{Connors}, A.~{Deerenberg}, R.~{Diehl}, A.~{von Dordrecht},
  J.~W. {den Herder}, W.~{Hermsen}, M.~{Kippen}, L.~{Kuiper}, G.~{Lichti},
  J.~{Lockwood}, J.~{Macri}, M.~{McConnell}, D.~{Morris}, R.~{Much}, J.~{Ryan},
  G.~{Simpson}, M.~{Snelling}, G.~{Stacy}, H.~{Steinle}, A.~{Strong}, B.~N.
  {Swanenburg}, B.~{Taylor}, C.~{de Vries}, and C.~{Winkler}.
\newblock {Instrument Description and Performance of the Imaging Gamma-Ray
  Telescope COMPTEL aboard the Compton Gamma-Ray Observatory}.
\newblock {\em \apjs}, 86:657, June 1993.

\bibitem{1993ApJS...86..693J}
W.~N. {Johnson}, R.~L. {Kinzer}, J.~D. {Kurfess}, M.~S. {Strickman}, W.~R.
  {Purcell}, D.~A. {Grabelsky}, M.~P. {Ulmer}, D.~A. {Hillis}, G.~V. {Jung},
  and R.~A. {Cameron}.
\newblock {The Oriented Scintillation Spectrometer Experiment: Instrument
  Description}.
\newblock {\em \apjs}, 86:693, June 1993.

\bibitem{4333363}
V.~Schonfelder, R.~Diehl, G.~G. Lichti, H.~Steinle, B.~N. Swanenburg, A.~J.~M.
  Deerenberg, H.~Aarts, J.~Lockwood, W.~Webber, J.~Macri, J.~Ryan, G.~Simpson,
  B.~G. Taylor, K.~Bennett, and M.~Snelling.
\newblock The imaging compton telescope comptel on the gamma ray observatory.
\newblock {\em IEEE Transactions on Nuclear Science}, 31(1):766--770, 1984.

\bibitem{2009ApJ...702..791M}
C.~{Meegan} et~al.
\newblock {\em \apj}, 702:791--804, September 2009.

\bibitem{QUARATI2013596}
F.G.A. Quarati, P.~Dorenbos, J.~{van der Biezen}, Alan Owens, M.~Selle,
  L.~Parthier, and P.~Schotanus.
\newblock Scintillation and detection characteristics of high-sensitivity cebr3
  gamma-ray spectrometers.
\newblock {\em Nuclear Instruments and Methods in Physics Research Section A:
  Accelerators, Spectrometers, Detectors and Associated Equipment},
  729:596--604, 2013.

\bibitem{1239275}
K.S. Shah, J.~Glodo, M.~Klugerman, W.W. Moses, S.E. Derenzo, and M.J. Weber.
\newblock Labr/sub 3/:ce scintillators for gamma ray spectroscopy.
\newblock In {\em 2002 IEEE Nuclear Science Symposium Conference Record},
  volume~1, pages 92--95 vol.1, 2002.

\bibitem{10.1117/12.830016}
N.~J. Cherepy, B.~W. Sturm, O~B. Drury, T.~A. Hurst, S.~A. Sheets, L.~E. Ahle,
  C.~K. Saw, M.~A. Pearson, S.~A. Payne, A.~Burger, L.~A. Boatner, J.~O. Ramey,
  E.~V. van Loef, J.~Glodo, R.~Hawrami, W.~M. Higgins, K.~S. Shah, and W.~W.
  Moses.
\newblock {SrI[sub]2[/sub] scintillator for gamma ray spectroscopy}.
\newblock In Ralph~B. James, Larry~A. Franks, and Arnold Burger, editors, {\em
  Hard X-Ray, Gamma-Ray, and Neutron Detector Physics XI}, volume 7449, pages
  100 -- 105. International Society for Optics and Photonics, SPIE, 2009.

\bibitem{10.1117/12.2528073}
Lee~J. Mitchell, Bernard~F. Phlips, Richard~S. Woolf, Theodore~T. Finne, and
  W.~Neil Johnson.
\newblock {Strontium iodide radiation instrumentation II (SIRI-2)}.
\newblock In Oswald~H. Siegmund, editor, {\em UV, X-Ray, and Gamma-Ray Space
  Instrumentation for Astronomy XXI}, volume 11118, pages 122 -- 139.
  International Society for Optics and Photonics, SPIE, 2019.

\bibitem{osti_1096473}
F.~Patrick Doty, Xiao~Wang Zhou, Pin Yang, and Mark~A Rodriguez.
\newblock Elpasolite scintillators.

\bibitem{4545124}
Jarek Glodo, William~M. Higgins, Edgar V.~D. van Loef, and Kanai~S. Shah.
\newblock Scintillation properties of 1 inch <formula
  formulatype="inline"><tex>${\rm cs}_{2}{\rm liycl}_{6}{:}{\rm
  ce}$</tex></formula> crystals.
\newblock {\em IEEE Transactions on Nuclear Science}, 55(3):1206--1209, 2008.

\bibitem{SHIRWADKAR2011268}
Urmila Shirwadkar, Jarek Glodo, Edgar~V. {van Loef}, Rastgo Hawrami,
  Sharmishtha Mukhopadhyay, Alexei Churilov, William~M. Higgins, and Kanai~S.
  Shah.
\newblock Scintillation properties of cs2lilabr6 (cllb) crystals with varying
  ce3+ concentration.
\newblock {\em Nuclear Instruments and Methods in Physics Research Section A:
  Accelerators, Spectrometers, Detectors and Associated Equipment},
  652(1):268--270, 2011.
\newblock Symposium on Radiation Measurements and Applications (SORMA) XII
  2010.

\bibitem{10.1117/12.2060204}
Paul~P. Guss, Thomas~G. Stampahar, Sanjoy Mukhopadhyay, Alexander Barzilov, and
  Amber Guckes.
\newblock {Scintillation properties of a Cs2LiLa(Br6)90
  crystal}.
\newblock In Gary~P. Grim and H.~Bradford Barber, editors, {\em Radiation
  Detectors: Systems and Applications XV}, volume 9215, pages 27 -- 41.
  International Society for Optics and Photonics, SPIE, 2014.

\bibitem{doi:10.1021/acs.cgd.7b00583}
R.~Hawrami, E.~Ariesanti, H.~Wei, J.~Finkelstein, J.~Glodo, and K.~Shah.
\newblock Tl2liycl6: Large diameter, high performing dual mode scintillator.
\newblock {\em Crystal Growth \& Design}, 17(7):3960--3964, 2017.

\bibitem{DOLYMPIA2012140}
N.~D'Olympia, P.~Chowdhury, C.J. Guess, T.~Harrington, E.G. Jackson,
  S.~Lakshmi, C.J. Lister, J.~Glodo, R.~Hawrami, K.~Shah, and U.~Shirwadkar.
\newblock Optimizing cs2liycl6 for fast neutron spectroscopy.
\newblock {\em Nuclear Instruments and Methods in Physics Research Section A:
  Accelerators, Spectrometers, Detectors and Associated Equipment},
  694:140--146, 2012.

\bibitem{Hutcheson2021}
A.~L. Hutcheson, B.~F. Phlips, W.~N. Johnson, L.~J. Mitchell, R.~S. Perea,
  J.~M. Wolf, M.~V. Johnson-Rambert, J.~M. Davis, and T.~T. Finne.
\newblock {Neutron Radiation Detection Instrument (NeRDI)}.
\newblock In {\em IEEE Nuclear Science Symposium and Medical Imaging
  Conference}, pages N--26--02, October 2021.

\bibitem{Kamada_2011}
K~Kamada, T~Yanagida, J~Pejchal, M~Nikl, T~Endo, K~Tsutumi, Y~Fujimoto,
  A~Fukabori, and A~Yoshikawa.
\newblock Scintillator-oriented combinatorial search in ce-doped
  (y,gd)3(ga,al)5o12multicomponent garnet compounds.
\newblock {\em Journal of Physics D: Applied Physics}, 44(50):505104, dec 2011.

\bibitem{GLODO2017285}
Jarek Glodo, Yimin Wang, Ryan Shawgo, Charles Brecher, Rastgo~H. Hawrami,
  Joshua Tower, and Kanai~S. Shah.
\newblock New developments in scintillators for security applications.
\newblock {\em Physics Procedia}, 90:285--290, 2017.
\newblock Conference on the Application of Accelerators in Research and
  Industry, CAARI 2016, 30 October – 4 November 2016, Ft. Worth, TX, USA.

\bibitem{10.1117/12.2598588}
Lee~J. Mitchell, Bernard~F. Phlips, Richard~S. Woolf, Theodore~T. Finne,
  Anthony~L. Hutcheson, W.~Neil Johnson, Mary Johnson-Rambert, and Rose Perea.
\newblock {GAGG Radiation Instrumentation (GARI)}.
\newblock In Oswald~H. Siegmund, editor, {\em UV, X-Ray, and Gamma-Ray Space
  Instrumentation for Astronomy XXII}, volume 11821, pages 38 -- 50.
  International Society for Optics and Photonics, SPIE, 2021.

\bibitem{PHOS:Shaw2003}
S.~E. {Shaw}, M.~J. {Westmore}, A.~J. {Bird}, A.~J. {Dean}, C.~{Ferguson},
  R.~{Gurriaran}, J.~J. {Lockley}, and D.~R. {Willis}.
\newblock {A mass model for estimating the gamma ray background of the Burst
  and Transient Source Experiment}.
\newblock {\em \aap}, 398:391--402, January 2003.

\bibitem{PHOS:Meegan2009}
Charles {Meegan}, Giselher {Lichti}, P.~N. {Bhat}, Elisabetta {Bissaldi},
  Michael~S. {Briggs}, Valerie {Connaughton}, Roland {Diehl}, Gerald {Fishman},
  Jochen {Greiner}, Andrew~S. {Hoover}, Alexander~J. {van der Horst}, Andreas
  {von Kienlin}, R.~Marc {Kippen}, Chryssa {Kouveliotou}, Sheila {McBreen},
  W.~S. {Paciesas}, Robert {Preece}, Helmut {Steinle}, Mark~S. {Wallace},
  Robert~B. {Wilson}, and Colleen {Wilson-Hodge}.
\newblock {The Fermi Gamma-ray Burst Monitor}.
\newblock {\em \apj}, 702(1):791--804, September 2009.

\bibitem{PHOS:Fishman1989}
William~S. {Paciesas}, Geoffrey~N. {Pendleton}, J.~P. {Lestrade}, Gerald~J.
  {Fishman}, Charles~A. {Meegan}, Robert~B. {Wilson}, Thomas~A. {Parnell},
  Robert~A. {Austin}, Jr. {Berry}, Fred~A., John~M. {Horack}, and Scott~D.
  {Storey}.
\newblock {Performance of the large-area detectors for the Burst and Transient
  Source Experiment on the Gamma Ray Observatory}.
\newblock In Charles~J. {Hailey} and Oswald H.~W. {Siegmund}, editors, {\em
  EUV, X-Ray, and Gamma-Ray Instrumentation for Astronomy and Atomic Physics},
  volume 1159 of {\em Society of Photo-Optical Instrumentation Engineers (SPIE)
  Conference Series}, pages 156--0, November 1989.

\bibitem{PHOS:Bissaldi2009}
E.~{Bissaldi}, A.~{von Kienlin}, G.~{Lichti}, H.~{Steinle}, P.~N. {Bhat}, M.~S.
  {Briggs}, G.~J. {Fishman}, A.~S. {Hoover}, R.~M. {Kippen}, M.~{Krumrey},
  M.~{Gerlach}, V.~{Connaughton}, R.~{Diehl}, J.~{Greiner}, A.~J. {van der
  Horst}, C.~{Kouveliotou}, S.~{McBreen}, C.~A. {Meegan}, W.~S. {Paciesas},
  R.~D. {Preece}, and C.~A. {Wilson-Hodge}.
\newblock {Ground-based calibration and characterization of the Fermi gamma-ray
  burst monitor detectors}.
\newblock {\em Experimental Astronomy}, 24(1-3):47--88, May 2009.

\bibitem{PHOS:Matteson1978}
J.~L. Matteson.
\newblock {The UCSD/MIT hard X-ray and low energy gamma-ray experiment for
  HEAO-1 - Design and early results}.
\newblock {\em Proc. AIAA 16th Aerospace Sciences Meeting}, (35), January 1978.

\bibitem{PHOS:Rothschild1998}
R.~E. {Rothschild}, P.~R. {Blanco}, D.~E. {Gruber}, W.~A. {Heindl}, D.~R.
  {MacDonald}, D.~C. {Marsden}, M.~R. {Pelling}, L.~R. {Wayne}, and P.~L.
  {Hink}.
\newblock {In-Flight Performance of the High-Energy X-Ray Timing Experiment on
  the Rossi X-Ray Timing Explorer}.
\newblock {\em \apj}, 496(1):538--549, March 1998.

\bibitem{PHOS:Svertilov2017}
S.~I. Svertilov, M.~I. Panasyuk, V.~V. Bogomolov, A.~M. Amelushkin, V.~O.
  Barinova, V.~I. Galkin, A.~F. Iyudin, E.~A. Kuznetsova, A.~V. Prokhorov,
  V.~L. Petrov, G.~V. Rozhkov, I.~V. Yashin, E.~S. Gorbovskoy, V.~M. Lipunov,
  I.~H. Park, S.~Jeong, and M.~B. Kim.
\newblock {Wide-Field Gamma-Spectrometer BDRG: GRB Monitor On-Board the
  Lomonosov Mission}.
\newblock {\em Space Science Reviews}, 214(1):8, November 2017.

\bibitem{PHOS:Zhang2020}
Shuang-Nan {Zhang}, TiPei {Li}, FangJun {Lu}, LiMing {Song}, YuPeng {Xu},
  CongZhan {Liu}, Yong {Chen}, XueLei {Cao}, QingCui {Bu}, Zhi {Chang}, Gang
  {Chen}, Li~{Chen}, TianXiang {Chen}, YiBao {Chen}, YuPeng {Chen}, Wei {Cui},
  WeiWei {Cui}, JingKang {Deng}, YongWei {Dong}, YuanYuan {Du}, MinXue {Fu},
  GuanHua {Gao}, He~{Gao}, Min {Gao}, MingYu {Ge}, YuDong {Gu}, Ju~{Guan}, Can
  {Gungor}, ChengCheng {Guo}, DaWei {Han}, Wei {Hu}, Yue {Huang}, Jia {Huo},
  ShuMei {Jia}, LuHua {Jiang}, WeiChun {Jiang}, Jing {Jin}, YongJie {Jin}, Bing
  {Li}, ChengKui {Li}, Gang {Li}, MaoShun {Li}, Wei {Li}, Xian {Li}, XiaoBo
  {Li}, XuFang {Li}, YanGuo {Li}, ZiJian {Li}, ZhengWei {Li}, XiaoHua {Liang},
  JinYuan {Liao}, GuoQing {Liu}, HongWei {Liu}, ShaoZhen {Liu}, XiaoJing {Liu},
  Yuan {Liu}, YiNong {Liu}, Bo~{Lu}, XueFeng {Lu}, Tao {Luo}, Xiang {Ma}, Bin
  {Meng}, Yi~{Nang}, JianYin {Nie}, Ge~{Ou}, JinLu {Qu}, Na~{Sai}, RenCheng
  {Shang}, GuoHong {Shen}, Liang {Sun}, Ying {Tan}, Lian {Tao}, YouLi {Tuo},
  Chen {Wang}, ChunQin {Wang}, GuoFeng {Wang}, HuanYu {Wang}, Juan {Wang},
  WenShuai {Wang}, YuSa {Wang}, XiangYang {Wen}, BaiYang {Wu}, BoBing {Wu}, Mei
  {Wu}, GuangCheng {Xiao}, ShaoLin {Xiong}, LinLi {Yan}, JiaWei {Yang}, Sheng
  {Yang}, YanJi {Yang}, QiBin {Yi}, Bin {Yuan}, AiMei {Zhang}, ChunLei {Zhang},
  ChengMo {Zhang}, Fan {Zhang}, HongMei {Zhang}, Juan {Zhang}, Qiang {Zhang},
  ShenYi {Zhang}, Shu {Zhang}, Tong {Zhang}, WanChang {Zhang}, Wei {Zhang},
  WenZhao {Zhang}, Yi~{Zhang}, YiFei {Zhang}, YongJie {Zhang}, Yue {Zhang},
  Zhao {Zhang}, Zhi {Zhang}, ZiLiang {Zhang}, HaiSheng {Zhao}, XiaoFan {Zhao},
  ShiJie {Zheng}, JianFeng {Zhou}, YuXuan {Zhu}, Yue {Zhu}, RenLin {Zhuang},
  and {Insight-HXMT Team}.
\newblock {Overview to the Hard X-ray Modulation Telescope (Insight-HXMT)
  Satellite}.
\newblock {\em Science China Physics, Mechanics, and Astronomy}, 63(4):249502,
  February 2020.

\bibitem{PHOS:von_Kienlin2020}
A.~von Kienlin, C.~A. Meegan, W.~S. Paciesas, P.~N. Bhat, E.~Bissaldi, M.~S.
  Briggs, E.~Burns, W.~H. Cleveland, M.~H. Gibby, M.~M. Giles, A.~Goldstein,
  R.~Hamburg, C.~M. Hui, D.~Kocevski, B.~Mailyan, C.~Malacaria, S.~Poolakkil,
  R.~D. Preece, O.~J. Roberts, P.~Veres, and C.~A. Wilson-Hodge.
\newblock The fourth fermi-{GBM} gamma-ray burst catalog: A decade of data.
\newblock {\em The Astrophysical Journal}, 893(1):46, apr 2020.

\bibitem{PHOS:Poolakkil2021}
S.~Poolakkil, R.~Preece, C.~Fletcher, A.~Goldstein, P.~N. Bhat, E.~Bissaldi,
  M.~S. Briggs, E.~Burns, W.~H. Cleveland, M.~M. Giles, C.~M. Hui, D.~Kocevski,
  S.~Lesage, B.~Mailyan, C.~Malacaria, W.~S. Paciesas, O.~J. Roberts, P.~Veres,
  A.~von Kienlin, and C.~A. Wilson-Hodge.
\newblock The fermi-{GBM} gamma-ray burst spectral catalog: 10 yr of data.
\newblock {\em The Astrophysical Journal}, 913(1):60, may 2021.

\bibitem{PHOS:Pendleton1999}
Geoffrey~N. {Pendleton}, Michael~S. {Briggs}, R.~Marc {Kippen}, William.~S.
  {Paciesas}, Mark {Stollberg}, Pete {Woods}, Charles~A. {Meegan}, Gerald~J.
  {Fishman}, Mike~L. {McCollough}, and Valerie {Connaughton}.
\newblock {The Structure and Evolution of LOCBURST: The BATSE Burst Location
  Algorithm}.
\newblock {\em \apj}, 512(1):362--376, February 1999.

\bibitem{PHOS:Goldstein2020}
A.~Goldstein, C.~Fletcher, P.~Veres, M.~S. Briggs, W.~H. Cleveland, M.~H.
  Gibby, C.~M. Hui, E.~Bissaldi, E.~Burns, R.~Hamburg, A.~von Kienlin,
  D.~Kocevski, B.~Mailyan, C.~Malacaria, W.~S. Paciesas, O.~J. Roberts, and
  C.~A. Wilson-Hodge.
\newblock Evaluation of automated fermi {GBM} localizations of gamma-ray
  bursts.
\newblock {\em The Astrophysical Journal}, 895(1):40, may 2020.

\bibitem{PHOS:Bloser2013}
Peter~F. Bloser, Jason Legere, Christopher Bancroft, Mark~L. McConnell,
  James~M. Ryan, and Nathan Schwadron.
\newblock {Scintillator gamma-ray detectors with silicon photomultiplier
  readouts for high-energy astronomy}.
\newblock In Oswald~H. Siegmund, editor, {\em UV, X-Ray, and Gamma-Ray Space
  Instrumentation for Astronomy XVIII}, volume 8859, pages 66 -- 73.
  International Society for Optics and Photonics, SPIE, 2013.

\bibitem{PHOS:Guss2010}
Paul Guss, Michael Reed, Ding Yuan, Matthew Cutler, Christopher Contreras, and
  Denis Beller.
\newblock {Comparison of CeBr3 with LaBr3:Ce, LaCl3:Ce, and NaI:Tl detectors}.
\newblock In Arnold Burger, Larry~A. Franks, and Ralph~B. James, editors, {\em
  Hard X-Ray, Gamma-Ray, and Neutron Detector Physics XII}, volume 7805, pages
  93 -- 108. International Society for Optics and Photonics, SPIE, 2010.

\bibitem{PHOS:Knoll2000}
Glenn~F. Knoll.
\newblock {\em {Radiation Detection and Measurement, 3rd ed.}}
\newblock John Wiley and Sons, New York, 3rd edition edition, 2000.

\bibitem{Grove2010}
J.~Eric {Grove} and W.~Neil {Johnson}.
\newblock {The calorimeter of the Fermi Large Area Telescope}.
\newblock In {\em Space Telescopes and Instrumentation 2010: Ultraviolet to
  Gamma Ray}, volume 7732 of {\em Society of Photo-Optical Instrumentation
  Engineers (SPIE) Conference Series}, page 77320J, July 2010.

\bibitem{McEnery2020}
J.~{McEnery} and {AMEGO Team}.
\newblock {All Sky Medium Energy Gamma-ray Observatory (AMEGO): Exploring the
  Extreme Multimessenger Universe}.
\newblock In {\em American Astronomical Society Meeting Abstracts}, American
  Astronomical Society Meeting Abstracts, page 372.15, January 2020.

\bibitem{Mitchell2020}
Lee~J. {Mitchell}, Bernard {Phlips}, W.~Neil {Johnson}, Mary {Johnson-Rambert},
  Anika~N. {Kansky}, and Richard {Woolf}.
\newblock {Radiation damage assessment of SensL SiPMs}.
\newblock {\em arXiv e-prints}, page arXiv:2003.08213, March 2020.

\bibitem{Smith2019}
J.~{Smith}.
\newblock {BurstCube: Mission Concept, Performance, and Status}.
\newblock In {\em 36th International Cosmic Ray Conference (ICRC2019)},
  volume~36 of {\em International Cosmic Ray Conference}, page 604, July 2019.

\bibitem{Wen2019}
Jiaxing {Wen}, Xiangyun {Long}, Xutao {Zheng}, Yu~{An}, Zhengyang {Cai}, Jirong
  {Cang}, Yuepeng {Che}, Changyu {Chen}, Liangjun {Chen}, Qianjun {Chen}, Ziyun
  {Chen}, Yingjie {Cheng}, Litao {Deng}, Wei {Deng}, Wenqing {Ding}, Hangci
  {Du}, Lian {Duan}, Quan {Gan}, Tai {Gao}, Zhiying {Gao}, Wenbin {Han}, Yiying
  {Han}, Xinbo {He}, Xinhao {He}, Long {Hou}, Fan {Hu}, Junling {Hu},
  Bo~{Huang}, Dongyang {Huang}, Xuefeng {Huang}, Shihai {Jia}, Yuchen {Jiang},
  Yifei {Jin}, Ke~{Li}, Siyao {Li}, Yurong {Li}, Jianwei {Liang}, Yuanyuan
  {Liang}, Wei {Lin}, Chang {Liu}, Gang {Liu}, Mengyuan {Liu}, Rui {Liu},
  Tianyu {Liu}, Wanqiang {Liu}, Di'an {Lu}, Peiyibin {Lu}, Zhiyong {Lu}, Xiyu
  {Luo}, Sizheng {Ma}, Yuanhang {Ma}, Xiaoqing {Mao}, Yanshan {Mo}, Qiyuan
  {Nie}, Shuiyin {Qu}, Xiaolong {Shan}, Gengyuan {Shi}, Weiming {Song}, Zhigang
  {Sun}, Xuelin {Tan}, Songsong {Tang}, Mingrui {Tao}, Boqin {Wang}, Yue
  {Wang}, Zhiang {Wang}, Qiaoya {Wu}, Xuanyi {Wu}, Yuehan {Xia}, Hengyuan
  {Xiao}, Wenjin {Xie}, Dacheng {Xu}, Rui {Xu}, Weili {Xu}, Longbiao {Yan},
  Shengyu {Yan}, Dongxin {Yang}, Hang {Yang}, Haoguang {Yang}, Yi-Si {Yang},
  Yifan {Yang}, Lei {Yao}, Huan {Yu}, Yangyi {Yu}, Aiqiang {Zhang}, Bingtao
  {Zhang}, Lixuan {Zhang}, Maoxing {Zhang}, Shen {Zhang}, Tianliang {Zhang},
  Yuchong {Zhang}, Qianru {Zhao}, Ruining {Zhao}, Shiyu {Zheng}, Xiaolong
  {Zhou}, Runyu {Zhu}, Yu~{Zou}, Peng {An}, Yifu {Cai}, Hongbing {Chen}, Zigao
  {Dai}, Yizhong {Fan}, Changqing {Feng}, Hua {Feng}, He~{Gao}, Liang {Huang},
  Mingming {Kang}, Lixin {Li}, Zhuo {Li}, Enwei {Liang}, Lin {Lin}, Qianqian
  {Lin}, Congzhan {Liu}, Hongbang {Liu}, Xuewen {Liu}, Yinong {Liu}, Xiang
  {Lu}, Shude {Mao}, Rongfeng {Shen}, Jing {Shu}, Meng {Su}, Hui {Sun}, Pak-Hin
  {Tam}, Chi-Pui {Tang}, Yang {Tian}, Fayin {Wang}, Jianjun {Wang}, Wei {Wang},
  Zhonghai {Wang}, Jianfeng {Wu}, Xuefeng {Wu}, Shaolin {Xiong}, Can {Xu},
  Jiandong {Yu}, Wenfei {Yu}, Yunwei {Yu}, Ming {Zeng}, Zhi {Zeng}, Bin-Bin
  {Zhang}, Bing {Zhang}, Zongqing {Zhao}, Rong {Zhou}, and Zonghong {Zhu}.
\newblock {GRID: a student project to monitor the transient gamma-ray sky in
  the multi-messenger astronomy era}.
\newblock {\em Experimental Astronomy}, 48(1):77--95, August 2019.

\bibitem{Vestrand2019}
T.~{Vestrand}.
\newblock {The Micro Astrophysical MeV Background Observatory (MAMBO)}.
\newblock In {\em 36th International Cosmic Ray Conference (ICRC2019)},
  volume~36 of {\em International Cosmic Ray Conference}, page 608, July 2019.

\bibitem{Link2019}
J.~{Link}.
\newblock {Silicon Photomultiplier use in Particle Astrophysics and
  Heliophysics Missions}.
\newblock In {\em 36th International Cosmic Ray Conference (ICRC2019)},
  volume~36 of {\em International Cosmic Ray Conference}, page~96, July 2019.

\bibitem{Ulyanov2020}
Alexei {Ulyanov}, David {Murphy}, Joseph {Mangan}, Viyas {Gupta}, Wojciech
  {Hajdas}, Daithi {de Faoite}, Brian {Shortt}, Lorraine {Hanlon}, and Sheila
  {McBreen}.
\newblock {Radiation damage study of SensL J-series silicon photomultipliers
  using 101.4 MeV protons}.
\newblock {\em Nuclear Instruments and Methods in Physics Research A},
  976:164203, October 2020.

\bibitem{Tamer2020}
Tamer {Tolba}, Dieter {Grzonka}, Thomas {Sefzick}, and James {Ritman}.
\newblock {Study of Silicon Photomultiplier Radiation Hardness with the JULIC
  Cyclotron}.
\newblock {\em arXiv e-prints}, page arXiv:2003.04933, March 2020.

\bibitem{Bartlett2020}
K.~D. {Bartlett}, D.~D.~S. {Coupland}, D.~T. {Beckman}, and K.~E. {Mesick}.
\newblock {Proton irradiation damage and annealing effects in ON Semiconductor
  J-series silicon photomultipliers}.
\newblock {\em Nuclear Instruments and Methods in Physics Research A},
  969:163957, July 2020.

\bibitem{Krizmanic:2020shl}
John~F. Krizmanic.
\newblock {Space-based Extensive Air Shower Optical Cherenkov and Fluorescence
  Measurements using SiPM Detectors in context of POEMMA}.
\newblock {\em Nucl. Instrum. Meth. A}, 985:164614, 2021.

\bibitem{Otte:2018cnv}
Adam~Nepomuk Otte, Thanh Nguyen, and Joel Stansbury.
\newblock {Locating the avalanche structure and the origin of breakdown
  generating charge carriers in silicon photomultipliers by using the bias
  dependent breakdown probability}.
\newblock {\em Nucl. Instrum. Meth. A}, 916:283--289, 2019.

\bibitem{Krizmanic:1999gf}
J.~F. Krizmanic.
\newblock {Performance of the Orbiting Wide-Angle Light collector
  (OWL/AirWatch) experiment via Monte Carlo simulation}.
\newblock In {\em {26th International Cosmic Ray Conference}}, 1999.

\bibitem{Cummings:2020ycz}
A.~L. Cummings, R.~Aloisio, and J.~F. Krizmanic.
\newblock {Modeling of the Tau and Muon Neutrino-induced Optical Cherenkov
  Signals from Upward-moving Extensive Air Showers}.
\newblock {\em Phys. Rev. D}, 103(4):043017, 2021.

\bibitem{Meier1991}
R.~R. {Meier}.
\newblock {Ultraviolet spectroscopy and remote sensing of the upper
  atmosphere}.
\newblock {\em Space Science Reviews}, 58:1--185, 1991.

\bibitem{Shepherd2006}
Gordon~G. {Shepherd}, Young-Min {Cho}, Guiping {Liu}, Marianna~G. {Shepherd},
  and Raymond~G. {Roble}.
\newblock {Airglow variability in the context of the global mesospheric
  circulation}.
\newblock {\em Journal of Atmospheric and Solar-Terrestrial Physics},
  68:2000--2011, December 2006.

\bibitem{2003A&A...407.1157H}
R.~W. {Hanuschik}.
\newblock {A flux-calibrated, high-resolution atlas of optical sky emission
  from UVES}.
\newblock {\em Astronomy and Astrophysics}, 407:1157--1164, Sep 2003.

\bibitem{2006JGRA..11112307C}
Philip~C. {Cosby}, Brian~D. {Sharpee}, Tom~G. {Slanger}, David~L. {Huestis},
  and Reinhard~W. {Hanuschik}.
\newblock {High-resolution terrestrial nightglow emission line atlas from
  UVES/VLT: Positions, intensities, and identifications for 2808 lines at
  314-1043 nm}.
\newblock {\em Journal of Geophysical Research (Space Physics)},
  111(A12):A12307, Dec 2006.

\bibitem{geant41}
S.~Agostinelli et~al.
\newblock Geant4—a simulation toolkit.
\newblock {\em Nuclear Instruments and Methods in Physics Research Section A:
  Accelerators, Spectrometers, Detectors and Associated Equipment},
  506(3):250--303, 2003.

\bibitem{geant42}
J.~et~al. Allison.
\newblock Geant4 developments and applications.
\newblock {\em IEEE Transactions on Nuclear Science}, 53(1):270--278, 2006.

\bibitem{mcnp}
Michael~Evan Rising, Forrest~B. Brown, Jackie~Renee Salazar, and Jeremy~Ed
  Sweezy.
\newblock Overview of the mcnp6® sqa plan and requirements [memorandum].
\newblock 8 2020.

\bibitem{corsika}
D.~et~al. {Heck}.
\newblock {\em {CORSIKA: a Monte Carlo code to simulate extensive air
  showers.}}
\newblock 1998.

\bibitem{sword}
Wade Duvall, Bernard Phlips, Anthony Hutcheson, Ryan Cordes, Joseph Hartsell,
  and Mark Strickman.
\newblock Improving radiation transport simulation capabilities for nuclear
  threat detection using sword.
\newblock In {\em 2019 IEEE International Symposium on Technologies for
  Homeland Security (HST)}, pages 1--6, 2019.

\bibitem{siri}
Lee J.~Mitchell et~al.
\newblock {Strontium iodide radiation instrumentation II (SIRI-2)}.
\newblock In Oswald~H. Siegmund, editor, {\em UV, X-Ray, and Gamma-Ray Space
  Instrumentation for Astronomy XXI}, volume 11118, pages 122 -- 139.
  International Society for Optics and Photonics, SPIE, 2019.

\bibitem{gamma_detector}
Richard S.~Woolf et~al.
\newblock {Novel inorganic scintillators for future space-based solar gamma-ray
  and neutron research}.
\newblock In Ralph~B. James, Arnold Burger, and Stephen~A. Payne, editors, {\em
  Hard X-Ray, Gamma-Ray, and Neutron Detector Physics XXI}, volume 11114, pages
  7 -- 24. International Society for Optics and Photonics, SPIE, 2019.

\bibitem{glowbug}
J.~E. Grove et~al.
\newblock {Glowbug, a Low-Cost, High-Sensitivity Gamma-Ray Burst Telescope}.
\newblock In {\em {71st Yamada Conference: Gamma-ray Bursts in the
  Gravitational Wave Era 2019}}, 9 2020.

\bibitem{nustar}
Takao et~al. {Kitaguchi}.
\newblock {Spectral calibration and modeling of the NuSTAR CdZnTe pixel
  detectors}.
\newblock In {\em Society of Photo-Optical Instrumentation Engineers (SPIE)
  Conference Series}, volume 8145 of {\em Society of Photo-Optical
  Instrumentation Engineers (SPIE) Conference Series}, page 814507, September
  2011.

\bibitem{cosi}
John A. et~al. Tomsick.
\newblock {The Compton Spectrometer and Imager}.
\newblock 8 2019.

\bibitem{amego}
Regina et~al. Caputo.
\newblock {All-sky Medium Energy Gamma-ray Observatory: Exploring the Extreme
  Multimessenger Universe}.
\newblock 7 2019.

\bibitem{robast}
Akira Okumura, Koji Noda, and Cameron Rulten.
\newblock Robast: Development of a root-based ray-tracing library for
  cosmic-ray telescopes and its applications in the cherenkov telescope array.
\newblock {\em Astroparticle Physics}, 76:38--47, Mar 2016.

\bibitem{root}
Rene~Brun et~al.
\newblock Root — an object oriented data analysis framework.
\newblock {\em Nuclear Instruments and Methods in Physics Research Section A:
  Accelerators, Spectrometers, Detectors and Associated Equipment},
  389(1):81--86, 1997.
\newblock New Computing Techniques in Physics Research V.

\bibitem{simtel}
Konrad Bernlohr.
\newblock {Simulation of Imaging Atmospheric Cherenkov Telescopes with CORSIKA
  and sim\_telarray}.
\newblock {\em Astropart. Phys.}, 30:149--158, 2008.

\bibitem{zemax}
J.~A. et~al. Aguilar.
\newblock {Design, optimization and characterization of the light concentrators
  of the single-mirror small size telescopes of the Cherenkov Telescope Array}.
\newblock {\em Astropart. Phys.}, 60:32--40, 2015.

\bibitem{postnikov2019gamma}
EB~Postnikov, AP~Kryukov, SP~Polyakov, DA~Shipilov, and DP~Zhurov.
\newblock Gamma/hadron separation in imaging air cherenkov telescopes using
  deep learning libraries tensorflow and pytorch.
\newblock In {\em Journal of Physics: Conference Series}, volume 1181, page
  012048. IOP Publishing, 2019.

\bibitem{capistran2021use}
T~Capistr{\'a}n, KL~Fan, JT~Linnemann, I~Torres, PM~Parkinson, and PLH Yu.
\newblock Use of machine learning for gamma/hadron separation with hawc.
\newblock {\em arXiv preprint arXiv:2108.00112}, 2021.

\bibitem{Kaur_2019}
Amanpreet Kaur, Abraham~D. Falcone, Michael~D. Stroh, Jamie~A. Kennea, and
  Elizabeth~C. Ferrara.
\newblock Classification of new x-ray counterparts for fermi unassociated
  gamma-ray sources using the swift x-ray telescope.
\newblock {\em The Astrophysical Journal}, 887(1):18, dec 2019.

\bibitem{Saz_Parkinson_2016}
P.~M.~Saz Parkinson, H.~Xu, P.~L.~H. Yu, D.~Salvetti, M.~Marelli, and A.~D.
  Falcone.
\newblock {CLASSIFICATION} {AND} {RANKING} {OFFERMILAT} {GAMMA}-{RAY} {SOURCES}
  {FROM} {THE} 3fgl {CATALOG} {USING} {MACHINE} {LEARNING} {TECHNIQUES}.
\newblock {\em The Astrophysical Journal}, 820(1):8, mar 2016.

\bibitem{10.1093/mnras/staa166}
Shengda Luo, Alex~P Leung, C~Y Hui, and K~L Li.
\newblock {An investigation on the factors affecting machine learning
  classifications in gamma-ray astronomy}.
\newblock {\em Monthly Notices of the Royal Astronomical Society},
  492(4):5377--5390, 01 2020.

\bibitem{Tavani2009}
M.~Tavani et~al.
\newblock {The AGILE Mission}.
\newblock {\em Astron. Astrophys.}, 502:995--1013, 2009.

\bibitem{Atwood2009}
W.~B. Atwood, A.~A. Abdo, M.~Ackermann, W.~Althouse, B.~Anderson, M.~Axelsson,
  L.~Baldini, J.~Ballet, D.~L. Band, G.~Barbiellini, J.~Bartelt, D.~Bastieri,
  B.~M. Baughman, K.~Bechtol, D.~B{\'{e}}d{\'{e}}r{\`{e}}de, F.~Bellardi,
  R.~Bellazzini, B.~Berenji, G.~F. Bignami, D.~Bisello, E.~Bissaldi, R.~D.
  Blandford, E.~D. Bloom, J.~R. Bogart, E.~Bonamente, J.~Bonnell, A.~W.
  Borgland, A.~Bouvier, J.~Bregeon, A.~Brez, M.~Brigida, P.~Bruel, T.~H.
  Burnett, G.~Busetto, G.~A. Caliandro, R.~A. Cameron, P.~A. Caraveo,
  S.~Carius, P.~Carlson, J.~M. Casandjian, E.~Cavazzuti, M.~Ceccanti,
  C.~Cecchi, E.~Charles, A.~Chekhtman, C.~C. Cheung, J.~Chiang, R.~Chipaux,
  A.~N. Cillis, S.~Ciprini, R.~Claus, J.~Cohen-Tanugi, S.~Condamoor, J.~Conrad,
  R.~Corbet, L.~Corucci, L.~Costamante, S.~Cutini, D.~S. Davis, D.~Decotigny,
  M.~Deklotz, C.~D. Dermer, A.~{De Angelis}, S.~W. Digel, E.~{Do Couto E
  Silva}, P.~S. Drell, R.~Dubois, D.~Dumora, Y.~Edmonds, D.~Fabiani,
  C.~Farnier, C.~Favuzzi, D.~L. Flath, P.~Fleury, W.~B. Focke, S.~Funk,
  P.~Fusco, F.~Gargano, D.~Gasparrini, N.~Gehrels, F.~X. Gentit, S.~Germani,
  B.~Giebels, N.~Giglietto, P.~Giommi, F.~Giordano, T.~Glanzman, G.~Godfrey,
  I.~A. Grenier, M.~H. Grondin, J.~E. Grove, L.~Guillemot, S.~Guiriec,
  G.~Haller, A.~K. Harding, P.~A. Hart, E.~Hays, S.~E. Healey, M.~Hirayama,
  L.~Hjalmarsdotter, R.~Horn, R.~E. Hughes, G.~J{\'{o}}hannesson, G.~Johansson,
  A.~S. Johnson, R.~P. Johnson, T.~J. Johnson, W.~N. Johnson, T.~Kamae,
  H.~Katagiri, J.~Kataoka, A.~Kavelaars, N.~Kawai, H.~Kelly, M.~Kerr,
  W.~Klamra, J.~Kn{\"{o}}dlseder, M.~L. Kocian, N.~Komin, F.~Kuehn, M.~Kuss,
  D.~Landriu, L.~Latronico, B.~Lee, S.~H. Lee, M.~Lemoine-Goumard, A.~M.
  Lionetto, F.~Longo, F.~Loparco, B.~Lott, M.~N. Lovellette, P.~Lubrano, G.~M.
  Madejski, A.~Makeev, B.~Marangelli, M.~M. Massai, M.~N. Mazziotta, J.~E.
  McEnery, N.~Menon, C.~Meurer, P.~F. Michelson, M.~Minuti, N.~Mirizzi,
  W.~Mitthumsiri, T.~Mizuno, A.~A. Moiseev, C.~Monte, M.~E. Monzani,
  E.~Moretti, A.~Morselli, I.~V. Moskalenko, S.~Murgia, T.~Nakamori,
  S.~Nishino, P.~L. Nolan, J.~P. Norris, E.~Nuss, M.~Ohno, T.~Ohsugi,
  N.~Omodei, E.~Orlando, J.~F. Ormes, A.~Paccagnella, D.~Paneque, J.~H.
  Panetta, D.~Parent, M.~Pearce, M.~Pepe, A.~Perazzo, M.~Pesce-Rollins,
  P.~Picozza, L.~Pieri, M.~Pinchera, F.~Piron, T.~A. Porter, L.~Poupard,
  S.~Rain{\`{o}}, R.~Rando, E.~Rapposelli, M.~Razzano, A.~Reimer, O.~Reimer,
  T.~Reposeur, L.~C. Reyes, S.~Ritz, L.~S. Rochester, A.~Y. Rodriguez, R.~W.
  Romani, M.~Roth, J.~J. Russell, F.~Ryde, S.~Sabatini, H.~F.W. Sadrozinski,
  D.~Sanchez, A.~Sander, L.~Sapozhnikov, P.~M.Saz Parkinson, J.~D. Scargle,
  T.~L. Schalk, G.~Scolieri, C.~Sgr{\`{o}}, G.~H. Share, M.~Shaw,
  T.~Shimokawabe, C.~Shrader, A.~Sierpowska-Bartosik, E.~J. Siskind, D.~A.
  Smith, P.~D. Smith, G.~Spandre, P.~Spinelli, J.~L. Starck, T.~E. Stephens,
  M.~S. Strickman, A.~W. Strong, D.~J. Suson, H.~Tajima, H.~Takahashi,
  T.~Takahashi, T.~Tanaka, A.~Tenze, S.~Tether, J.~B. Thayer, J.~G. Thayer,
  D.~J. Thompson, L.~Tibaldo, O.~Tibolla, D.~F. Torres, G.~Tosti, A.~Tramacere,
  M.~Turri, T.~L. Usher, N.~Vilchez, V.~Vitale, P.~Wang, K.~Watters, B.~L.
  Winer, K.~S. Wood, T.~Ylinen, and M.~Ziegler.
\newblock {The large area telescope on the fermi gamma-ray space telescope
  mission}.
\newblock {\em Astrophysical Journal}, 697(2), 2009.

\bibitem{Lucarelli2020}
F.~{Lucarelli} and C.~{Pittori}.
\newblock {The AGILE Gamma-Ray Legacy Archive and the User-Friendly AGILE-LV3
  Web Tool Integrated in the ASI-SSDC MWL Environment}.
\newblock In R.~{Pizzo}, E.~R. {Deul}, J.~D. {Mol}, J.~{de Plaa}, and
  H.~{Verkouter}, editors, {\em Astronomical Data Analysis Software and Systems
  XXIX}, volume 527 of {\em Astronomical Society of the Pacific Conference
  Series}, page~33, January 2020.

\bibitem{2022arXiv220111184F}
{Fermi-LAT collaboration}, {:}, Soheila {Abdollahi}, Fabio {Acero}, Luca
  {Baldini}, Jean {Ballet}, Denis {Bastieri}, Ronaldo {Bellazzini}, Bijan
  {Berenji}, Alessandra {Berretta}, Elisabetta {Bissaldi}, Roger~D.
  {Blandford}, Elliott {Bloom}, Raffaella {Bonino}, Ari {Brill}, Richard~J.
  {Britto}, Philippe {Bruel}, Toby~H. {Burnett}, Sara {Buson}, Rob~A.
  {Cameron}, Regina {Caputo}, Patrizia~A. {Caraveo}, Daniel {Castro}, Sylvain
  {Chaty}, Teddy~C. {Cheung}, Graziano {Chiaro}, Nicolo {Cibrario}, Stefano
  {Ciprini}, Javier {Coronado-Blazquez}, Milena {Crnogorcevic}, Sara {Cutini},
  Filippo {D'Ammando}, Salvatore {De Gaetano}, Seth~W. {Digel}, Niccolo {Di
  Lalla}, Feraol~F. {Dirirsa}, Leonardo {Di Venere}, Alberto {Dominguez},
  Vandad {Fallah Ramazani}, Stephen~J. {Fegan}, Elizabeth~C. {Ferrara}, Alessio
  {Fiori}, Henrike {Fleischhack}, Anna {Franckowiak}, Yasushi {Fukazawa},
  Stefan {Funk}, Piergiorgio {Fusco}, Giorgio {Galanti}, Viviana {Gammaldi},
  Fabio {Gargano}, Simone {Garrappa}, Dario {Gasparrini}, Federica {Giacchino},
  Nico {Giglietto}, Francesco {Giordano}, Marcello {Giroletti}, Thomas
  {Glanzman}, David {Green}, Isabelle~A. {Grenier}, Marie-Helene {Grondin},
  Lucas {Guillemot}, Sylvain {Guiriec}, Michael {Gustafsson}, Alice~K.
  {Harding}, Liz {Hays}, John~W. {Hewitt}, Deirdre {Horan}, Xian {Hou},
  Gudlaugur {Johannesson}, Christopher~M. {Karwin}, Taishu {Kayanoki},
  Matthew~T. {Kerr}, Michael {Kuss}, David {Landriu}, Stefan {Larsson}, Luca
  {Latronico}, Marianne {Lemoine-Goumard}, Jian {Li}, Ioannis {Liodakis},
  Francesco {Longo}, Francesco {Loparco}, Benoit {Lott}, Pasquale {Lubrano},
  Simone {Maldera}, Dmitry {Malyshev}, Alberto {Manfreda}, Guillem
  {Marti-Devesa}, Mario~N. {Mazziotta}, Isabella {Mereu}, Manuel {Meyer},
  Peter~F. {Michelson}, Nestor {Mirabal}, Warit {Mitthumsiri}, Tsunefumi
  {Mizuno}, Alex~A. {Moiseev}, Maria~E. {Monzani}, Aldo {Morselli}, Igor~V.
  {Moskalenko}, Michela {Negro}, Eric {Nuss}, Nicola {Omodei}, Monica
  {Orienti}, Elena {Orlando}, David {Paneque}, Zhiyuan {Pei}, Jeremy~S.
  {Perkins}, Massimo {Persic}, Melissa {Pesce-Rollins}, Vahe {Petrosian},
  Roberta {Pillera}, Helen {Poon}, Troy~A. {Porter}, Giacomo {Principe}, Silvia
  {Raino}, Riccardo {Rando}, Bindu {Rani}, Massimiliano {Razzano}, Soebur
  {Razzaque}, Anita {Reimer}, Olaf {Reimer}, Thierry {Reposeur}, Miguel~A.
  {Sanchez-Conde}, Pablo~M. {Saz Parkinson}, Lorenzo {Scotton}, Davide
  {Serini}, Carmelo {Sgro}, Eric~J. {Siskind}, David~A. {Smith}, Gloria
  {Spandre}, Paolo {Spinelli}, Kohei {Sueoka}, Dan~J. {Suson}, Hiro {Tajima},
  Dongguen {Tak}, Jana~B. {Thayer}, David~J. {Thompson}, Diego~F. {Torres},
  Eleonora {Troja}, Janeth {Valverde}, Kent {Wood}, and Gabrijela {Zaharijas}.
\newblock {Incremental Fermi Large Area Telescope Fourth Source Catalog}.
\newblock {\em arXiv e-prints}, page arXiv:2201.11184, January 2022.

\bibitem{2020ApJ...905...76A}
A.~{Albert}, R.~{Alfaro}, C.~{Alvarez}, J.~R.~Angeles {Camacho}, J.~C.
  {Arteaga-Vel{\'a}zquez}, K.~P. {Arunbabu}, D.~{Avila Rojas}, H.~A. {Ayala
  Solares}, V.~{Baghmanyan}, E.~{Belmont-Moreno}, S.~Y. {BenZvi},
  C.~{Brisbois}, K.~S. {Caballero-Mora}, T.~{Capistr{\'a}n},
  A.~{Carrami{\~n}ana}, S.~{Casanova}, U.~{Cotti}, S.~{Couti{\~n}o de
  Le{\'o}n}, E.~{De la Fuente}, R.~{Diaz Hernandez}, L.~{Diaz-Cruz}, B.~L.
  {Dingus}, M.~A. {DuVernois}, M.~{Durocher}, J.~C. {D{\'\i}az-V{\'e}lez},
  R.~W. {Ellsworth}, K.~{Engel}, C.~{Espinoza}, K.~L. {Fan}, K.~{Fang},
  M.~Fern{\'a}ndez {Alonso}, H.~{Fleischhack}, N.~{Fraija},
  A.~{Galv{\'a}n-G{\'a}mez}, D.~{Garcia}, J.~A. {Garc{\'\i}a-Gonz{\'a}lez},
  F.~{Garfias}, G.~{Giacinti}, M.~M. {Gonz{\'a}lez}, J.~A. {Goodman}, J.~P.
  {Harding}, S.~{Hernandez}, J.~{Hinton}, B.~{Hona}, D.~{Huang},
  F.~{Hueyotl-Zahuantitla}, P.~{H{\"u}ntemeyer}, A.~{Iriarte},
  A.~{Jardin-Blicq}, V.~{Joshi}, D.~{Kieda}, A.~{Lara}, W.~H. {Lee},
  H.~{Le{\'o}n Vargas}, J.~T. {Linnemann}, A.~L. {Longinotti}, G.~{Luis-Raya},
  J.~{Lundeen}, R.~{L{\'o}pez-Coto}, K.~{Malone}, V.~{Marandon}, O.~{Martinez},
  I.~{Martinez-Castellanos}, J.~{Mart{\'\i}nez-Castro}, J.~A. {Matthews},
  P.~{Miranda-Romagnoli}, J.~A. {Morales-Soto}, E.~{Moreno}, M.~{Mostaf{\'a}},
  A.~{Nayerhoda}, L.~{Nellen}, M.~{Newbold}, M.~U. {Nisa},
  R.~{Noriega-Papaqui}, L.~{Olivera-Nieto}, N.~{Omodei}, A.~{Peisker},
  Y.~{P{\'e}rez Araujo}, E.~G. {P{\'e}rez-P{\'e}rez}, Z.~{Ren}, C.~D. {Rho},
  C.~{Rivi{\`e}re}, D.~{Rosa-Gonz{\'a}lez}, E.~{Ruiz-Velasco}, H.~{Salazar},
  F.~{Salesa Greus}, A.~{Sandoval}, M.~{Schneider}, H.~{Schoorlemmer},
  F.~{Serna}, G.~{Sinnis}, A.~J. {Smith}, R.~W. {Springer}, P.~{Surajbali},
  K.~{Tollefson}, I.~{Torres}, R.~{Torres-Escobedo}, T.~N. {Ukwatta},
  F.~{Ure{\~n}a-Mena}, T.~{Weisgarber}, F.~{Werner}, E.~{Willox}, A.~{Zepeda},
  H.~{Zhou}, C.~{de Le{\'o}n}, J.~D. {{\'A}lvarez}, and {HAWC Collaboration}.
\newblock {3HWC: The Third HAWC Catalog of Very-high-energy Gamma-Ray Sources}.
\newblock {\em \apj}, 905(1):76, December 2020.

\bibitem{2012ASPC..466..167K}
O.~{Kargaltsev}, G.~G. {Pavlov}, and M.~{Durant}.
\newblock {Pulsar Wind Nebulae from X-rays to VHE {\ensuremath{\gamma}}-rays}.
\newblock In W.~{Lewandowski}, O.~{Maron}, and J.~{Kijak}, editors, {\em
  Electromagnetic Radiation from Pulsars and Magnetars}, volume 466 of {\em
  Astronomical Society of the Pacific Conference Series}, page 167, December
  2012.

\bibitem{2018A&A...612A...2H}
{H.~E.~S.~S. Collaboration}, H.~{Abdalla}, A.~{Abramowski}, F.~{Aharonian},
  F.~{Ait Benkhali}, A.~G. {Akhperjanian}, T.~{Andersson}, E.~O. {Ang{\"u}ner},
  M.~{Arrieta}, P.~{Aubert}, M.~{Backes}, A.~{Balzer}, M.~{Barnard},
  Y.~{Becherini}, J.~{Becker Tjus}, D.~{Berge}, S.~{Bernhard},
  K.~{Bernl{\"o}hr}, R.~{Blackwell}, M.~{B{\"o}ttcher}, C.~{Boisson},
  J.~{Bolmont}, P.~{Bordas}, J.~{Bregeon}, F.~{Brun}, P.~{Brun}, M.~{Bryan},
  T.~{Bulik}, M.~{Capasso}, J.~{Carr}, S.~{Carrigan}, S.~{Casanova},
  M.~{Cerruti}, N.~{Chakraborty}, R.~{Chalme-Calvet}, R.~C.~G. {Chaves},
  A.~{Chen}, J.~{Chevalier}, M.~{Chr{\'e}tien}, S.~{Colafrancesco},
  G.~{Cologna}, B.~{Condon}, J.~{Conrad}, C.~{Couturier}, Y.~{Cui}, I.~D.
  {Davids}, B.~{Degrange}, C.~{Deil}, J.~{Devin}, P.~{deWilt}, L.~{Dirson},
  A.~{Djannati-Ata{\"\i}}, W.~{Domainko}, A.~{Donath}, L.~O.~'C. {Drury},
  G.~{Dubus}, K.~{Dutson}, J.~{Dyks}, T.~{Edwards}, K.~{Egberts}, P.~{Eger},
  J.~P. {Ernenwein}, S.~{Eschbach}, C.~{Farnier}, S.~{Fegan}, M.~V.
  {Fernandes}, A.~{Fiasson}, G.~{Fontaine}, A.~{F{\"o}rster}, S.~{Funk},
  M.~{F{\"u}{\ss}ling}, S.~{Gabici}, M.~{Gajdus}, Y.~A. {Gallant},
  T.~{Garrigoux}, G.~{Giavitto}, B.~{Giebels}, J.~F. {Glicenstein},
  D.~{Gottschall}, A.~{Goyal}, M.~H. {Grondin}, D.~{Hadasch}, J.~{Hahn},
  M.~{Haupt}, J.~{Hawkes}, G.~{Heinzelmann}, G.~{Henri}, G.~{Hermann},
  O.~{Hervet}, A.~{Hillert}, J.~A. {Hinton}, W.~{Hofmann}, C.~{Hoischen},
  M.~{Holler}, D.~{Horns}, A.~{Ivascenko}, A.~{Jacholkowska}, M.~{Jamrozy},
  M.~{Janiak}, D.~{Jankowsky}, F.~{Jankowsky}, M.~{Jingo}, T.~{Jogler},
  L.~{Jouvin}, I.~{Jung-Richardt}, M.~A. {Kastendieck}, K.~{Katarzy{\'n}ski},
  U.~{Katz}, D.~{Kerszberg}, B.~{Kh{\'e}lifi}, M.~{Kieffer}, J.~{King},
  S.~{Klepser}, D.~{Klochkov}, W.~{Klu{\'z}niak}, D.~{Kolitzus}, Nu. {Komin},
  K.~{Kosack}, S.~{Krakau}, M.~{Kraus}, F.~{Krayzel}, P.~P. {Kr{\"u}ger},
  H.~{Laffon}, G.~{Lamanna}, J.~{Lau}, J.~P. {Lees}, J.~{Lefaucheur},
  V.~{Lefranc}, A.~{Lemi{\`e}re}, M.~{Lemoine-Goumard}, J.~P. {Lenain},
  E.~{Leser}, T.~{Lohse}, M.~{Lorentz}, R.~{Liu}, R.~{L{\'o}pez-Coto},
  I.~{Lypova}, V.~{Marandon}, A.~{Marcowith}, C.~{Mariaud}, R.~{Marx},
  G.~{Maurin}, N.~{Maxted}, M.~{Mayer}, P.~J. {Meintjes}, M.~{Meyer}, A.~M.~W.
  {Mitchell}, R.~{Moderski}, M.~{Mohamed}, L.~{Mohrmann}, K.~{Mor{\r{a}}},
  E.~{Moulin}, T.~{Murach}, M.~{de Naurois}, F.~{Niederwanger}, J.~{Niemiec},
  L.~{Oakes}, P.~{O'Brien}, H.~{Odaka}, S.~{{\"O}ttl}, S.~{Ohm}, E.~{de O{\~n}a
  Wilhelmi}, M.~{Ostrowski}, I.~{Oya}, M.~{Padovani}, M.~{Panter}, R.~D.
  {Parsons}, M.~{Paz Arribas}, N.~W. {Pekeur}, G.~{Pelletier}, C.~{Perennes},
  P.~O. {Petrucci}, B.~{Peyaud}, S.~{Pita}, H.~{Poon}, D.~{Prokhorov},
  H.~{Prokoph}, G.~{P{\"u}hlhofer}, M.~{Punch}, A.~{Quirrenbach}, S.~{Raab},
  A.~{Reimer}, O.~{Reimer}, M.~{Renaud}, R.~{de los Reyes}, F.~{Rieger},
  C.~{Romoli}, S.~{Rosier-Lees}, G.~{Rowell}, B.~{Rudak}, C.~B. {Rulten},
  V.~{Sahakian}, D.~{Salek}, D.~A. {Sanchez}, A.~{Santangelo}, M.~{Sasaki},
  R.~{Schlickeiser}, F.~{Sch{\"u}ssler}, A.~{Schulz}, U.~{Schwanke},
  S.~{Schwemmer}, M.~{Settimo}, A.~S. {Seyffert}, N.~{Shafi}, I.~{Shilon},
  R.~{Simoni}, H.~{Sol}, F.~{Spanier}, G.~{Spengler}, F.~{Spies},
  {\L}.~{Stawarz}, R.~{Steenkamp}, C.~{Stegmann}, F.~{Stinzing}, K.~{Stycz},
  I.~{Sushch}, J.~P. {Tavernet}, T.~{Tavernier}, A.~M. {Taylor}, R.~{Terrier},
  L.~{Tibaldo}, D.~{Tiziani}, M.~{Tluczykont}, C.~{Trichard}, R.~{Tuffs},
  Y.~{Uchiyama}, K.~{Valerius}, D.~J. {van der Walt}, C.~{van Eldik}, B.~{van
  Soelen}, G.~{Vasileiadis}, J.~{Veh}, C.~{Venter}, A.~{Viana}, P.~{Vincent},
  J.~{Vink}, F.~{Voisin}, H.~J. {V{\"o}lk}, T.~{Vuillaume}, Z.~{Wadiasingh},
  S.~J. {Wagner}, P.~{Wagner}, R.~M. {Wagner}, R.~{White}, A.~{Wierzcholska},
  P.~{Willmann}, A.~{W{\"o}rnlein}, D.~{Wouters}, R.~{Yang}, V.~{Zabalza},
  D.~{Zaborov}, M.~{Zacharias}, A.~A. {Zdziarski}, A.~{Zech}, F.~{Zefi},
  A.~{Ziegler}, and N.~{{\.Z}ywucka}.
\newblock {The population of TeV pulsar wind nebulae in the H.E.S.S. Galactic
  Plane Survey}.
\newblock {\em \aap}, 612:A2, April 2018.

\bibitem{2009Natur.458..607A}
O.~{Adriani}, G.~C. {Barbarino}, G.~A. {Bazilevskaya}, R.~{Bellotti},
  M.~{Boezio}, E.~A. {Bogomolov}, L.~{Bonechi}, M.~{Bongi}, V.~{Bonvicini},
  S.~{Bottai}, A.~{Bruno}, F.~{Cafagna}, D.~{Campana}, P.~{Carlson},
  M.~{Casolino}, G.~{Castellini}, M.~P. {de Pascale}, G.~{de Rosa}, N.~{de
  Simone}, V.~{di Felice}, A.~M. {Galper}, L.~{Grishantseva}, P.~{Hofverberg},
  S.~V. {Koldashov}, S.~Y. {Krutkov}, A.~N. {Kvashnin}, A.~{Leonov},
  V.~{Malvezzi}, L.~{Marcelli}, W.~{Menn}, V.~V. {Mikhailov}, E.~{Mocchiutti},
  S.~{Orsi}, G.~{Osteria}, P.~{Papini}, M.~{Pearce}, P.~{Picozza}, M.~{Ricci},
  S.~B. {Ricciarini}, M.~{Simon}, R.~{Sparvoli}, P.~{Spillantini}, Y.~I.
  {Stozhkov}, A.~{Vacchi}, E.~{Vannuccini}, G.~{Vasilyev}, S.~A. {Voronov},
  Y.~T. {Yurkin}, G.~{Zampa}, N.~{Zampa}, and V.~G. {Zverev}.
\newblock {An anomalous positron abundance in cosmic rays with energies
  1.5-100GeV}.
\newblock {\em \nat}, 458(7238):607--609, April 2009.

\bibitem{PhysRevLett.110.141102}
M.~Aguilar, G.~Alberti, B.~Alpat, A.~Alvino, G.~Ambrosi, K.~Andeen,
  H.~Anderhub, L.~Arruda, P.~Azzarello, A.~Bachlechner, F.~Barao, B.~Baret,
  A.~Barrau, L.~Barrin, A.~Bartoloni, L.~Basara, A.~Basili, L.~Batalha,
  J.~Bates, R.~Battiston, J.~Bazo, R.~Becker, U.~Becker, M.~Behlmann,
  B.~Beischer, J.~Berdugo, P.~Berges, B.~Bertucci, G.~Bigongiari, A.~Biland,
  V.~Bindi, S.~Bizzaglia, G.~Boella, W.~de~Boer, K.~Bollweg, J.~Bolmont,
  B.~Borgia, S.~Borsini, M.~J. Boschini, G.~Boudoul, M.~Bourquin, P.~Brun,
  M.~Bu\'enerd, J.~Burger, W.~Burger, F.~Cadoux, X.~D. Cai, M.~Capell,
  D.~Casadei, J.~Casaus, V.~Cascioli, G.~Castellini, I.~Cernuda, F.~Cervelli,
  M.~J. Chae, Y.~H. Chang, A.~I. Chen, C.~R. Chen, H.~Chen, G.~M. Cheng, H.~S.
  Chen, L.~Cheng, N.~Chernoplyiokov, A.~Chikanian, E.~Choumilov, V.~Choutko,
  C.~H. Chung, C.~Clark, R.~Clavero, G.~Coignet, V.~Commichau, C.~Consolandi,
  A.~Contin, C.~Corti, M.~T. Costado~Dios, B.~Coste, D.~Crespo, Z.~Cui, M.~Dai,
  C.~Delgado, S.~Della~Torre, B.~Demirkoz, P.~Dennett, L.~Derome, S.~Di~Falco,
  X.~H. Diao, A.~Diago, L.~Djambazov, C.~D\'{\i}az, P.~von Doetinchem, W.~J.
  Du, J.~M. Dubois, R.~Duperay, M.~Duranti, D.~D'Urso, A.~Egorov, A.~Eline,
  F.~J. Eppling, T.~Eronen, J.~van Es, H.~Esser, A.~Falvard, E.~Fiandrini,
  A.~Fiasson, E.~Finch, P.~Fisher, K.~Flood, R.~Foglio, M.~Fohey, S.~Fopp,
  N.~Fouque, Y.~Galaktionov, M.~Gallilee, L.~Gallin-Martel, G.~Gallucci,
  B.~Garc\'{\i}a, J.~Garc\'{\i}a, R.~Garc\'{\i}a-L\'opez,
  L.~Garc\'{\i}a-Tabares, C.~Gargiulo, H.~Gast, I.~Gebauer, S.~Gentile,
  M.~Gervasi, W.~Gillard, F.~Giovacchini, L.~Girard, P.~Goglov, J.~Gong,
  C.~Goy-Henningsen, D.~Grandi, M.~Graziani, A.~Grechko, A.~Gross, I.~Guerri,
  C.~de~la Gu\'{\i}a, K.~H. Guo, M.~Habiby, S.~Haino, F.~Hauler, Z.~H. He,
  M.~Heil, J.~Heilig, R.~Hermel, H.~Hofer, Z.~C. Huang, W.~Hungerford,
  M.~Incagli, M.~Ionica, A.~Jacholkowska, W.~Y. Jang, H.~Jinchi, M.~Jongmanns,
  L.~Journet, L.~Jungermann, W.~Karpinski, G.~N. Kim, K.~S. Kim, Th. Kirn,
  R.~Kossakowski, A.~Koulemzine, O.~Kounina, A.~Kounine, V.~Koutsenko, M.~S.
  Krafczyk, E.~Laudi, G.~Laurenti, C.~Lauritzen, A.~Lebedev, M.~W. Lee, S.~C.
  Lee, C.~Leluc, H.~Le\'on~Vargas, V.~Lepareur, J.~Q. Li, Q.~Li, T.~X. Li,
  W.~Li, Z.~H. Li, P.~Lipari, C.~H. Lin, D.~Liu, H.~Liu, T.~Lomtadze, Y.~S. Lu,
  S.~Lucidi, K.~L\"ubelsmeyer, J.~Z. Luo, W.~Lustermann, S.~Lv, J.~Madsen,
  R.~Majka, A.~Malinin, C.~Ma\~n\'a, J.~Mar\'{\i}n, T.~Martin,
  G.~Mart\'{\i}nez, F.~Masciocchi, N.~Masi, D.~Maurin, A.~McInturff,
  P.~McIntyre, A.~Menchaca-Rocha, Q.~Meng, M.~Menichelli, I.~Mereu,
  M.~Millinger, D.~C. Mo, M.~Molina, P.~Mott, A.~Mujunen, S.~Natale, P.~Nemeth,
  J.~Q. Ni, N.~Nikonov, F.~Nozzoli, P.~Nunes, A.~Obermeier, S.~Oh, A.~Oliva,
  F.~Palmonari, C.~Palomares, M.~Paniccia, A.~Papi, W.~H. Park, M.~Pauluzzi,
  F.~Pauss, A.~Pauw, E.~Pedreschi, S.~Pensotti, R.~Pereira, E.~Perrin,
  G.~Pessina, G.~Pierschel, F.~Pilo, A.~Piluso, C.~Pizzolotto, V.~Plyaskin,
  J.~Pochon, M.~Pohl, V.~Poireau, S.~Porter, J.~Pouxe, A.~Putze, L.~Quadrani,
  X.~N. Qi, P.~G. Rancoita, D.~Rapin, Z.~L. Ren, J.~S. Ricol, E.~Riihonen,
  I.~Rodr\'{\i}guez, U.~Roeser, S.~Rosier-Lees, L.~Rossi, A.~Rozhkov, D.~Rozza,
  A.~Sabellek, R.~Sagdeev, J.~Sandweiss, B.~Santos, P.~Saouter, M.~Sarchioni,
  S.~Schael, D.~Schinzel, M.~Schmanau, G.~Schwering, A.~Schulz~von Dratzig,
  G.~Scolieri, E.~S. Seo, B.~S. Shan, J.~Y. Shi, Y.~M. Shi, T.~Siedenburg,
  R.~Siedling, D.~Son, F.~Spada, F.~Spinella, M.~Steuer, K.~Stiff, W.~Sun,
  W.~H. Sun, X.~H. Sun, M.~Tacconi, C.~P. Tang, X.~W. Tang, Z.~C. Tang, L.~Tao,
  J.~Tassan-Viol, Samuel C.~C. Ting, S.~M. Ting, C.~Titus, N.~Tomassetti,
  F.~Toral, J.~Torsti, J.~R. Tsai, J.~C. Tutt, J.~Ulbricht, T.~Urban,
  V.~Vagelli, E.~Valente, C.~Vannini, E.~Valtonen, M.~Vargas~Trevino,
  S.~Vaurynovich, M.~Vecchi, M.~Vergain, B.~Verlaat, C.~Vescovi, J.~P. Vialle,
  G.~Viertel, G.~Volpini, D.~Wang, N.~H. Wang, Q.~L. Wang, R.~S. Wang, X.~Wang,
  Z.~X. Wang, W.~Wallraff, Z.~L. Weng, M.~Willenbrock, M.~Wlochal, H.~Wu, K.~Y.
  Wu, Z.~S. Wu, W.~J. Xiao, S.~Xie, R.~Q. Xiong, G.~M. Xin, N.~S. Xu, W.~Xu,
  Q.~Yan, J.~Yang, M.~Yang, Q.~H. Ye, H.~Yi, Y.~J. Yu, Z.~Q. Yu, S.~Zeissler,
  J.~G. Zhang, Z.~Zhang, M.~M. Zhang, Z.~M. Zheng, H.~L. Zhuang, V.~Zhukov,
  A.~Zichichi, P.~Zuccon, and C.~Zurbach.
\newblock First result from the alpha magnetic spectrometer on the
  international space station: Precision measurement of the positron fraction
  in primary cosmic rays of 0.5--350 gev.
\newblock {\em Phys. Rev. Lett.}, 110:141102, Apr 2013.

\bibitem{2016A&A...586A..84S}
T.~{Siegert}, R.~{Diehl}, G.~{Khachatryan}, M.~G.~H. {Krause},
  F.~{Guglielmetti}, J.~{Greiner}, A.~W. {Strong}, and X.~{Zhang}.
\newblock {Gamma-ray spectroscopy of positron annihilation in the Milky Way}.
\newblock {\em \aap}, 586:A84, February 2016.

\bibitem{2015ICRC...34...14C}
M.~{Cirelli}.
\newblock {Dark matter phenomena}.
\newblock In {\em 34th International Cosmic Ray Conference (ICRC2015)},
  volume~34 of {\em International Cosmic Ray Conference}, page~14, July 2015.

\bibitem{PhysRevD.96.103013}
Dan Hooper, Ilias Cholis, Tim Linden, and Ke~Fang.
\newblock Hawc observations strongly favor pulsar interpretations of the
  cosmic-ray positron excess.
\newblock {\em Phys. Rev. D}, 96:103013, Nov 2017.

\bibitem{2009PhRvD..80f3005M}
Dmitry {Malyshev}, Ilias {Cholis}, and Joseph {Gelfand}.
\newblock {Pulsars versus dark matter interpretation of ATIC/PAMELA}.
\newblock {\em \prd}, 80(6):063005, September 2009.

\bibitem{2012A&A...548A..46H}
{H.~E.~S.~S. Collaboration}, A.~{Abramowski}, F.~{Acero}, F.~{Aharonian}, A.~G.
  {Akhperjanian}, G.~{Anton}, S.~{Balenderan}, A.~{Balzer}, A.~{Barnacka},
  Y.~{Becherini}, J.~{Becker}, K.~{Bernl{\"o}hr}, E.~{Birsin}, J.~{Biteau},
  A.~{Bochow}, C.~{Boisson}, J.~{Bolmont}, P.~{Bordas}, J.~{Brucker},
  F.~{Brun}, P.~{Brun}, T.~{Bulik}, I.~{B{\"u}sching}, S.~{Carrigan},
  S.~{Casanova}, M.~{Cerruti}, P.~M. {Chadwick}, A.~{Charbonnier}, R.~C.~G.
  {Chaves}, A.~{Cheesebrough}, G.~{Cologna}, J.~{Conrad}, C.~{Couturier},
  M.~{Dalton}, M.~K. {Daniel}, I.~D. {Davids}, B.~{Degrange}, C.~{Deil}, H.~J.
  {Dickinson}, A.~{Djannati-Ata{\"\i}}, W.~{Domainko}, L.~O'C. {Drury},
  G.~{Dubus}, K.~{Dutson}, J.~{Dyks}, M.~{Dyrda}, K.~{Egberts}, P.~{Eger},
  P.~{Espigat}, L.~{Fallon}, C.~{Farnier}, S.~{Fegan}, F.~{Feinstein}, M.~V.
  {Fernandes}, A.~{Fiasson}, G.~{Fontaine}, A.~{F{\"o}rster},
  M.~{F{\"u}{\ss}ling}, M.~{Gajdus}, Y.~A. {Gallant}, T.~{Garrigoux},
  H.~{Gast}, L.~{G{\'e}rard}, B.~{Giebels}, J.~F. {Glicenstein},
  B.~{Gl{\"u}ck}, D.~{G{\"o}ring}, M.~H. {Grondin}, S.~{H{\"a}ffner}, J.~D.
  {Hague}, J.~{Hahn}, D.~{Hampf}, J.~{Harris}, M.~{Hauser}, S.~{Heinz},
  G.~{Heinzelmann}, G.~{Henri}, G.~{Hermann}, A.~{Hillert}, J.~A. {Hinton},
  W.~{Hofmann}, P.~{Hofverberg}, M.~{Holler}, D.~{Horns}, A.~{Jacholkowska},
  C.~{Jahn}, M.~{Jamrozy}, I.~{Jung}, M.~A. {Kastendieck},
  K.~{Katarzy{\'n}ski}, U.~{Katz}, S.~{Kaufmann}, B.~{Kh{\'e}lifi},
  D.~{Klochkov}, W.~{Klu{\'z}niak}, T.~{Kneiske}, Nu. {Komin}, K.~{Kosack},
  R.~{Kossakowski}, F.~{Krayzel}, H.~{Laffon}, G.~{Lamanna}, J.~P. {Lenain},
  D.~{Lennarz}, T.~{Lohse}, A.~{Lopatin}, C.~C. {Lu}, V.~{Marandon},
  A.~{Marcowith}, J.~{Masbou}, G.~{Maurin}, N.~{Maxted}, M.~{Mayer}, T.~J.~L.
  {McComb}, M.~C. {Medina}, J.~{M{\'e}hault}, U.~{Menzler}, R.~{Moderski},
  M.~{Mohamed}, E.~{Moulin}, C.~L. {Naumann}, M.~{Naumann-Godo}, M.~{de
  Naurois}, D.~{Nedbal}, D.~{Nekrassov}, N.~{Nguyen}, B.~{Nicholas},
  J.~{Niemiec}, S.~J. {Nolan}, S.~{Ohm}, E.~{de O{\~n}a Wilhelmi}, B.~{Opitz},
  M.~{Ostrowski}, I.~{Oya}, M.~{Panter}, M.~{Paz Arribas}, N.~W. {Pekeur},
  G.~{Pelletier}, J.~{Perez}, P.~O. {Petrucci}, B.~{Peyaud}, S.~{Pita},
  G.~{P{\"u}hlhofer}, M.~{Punch}, A.~{Quirrenbach}, M.~{Raue}, A.~{Reimer},
  O.~{Reimer}, M.~{Renaud}, R.~{de los Reyes}, F.~{Rieger}, J.~{Ripken},
  L.~{Rob}, S.~{Rosier-Lees}, G.~{Rowell}, B.~{Rudak}, C.~B. {Rulten},
  V.~{Sahakian}, D.~A. {Sanchez}, A.~{Santangelo}, R.~{Schlickeiser},
  A.~{Schulz}, U.~{Schwanke}, S.~{Schwarzburg}, S.~{Schwemmer}, F.~{Sheidaei},
  J.~L. {Skilton}, H.~{Sol}, G.~{Spengler}, {\L}.~{Stawarz}, R.~{Steenkamp},
  C.~{Stegmann}, F.~{Stinzing}, K.~{Stycz}, I.~{Sushch}, A.~{Szostek}, J.~P.
  {Tavernet}, R.~{Terrier}, M.~{Tluczykont}, K.~{Valerius}, C.~{van Eldik},
  G.~{Vasileiadis}, C.~{Venter}, A.~{Viana}, P.~{Vincent}, H.~J. {V{\"o}lk},
  F.~{Volpe}, S.~{Vorobiov}, M.~{Vorster}, S.~J. {Wagner}, M.~{Ward},
  R.~{White}, A.~{Wierzcholska}, M.~{Zacharias}, A.~{Zajczyk}, A.~A.
  {Zdziarski}, A.~{Zech}, and H.~S. {Zechlin}.
\newblock {Identification of HESS J1303-631 as a pulsar wind nebula through
  {\ensuremath{\gamma}}-ray, X-ray, and radio observations}.
\newblock {\em \aap}, 548:A46, December 2012.

\bibitem{2015ApJ...803L..27B}
Slavko {Bogdanov} and Jules~P. {Halpern}.
\newblock {Identification of the High-energy Gamma-Ray Source 3FGL J1544.6-1125
  as a Transitional Millisecond Pulsar Binary in an Accreting State}.
\newblock {\em \apjl}, 803(2):L27, April 2015.

\bibitem{2019ApJ...875..107H}
Jeremy {Hare}, Igor {Volkov}, Oleg {Kargaltsev}, George {Younes}, and Blagoy
  {Rangelov}.
\newblock {XMM-Newton and Chandra Observations of the Unidentified Fermi-LAT
  Source 3FGL J1016.5-6034: A Young Pulsar with a Nebula?}
\newblock {\em \apj}, 875(2):107, April 2019.

\bibitem{2005A&A...442L..25A}
F.~A. {Aharonian}, A.~G. {Akhperjanian}, A.~R. {Bazer-Bachi}, M.~{Beilicke},
  W.~{Benbow}, D.~{Berge}, K.~{Bernl{\"o}hr}, C.~{Boisson}, O.~{Bolz},
  V.~{Borrel}, I.~{Braun}, F.~{Breitling}, A.~M. {Brown}, P.~M. {Chadwick},
  L.~M. {Chounet}, R.~{Cornils}, L.~{Costamante}, B.~{Degrange}, H.~J.
  {Dickinson}, A.~{Djannati-Ata{\"\i}}, L.~{O'C. Drury}, G.~{Dubus},
  D.~{Emmanoulopoulos}, P.~{Espigat}, F.~{Feinstein}, G.~{Fontaine},
  Y.~{Fuchs}, S.~{Funk}, Y.~A. {Gallant}, B.~{Giebels}, S.~{Gillessen}, J.~F.
  {Glicenstein}, P.~{Goret}, C.~{Hadjichristidis}, M.~{Hauser},
  G.~{Heinzelmann}, G.~{Henri}, G.~{Hermann}, J.~A. {Hinton}, W.~{Hofmann},
  M.~{Holleran}, D.~{Horns}, A.~{Jacholkowska}, O.~C. {de Jager},
  B.~{Kh{\'e}lifi}, Nu. {Komin}, A.~{Konopelko}, I.~J. {Latham}, R.~{Le
  Gallou}, A.~{Lemi{\`e}re}, M.~{Lemoine-Goumard}, N.~{Leroy}, T.~{Lohse},
  J.~M. {Martin}, O.~{Martineau-Huynh}, A.~{Marcowith}, C.~{Masterson},
  T.~J.~L. {McComb}, M.~{de Naurois}, S.~J. {Nolan}, A.~{Noutsos}, K.~J.
  {Orford}, J.~L. {Osborne}, M.~{Ouchrif}, M.~{Panter}, G.~{Pelletier},
  S.~{Pita}, G.~{P{\"u}hlhofer}, M.~{Punch}, B.~C. {Raubenheimer}, M.~{Raue},
  J.~{Raux}, S.~M. {Rayner}, A.~{Reimer}, O.~{Reimer}, J.~{Ripken}, L.~{Rob},
  L.~{Rolland}, G.~{Rowell}, V.~{Sahakian}, L.~{Saug{\'e}}, S.~{Schlenker},
  R.~{Schlickeiser}, C.~{Schuster}, U.~{Schwanke}, M.~{Siewert}, H.~{Sol},
  D.~{Spangler}, R.~{Steenkamp}, C.~{Stegmann}, J.~P. {Tavernet}, R.~{Terrier},
  C.~G. {Th{\'e}oret}, M.~{Tluczykont}, G.~{Vasileiadis}, C.~{Venter},
  P.~{Vincent}, H.~J. {V{\"o}lk}, and S.~J. {Wagner}.
\newblock {A possible association of the new VHE {\ensuremath{\gamma}}-ray
  source HESS J1825-137 with the pulsar wind nebula G18.0-0.7}.
\newblock {\em \aap}, 442(3):L25--L29, November 2005.

\bibitem{2013arXiv1305.2552K}
Oleg {Kargaltsev}, Blagoy {Rangelov}, and George~G. {Pavlov}.
\newblock {Gamma-ray and X-ray Properties of Pulsar Wind Nebulae and
  Unidentified Galactic TeV Sources}.
\newblock {\em arXiv e-prints}, page arXiv:1305.2552, May 2013.

\bibitem{2014A&A...562A..40H}
{H.~E.~S.~S. Collaboration}, A.~{Abramowski}, F.~{Aharonian}, F.~{Ait
  Benkhali}, A.~G. {Akhperjanian}, E.~{Ang{\"u}ner}, G.~{Anton},
  S.~{Balenderan}, A.~{Balzer}, A.~{Barnacka}, Y.~{Becherini}, J.~{Becker
  Tjus}, K.~{Bernl{\"o}hr}, E.~{Birsin}, E.~{Bissaldi}, J.~{Biteau},
  M.~{B{\"o}ttcher}, C.~{Boisson}, J.~{Bolmont}, P.~{Bordas}, J.~{Brucker},
  F.~{Brun}, P.~{Brun}, T.~{Bulik}, S.~{Carrigan}, S.~{Casanova}, M.~{Cerruti},
  P.~M. {Chadwick}, R.~{Chalme-Calvet}, R.~C.~G. {Chaves}, A.~{Cheesebrough},
  M.~{Chr{\'e}tien}, S.~{Colafrancesco}, G.~{Cologna}, J.~{Conrad},
  C.~{Couturier}, Y.~{Cui}, M.~{Dalton}, M.~K. {Daniel}, I.~D. {Davids},
  B.~{Degrange}, C.~{Deil}, P.~{deWilt}, H.~J. {Dickinson},
  A.~{Djannati-Ata{\"\i}}, W.~{Domainko}, L.~O'C. {Drury}, G.~{Dubus},
  K.~{Dutson}, J.~{Dyks}, M.~{Dyrda}, T.~{Edwards}, K.~{Egberts}, P.~{Eger},
  P.~{Espigat}, C.~{Farnier}, S.~{Fegan}, F.~{Feinstein}, M.~V. {Fernandes},
  D.~{Fernandez}, A.~{Fiasson}, G.~{Fontaine}, A.~{F{\"o}rster},
  M.~{F{\"u}{\ss}ling}, M.~{Gajdus}, Y.~A. {Gallant}, T.~{Garrigoux},
  G.~{Giavitto}, B.~{Giebels}, J.~F. {Glicenstein}, M.~H. {Grondin},
  M.~{Grudzi{\'n}ska}, S.~{H{\"a}ffner}, J.~{Hahn}, J.~{Harris},
  G.~{Heinzelmann}, G.~{Henri}, G.~{Hermann}, O.~{Hervet}, A.~{Hillert}, J.~A.
  {Hinton}, W.~{Hofmann}, P.~{Hofverberg}, M.~{Holler}, D.~{Horns},
  A.~{Jacholkowska}, C.~{Jahn}, M.~{Jamrozy}, M.~{Janiak}, F.~{Jankowsky},
  I.~{Jung}, M.~A. {Kastendieck}, K.~{Katarzy{\'n}ski}, U.~{Katz},
  S.~{Kaufmann}, B.~{Kh{\'e}lifi}, M.~{Kieffer}, S.~{Klepser}, D.~{Klochkov},
  W.~{Klu{\'z}niak}, T.~{Kneiske}, D.~{Kolitzus}, Nu. {Komin}, K.~{Kosack},
  S.~{Krakau}, F.~{Krayzel}, P.~P. {Kr{\"u}ger}, H.~{Laffon}, G.~{Lamanna},
  J.~{Lefaucheur}, A.~{Lemi{\`e}re}, M.~{Lemoine-Goumard}, J.~P. {Lenain},
  D.~{Lennarz}, T.~{Lohse}, A.~{Lopatin}, C.~C. {Lu}, V.~{Marandon},
  A.~{Marcowith}, R.~{Marx}, G.~{Maurin}, N.~{Maxted}, M.~{Mayer}, T.~J.~L.
  {McComb}, J.~{M{\'e}hault}, P.~J. {Meintjes}, U.~{Menzler}, M.~{Meyer},
  R.~{Moderski}, M.~{Mohamed}, E.~{Moulin}, T.~{Murach}, C.~L. {Naumann},
  M.~{de Naurois}, J.~{Niemiec}, S.~J. {Nolan}, L.~{Oakes}, S.~{Ohm}, E.~{de
  O{\~n}a Wilhelmi}, B.~{Opitz}, M.~{Ostrowski}, I.~{Oya}, M.~{Panter}, R.~D.
  {Parsons}, M.~{Paz Arribas}, N.~W. {Pekeur}, G.~{Pelletier}, J.~{Perez},
  P.~O. {Petrucci}, B.~{Peyaud}, S.~{Pita}, H.~{Poon}, G.~{P{\"u}hlhofer},
  M.~{Punch}, A.~{Quirrenbach}, S.~{Raab}, M.~{Raue}, A.~{Reimer}, O.~{Reimer},
  M.~{Renaud}, R.~{de los Reyes}, F.~{Rieger}, L.~{Rob}, C.~{Romoli},
  S.~{Rosier-Lees}, G.~{Rowell}, B.~{Rudak}, C.~B. {Rulten}, V.~{Sahakian},
  D.~A. {Sanchez}, A.~{Santangelo}, R.~{Schlickeiser}, F.~{Sch{\"u}ssler},
  A.~{Schulz}, U.~{Schwanke}, S.~{Schwarzburg}, S.~{Schwemmer}, H.~{Sol},
  G.~{Spengler}, F.~{Spies}, {\L}.~{Stawarz}, R.~{Steenkamp}, C.~{Stegmann},
  F.~{Stinzing}, K.~{Stycz}, I.~{Sushch}, A.~{Szostek}, J.~P. {Tavernet},
  T.~{Tavernier}, A.~M. {Taylor}, R.~{Terrier}, M.~{Tluczykont}, C.~{Trichard},
  K.~{Valerius}, C.~{van Eldik}, B.~{van Soelen}, G.~{Vasileiadis},
  C.~{Venter}, A.~{Viana}, P.~{Vincent}, H.~J. {V{\"o}lk}, F.~{Volpe},
  M.~{Vorster}, T.~{Vuillaume}, S.~J. {Wagner}, P.~{Wagner}, M.~{Ward},
  M.~{Weidinger}, Q.~{Weitzel}, R.~{White}, A.~{Wierzcholska}, P.~{Willmann},
  A.~{W{\"o}rnlein}, D.~{Wouters}, V.~{Zabalza}, M.~{Zacharias}, A.~{Zajczyk},
  A.~A. {Zdziarski}, A.~{Zech}, and H.~S. {Zechlin}.
\newblock {HESS J1818-154, a new composite supernova remnant discovered in TeV
  gamma rays and X-rays}.
\newblock {\em \aap}, 562:A40, February 2014.

\bibitem{2019ApJ...884...93C}
R.~H.~D. {Corbet}, L.~{Chomiuk}, M.~J. {Coe}, J.~B. {Coley}, G.~{Dubus}, P.~G.
  {Edwards}, P.~{Martin}, V.~A. {McBride}, J.~{Stevens}, J.~{Strader}, and
  L.~J. {Townsend}.
\newblock {Discovery of the Galactic High-mass Gamma-Ray Binary 4FGL
  J1405.1-6119}.
\newblock {\em \apj}, 884(1):93, October 2019.

\bibitem{2006ApJS..163..344K}
D.~L. {Kaplan}, B.~M. {Gaensler}, S.~R. {Kulkarni}, and P.~O. {Slane}.
\newblock {An X-Ray Search for Compact Central Sources in Supernova Remnants.
  II. Six Large-Diameter SNRs}.
\newblock {\em \apjs}, 163(2):344--371, April 2006.

\bibitem{2012ApJ...756...27L}
Dacheng {Lin}, Natalie~A. {Webb}, and Didier {Barret}.
\newblock {Classification of X-Ray Sources in the XMM-Newton Serendipitous
  Source Catalog}.
\newblock {\em \apj}, 756(1):27, September 2012.

\bibitem{2012arXiv1201.0490P}
Fabian {Pedregosa}, Ga{\"e}l {Varoquaux}, Alexandre {Gramfort}, Vincent
  {Michel}, Bertrand {Thirion}, Olivier {Grisel}, Mathieu {Blondel}, Andreas
  {M{\"u}ller}, Joel {Nothman}, Gilles {Louppe}, Peter {Prettenhofer}, Ron
  {Weiss}, Vincent {Dubourg}, Jake {Vanderplas}, Alexandre {Passos}, David
  {Cournapeau}, Matthieu {Brucher}, Matthieu {Perrot}, and {\'E}douard
  {Duchesnay}.
\newblock {Scikit-learn: Machine Learning in Python}.
\newblock {\em arXiv e-prints}, page arXiv:1201.0490, January 2012.

\bibitem{2020AAS...23515405E}
I.~N. {Evans}, F.~A. {Primini}, J.~B. {Miller}, J.~D. {Evans}, C.~E. {Allen},
  C.~S. {Anderson}, G.~{Becker}, J.~A. {Budynkiewicz}, D.~{Burke}, J.~C.
  {Chen}, F.~{Civano}, R.~{D'Abrusco}, S.~M. {Doe}, G.~{Fabbiano}, J.~{Martinez
  Galarza}, II~{Gibbs}, D.~G., K.~J. {Glotfelty}, D.~E. {Graessle}, Jr.
  {Grier}, J.~D., R.~M. {Hain}, D.~M. {Hall}, P.~N. {Harbo}, J.~C. {Houck},
  J.~L. {Lauer}, O.~{Laurino}, N.~P. {Lee}, M.~L. {McCollough}, J.~C.
  {McDowell}, W.~{McLaughlin}, D.~L. {Morgan}, A.~E. {Mossman}, D.~T. {Nguyen},
  J.~S. {Nichols}, M.~A. {Nowak}, C.~{Paxson}, M.~{Perdikeas}, D.~A. {Plummer},
  A.~H. {Rots}, A.~L. {Siemiginowska}, B.~A. {Sundheim}, S.~{Thong}, M.~S.
  {Tibbetts}, D.~W. {Van Stone}, S.~L. {Winkelman}, and P.~{Zografou}.
\newblock {The Chandra Source Catalog {\textemdash} A Billion X-ray Photons}.
\newblock In {\em American Astronomical Society Meeting Abstracts \#235},
  volume 235 of {\em American Astronomical Society Meeting Abstracts}, page
  154.05, January 2020.

\bibitem{2020A&A...641A.136W}
N.~A. {Webb}, M.~{Coriat}, I.~{Traulsen}, J.~{Ballet}, C.~{Motch}, F.~J.
  {Carrera}, F.~{Koliopanos}, J.~{Authier}, I.~{de la Calle}, M.~T. {Ceballos},
  E.~{Colomo}, D.~{Chuard}, M.~{Freyberg}, T.~{Garcia}, M.~{Kolehmainen},
  G.~{Lamer}, D.~{Lin}, P.~{Maggi}, L.~{Michel}, C.~G. {Page}, M.~J. {Page},
  J.~V. {Perea-Calderon}, F.~X. {Pineau}, P.~{Rodriguez}, S.~R. {Rosen},
  M.~{Santos Lleo}, R.~D. {Saxton}, A.~{Schwope}, L.~{Tom{\'a}s}, M.~G.
  {Watson}, and A.~{Zakardjian}.
\newblock {The XMM-Newton serendipitous survey. IX. The fourth XMM-Newton
  serendipitous source catalogue}.
\newblock {\em \aap}, 641:A136, September 2020.

\bibitem{2021RNAAS...5..102Y}
Hui {Yang}, Jeremy {Hare}, Igor {Volkov}, and Oleg {Kargaltsev}.
\newblock {Visualizing Multiwavelength Properties of Classified X-Ray Sources
  from Chandra Source Catalog}.
\newblock {\em Research Notes of the American Astronomical Society}, 5(5):102,
  May 2021.

\bibitem{2016ApJ...821...54S}
E.~{Sonbas}, B.~{Rangelov}, O.~{Kargaltsev}, K.~S. {Dhuga}, J.~{Hare}, and
  I.~{Volkov}.
\newblock {X-Ray Sources in the Dwarf Spheroidal Galaxy Draco}.
\newblock {\em \apj}, 821(1):54, April 2016.

\bibitem{2017ApJ...839...59P}
Thomas~G. {Pannuti}, Jeonghee {Rho}, Oleg {Kargaltsev}, Blagoy {Rangelov},
  Alekzander~R. {Kosakowski}, P.~Frank {Winkler}, Jonathan~W. {Keohane}, Jeremy
  {Hare}, and Sonny {Ernst}.
\newblock {CTIO, ROSAT HRI, and Chandra ACIS Observations of the Archetypical
  Mixed-morphology Supernova Remnant W28 (G6.4-0.1)}.
\newblock {\em \apj}, 839(1):59, April 2017.

\bibitem{2016ApJ...816...52H}
Jeremy {Hare}, Blagoy {Rangelov}, Eda {Sonbas}, Oleg {Kargaltsev}, and Igor
  {Volkov}.
\newblock {Multi-wavelength Study of HESS J1741-302}.
\newblock {\em \apj}, 816(2):52, January 2016.

\bibitem{2017ApJ...841...81H}
Jeremy {Hare}, Oleg {Kargaltsev}, George~G. {Pavlov}, Blagoy {Rangelov}, and
  Igor {Volkov}.
\newblock {Chandra Observations of the Field Containing HESS J1616-508}.
\newblock {\em \apj}, 841(2):81, June 2017.

\bibitem{2020ApJ...901..157K}
Noel {Klingler}, Hui {Yang}, Jeremy {Hare}, Oleg {Kargaltsev}, George~G.
  {Pavlov}, and Bettina {Posselt}.
\newblock {Chandra Monitoring of the J1809-1917 Pulsar Wind Nebula and Its
  Field}.
\newblock {\em \apj}, 901(2):157, October 2020.

\bibitem{2021A&A...647A...1P}
P.~{Predehl}, R.~{Andritschke}, V.~{Arefiev}, V.~{Babyshkin}, O.~{Batanov},
  W.~{Becker}, H.~{B{\"o}hringer}, A.~{Bogomolov}, T.~{Boller}, K.~{Borm},
  W.~{Bornemann}, H.~{Br{\"a}uninger}, M.~{Br{\"u}ggen}, H.~{Brunner},
  M.~{Brusa}, E.~{Bulbul}, M.~{Buntov}, V.~{Burwitz}, W.~{Burkert}, N.~{Clerc},
  E.~{Churazov}, D.~{Coutinho}, T.~{Dauser}, K.~{Dennerl}, V.~{Doroshenko},
  J.~{Eder}, V.~{Emberger}, T.~{Eraerds}, A.~{Finoguenov}, M.~{Freyberg},
  P.~{Friedrich}, S.~{Friedrich}, M.~{F{\"u}rmetz}, A.~{Georgakakis},
  M.~{Gilfanov}, S.~{Granato}, C.~{Grossberger}, A.~{Gueguen}, P.~{Gureev},
  F.~{Haberl}, O.~{H{\"a}lker}, G.~{Hartner}, G.~{Hasinger}, H.~{Huber},
  L.~{Ji}, A.~v. {Kienlin}, W.~{Kink}, F.~{Korotkov}, I.~{Kreykenbohm},
  G.~{Lamer}, I.~{Lomakin}, I.~{Lapshov}, T.~{Liu}, C.~{Maitra},
  N.~{Meidinger}, B.~{Menz}, A.~{Merloni}, T.~{Mernik}, B.~{Mican}, J.~{Mohr},
  S.~{M{\"u}ller}, K.~{Nandra}, V.~{Nazarov}, F.~{Pacaud}, M.~{Pavlinsky},
  E.~{Perinati}, E.~{Pfeffermann}, D.~{Pietschner}, M.~E. {Ramos-Ceja},
  A.~{Rau}, J.~{Reiffers}, T.~H. {Reiprich}, J.~{Robrade}, M.~{Salvato},
  J.~{Sanders}, A.~{Santangelo}, M.~{Sasaki}, H.~{Scheuerle}, C.~{Schmid},
  J.~{Schmitt}, A.~{Schwope}, A.~{Shirshakov}, M.~{Steinmetz}, I.~{Stewart},
  L.~{Str{\"u}der}, R.~{Sunyaev}, C.~{Tenzer}, L.~{Tiedemann},
  J.~{Tr{\"u}mper}, V.~{Voron}, P.~{Weber}, J.~{Wilms}, and V.~{Yaroshenko}.
\newblock {The eROSITA X-ray telescope on SRG}.
\newblock {\em \aap}, 647:A1, March 2021.

\bibitem{2020PASP..132c5001L}
M.~{Lacy}, S.~A. {Baum}, C.~J. {Chandler}, S.~{Chatterjee}, T.~E. {Clarke},
  S.~{Deustua}, J.~{English}, J.~{Farnes}, B.~M. {Gaensler}, N.~{Gugliucci},
  G.~{Hallinan}, B.~R. {Kent}, A.~{Kimball}, C.~J. {Law}, T.~J.~W. {Lazio},
  J.~{Marvil}, S.~A. {Mao}, D.~{Medlin}, K.~{Mooley}, E.~J. {Murphy},
  S.~{Myers}, R.~{Osten}, G.~T. {Richards}, E.~{Rosolowsky}, L.~{Rudnick},
  F.~{Schinzel}, G.~R. {Sivakoff}, L.~O. {Sjouwerman}, R.~{Taylor}, R.~L.
  {White}, J.~{Wrobel}, H.~{Andernach}, A.~J. {Beasley}, E.~{Berger},
  S.~{Bhatnager}, M.~{Birkinshaw}, G.~C. {Bower}, W.~N. {Brandt}, S.~{Brown},
  S.~{Burke-Spolaor}, B.~J. {Butler}, J.~{Comerford}, P.~B. {Demorest},
  H.~{Fu}, S.~{Giacintucci}, K.~{Golap}, T.~{G{\"u}th}, C.~A. {Hales},
  R.~{Hiriart}, J.~{Hodge}, A.~{Horesh}, {\v{Z}}.~{Ivezi{\'c}}, M.~J. {Jarvis},
  A.~{Kamble}, N.~{Kassim}, X.~{Liu}, L.~{Loinard}, D.~K. {Lyons},
  J.~{Masters}, M.~{Mezcua}, G.~A. {Moellenbrock}, T.~{Mroczkowski},
  K.~{Nyland}, C.~P. {O'Dea}, S.~P. {O'Sullivan}, W.~M. {Peters}, K.~{Radford},
  U.~{Rao}, J.~{Robnett}, J.~{Salcido}, Y.~{Shen}, A.~{Sobotka}, S.~{Witz},
  M.~{Vaccari}, R.~J. {van Weeren}, A.~{Vargas}, P.~K.~G. {Williams}, and
  I.~{Yoon}.
\newblock {The Karl G. Jansky Very Large Array Sky Survey (VLASS). Science Case
  and Survey Design}.
\newblock {\em \pasp}, 132(1009):035001, March 2020.

\bibitem{2019PASP..131a8002B}
Eric~C. {Bellm}, Shrinivas~R. {Kulkarni}, Matthew~J. {Graham}, Richard
  {Dekany}, Roger~M. {Smith}, Reed {Riddle}, Frank~J. {Masci}, George {Helou},
  Thomas~A. {Prince}, Scott~M. {Adams}, C.~{Barbarino}, Tom {Barlow}, James
  {Bauer}, Ron {Beck}, Justin {Belicki}, Rahul {Biswas}, Nadejda
  {Blagorodnova}, Dennis {Bodewits}, Bryce {Bolin}, Valery {Brinnel}, Tim
  {Brooke}, Brian {Bue}, Mattia {Bulla}, Rick {Burruss}, S.~Bradley {Cenko},
  Chan-Kao {Chang}, Andrew {Connolly}, Michael {Coughlin}, John {Cromer},
  Virginia {Cunningham}, Kishalay {De}, Alex {Delacroix}, Vandana {Desai},
  Dmitry~A. {Duev}, Gwendolyn {Eadie}, Tony~L. {Farnham}, Michael {Feeney},
  Ulrich {Feindt}, David {Flynn}, Anna {Franckowiak}, S.~{Frederick},
  C.~{Fremling}, Avishay {Gal-Yam}, Suvi {Gezari}, Matteo {Giomi}, Daniel~A.
  {Goldstein}, V.~Zach {Golkhou}, Ariel {Goobar}, Steven {Groom}, Eugean
  {Hacopians}, David {Hale}, John {Henning}, Anna Y.~Q. {Ho}, David {Hover},
  Justin {Howell}, Tiara {Hung}, Daniela {Huppenkothen}, David {Imel},
  Wing-Huen {Ip}, {\v{Z}}eljko {Ivezi{\'c}}, Edward {Jackson}, Lynne {Jones},
  Mario {Juric}, Mansi~M. {Kasliwal}, S.~{Kaspi}, Stephen {Kaye}, Michael S.~P.
  {Kelley}, Marek {Kowalski}, Emily {Kramer}, Thomas {Kupfer}, Walter {Landry},
  Russ~R. {Laher}, Chien-De {Lee}, Hsing~Wen {Lin}, Zhong-Yi {Lin}, Ragnhild
  {Lunnan}, Matteo {Giomi}, Ashish {Mahabal}, Peter {Mao}, Adam~A. {Miller},
  Serge {Monkewitz}, Patrick {Murphy}, Chow-Choong {Ngeow}, Jakob {Nordin},
  Peter {Nugent}, Eran {Ofek}, Maria~T. {Patterson}, Bryan {Penprase}, Michael
  {Porter}, Ludwig {Rauch}, Umaa {Rebbapragada}, Dan {Reiley}, Mickael
  {Rigault}, Hector {Rodriguez}, Jan {van Roestel}, Ben {Rusholme}, Jakob {van
  Santen}, S.~{Schulze}, David~L. {Shupe}, Leo~P. {Singer}, Maayane~T.
  {Soumagnac}, Robert {Stein}, Jason {Surace}, Jesper {Sollerman}, Paula
  {Szkody}, F.~{Taddia}, Scott {Terek}, Angela {Van Sistine}, Sjoert {van
  Velzen}, W.~Thomas {Vestrand}, Richard {Walters}, Charlotte {Ward}, Quan-Zhi
  {Ye}, Po-Chieh {Yu}, Lin {Yan}, and Jeffry {Zolkower}.
\newblock {The Zwicky Transient Facility: System Overview, Performance, and
  First Results}.
\newblock {\em \pasp}, 131(995):018002, January 2019.

\bibitem{2019ApJ...873..111I}
{\v{Z}}eljko {Ivezi{\'c}}, Steven~M. {Kahn}, J.~Anthony {Tyson}, Bob {Abel},
  Emily {Acosta}, Robyn {Allsman}, David {Alonso}, Yusra {AlSayyad}, Scott~F.
  {Anderson}, John {Andrew}, James Roger~P. {Angel}, George~Z. {Angeli}, Reza
  {Ansari}, Pierre {Antilogus}, Constanza {Araujo}, Robert {Armstrong}, Kirk~T.
  {Arndt}, Pierre {Astier}, {\'E}ric {Aubourg}, Nicole {Auza}, Tim~S.
  {Axelrod}, Deborah~J. {Bard}, Jeff~D. {Barr}, Aurelian {Barrau}, James~G.
  {Bartlett}, Amanda~E. {Bauer}, Brian~J. {Bauman}, Sylvain {Baumont}, Ellen
  {Bechtol}, Keith {Bechtol}, Andrew~C. {Becker}, Jacek {Becla}, Cristina
  {Beldica}, Steve {Bellavia}, Federica~B. {Bianco}, Rahul {Biswas}, Guillaume
  {Blanc}, Jonathan {Blazek}, Roger~D. {Blandford}, Josh~S. {Bloom}, Joanne
  {Bogart}, Tim~W. {Bond}, Michael~T. {Booth}, Anders~W. {Borgland}, Kirk
  {Borne}, James~F. {Bosch}, Dominique {Boutigny}, Craig~A. {Brackett}, Andrew
  {Bradshaw}, William~Nielsen {Brandt}, Michael~E. {Brown}, James~S. {Bullock},
  Patricia {Burchat}, David~L. {Burke}, Gianpietro {Cagnoli}, Daniel
  {Calabrese}, Shawn {Callahan}, Alice~L. {Callen}, Jeffrey~L. {Carlin},
  Erin~L. {Carlson}, Srinivasan {Chandrasekharan}, Glenaver {Charles-Emerson},
  Steve {Chesley}, Elliott~C. {Cheu}, Hsin-Fang {Chiang}, James {Chiang}, Carol
  {Chirino}, Derek {Chow}, David~R. {Ciardi}, Charles~F. {Claver}, Johann
  {Cohen-Tanugi}, Joseph~J. {Cockrum}, Rebecca {Coles}, Andrew~J. {Connolly},
  Kem~H. {Cook}, Asantha {Cooray}, Kevin~R. {Covey}, Chris {Cribbs}, Wei {Cui},
  Roc {Cutri}, Philip~N. {Daly}, Scott~F. {Daniel}, Felipe {Daruich}, Guillaume
  {Daubard}, Greg {Daues}, William {Dawson}, Francisco {Delgado}, Alfred
  {Dellapenna}, Robert {de Peyster}, Miguel {de Val-Borro}, Seth~W. {Digel},
  Peter {Doherty}, Richard {Dubois}, Gregory~P. {Dubois-Felsmann}, Josef
  {Durech}, Frossie {Economou}, Tim {Eifler}, Michael {Eracleous}, Benjamin~L.
  {Emmons}, Angelo {Fausti Neto}, Henry {Ferguson}, Enrique {Figueroa}, Merlin
  {Fisher-Levine}, Warren {Focke}, Michael~D. {Foss}, James {Frank}, Michael~D.
  {Freemon}, Emmanuel {Gangler}, Eric {Gawiser}, John~C. {Geary}, Perry {Gee},
  Marla {Geha}, Charles J.~B. {Gessner}, Robert~R. {Gibson}, D.~Kirk {Gilmore},
  Thomas {Glanzman}, William {Glick}, Tatiana {Goldina}, Daniel~A. {Goldstein},
  Iain {Goodenow}, Melissa~L. {Graham}, William~J. {Gressler}, Philippe {Gris},
  Leanne~P. {Guy}, Augustin {Guyonnet}, Gunther {Haller}, Ron {Harris},
  Patrick~A. {Hascall}, Justine {Haupt}, Fabio {Hernandez}, Sven {Herrmann},
  Edward {Hileman}, Joshua {Hoblitt}, John~A. {Hodgson}, Craig {Hogan},
  James~D. {Howard}, Dajun {Huang}, Michael~E. {Huffer}, Patrick {Ingraham},
  Walter~R. {Innes}, Suzanne~H. {Jacoby}, Bhuvnesh {Jain}, Fabrice {Jammes},
  M.~James {Jee}, Tim {Jenness}, Garrett {Jernigan}, Darko {Jevremovi{\'c}},
  Kenneth {Johns}, Anthony~S. {Johnson}, Margaret W.~G. {Johnson}, R.~Lynne
  {Jones}, Claire {Juramy-Gilles}, Mario {Juri{\'c}}, Jason~S. {Kalirai},
  Nitya~J. {Kallivayalil}, Bryce {Kalmbach}, Jeffrey~P. {Kantor}, Pierre
  {Karst}, Mansi~M. {Kasliwal}, Heather {Kelly}, Richard {Kessler}, Veronica
  {Kinnison}, David {Kirkby}, Lloyd {Knox}, Ivan~V. {Kotov}, Victor~L.
  {Krabbendam}, K.~Simon {Krughoff}, Petr {Kub{\'a}nek}, John {Kuczewski}, Shri
  {Kulkarni}, John {Ku}, Nadine~R. {Kurita}, Craig~S. {Lage}, Ron {Lambert},
  Travis {Lange}, J.~Brian {Langton}, Laurent {Le Guillou}, Deborah {Levine},
  Ming {Liang}, Kian-Tat {Lim}, Chris~J. {Lintott}, Kevin~E. {Long}, Margaux
  {Lopez}, Paul~J. {Lotz}, Robert~H. {Lupton}, Nate~B. {Lust}, Lauren~A.
  {MacArthur}, Ashish {Mahabal}, Rachel {Mandelbaum}, Thomas~W. {Markiewicz},
  Darren~S. {Marsh}, Philip~J. {Marshall}, Stuart {Marshall}, Morgan {May},
  Robert {McKercher}, Michelle {McQueen}, Joshua {Meyers}, Myriam {Migliore},
  Michelle {Miller}, David~J. {Mills}, Connor {Miraval}, Joachim {Moeyens},
  Fred~E. {Moolekamp}, David~G. {Monet}, Marc {Moniez}, Serge {Monkewitz},
  Christopher {Montgomery}, Christopher~B. {Morrison}, Fritz {Mueller}, Gary~P.
  {Muller}, Freddy {Mu{\~n}oz Arancibia}, Douglas~R. {Neill}, Scott~P.
  {Newbry}, Jean-Yves {Nief}, Andrei {Nomerotski}, Martin {Nordby}, Paul
  {O'Connor}, John {Oliver}, Scot~S. {Olivier}, Knut {Olsen}, William
  {O'Mullane}, Sandra {Ortiz}, Shawn {Osier}, Russell~E. {Owen}, Reynald
  {Pain}, Paul~E. {Palecek}, John~K. {Parejko}, James~B. {Parsons}, Nathan~M.
  {Pease}, J.~Matt {Peterson}, John~R. {Peterson}, Donald~L. {Petravick}, M.~E.
  {Libby Petrick}, Cathy~E. {Petry}, Francesco {Pierfederici}, Stephen
  {Pietrowicz}, Rob {Pike}, Philip~A. {Pinto}, Raymond {Plante}, Stephen
  {Plate}, Joel~P. {Plutchak}, Paul~A. {Price}, Michael {Prouza}, Veljko
  {Radeka}, Jayadev {Rajagopal}, Andrew~P. {Rasmussen}, Nicolas {Regnault},
  Kevin~A. {Reil}, David~J. {Reiss}, Michael~A. {Reuter}, Stephen~T. {Ridgway},
  Vincent~J. {Riot}, Steve {Ritz}, Sean {Robinson}, William {Roby}, Aaron
  {Roodman}, Wayne {Rosing}, Cecille {Roucelle}, Matthew~R. {Rumore}, Stefano
  {Russo}, Abhijit {Saha}, Benoit {Sassolas}, Terry~L. {Schalk}, Pim
  {Schellart}, Rafe~H. {Schindler}, Samuel {Schmidt}, Donald~P. {Schneider},
  Michael~D. {Schneider}, William {Schoening}, German {Schumacher}, Megan~E.
  {Schwamb}, Jacques {Sebag}, Brian {Selvy}, Glenn~H. {Sembroski}, Lynn~G.
  {Seppala}, Andrew {Serio}, Eduardo {Serrano}, Richard~A. {Shaw}, Ian
  {Shipsey}, Jonathan {Sick}, Nicole {Silvestri}, Colin~T. {Slater}, J.~Allyn
  {Smith}, R.~Chris {Smith}, Shahram {Sobhani}, Christine {Soldahl}, Lisa
  {Storrie-Lombardi}, Edward {Stover}, Michael~A. {Strauss}, Rachel~A.
  {Street}, Christopher~W. {Stubbs}, Ian~S. {Sullivan}, Donald {Sweeney},
  John~D. {Swinbank}, Alexander {Szalay}, Peter {Takacs}, Stephen~A. {Tether},
  Jon~J. {Thaler}, John~Gregg {Thayer}, Sandrine {Thomas}, Adam~J. {Thornton},
  Vaikunth {Thukral}, Jeffrey {Tice}, David~E. {Trilling}, Max {Turri}, Richard
  {Van Berg}, Daniel {Vanden Berk}, Kurt {Vetter}, Francoise {Virieux},
  Tomislav {Vucina}, William {Wahl}, Lucianne {Walkowicz}, Brian {Walsh},
  Christopher~W. {Walter}, Daniel~L. {Wang}, Shin-Yawn {Wang}, Michael
  {Warner}, Oliver {Wiecha}, Beth {Willman}, Scott~E. {Winters}, David
  {Wittman}, Sidney~C. {Wolff}, W.~Michael {Wood-Vasey}, Xiuqin {Wu}, Bo~{Xin},
  Peter {Yoachim}, and Hu~{Zhan}.
\newblock {LSST: From Science Drivers to Reference Design and Anticipated Data
  Products}.
\newblock {\em \apj}, 873(2):111, March 2019.

\bibitem{2021AJ....162...94S}
Paula {Szkody}, Claire {Olde Loohuis}, Brad {Koplitz}, Jan {van Roestel},
  Brooke {Dicenzo}, Anna Y.~Q. {Ho}, Lynne~A. {Hillenbrand}, Eric~C. {Bellm},
  Richard {Dekany}, Andrew~J. {Drake}, Dmitry~A. {Duev}, Matthew~J. {Graham},
  Mansi~M. {Kasliwal}, Ashish~A. {Mahabal}, Frank~J. {Masci}, James~D. {Neill},
  Reed {Riddle}, Benjamin {Rusholme}, Jesper {Sollerman}, and Richard
  {Walters}.
\newblock {Cataclysmic Variables in the Second Year of the Zwicky Transient
  Facility}.
\newblock {\em \aj}, 162(3):94, September 2021.

\bibitem{2021AJ....161..267V}
Jan {van Roestel}, Dmitry~A. {Duev}, Ashish~A. {Mahabal}, Michael~W.
  {Coughlin}, Przemek {Mr{\'o}z}, Kevin {Burdge}, Andrew {Drake}, Matthew~J.
  {Graham}, Lynne {Hillenbrand}, Eric~C. {Bellm}, Thomas {Kupfer}, Alexandre
  {Delacroix}, C.~{Fremling}, V.~Zach {Golkhou}, David {Hale}, Russ~R. {Laher},
  Frank~J. {Masci}, Reed {Riddle}, Philippe {Rosnet}, Ben {Rusholme}, Roger
  {Smith}, Maayane~T. {Soumagnac}, Richard {Walters}, Thomas~A. {Prince}, and
  S.~R. {Kulkarni}.
\newblock {The ZTF Source Classification Project. I. Methods and
  Infrastructure}.
\newblock {\em \aj}, 161(6):267, June 2021.

\bibitem{2020arXiv200409213K}
J{\"u}rgen {Kn{\"o}dlseder}.
\newblock {The Cherenkov Telescope Array}.
\newblock {\em arXiv e-prints}, page arXiv:2004.09213, April 2020.

\bibitem{Goble_2014}
Carole Goble.
\newblock Better software, better research.
\newblock {\em IEEE Internet Computing}, 18(5):4--8, 2014.

\bibitem{Mushotzky_2011}
R.~{Mushotzky}.
\newblock What do astronomers really want?
\newblock In I.~N. {Evans}, A.~{Accomazzi}, D.~J. {Mink}, and A.~H. {Rots},
  editors, {\em Astronomical Data Analysis Software and Systems XX}, volume 442
  of {\em Astronomical Society of the Pacific Conference Series}, page 235,
  July 2011.

\bibitem{momcheva_software_2015}
I.~Momcheva and E.~Tollerud.
\newblock Software use in astronomy: an informal survey.

\bibitem{dicosmo:hal-01590958}
Roberto Di~Cosmo and Stefano Zacchiroli.
\newblock {Software Heritage: Why and How to Preserve Software Source Code}.
\newblock In {\em {iPRES 2017 - 14th International Conference on Digital
  Preservation}}, pages 1--10, Kyoto, Japan, September 2017.

\bibitem{EONR_Zenodo}
{European Organization For Nuclear Research} and {OpenAIRE}.
\newblock Zenodo, 2013.
\newblock https://doi.org/10.25495/7gxk-rd71.

\bibitem{Allen_scicodes}
Alice {Allen}.
\newblock {SciCodes: Astronomy Research Software and Beyond}.
\newblock {\em arXiv e-prints}, page arXiv:2111.14278, November 2021.

\bibitem{DOECODE_2017}
Neal Ensor, Shelby Stooksbury, Andrew Smith, Lorrie~Apple Johnson, Lance
  Vowell, Mark Martin, Mike Hensley, Darel Finkbeiner, Carly Robinson, Katie
  Knight, Josh Nelson, Lynn Davis, Ian Lee, Crystal Sherline, Thomas Welsch,
  Jay~Jay Billings, Mary~Beth West, Tim Sowers, and Amber Watson.
\newblock Doe code, 2017.
\newblock https://doi.org/10.11578/dc.20171031.3.

\bibitem{ORNL_DAAC}
ORNL.
\newblock {Oak Ridge National Laboratory Distributed Active Archive Center},
  2013.

\bibitem{Remy2021}
Quentin Remy, L.~Tibaldo, F.~Acero, M.~Fiori, J.~Knödlseder, B.~Olmi, and
  P.~Sharma.
\newblock {Survey of the Galactic Plane with the Cherenkov Telescope Array}.
\newblock {\em PoS}, ICRC2021:886, 2021.

\bibitem{Mohrmann2019}
{Mohrmann, L.}, {Specovius, A.}, {Tiziani, D.}, {Funk, S.}, {Malyshev, D.},
  {Nakashima, K.}, and {van Eldik, C.}
\newblock Validation of open-source science tools and background model
  construction in astronomy.
\newblock {\em A\&A}, 632:A72, 2019.

\bibitem{OliveraNieto2021}
Laura Olivera-Nieto, Vikas Joshi, Harm Schoorlemmer, Axel Donath, Anushka~Udara
  Abeysekara, Andrea Albert, Ruben Alfaro, César Alvarez, Juan de~Dios
  Álvarez Romero, José~Roberto Angeles~Camacho, Juan~Carlos
  Arteaga~Velazquez, Arun~Babu Kollamparambil, Daniel~Omar Avila~Rojas,
  Hugo~Alberto Ayala~Solares, Rishi Babu, Vardan Baghmanyan, Ahron~S. Barber,
  Josefa Becerra~Gonzalez, Ernesto Belmont-Moreno, Segev BenZvi, David Berley,
  Chad Brisbois, Karen~S. Caballero~Mora, Tomás Capistrán, Alberto
  Carramiñana, Sabrina Casanova, Oscar Chaparro-Amaro, Umberto Cotti, Jorge
  Cotzomi, Sara Coutiño~de Leon, Eduardo de~la Fuente, Cederik~León de~León,
  Lorenzo Diaz, Raquel Diaz~Hernandez, Juan~Carlos Díaz~Vélez, Brenda Dingus,
  Mora Durocher, Michael DuVernois, Robert Ellsworth, Kristi Engel,
  María~Catalina Espinoza~Hernández, Jason Fan, Ke~Fang, Mateo
  Fernandez~Alonso, Brian Fick, Henrike Fleischhack, Jorge~Luis Flores,
  Nissim~Illich Fraija, Diego Garcia~Aguilar, Jose~Andres Garcia-Gonzalez,
  Jose~Luis García-Luna, Guillermo García-Torales, Fernando Garfias, Gwenael
  Giacinti, Hazal Goksu, Maria~Magdalena González, Jordan~A. Goodman,
  J.~Patrick Harding, Sergio Hernández~Cadena, Ian Herzog, Jim Hinton, Binita
  Hona, Dezhi Huang, Filiberto Hueyotl-Zahuantitla, Michelle Hui, Brian
  Humensky, Petra Hüntemeyer, Arturo Iriarte, Armelle Jardin-Blicq, Hannah
  Jhee, David Kieda, Gerd~J. Kunde, Samridha Kunwar, Alejandro Lara, Jason LEE,
  William~H. Lee, Dirk Lennarz, Hermes Leon~Vargas, Jim Linnemann, Anna~Lia
  Longinotti, Ruben Lopez-Coto, Gilgamesh Luis-Raya, Joe Lundeen, Kelly Malone,
  Vincent Marandon, Oscar MARTINEZ, Israel Martinez~Castellanos, Humberto
  Martínez~Huerta, Jesús Martínez-Castro, John Matthews, Julie McEnery,
  Pedro Miranda-Romagnoli, Jorge~Antonio Morales~Soto, Eduardo Moreno~Barbosa,
  Miguel Mostafa, Amid Nayerhoda, Lukas Nellen, Michael Newbold, Mehr~Un Nisa,
  Roberto Noriega-Papaqui, Nicola Omodei, Alison Peisker, Yunior Pérez~Araujo,
  Eucario~Gonzalo Pérez~Pérez, Chang~Dong Rho, Colas Rivière, Daniel
  Rosa-Gonzalez, Edna Ruiz-Velasco, James Ryan, Humberto~Ibarguen Salazar,
  Francisco Salesa~Greus, Andrés Sandoval, Michael Schneider, José
  Serna-Franco, Gus Sinnis, Andrew~James Smith, Wayne~Robert Springer, Pooja
  Surajbali, Ignacio Taboada, Meghan Tanner, Kirsten Tollefson, Ibrahim Torres,
  Ramiro Torres~Escobedo, Rhiannon Turner, Fernando Ureña-Mena, Luis
  Villaseñor, Xiaojie Wang, Ian~James Watson, Thomas Weisgarber, Felix Werner,
  Elijah Willox, Joshua Wood, Gaurang Yodh, Arnulfo Zepeda, and Hao Zhou.
\newblock {Standardized formats for gamma-ray analysis applied to HAWC
  observatory data}.
\newblock {\em PoS}, ICRC2021:727, 2021.

\bibitem{Nigro2019}
{Nigro, C.}, {Deil, C.}, {Zanin, R.}, {Hassan, T.}, {King, J.}, {Ruiz, J. E.},
  {Saha, L.}, {Terrier, R.}, {Br\"ugge, K.}, {N\"othe, M.}, {Bird, R.}, {Lin,
  T. T. Y.}, {Aleksi\'{}c, J.}, {Boisson, C.}, {Contreras, J. L.}, {Donath,
  A.}, {Jouvin, L.}, {Kelley-Hoskins, N.}, {Khelifi, B.}, {Kosack, K.}, {Rico,
  J.}, and {Sinha, A.}
\newblock Towards open and reproducible multi-instrument analysis in gamma-ray
  astronomy.
\newblock {\em A\&A}, 625:A10, 2019.

\bibitem{2021ApJS..256...12A}
M.~{Ajello}, W.~B. {Atwood}, M.~{Axelsson}, R.~{Bagagli}, M.~{Bagni},
  L.~{Baldini}, D.~{Bastieri}, F.~{Bellardi}, R.~{Bellazzini}, E.~{Bissaldi},
  E.~D. {Bloom}, R.~{Bonino}, J.~{Bregeon}, A.~{Brez}, P.~{Bruel},
  R.~{Buehler}, S.~{Buson}, R.~A. {Cameron}, P.~A. {Caraveo}, E.~{Cavazzuti},
  M.~{Ceccanti}, S.~{Chen}, C.~C. {Cheung}, S.~{Ciprini}, I.~{Cognard},
  J.~{Cohen-Tanugi}, S.~{Cutini}, F.~{D'Ammando}, P.~{de la Torre Luque},
  F.~{de Palma}, S.~W. {Digel}, F.~{Dirirsa}, N.~{Di Lalla}, L.~{Di Venere},
  A.~{Dom{\'\i}nguez}, D.~{Fabiani}, E.~C. {Ferrara}, A.~{Fiori}, G.~{Foglia},
  Y.~{Fukazawa}, P.~{Fusco}, F.~{Gargano}, D.~{Gasparrini}, M.~{Giroletti},
  T.~{Glanzman}, D.~{Green}, S.~{Griffin}, M.~H. {Grondin}, J.~E. {Grove},
  L.~{Guillemot}, S.~{Guiriec}, M.~{Gustafsson}, E.~{Hays}, D.~{Horan},
  G.~{J{\'o}hannesson}, T.~J. {Johnson}, T.~{Kamae}, M.~{Kerr}, M.~{Kuss},
  S.~{Larsson}, L.~{Latronico}, M.~{Lemoine-Goumard}, J.~{Li}, I.~{Liodakis},
  F.~{Longo}, F.~{Loparco}, M.~N. {Lovellette}, P.~{Lubrano}, S.~{Maldera},
  A.~{Manfreda}, G.~{Mart{\'\i}-Devesa}, M.~N. {Mazziotta}, N.~{Menon},
  I.~{Mereu}, M.~{Meyer}, P.~F. {Michelson}, M.~{Minuti}, W.~{Mitthumsiri},
  T.~{Mizuno}, M.~{Mongelli}, M.~E. {Monzani}, I.~V. {Moskalenko}, M.~{Negro},
  E.~{Nuss}, R.~{Ojha}, M.~{Orienti}, E.~{Orlando}, A.~{Paccagnella}, V.~S.
  {Paliya}, D.~{Paneque}, Z.~{Pei}, J.~S. {Perkins}, M.~{Pesce-Rollins},
  M.~{Pinchera}, F.~{Piron}, H.~{Poon}, T.~A. {Porter}, R.~{Primavera},
  G.~{Principe}, J.~L. {Racusin}, S.~{Rain{\`o}}, R.~{Rando}, B.~{Rani},
  E.~{Rapposelli}, M.~{Razzano}, S.~{Razzaque}, A.~{Reimer}, O.~{Reimer}, J.~J.
  {Russell}, N.~{Saggini}, P.~M. {Saz Parkinson}, N.~{Scolieri}, D.~{Serini},
  C.~{Sgr{\`o}}, E.~J. {Siskind}, D.~A. {Smith}, G.~{Spandre}, P.~{Spinelli},
  D.~J. {Suson}, H.~{Tajima}, J.~G. {Thayer}, D.~J. {Thompson}, L.~{Tibaldo},
  D.~F. {Torres}, G.~{Tosti}, J.~{Valverde}, L.~{Vigiani}, and G.~{Zaharijas}.
\newblock {Fermi Large Area Telescope Performance after 10 Years of Operation}.
\newblock {\em \apjs}, 256(1):12, September 2021.

\bibitem{2017ICRC...35..824W}
M.~{Wood}, R.~{Caputo}, E.~{Charles}, M.~{Di Mauro}, J.~{Magill}, J.~S.
  {Perkins}, and {Fermi-LAT Collaboration}.
\newblock {Fermipy: An open-source Python package for analysis of Fermi-LAT
  Data}.
\newblock In {\em 35th International Cosmic Ray Conference (ICRC2017)}, volume
  301 of {\em International Cosmic Ray Conference}, page 824, January 2017.

\bibitem{2017AIPC.1792g0016P}
Giacomo {Principe} and Dmitry {Malyshev}.
\newblock {Point source detection and flux determination with PGWave}.
\newblock In {\em 6th International Symposium on High Energy Gamma-Ray
  Astronomy}, volume 1792 of {\em American Institute of Physics Conference
  Series}, page 070016, January 2017.

\bibitem{2018A&A...618A..22P}
G.~{Principe}, D.~{Malyshev}, J.~{Ballet}, and S.~{Funk}.
\newblock {The first catalog of Fermi-LAT sources below 100 MeV}.
\newblock {\em \aap}, 618:A22, October 2018.

\bibitem{2013A&A...558A..33A}
{Astropy Collaboration}, Thomas~P. {Robitaille}, Erik~J. {Tollerud}, Perry
  {Greenfield}, Michael {Droettboom}, Erik {Bray}, Tom {Aldcroft}, Matt
  {Davis}, Adam {Ginsburg}, Adrian~M. {Price-Whelan}, Wolfgang~E. {Kerzendorf},
  Alexander {Conley}, Neil {Crighton}, Kyle {Barbary}, Demitri {Muna}, Henry
  {Ferguson}, Fr{\'e}d{\'e}ric {Grollier}, Madhura~M. {Parikh}, Prasanth~H.
  {Nair}, Hans~M. {Unther}, Christoph {Deil}, Julien {Woillez}, Simon
  {Conseil}, Roban {Kramer}, James E.~H. {Turner}, Leo {Singer}, Ryan {Fox},
  Benjamin~A. {Weaver}, Victor {Zabalza}, Zachary~I. {Edwards}, K.~{Azalee
  Bostroem}, D.~J. {Burke}, Andrew~R. {Casey}, Steven~M. {Crawford}, Nadia
  {Dencheva}, Justin {Ely}, Tim {Jenness}, Kathleen {Labrie}, Pey~Lian {Lim},
  Francesco {Pierfederici}, Andrew {Pontzen}, Andy {Ptak}, Brian {Refsdal},
  Mathieu {Servillat}, and Ole {Streicher}.
\newblock {Astropy: A community Python package for astronomy}.
\newblock {\em \aap}, 558:A33, October 2013.

\bibitem{2020NatMe..17..261V}
Pauli {Virtanen}, Ralf {Gommers}, Travis~E. {Oliphant}, Matt {Haberland}, Tyler
  {Reddy}, David {Cournapeau}, Evgeni {Burovski}, Pearu {Peterson}, Warren
  {Weckesser}, Jonathan {Bright}, St{\'e}fan~J. {van der Walt}, Matthew
  {Brett}, Joshua {Wilson}, K.~Jarrod {Millman}, Nikolay {Mayorov}, Andrew
  R.~J. {Nelson}, Eric {Jones}, Robert {Kern}, Eric {Larson}, C.~J. {Carey},
  {\.I}lhan {Polat}, Yu~{Feng}, Eric~W. {Moore}, Jake {VanderPlas}, Denis
  {Laxalde}, Josef {Perktold}, Robert {Cimrman}, Ian {Henriksen}, E.~A.
  {Quintero}, Charles~R. {Harris}, Anne~M. {Archibald}, Ant{\^o}nio~H.
  {Ribeiro}, Fabian {Pedregosa}, Paul {van Mulbregt}, and {SciPy 1. 0
  Contributors}.
\newblock {SciPy 1.0: fundamental algorithms for scientific computing in
  Python}.
\newblock {\em Nature Methods}, 17:261--272, February 2020.

\bibitem{2018AJ....156..123A}
{Astropy Collaboration}, A.~M. {Price-Whelan}, B.~M. {Sip{\H{o}}cz}, H.~M.
  {G{\"u}nther}, P.~L. {Lim}, S.~M. {Crawford}, S.~{Conseil}, D.~L. {Shupe},
  M.~W. {Craig}, N.~{Dencheva}, A.~{Ginsburg}, J.~T. {VanderPlas}, L.~D.
  {Bradley}, D.~{P{\'e}rez-Su{\'a}rez}, M.~{de Val-Borro}, T.~L. {Aldcroft},
  K.~L. {Cruz}, T.~P. {Robitaille}, E.~J. {Tollerud}, C.~{Ardelean},
  T.~{Babej}, Y.~P. {Bach}, M.~{Bachetti}, A.~V. {Bakanov}, S.~P. {Bamford},
  G.~{Barentsen}, P.~{Barmby}, A.~{Baumbach}, K.~L. {Berry}, F.~{Biscani},
  M.~{Boquien}, K.~A. {Bostroem}, L.~G. {Bouma}, G.~B. {Brammer}, E.~M. {Bray},
  H.~{Breytenbach}, H.~{Buddelmeijer}, D.~J. {Burke}, G.~{Calderone}, J.~L.
  {Cano Rodr{\'\i}guez}, M.~{Cara}, J.~V.~M. {Cardoso}, S.~{Cheedella},
  Y.~{Copin}, L.~{Corrales}, D.~{Crichton}, D.~{D'Avella}, C.~{Deil},
  {\'E}.~{Depagne}, J.~P. {Dietrich}, A.~{Donath}, M.~{Droettboom}, N.~{Earl},
  T.~{Erben}, S.~{Fabbro}, L.~A. {Ferreira}, T.~{Finethy}, R.~T. {Fox}, L.~H.
  {Garrison}, S.~L.~J. {Gibbons}, D.~A. {Goldstein}, R.~{Gommers}, J.~P.
  {Greco}, P.~{Greenfield}, A.~M. {Groener}, F.~{Grollier}, A.~{Hagen},
  P.~{Hirst}, D.~{Homeier}, A.~J. {Horton}, G.~{Hosseinzadeh}, L.~{Hu}, J.~S.
  {Hunkeler}, {\v{Z}}.~{Ivezi{\'c}}, A.~{Jain}, T.~{Jenness}, G.~{Kanarek},
  S.~{Kendrew}, N.~S. {Kern}, W.~E. {Kerzendorf}, A.~{Khvalko}, J.~{King},
  D.~{Kirkby}, A.~M. {Kulkarni}, A.~{Kumar}, A.~{Lee}, D.~{Lenz}, S.~P.
  {Littlefair}, Z.~{Ma}, D.~M. {Macleod}, M.~{Mastropietro}, C.~{McCully},
  S.~{Montagnac}, B.~M. {Morris}, M.~{Mueller}, S.~J. {Mumford}, D.~{Muna},
  N.~A. {Murphy}, S.~{Nelson}, G.~H. {Nguyen}, J.~P. {Ninan}, M.~{N{\"o}the},
  S.~{Ogaz}, S.~{Oh}, J.~K. {Parejko}, N.~{Parley}, S.~{Pascual}, R.~{Patil},
  A.~A. {Patil}, A.~L. {Plunkett}, J.~X. {Prochaska}, T.~{Rastogi}, V.~{Reddy
  Janga}, J.~{Sabater}, P.~{Sakurikar}, M.~{Seifert}, L.~E. {Sherbert},
  H.~{Sherwood-Taylor}, A.~Y. {Shih}, J.~{Sick}, M.~T. {Silbiger},
  S.~{Singanamalla}, L.~P. {Singer}, P.~H. {Sladen}, K.~A. {Sooley},
  S.~{Sornarajah}, O.~{Streicher}, P.~{Teuben}, S.~W. {Thomas}, G.~R.
  {Tremblay}, J.~E.~H. {Turner}, V.~{Terr{\'o}n}, M.~H. {van Kerkwijk}, A.~{de
  la Vega}, L.~L. {Watkins}, B.~A. {Weaver}, J.~B. {Whitmore}, J.~{Woillez},
  V.~{Zabalza}, and {Astropy Contributors}.
\newblock {The Astropy Project: Building an Open-science Project and Status of
  the v2.0 Core Package}.
\newblock {\em \aj}, 156(3):123, September 2018.

\bibitem{2015ICRC...34..789D}
A.~{Donath}, C.~{Deil}, M.~Paz {Arribas}, J.~{King}, E.~{Owen}, R.~{Terrier},
  I.~{Reichardt}, J.~{Harris}, R.~{Buehler}, and S.~{Klepser}.
\newblock {Gammapy: An open-source Python package for gamma-ray astronomy}.
\newblock In {\em 34th International Cosmic Ray Conference (ICRC2015)},
  volume~34 of {\em International Cosmic Ray Conference}, page 789, July 2015.

\bibitem{2007CSE.....9...90H}
John~D. {Hunter}.
\newblock {Matplotlib: A 2D Graphics Environment}.
\newblock {\em Computing in Science and Engineering}, 9(3):90--95, May 2007.

\bibitem{2022arXiv220105567A}
Fabio {Acero}, Marianne {Lemoine-Goumard}, and Jean {Ballet}.
\newblock {Characterization of the GeV emission from the Kepler supernova
  remnant}.
\newblock {\em arXiv e-prints}, page arXiv:2201.05567, January 2022.

\bibitem{2020A&A...640A..76P}
G.~{Principe}, A.~M.~W. {Mitchell}, S.~{Caroff}, J.~A. {Hinton}, R.~D.
  {Parsons}, and S.~{Funk}.
\newblock {Energy dependent morphology of the pulsar wind nebula HESS J1825-137
  with Fermi-LAT}.
\newblock {\em \aap}, 640:A76, August 2020.

\bibitem{2020A&A...635A.185P}
G.~{Principe}, G.~{Migliori}, T.~J. {Johnson}, F.~{D'Ammando}, M.~{Giroletti},
  M.~{Orienti}, C.~{Stanghellini}, G.~B. {Taylor}, E.~{Torresi}, and C.~C.
  {Cheung}.
\newblock {NGC 3894: a young radio galaxy seen by Fermi-LAT}.
\newblock {\em \aap}, 635:A185, March 2020.

\bibitem{2021ApJ...911L..11E}
{EHT MWL Science Working Group}, J.~C. {Algaba}, J.~{Anczarski}, K.~{Asada},
  M.~{Balokovi{\'c}}, S.~{Chandra}, Y.~Z. {Cui}, A.~D. {Falcone},
  M.~{Giroletti}, C.~{Goddi}, K.~{Hada}, D.~{Haggard}, S.~{Jorstad}, A.~{Kaur},
  T.~{Kawashima}, G.~{Keating}, J.~Y. {Kim}, M.~{Kino}, S.~{Komossa}, E.~V.
  {Kravchenko}, T.~P. {Krichbaum}, S.~S. {Lee}, R.~S. {Lu}, M.~{Lucchini},
  S.~{Markoff}, J.~{Neilsen}, M.~A. {Nowak}, J.~{Park}, G.~{Principe},
  V.~{Ramakrishnan}, M.~T. {Reynolds}, M.~{Sasada}, S.~S. {Savchenko}, K.~E.
  {Williamson}, {Event Horizon Telescope Collaboration}, Kazunori {Akiyama},
  Antxon {Alberdi}, Walter {Alef}, Richard {Anantua}, Rebecca {Azulay},
  Anne-Kathrin {Baczko}, David {Ball}, John {Barrett}, Dan {Bintley},
  Bradford~A. {Benson}, Lindy {Blackburn}, Raymond {Blundell}, Wilfred
  {Boland}, Katherine~L. {Bouman}, Geoffrey~C. {Bower}, Hope {Boyce}, Michael
  {Bremer}, Christiaan~D. {Brinkerink}, Roger {Brissenden}, Silke {Britzen},
  Avery~E. {Broderick}, Dominique {Broguiere}, Thomas {Bronzwaer}, Do-Young
  {Byun}, John~E. {Carlstrom}, Andrew {Chael}, Chi-Kwan {Chan}, Shami
  {Chatterjee}, Koushik {Chatterjee}, Ming-Tang {Chen}, Yongjun {Chen}, Paul~M.
  {Chesler}, Ilje {Cho}, Pierre {Christian}, John~E. {Conway}, James~M.
  {Cordes}, Thomas~M. {Crawford}, Geoffrey~B. {Crew}, Alejandro {Cruz-Osorio},
  Jordy {Davelaar}, Mariafelicia {de Laurentis}, Roger {Deane}, Jessica
  {Dempsey}, Gregory {Desvignes}, Jason {Dexter}, Sheperd~S. {Doeleman},
  Ralph~P. {Eatough}, Heino {Falcke}, Joseph {Farah}, Vincent~L. {Fish},
  Ed~{Fomalont}, H.~Alyson {Ford}, Raquel {Fraga-Encinas}, Per {Friberg},
  Christian~M. {Fromm}, Antonio {Fuentes}, Peter {Galison}, Charles~F.
  {Gammie}, Roberto {Garc{\'\i}a}, Olivier {Gentaz}, Boris {Georgiev}, Roman
  {Gold}, Jos{\'e}~L. {G{\'o}mez}, Arturo~I. {G{\'o}mez-Ruiz}, Minfeng {Gu},
  Mark {Gurwell}, Michael~H. {Hecht}, Ronald {Hesper}, Luis~C. {Ho}, Paul {Ho},
  Mareki {Honma}, Chih-Wei~L. {Huang}, Lei {Huang}, David~H. {Hughes}, Shiro
  {Ikeda}, Makoto {Inoue}, Sara {Issaoun}, David~J. {James}, Buell~T.
  {Jannuzi}, Michael {Janssen}, Britton {Jeter}, Wu~{Jiang}, Alejandra
  {Jim{\'e}nez-Rosales}, Michael~D. {Johnson}, Taehyun {Jung}, Mansour
  {Karami}, Ramesh {Karuppusamy}, Mark {Kettenis}, Dong-Jin {Kim}, Jongsoo
  {Kim}, Junhan {Kim}, Jun~Yi {Koay}, Yutaro {Kofuji}, Patrick~M. {Koch}, Shoko
  {Koyama}, Michael {Kramer}, Carsten {Kramer}, Cheng-Yu {Kuo}, Tod~R. {Lauer},
  Aviad {Levis}, Yan-Rong {Li}, Zhiyuan {Li}, Michael {Lindqvist}, Rocco
  {Lico}, Greg {Lindahl}, Jun {Liu}, Kuo {Liu}, Elisabetta {Liuzzo}, Wen-Ping
  {Lo}, Andrei~P. {Lobanov}, Laurent {Loinard}, Colin {Lonsdale}, Nicholas~R.
  {MacDonald}, Jirong {Mao}, Nicola {Marchili}, Daniel~P. {Marrone}, Alan~P.
  {Marscher}, Iv{\'a}n {Mart{\'\i}-Vidal}, Satoki {Matsushita}, Lynn~D.
  {Matthews}, Lia {Medeiros}, Karl~M. {Menten}, Izumi {Mizuno}, Yosuke
  {Mizuno}, James~M. {Moran}, Kotaro {Moriyama}, Monika {Moscibrodzka},
  Cornelia {M{\"u}ller}, Gibwa {Musoke}, Alejandro~Mus {Mej{\'\i}as}, Hiroshi
  {Nagai}, Neil~M. {Nagar}, Masanori {Nakamura}, Ramesh {Narayan}, Gopal
  {Narayanan}, Iniyan {Natarajan}, Antonios {Nathanail}, Roberto {Neri},
  Chunchong {Ni}, Aristeidis {Noutsos}, Hiroki {Okino}, H{\'e}ctor {Olivares},
  Gisela~N. {Ortiz-Le{\'o}n}, Tomoaki {Oyama}, Feryal {{\"O}zel}, Daniel C.~M.
  {Palumbo}, Nimesh {Patel}, Ue-Li {Pen}, Dominic~W. {Pesce}, Vincent
  {Pi{\'e}tu}, Richard {Plambeck}, Aleksandar {Popstefanija}, Oliver {Porth},
  Felix~M. {P{\"o}tzl}, Ben {Prather}, Jorge~A. {Preciado-L{\'o}pez}, Dimitrios
  {Psaltis}, Hung-Yi {Pu}, Ramprasad {Rao}, Mark~G. {Rawlings}, Alexander~W.
  {Raymond}, Luciano {Rezzolla}, Angelo {Ricarte}, Bart {Ripperda}, Freek
  {Roelofs}, Alan {Rogers}, Eduardo {Ros}, Mel {Rose}, Arash {Roshanineshat},
  Helge {Rottmann}, Alan~L. {Roy}, Chet {Ruszczyk}, Kazi L.~J. {Rygl}, Salvador
  {S{\'a}nchez}, David {S{\'a}nchez-Arguelles}, Tuomas {Savolainen}, F.~Peter
  {Schloerb}, Karl-Friedrich {Schuster}, Lijing {Shao}, Zhiqiang {Shen}, Des
  {Small}, Bong~Won {Sohn}, Jason {Soohoo}, He~{Sun}, Fumie {Tazaki},
  Alexandra~J. {Tetarenko}, Paul {Tiede}, Remo P.~J. {Tilanus}, Michael
  {Titus}, Kenji {Toma}, Pablo {Torne}, Tyler {Trent}, Efthalia {Traianou},
  Sascha {Trippe}, Ilse {van Bemmel}, Huib~Jan {van Langevelde}, Daniel~R. {van
  Rossum}, Jan {Wagner}, Derek {Ward-Thompson}, John {Wardle}, Jonathan
  {Weintroub}, Norbert {Wex}, Robert {Wharton}, Maciek {Wielgus}, George~N.
  {Wong}, Qingwen {Wu}, Doosoo {Yoon}, Andr{\'e} {Young}, Ken {Young}, Ziri
  {Younsi}, Feng {Yuan}, Ye-Fei {Yuan}, J.~Anton {Zensus}, Guang-Yao {Zhao},
  Shan-Shan {Zhao}, {Fermi Large Area Telescope Collaboration}, G.~{Principe},
  M.~{Giroletti}, F.~{D'Ammando}, M.~{Orienti}, {H.~E.~S.~S. Collaboration},
  H.~{Abdalla}, R.~{Adam}, F.~{Aharonian}, F.~Ait {Benkhali}, E.~O.
  {Ang{\"u}ner}, C.~{Arcaro}, C.~{Armand}, T.~{Armstrong}, H.~{Ashkar},
  M.~{Backes}, V.~{Baghmanyan}, V.~{Barbosa Martins}, A.~{Barnacka},
  M.~{Barnard}, Y.~{Becherini}, D.~{Berge}, K.~{Bernl{\"o}hr}, B.~{Bi},
  M.~{B{\"o}ttcher}, C.~{Boisson}, J.~{Bolmont}, M.~De~Bony {de Lavergne},
  M.~{Breuhaus}, F.~{Brun}, P.~{Brun}, M.~{Bryan}, M.~{B{\"u}chele},
  T.~{Bulik}, T.~{Bylund}, S.~{Caroff}, A.~{Carosi}, S.~{Casanova}, T.~{Chand},
  A.~{Chen}, G.~{Cotter}, M.~{Cury{\l}o}, J.~{Damascene Mbarubucyeye}, I.~D.
  {Davids}, J.~{Davies}, C.~{Deil}, J.~{Devin}, P.~{Dewilt}, L.~{Dirson},
  A.~{Djannati-Ata{\"\i}}, A.~{Dmytriiev}, A.~{Donath}, V.~{Doroshenko},
  C.~{Duffy}, J.~{Dyks}, K.~{Egberts}, F.~{Eichhorn}, S.~{Einecke}, G.~{Emery},
  J.~P. {Ernenwein}, K.~{Feijen}, S.~{Fegan}, A.~{Fiasson}, G.~Fichet {de
  Clairfontaine}, G.~{Fontaine}, S.~{Funk}, M.~{F{\"u}{\ss}ling}, S.~{Gabici},
  Y.~A. {Gallant}, G.~{Giavitto}, L.~{Giunti}, D.~{Glawion}, J.~F.
  {Glicenstein}, D.~{Gottschall}, M.~H. {Grondin}, J.~{Hahn}, M.~{Haupt},
  G.~{Hermann}, J.~A. {Hinton}, W.~{Hofmann}, C.~{Hoischen}, T.~L. {Holch},
  M.~{Holler}, M.~{H{\"o}rbe}, D.~{Horns}, D.~{Huber}, M.~{Jamrozy},
  D.~{Jankowsky}, F.~{Jankowsky}, A.~{Jardin-Blicq}, V.~{Joshi},
  I.~{Jung-Richardt}, E.~{Kasai}, M.~A. {Kastendieck}, K.~{Katarzy{\'n}ski},
  U.~{Katz}, D.~{Khangulyan}, B.~{Kh{\'e}lifi}, S.~{Klepser},
  W.~{Klu{\'z}niak}, Nu. {Komin}, R.~{Konno}, K.~{Kosack}, D.~{Kostunin},
  M.~{Kreter}, G.~{Lamanna}, A.~{Lemi{\`e}re}, M.~{Lemoine-Goumard}, J.~P.
  {Lenain}, C.~{Levy}, T.~{Lohse}, I.~{Lypova}, J.~{Mackey}, J.~{Majumdar},
  D.~{Malyshev}, D.~{Malyshev}, V.~{Marandon}, P.~{Marchegiani},
  A.~{Marcowith}, A.~{Mares}, G.~{Mart{\'\i}-Devesa}, R.~{Marx}, G.~{Maurin},
  P.~J. {Meintjes}, M.~{Meyer}, R.~{Moderski}, M.~{Mohamed}, L.~{Mohrmann},
  A.~{Montanari}, C.~{Moore}, P.~{Morris}, E.~{Moulin}, J.~{Muller},
  T.~{Murach}, K.~{Nakashima}, A.~{Nayerhoda}, M.~{de Naurois}, H.~{Ndiyavala},
  F.~{Niederwanger}, J.~{Niemiec}, L.~{Oakes}, P.~{O'Brien}, H.~{Odaka},
  S.~{Ohm}, L.~{Olivera-Nieto}, E.~{de Ona Wilhelmi}, M.~{Ostrowski},
  M.~{Panter}, S.~{Panny}, R.~D. {Parsons}, G.~{Peron}, B.~{Peyaud}, Q.~{Piel},
  S.~{Pita}, V.~{Poireau}, A.~Priyana {Noel}, D.~A. {Prokhorov}, H.~{Prokoph},
  G.~{P{\"u}hlhofer}, M.~{Punch}, A.~{Quirrenbach}, R.~{Rauth},
  P.~{Reichherzer}, A.~{Reimer}, O.~{Reimer}, Q.~{Remy}, M.~{Renaud},
  F.~{Rieger}, L.~{Rinchiuso}, C.~{Romoli}, G.~{Rowell}, B.~{Rudak},
  E.~{Ruiz-Velasco}, V.~{Sahakian}, S.~{Sailer}, D.~A. {Sanchez},
  A.~{Santangelo}, M.~{Sasaki}, M.~{Scalici}, H.~M. {Schutte}, U.~{Schwanke},
  S.~{Schwemmer}, M.~{Seglar-Arroyo}, M.~{Senniappan}, A.~S. {Seyffert},
  N.~{Shafi}, K.~{Shiningayamwe}, R.~{Simoni}, A.~{Sinha}, H.~{Sol},
  A.~{Specovius}, S.~{Spencer}, M.~{Spir-Jacob}, {\L}.~{Stawarz}, L.~{Sun},
  R.~{Steenkamp}, C.~{Stegmann}, S.~{Steinmassl}, C.~{Steppa}, T.~{Takahashi},
  T.~{Tavernier}, A.~M. {Taylor}, R.~{Terrier}, D.~{Tiziani}, M.~{Tluczykont},
  L.~{Tomankova}, C.~{Trichard}, M.~{Tsirou}, R.~{Tuffs}, Y.~{Uchiyama}, D.~J.
  {van der Walt}, C.~{van Eldik}, C.~{van Rensburg}, B.~{van Soelen},
  G.~{Vasileiadis}, J.~{Veh}, C.~{Venter}, P.~{Vincent}, J.~{Vink}, H.~J.
  {V{\"o}lk}, T.~{Vuillaume}, Z.~{Wadiasingh}, S.~J. {Wagner}, J.~{Watson},
  F.~{Werner}, R.~{White}, A.~{Wierzcholska}, Yu~Wun {Wong}, A.~{Yusafzai},
  M.~{Zacharias}, R.~{Zanin}, D.~{Zargaryan}, A.~A. {Zdziarski}, A.~{Zech},
  S.~J. {Zhu}, J.~{Zorn}, S.~{Zouari}, N.~{{\.Z}ywucka}, {MAGIC Collaboration},
  V.~A. {Acciari}, S.~{Ansoldi}, L.~A. {Antonelli}, A.~Arbet {Engels},
  M.~{Artero}, K.~{Asano}, D.~{Baack}, A.~{Babi{\'c}}, A.~{Baquero}, U.~Barres
  {de Almeida}, J.~A. {Barrio}, J.~{Becerra Gonz{\'a}lez}, W.~{Bednarek},
  L.~{Bellizzi}, E.~{Bernardini}, M.~{Bernardos}, A.~{Berti}, J.~{Besenrieder},
  W.~{Bhattacharyya}, C.~{Bigongiari}, A.~{Biland}, O.~{Blanch}, G.~{Bonnoli},
  {\v{Z}}.~{Bo{\v{s}}njak}, G.~{Busetto}, R.~{Carosi}, G.~{Ceribella},
  M.~{Cerruti}, Y.~{Chai}, A.~{Chilingarian}, S.~{Cikota}, S.~M. {Colak},
  E.~{Colombo}, J.~L. {Contreras}, J.~{Cortina}, S.~{Covino}, G.~{D'Amico},
  V.~{D'Elia}, P.~{da Vela}, F.~{Dazzi}, A.~{de Angelis}, B.~{de Lotto},
  M.~{Delfino}, J.~{Delgado}, C.~{Delgado Mendez}, D.~{Depaoli}, F.~{di
  Pierro}, L.~{di Venere}, E.~{Do Souto Espi{\~n}eira}, D.~{Dominis Prester},
  A.~{Donini}, D.~{Dorner}, M.~{Doro}, D.~{Elsaesser}, V.~Fallah {Ramazani},
  A.~{Fattorini}, G.~{Ferrara}, M.~V. {Fonseca}, L.~{Font}, C.~{Fruck},
  S.~{Fukami}, R.~J. {Garc{\'\i}a L{\'o}pez}, M.~{Garczarczyk}, S.~{Gasparyan},
  M.~{Gaug}, N.~{Giglietto}, F.~{Giordano}, P.~{Gliwny}, N.~{Godinovi{\'c}},
  J.~G. {Green}, D.~{Green}, D.~{Hadasch}, A.~{Hahn}, L.~{Heckmann},
  J.~{Herrera}, J.~{Hoang}, D.~{Hrupec}, M.~{H{\"u}tten}, T.~{Inada},
  S.~{Inoue}, K.~{Ishio}, Y.~{Iwamura}, I.~{Jim{\'e}nez}, J.~{Jormanainen},
  L.~{Jouvin}, Y.~{Kajiwara}, M.~{Karjalainen}, D.~{Kerszberg}, Y.~{Kobayashi},
  H.~{Kubo}, J.~{Kushida}, A.~{Lamastra}, D.~{Lelas}, F.~{Leone},
  E.~{Lindfors}, S.~{Lombardi}, F.~{Longo}, R.~{L{\'o}pez-Coto},
  M.~{L{\'o}pez-Moya}, A.~{L{\'o}pez-Oramas}, S.~{Loporchio}, B.~{Machado de
  Oliveira Fraga}, C.~{Maggio}, P.~{Majumdar}, M.~{Makariev}, M.~{Mallamaci},
  G.~{Maneva}, M.~{Manganaro}, K.~{Mannheim}, L.~{Maraschi}, M.~{Mariotti},
  M.~{Mart{\'\i}nez}, D.~{Mazin}, S.~{Menchiari}, S.~{Mender},
  S.~{Mi{\'c}anovi{\'c}}, D.~{Miceli}, T.~{Miener}, M.~{Minev}, J.~M.
  {Miranda}, R.~{Mirzoyan}, E.~{Molina}, A.~{Moralejo}, D.~{Morcuende},
  V.~{Moreno}, E.~{Moretti}, V.~{Neustroev}, C.~{Nigro}, K.~{Nilsson},
  K.~{Nishijima}, K.~{Noda}, S.~{Nozaki}, Y.~{Ohtani}, T.~{Oka},
  J.~{Otero-Santos}, S.~{Paiano}, M.~{Palatiello}, D.~{Paneque}, R.~{Paoletti},
  J.~M. {Paredes}, L.~{Pavleti{\'c}}, P.~{Pe{\~n}il}, C.~{Perennes},
  M.~{Persic}, P.~G.~Prada {Moroni}, E.~{Prandini}, C.~{Priyadarshi},
  I.~{Puljak}, W.~{Rhode}, M.~{Rib{\'o}}, J.~{Rico}, C.~{Righi},
  A.~{Rugliancich}, L.~{Saha}, N.~{Sahakyan}, T.~{Saito}, S.~{Sakurai},
  K.~{Satalecka}, F.~G. {Saturni}, B.~{Schleicher}, K.~{Schmidt},
  T.~{Schweizer}, J.~{Sitarek}, I.~{{\v{S}}nidari{\'c}}, D.~{Sobczynska},
  A.~{Spolon}, A.~{Stamerra}, D.~{Strom}, M.~{Strzys}, Y.~{Suda},
  T.~{Suri{\'c}}, M.~{Takahashi}, F.~{Tavecchio}, P.~{Temnikov},
  T.~{Terzi{\'c}}, M.~{Teshima}, L.~{Tosti}, S.~{Truzzi}, A.~{Tutone},
  S.~{Ubach}, J.~{van Scherpenberg}, G.~{Vanzo}, M.~{Vazquez Acosta},
  S.~{Ventura}, V.~{Verguilov}, C.~F. {Vigorito}, V.~{Vitale}, I.~{Vovk},
  M.~{Will}, C.~{Wunderlich}, D.~{Zari{\'c}}, {VERITAS Collaboration}, C.~B.
  {Adams}, W.~{Benbow}, A.~{Brill}, M.~{Capasso}, J.~L. {Christiansen}, A.~J.
  {Chromey}, M.~K. {Daniel}, M.~{Errando}, K.~A. {Farrell}, Q.~{Feng}, J.~P.
  {Finley}, L.~{Fortson}, A.~{Furniss}, A.~{Gent}, C.~{Giuri}, T.~{Hassan},
  O.~{Hervet}, J.~{Holder}, G.~{Hughes}, T.~B. {Humensky}, W.~{Jin},
  P.~{Kaaret}, M.~{Kertzman}, D.~{Kieda}, S.~{Kumar}, M.~J. {Lang}, M.~{Lundy},
  G.~{Maier}, P.~{Moriarty}, R.~{Mukherjee}, D.~{Nieto}, M.~{Nievas-Rosillo},
  S.~{O'Brien}, R.~A. {Ong}, A.~N. {Otte}, S.~{Patel}, K.~{Pfrang}, M.~{Pohl},
  R.~R. {Prado}, E.~{Pueschel}, J.~{Quinn}, K.~{Ragan}, P.~T. {Reynolds},
  D.~{Ribeiro}, G.~T. {Richards}, E.~{Roache}, C.~{Rulten}, J.~L. {Ryan},
  M.~{Santander}, G.~H. {Sembroski}, R.~{Shang}, A.~{Weinstein}, D.~A.
  {Williams}, T.~J. {Williamson}, {Eavn Collaboration}, Tomoya {Hirota}, Lang
  {Cui}, Kotaro {Niinuma}, Hyunwook {Ro}, Nobuyuki {Sakai}, Satoko
  {Sawada-Satoh}, Kiyoaki {Wajima}, Na~{Wang}, Xiang {Liu}, and Yoshinori
  {Yonekura}.
\newblock {Broadband Multi-wavelength Properties of M87 during the 2017 Event
  Horizon Telescope Campaign}.
\newblock {\em \apjl}, 911(1):L11, April 2021.

\bibitem{2021MNRAS.507.4564P}
G.~{Principe}, L.~{Di Venere}, M.~{Orienti}, G.~{Migliori}, F.~{D'Ammando},
  M.~N. {Mazziotta}, and M.~{Giroletti}.
\newblock {Gamma-ray emission from young radio galaxies and quasars}.
\newblock {\em \mnras}, 507(3):4564--4583, November 2021.

\bibitem{2021ApJS..257...37P}
Vaidehi~S. {Paliya}, M.~{B{\"o}ttcher}, Mark {Gurwell}, and C.~S. {Stalin}.
\newblock {On the Origin of Gamma-Ray Flares from Bright Fermi Blazars}.
\newblock {\em \apjs}, 257(2):37, December 2021.

\bibitem{2021arXiv210903548P}
Giacomo {Principe}, Nicola {Omodei}, Niccol{\`o} {Di Lalla}, Leonardo {Di
  Venere}, and Francesco {Longo}.
\newblock {Hunting the gamma-ray emission from Fast Radio Burst with
  Fermi-LAT}.
\newblock {\em arXiv e-prints}, page arXiv:2109.03548, September 2021.

\bibitem{2021ApJS..256...13B}
L.~{Baldini}, J.~{Ballet}, D.~{Bastieri}, J.~{Becerra Gonzalez},
  R.~{Bellazzini}, A.~{Berretta}, E.~{Bissaldi}, R.~D. {Blandford}, E.~D.
  {Bloom}, R.~{Bonino}, E.~{Bottacini}, P.~{Bruel}, S.~{Buson}, R.~A.
  {Cameron}, P.~A. {Caraveo}, E.~{Cavazzuti}, S.~{Chen}, G.~{Chiaro},
  D.~{Ciangottini}, N.~{Cibario}, S.~{Ciprini}, P.~{Cristarella Orestano},
  M.~{Crnogorcevic}, S.~{Cutini}, F.~{D'Ammando}, P.~{de la Torre Luque},
  F.~{de Palma}, S.~W. {Digel}, N.~{Di Lalla}, F.~{Dirirsa}, L.~{Di Venere},
  A.~{Dom{\'\i}nguez}, A.~{Fiori}, H.~{Fleischhack}, A.~{Franckowiak},
  Y.~{Fukazawa}, S.~{Funk}, P.~{Fusco}, F.~{Gargano}, D.~{Gasparrini},
  S.~{Germani}, N.~{Giglietto}, F.~{Giordano}, M.~{Giroletti}, D.~{Green},
  I.~A. {Grenier}, S.~{Griffin}, S.~{Guiriec}, M.~{Gustafsson}, J.~W. {Hewitt},
  D.~{Horan}, R.~{Imazawa}, G.~{J{\'o}hannesson}, M.~{Kerr}, D.~{Kocevski},
  M.~{Kuss}, S.~{Larsson}, L.~{Latronico}, J.~{Li}, I.~{Liodakis}, F.~{Longo},
  F.~{Loparco}, M.~N. {Lovellette}, P.~{Lubrano}, S.~{Maldera}, A.~{Manfreda},
  G.~{Mart{\'\i}-Devesa}, H.~{Matake}, M.~N. {Mazziotta}, I.~{Mereu},
  M.~{Meyer}, N.~{Mirabal}, W.~{Mitthumsiri}, T.~{Mizuno}, M.~E. {Monzani},
  A.~{Morselli}, I.~V. {Moskalenko}, S.~{Nagasawa}, M.~{Negro}, R.~{Ojha},
  M.~{Orienti}, E.~{Orlando}, M.~{Palatiello}, V.~{Paliya}, D.~{Paneque},
  Z.~{Pei}, M.~{Persic}, M.~{Pesce-Rollins}, V.~{Petrosian}, H.~{Poon}, T.~A.
  {Porter}, G.~{Principe}, J.~L. {Racusin}, S.~{Rain{\`o}}, R.~{Rando},
  B.~{Rani}, M.~{Razzano}, S.~{Razzaque}, A.~{Reimer}, O.~{Reimer}, P.~M. {Saz
  Parkinson}, L.~{Scotton}, D.~{Serini}, C.~{Sgr{\`o}}, E.~J. {Siskind},
  G.~{Spandre}, P.~{Spinelli}, D.~J. {Suson}, H.~{Tajima}, D.~{Tak}, D.~F.
  {Torres}, G.~{Tosti}, E.~{Troja}, K.~{Wood}, M.~{Yassine}, G.~{Zaharijas},
  and {Fermi-LAT Collaboration}.
\newblock {Catalog of Long-term Transient Sources in the First 10 yr of
  Fermi-LAT Data}.
\newblock {\em \apjs}, 256(1):13, September 2021.

\bibitem{2019ApJ...877...39M}
Manuel {Meyer}, Jeffrey~D. {Scargle}, and Roger~D. {Blandford}.
\newblock {Characterizing the Gamma-Ray Variability of the Brightest Flat
  Spectrum Radio Quasars Observed with the Fermi LAT}.
\newblock {\em \apj}, 877(1):39, May 2019.

\bibitem{2021NatAs...5..510S}
Robert {Stein}, Sjoert~van {Velzen}, Marek {Kowalski}, Anna {Franckowiak}, Suvi
  {Gezari}, James C.~A. {Miller-Jones}, Sara {Frederick}, Itai {Sfaradi},
  Michael~F. {Bietenholz}, Assaf {Horesh}, Rob {Fender}, Simone {Garrappa},
  Tom{\'a}s {Ahumada}, Igor {Andreoni}, Justin {Belicki}, Eric~C. {Bellm},
  Markus {B{\"o}ttcher}, Valery {Brinnel}, Rick {Burruss}, S.~Bradley {Cenko},
  Michael~W. {Coughlin}, Virginia {Cunningham}, Andrew {Drake}, Glennys~R.
  {Farrar}, Michael {Feeney}, Ryan~J. {Foley}, Avishay {Gal-Yam}, V.~Zach
  {Golkhou}, Ariel {Goobar}, Matthew~J. {Graham}, Erica {Hammerstein}, George
  {Helou}, Tiara {Hung}, Mansi~M. {Kasliwal}, Charles~D. {Kilpatrick}, Albert
  K.~H. {Kong}, Thomas {Kupfer}, Russ~R. {Laher}, Ashish~A. {Mahabal}, Frank~J.
  {Masci}, Jannis {Necker}, Jakob {Nordin}, Daniel~A. {Perley}, Mickael
  {Rigault}, Simeon {Reusch}, Hector {Rodriguez}, C{\'e}sar {Rojas-Bravo}, Ben
  {Rusholme}, David~L. {Shupe}, Leo~P. {Singer}, Jesper {Sollerman}, Maayane~T.
  {Soumagnac}, Daniel {Stern}, Kirsty {Taggart}, Jakob {van Santen}, Charlotte
  {Ward}, Patrick {Woudt}, and Yuhan {Yao}.
\newblock {A tidal disruption event coincident with a high-energy neutrino}.
\newblock {\em Nature Astronomy}, 5:510--518, February 2021.

\bibitem{2020PhRvL.124w1101M}
Manuel {Meyer}, Tanja {Petrushevska}, and {Fermi-LAT Collaboration}.
\newblock {Search for Axionlike-Particle-Induced Prompt {\ensuremath{\gamma}}
  -Ray Emission from Extragalactic Core-Collapse Supernovae with the Fermi
  Large Area Telescope}.
\newblock {\em \prl}, 124(23):231101, June 2020.

\bibitem{2020PhRvD.102j3010D}
Mattia {Di Mauro}, Martin {Stref}, and Francesca {Calore}.
\newblock {Investigating the detection of dark matter subhalos as extended
  sources with Fermi-LAT}.
\newblock {\em \prd}, 102(10):103010, November 2020.

\bibitem{2015arXiv150708343V}
Giacomo Vianello et~al.
\newblock {The Multi-Mission Maximum Likelihood framework (3ML)}.
\newblock In {\em Proceedings of the 34th International Cosmic Ray Conference
  (PoS)}, 2015.

\end{thebibliography}
